\documentclass[12]{report}
\usepackage{amsmath}
\usepackage{color}
\def\dd{{\rm d}}\def\half{{\textstyle\frac12}}
\usepackage{amssymb}

\definecolor{red}{rgb}{1,0,0}
\definecolor{blue}{rgb}{0,0,1}
\definecolor{green}{rgb}{0,1,0}
\definecolor{black}{rgb}{0,0,0}
\definecolor{yellow}{rgb}{1,1,0}
\definecolor{mdwblue}{rgb}{0.2,0.2,0.6}
\definecolor{gray}{rgb}{0.7,0.7,0.7}
\definecolor{darkgreen}{rgb}{0.2,0.7,0.2}
\begin{document}
\title{\textbf{Foundations of a theory of quantum gravity}}
 \author{Johan Noldus\footnote{Johan.Noldus@gmail.com}} \maketitle
\tableofcontents
\chapter{Introduction}
This adventure started out as a paper\footnote{Disclaimer: whenever I use the word God, I do so in a highly theoretical sense as a completion of Platonic space, a class of prefixed global reference frames, a unique superobserver\ldots  By no means do I intend to say something about the contemporary sociological concept of God albeit I would assert this to be emergent and not fundamental in a deeper sense and therefore  apologize to any religious reader who might feel upset about my usage of this word.}, but soon it grew considerably in size and there was no choice left anymore but to present it as a full blown book written in a style which is intermediate between that of an original research paper and that of a book.  More precisely, I opted for a style which is somewhat between the historical and axiomatic approach and this manuscript can therefore be read from different perspectives depending upon the knowledge and skills of the reader.  Since quantum gravity is more than a technical problem, the mandatory sections constitute the introduction as well as the technical and axiomatic framework of sections seven till eleven.  However, the reader who is also interested in the philosophical aspects as well as a general overview of the problem is advised to study sections two and three as well.  The critical reader who is not willing to take any statement for granted should include also sections four till six, since these are somewhat of a transitional nature closing the gap between the conservative initial point of view and the new theory developed later on.  Lecturing about this work made me aware that there is also a more direct way to arrive in Rome and for that very reason, this introduction is also split into two parts.  The first one takes the conservative point of view as it is done by the very large majority of researchers which necessitates a careful and precise way of phrasing the content; the second approach however is more bold and direct but goes, in my humble opinion, much more economic to the heart of the matter.  I believe that the variety of presenting the same material in this introduction will allow the reader to choose which way he prefers to follow.  \\* \\*  Let me also say from the outset what this book achieves and what it leaves as open issues, where the last phrase is to be understood in the sense that these issues are technically open but that the successful realization of them is motivated to some extend.  As is always the case in science, the judgment of whether an argument is compelling or not depends upon the history and experience of the beholder and I certainly do not claim to be the oracle of Delphi in this regard.  However, I deem these conjectures to be utterly reasonable and received no serious signs of doubt from those people I actually explained the content to in detail and who understood the material.  If there were no major open \emph{technical} issues anymore, at least the mathematical side of the theory would be fully specified and detailed leaving merely the duty to match experiment, something which remains at this point to be done.  For example verifying the post Newtonian expansion as well as the emergence of QED are mandatory tasks.  Nevertheless, let us start by the achievements: a new class of gravitational theories is presented which naturally incorporate a novel relativistic quantum theory whose formulation is entirely \emph{local} on spacetime in a way which is identical to Einstein's original formulation of the theory of relativity.  This means that we dismiss global Hamiltonian and path integral approaches to quantum mechanics and that causality, amongst other things, is an emergent property instead of a fundamental principle.  Moreover, it is shown that on Minkowski, ordinary free quantum field theory is the ``natural'' limit of our theory in the absence of interactions.  In section ten, we started the study of how the free theory behaves in a nontrivial gravitational background with a global spatial rotational symmetry and we impose natural boundary conditions at infinity for the quantum theory.  Some of the main realizations, however, are that (a) we have a full nonperturbative formulation of a candidate theory of quantum gravity (b) we present a solution to the question of ``Where the collapse takes place?'' (c) our laws have a local four dimensional formulation which allows for a consistent treatment of singularities (d) we present a natural class of local physical observables (e) we give a natural interpretation to the Weinberg-Witten theorem and circumvent as well Haag's theorem as the Coleman-Mandula no-go argument (f) the local vacuum states are dynamically determined (g) we have shown that Newton's law and free Quantum Field Theory emerge in the suitable limits assuming natural boundary conditions.  To the best knowledge of the author, string theory managed to solve (g) as well as some form of the post-Newtonian limit; however (a) till (c) are certainly open issues in that approach.  An emergent virtue is that ``new'' and more general mathematical concepts and techniques enter the formulation and sections seven and eleven are entirely devoted to the introduction of these tools.  Obviously, a lot of work will have to be done before these new mathematical gadgets are understood at an appropriate level but that is nothing extraneous to other approaches.  In all honesty, I believe it is quite remarkable that someone can offer a complete new quantum theory, based upon totally different principles, which appears to have the right limits ninety years after birth of that same theoretical framework.  \\* \\*   Certainly, these promises must arouse some skepticism and also I did not believe much of it in the beginning. However, as the work evolved, the inner coherence became stronger which reassured me that it was not all utter nonsense.  At this point, it might be opportune to make some philosophical remarks as to why it is not very surprising that a singleton comes up with some novel ideas regarding this old problem which opposes modern culture.  As an act of wisdom and cowardry resulting from the fear of a potential downfall on the sales ranks of this work, I shall refrain from doing so hoping that the intelligent reader understands what I am talking about.  Let us now take the historical and conservative approach and say how the argumentation and gist behind this work is subdivided contentwise.  Sections two and three have a rather special place in this work and reflect more my own way of thinking than anything else.  Nevertheless, a reader who would finish the entire book might have the feeling that somehow these two sections already contained some of the main seeds of the later construction, albeit in a somewhat hidden form which is, at least, my humble intention.  Why are sections of a ``revisional'' and ``philosophical'' nature important?  Well, they reflect how one thinks about contemporary science; what its main lessons are, where reside the most important shortcomings and what logical gaps might imply a very different worldview which in turn generates new mathematics, hence new physics and the cycle starts again.  In spite of the even more radical character of the end product, I decided not to change any word here because I want to convey that many cycles can lead to very different conclusions, but one has to go one ``rotation'' at a time.  Especially the role of  consciousness in physics, historically stressed by Von Neumann, and more recently revived by Penrose and others gets a more central place in the theory and as an amateur philosopher, I spur some resemblance to monism.  In the third chapter, the focus is changed from relativity and quantum mechanics to quantum gravity; this chapter will contain technical arguments as well as metaphysical ones.  I realize that this is a rather unconventional course of action for a physics book but sometimes it is good to be liberated from too restrictive formal rules.  \\* \\*
Considering this philosophical and physical input, it requires a novel idea to save manifest background independence in the sense that we demand a well defined representation of the group of coordinate transformations as well as a covariant (hence dynamical) procedure for fixing the \textit{local} vacuum state and particle interpretation.  Loop quantum gravity certainly tries to construct this representation as well as vacuum state however unsuccessfully so far and the issue of a particle interpretation is nonexistent apart from some naive attempts trying to identify particles with knot like configurations in the spin network basis states.  String theory follows a more conventional approach, however, to my knowledge the issue of the vacuum state has not recieved any answer.  A radical new construction is presented in chapter eight which allows for a treatment of all these issues which appears to be consistent so far.  However, these ideas are highly nontrivial if you look through conservative glasses and in chapters four and five, we present a representation in terms of background dependent physics.  The germs of this theory, that is the kinematical setting and classical dynamics, are presented in chapter four.  Here, I study a novel type of background dependent dynamics which resembles the Polyakov action but with the important difference that the worldsheet metric is not a dynamical variable.  Therefore, we do not have to consider the Virasoro constraints and a kinematical volume constraint is put in by hand.  The motivation for committing this ugly crime comes from the technical idea that inverting a metric becomes an analytic operation if one does not have to divide trough its determinant (in either volume).  The problems of causality and ``localizability'' are discussed and an old idea of how to retrieve matter from such framework is revived (just consider the Einstein-Cartan equations to be an identity).  It turns out that Quantum Field Theory generates local degrees of freedom which are not present classically because the curvature tensor may be nonvanishing depending upon the type of Wick ordering one considers (something which one may call a quantum anomaly).  However, this theory cannot be rescued but trying to do so lead me to work done in chapter six which by itself formed an important corner stone for the ideas presented later on.  A philosophical principle, which constitutes the very core of the reasoning behind that later work, is that there is no point in axiomatizing based upon representation prejudices.  Indeed, all inequivalent representations should be investigated and therefore one should only \textsl{try} to formalize physical principles.  There are plenty of examples in the literature of the first kind of activity: (a) the old Wightmann axioms (and more recently Wald's) of Quantum Field Theory (b) the work of Piron on some possible extensions of Quantum Mechanics (c) General Relativity as the Einstein equations (d) Dirac's fermion theory (e) Weinberg's analysis of the implications of first principles of Quantum Field Theory \cite{Weinberg} even if this work is by far superior to anything else in literature (and was actually the key motivator for my ideas).  Indeed, the philosophical ideas explained in sections two and three do not change later on, only the mathematical representation does.  In other words, this work is written in the old spirit of natural philosophy complemented with novel mathematical techniques exceeding the current use in mainstream physics.  Valuable inspiration for these ideas originated from literature on quantum group theory, Von Neumann algebras, measure theory, Krein spaces, operator theory and many other branches of mathematics.  \\* \\* Chapter five starts with a general discussion about interpretational subtleties in quantum physics regarding observables which do not commute with the Hamiltonian and give rise to fairly complicated interpretations of pretty simple dynamical systems.   Consequently, we apply this idea to the simple theory proposed in chapter three and, as said previously, define observed matter though calculation of relevant tensors in Einstein-Cartan theory.  Those observables are highly nonlinear and noncommuting with the Hamiltonian and it could be hoped that the probability of decay for their low energy eigenstates on the time scale of observation is sufficiently low for no inconsistencies to arise.  However, computation of the metric tensor and (anti)commutation relations thereof leads to unwanted infinties which I try to dissolve through a modification of the quantization procedure and particle statistics.  This leads to a split in the content of the chapter where on one on side the question of statistics is readressed and on the other the ``quantization'' of our preliminary theory is continued.  I have decided to move the reinvestigation of the spin-statistics theorem, which is justified because Minkowski causality is not a valid assumption anymore, to a separate appendix in order to improve the general readership of this chapter.  The outcome of this investigation is rather surprising since a consistent quantization of our theory (that is one without normal ordering infinities at fourth order) does not only require spin $\frac{1}{2}$ Clifford particles, but we must also allow for negative energies.  The latter cannot be replaced by negative norm, positive energy bosons as such particles would not cancel out the infinities in the Hamiltonian as well as the commutation relations of the metric.  Given the importance of the Clifford numbers in this procedure, it is logical to study Clifford valued actions and quantize them; a study which is initiated in chapter six.  Here, a trade off between negative energies and negative probabilities occurs and the resulting particles have genuinely different transformation properties under the Poincar\'e group than is allowed for by the analysis of Wigner \cite{Weinberg}.  Given that we have to work on indefinite Hilbert spaces, the spin statistics connection vanishes and we shall have better things to say about that later on.  All this requires a first extension of Quantum Field Theory, that is one must study representation theory on indefinite Hilbert spaces and construct a consistent local and causal interpretation.  At the same time, one might investigate the possibility of negative energies and study if this theory is really as screwed as most people believe.  Although the quantization scheme in chapter six is the first example in the literature where negative probabilities are mandatory, since without them negative energy spin $\frac{1}{2}$ particles would have to be bosons, indefinite Hilbert spaces have shown up in history on several other occasions such as Gupta-Bleuler quantization of gauge theories.  Moreover, negative probabilities allow one to sidestep the famous Weinberg-Witten theorem, which states that there exists no theory with a Lorentz covariant energy momentum tensor containing massless spin two particles.  There are plenty of other means for achieving this goal such as allowing for fat gravitons, or one might dismiss gravitons and recuperate the Newtonian gravitational force from virtual particle interactions\footnote{I acknowledge useful private correspondence with Alejandro Jenkins about the Weinberg-Witten theorem although he would not morally agree with all conclusions I draw here \cite{Jenkins}.}.  Anyhow, all above results strongly indicate that indefinite Hilbert spaces do not only allow for a broader class of phenomena, but appear also to be necessary for quantum gravity.  There is still another way of looking at the Weinberg-Witten theorem which does not seem to have been appreciated too much which is simply accepting its conclusion: that is, gravitons do not gravitate directly (they do nevertheless indirectly through interaction with matter)!  This must appear nutty for someone who thinks in the conventional way about how gravitons arise (through quantization of a classical field theory), but as will become clear in chapter eight, it is completely consistent and physical within the new framework.  Therefore, in my mind, we are left with essentially two possibilities : (a) gravitons on Nevanlinna spaces which do gravitate and (b) non-gravitating gravitons (on Nevanlinna spaces or not).  In sections seven and eight, we will come to the conclusion that option (b) on Clifford-Nevanlinna modules is the right way to go\footnote{To add to the reader's confusion, these non-gravitating gravitons can nevertheless scatter in a non-trivial way.}.  In a nutshell, chapter four is a fairly ordinary analysis of a simple theory which realizes the ideas of chapter three in a straightforward way, while sections five and six are of a transitional nature; the ``real'' theory starts to be developed from chapter seven onwards.  \\* \\*
So, chapter seven paves the way for a future study of representation theory of the Poincar\'e group on infinite dimensional Clifford-Nevanlinna modules which is an even wider first generalization of Quantum Field Theory.  For starters, I was quite unhappy with the definition of Nevanlinna spaces by Krein and Jadczyk and decided to rigorously construct my own concept; the latter is a lot more advanced and relates to concepts such as an observer dependent topology.  The definition suggests an even wider generalization to non-associative structures we baptise to be kroups, as opposed to groupoids and semi-groups.  The construction of a rigorous definition of a Nevanlinna space constitutes the main body of the chapter as it currently stands while the study of \textit{finite dimensional} Clifford-Nevanlinna modules and a suitable spectral theorem thereon is its primary stages.  We learn for now that an Hermitian operator allows for \emph{many} (approximate) decompositions of several inequivalent types, each with their own probability interpretation, but as it stands no general theorem is formulated.  These preliminary results suggest such an interpretational ``revolution'' that is legitimate to spend many pages spend to it.  The interpretation needs to be further worked out and generalization towards the infinite dimensional context needs to be made prior to studying representation theory of the Poincar\'e group. \\* \\*  The dynamics presented in chapter eight incorporates the idea of a quantum bundle in which the unitary relators form a group locally, but only have a kroup structure globally.  As mentioned there, I foresee the possibility for a slight generalization of this to kroups with a special kind of connectedness property but I feel it would be hard, if not impossible, to construct a dynamics while assuming only a general kroup structure to hold.  Hard computations will have to show whether the ``postulate'' of a local group structure can be sustained, otherwise one would have to give up associativity even locally;  this is one of the issues I still need to adress in sections nine and ten, but this book is not going to give a final answer to this question.  The second idea consists in putting free Quantum Field Theory on the tangent bundle instead of on spacetime itself: the physical and mathematical ideas behind this are nontrivial and I go through a great deal to explain them properly.  Moreover, the setting discussed here is just a special case of an even much wider class of possibilities and only future work can tell to which extend our limitations are justified.  The third idea deals with a totally nonperturbative treatment of particle interactions; particles originate from ultralocal ``hidden variables'' living on tangent space and the relators between those hidden variables are subject of the real dynamical content.  In this sense, our approach is radically quantum and many ideas are natural continuations of suggestions made, even as early, by Von Neumann, Wigner and Heisenberg.  We dismiss the path integral as a step back in the natural evolution of quantum theory in the sense that it hinges too close on concepts involving a classical reality and it is moreover not as relativistic as one would like it to be.  Indeed, as mentioned previously, our theory really has a \emph{local} formulation and global considerations like hypersurfaces, action principles with ill defined integration over noncompact spacetimes definitively belong to the past.  Not only do the laws have a local formulation on spacetime, also the probability interpretation and state of the universe have a mere local meaning.  It would be too much to simply explain these things at this point, but let me say that (a) a boundary value point of view is more natural for the theory of gravitation than initial values are (b) the holographic principle is directly reflected in the quantum and geometry theory.  Many of the philosophical implications (which were not foreseen in chapter three) would simply be too mind stretching to explain without any understanding of the mathematical formalism and the chapter finishes with a more in depth discussion where physics could go from thereon.  For all these reasons, I believe it is not a good idea to start at chapter eight or just even chapter six for that matter.  Chapters nine and ten, which are currently under construction, will deal with phenomenology as well as some representation theory of the Poincar\'e group on Hilbert spaces in which an infinite number of copies of the same particles are allowed for.  The latter involve a length scale which has to be sufficiently large so that the corresponding violations of the Pauli principle do not lead to conflicts with observation.  Chapter nine in particular will deal with corrections to the Hawking effect as calculated in our novel quantum theory.  A full mathematical investigation of integrability of the equations of motion is, as said previously, not treated in this work for the understandable reason that it would take too much work to fill all the gaps.  Chapter eleven is meant as a teaser and provides an even wider mathematical implementation of the physical principles we enunciated before; a novel and universal concept developed in that direction is the notion of a quantum manifold.  This concludes the overview from the conservative vantage point of view.  \\* \\*
As an alternative way of reaching similar conclusions and of deepening ones understanding of the physical principles which go into the theory, let me present an exercise which is seldomly made but can have an illuminating effect after one has gone through all the painful derivations.  That is, I shall first present the known principles behind Quantum Field Theory and General Relativity and comment upon which ones are to remain there and which should be the approximate result of a computation in weak gravitational fields instead of a fundamental law of nature.  The physical principles behind Quantum Field Theory are (a) locality (b) Poincar\'e covariance (c) causality, in the sense that spacelike separated observables commute, (d) positive energies (e) the statistics assumption (f) cluster decomposition principle and the technical assumption made is that all representations should be on separable Hilbert spaces.  Of course, some of these principles can be exchanged such as the statistics assumption which follows from the existence of a well defined number operator, Poincar\'e covariance and a relative isotropy condition while ignoring parastatistics.  Now, there is no doubt that all these physical restrictions should apply in case all interactions are shut off, but there are no good indications for the technical requirements.  Indeed, positive probability is tightened to the straightforward Born rule, but the latter can be extended to representations on Nevanlinna space; likewise, it is rather unnatural that the representation space should be separable since it is impossible to describe the situation with an infinite number of particles which should be allowed, in principle, if one is describing the whole universe.  However, this puts doubt on the principle of causality since the spin statistics theorem fails if \emph{any} one of the above restrictions is dropped; replacing causality by spin statistics as a fundamental principle of nature appears a better thing to do since the implication of causality would be much more robust (that is, not depend upon any of these technical assumptions).  Another argument which leads to this conclusion is the desire to have a truly local, four dimensional formulation of quantum interactions; in that case, the commutation relations cannot be implemented since they depend upon a global apriori notion of spatiality.  For quantum gravity therefore, we demand that the \emph{interactions} satisfy laws which have a local formulation, are covariant under \emph{local} Lorentz transformations and are ``locally unitary''.  The free theory on the other hand should obey locality, Poincar\'e covariance, spin statistics, positive energies and cluster decomposition; the reader notices that we dropped the technical requirements as well as the statistics assumption.  To merge these views, the free theory should live on the tangent bundle and the representation of the Poincar\'e group should live on the tangent plane and not on spacetime.  This means that the translation symmetry of the free theory is broken by means of the interactions which single out a preferred origin.  \\* \\*
On the side of Relativity, the main principles are (a) locality (b) background independence (c) local Lorentz covariance (d) general covariance (e) the equivalence of gravitational and inertial mass.  Except for the last principle, all the latter are mathematically well defined and there is no reason to abandon them in a theory of quantum gravity and one has the choice whether to make the gravitational theory locally Lorentz covariant or locally Poincar\'e covariant (it does not really matter).  However (e) is something which should only hold in the linearization of the theory around a Minkowski background and current work reveals it does not hold if nonlinear corrections are taken into account.  From all the above, it follows that if one probes the world at small distance scales, the theory should become free and therefore asymptotic freedom is build into the construction right from the start.  These constitute the very foundations upon which the construction in chapter eight hinges and we have more to say about these things in the course of this book.  Most attention however is spend to the principle of locality which \emph{appears} to necessitate the framework of classical abelian manifolds.  However, there is a small caveat here and in section eleven we show how the standard locality concept can be canonically lifted to non-abelian manifolds.  This is an extremely strong result since it allows for the construction of a ``universal'' differential calculus where the ambiguity in the derivative operators originates from a quantum connection.  We shall not further treat this construction in this book since I feel that the more conservative theory is already more than complex enough to start with.           
\chapter{On quantum mechanics and relativity}   
My first reaction when learning about quantum mechanics was that this could not be and that eventually quantum theory would prove to be an excellent approach to an otherwise deterministic theory.  This (local) realist stance remained with me for a long time even in spite of Bell's theorem which strictly speaking doesn't prove anything since it assumes a nondetermistic feature of nature, namely ``free will''.  This has recently been pointed out again by 't Hooft \cite{'t Hooft} and resulted in a debate with Conway and Kochen \cite{Conway1} \cite{Conway2}.  Indeed, the textbook case for quantum mechanics is rather weak, first of all do you need to assume a two fold level of reality, the classical observer and the quantum system under consideration, but moreover is the dynamics presented as a procedure applied to a classical system.  This is certainly so in the Dirac quantization scheme where classically meaningless Poisson brackets get promoted to physical statements about the quantum world; this situation, however, is already considerably improved upon - but not completely erased - in ``the'' path integral formulation.  In that sense quantum mechanics is not even a theory, rather an algorithm, and the only argument in favor of it is that it manages to produce accurate outcomes of experiments.  This is of course a very strong indication that something about it must be right but as long as we do not understand quantum mechanics ``an sich'' the situation is theoretically rather unsatisfying.  That is, until we figure out \emph{why} nature would prefer some of its ideas, the theorist must remain skeptical and open to alternatives.  Before I proceed, let me stress that I am an unashamed realist in the sense that I believe some stuff to exist, but the question is what does and how it connects to our observations.  Indeed, suppose you want to make a theory for the universe, then your Platonic objects might be a fixed four manifold $\mathcal{M}$ and the \emph{definition} of a Lorentzian metric, i.e. a symmetric covariant two tensor with signature $(- + + +)$ or you might want to be more ambitious and take as Platonic object the definition of a causal set.  Now, classical mechanics corresponds to a single universe which we need to find out by specifying initial conditions and by proposing a certain dynamics.  The view on this procedure is rather limited since it allows only for globally hyperbolic universes and wouldn't allow us to think of black holes while we clearly can do that within general relativity.  There, the Einstein equations should be thought of as a constraint on the universe and the initial value point of view must be entirely dropped.  This leads one to propose that classical mechanics could be thought of as a singular probability measure with support on one Lorentzian metric on the space of all Lorentzian metrics on $\mathcal{M}$.  Specifying which measures $\mu$ of that type are allowed is equivalent to formulating a dynamics; in that respect a single measure unifies the idea of ``initial values'' with the dynamics and putting physical demands on $\mu$ would constrain as well the kinematics and dynamics at the same time.  A first, albeit limited, generalization of this would consist in studying nonsingular probability measures.  This can give rise to a genuine stochastic \emph{dynamics} with fixed initial boundary conditions such as happens in the Sorkin Rideout-dynamics for causal sets \cite{Sorkin1} \cite{Rideout} and does not need to be limited to measures expressing lack of knowledge of the initial data.             
One recognizes that this is already a higher form of physics since it involves the entire space of representations (usually called histories) of the Platonic theory.  ``Quantum mechanics'' is another generalization of this idea which contains the latter as a special case; actually as Sorkin noticed, it is the next alternative in an infinite series of theories expressing higher types of correlations between alternate histories \cite{Sorkin2} \cite{Sorkin3}.   More precisely, assume the space of histories is equipped with a topology and its subsequent sigma algebra $\Sigma$, then a function $\mu : \Sigma \rightarrow \mathbb{R}_{+}$ is said to be a measure of order $n-1$ if for every $n$ tuple of disjoint elements $A_i \in \Sigma$, $\mu$ satisfies
$$\mu \left( A_1 \cup A_2 \ldots A_n \right) - \sum \mu \left( \textrm{n-1 tuples}  \right) + \sum \mu \left( \textrm{n-2 tuples}  \right) \ldots + (-1)^{n-1} \sum \mu(A_i) = 0.$$
Sorkin's generalization of quantum mechanics deals with measures of order $2$.  One can show that this implies the existence of a real valued function $I(A,B)$ satisfying for $A$ and $B$ disjoint $$I(A \cup B, C) = I(A,C) + I(B,C)$$ and $$\mu(A) = I(A,A).$$  This ties actually with the decoherence functional approach developed by amongst others Dowker and Halliwell \cite{Dowker}.  A decoherence functional is a complex valued function $D$ on $\Sigma \times \Sigma$ satisfying $D(A,B) = \overline{D(B,A)}$, $D(A \cup B,C) = D(A,C) + D(B,C)$ and for any $n$ and n-tuple $A_i$, the matrix $D(A_i,A_j)$ is positive definite; $I(A,B)$ can be thought of as the real part of $D(A,B)$.  The way all these notions tie with the ordinary path integral is as follows :
$$D(A,B) = \int_{\gamma \in A, \chi \in B} D \gamma D \chi e^{i(S(\gamma) - S(\chi))} \delta(\gamma(T),\chi(T))$$ where $T$ is a so called truncation time and $S$ is the ordinary action.  A constrained history $A$ is equivalent to the insertion of a (possibly distributional) operator in the Hamiltonian formalism.  In this language, there is no room for operators and Hilbert spaces (just as in the path integral language) and one needs to figure out an (objective) interpretation based upon the measure alone.  Likewise, the measurement problem in quantum mechanics needs to find a translation and resolution in this language.  A promising framework for such interpretation has recently been proposed by Sorkin and Gudder \cite{Sorkin4} \cite{Sorkin5} \cite{Gudder}.  The approach I will take later on is based upon a much more sophisticated operational formalism and is likewise genuinely quantum in the sense that it does not start from a classical action principle.  But the unification of the ``state'' and ``action'' in a single measure of order $3$ is certainly a nice idea which is also capable of encapsulating topology change in quantum gravity, as is our formulation of the quantum laws enunciated in chapter eight.  In this framework, one recognizes that ``quantum mechanics'' is a higher order theory than classical mechanics is which in a certain sense respects more the Platonic world because it expresses pairwise relations between measurable sets of representations.  However, the above discussion also puts into doubt the universality of quantum theory as a theory of nature and a three split experiment has been devised to verify if nature does not entail higher order correlations \cite{three}.  Let me mention here that all my comments concerning quantum gravity below also apply to these higher order theories.  \\* \\*
The traditional physicist might now object that the Hilbert space framework with a well defined Hamiltonian or a more traditional path integral point of view ensures a \emph{unitary} dynamics or at least a unitary scattering matrix.  In the above interpretational framework, there is nothing which automatically ensures unitarity and one is left with the task of constructing theories in which the breakdown of unitarity is sufficiently small such that no reasonable contradiction with observation arises \cite{Sorkin6}.  The acceptance of a lack of unitarity mainly stems from two different observations : (a) unitarity is not a logical requirement to have a consistent probability interpretation (b) Hawking radiation seems to suggest a violation of unitarity in quantum gravity albeit the opinions upon that are rather divided \cite{Wald1} \cite{Sorkin7} \cite{'t Hooft1} \cite{'t Hooft2} \cite{Susskind}.  The dynamics I am about to propose in the fourth and fifth chapter is not unitary either due to a novel implementation of the commutation relations.  In the usual path integral formulation, the measure $\mu$ is split into an infinite dimensional Lebesgue measure and the exponential of the action.  The Lebesgue measure does however not exist and to make it precise, one has to start with a theory on a finite lattice and take the thermodynamic and continuum limit (in the right order) later while renormalizing at the same moment \cite{Salmhofer}.  The same can be understood in -say- free Klein Gordon field theory starting from the Hamiltonian Fock space quantization.  It might be an instructive exercise to explicitly construct formal ``field'' $\psi$ and ``field momentum'' $\pi$ eigenstates on the Fock space and calculate their inner products.  Both operators are defined in a distributional sense (as a limit of bounded operators corresponding to a momentum cutoff) and the domain $\mathcal{D}$ of $\psi$ is defined as the set of all vectors $v$ in Fock space such that $\lim_{L \rightarrow \infty} \psi_L (v)$ is well defined (where the $\psi_L$ are the cutoff operators) as a distribution\footnote{This construction works by taking the scalar product with states constructed from Schwartz functions on momentum space or Cartesian products thereof.}.  Hence, we may define the resolvent of these operators as the set of all $\lambda \in \mathbb{C}$ such that there exists a sequence of bounded operators $\phi^{\lambda}_L : \psi_L \left(\mathcal{D}_L \right) \rightarrow \mathcal{D}_L$ such that $\phi^{\lambda}_L \circ (\psi_L - \lambda 1)= 1_{\mathcal{D}_L}$ and $(\psi_L - \lambda 1) \circ \phi^{\lambda}_L = 1_{\psi_L \left( \mathcal{D}_L \right)}$ and the sequence $\phi^{\lambda}_{L}$ converges to a well defined distribution mapping $\psi \left(\mathcal{D} \right)$ to $\mathcal{D}$.  The spectrum is then defined, as usual, as the complement of the resolvent.  \\* \\*
Let me now come to the issue of quantum mechanical measurement and discuss that from a traditional point of view, I will not dwell here on the anhomomorphic logic proposed by Sorkin.  I shall on purpose refrain from jumping immediately to the conclusions I will reach and postpone this for later since a universe with classical observers does not force me yet to take such radical point of view.  Instead, I will merely comment on the existing interpretations and point out their weaknesses.  Let us first start with the Copenhagen point of view which -in a precise sense- is the cleanest but also the one which is the least suitable.  I don't have to explain the textbook version of the U(nitary)/R(eduction) process here but shall merely comment on the underlying assumptions.  First of all, one assumes classical observers which move around in a fixed geometry and study a quantum system.  Although the observer is in one world, the system under study is in many worlds at the same time (where the number of worlds depends upon some pointer basis), but somehow the observer is \emph{aware} of the simultaneity of these worlds and collapses the system at \emph{regular times} and not only when ``macroscopically distinct'' alternatives in the state occur.  This regular collapse time is probably related to an internal clock associated to the observer's brain activity which has not been taken into account in the description.  It just happens to be so that the time scale at which different motions of \emph{most} macroscopic objects occur is comparable to the observer's internal clock.  This is however not always so: a bullet shot from a gun is a clear counterexample and although it is definetly a classical object, one could argue that it behaves somewhat quantum mechanically with respect to the observer (in the sense that we can only locate it when it hits a macroscopic body).  Moreover, when \emph{interpreting} the ``macroscopic'' situation at hand we always make use of Cournot's principle (the author has learned about this by Rafael Sorkin) which says that if something is extremely unlikely to occur, it actually \emph{never} does.  To understand this, consider the following situation: the quantum mechanical situation at hand are two measurement apparati, one to the left of the observer and another to the right (with respect to some spatial axis).  Now, initially, the left one is in the ``up'' state and the right one in the ``down'' state; a ``consciousness time $\delta t$'' later the observer sees an apparatus to the left in the ``down'' state and one to the right in the ``up'' state.  The conclusion he draws from this is that both apparati stayed where they were but simply changed their state.  However, there exists an extremely tiny possibility that both apparati simply switched position; the fact that we \emph{never} infer this logical possibility is precisely Cournot's principle.  This implies for example that we could disagree with a ``conscious being'' -call it X- with a much higher state of awareness (but not necessarily intelligence) in very few cases since they actually might see the apparati moving around.  Is there a contradiction or does it mean that both these conscious beings live in a different world?  No, there is no such thing of a kind!  Imagine that at time zero, both observers have a ``consciousness moment'' and X has a ``consciousness interval'' $\delta t/n$ while the human has $\delta t$.  Initially, the \emph{objective} state will be :
$$ | \textrm{X sees left up and right down} ! \rangle | \textrm{left up} \rangle |\textrm{right down} \rangle .$$
At an intermediate time, the consciousness of X (which we indicate by !) has made a choice (which we could interpret as a collapse but do not need to do so):
$$ | \textrm{X: left and right are switching position} ! \rangle | \textrm{left moving to right: up} \rangle  |\textrm{right moving to left: down} \rangle  + $$ $$ | \textrm{corrections showing that X makes the wrong interpretation and X's consciousness could also be here} \rangle.$$
Of course, this is just an intermediate state and the human doesn't know about it; in a quantum description incoorporating the human observer, the latter would be in a state $|\textrm{ignorant : no perception made} \rangle$ in almost \emph{all} universes in which he or she is residing.  Then, when the human finally makes his measurement, he is left with the state:
$$ | \textrm{X sees right up and left down but knows both apparati switched position } !\rangle | \textrm{right: up}\rangle |\textrm{left: down} \rangle $$
but he will nevertheless make the \emph{ab initio} interpretation that left and right switched state.  However, communication with X can make him revise his point of view at a later moment even if X has a objective much smaller chance to be wrong too.  Of course this works only well if X is still ``macroscopic'' in a reasonable sense; clearly electrons cannot have a higher awareness state than humans since we would never see an interference pattern in a double slit experiment.  Does this mean that electrons have no awareness?  Not necessarily, their awareness time intervals \emph{could} simply be gigantically large which is the very premise behind the collapse interpretation by Ghirardi, Rimini and Weber \cite{Tumulka}.  There, ``macroscopic'' bodies have a much shorter collapse time since they contain a gigantic number of elementary particles; this would give a fairly linear relation between awareness time and mass of an observer which is I believe not true per se.  The attentive reader might infer at this moment that I did not assign any conscious perception to the measurement apparati in the above experiment; I should have done that but it would only have made the situation more complex while letting the conclusion remain identical. \\* \\*
Let me end this discussion by one further observation : the fact that the classical observer manages at time zero to distinguish between both measurement apparati is because he has a classical reference frame at his disposal and he can make at all awareness times the distinction between left and right.  Before we proceed to the Everett interpretation, let me comment on a distinction between two quantum mechanical descriptions of reality which is rarely stressed but will become important later on.  Consider for simplicity a superposition of two particle states (it works generally for a superposition of n-particle states, but not for superpositions of n with m particle states) in ordinary free Klein Gordon field theory:
$$ \Psi  = (a^{\dagger}_1 a^{\dagger}_2 + a^{\dagger}_3 a^{\dagger}_4 ) | 0 \rangle $$           
Suppose that $a^{\dagger}_1$ and $a^{\dagger}_3$ correspond to left movers with respect to (a fixed axis) the classical observers inertial frame and $a^{\dagger}_2$, $a^{\dagger}_4$ to right movers.  Now, in general, for a theory with $n$ particles and $m$ superpositions I can make $(n!)^m$ identifications (at least if no internal symmetries are present) which is much larger than the usual $n!$ considered in the literature.  In our case there are four distinct worlds : denoting with P1,P2 particle 1,2 we have $$\{ (P1 = 13, P2=24), (P1 = 14, P2=23),   
(P1 = 23, P2=14), (P1 = 24, P2=13) \}$$ and the actual different wavefunctions are all \emph{physically} the same.  However, this is not the end of the story : in the multiparticle wavefunction formalism, one introduces the tensor Hilbert space construction and pretends as if every particle has a separate position, momentum and angular momentum operator which in a sense is totally wrong since there exists only one field operator.  This allows one in principle to ask a question like ``Where is particle i on average?'' or ``What is the average distance between particle 1 and 2?'', these are well defined questions since one has a fixed notion of locality and metricity due to the classical observer.  Even more, one could pretend collapsing particle 1 !  However quantum field theory does not allow for this and only ``localized'' questions are possible; the latter do not depend upon the labeling of course since they only feel left or right and not 1 and 2.  Nevertheless, the classical observers \emph{awareness} (or consciousness) seems to make exactly such labeling for ``macroscopic'' many particle systems in different universes by Cournot's principle as argued before (!).  Now, there is an edge to this : it is quite safe to assume that on the awareness scales of the human observer, very few if no particles inside the apparatus are created and/or annihilated.  Therefore, the ``labeling'' on the apparati induces a labeling on the elementary particles composing it, we shall come back to this in the next chapter. It is clear that such labeling is nothing physical (a priori) and does not belong to the world of material variables, but it is nevertheless necessary to make sense out of the multiverse.  This seems to have been contemplated by various physicists in the course of history amongst which Wolfgang Pauli and Eugene Wigner \cite{d'Espagnat} albeit I do not know if they meant it in the precise sense I elaborated upon above (I am not a historian of science). \\* \\*
In the Everett interpretation you initially take the point of view of God and write down the state of the Universe undergoing a unitary dynamics.  Now as said before, different ``macroscopic'' states in different universes get identified by means of an infinite collection of ``quasilocal'' consciousnesses.  A first remark is that any ``theory'' which produces such identifications has to rely upon a notion of ``God'' or a non materialistic superobserver which observes the entire dynamical scene with respect to his reference frame and classical measure stick.  Now, such assignment of quasilocal ``consciousness'' by God is of vital importance, since it \emph{determines} each observers notion of localization and causation.  Now, in contrast to the ``standard'' Everett interpretation, I assume God will be that kind to make all quasilocal consciousnesses go to the same material reality.  It is clear that this point of view is filled with tremendous philosophical difficulties (amongst which the fact that our individual free will would be an illusion, only God possesses it) and we shall come back to this in the next chapter.  However, let me stress that, no matter how difficult this point of view is to swallow, there is as yet no practical problem in the sense that we -as humans- are still capable to (approximately) trace God's steps (in retrospect) since the entire setup remains \emph{computable}; that is we are not only allowed to read God's mind but we can also exactly solve the (unitary) dynamics the good man had in mind.  For a theory of quantum gravity in which the superposition principle applies to spacetime itself, this problem becomes more severe since now God no longer disposes of his fixed classical measure stick.  Therefore, one must wonder what will happen to localization and causation of the human observers in such framework.  Actually, this concern has also been expressed by 't Hooft \cite{'t Hooft3} \cite{'t Hooft4} to me when I was a post doc in Utrecht, I am not sure if it this precise statement that he had in mind, but at least that is how I understood it.  I will elaborate on these issues further in the following chapters and propose a radical resolution to this problem which effectively will be a ``new kind of physics'' with a well defined \emph{unity} principle between consciousness and materialism.  However, as far as it stands, I do not (yet) need to develop a dynamical theory of consciousness.  Now, we will turn to general relativity and show that similar problems also appear there at the surface. \\* \\*
Einstein's theory of General Relativity is, although only a second order theory, another pillar of modern physics.  I learned relativity for the first time in 1998 as a fresh graduate student and remember very well being struck with its conceptual coherence and mathematical beauty, this truly was physics of a very different order than anything I had learned so far; for excellent textbooks see \cite{Wald2} \cite{Hawking1} \cite{Beem}.  I think every beginning relativity student always first studies the Lagrangian formulation of Relativity and only comes to the Hamiltonian formulation at a later stage which forces one to study constrained Hamiltonian systems and the Dirac algorithm, see for instance \cite{Thiemann} \cite{Henneaux} for excellent reviews on that topic.  However, something which struck me immediately about the theory was that, just like the beginning of the Everett interpretation of quantum mechanics, it was a theory of the universe from God's perspective; it for sure was no theory of conscious ``entities'' living inside the universe.  This problem, downside or whatever you might want to call it of Einstein's beautiful concepts has many technical faces and I shall elaborate upon them in great detail (while probably omitting many references of people who have worked on this).  The upshot of the discussion will be however that the theory of Relativity needs to have an extension, not necessarily changing its large scale physics but complementing its interpretation, just like the ``quasilocal'' consciousnesses did it for Everett's theory.  The entire story starts of course with the observation that we live within the universe and make \emph{(quasi) local} observations.  Such observables are not diffeomorphism invariant (in the Lagrangian formalism) or do not commute with the constraints on shell (in the Hamiltonian formalism) so they are not what is called Dirac observables.  That is, they are not gauge invariant; this suggests that the beautiful principle of gauge invariant observables which worked so well for (real) gauge theories (that is for connections living in a fiber bundle over space-time) fails for gravity.  Undoubtedly, there exist people (and I know some of them) who maintain that Dirac observables are the only legitimate objects in Relativity.  Dirac observables are by definition nonlocal since they do not discriminate against points in spacetime and it is fairly obvious that there exist only a countable infinite number $\aleph_0$ of them.  Making sure they are well defined, one must either work with compact universes or universes with severe asymptotic conditions and sufficiently rapid fall off behavior of the fields towards infinity.  Now, the hope that these people seem to have is that by measuring a sufficient number of them, one is able to reconstruct the universe ``locally'' with a sufficient degree of accuracy and uniqueness.  Of course, this mapping can never be exact since there are an $\aleph_1$ local degrees of freedom in gravitational physics.  I would say that such hopes have a probability of coming out which is far inferior to the probability of winning the lottery.  Moreover, and far more important than this, this is simply not the way we make observations (I shall come back to this in the next chapter).   A more useful point of view consists in admitting that we are allowed to make local \emph{scalar} observations such as -say- the Ricci scalar $R(x)$.  The importance of these so called ``partial'' observables has been stressed many times by authors such as Karel Kuchar and more recently Carlo Rovelli \cite{Rovelli1} \cite{Rovelli2}.  Partial observables will not be generically useful to obtain \emph{predictions} of the theory, Minkowski or de Sitter are classic examples of such stubborn space-times.  I perceive the main distinction between both types of reasoning as follows : (a) a Dirac observable does not care about the initial conditions of the universe at all, it is a formal expression in terms of the dynamical variables which should work for all universes (which are solutions to the field equations) (b) a partial observable hinges upon the initial conditions, the universe must be ``well chosen'' for partial observables to be useful.  Relativists have therefore tried for a long time to construct \emph{physical} coordinate systems using scalar invariants hoping that these would fully determine the event at hand.  Obviously, this puts a constraint on the possible universes and the relativist might conclude he had to limit the solutions of Einstein's equations to those guys.  This would be a classical solution for the problem that a single observer making local observations would immediately know where he is on a local map of the universe and therefore also know his future.  In a fully deterministic theory, the future exists and it would be rather contradictory that a single observer in the universe with a local map, containing information beyond his current event horizon, at his disposal would not find his place even though such information should be embedded in the theory.  Smolin has given such relational ideas a name: he baptized these universes as having the Leibniz property by which \emph{I} mean that every point distinguishes itself by the \emph{local} matter and geometry configuration\footnote{One could assume the somewhat weaker notion that two points should distinguish themselves on basis of their past lightcone: this would be the point of view taken by someone adhering to the notion of Einstein causality instead of local causality (meaning that nonlocal signals from the past can reach an observer).  However, I find this notion contrived and my conclusions about the quantum mechanical use of this idea remain identical.}.  However, even Leibniz universes wouldn't save us quantum mechanically as I shall elaborate upon in the next chapter.  \\* \\*
There is still a third class of people who have given up the idea that we live \emph{inside} the universe, instead they live on a conformal boundary which is held fixed (in the variation of Einstein's equations).  One must wonder why these collegues are willing to voluntarily leave us and live on one of the two null boundaries of a topologically trivial observable universe.  As far as I understand, there are several reasons for this : (a) first of all, the asymptotic universe is held fixed even in the quantum world so classical observers survive there without any quantum trouble (b) there exist well defined asymptotic spacetime symmetries (an asymptotic Poincar\'e (semi) group) which allow for unambiguous particle notions  \cite{Szabados1} \cite{Szabados2} (c) the non-dynamical character of asymptotic infinity allows for the existence of a nonzero ADM Hamiltonian (in contrast to what happens in the bulk) (d) the $S$ - matrix philosophy (realistically) merely allows one to define asymptotic particle states.  It may be that I have forgotten another reason, but four of them already seem sufficient.  What is there to argue against such philosophy?  Well if you look at the bulk of the universe like a black box, then you obviously lose \emph{ab initio} all trace of the local and causal relations within the universe (which becomes utterly clear if you stick to a strong form of the holographic principle).  That doesn't mean you will not be able to restore these eventually, but I wish these mighty sailors ``good luck'' in doing so.  But a more trivial remark would be that we live inside the universe and we are as quantum mechanical as anything else.  \\* \\*
Until now, we arrived at the conclusion that keeping up with relativity's philosophy would ``force'' us to live in a Leibniz universe: this would lead to the staggering conclusion that no two electrons are \emph{physically} identical (and one electron would constantly change) which might very well be \emph{logically} possible.   Indeed our current apparati are living on a scale which is about $10^{10}$ larger than those of electrons and could never distinguish any of them, just like the human eye cannot distinguish between ants.  I believe this to be true and will come back to this point in the following chapter.  To appreciate the kind of paradoxes which arise when not living in a Leibniz universe one might consider a universe which had a beginning and two identical humans living in it having identical perceptions up to their clock time $t$; from that moment on the event horizons start to differ.  Suppose that both humans have a map of a chunk of the universe they are actually living in containing information beyond the event horiza at time $t$.  It is clear that up to time $t$, none of these humans could actually predict what their future is (indeed their future would be with probability 1/2 the one of either human) in spite of the fact that they definitely can be identified with one of both observers in the universe.  A good friend of mine does not think this is a problem, I disagree with him.  Given the above, how can these humans speak about particle observations \emph{inside} the universe in the context of general relativity and what type of particles might we expect to see?  This is a very difficult question and one of the main reasons some people are happy to reside at the asymptotic boundary.  This question is identical to finding ``quasi local'' representations of the Poincar\'e group and people have gone through a great deal of pain to give meaning to this.  Let me confess straight forward that I have not followed up the developments on this interesting topic myself so I limit myself here to cite some valuable recourses and some surprising results.  As far as I know, on the notion of quasi local mass and angular momentum, the specialist at hand is the Hungarian relativist Laszlo Szabados and I refer the reader to his Living Reviews article in Relativity \cite{Szabados3} for the main content and further references.  Concerning the type of particle one might find in \emph{classical} relativity, one has of course the well known spin 2 graviton, but also geons \cite{Marolf} \cite{Anderson} and spin 1 particles \cite{Canfora} occur.  \\* \\*
This concludes my \emph{necessarily} somewhat idiosyncratic view of these two jewels of modern physics; I think it is fair to say that up till now I haven't said anything controversial, everything being well documented in the literature.  I have done my best to expose the necessity of at least two types of consciousness in quantum physics as it stands now: a ``quasilocal one'' associated to ``macroscopic'' observers and a global one making sure all quasilocal consciousnesses choose the same physical \emph{reality} (this is not in conflict with the laws of special relativity as one may think at first since these apply to the physical world).  As I have pointed out, relativity either necessitates the introduction of consciouness too (a preferred physical gauge choice) \emph{or} requires something like the Leibniz universe.  In the next chapter I will argue that the second option is not viable: the argument will be lenghty and will consist out of technical arguments as well as metaphysical considerations (where technical arguments cannot settle the matter).  As always, I will start with the former arguments since they distinguish the real metaphysical issues from the unnecessary ones.  Therefore, I believe, it is good for any reader to at least embark on this third chapter; if he or she grows grey hairs from so much metaphysical thoughts he or she may decide for themselves that they want to see the real beef while they prefer to remain somewhat ignorant of the interpretational issues.  This is acceptable and that reader is at this point referred to chapter four; it is not ``allowed'' to go straight ahead to chapter four and conclude that you do not like its \emph{philosophy} while refusing to study chapter three.  That, after all, would not constitute a fair judgement.  With this warning in mind, let me proceed to the third chapter.
\chapter{On the issue of consciousness in physics}  
It is always great ``fun'' when a philosopher of science tries to say something meaningful about physics to professional physicists; likewise I can imagine that the well intended but uneducated metaphysical thoughts of a physicist might cause some mild form of amusement to professionals of the ``other camp''.  It is very likely that many of the arguments I will spell out below have already appeared in journals about metaphysics and it is almost equally likely that rebuttals of all sorts have been construed.  I must confess I am ignorant about this and the lack of resource material below is the consequence of this ignorance supplemented with a manifest lack of time to delve into that unknown literature.  If a philosopher of science would read this book and recognize some of its arguments, this person is more than welcome to inform the author about references and alternative ideas; I will consequently refer to ideas properly and expand upon the arguments.  Nevertheless, I am not a complete ignoramus about metaphysics either; when being a teenager of about 15 years old, I was reading intensely books of Jung on archetypes and books of Freud on ``depth psychology''.  This has somewhat remained with me over the years and it is of some considerable personal joy to see that some of these ideas actually become useful.  The structure of this chapter will be as follows : first I will expose in detail how I see measurement in \emph{my} version of the consciousness interpretation and address at the same time some issues posed by Roger Penrose in this regard \cite{Penrose1}.  Second, I will expand upon the difficulties imposed by quantum gravity : the discussion of that topic will involve some work I have done in the past.  Third, I will try to argue why the very ontology change presented in this book is the \emph{only} reasonable way out: this will involve technical and philosophical arguments presented earlier in the discussion.  The next chapter will then deal with a particular representation of these ideas but I certainly do not claim it is the only possible consistent one!  Comments upon that will follow in chapter six.  \\* \\*
Let me return now to the comments I spelled out regarding the ``standard'' Everett interpretation and give my resolution to (a part of) this problem.  There are, in my mind, four main comments on the Everett interpretation: (a) strictly speaking, there are ``quasilocal'' consciousnesses associated to ``macroscopic'' observers in every world : how do these quasilocal consciousnesses know where they should be in these different worlds?  The only answer this interpretation can give is that God must have put them there; this author believes this gives too much credit to God and I shall solve this issue at the end of this chapter. (b) These ``quasilocal consciousnesses'' can bifurcate into different physical realities ; here the Everettians do not want to involve God or a collapse of the wave function since that would entail some nonlocality which they believe to be in conflict with (special) relativity.  I will argue that such nonlocality is unavoidable but is certainly not in conflict with relativity since only ``mental'' processes are involved here, nothing physical is happening (a priori).  Making a ``genericity'' assumption suffices to have a consistent interpretation without any need for God or a global ``consciousness'' (c)  Only ``living creatures'' have consciousness: I see no reason for this and as suggested before, I believe everything to have a \emph{seed} of consciousness but clearly there must be some relation between the energy scale of the material object and the timescale of awareness so that no contradictions arise.  (d) Somehow God knows about the relationship between the time scale of the observers ``consciousness'' which is a \emph{physical} property and the energy/mass scale of it; again this makes the free will of my ``quasilocal'' consciousness a complete illusion and gives all of it to the Good Lord.  Clearly, as I have emphasized before,  a consistent interpretation of consciousness \emph{requires} a notion of \emph{reality}, such a thing being only present in path integral like formulations.  Indeed, Penrose suggested that the lack of a preferred ontological basis prevented the Everett interpretation from solving the cat problem \cite{Penrose1} p. 807 and therefore this is something which has to be added to quantum mechanics.  However, the underlying reality which exists and of which one is conscious is not one which can be directly accessed through measurements; so the problem Roger was referring to is why we have only conscious experiences of certain types of superpositions of these fundamental realities while we certainly can be conscious about the existence of them separately.  The suggestion I would like to make is that the only superpositions which can be acessed are the eigenstates of the quasi local Hamiltonian of the observer (which can be constructed from the total Hamiltonian by simply ignoring the interactions with all degrees of freedom which do not belong to the material body).  I will comment in chapter eight how this could be done (notice that this is a first imposition of the idea ``mind-matter'' unity: the dynamics of matter is telling you what the ``mind'' can only access).  My comments regarding point (c) also solve a ``problem'' Penrose suggested concerning the non classicality of planets where no conscious beings should live (see \cite{Penrose1} p. 806).  Also, Roger suggests an ``objective reduction'' of the state should take place in the sense that all conscious beings should be in the same physical universe (in other words: to \emph{consciously} agree more or less on outcomes of experiments), I totally agree with him.  Now let me explain why this is not in conflict with the notion of free will, but first let me clarify how I see ``quasi local consciounesses'' make ``quasi local'' measurements.  Let $\Psi$ be the state of the universe written down in the ontological orthogonal \emph{local} basis\footnote{For example in Klein Gordon field theory on finite lattice such a local state might be given by $|a,x \rangle$ where $a$ is a real number indicating the value of the field and $x$ is a spatial point in the lattice; the scalar product is given by $ \langle a,x | b,y \rangle  = \delta(a - b) \delta_{x,y}$ where the first ``delta'' is the Dirac delta ``function'' and the second one is the Kronecker delta.} constituting the realities in the path integral formulation.   The reality as seen by a quasi local conscious observer is not given by $\Psi$ but can be constructed from $\Psi$ by inserting the \emph{quasi local} identity operator written down in terms of the irreducible projection operators coming from the spectral decomposition of the quasi local Hamiltonian.  That is, the ontology for the quasilocal observer is given by $$ \sum_{\alpha \in \sigma ( H_{\textrm{loc}}) } |\chi_{\alpha} \rangle \langle \chi_{\alpha} | \Psi.$$  Notice that performing this change of basis only requires knowledge of $\Psi$ concerning the quasi-local degrees of freedom and nothing more.  However, a slight nonlocality has to enter the argument (but this is much less harmful than in for instance the decoherence interpretation), that is the observer's consciousness has to be aware of the existence of all possible universes to which the \emph{same} quasi local state couples.  It does not however need to be aware of any details of these states, just that they are there, that they are \emph{orthonormal} by construction and the amplitude $\lambda$ which is carried by this state.  This is nothing out of the ordinary since $$ \lambda | \chi_{\alpha} \rangle  | \textrm{pure state for rest of universe in ontological basis} \rangle $$ can be written as $$  ( \lambda | \chi_{\alpha} \rangle ) | \textrm{pure state for rest of universe in ontological variables} \rangle $$ and $$ | \chi_{\alpha} \rangle (\lambda | \textrm{pure state for rest of universe in ontological variables} \rangle ) $$  so the reality of $\lambda$ can shift through any part of the state.   Therefore, the only thing the local observer can be aware of is the probability associated to each state $| \chi_{\alpha} \rangle$, that is $$\sum_{ \textrm{all pure states} \, \beta \, \textrm{for the rest of the universe coupling to} \, \chi_{\alpha} } |\lambda_{\beta} |^2$$ which is nothing but the diagonal of the density matrix in the decoherence interpretation.  There however, the knowledge of the \emph{off} diagonal elements is constrained by more details about the rest of the universe (for example that one pure state for the rest of the universe couples to two different $\chi_{\alpha}$).  There is no way any quasi local observer could be conscious about this, so the density matrix is quite an unrealistic construct.  This gives the quasi-local Born rule for each observer and after observing a $|\chi_{\alpha}\rangle$, the $\Psi$ gets objectively reduced to the renormalized $|\chi_{\alpha} \rangle \langle \chi_{\alpha} | \Psi$ which again can be rewritten in the realist ontology.  The attentive reader could have noticed that the entire construction can be easily adapted to the case where the spectrum of the local Hamiltonian is degenerate.  In general, suppose observer one measures at time $t_1$ and observer two measures at $t_2 > t_1$; the second observer's consciousness is not going to be aware of the damage done by observer one and would feel that it makes an entirely free choice even if the latter really is constrained by the measurement act of the first observer.  Finally, let me comment why this is not in conflict with the notion of free will; it is clear that the measurement act of different observers can only affect that part of the brain which deals with external sensory impulses.  The other part, which is connected to self-awareness and the internal thought process remains largely unaffected so every observer's consciouness still has a large portion of free will left even though the sensory impulses about the outside world are constrained.  Let me now turn to the problems imposed by quantum gravity. \\* \\*
One might fear at first that in the multiverse notions of locality and causality do get screwed up since it is possible to imagine situations with multiple identical observers which could only be distinguished by God through their relations with the environment.  However, this is only \emph{one} time a problem and once a collapse to a classical world at ``macroscopic level'' has been made, all identical observers cannot be confused anymore by Cournot's principle and they happily satisfy quasi local laws (on the scale set by the human eye).  Quantum gravity however adds an important twist to this and the problem resides in what is usually meant with background independence (or equivalently quantum diffeomorphism invariance).  I shall come to the conclusion that background independence cannot be upheld at the quantum level but \emph{must} be a property of the classical limit; an emergent symmetry as to speak.  This idea has been suggested before by amongst others 't Hooft \cite{'t Hooft5}.  This fact might have been obvious to certain string theorists, but I shall present the short argumentation in order not to insult my background independent friends.  Moreover, I am definetly not happy with the kind of background dependence string theory suggests, since this corresponds again to a preferred frame for some classical observer and I shall dispose of this artifact later on.  I have already discussed certain ``consciousness'' issues which arise in general relativity when not restricting to Leibniz universes; the trouble however now is that one certainly cannot restrict to Leibniz universes when taking the path integral since this would violate the Markov property of the transfer matrix\footnote{In the strong form where points must distinguish themselves locally, this is obvious because ``gluing'' two such universes can destroy this property.  In the weaker version, the problem is that -although the gluing of two Leibniz universes is certainly again a Leibniz universe- not \emph{all} Leibniz universes can be obtained from such gluings.} (one does not necessarily have to sum over all possible universes but the ``gluing conditions'' certainly must be consistent).  Therefore, the idea of Leibniz universes is worthless at the ``quantum'' level where the reader will notice that I interpreted quantum in a broader sense since the strict quantum physicist would say we have to sum over them all (while I consider the notion of ``consistent classes'' as the only restriction as long as such restriction is \emph{physically} motivated).  This broader view on the path integral is for example also assumed in the framework of causal dynamical triangulations (where each universe has a preferred kinematical time coordinate).  Consider first a fixed (say without any symmetries) classical spacetime with some matter and gauge fields on it and perform the path integral with respect to those fields (we assume that the full path integral - including gravitational fields - can be split in this way, which always is the case if you would consider dynamical lattice theories with a finite cutoff and fields on the vertices and edges of the lattice).  Now, within this fixed spacetime ``a point'' has an \emph{absolute} meaning since we assumed no Killing fields to exist and the quantum dynamics certainly depends upon the causal relations of that point with other points in the rest of the universe.  Therefore, if one would consider shifting points around in this spacetime and compute observables taking in account the shifting, one would have to be extremely careful in doing so since moving points around in a rather wild fashion would destroy completely causality and ``localizability''.  I prefer the use here of ``localizability'' over locality since strictly speaking locality is never lost since it only deals with infinitesimal relationships between points; localizability however deals with metric relations between two points and they actually form the basis with respect to which we observe the world.  Making points ``fuzzy'' in this way actually smears out a bit the light cone and would allow for mild violations of causality.  Now, we are going to do the gravitational path integral - actually we only need to sum over three different spacetimes to understand where the trouble is - in a \emph{background independent} way.  Of course, even background independent physicists do not negate the importance of ``localizability'' (which is a dynamical notion) and causality.  Imposing a gauge condition and calculating \emph{gauge dependent} variables (that is \emph{partial observables}) might solve our problem although gravity has the nasty habit of not letting itself be gauge fixed.  The philosophy behind this is rather doubtful too: you start from a diffeomorphism invariant theory, break it in an arbitrary fashion and then assign physical meaning to this procedure.  It would be much neater if this breaking of diffeomorphism invariance would be spontanious in the sense of Higgs fields performing the crime, see 't Hooft \cite{'t Hooft6} for an interesting suggestion.  However, this is not what our background independent friends have in mind, they genuinely think that localizability and causality can be restored without appeal to anything of the kind I mentioned before.  Let me give an argument why this will not work: consider three \emph{generic} spacetimes $\mathcal{S}_i$ of the same topology (to facilitate the argument).  The task at hand is to \emph{identify} events in different space-times in a \emph{physical} way - that means the identification only depends upon the geometries $\mathcal{S}_i$ and not on some gauge.  Of course this is a terribly non-local question which is \emph{not} computable (for continuum spacetimes) so such identification would be in God's hands.  Let me first comment that while you can construct such criteria mathematically \cite{Noldus1} \cite{Noldus2} \cite{Noldus3}, the identification itself, is generally not known and if it would exist, it would be generically not unique.  However, let us not be frightened by this and suppose there would exist such unique diffeomorphism between any two spacetimes (one can relax the diffeomorphism criterion but it would only make things worse, not better): that is, we can construct diffeomorphisms $\psi_{12}, \psi_{13}$ and $\psi_{23}$ where for example $\psi_{21} = \psi_{12}^{-1}$.  Now, the identification criterion must satifisfy symmetry and transitivity otherwise it would single out one or more backgrounds.  The problem arises from the fact that generically the cycle $\psi \equiv \psi_{31} \psi_{23} \psi_{12}$ is not the identity diffeomorphism.  Applying the above criteria consistently implies that one has an infinite number of identifications on $\mathcal{S}_1$ given by $\psi^n$ where $n$ is an integer number.  As if this were not bad enough, this pattern grows at least exponentially in complexity in terms of the number of spacetimes considered.  Hence, it is quite reasonable to assume that in the end all possible identifications within $\mathcal{S}_1$ have to be made screwing totally localizability and causality.  Within theories such as the causal dynamical triangulations approach, the situation appears to be a bit better since one disposes of a prefferred slicing, but the above argument works as devastating there as it does in the more generic case.  Therefore, researchers (in that particular approach amongst others) have voiced the opinion that \emph{pure} gravity will be diffeomorphism invariant (at least with respect to the spatial diffeomorphisms) but once you include matter this invariance is broken.  This is very unsatisfying for at least two reasons : (a) no mechanism for breaking diffeomorphism invariance is presented but merely a kinematical labeling is chosen by hand (b) a unified theory should not make a distinction between gravitational degrees of freedom and other stuff floating around in the universe, at least string theory satisfies that criterion.  Sometimes, I jokingly ask to some loopy friends if they already made sense out of a superposition of spin network states: I am afraid they never will -at least not without breaking diffeomorphism invariance.  \\* \\*
Philosophically, the drama is complete: while God still had at least a theory at his disposal to put the observer's consciousness in different worlds in Quantum Field Theory on a fixed background, here he must proceed by ``random'' identifications.  This is simply unacceptable for a physicist and it leads me to the conclusion that background dependence must be an ingredient of ``quantum gravity''.  All arguments given up till now have a common denominator causing all the trouble: that is, as well quantum theory as general relativity take the point of view of God and are not in any sense theories for ``conscious'' beings living \emph{inside} the universe.  I will show in the next chapter that \emph{assuming} this metaphysical insight from the beginning allows one to construct a class of theories which has apparently no problems with localizability and causality at all; God is expelled from this worldview and I side completely with Pierre-Simon Laplace on this issue.  We proceed by explaining the metaphysical input behind this ``new view on physics''. \\* \\*
The theory I will start to explain now is the simplest of its kind and the reader who is eager to object and propose generalizations will find what he or she is looking for in chapter eight.  At all steps, I will implement the idea of ``mind-matter'' unity which should be at the core of any theory for ``beings'' inside the universe.  As 't Hooft has repeatedly argued, no generally covariant theory can single out Minkowski as its vacuum state and the point of view taken here is that Minkowski is not the physical vacuum (which it cannot be due to geometric vacuum fluctuations\footnote{It is also in a sense the physical vacuum because the vacuum fluctuations are defined with respect to it.}) but it is the ``mental'' vacuum.  This idea is certainly not new and indeed the suggestion that geometry is all in the mind has been made by -amongst others- Lasenby, Doran and Gull \cite{Lasenby} who constructed a new gauge theory of gravity starting from Minkowski.  I remember being impressed with this idea when I was a young PhD student; unfortunately my promoter at that time, a hard core relativist, was less so.  Since the term ``mental'' vacuum must sound a bit weird, let me explain precisely what I mean by it and how the idea of ``mind-matter'' unity is incorporated in it.  Primary to any theory is the notion of space and time, but neither have to be thought of as physical space and time but as a Platonic notion which carries in itself the \emph{potentiality} for dynamical space and time to arise.  Space is filled with points (labbeled by coordinates $x^{i}$), which we shall call monads; actually the primary notion in empty space-time is not given by the monads but by their worldlines (given by the time lines of constant $x^{i}$).  Why is this so?  In order not to construct an eather theory, these monads cannot carry any energy (the entropy -as the temperature- of the vacuum is exactly zero) so they cannot be ``conscious'' of time also.  Moreover, they have no consciousness at all regarding the \emph{other} monads and are therefore timeless and spaceless (notice the ``mind-matter'' unity here), the only thing a monad is ``aware'' of is its identity (which can be encoded in the theory by its coordinates).  Nevertheless, a notion of time is necessary to create the potentiality for time to arise dynamically.  The idea is that notions of space and time arise dynamically due to \emph{relative} changes in the relations between the monads.  Logically, this requires one to add Platonic relations between the atoms even if they are not aware of them in the ``mental'' vacuum state.  There is \emph{no} choice in the relations one can impose since the potentiality for each monad at every moment in time ``t'' must be the same and therefore no a priori direction should exist in space (since that would favour potential excitations between monads in that direction).  That is, space must be homogeneous and isotropic and time invariant.  Moreover, space and time should be decoupled from one and another meaning that the $dx^i dt$ terms in the metric must vanish.  Also, time must be linear otherwise the notion of ``change in time'' would not be time translation invariant - which cannot be for the vacuum state (again, a nonlinear time could and, generically, will arise dynamically).  This leaves us with two scale factors: one overall conformal constant (which induces a renormalization of the coupling constants) and one ``velocity'' to be freely chosen (the metric cannot be Euclidean since there would be no distinction between space and time).  Personally, this derivation is much easier and certainly more convincing than a more traditional one which requires more assumptions \cite{Leblond}.  Our monads are what I would call -to use a term invented by Karel Kuchar- perennials: they exist forever and cannot be destroyed or created.  This leads us to a philosophy which is closer to the ``multi-particle'' wave function, than the one of quantum field theory as discussed in the previous chapter.  Notice that there is no conflict between this point of view and the possibility for particle creation and annihilation in quantum field theory since what we call particles are collective excitations of the ``monads of space'' which clearly can be created as well as die out (the ``perennials'' in Quantum Field Theory are the space-coordinates).  My view is that the more complex the pattern of excitations becomes, the more complex the (quasi-local) notion of consciousness grows.  This requires a theory of consciousness which runs parallel to the material world, contradictory to the view of Penrose \cite{Penrose2} who regards consciousness as an emergent property of a theory of quantum gravity. \\* \\*
Since the dynamics is about a change in relations between the monads of space (inducing mass, energy, dynamical space-time and consciousness) there is no problem with a multiverse whatsoever.  Actually, \emph{one} monad of space will not feel it is in a multiverse at all; it could however \emph{potentially} (but no such thing will happen since one monad doesn't carry any energy) see the other monads in a superposition \emph{relative} to itself.  This is an entirely democratic view and doesn't distinguish anything and/or anybody.  ``Macroscopic'' objects (an excitation of certain monads) will ``feel'' nevertheless that they are in different universes since their \emph{internal} state can be in a superposition too; this clearly requires a higher notion of awareness and the necessity for a thing of this kind is a guideline for the construction of a theory of consciousness.  Does this mean that a ``macroscopic'' observer can decipher the internal code of say an electron?  Although this question can be asked in principle within this construction, I think it is unlikely that the eigenstates of the quasi-local Hamiltonian of the observers' body (brain) will contain any such information.  At this point, it is also clear what I meant with the quasi-local Hamiltonian, since this one is now expressed in terms of the monads of space constituting the observer.  Let me now discuss the issues of localizability and causality in this framework.  I think localizability is limited to the extend of the observer's (note that our observers don't have to be macroscopic per se) body, the outside world being encoded in the quantum state of the observer's brain\footnote{The way a consciousness can be aware of a body being localized despite of the fact that it occurs in a superposition would be by having insight into the relations between the constituting atoms in different universes.  The \emph{statistics} of these relations should define a notion of localizability in the multiverse and vice versa, this notion of localizability should have an impact on the consciousness.}.   That is certainly consistent with the fact that we can never ask the code of the observed phenomena but merely the physical characteristics of it.  As far as causality is concerned, the reader will learn in sections four and five that causality is a dynamical property (even though we start out from Minkowski) and by definition our \emph{physical} particles always travel on timelike or null curves.  A nonzero space-time curvature will also emerge dynamically suggesting that Einstein's theory of relativity may be present and I actually will \emph{prove} that quantized Einstein Cartan theory is a part of our theory - in a most unexpected way.  I think it is utterly clear that the viewpoint enunciated above is very different from as well the relativist's as particle physicist's view on Minkowski and therefore the dynamics will also deviate from those theories.  Now, we will show that this is indeed the case.
\chapter{A new kind of dynamics: a prelude} 
Technically, we start from Minkowski spacetime and see dynamics as a change in relations between the monads thereof.  Thinking in terms of a path integral, all such changes in relations should satisfy a few criteria: (a) it is almost everywhere differentiable (asking it to be differentiable everywhere is in conflict with the Markov property of the transfer matrix) (b) it must respect the ``mental'' notion of space and preserve the ``mental'' space-time volume and orientation (cfr. unimodular gravity).  In other words, we have an almost everywhere differentiable homeomorphism $X^{\mu}(x^{\alpha},t)$ such that $$ \eta_{\mu \nu} \partial_{\alpha} X^{\mu} \partial_{\beta}X^{\nu} $$ is a metric of Lorentzian signature.  Actually, since only the \emph{physical} properties of Minkowski spacetime count -and not a particular embedding- we work with the Poincar\'e equivalence classes $[ X^{\mu}(x^{\alpha},t) ]$.  So, instead of having a unitary representation of the Poincar\'e semi-group on the Hilbert space spanned by distributional states associated to spacelike embeddings $| X^{\mu}(x^{\alpha}) \rangle$ defined by $U(\Lambda, a)| X^{\mu}(x^{\alpha}) \rangle  = | \Lambda^{\mu}_{\nu} X^{\nu}(x^{\alpha}) + a^{\mu} \rangle$, we regard all these states as being the same one (where $\Lambda$ is an ortochronous Lorentz transformation).  This allows for much more interference since the norm of $$\lambda | X^{\mu}(x^{\alpha}) \rangle + \kappa | X'^{\mu}(x^{\alpha}) \rangle $$ is generally $(|\lambda|^2 + |\kappa|^2) \infty $ while -when X and X' belong to the same equivalence class- this becomes $ (|\lambda|^2 + |\kappa|^2 + 2 Re ( \overline{\lambda} \kappa ) ) \infty$.  We demanded that the space-time orientation should be preserved which breaks time $T$ and space $S$ reversal (of course, time and space reversal won't be broken a priori in terms of the labels of our space-time atoms) but preserves $ST$.  Actually, this is not yet sufficient but its origin can be traced back to problems which arise when one must glue the future boundary of a ``cobordism'' to the past boundary of a ``cobordism'' running backwards in time.  The result is of course not a homeomorphism anymore, however to fully exclude such possibility one must also break $ST$; therefore one demands that $$ \partial_t X^{\mu} (x^{\alpha},t) n_{\mu}(X^{\nu} (x^{\alpha},t)) < 0$$ where $ n^{\mu}(X^{\nu} (x^{\alpha},t))$ is the future pointing unit normal\footnote{An explicit formula for $n^{\mu}$ is given by the normalization of $ - \epsilon^{\mu}_{\nu \alpha \beta} \partial_1 X^{\nu} \partial_2 X^{\alpha} \partial_3 X^{\beta}$ satisfying $\epsilon_{T \nu \alpha \beta} \partial_1 X^{\nu} \partial_2 X^{\alpha} \partial_3 X^{\beta} > 0$.} to the hypersurface of constant $t$.  This induces also an orientation on ``space'' and the above requirement provides a space-orientation preserving homeomorphism.  Now, it is easy to see that everything is consistent and the gluing of two equivalence classes of ``cobordisms'' provides a unique equivalence class.  By a ``cobordism'' I mean a mapping $X^{\mu}(x^{\alpha},t)$ restricted to some time interval $\mathbb{R}^3 \times [t_0, t_1]$ with $t_1 > t_0$ and gluing happens between some $[ X^{\mu}(x^{\alpha},t), \mathbb{R}^3 \times [t_0 , t_1] ]$ and $[ X'^{\mu}(x^{\alpha},t), \mathbb{R}^3 \times [t_1 , t_2] ]$ in case $[ X^{\mu}(x^{\alpha},t_1)] = [ X'^{\mu}(x^{\alpha},t_1)]$.  Another way of saying what we are doing so far is that the representation of the Poincar\'e (semi) group on target space is trivial; that is all physical states are manifestly Poincar\'e and $S,T$ and $ST$ invariant.  \\* \\*
Now, it is probably impossible to write down a classical dynamics such that above constraints are preserved under the equations of motion, so we must keep in mind that the classical starting point which we shall assume is not the ``classical limit'' of the quantum theory; the latter will probably be much more complicated.  It is a good exercise to figure out what the constraints do for the easiest theory one could imagine.  The latter is given by the action:
$$ - \alpha' \int d^{D+1}x \, \partial_{\alpha} X^{\mu} \partial_{\beta} X^{\nu} \eta^{\alpha \beta} \eta_{\mu \nu} $$ which we shall study in $1+1$ dimensions (since the analysis simplifies considerably there).  This action has a remarkable property regarding infinitesimal \emph{unimodular} perturbations of the ``mental'' frame 
$$X^{\mu}  = x^{\mu} + \epsilon f^{\mu}(x^{\nu})$$ where $\epsilon$ is an infinitesimal number and we slightly abused notation by identifying the Lorentz indices of the ``relational'' and ``mental'' Lorentz group.  Not only remains the action stationary under such perturbations, but also the action \emph{density} does.  This may be considered a sign of stability of the vacuum.  Indeed, the volume constraint requires that $\partial_{\mu} f^{\mu}  = 0$ while the perturbation on the action density transforms\footnote{The following actually reveals that the bare cosmological constant has to be set to the value $\alpha' (D+1)$.} as $$ - \alpha' (D+1) \rightarrow -\alpha' (D+1 + 2 \epsilon \partial_{\mu} f^{\mu}).$$  Let me stress an undesirable feature of the standard quantization procedure which is immediately clear.  I have argued above that only Poincar\'e invariant (with respect to the ``relational'' group) states and operators have a physical meaning while the ``ordinary'' momentum $P_{\mu} = 2 \alpha' \partial_t X_{\mu}$ transforms \emph{covariantly} under these Poincar\'e transformations.  Moreover, we stressed that our monads should be seen as particles and not fields, so one would expect the momentum to be Lorentz covariant under the ``mental'' Lorentz transformations; indeed, this observation will serve as a guideline for our alternative ``quantization'' procedure.  \\* \\*    
Obviously, the constraints break the classical superposition principle and the space of solutions splits into ten sectors which ,quantum mechanically, can live together happily due to non-commutativity.  Moreover, the particle interpretation we shall adhere to is very different from the one suggested by string theory, we will jump a bit ahead in time and enunciate our novel point of view here (this paragraph is merely a lengthy introduction to motivate where we go to).  Up till now, I have said that ``macroscopic'' localized (in the space of monads) configurations have the ability to grow in consciousness.  Now, by this, I do not want to say at all that a macroscopic configuration cannot ``travel'' on the monads, that is change the internal constitution of its labeling and, while doing so, all alternate possibilities are immediately eliminated.  However, elementary particles such as electrons do not have this property (and it is here that I propose a radical reinterpretation); that is, once an electron starts bifurcating into different universes no ``identity information'' is transferred (and therefore no collapse of the wavefunction happens), all these \emph{new} electrons being carried by different monads are genuine different identities.  That is, electrons do not ``travel'', they constantly get annihilated and created which is a rather logical point of view since the correlations between the different universes where ``one electron'' can go to are infinitely weaker than the correlations between the different universes for macroscopic observers\footnote{To put it clear; electrons carried by some monads do not develop consciousness since the same monads are not excited in almost all different universes (!).  It is precisely this ``consistency'' through different universes which allows for consciousness to grow.  A ``localized'' electron however (by means of some electromagnetic field) can develop some ``awareness''.}.  When the observer's consciousness will collapse his brain observing an alternative of ``macroscopic'' states, the infinity  (which do not correlate to this brain wave) of other electrons (in different universes) disappears too.  So, the traditional point of view, which tells you that the \emph{same} electron is traveling different paths cannot be upheld, since it would effectively have to be conscious about them all which contradicts everything we said up till now. \\* \\*  Let us continue by examining the classical constrained theory in $1+1$ dimensions.  Although $1+1$ dimensions allows for explicit calculations, it is also a rather peculiar dimension for our proposal.  Indeed, the volume constraint will break the space reversal symmetry which in $1+1$ dimensions coincides with the isotropy of space.  Therefore, the classical theory will break into two sectors corresponding to left and right moving waves (which does not occur in higher dimensions).  Comparing the ``germs'' of our theory with Polyakov theory, one is struck by two differences: (a) in Polyakov theory, background independence of the ``mental'' frame (taken together with conformal invariance) produces two local independent hard constraints, while here only one hard constraint is present (the volume constraint) (b) we have (so far) two soft constraints (orientability and spacelike character) which will cause a division in parameter space but not eliminate any local degrees of freedom.  We shall first work out the premises of our construction as they are stated so far and examine their physical properties.  Next we work out the Virasoro constraints nonperturbatively (I am unaware of such analysis in the literature), compare both results and possibly suggest improvements. \\* \\*
It is well known \cite{Witten} that the unconstrained solutions to the massless d'Alembert equation in $1+1$ dimensions can be written as
$$ X^{\mu}(t,x) = f^{\mu}(t+x) + g^{\mu}(t-x)$$  which automatically precludes the existence of bound states (but they will appear in $3+1$ dimensions).  The volume constraint $$1 = \partial_{[1} X^1 \partial_{2 ]} X^2 $$ becomes $$\frac{1}{2} = f'^2(t+x) \, g'^1(t-x) - f'^1(t+x)g'^2(t-x).$$  Further analysis reveals there are essentially two different cases and the solutions are given by:   
\begin{eqnarray*}
X^1(t,x) & = & f(t + x) + \frac{1}{2 \kappa} (t - x) \\*
X^2(t,x) & = & a + \gamma X^1(t,x) + \kappa (t + x) \\*
\end{eqnarray*} or
\begin{eqnarray*}
X^1(t,x) & = & g(t - x) - \frac{1}{2 \kappa} (t + x) \\*
X^2(t,x) & = & b + \gamma X^1(t,x) + \kappa (t - x) \\*
\end{eqnarray*} where $a,b, \kappa, \gamma$ are constants and $f$,$g$ any functions\footnote{The reader should notice the mild breaking of space reversal invariance $x \rightarrow -x$ in the ``mental'' frame.}.  Notice that at this point, we have effectively eliminated a left and right moving ``polarization degree'' of freedom when comparing the constrained solution space to the full solution space of the d'Alembertian equation.  It is instructive to realize that the Virasoro constraint $\partial_t X^{\mu} \partial_x X_{\mu} = 0$ on the left moving sector leads to $\kappa = 0$ and $\gamma = \pm 1$ (the critical points of our theory), which is excluded from our solution space as we shall see in a moment.  It remains to impose the spatial character of slices of constant $t$ as well as time orientability; that is, $\partial_x X^2(t,x) > 0$ and $$\partial_x X^{\mu}(t,x) \partial_x X_{\mu}(t,x) > 0$$ (we shall only examine the ``left moving sector'', leaving the other one to the reader).  There are several distinctions one must make, based upon the value of $\gamma$.  If $\gamma > 1$, then both conditions give $$f'(t+x) > \max \{ \frac{1}{2 \kappa} - \frac{\kappa}{(\gamma - 1)}, \frac{1}{2 \kappa} - \frac{\kappa}{(\gamma + 1)} \}.$$  $\gamma = 1$ is a critical point, since then $$f'(t+x) > \frac{1}{2 \kappa} - \frac{\kappa}{2} \, \textrm{and} \, \kappa > 0.$$  For $1 > \gamma > -1$, one obtains that $\kappa > 0$ and $$\frac{1}{2 \kappa} - \frac{\kappa}{(1 + \gamma)} < f'(t+x) < \frac{1}{2 \kappa} + \frac{\kappa}{(1 - \gamma)}.$$   $\gamma = -1$ is another critical point and the restrictions are $\kappa > 0$ and $$f'(t+x) < \frac{1}{2 \kappa} + \frac{\kappa}{2}.$$  Finally for $\gamma < - 1$, the relations are $$f'(t+x) < \min \{ \frac{1}{2 \kappa} + \frac{\kappa}{(1 - \gamma)},  \frac{1}{2 \kappa} + \frac{\kappa}{(- 1 - \gamma)} \}.$$  Therefore, in total, we have ten distinct classical sectors, at the transition between sectors a discontinuity occurs which can only be cured in the quantum theory. \\* \\*
Having arrived at this point, it is now opportune to spell out some remarks regarding the notion of causality.  From the classical string point of view, it is by no means guaranteed that the points of constant $x$ must move on timelike curves (and they generically won't).  Everything the classical string point of view is concerned about is the center of mass momentum and even that one is not always timelike (as is well known, the closed string theory contains tachyons).  Now one can wonder whether \emph{imposing} such constraint upon the theory will solve the causality problem in my approach (classically).  The answer is that it will by any reasonable definition of a particle, and we shall illustrate this at the end of this chapter.  However, in classical relativity where one would associate particles to ``distinguished'' geometrical excitations such as geons, this is not guaranteed to be the case.  It is by no means so that a well defined center of mass should move on timelike curves although one would suspect it to be the case for ``macroscopic'' bodies (references and progress in the literature will be discussed later).  The latter concern is valid for any approach which considers the same dynamical variables to cause a dynamical geometry as well as a particle spectrum.  However, even if one ``observer'' would move on a spacelike curve, there is not necessarily yet an operational problem of causality and the whole issue entangles with what one means by ``quasi-local observations''.  To appreciate this, consider Minkowski spacetime in $1+1$ dimensions and two observers; number one moving on $x=0$ to the future and another one moving to the right on $t=0$.  Moreover, we assume they can communicate by means of null particles (photons).  Observer one will have a local frame at his disposal $\partial_t, \partial_x$ where $\partial_t$ is interpreted as time and $\partial_x$ as space.  For the second observer, the role of both vectors is switched (actually, his notion of space will coincide with $- \partial_t$) and the only \emph{mathematical} curiosity which occurs is that his spacelike vector will have a negative ``norm'' and his time vector a positive one.  It is clear that $1$ can only communicate with $2$ up to $t=0$ and he will receive responses only after that moment.  The funny thing however is that the order in which he shall receive responses is exactly the reversal of the order in which he has sent the messages (so who comes last gets first served).  Anyway suppose $1$ sends prior to $t = 0$ a message to $2$ given by the vector $(k,\pm k)$ where $k > 0$, then observer $2$ will perceive this as an incoming photon with energy $\pm k$ and momentum $k$ (in the Dirac interpretation, he might perceive a negative energy photon as a hole in the sea of photons), but there is no contradiction whatsoever since the photon will still be perceived to move with the same speed of light as it does for observer $1$; the signature of the metric is merely a theoretical convention which cannot have any operational consequence - after all we do not say either that the time interval between two events is negative.  I realize that this ``simplistic'' reasoning is only valid in $1+1$ dimensions and one would naively expect observer $2$ to see ``tachyons'' in higher dimensions.  I stress naively since I believe this not to be true at all; certainly not for a timelike observer receiving ``tachyons''.  The dynamical picture I have in mind is the following: with a high probability (in the multiverse sense) the tachyon will termalize with the apparatus causing a ``macroscopic'' response within the apparatus (human).  I conjecture that the eigenstates of the quasi local Hamiltonian will only contain information regarding the center of mass motion of these flows and some other coarse grained properties (which certainly doesn't mean that our brain is a simple thing since there are zillions of neurons firing classically at the same time).  The latter will almost certainly be timelike and therefore tachyonic particles are never observed even if they exist.  Anyway, I just wanted to say the problem is much more difficult than is usually thought. \\* \\*
We now return to standard string theory and solve the Virasoro constraints classically.  After having done that, we will try to understand the causality problem in both approaches.  Obviously, the Virasoro constraints will not impose any asymmetry between the left moving and right moving sectors and the resulting conditions are
\begin{eqnarray*} (f'^{1}(t+x))^2 & = & (f'^2(t+x))^2 \\*
(g'^{1}(t-x))^2 & = & (g'^{2}(t-x))^2 \\*
\end{eqnarray*}
As said before, this puts $\gamma = \pm 1$ and $\kappa = 0$.  Restricting ourselves to the sector where globally
\begin{eqnarray*} f'^{2}(t+x) & = & \alpha f'^1(t+x) \\*
g'^{2}(t-x) & = & \beta g'^{1}(t-x) \\*
\end{eqnarray*}             
with $\alpha, \beta = \pm 1$ (one could still identify the ``linear momenta'' as happens in open string theory by allowing for nontrivial variations of the linear part of $f$ and $g$ at infinity).  However in the bulk, one has an energy given\footnote{We put here $\alpha' = \frac{1}{2}$ and define the energy observed by an observer in \emph{target} space $T^{\mu}$ as $- P_{\mu} T^{\mu}$.  The momentum, with respect to the spatial part of the tetrad $E_i$ defined by that observer is given by $P_{\mu} E_i^{\mu}$.} by $f'(t+x) + g'(t-x)$, momentum $\alpha f'(t+x) + \beta g'(t-x)$ and the ``norm squared'' is given by
$$ \partial_t X^{\mu} \partial_t X_{\mu} = 2 (\alpha \beta - 1) f'(t+x)g'(t-x).$$
Trivially, if $\alpha = \beta$, only null particles are allowed with positive and/or negative energy.  For $\alpha \beta = -1$, as well tachyons as massive particles exist with any energy.  Within this context, four different sectors exist classically (strictly speaking there is an infinite number of them since at each point where -say- $f' = 0 = f''$, $\alpha$ could switch sign)\footnote{The reader notices that also here, the superposition principle is broken.}.  In our constrained ansatz one obtains that for $\kappa > 0$, the solutions with positive energy and timelike momentum can reside in each sector for $\gamma < 1$.  The conditions for $-1 < \gamma < 1$ are $$\frac{1}{2 \kappa} + \frac{\kappa}{(1 - \gamma)} > f'(t+x) > \max \{ - \frac{1}{2 \kappa} + \frac{\kappa}{(1 - \gamma)},  \frac{1}{2 \kappa} - \frac{\kappa}{(1 + \gamma)} \}$$ and for $\gamma = -1$ one obtains $$ - \frac{\kappa}{2} < f'(t+x) < \frac{1}{2 \kappa} + \frac{\kappa}{2}$$  The reader can work out the case $\gamma < -1$.  Let me stress that the previous discussion took the point of view of string theory with an observer living in target space.  This point of view is entirely ``wrong'', the observer as well as all other particles resides in ``mental'' space.  To appreciate this, the reader must realize that the action we started from has \emph{two} Lorentz groups, one on ``mental'' space and another one on target space.  At the beginning of this chapter, we have killed of the latter one by demanding that its representations on Hilbert space were trivial: only Poincar\'e invariant properties on target space are allowed for.  This implies that the correct momentum has to be calculated by applying Noether's theorem on mental space.  Before I address this issue in greater depth, let us see what it gives for our particular approach and what this has to do with the operational notion of causality I have spoken about before.  The energy momentum tensor on mental space is given by\footnote{Note that the latter is constrained to vanish in standard string theory.}: 
$$T_{\alpha \beta} = \partial_{\alpha} X^{\mu} \partial_{\beta} X_{\mu} - \frac{1}{2} \eta_{\alpha \beta} \partial_{\gamma} X^{\mu} \partial^{\gamma} X_{\mu} + \frac{(D - 1)}{2} \eta_{\alpha \beta} $$ and ``the'' spin tensor by 
$$S^{\kappa \gamma}_{\alpha} = \partial_{\alpha}X^{\mu} \partial^{[ \kappa} X_{\mu} x^{\gamma ]} - \frac{1}{2} \delta_{\alpha}^{[ \kappa} x^{\gamma ]} \partial_{\beta} X^{\mu} \partial^{\beta} X_{\mu} + \frac{(D-1)}{2} \delta_{\alpha}^{[ \kappa} x^{\gamma ]}.$$  Note that the ``constant'' terms at the end of each expression make the energy momentum and spin vanish for inertial embeddings as is mandatory.  Of course, the spin tensor depends upon the origin of our coordinates - as it always does; therefore this cannot be the physical spin tensor and we shall solve this issue later on.  The energy-momentum vector does not suffer from such problem and is given (in $1+1$ dimensions) by:
\begin{eqnarray*}
P_x & = & \partial_{t} X^{\mu} \partial_{x} X_{\mu} \\*
P_t & = & \frac{1}{2} \left( \partial_{t} X^{\mu} \partial_{t} X_{\mu} + \partial_{x} X^{\mu} \partial_{x} X_{\mu} \right)
\end{eqnarray*} and their integral over space is conserved in time.  What do we have so far?  We have a classical theory of excitations of monads.  The $X^{\mu}$ are to be thought of as hidden variables since observables are non-linear, Poincar\'e invariant (on target space) functions of them.  The monads of space certainly get excited since relations between them change and they can acquire a dynamical energy, momentum and angular momentum.  However, since no nontrivial space-time curvature is present, no physical particles can be present.  We have discussed the problem of how to define particles in a relational context previously and now we shall solve this question.  Let me stress that in general relativity no canonical procedure exists; here, the ``mental'' frame comes to our rescue.  What I want to say is that the Einstein tensor \emph{defines} particles up to a constant (which depends on how we define the notion of mass relative to our geometrical units).  In other words, the Einstein equations are true by definition (it is a tautology), there is nothing to prove in a theory which studies relations between fundamental monads.  How is this so?  As I have explained several times, excitations of our monads cannot serve as a definition for a particle, so it appears that we have used up all conservation laws in our theory.  Since particles are composites of the geometry defined by our monads and we want conservation laws for some particle energy momentum tensor, our only option is to construct this object bottoms up from the dynamical spacetime metric.  It is here that general covariance comes into play and the easiest such tensor is the Einstein tensor with a cosmological constant.  As I said, classically nothing happens in our theory; non-trivial spacetime curvature and therefore quantum particles are purely quantum mechanical effects which are due to the commutation relations between the $g_{0\alpha}$ and spatial geometry $g_{kl}$ induced by our hidden variables.  This is an example where quantum mechanics \emph{generates} local degrees of freedom which are not present in the classical description.  Indeed, non-commutativity does not allow for a simple counting of degrees of freedom as occurs in the commutative case: in such a theory, mass and geometry literally arise out of nothing. \\* \\*  One can certainly define a suitable Einstein tensor quantum mechanically, but it is unreasonable to expect that the Bianchi identities will hold at that level; they must emerge in some classical limit.  The form of general covariance we have is \emph{classical} even for the quantum operators; there is no such thing as quantum diffeomorphism invariance as explained before.  Of course, it remains to \emph{prove} that particles defined as such will satisfy reasonable energy conditions (at least in some classical limit) and obey a satisfactory notion of operational causality.  Also, macroscopic objects of the size of comets, planets and stars should move more or less on geodesics of the dynamical spacetime metric.  These questions are still subject of study within the context of general relativity and recent work by, amongs others, Bob Wald has been performed on these issues.  A part of the causality and energy problem could be solved ``by hand'' for microscopic particles by imposing the \emph{dominant} energy condition\footnote{Meaning that $-T^{\alpha \beta} v_{\beta}$ is a future pointing timelike vector for any timelike vector $v^{\alpha}$.}  (so one could try to prove if this one holds for our particles).  The difficulty of the problem that planets should move on geodesics is of an entirely different order, but I believe it to be true (if one defines planets purely within the context of general relativity, I would think they either become unstable after a relatively small timescale or collapse to a black hole - at least this is what appears to happen to geons\footnote{This shouldn't come as a surprise to anyone, we know for a century by now that stability of matter is a quantum mechanical property.} \cite{Cooperstock} \cite{Thorne}).  Geodesics in a sense are the path of least resistance through the geometry; now, I think it must be a deep hidden property of the Einstein tensor that large scale geometric excitations are following this path, this is a majestic back-coupling of the geometry to itself indeed.  It is also exactly what happens in thermodynamics of real life phenomena: currents inside a gas move into the direction where the density of the gas is the lowest (hence, the amount of collisions is minimized which is the way of least resistance), individual particles certainly don't.  Recently, people have been working however on a logical ``converse'' of this question: if we start from the action of a free particle and take into account gravitational self effects, is the particle still going to move on a geodesic and obey causal laws with respect to the full dynamical metric?  The answer to both questions appears to be no \cite{Wald3} although deviations from the geodesic path are expected to be small.  On the other hand, our questions could receive a better quantum mechanical answer than classical physics might be able to provide; this is left for future investigations.  The cosmological constant is fixed by demanding that the expectation value of the energy and momentum operators for the physical particles with respect to the ``ontological'' vacuum (for our ``beables'') vanish; in that sense it is a pure renormalization constant.  I mentioned previously that we would recuperate Einstein Cartan theory and indeed, the commutation relations will induce a nonzero torsion and hence spin.  Let me jump a bit ahead now and express what can be expected from such framework.  Since our particle energy, momentum and spin correspond to nonlinear but analytic expressions in terms of the ``beable'' operators, one can reasonably expect to observe discrete spectra even if the latter operators have a continuous spectrum.  As is well known, traditional string theory has this salient feature due to periodic boundary conditions (at least for the mass operator); no such thing exists here, but our notion of particles is drastically different so we might recover this virtue in a totally different way.  We will spell out more details in the next chapter.  \\* \\*
I have promised to study the causality question classically and by doing so I must define particles in a different way.  The reader understands now that this implies I must ``cheat'' a bit (since classically no particles exist), but nevertheless the exercise is instructive.  Let us define $$g_{\alpha \beta} = \eta_{\mu \nu}
\partial_{\alpha} X^{\mu} \partial_{\beta} X^{\nu}$$ then what we should be calculating is either $$g_{\alpha \beta} P^{\alpha} P^{\beta}$$ where the raising of the indices on the $P$'s is done by $\eta^{\alpha \beta}$, or
$$g^{\alpha \beta} P_{\alpha} P_{\beta}.$$  The reader notices the ambiguity in the definition of the ``physical'' norm squared of the monad's energy-momentum vector; this already indicates that this is not a good concept.  Such problem does not occur of course for the energy-momentum of our particles.  Nevertheless, we shall compute another expression (which is unambiguous) given by
$$\eta^{\alpha \beta} P_{\alpha} P_{\beta}.$$  In $1+1$ dimensions, this expression simplifies due to the volume constraint\footnote{This is a peculiar feature of $1+1$.} to
$$ -\frac{1}{4} \left( \partial_t X^{\mu} \partial_t X_{\mu} - \partial_x X^{\mu} \partial_x X_{\mu} \right)^2 + 1.$$ 
Hence, all beable tachyons in the theory have a norm squared smaller than $1$.  This is in sharp contrast with the ``stringy'' definition given by:
$$ \partial_t X^{\mu} \partial_t X_{\mu}.$$
Here, one can easily see that within the left moving sector and for $\gamma =1$ the above expression reduces to 
$$2 \kappa f'(t+x) + \kappa^2 + 1$$ where $\kappa > 0$ and  $f'(t+x) > \frac{1}{2 \kappa} - \frac{\kappa}{2}$.  It is clear that this can grow unboundedly to $+ \infty$.  Now, the classical issue I want to discuss is that if monads\footnote{For example, one can calculate that for $\gamma > 1$, $\kappa > 0$ and $f'(t+x) \geq \frac{1}{2 \kappa}$ all atoms move on future oriented timelike curves.} are moving on future pointing timelike curves; any reasonable definition of a particle will obey this property.  Usually, a particle is thought of as the top of a bump in the spatial geometry; the latter is given by
$$ \partial_x X^{\mu} \partial_x X_{\mu} = - (f'(t+x) - \frac{1}{2 \kappa})^2 + (\gamma f'(t+x) - \frac{\gamma}{2 \kappa} + \kappa)^2 $$ so any bump will satisfy an equation of the form $t+x = a$ and therefore moves on a null geodesic with respect to the background metric.    
\chapter{Quantum physics is crazy, but to what extend ?}
A famous physicist once said that it is not the right question whether your theory is too crazy, but whether it is crazy enough.  Indeed, the difference between a crank's paradise and the seeminly unlimited possibilities of quantum physics is very subtle and one may wonder whether labeling the crank as cranky is really not a matter of taste or ``proper'' conduct, rather than anything else.  The attentive reader may have discovered in the previous chapter a hidden suggestion: that is, the quantum Hamiltonian of our hidden variable theory corresponds to a free theory and macroscopic nonlinearity of what we call particle energy and momentum is just how we ask questions about the world.  Does this mean that we shall ``perceive'' a non-free (or even nonlinear) dynamics for what we call particles?  I think it does and I shall illustrate this point here by calculating a concrete example.  Notice upfront that the particle observables which we defined in the previous chapter are still local; in reality, when asking questions about the real world we shall deal with non-local observables.  The upshot of this chapter will be that a theory of quantum gravity cannot function in this way and that dynamical nonlinearity is a vital ingredient implying the role of consciousness in the physical world is rather limited.  The ontological theory is the free theory of one particle with Hamiltonian given by:
$$H = \frac{1}{2 m} P^2$$ where $[ X, P] = i \hbar$ (we shall restore units here to make realistic estimations later on).  Suppose our particle Hamiltonian by definition is given by $$H' = \frac{1}{2 m} P^2 + \frac{m \omega^2}{2} X^2$$ that is, by the harmonic oscillator with spring constant $\omega$.  Of course $H'$ is not conserved and its Heisenberg equation can be written as
$$\partial_t H' =  \frac{\omega^2}{2} (PX + XP).$$
The observer measures the bound states of $H'$ and not those of $H$ so the question is whether he will observe a nonstationary dynamics.  The answer is that for reasonable values of $\omega$ and of perception time $\Delta t$, the lowest energy states of $H'$ will remain stationary with a probability higher than $99 \%$.  Even in the extreme rare case a transition may occur then its energy difference would be too low for it being observed.  However, if one would increase the value of $\omega$ and keep $\Delta t$ fixed, even the lowest states would become nonstationary.  In case the observer would notice this violation of energy, he would attribute it to gravitational effects by definition.  The conclusion is that in the first case the observer would be fooled into believing that the correct Hamiltonian is indeed $H'$, $H'$ and $H$ are operationally indistinguishable.  In the second case, he is looking at the world at high energies relative to his own awareness time, and at such scales physics is no longer conservative (at least if there is no ``being'' with a higher consciousness (lower $\Delta t$) than the observer performing the reduction).  Classically, this is impossible since observation does not affect the system; quantum mechanically however, we are saved by $R$ and the superposition principle.  What are reasonable estimates for $\Delta t$?  There are sources that the human brain would have a consciousness timescale of $10^{-12}$ seconds \cite{Georgiev} and I believe an upper bound to the timescale between two $R$ processes in high energy experiments -such as occuring in Cern- to have a magnitude of $10^{-7}$ seconds.  Of course, the calculation of the scattering matrix runs over an infinite time interval, but the above might explain why perfect correspondence is achieved at the two loop level (where fairly elementary scattering processes occur) but also why a nonperturbative summing over higher loops gives divergent results.  I would claim that within the energy scale we are doing experiments so far, the difference in evolution between the free beable Hamiltonian and the particle Hamiltonian (evolving the \emph{particle} states) is so small that it will practically never happen that physical energy-momentum is not preserved.  We now show the validity of these claims by a concrete calculation in the above setup.  As always, it is convenient to introduce the ladder operators:
\begin{eqnarray*}
a & = & \sqrt{\frac{m \omega}{2 \hbar}} \left( X + \frac{i}{m \omega} P \right) \\*
a^{\dag} & = & \sqrt{\frac{m \omega}{2 \hbar}} \left( X - \frac{i}{m \omega} P \right) \\*
\end{eqnarray*} and $[ a ,a^{\dag} ] = 1$.  As is well known
$$H' = \hbar \omega ( a^{\dag} a + \frac{1}{2} )$$ and 
$$H = - \frac{\hbar \omega ( a - a^{\dag} )^2}{4}.$$
Let us calculate the time evolution of the first excited state $a^{\dag} | 0 >$ order by order in perturbation theory.  The first order corrections to the free evolution of $e^{- \frac{i \omega t ( a - a^{\dag})^2}{4}}  a^{\dag} | 0 >$ are 
$$ (1 + \frac{3 i \omega t}{4}) a^{\dag} | 0 > - \frac{i \omega t}{4} (a^{\dag})^3 | 0 >$$ and its norm squared is given by $1 +\frac{15 t^2 \omega^2}{16}$.  The probability for staying in $a^{\dag}| 0>$ is $$1 - \frac{6 t^2 \omega^2}{16}$$ and the transition probability to $\frac{1}{\sqrt{6}} (a^{\dag})^3 |0 >$ is therefore      
$$\frac{6 t^2 \omega^2}{16}.$$  Now, for $t\omega \ll 1$ this result is perturbatively stable.  Indeed, taking into account the second order corrections modifies the probability to stay in $a^{\dag}| 0>$ to $$1 - \frac{6 t^2 \omega^2}{16} - \frac{672 t^4 \omega^4}{1024}.$$  Now, to obtain the localization property of our particle in this first excited state, we calculate the standard deviation.  The latter turns out to be given by
$$\Delta X = \sqrt{\frac{6 \hbar}{4 m \omega}}$$ and for an electron the order of magnitude is
$$\Delta X = \sqrt{\frac{10^{-4}}{\omega}}.$$  Hence, for $t = 10^{-7}$ and $\omega = 10^6$, we can localize the particle within a radius of $10^{-5}$ meters and the probability for it staying in this state would be around $1 - \frac{6}{1600} \sim 1$ by Cournot's principle.  However, in our model, we have made the assumption that the beable mass was equal to the particle mass.  This does not need to be and one might expect the beable mass $M$ to be much higher (since the theory becomes only free at very high energies).  The formula for $\Delta X$ remains the same but the probability gets renormalized by $$1 - \frac{6 t^2 m^2 \omega^2}{16 M^2}.$$ Now, it is possible to obtain much better localization properties; suppose $\frac{m}{M} = 10^{-4}$, then one may choose $\omega = 10^{10}$ to have the same probabilities, but $\Delta X$ now becomes $10^{-7}$ meters.  As I told energy eigenstates of the very high end of $H'$ will become unstable, but then gravitation comes into play.  Let me summarize these results again; the observer doesn't know what the fundamental theory is, neither does he know what his operators are which he measures.  But he turns out to measure discrete spectra which are very stable up to relatively high energies, so he is fooled to believe that the \emph{beable} Hamiltonian must be one with bound states since he readily identifies the underlying reality with the results of his measurements.  Moreover, deviations at high energies may cause him to believe that the theory gets more \emph{complicated} at higher energies, while it is actually the reverse.  This phenomenon has been well documented since the $1970$ ties when asymptotic freedom in the theory of strong interactions was discovered.  This must be extremely hard to swallow for the classical physicist who is used to believe that things simply are what they are.  Here, the reality depends upon the time scale of observation and the questions nature allows one to ask; it could be that in this free theory, the reason for internal stability of \emph{macroscopic} objects is due to the minute awareness time.  If this were true, then the interplay between consciousness and materialism would be much stronger than anything a classical physicist could ever imagine: it would mean the end of classical physics as we know it.  Before we come to the quantization of our theory, we will have to make another crucial observation. \\* \\*   
This observation will reduce three problems to a single one; that is, it unifies (a) (a part of the) causality problem (b) problem of negative energies (c) spin statistics relations.  We will show that the ``quantization'' of this bosonic theory is inconsistent and the action needs to be extended providing for fermionic degrees of freedom.  That will give rise to a mixed statistics requiring negative energy monadic degrees of freedom for the theory to be consistent at fourth order.   This will be the main content of the chapter and the issue of how to satisfy (a) and (b) is dealt with in the same manner as the strategy which historically lead to the discovery of the spin-statistics relation.  That is, I don't know what the relations between the beable ``position'' and ``momentum'' are but I will constrain them by demanding that the ``dominant energy'' condition\footnote{I put these words between parentheses because I have to define what the dominant energy condition is at the quantum level.} at the level of particles is satisfied
(if these constraints are not severe enough, this might lead to inequivalent theories).  This will probably imply that as well my beables as particles have a mixed form of statistics; the reader should notice that the statistics a particle satisfies here is decided upon at the level of the state-vector \emph{and} at the level of the operators.  Of course, the burden of proof of the spin-statistics theorem is still upon us.  Concretely, I will follow the same steps as are usual in quantum field theory; that is, I shall solve the d'Alembertian equations of motion and impose the constraints at the level of the beables \emph{but} I shall leave the usual commutation relations between the mode operators completely unspecified\footnote{There is no reason to impose that our beables should satisfy some notion of causality with respect to the Minkowski background, that would be entirely unphysical.}.  Then, I will try to construct the Einstein tensor and impose the dominant energy condition.  Next, I will try to ``guess'' clever relations between the mode operators such that the above is satisfied.  This will define my statistics; it is entirely logical and very quantum mechanical, it is just not the usual procedure people would think about.  Note that my reasoning is extremely tight and the survival of the proposal as it stands crucially depends upon a reasonable form of the spin-statistics theorem.  It is instructive at this point to guess what relations one might expect the beable operators to satisfy: suppose the beable mode operators are given by $c = \alpha a + \beta b$ (or equivalently $d = \alpha a + \beta b^{\dag}$) where $a$ is bosonic, $b$ fermionic and $a$ (or dagger) and $b$ (or dagger) commute.  Notice that if $|\alpha| = |\beta|$ the commutator $[d,d^{\dag}] = 2 b^{\dag} b$ has no central part; hence no normal ordering infinities arise here at least if one orders with respect to $d,d^{\dag}$.  However, this would not solve our normal ordering problem since the latter is formulated with respect to the $a$'s and $b$'s and not the $d$'s.  A genuine mechanism would consist in allowing for $X^{\mu}$ to become complex quantum mechanically (there is no general principle which dictates that real classical fields should remain so at the quantum level-all what is required is that observables are Hermitian) and consider operators of the kind $c= a + ib, d = a^{\dag} + ib^{\dag}$ then $\mathcal{N} \{c,d \} = 2 a^{\dag} a + 2i( ab^{\dag} + ba^{\dag})$ where $\mathcal{N}$ denotes normal ordering.  One notices here that in such theory, free fermion operators do not occur.  Such thing would require the commutator $[c,d]$ and the normal ordered expression $\mathcal{N} [c,d] = 2 b^{\dag} b$ (and therefore doesn't contain any central term either).  Remark also that taking the real part of such expressions would kill off any term with an odd number of fermions - this is precisely what we need.  Notice that the relations below also hold with $c^{\dag}$ replaced by $d$.  Therefore, this could be a truely unified mechanism replacing the duality of supersymmetry with the enormous virtue that no Grassmann directions need to be added to space-time.  Let us find out support for this idea by calculating the total canonical beable momentum in this approach.  That is, the quantization of $ - \int d^3x \, \partial_t X^{\mu} \partial_{x^{j}} X_{\mu}$ leads to $ -\frac{1}{2} \, \int d^3x \, \mathcal{S} \, \partial_t X^{\mu} \partial_{x^{j}} X_{\mu} + hc$ in the quantum theory, where $hc$ denotes hermitian conjugate.  Write $X^{\mu} = a^{\mu} + p^{\mu}_{\alpha} x^{\alpha} + \frac{1}{\kappa} \int \frac{d^3 \vec{k}}{\sqrt{k}} \, \left[ e^{-i(kt - \vec{k}\dot \vec{x})}c^{\mu}_{\vec{k}} + e^{i(kt - \vec{k}\dot \vec{x})}d^{\mu}_{\vec{k}} \right] $ and as previously $c^{\mu}_{\vec{k}} = a^{\mu}_{\vec{k}} + i b^{\mu}_{\vec{k}}$ where $\left[ a^{\mu}_{\vec{k}}, a^{\nu \, \dag}_{\vec{l}} \right] = \delta(\vec{k} - \vec{l}) \eta^{\mu \nu}$, $\{ b^{\mu}_{\vec{k}}, b^{\nu \, \dag}_{\vec{l}} \} = \delta(\vec{k} - \vec{l}) \eta^{\mu \nu}$ and the rest of the $a$'s commute with one and another and all the $b$'s and the $b$'s anticommute amongst one and another.  The $p^{\mu}_{\alpha}$ commute amongst themselves and with the $a$'s and $b$'s and are constrained to satify $p^{\mu}_{t} p_{\mu \beta} = 0$ at least for $\beta \neq 1$.  Hence, one can caculate that the momentum equals
$$ \frac{1}{4 \kappa^2} \int d^3\vec{k} \, \left[ k_j e^{-2ikt} \{ c^{\mu}_{\vec{k}} , c_{- \vec{k} \, \mu} \} + k_j e^{2ikt} \{ d^{\mu}_{\vec{k}} , d_{- \vec{k} \, \mu} \} + 2 k_j \{ c^{\mu}_{\vec{k}} , d_{\vec{k} \, \mu} \} \right] + hc.$$  The first two expressions vanish due to assymetry under $\vec{k} \rightarrow - \vec{k}$ and therefore, the formula reduces to $$ \frac{1}{2 \kappa^2} \int d^3\vec{k} \, k_j \left( 2 a^{\mu \, \dag}_{\vec{k}} a_{\vec{k} \, \mu} + 2i \left( b^{\mu \, \dag}_{\vec{k}} a_{\vec{k} \, \mu} + a^{\mu \, \dag}_{\vec{k}} b_{\vec{k} \, \mu} \right) \right) + hc.$$  Here one notices that the contributions from the vacuum have cancelled out (this was not a problem for the momentum operator but definetly for the energy operator) and the formula reduces to
$$ \frac{2}{\kappa^2} \int d^3 \vec{k} \, k_j  a^{\mu \, \dag}_{\vec{k}} a_{\vec{k} \, \mu} $$ which is the usual expression.  Strange enough, the fermions have vanished from this expression.  To understand this better, let us calculate the Hamiltonian and see if the infinite (bosonic) vacuum contribution is cancelled.  One calculates that
$$ \frac{1}{4} \int d^3x \, \left[ \partial_t X^{\mu}\partial_t X_{\mu} + \partial_{x^j} X^{\mu}\partial_{x^j} X_{\mu} \right] + hc$$ equals 
$$\frac{2}{\kappa^2} \int d^3 \vec{k} \, k a^{\mu \, \dag}_{\vec{k}} a_{\mu \, \vec{k}}$$ and therefore the fermions do what they are expected to do.  So, we have to conclude that our fermions do a magnificent job in canceling infinities but are nevertheless energy and momentumless.  Does this mean that fermions cannot be ``physical'' perennials?  Let me explain why this is not so: first remark that all creation operators have to be Lorentz invariant with respect to the $\mu$ indices.  Indeed, we have explained this in detail at the beginning of the previous chapter; therefore only creation operators of the kind $a^{\mu \, \dag}_{\vec{k}} a^{\dag}_{\vec{l} \, \mu}$, $a^{\mu \, \dag}_{\vec{k}} b^{\dag}_{\vec{l} \, \mu}$, $b^{\mu \, \dag}_{\vec{k}} b^{\dag}_{\vec{l} \, \mu}$ or $ \epsilon_{\mu_1 \mu_2 \mu_3 \mu_4} b^{\mu_1 \, \dag}_{\vec{k_1}}b^{\mu_2 \, \dag}_{\vec{k_2}}b^{\mu_3 \, \dag}_{\vec{k_3}}b^{\mu_4 \, \dag}_{\vec{k_4}}$ are to be used to construct beable states.  Consider for example the operator $a^{\mu \, \dag}_{\vec{k}} b^{\dag}_{\vec{l} \, \mu}$, the latter is of the fermionic type since $(a^{\mu \, \dag}_{\vec{k}} b^{\dag}_{\vec{l} \, \mu})^2 = 0$.  So, does this operator produce a single fermion or a boson and fermion separately?  There are two independent reasons to prefer the former option: (a) the separate entities do not form beables, therefore such operator should be regarded as one entity (b) in quantum field theory fermions are ``dressed'' with bosons to give them a dynamical mass, neverteless we still call the outcome of this process a fermion.  It is now easy to see that the fermion created by $a^{\mu \, \dag}_{\vec{k}} b^{\dag}_{\vec{l} \, \mu}$ has energy $k$ and momentum $\vec{k}$ which is of course still strange since $\vec{l}$ is irrelevant (and therefore our fermions do not satisfy the physical Pauli exclusion principle at this moment - strictly speaking they do, but alas $\vec{l}$ is not observable yet).  Later on, at the level of particles, when fermions shall become dynamical due to the contorsion tensor, this momentum vector $\vec{l}$ will of course matter.  Likewise, by considering higher composite operators, one can construct massive fermions; there is probably no spin statistics theorem at this level.  It is furthermore important to notice that $X^{\mu}(t,x) \neq e^{iHt} X^{\mu}(0,x)e^{-iHt}$, so the normal Hamiltonian evolution picture does not hold here anymore.  It is in this precise sense that our ``quantum theory'' is not strictly quantum; however, as elaborated upon previously, the physical predictions of the bosonic sector of this theory do obey this unitary evolution and this is the only thing that matters (so our quantization procedure is a very subtle deviation of what is usually done).  Therefore, on the level of observables, one may conclude that the ``new'' Hamiltonian evolution coincides with the ``old'' one; hence the novel dynamics at the level of beables is somehow forgotten at the level of observables.  However, one may wonder now whether defining $X^{\mu}(t,\vec{x}) = e^{iHt} X^{\mu}(0,\vec{x})e^{-iHt}$ and substituting this expression in our field formula for the original Hamiltonian density yields the same density as we arrived at.  The answer is no and the reader may verify that by doing so, the original infinite vacuum density shows up again.  This suggests in my view that the original $X^{\mu}(t,\vec{x})$ are hidden variables with well defined observables and ``Hamiltonian'' $H$; once the latter has been constructed, one should forget the original prescription of the observables in terms of these hidden variables and work with them as such.  This is consistent since the observables satisfy the Heisenberg equation with respect to $H$.  Looking for alternative prescriptions for energy and momentum by mixing $X^{\mu}$ with $X^{\mu \, \dag}$ is forbidden since these reintroduce the familiar problems with the beable Hamiltonian (the reader is invited to show this explicitely).  A more detailed computation reveals that the $b,b^{\dag}$ operators also vanish from the \emph{local} energy and momentum and therefore causality is preserved in the standard way.  More in general, the entire Poincar\'e group is the standard bosonic one and therefore Poincar\'e invariance and causality are preserved in this alternative quantization scheme.  The reader may compute that $$S^{rs} = \frac{4i}{\kappa^2} \int d^3\vec{k} \, a^{\dag \, \mu}_{\vec{k}}k_{[r} \partial^{\vec{k}}_{s]}a_{\vec{k} \, \mu}$$ and $$S^{0j} =\frac{2i}{\kappa^2} \int d^3 \vec{k} \, k \partial_j a^{\dag \, \mu}_{\vec{k}}a_{\vec{k} \, \mu}.$$  The vacuum problem however is solved and this is exactly what supersymmetry normally does for you (the appearant price to pay is a bunch of unmeasurable fermions at the beable level). \\* \\*  Our previous ansatz satisfies 
$$[ c, [c , [c, c^{\dag}]]] = 0 $$ and $$[ c^{\dag}, [c^{\dag} , [c, c^{\dag}]]] = 0.$$  and as we will show later, mode operators satisfying these relations are not sufficient to make the theory well defined\footnote{One notices that these relations are clearly of higher order than they usually are.  The commutator is still preferred over the anti-commutator since the relations between $a$ and $b$ break the symmetry between both brackets.}.  We are now in a position to make some preliminary steps towards the ``quantization'' of the theory.  When quantizing a classical theory, it sometimes happens that a classical symmetry goes havoc.  In the case of bosonic string theory, this might happen to the conformal symmetry in $D = 26$ if the background geometry does not satisfy covariant equations which contain higher order corrections in the string coupling constant to the vacuum Einstein equations.  The critical dimension originates from the demand that the Virasoro algebra contains no central extension, the latter can occur because the Lie-algebra of normal ordered symmetry generators can differ from the original classical algebra.  The Einstein tensor however, is a beast of a totally different category, even classically there is a not an \emph{entire} analytic function expressing the inverse of a matrix\footnote{However, locally in matrix space, analytic expressions can be found - but this is meaningless for quantum mechanics.}.  However, our volume constraint comes here to the rescue, and indeed the lack of (entire) analycity is magically transferred into a polynomial expression of finite degree.  This is the main technical reason why we assumed this constraint to hold: it actually makes our computations possible quantum mechanically.  Of course, one has to be very careful now in defining the right \emph{product} so that the inverse property and the polynomial expression happily marry together at the quantum level.  However, the ``breaking'' of a symmetry I wish to talk about here is not exactly the same as is meant in the context of string theory.  That is, the coordinate covariance of the Einstein tensor is a \emph{classical} symmetry and remains at best so at the quantum level (just as happens to gauge invariance).    
\\* \\*
Classically, the inverse matrix of $\left( \frac{\partial X^{\mu}}{\partial x^{\alpha}} \right)$ is given by
$$\frac{\partial x^{\alpha}}{\partial X^{\mu}} = - \frac{1}{3!} \epsilon_{\mu \nu \gamma \delta} \epsilon^{\alpha \beta \kappa \zeta} \frac{\partial X^{\nu}}{\partial x^{\beta}} \frac{\partial X^{\gamma}}{\partial x^{\kappa}} \frac{\partial X^{\delta}}{\partial x^{\zeta}}$$ and therefore the inverse metric reads as
$$g^{\alpha \beta} = \frac{\partial x^{\alpha}}{\partial X^{\mu}} \eta^{\mu \nu} \frac{\partial x^{\beta}}{\partial X^{\nu}} = \frac{1}{3!} \sum_{\mu_i \neq \mu_j \neq \mu \, i \neq j} \eta^{\mu \mu} \epsilon^{\alpha \kappa \gamma \zeta} \epsilon^{\beta \pi \chi \delta} \frac{\partial X^{\mu_1}}{\partial x^{\kappa}} \frac{\partial X^{\mu_2}}{\partial x^{\gamma}}\frac{\partial X^{\mu_3}}{\partial x^{\zeta}}\frac{\partial X^{\mu_1}}{\partial x^{\pi}}\frac{\partial X^{\mu_2}}{\partial x^{\chi}}\frac{\partial X^{\mu_3}}{\partial x^{\delta}}.$$  Obviously, the Ricci tensor consists of terms of $4$'th and $8$'th order in the $\frac{\partial X^{\mu}}{\partial x^{\beta}}$ and second derivatives while the Ricci scalar term in the Einstein action contains terms of $10$'th and $14$'th order respectively.  This is an infinite simplification with respect to the situation in standard perturbation theory.  When quantizing, one needs to be careful (a) about preserving the above properties (b) having Hermitian operators.  Define $\mathcal{S} \, \partial_{\beta_1} X^{\mu_1} \ldots \partial_{\beta_n} X^{\mu_n}$ as $$ \mathcal{S} \, \partial_{\beta_1} X^{\mu_1} \ldots \partial_{\beta_n} X^{\mu_n} = \frac{1}{2} \left( \frac{1}{n!} \sum_{\sigma \in S_n} \partial_{\beta_{\sigma(1)}} X^{\mu_{\sigma(1)}} \ldots \partial_{\beta_{\sigma(n)}} X^{\mu_{\sigma(n)}} + \textrm{hc} \right)$$ and extend this definition by linearity\footnote{$\mathcal{S}$ is the usual symmetrization: note that the definition of $\mathcal{S}$ trivially extends if higher order derivatives are included (we shall need this extension later on).}; then the volume constraint becomes $$ - \mathcal{S} \, \frac{1}{4!} \epsilon_{\mu \nu \gamma \delta} \epsilon^{\alpha \beta \kappa \zeta} \frac{\partial X^{\mu}}{\partial x^{\alpha}} \frac{\partial X^{\nu}}{\partial x^{\beta}} \frac{\partial X^{\gamma}}{\partial x^{\kappa}} \frac{\partial X^{\delta}}{\partial x^{\zeta}} = 1.$$  Similarly, the inverse ``coordinate transformation'' is given by
$$\frac{\partial x^{\alpha}}{\partial X^{\mu}} = - \mathcal{S} \, \frac{1}{3!} \epsilon_{\mu \nu \gamma \delta} \epsilon^{\alpha \beta \kappa \zeta} \frac{\partial X^{\nu}}{\partial x^{\beta}} \frac{\partial X^{\gamma}}{\partial x^{\kappa}} \frac{\partial X^{\delta}}{\partial x^{\zeta}}.$$  One can define the symmetric $\star$ product between two monomials $\partial_{\beta_1} X^{\mu_1} \ldots \partial_{\beta_n} X^{\mu_n}$ and $\partial_{\gamma_1} X^{\nu_1} \ldots \partial_{\gamma_m} X^{\nu_m}$ as $$\partial_{\beta_1} X^{\mu_1} \ldots \partial_{\beta_n} X^{\mu_n} \star \partial_{\gamma_1} X^{\nu_1} \ldots \partial_{\gamma_m} X^{\nu_m} = \mathcal{S} \, \partial_{\beta_1} X^{\mu_1} \ldots \partial_{\beta_n} X^{\mu_n}\partial_{\gamma_1} X^{\nu_1} \ldots \partial_{\gamma_m} X^{\nu_m}.$$  
Therefore, one obtains that  
$$\frac{\partial x^{\alpha}}{\partial X^{\mu}} \star \frac{\partial X^{\mu}}{\partial x^{\beta}} = \delta_{\beta}^{\alpha}$$ and clearly, the symmetric  $\star$ product is associative on the \emph{free} algebra generated by all derivatives of the $X^{\mu}$.  As it stands, however, this product is incompatible with the volume constraint.  In order to keep the definition as canonical as possible (that is to avoid ambiguities) we restrict to expressions which are scalar in the $\mu$ coordinates and (pseudo) tensors in the $\alpha$ coordinates.  Hence, we \emph{define} that for a string of products, maximal reduction by means of the volume constraint has to be made; clearly, an operation of this kind preserves the transformation properties of the original expression since the volume constraint is a scalar (with respect to unimodular diffeomorphisms).  We extend therefore the definition of $\star$ by imposing that $A \star 1 = A$ where $A$ is any (pseudo) tensor in the derivatives of the fields (notice that this also redefines $\mathcal{S}$, the reader is invited to fill in the details).  Although this definition is straightforward for the volume constraint itself, one can forget about this when derivatives of the volume constraint appear.  For example, consider the expression
$$\epsilon_{\nu_1 \nu_2 \nu_3 \nu_4} \epsilon^{\kappa \alpha \beta \delta} \mathcal{S} \left( \frac{\partial^2 X^{\nu_1}}{\partial x^{\gamma} \partial x^{\kappa}} \frac{\partial X^{\nu_2}}{\partial x^{\alpha}} \frac{\partial X^{\nu_3}}{\partial x^{\beta}}\frac{\partial X^{\nu_4}}{\partial x^{\delta}} \frac{\partial X^{\mu}}{\partial x^{\zeta}} \frac{\partial X_{\mu}}{\partial x^{\xi}} \right).$$  Strictly speaking one should say that this is equal to $$ - \epsilon_{\nu_1 \nu_2 \nu_3 \nu_4} \epsilon^{\kappa \alpha \beta \delta} \mathcal{S} \left( \frac{\partial X^{\nu_1}}{\partial x^{\kappa}} \frac{\partial}{\partial x^{\gamma}} \left( \frac{\partial X^{\nu_2}}{\partial x^{\alpha}} \frac{\partial X^{\nu_3}}{\partial x^{\beta}}\frac{\partial X^{\nu_4}}{\partial x^{\delta}} \right) \frac{\partial X^{\mu}}{\partial x^{\zeta}} \frac{\partial X_{\mu}}{\partial x^{\xi}} \right)$$  although this would be blatant nonsense\footnote{They would certainly not be for ordinary bosonic quantization.} for our original definition of $\mathcal{S}$.  Also, a move of the above type might simplify things considerably if one would take an expression like $$\epsilon_{\mu_1 \mu_2 \mu_3 \mu_4} \epsilon_{\nu_1 \nu_2 \nu_3 \nu_4} \epsilon^{\kappa \alpha \beta \delta} \mathcal{S} \left( \frac{\partial X^{\mu_1}}{\partial x^{\psi}}\frac{\partial X^{\mu_2}}{\partial x^{\zeta}}\frac{\partial X^{\mu_3}}{\partial x^{\xi}} \frac{\partial^2 X^{\mu_4}}{\partial x^{\gamma} \partial x^{\kappa}} \frac{\partial X^{\nu_1}}{\partial x^{\alpha}} \frac{\partial X^{\nu_2}}{\partial x^{\beta}}\frac{\partial X^{\nu_3}}{\partial x^{\delta}} \frac{\partial^2 X^{\nu_4}}{\partial x^{\pi} \partial x^{\chi}} \right).$$  The latter would ``reduce'' to $$ - 2 \epsilon_{\psi \zeta \xi \kappa} \epsilon_{\nu_1 \nu_2 \nu_3 \nu_4} \epsilon^{\kappa \alpha \beta \delta} \mathcal{S} \left(    \frac{\partial}{\partial x^{\gamma}} \left( \frac{\partial X^{\nu_1}}{\partial x^{\alpha}} \frac{\partial X^{\nu_2}}{\partial x^{\beta}}\frac{\partial X^{\nu_3}}{\partial x^{\delta}}  \right) \frac{\partial^2 X^{\nu_4}}{\partial x^{\pi} \partial x^{\chi}} \right)$$ which is a substantial simplification.  Therefore, what should we do with expressions allowing equivalences by means of derivatives of the volume constraint which are not identities?  One argument would be to avoid them.  Another position would be that if a true simplification happens (such as in the latter example), then one should define it by this expression and cook up something else otherwise (such as in the former example).  A final idea would be to average the expressions over all ``equivalent'' ones; this would certainly not allow for such drastic reduction as in the previous example. \\* \\*  For now, we split the ``metric'' into two parts:
$$g_{\alpha \beta} = \mathcal{S} \, \frac{\partial X^{\mu}}{\partial x^{\alpha}} \frac{\partial X^{\nu}}{\partial x^{\beta}} \eta_{\mu \nu}$$ and
$$B_{\alpha \beta} = \frac{i}{2} \left( \frac{\partial X^{\mu}}{\partial x^{\alpha}} \frac{\partial X^{\nu}}{\partial x^{\beta}} -  \frac{\partial X^{\mu}}{\partial x^{\beta}} \frac{\partial X^{\nu}}{\partial x^{\alpha}}\right) \eta_{\mu \nu}$$ the symmetric metric and antisymmetric tensor respectively.  One computes that $g_{\alpha \beta}$ is a pure bosonic quantity and its formula is given by 
\begin{eqnarray*} g_{\alpha \beta}& = & \frac{1}{\kappa^2} \int \frac{d^3 \vec{k} \, d^3 \vec{l}}{\sqrt{k} \sqrt{l}} \, -k_{( \alpha} l_{\beta )}  e^{-i((k+l)t - ( \vec{k} + \vec{l} ). \vec{x})}  a^{\mu}_{\vec{k}} a_{\vec{l} \, \mu}   - k_{( \alpha} l_{\beta )}  e^{i((k+l)t - ( \vec{k} + \vec{l} ). \vec{x})}  a^{\mu \, \dag}_{\vec{k}} a^{\dag}_{\vec{l} \, \mu} \\*
& + & k_{( \alpha} l_{\beta )}  e^{-i((k-l)t - ( \vec{k} - \vec{l} ). \vec{x})}  a^{\mu \, \dag}_{\vec{l}} a_{\vec{k} \, \mu} + k_{( \alpha} l_{\beta )}  e^{i((k-l)t - ( \vec{k} - \vec{l} ). \vec{x})}  a^{\mu \, \dag}_{\vec{k}} a_{\vec{l} \, \mu}. \end{eqnarray*}  The commutator of two metric operators reads:
\begin{eqnarray*}   
\left[ g_{\alpha \beta}(t,\vec{x}),g_{\gamma \delta}(s,\vec{y}) \right] & = & \frac{8}{\kappa^4} \int \frac{d^3 \vec{k} \, d^3 \vec{l}}{k l} k_{( \alpha} l_{\beta)} k_{(\gamma} l_{\delta)} e^{-i((k+l)(t-s) - (\vec{k} + \vec{l}).(\vec{x} - \vec{y}))} \\* & - & \frac{4}{\kappa^4} \int \frac{d^3 \vec{k} \, d^3 \vec{l} \, d^3 \vec{q}}{\sqrt{k} l \sqrt{q}} k_{( \alpha} l_{\beta)} l_{(\gamma} q_{\delta)} e^{-i(kt -qs - \vec{k}.\vec{x} + \vec{q}.\vec{y})} \\* & & \left( e^{i(l(t-s) - \vec{l}.(\vec{x} - \vec{y}))} -  e^{-i(l(t-s) - \vec{l}.(\vec{x} - \vec{y}))}   \right)a^{\dag \, \mu}_{\vec{q}}a_{\vec{k} \, \mu} \\* & + & \frac{4}{\kappa^4} \int \frac{d^3 \vec{k} \, d^3 \vec{l} \, d^3 \vec{q}}{\sqrt{k} l \sqrt{q}} k_{( \alpha} l_{\beta)} l_{(\gamma} q_{\delta)} e^{-i(kt + qs - \vec{k}.\vec{x} - \vec{q}.\vec{y})} \\* 
& & \left( e^{i(l(t-s) - \vec{l}.(\vec{x} - \vec{y}))} -  e^{-i(l(t-s) - \vec{l}.(\vec{x} - \vec{y}))}\right)a^{\mu}_{\vec{k}}a_{\vec{q} \, \mu} \\*
& - & \textrm{hc}. \\*
\end{eqnarray*}  Obviously, $\left[ g_{ij}(t,\vec{x}),g_{rs}(t,\vec{y}) \right] = 0$ for space indices $i,j,r,s$ and one calculates that 
\begin{eqnarray*} \left[ g_{j0}(t,\vec{x}),g_{rs}(t,\vec{y}) \right] & =  & \frac{2i}{\kappa^2} \left( \partial_r \delta(\vec{x} - \vec{y}) g_{js}(t,\vec{x},\vec{y}) + \partial_s \delta(\vec{x} - \vec{y})g_{jr}(t,\vec{x},\vec{y}) \right)\\* & - & \frac{8i}{\kappa^4} \left( \left( \int \frac{d^3 \vec{k}}{k} k_j k_r \right)  \partial_s \delta(\vec{x} - \vec{y}) + \left( \int \frac{d^3 \vec{k}}{k} k_j k_s \right)  \partial_r \delta(\vec{x} - \vec{y}) \right) \end{eqnarray*} where unfortunately, an infinite central extension appears (it seems we were not clever enough yet and not all renormalization problems have been cured); the reader is invited to figure out the definition of $g_{jr}(t,\vec{x},\vec{y})$.  The reader may furthermore compute that
$$\left[ g_{j0}(t,\vec{x}), g_{r0}(t,\vec{y})\right] = \frac{2i}{\kappa^2}\left( \partial_r \delta(\vec{x}-\vec{y})g_{j0}(t,\vec{x},\vec{y}) + \partial_j \delta(\vec{x}-\vec{y}) g_{0r}(t,\vec{x},\vec{y}) \right)$$ (note that all partial derivatives are here with respect to $\vec{x}$) and $$\left[ g_{00}(t,\vec{x}), g_{r0}(t,\vec{y})\right] = \frac{4i}{\kappa^2}\partial_r \delta(\vec{x} - \vec{y})g_{00}(t,\vec{x},\vec{y}) - \frac{16i}{\kappa^4} \partial_r \delta(\vec{x}-\vec{y}) \int d^3 \vec{k} \, k.$$  Finally, $$\left[ g_{00}(t,\vec{x}), g_{rs}(t,\vec{y})\right] = \frac{4i}{\kappa^2}\left( \partial_r \delta(\vec{x}-\vec{y})g_{0s}(t,\vec{x},\vec{y}) + \partial_s \delta(\vec{x}-\vec{y})g_{0r}(t,\vec{x},\vec{y}) \right).$$  Suppose $(t,\vec{x})$ and $(s,\vec{y})$ are spatially separated; that is, there exists a Lorentz transformation $\Lambda$ such that $x^{\mu} - y^{\mu} = \Lambda^{\mu}_{\,\, \nu} (x'^{\nu} - y'^{\nu})$ where $x' - y' = (0, \vec{x}' - \vec{y}')$ then $$\left[ g_{\alpha \beta}(t,\vec{x}),g_{\gamma \delta}(s,\vec{y}) \right] = \Lambda_{\alpha}^{\,\, \alpha'}\Lambda_{\beta}^{\,\, \beta'}\Lambda_{\gamma}^{\,\, \gamma'}\Lambda_{\delta}^{\,\, \delta'} \left[ g_{\alpha' \beta'}(t',\vec{x}'),g_{\gamma' \delta'}(t',\vec{y}') \right]$$ as follows straight from the definition of $g_{\alpha \beta}$.  Obviously, Minkowski causality holds with the caveat that, if two points coincide, the commutation property can fail.  The Levi-Civita connection
\begin{eqnarray*} \Gamma^{\alpha}_{\beta \gamma}  & = & \frac{1}{3!} \epsilon_{\mu_1 \mu_2 \mu_3 \mu_4} \epsilon^{\alpha \kappa_1 \delta_1 \xi_1} \mathcal{S} \left( \frac{\partial X^{\mu_1}}{\partial x^{\kappa_1}}\frac{\partial X^{\mu_2}}{\partial x^{\delta_1}}\frac{\partial X^{\mu_3}}{\partial x^{\xi_1}} \frac{\partial^2 X^{\mu_4}}{\partial x^{\gamma} \partial x^{\beta}} \right) \end{eqnarray*} is of order $4$ in the $X^{\mu}$ and satisfies $\partial_{\alpha} g_{\beta \gamma} - \Gamma^{\kappa}_{\alpha \beta} \star g_{\kappa \gamma} - \Gamma^{\kappa}_{\alpha \gamma} \star g_{\beta \kappa} = 0$ where the connection star products are computed to be
\begin{eqnarray*}
\Gamma^{\kappa}_{\alpha \beta} \star g_{\kappa \gamma} & = & \frac{1}{3!} \epsilon_{\mu_1 \mu_2 \mu_3 \mu_4} \epsilon^{\kappa \kappa_1 \delta_1 \xi_1} \mathcal{S} \left( \frac{\partial X^{\mu_1}}{\partial x^{\kappa_1}}\frac{\partial X^{\mu_2}}{\partial x^{\delta_1}}\frac{\partial X^{\mu_3}}{\partial x^{\xi_1}} \frac{\partial^2 X^{\mu_4}}{\partial x^{\alpha} \partial x^{\beta}} \right) \, \star \, \mathcal{S} \left( \frac{\partial X^{\nu}}{\partial x^{\kappa}} \frac{\partial X_{\nu}}{\partial x^{\gamma}} \right) \\*
& = &  \frac{1}{3!} \epsilon_{\mu_1 \mu_2 \mu_3 \mu_4} \epsilon^{\kappa \kappa_1 \delta_1 \xi_1} \mathcal{S} \left( \mathcal{S} \left( \frac{\partial X^{\nu}}{\partial x^{\kappa}} \frac{\partial X^{\mu_1}}{\partial x^{\kappa_1}}\frac{\partial X^{\mu_2}}{\partial x^{\delta_1}}\frac{\partial X^{\mu_3}}{\partial x^{\xi_1}} \right) \frac{\partial^2 X^{\mu_4}}{\partial x^{\alpha} \partial x^{\beta}} \frac{\partial X_{\nu}}{\partial x^{\gamma}} \right) \\* 
& = & \mathcal{S} \left( \frac{\partial X^{\nu}}{\partial x^{\alpha} \partial x^{\beta}} \frac{\partial X_{\nu}}{\partial x^{\gamma}} \right).  
\end{eqnarray*}  Here, we notice again that in the definition of the star product a maximal reduction has taken place by using the volume constraint.  As explained previously, the reduction preserves the correct transformation properties under unimodular coordinate transformations; this is because the volume constraint is a scalar invariant.  Within the context where $X^{\mu}$ would be subtly complex, one can propose two definitions : $$B_{\alpha \beta} =  \frac{i}{2} \frac{\partial X^{\mu}}{\partial x^{[ \alpha}} \frac{\partial X_{\mu}}{\partial x^{\beta]}} + hc$$ or without the '$i$' in front $$B'_{\alpha \beta} =  \frac{1}{2} \frac{\partial X^{\mu}}{\partial x^{[ \alpha}} \frac{\partial X_{\mu}}{\partial x^{\beta]}} + hc.$$ Continuing the previous calculations, these expressions reduce to
$$ B'_{\alpha \beta} = \frac{1}{\kappa^2} \int \frac{d^3 \vec{k} \, d^3 \vec{l}}{\sqrt{k} \sqrt{l}} \, \left[ k_{ [ \alpha } l_{\beta ]} \, e^{-i((k-l)t - ( \vec{k} - \vec{l} ). \vec{x})}  a^{\mu \, \dag}_{\vec{l}} a_{\vec{k} \, \mu}  + k_{[ \alpha } l_{\beta ]} \, e^{i((k-l)t - (\vec{k} - \vec{l}).\vec{x})} a^{\mu \, \dag}_{\vec{k}} a_{\vec{l} \, \mu} \right]$$ and
\begin{eqnarray*}
B_{\alpha \beta} & = & \frac{1}{\kappa^2} \int \frac{d^3 \vec{k} \, d^3 \vec{l}}{\sqrt{k} \sqrt{l}} \,  - i k_{ [ \alpha } l_{\beta ]} e^{-i((k+l)t - ( \vec{k} + \vec{l} ). \vec{x})}  b^{\mu}_{\vec{k}} b_{\vec{l} \, \mu}  - i k_{ [ \alpha } l_{\beta ]} e^{i((k+l)t - ( \vec{k} + \vec{l} ). \vec{x})}  b^{\mu \, \dag}_{\vec{k}} b^{\dag}_{\vec{l} \, \mu}  \\*
& + &  ik_{ [ \alpha } l_{\beta ]} e^{-i((k-l)t - ( \vec{k} - \vec{l} ). \vec{x})} b^{\mu \, \dag}_{\vec{l}} b_{\vec{k} \, \mu} - ik_{ [ \alpha } l_{\beta ]} e^{i((k-l)t - ( \vec{k} - \vec{l} ). \vec{x})} b^{\mu \, \dag}_{\vec{k}} b_{\vec{l} \, \mu}.   
\end{eqnarray*} A remarkable property of both tensors is that their integral over space vanishes; at least, we have already local Fermi densities in $B_{\alpha \beta}$ and this gives hope for work to come.  One should note that the $B_{\alpha \beta}$ do not satisfy the Heisenberg equations with respect to $H$ and its time dependence is therefore explicit.  It is as if a background field has been switched on at the level of $H$ which was really hidden in the original beables.  Indeed, this is the only way in which our fermions can become propagating entities.  The reader is invited to work out the causality properties of both tensors.  As explained previously, the problem is that I do not know so far the correct normal ordering scheme since the (anti)commutation relations are not fixed yet.  Probably, no normal ordering problems arise if one considers mixed statistics and complex fields.  This would be preferable since renormalization could endanger covariance of the Einstein tensor.  At least, we have understood now that this holds for operators with less than four products of the derivatives of the $X^{\mu}$.  Let us now derive a higher form of statistics; we shall be conservative here and assume exact Lorentz invariance to hold.  Suppose only one type of particle is allowed for and we work in the language of creation $a^{\dag}$ and annihilation operators $a$, then the only reasonable equation one can write down is of the form
$$aa^{\dag} + \alpha a^{\dag}a = \beta 1$$ where $\alpha, \beta$ are real numbers.  Indeed, the meaning of $a$ is that it eats away a particle created by $a^{\dag}$, therefore $aa^{\dag}|0 \rangle  \sim |0 \rangle$; hence - since $aa^{\dag}$ is a positive operator - the only terms we could add to this relation are of the form $a^{\dag \, n} a^{n}$ for $n > 1$ (since otherwise there would be terms \emph{raising} or preserving the degree of the original state).  However, within the context of quantum field theory, such terms would be responsible for highly singular operators which we reject so far (moreover, they would violate Lorentz invariance).  It is not necessary to impose that $\beta > 0$ since $\beta \leq 0$ does not automatically lead to \emph{physical} negative (or zero) norm states.  Furthermore we demand a from of simplicity (this is a generalization of our previous ansatz), that is $ad_{a}^n = ad_{a^{\dag}}^n = 0$ for some $n \geq 2$.  This leads to $\alpha = -1$ or $a^{\dag} a^n  = \frac{\beta}{\alpha + 1}a^{n-1}$; that is the usual ``bosonic'' statistics or something else.  One notices that the latter condition is always satisfied for $n \geq 3$ if $a^{n-1} = 0$ which is a trivial generalization of the Pauli principle (at least for one mode). 
There is one further observation we need to make and it concerns the construction of the number operator; suppose $a^3 = 0 \neq a^2$ then $\frac{1}{\beta}a^{\dag}a a^{\dag \, 2}| 0 \rangle = (1 - \alpha) a^{\dag \, 2} | 0 \rangle$; repairing this would ask for the addition of a fourth order operator $\frac{(1 + \alpha)}{\beta^2 (1 - \alpha)}a^{\dag \, 2}a^2$.  However, in Quantum Field Theory, such operator is ill defined and writing it in terms of a double integration will always yield nontrivial cross terms between different particles (of the same species) - dropping the continuum hypothesis therefore seems to allow for more complicated situations.  Hence, the existence of a number operator demands that $\left[ a^{\dag} a ,a^{\dag} \right] = \beta a^{\dag}$ implying that $\alpha = -1$ or $a^2 = 0$.  The hard question now is how the different modes should couple together: suppose one considers $a_{\vec{k} \, \lambda}$ and $a_{\vec{l} \, \lambda'}$ where $\vec{k}\neq \vec{l}$ are the usual momentum vectors and $\lambda, \lambda'$ are internal labels.  We shall restrict at first to the situation where the relations between both types of operators do not involve a third operator.  At first, one needs to make a distinction between different dimensions: (a) in case of $1+1$ dimensions, the null vectors $k$ and $l$ cannot be boosted into one and another and since we are free to break space reversal, there could be an asymmetry here (b) in the complementary case, this is not true anymore and therefore the relations need to be symmetric between $k$ and $l$.  Generically, it is true that $\left[ a^{\dag}_{\vec{k} \, \lambda}a_{\vec{k} \, \lambda}, a^{\dag}_{\vec{l} \, \lambda'} \right] =\left[ a^{\dag}_{\vec{k} \, \lambda}a_{\vec{k} \, \lambda}, a_{\vec{l} \, \lambda'} \right] = \left[ a^{\dag}_{\vec{l} \, \lambda'}a_{\vec{l} \, \lambda'}, a^{\dag}_{\vec{k} \, \lambda} \right] = \left[ a^{\dag}_{\vec{l} \, \lambda'}a_{\vec{l} \, \lambda'}, a_{\vec{k} \, \lambda} \right] = 0$ and we shall assume that different $a$'s do not interact (in other words, the number of creation and annihilation operators of both types remains the same - this is sensible in the context of Quantum Field Theory since allowing for the number of creation and annihilation operators to increase leads to highly singular operators).  Therefore, we are left with $$a^{\dag}_{\vec{k} \, \lambda}a^{\dag}_{\vec{l} \, \lambda'} = \kappa(k.l,\omega(k,l),\lambda,\lambda')a^{\dag}_{\vec{l} \, \lambda'}a^{\dag}_{\vec{k} \, \lambda}$$ and $$a_{\vec{k} \, \lambda}a^{\dag}_{\vec{l} \, \lambda'} = \kappa(k.l,\omega(k,l), \lambda,\lambda')^{-1}a^{\dag}_{\vec{l} \, \lambda'}a_{\vec{k} \, \lambda}$$ where $\omega$ exists only in $1+1$ dimensions (it is the volume form).  If one ignores the internal indices, then the formula still simplify: in case (a) it is easy to see that $\kappa(k.l, - \omega(k,l)) = \overline{\kappa(k.l, \omega(k,l))}$ and since $k.l = \omega(k,l)$ for $k$ leftmoving and $l$ rightmoving massless particles, we arrive at 
$\kappa(k.l, \omega(k,l)) = a(k.l)e^{i\omega(k,l)b(k.l)}$.  Demanding consistency requires that $\beta^2 = \langle a_1 a_2 a^{\dag}_2 a^{\dag}_1 \rangle = |\kappa(1,2)|^2 \langle a_2 a_1 a^{\dag}_1 a_2 \rangle = |\kappa(1,2)|^2 \beta^2$ and therefore $\kappa(k.l, \omega(k,l)) = e^{i\omega(k,l) \theta(k.l)}$ which is a slight generalization of the standard anyonic statistics, while if (b) holds the same reasoning implies that $\kappa(k.l) = \pm 1$, that is a generalization of ``Bose'' and bosonic parastatistics.  Even if one would not assume the number operator to be preserved, then one can show that consistency of the norm of two particle states (and assuming two relations instead of one - in the latter case one can define the so called quon statistics) leads to the previous results.  The same reasoning holds for $a^2 =0$ - $\alpha$ can be whatever-leading to a generalization of ``Fermi'' and Fermi parastatistics; this is interesting since picking $\beta = 1$ and $\alpha > 1$ allows for the ``Fermi'' terms to survive in our ansatz and have positive energy.  The question then is of course whether Lorentz invariance and causality will be preserved; we shall come back to this later on.  In the above, we made a rather well hidden assumption which is that we take the ordinary vector representation of the physical energy-momentum for granted.  There is one and only one other choice one could have made and that is sending $k \Rightarrow \displaystyle{\not}k$, that is the same vector in the Clifford algebra.  Then, in all dimensions, one has the following Lorentz covariant quantity $\displaystyle{\not}k\displaystyle{\not}l$ which has as well a symmetric as antisymmetric piece; this allows for a length factor $\theta(k.l)$ and asymmetry $\omega(k.l) = k_{\mu}l_{\nu}\frac{1}{2}\left(\gamma^{\mu}\gamma^{\nu} - \gamma^{\nu}\gamma^{\mu} \right)$.  We baptize the kind of particles obeying this statistics Cliffordons and it is clear that it contains ordinary (para)Bose, (para)Fermi and anyonic; given that this kind of statististics requires an extension of representation theory to Clifford modules, we discuss this further on in chapter seven - independent arguments for Cliffordons being provided in the next one.  Demanding $\kappa$ to be continuous (except in $0$) reduces to both cases globally - but I see no reason to impose this (our relations are not continuous anyway, so why not allow for another discontinuity).  It is now easy to show that for irreducible representations $D$ of the internal group, the general situation reduces to the latter.  Indeed, it follows from the first equation that
$$D_{\lambda}^{\,\, \gamma}D_{\lambda'}^{\,\, \gamma'}\kappa(k.l,\gamma,\gamma') =
D_{\lambda}^{\,\, \gamma}D_{\lambda'}^{\,\, \gamma'}\kappa(k.l,\lambda,\lambda')$$ for all $\lambda,\lambda',\gamma,\gamma'$ and therefore $\kappa(k.l, \lambda, \lambda') = \kappa(k.l)$.  We should still mention two things; that is the relations $a^{n}_{\vec{k} \, \lambda} = 0$ should be invariant with respect to the internal symmetries.  This leads to $$\mathcal{S} \, a_{\vec{k} \, \lambda_1} a_{\vec{k} \, \lambda_2} \ldots a_{\vec{k} \, \lambda_n} = 0$$ for all $\lambda_i$ and $\mathcal{S}$ denotes symmetrization.  Also, the relations $a_{\vec{k} \, \lambda}a^{\dag}_{\vec{k} \, \lambda'} + \alpha a^{\dag}_{\vec{k} \, \lambda'}a_{\vec{k} \, \lambda} = \beta 1 K_{\lambda \lambda'}$ should hold where $K_{\lambda \lambda'}$ is symmetric and invariant under the action of the internal group (usually, it is the defining element of the group such as $\eta_{\alpha \beta}$ for the Lorentz group).  In quantum field theory, this leads to the unified relation $$a_{\vec{k} \, \lambda} a^{\dag}_{\vec{l} \, \lambda'} + ( \left( \alpha + \kappa(k.l) \right) \delta_{\vec{k},\vec{l}} - \kappa(k.l) )a^{\dag}_{\vec{l} \, \lambda'} a_{\vec{k} \, \lambda} = \beta \delta(\vec{k} - \vec{l}) \eta_{\lambda \lambda'}$$ where $\delta_{\vec{k},\vec{l}}$ is the Kronecker delta.  We test here our new notion of Fermi statistics; that is $$b^{\mu}_{\vec{k}}b^{\dag \, \nu}_{\vec{l}} + (( \alpha - 1) \delta_{\vec{k},\vec{l}} + 1)b^{\dag \, \nu}_{\vec{l}}b^{\mu}_{\vec{k}} = \delta(\vec{k} - \vec{l})\eta^{\mu \nu}$$ and 
$b^{\mu}_{\vec{k}}b^{\nu}_{\vec{l}} + b^{\nu}_{\vec{l}} b^{\mu}_{\vec{k}} = 0$ so we don't allow for Fermi parastatistics.  The reader may now compute that for the local energy, momentum and angular momentum densities causality is obeyed and the generators of the translation group become
$$P^{\mu} = \frac{2}{\kappa^2} \int d^3 \vec{k} k^{\mu} \left( a^{\dag \, \mu}_{\vec{k}}a_{\vec{k} \, \mu} + \frac{(\alpha - 1)}{2} b^{\dag \, \mu}_{\vec{k}} b_{\vec{k} \, \mu} \right)$$ while the Lorentz algebra remains unaffected (that is the usual bosonic expression).  Also, the Heisenberg evolution is satisfied for the local ``physical'' beable observables and even for the beable operators $X^{\mu}(t,\vec{x})$ if and only if $\alpha = 3$.  I verified this theory by calculating all first and second order expressions and it appears to be entirely consistent.  The causality question reduces to the bosonic one and therefore Minkowski causality is preserved in this framework.  During the calculations, one has to generalize the definition of the the Dirac function, that is $\int dx f(x) \delta(x) = f(0)$ for any bounded function $f$.  Such definition however requires one to adapt the usual notion of integral.  To appreciate this, consider the expression $$\int dx dk e^{ikx} \delta_{k,l}$$ where $\delta$ is the Kronecker delta; performing first the $x$ integral and next the $k$ integral one gets $2 \pi \delta_{l,0}$ while taking the integral in the opposite order gives exactly zero.  The correct answer for the Lebesgue integral would be zero indeed but Fubini's theorem doesn't apply here.  Now, I believe the first answer to be correct and getting it out would require to define the integral by taking a sum over \emph{suprema} of the function instead of infima, like it occurs with the Lebesgue integral.  In order to appreciate how this would work, perform a lattice regularization with spacing $\epsilon$ and length $L$ so that $\frac{L}{2 \epsilon}$ is an integer number.  Discretize the delta function by changing it to $\delta_{n, \left[ \frac{L l}{2 \pi} \right]}$ where $n$ is the wave number and the square brackets indicate the integer part of this number.  Then our integral reduces to
$$ \epsilon \sum_{j = - \frac{L}{2 \epsilon}}^{\frac{L}{2 \epsilon} - 1} \frac{2 \pi}{L} \sum_{n = - \frac{L}{2 \epsilon}}^{\frac{L}{2 \epsilon} - 1} e^{\frac{i2\pi j n \epsilon}{L}}   \delta_{n, \left[ \frac{L l}{2 \pi} \right]}.$$  Taking the sum over $j$ gives $2\pi \delta_{0, \left[ \frac{L l}{2 \pi} \right]}$ which reduces in the limit $L \rightarrow \infty$ to $2\pi \delta_{l,0}$.  Obviously, for this integral, an equivalent of Fubini's theorem doesn't hold either.  Therefore, I believe the correct attitude one should take in Quantum Field Theory is to compute all expressions in a lattice approximation, compute any bracket or expectation value one wants and \emph{then} take the continuum and thermodynamic limit.  In Quantum Field Theory, there is an extra edge to this since the operators $b_{\vec{k}}$ are distributional; considering the smallest length scale $\epsilon$ and thermodynamic length $L$ (we put the system in a box), then one obtains that $b_{\vec{k}} = \left( \frac{L}{2 \pi} \right)^{\frac{3}{2}} b'_{\vec{n}}$ where $\vec{k} = \frac{2 \pi}{L}\vec{n}$ and $b'_{\vec{n}}$ satisfies either $\left[ b'_{\vec{n}}, b'^{\dag}_{\vec{m}} \right] = \delta_{\vec{n}, \vec{m}}$ or $\{ b'_{\vec{n}}, b'^{\dag}_{\vec{m}} \} = \delta_{\vec{n}, \vec{m}}$.  An integral we meet during the above computations is given by
$$\int \frac{d^3 \vec{k} \, d^3 \vec{l}}{\sqrt{k} \sqrt{l}} k l_r e^{i((k-l)t - (\vec{k} - \vec{l}).\vec{x})} b^{\dag}_{\vec{k}} b_{\vec{k}} \delta_{\vec{k},\vec{l}}$$ and going over to the lattice language as above reduces the whole thing to
$$\sum_{ n_j = - \frac{L}{2 \epsilon} }^{\frac{L}{2 \epsilon} - 1} \left( \frac{2  \pi}{L} \right)^{3} \frac{2 \pi n_r}{L} b'^{\dag}_{\vec{n}} b'_{\vec{n}}$$ which is a well defined non vanishing operator.  Taking the integral over $x$, gives $$ \epsilon^3 \sum_{x_k = - \frac{L}{2 \epsilon}}^{\frac{L}{2 \epsilon}-1}\sum_{ n_j = - \frac{L}{2 \epsilon} }^{\frac{L}{2 \epsilon} - 1} \left( \frac{2  \pi}{L} \right)^{3} \frac{2 \pi n_r}{L} b'^{\dag}_{\vec{n}} b'_{\vec{n}} = \sum_{ n_j = - \frac{L}{2 \epsilon} }^{\frac{L}{2 \epsilon} - 1} \left( 2  \pi \right)^{3} \frac{2 \pi n_r}{L} b'^{\dag}_{\vec{n}} b'_{\vec{n}}$$ which reduces to
$$(2\pi)^3 \sum_{ n_j = - \frac{L}{2 \epsilon} }^{\frac{L}{2 \epsilon} - 1} \left( \frac{2  \pi}{L} \right)^{3} \frac{2 \pi n_r}{L} b^{\dag}_{\vec{k}} b_{\vec{k}}$$ which is the usual integral.  Anyhow, we come to the surprising conclusion that Lorentz (but not Poincar\'e) covariance is preserved even though we have a very deviant statistics.  Actually, we have two Planck constants in our theory: $\hbar$ and $\frac{(\alpha - 1)\hbar}{2}$.  Demanding both to be equal reduces to $\alpha = 3$ or 
$$b^{\mu}_{\vec{k}}b^{\dag \, \nu}_{\vec{k}} + 3b^{\dag \, \nu}_{\vec{k}}b^{\mu}_{\vec{k}} = \delta(0)\eta^{\mu \nu}.$$  Let us study what happens when we relax one of our conditions in two ways : (a) define $b^{\mu}_{\vec{k}}b^{\dag \, \nu}_{\vec{l}} + 3b^{\dag \, \nu}_{\vec{l}}b^{\mu}_{\vec{k}} = \delta(\vec{k} - \vec{l})\eta^{\mu \nu}$ which is totally inconsistent since taking the conjugate of this relation brakes the symmetry between $\vec{k}$ and $\vec{l}$ (b) allow for Fermi parastatistics.  In case (b) however, Poincar\'e invariance is totally safe; when $1 < \alpha \neq 3$ one gets two copies of the Poincar\'e algebra, each with a different Planck constant.  Moreover, for $\alpha = 3$, the Heisenberg equations of all operators are satisfied: this is good news since now we can make the $b$ modes into truly propagating degrees of freedom.  In case $\kappa$ is not constant, fairly elementary calculations show that Minkowski causality is broken since different directions in spacetime ``commute'' in different ways.  The reader can verify this by calculating $\left[  P^j(t,\vec{x}), P^{k}(t, \vec{y}) \right]$.  To summarize, our conclusions are that we have two possible values of $\alpha$ compatible with Poincar\'e invariance, that is $\alpha = 1,3$.  In both cases one can consider Fermi parastatistics which will preserve Poincar\'e invariance but destroy Minkowski causality.  The latter fact is not necessarily a bad thing since we should find out how to define commutation relations with respect to the dynamical metric.  But it is my philosophy that the issue of renormalization of operators will guide us towards such notion.  For $\alpha = 1$ and Fermi parastatistics, the Heisenberg evolution is satisfied for physical operators; in case $\alpha = 3$ this also holds for the beable operators.  Although we just learned that some novel parastatistics is possible, by making $\kappa$ nonconstant, without violating Poincar\'e covariance, the latter does however introduces a relative momentum dependence and therefore we dismiss it as unphysical.  This leaves us with fairly elementary statistics which is rather well known to exist in the literature and the natural thing to do is apply our complexification trick and add bosons to ``imaginary'' fermions while still preserving Poincar\'e invariance and causality.  We shall shortly see we have to apply an almost isomorphic trick again, but the effect of it will be the ``destruction'' of Minkowski causality.  As explained before, this is what we \emph{want} since we insist upon causality being dynamical too.  Until now, there was no natural mechanism which violated it (breaking isotropy of space is no option) so in a sense, the infinities in the metric algebra had to show up.  Note that our ansatz so far contains a certain symmetry; that is, instead of $c = a + ib$ we could also have written $c = a + ib^{\dag}$ and all the results would have been the same (except that the Heisenberg equations would be broken).  The infinities at the level of the Hamiltonian might have also been killed by putting $c= a + ia'^{\dag}$ where $a'$ is bosonic, but this would come at the cost of an unbounded negative energy spectrum - however, the anomalies in the metric algebra would \emph{not} show up as an easy calculation confirms.  The latter view is consistent with the Heisenberg equations of motion; insisting upon a positive energy (as well as getting rid of the infinities) will destroy the Heisenberg equations and require two fermions besides these two bosons.  So here we are meeting our very first physical difficulty: should we allow for negative energies and preserve the Heisenberg equations while having a purely (para)bosonic theory, or should we insist upon positive energies only and have a theory with a symmetry between (para)bosons and (para)fermions but alas, the Heisenberg relations get destroyed (of course, our fermions remain strange).  I believe some theoretical bias is allowed here since the Heisenberg equations have proven their validity for a long while and moreover, it is clear that the negative mass solution is preferred from the mathematical perspective.  This suggests that we might want to dispose of our psychological prejudices and proceed by accepting what the theory and Poincar\'e invariance are telling us.  Negative mass can peacefully coexist with the equivalence principle as general relativity itself allows for negative mass solutions.  I cannot -moreover- stress enough that the negative energy particles live at the level of beables where energy operators are quadratic; at the level of particles our operators are at least quartic, so it may very well be that the physical energy is positive (just like $(-1)^2=1$).  Although negative energies solve these two renormalization problems at once, it does not solve all of them!  This can be easily seen by taking the anticommutator of the metric with itself, then the canceling (in case of the commutator) becomes an amplification.  There is no way we can solve this by means of bosons alone; therefore one might think of adding a pair of fermions just as we did before (that is a term of the form $b + ib'^{\dag}$ restoring symmetry between bosons and fermions), however this wouldn't help either since all fermionic terms would already vanish at the level of the metric tensor - only the cross terms $ab+ba$ and $-a'b'^{\dag} - b'^{\dag}a'$ would survive and these come all with the wrong sign in order to cancel the bosonic terms (again there is an amplification).  This conclusion also holds even if $a,b$ and $a',b'$ anticommute with one another instead of commuting.  Clearly, the complex numbers are insufficient to solve this problem: in other words, our fermions do not have the correct statistics.  It is kind of a miracle that a consistent quantization of a theory anyone would think of as bosonic forces one to go beyond the complex numbers and therefore consider the correct statistical properties of fermions. \\* \\*  The lack of a canonical definition of $\star$ (in the context of reducing expressions by means of the volume constraint) will put out its ugly head right here and we have to find better independent arguments as how to proceed.  It would be very bad indeed to let a quantization scheme depend upon a constraint one imposes to facilitate computations (amongst others).  Suppose we would apply the star product in the definition of the Riemann tensor, then a simple calculation reveals that our theory reduces to the classical case.  That is: the Riemann tensor is exactly zero and no quantum corrections are induced, something which is entirely unphysical as it would be equivalent to saying that $[X,P] = 0$ quantum mechanically.  This leads one to appreciate that connections are objects in their own right and therefore the Riemann tensor is uniquely defined as:
$$R^{\alpha}_{\gamma \delta \beta} = \partial_{[ \delta} \Gamma^{\alpha}_{\beta] \gamma} + \frac{1}{2} \left( \Gamma^{\alpha}_{[\delta | \kappa |}  \Gamma^{\kappa}_{\beta] \gamma} + \Gamma^{\kappa}_{[\beta |\gamma |} \Gamma^{\alpha}_{\delta ] \kappa} \right).$$  
For now, this is sufficient and the attentive reader knows now that the first thing to do is to reexamine the relation between spin and statistics.
\chapter{Towards a generalized spin statistics ``theorem'': mathematical preliminaries} 
As explained in the introduction, we cannot rely upon the standard spin statistics theorem anymore since the demand of Minkowski causality is not physical anymore.  By now, we have also discovered that the positive energy assumption is likely to be false; nevertheless we have two other physical conditions which are (a) the absence of infinities (b) the existence of a well defined number operator.  We have already treated the statistics side of the question and the result was that only (para)Bose, (para) Fermi and combinations of both are allowed for.  Moreover, the problem of infinities thought us that we have to introduce non commutative number algebra's: that is the correct spin needs to be assigned to particles with a particular statistics.  We shall explicitly work this idea out up till fourth order in our theory and study whether particles with spin $1/2$ need to be fermions.  This shall also make sure that they become propagating degrees of freedom: a problem we had with the ``wrong'' fermions we considered previously.  Since we would like to make this as much of a theorem as possible, we have to eliminate other possibilities such as (para)bosonic particles with spin $1/2$.  Indeed, we shall first assume this conservative attitude and show that it reasonably leads to contradictions\footnote{We leave it as a future exercise to close the tiny gaps left open here; doing so would lead us too much astray and it would probably not be of much value either.}.  The result of our investigation is quite ``revolutionary'' and will be treated in more detail in the next chapter; that is, spin statistics is more fundamental than causality is.  The first question one must ask is ``what are the correct non-commutative numbers?''.  Obviously, these must be the generators of the spin algebra since one must obtain covariant transformation properties under Lorentz transformations.  A mathematician would say that we need to work with the abstract Clifford algebra defined by $$e_{\mu} e_{\nu} + e_{\nu} e_{\mu} = 2 \eta_{\mu \nu} 1$$ and equipped with the reversion $\tilde{}$ defined by $$\tilde{e}_{\mu} = \eta_{\mu \mu} e_{\mu} = e_{0}e_{\mu}e_{0}.$$  Indeed, everything we do just depends upon this algebra and not upon a particular representation.  Nevertheless, I shall employ the physicists notation where the $e_ {\mu} = \gamma_{\mu}$ and $\tilde{} = \dag$.  As usual, we employ the Feynman slash notation where $\displaystyle{\not}p = p^{\mu}\gamma_{\mu}$.  From now on, our $X^{\mu}$ become Clifford valued operators and the usual bosonic part just gets a factor of $1$, therefore the scalar product should be redifined by taking the operation $\frac{1}{4} \textrm{Tr}$ since our representation is four dimensional (again, this can be abstractly expressed in terms of the Clifford algebra).  For a spin transformation $U$ one has that $U \gamma^0 = \gamma^0 U^{\dag \, -1}$ and the reader can take the inverse and/or Hermitean conjugate of this relation.  Since our conjugation should preserve the bosonic sector, $X^{\mu}(t,\vec{x})$ must transform as 
$$X^{\mu}(t',\vec{x}') = U X^{\mu}(t,\vec{x}) U^{-1}.$$  Writing $X^{\mu}$ as before gives
$$X^{\mu}(t,\vec{x}) =   \frac{1}{\sqrt{2}} \int \frac{d^3 \vec{k}}{\sqrt{k}} \left( e^{-i(kt - \vec{k}.\vec{x})} \alpha_{\vec{k}}a_{\vec{k}} + e^{i(kt - \vec{k}.\vec{x})} \beta_{\vec{k}}a^{\dag}_{\vec{k}} + \, \textrm{plus previous terms} \right)$$
where we take for now $a_{\vec{k}}$ to be (para)bosonic (later on we shall consider fermions).  We need to show that there exist no $\alpha_{\vec{k}}$ and $\beta_{\vec{k}}$ such that two conditions are satisfied : $$\alpha_{\vec{k}} \beta_{\vec{k}} + \beta_{\vec{k}} \alpha_{\vec{k}}$$ must square\footnote{Normally, one would expect this to be $2$.  However, as mentioned previously, the particle level is quartic in the operators and therefore we may allow for a ``nontraditional'' expression on the level of beables.} to a multiple of the identity and $$\left(\alpha_{\vec{k}}\alpha_{\vec{l}} + \alpha_{\vec{l}} \alpha_{\vec{k}} \right)\left(\beta_{\vec{k}}\beta_{\vec{l}} + \beta_{\vec{l}} \beta_{\vec{k}} \right) = - 1.$$  The first condition expresses positivity of the energy while the latter is the necessary condition for canceling the $$\{ a'a',a'^{\dag} a'^{\dag} \}$$ terms in the anticommutator of the metric with itself.  The above transformation properties imply that $$U \alpha_{\vec{k}} U^{-1} = \alpha_{\vec{k}'}$$ where $k'^{\mu} = \Lambda^{\mu}_{\nu} k^{\nu}$ and $U\gamma^{\mu}U^{-1} = (\Lambda^{-1})^{\mu}_{\nu}\gamma^{\nu}$.  Consider for simplicity $$\alpha_{\vec{k}} = \alpha 1 + \beta \gamma^5 + \kappa \displaystyle{\not}k + \delta \displaystyle{\not}k \gamma^5$$ where $\gamma^5 = i\epsilon_{\alpha \beta \kappa \delta} \gamma^{\alpha}\gamma^{\beta}\gamma^{\kappa}\gamma^{\delta}$ satisfying $(\gamma^{5})^2 = 1$ and $(\gamma^5)^{\dag} = \gamma^5$; then it is natural to take $$\beta_{\vec{k}} = \pm \gamma^{0} \alpha^{\dag}_{\vec{k}} \gamma^{0}$$ given that the mappings $$R_{\pm}: X^{\mu} \rightarrow \pm \gamma^{0}X^{\mu \,\dag}\gamma^{0}$$ are involutions.  That is $R_{\pm}^2 = 1$, however $R_{-}$ is preferred since it preserves the unity $R_{-}(1) = 1$.  Therefore, we shall demand that physics is $R_{-}$ symmetric: this leads to a generalization of the comlexification trick we considered in the previous chapter.  Anyway, we suppose for now that $X^{\mu}$ is real: that is $R_{-}(X^{\mu})=X^{\mu}$.  Hence, $$\beta_{\vec{k}} = \overline{\alpha}1 - \overline{\beta} \gamma^5 - \overline{\kappa} \displaystyle{\not}k - \overline{\delta} \displaystyle{\not}k \gamma^5$$ and a lengthy, though straightforward calulation reveals that
\begin{eqnarray*}
\left(\alpha_{\vec{k}}\alpha_{\vec{l}} + \alpha_{\vec{l}}\alpha_{\vec{k}}\right)
\left(\beta_{\vec{k}}\beta_{\vec{l}} + \beta_{\vec{l}}\beta_{\vec{k}}\right) & = & \left( 8 |\alpha|^2 \beta \overline{\delta} - 4(\alpha^2 + \beta^2 + (\kappa^2 - \delta^2)k.l)\overline{\alpha} \overline{\kappa} - hc \right)(\displaystyle{\not}k + \displaystyle{\not}l) \\*
& + & \left( 8 |\alpha|^2 \beta \overline{\kappa} - 4(\alpha^2 + \beta^2 + (\kappa^2 - \delta^2)k.l)\overline{\alpha} \overline{\delta} - hc \right)(\displaystyle{\not}k + \displaystyle{\not}l) \gamma^5 \\*
& + & \left( 8 |\alpha|^2 \delta \overline{\kappa}k.l - 8(\alpha^2 + \beta^2 + (\kappa^2 - \delta^2)k.l)\overline{\alpha} \overline{\beta} - hc \right)\gamma^5 \\*
& + & \left( 8 |\alpha|^2 (|\delta |^2 - |\kappa|^2)k.l - 16|\alpha|^2 |\beta|^2 + |2\alpha^2 + 2\beta^2 + 2(\kappa^2 - \delta^2)k.l|^2 \right)1.
\end{eqnarray*}        
Demanding this to be $-1$, leads to a few conditions\footnote{In an intermediate calculation, the alternative $\beta = r\alpha$ with $r$ real occurs but this condition keeps the product positive and is therefore excluded.}: (a) $\kappa = \pm \delta$ and (b) $\beta = e^{i\theta} \alpha$ where $\theta$ is a multiple of $\frac{\pi}{4}$ and $\kappa = r \alpha (1 \mp e^{i \theta})^2$.  Furthermore, one can show that a recurring pattern occurs with symmetry group $\mathbb{Z}_4$ and only $\frac{\pi}{4}$ and $\frac{\pi}{2}$ lead to different negative expressions which are $-8 |\alpha|^4$ and $-16| \alpha|^4$ respectively.  For these values, one computes the anticommutator $\{ \alpha_{\vec{k}}, \beta_{\vec{k}} \}$ and the latter equals
$$- |\alpha|^2 \left( 2 \sqrt{2} i \gamma^5 + 4ir \displaystyle{\not}k(1 \mp \sqrt{2}) \pm 4ri\displaystyle{\not}k \gamma^5 (1 \mp \sqrt{2}) \right)$$ for $\theta = \frac{\pi}{4}$ and a similar expression for $\theta = \frac{\pi}{2}$.  Clearly, this cannot be made equal to $2$; moreover, not even the square of it could realize to a multiple of the identity.  One might break the reality condition in the most brutal way by choosing totally different coefficients for $\beta_{\vec{k}}$.   The reader notices that this also leads to more general conditions and the computations become even much more lengthy.  \\* \\*
It is now clear that we should add a fermionic action principle to the standard bosonic one.  Several candidates do exist:
$$ S_1 = \beta \textrm{Tr} \, \int d^4x X^{\mu}\gamma^{\alpha} \partial_{\alpha} X_{\mu}$$ and
$$ S_2 = \beta \textrm{Tr} \, \int d^4x \gamma^{0 }X^{\mu \, \dag} \gamma^{0} \gamma^{\alpha} \partial_{\alpha} X_{\mu}.$$
The latter is the standard one, while the former is a truly novel description of Fermi particles.  Assuming that $X^{\mu}$ is real, that is $R_{\pm}X^{\mu} = X^{\mu}$, then $$R_{\pm} S_2 = \mp \frac{\overline{\beta}}{\beta} S_2$$ and $$R_{\pm}S_1 =  \mp \frac{\overline{\beta}}{\beta} \textrm{Tr} \, \int d^4x R_{\pm}X^{\mu} \gamma^{\alpha} \partial_{\alpha} R_{\pm} X_{\mu}.$$
Therefore, in both cases we arrive at the conclusion that $\beta$ is imaginary for $R_{+}$ and real for $R_{-}$.  Clearly, in order to have real energies, we must pick out $R_{+}$ as the reality mapping; more precisely, we want that the correct momentum operators are given by $i \partial_{\mu}$ and that $\displaystyle{\not}k$, with real $k_{\mu}$, is an Hermitian operator.  The first requirement comes from ordinary bosonic quantum mechanics (in a unified theory, the energy operator must be the same for bosons and fermions) while the second one says that the gamma matrices must have an Hermitian representation.  Note that this is not the case in quaternion quantum mechanics where the reality conditions on the quaternion elements do originate from an involution in a more complex way by separating the central piece of a quaternion from its vector and bivector part (the central piece is Hermitian while the vector part is not).  The relevant involution is equivalent to $R_{-}$ here and $\displaystyle{\not}k$ is therefore an anti-Hermitian operator in this formalism\footnote{A natural basis in $\mathcal{C}_{(1,3)}$ is given by using $\gamma^5$ for the pseudo vector and pseudo unit part as the reader may easily verify.}.  Indeed, the role of the imaginary unit in the real quaternion formalism is very subtle and the only reason to accept it is the virtue of working with a division algebra.  Since we do not have such luxury here, I see no natural argument to keep this split between the center and the rest of the algebra; that is we look at the comlex numbers as a whole and not as a real Clifford algebra.  Hence, we do not walk on this avenue but if the reader wishes to do so, he may decide otherwise. \\* \\* Therefore, we must say that the ``bosonic'' action is real by referring to $R_{-}$ and the ``fermionic'' action is real with respect to $R_{+}$.  This can be easily understood from the fact that $R_{\pm} X^{\mu}X^{\nu} = \mp X^{\nu}X^{\mu}$ so $R_{+}$ obeys anticommutation rules while $R_{-}$ satisfies commutation rules.  It is clear that $S_2$ is an anti-Hermitian action with respect to $\dag$ and therefore it is inconsistent.  Hence, do we have to accept that $S_1$ is our novel candidate for describing fermions?  A small inspection yields that it allows for very different solutions than the Dirac equation and therefore we did not find the correct action yet.  However, we have overlooked a reality condition $\widetilde{X^{\mu}}$ which is \emph{compatible} with the bosonic sector (something which one cannot achieve in the vector formalism) 
$$\widetilde{X^{\mu}} = \gamma^{0} \gamma^5 X^{\mu \, \dag} \gamma^{5 \, \dag} \gamma^{0 \, \dag}$$ and we shall explain later on why we used $\gamma^{5 \, \dag}$ instead of $\gamma^5$.  Indeed, $$\widetilde{X^{\mu}X^{\nu}} = \widetilde{X^{\nu}}\widetilde{X^{\mu}}$$ and \begin{eqnarray*}
\widetilde{1} & = & 1 \\*
\widetilde{\gamma^{\alpha}} & = & \gamma^{\alpha} \\*
\widetilde{\gamma^{\alpha} \gamma^{\beta}} & = & \gamma^{\beta} \gamma^{\alpha} \\*
\widetilde{\gamma^5} & = & - \gamma^{5 \, \dag} \\*
\widetilde{\gamma^{\alpha} \gamma^5} & = & - \gamma^{5 \, \dag} \gamma^{\alpha}. \\*
\end{eqnarray*} 
This leads one to suggest an action
$$S_3 = \pm i \, \textrm{Tr} \, \int d^4x \widetilde{X}^{\mu} \gamma^{\alpha} \partial_{\alpha} X_{\mu}$$ and a small computation yields that $S_3$ is hermitian with respect to $\dag$ and therefore also with respect to $\, \widetilde{}$.  One could add the following mass term
$$S_4 = \pm m\, \textrm{Tr} \, \int d^4x \left( \widetilde{X}^{\mu} X_{\mu} + \gamma^{0} X^{\mu \, \dag} \gamma^{0} X_{\mu} \right)$$
however, our new reality criterion cannot prevent one from considering spin-statistics violating actions (as well as its complex conjugate) such as
$$S_5 = \alpha'  \, \textrm{Tr} \, \int d^4 x \left( \gamma^{\alpha} \partial_{\alpha} X^{\mu} \gamma^{\beta} \partial_{\beta} X_{\mu} \right)$$                
where $\alpha'$ is a real constant.  Either, we must find new physical principles why these actions are excluded; otherwise, they exist and we must investigate the consequences.  The reader is invited to write down all other possible action principles in second order of the fields.  Let me briefly explain why we wrote $\gamma^{5 \, \dag}$ previously; this has to do with the very definition of $\gamma^{5}$.  That is, mathematicians who are in love with real Clifford algebra's have to define $\gamma'^{5} = - i \gamma^{5}$ and therefore $\widetilde{\gamma'^{5}} = \gamma'^{5}$.  In other words, our $\, \widetilde{}\,$ is equivalent to the definition of Hestenes \cite{Hestenes} and the reader may compute that for a spin transformation $U$, $\widetilde{U}U = U\widetilde{U}=1$.  We shall now second quantize the field theory defined by $S_3$, that is the massless case, and one may suspect that some clever regularization procedure is necessary since quantizing massless fermions is also quite tricky in the vector formalism (actually, usually it is avoided, see \cite{Peskin}). Obviously, the number of degrees of freedom for a real Clifford field equals $16$, eight in the odd sector and eight in the even sector.  This means there is twice as much information in our language than in the Dirac language which points in the direction of negative mass solutions.  We can solve the odd sector which is legitimate since -in the massless case- the odd and even sectors do not interact.  This is a severe problem since it leads to the conclusion that the (classical and quantum) beable Hamiltonian and momentum operators should be exactly zero in contrast to what occurs for the massless Dirac equation.  The reason is that the action density is odd valued (in the $\gamma$ matrices) and therefore the trace is exactly zero which really means that there is strictly speaking no equation of motion (at least if one does not consider variations of the fields which can belong to the \emph{even} sector).  Therefore, we are obliged to treat both sectors at once which is in conflict with the ideas of Hestenes \cite{Hestenes} who considers the even sector only.  Consider      
$$X^{\mu}(t,\vec{x}) = \int \frac{d^3 \vec{k}}{\sqrt{2k}} \, \left( \alpha_{\vec{k}} e^{-i(kt - \vec{k}.\vec{x})} + \beta_{\vec{k}} e^{i(kt - \vec{k}.\vec{x})} \right)$$
where we may impose at a later stage that $\widetilde{\alpha}_{\vec{k}} = \beta_{\vec{k}}$ and realize that the reality condition imposed on the field $X^{\mu}$ is not a problem \emph{per se}.  Obviously, $S_3$ leads to $$\gamma^{\alpha}\partial_{\alpha} X^{\mu}= 0$$ and therefore, on the odd sector, $\alpha_{\vec{k}} = \displaystyle{\not}u + \displaystyle{\not}v\gamma^5$ satisfies
$$\displaystyle{\not}k\displaystyle{\not}u + \displaystyle{\not}k\displaystyle{\not}v\gamma^5 = 0$$ whose solutions are given by $$\alpha_{\vec{k}} = a_{\vec{k}}\displaystyle{\not}k + b_{\vec{k}}\displaystyle{\not}k \gamma^5 + c_{\vec{k}}(\displaystyle{\not}n_1 - i\displaystyle{\not}n_2\gamma^{5}) + d_{\vec{k}}(\displaystyle{\not}n_2 + i\displaystyle{\not}n_1\gamma^{5})$$ where $(\vec{n}_j)^2 = 1$, $n_j.k = 0$, $n_j.l = 0$ where $l.l = 0$ and $k.l = -1$.  Moreover, orientability requires that $\epsilon_{\alpha \beta \gamma \delta}k^{\alpha}l^{\beta}n_1^{\gamma}n_2^{\delta} = -1$.  To interpret these states, consider $k^{\alpha}=(1,1,0,0)$ and choose $n^{\alpha}_1 = (0,0,1,0)$, $n^{\alpha}_2 = (0,0,0,1)$; then the helicity operator is given by $S = \frac{i}{2} \gamma^{2}\gamma^{3}$.  As is well known \cite{Weinberg}, the little group of $k$ is three dimensional and is determined by the helicity rotations 
$$R(\psi) = \left( \begin{array}{cccc}
1 & 0 & 0 & 0 \\
0 & 1 & 0 & 0 \\
0 & 0 & \cos\psi & - \sin\psi \\
0 & 0 & \sin\psi & \cos\psi 
\end{array} \right) 
$$ and the ``translations''
$$S(\alpha, \beta) = \left( \begin{array}{cccc}
1 - \zeta & \zeta & \alpha & \beta \\
- \zeta & 1 + \zeta &  \alpha  &  \beta \\
\alpha & -\alpha & 1 & 0 \\
\beta & -\beta & 0 & 1 
\end{array} \right) 
$$
where $\zeta = \frac{(\alpha^2 + \beta^2)}{2}$.  A generic element is therefore of the form $S(\alpha,\beta)R(\psi)$.  One computes the following remarkable identities:
\begin{eqnarray*}
\{ S, \displaystyle{\not}p_1 \} & = & 0 \\*
\{ S, \displaystyle{\not}p_2 \} & = &  0 \\*
\{ S, \displaystyle{\not}k \} & = & \displaystyle{\not}k \gamma^5 \\*
\{ S, \displaystyle{\not}k \gamma^5 \} & = & \displaystyle{\not}k \\*
\end{eqnarray*}    
and 
\begin{eqnarray*}
\left[ S, \displaystyle{\not}p_1 \right] & = & - i \displaystyle{\not}p_2 \\*
\left[ S, \displaystyle{\not}n_2 \right] & = &  i \displaystyle{\not}p_1 \\*
\left[ S, \displaystyle{\not}k \right] & = & 0 \\*
\left[ S, \displaystyle{\not}k \gamma^5 \right] & = & 0 \\*
\end{eqnarray*}         
where $\displaystyle{\not}p_1 = \displaystyle{\not}n_1 - i\displaystyle{\not}n_2\gamma^{5}$ and $\displaystyle{\not}p_2 = 
\displaystyle{\not}n_2 + i\displaystyle{\not}n_1\gamma^{5}$.  Although the commutator and anticommutator do distinguish different particles, simple left multiplication by $S$ -which equals one half of the sum of the commutator and anti-commutator- does not.  Indeed, all particles correspond to spin $\frac{1}{2}$ helicity states.  Therefore one may regard equations of the kind
$$S \left(\displaystyle{\not}p_1 - i\displaystyle{\not}p_2\right) = \frac{1}{2} \left(\displaystyle{\not}p_1 - i\displaystyle{\not}p_2\right)$$
as eigenvalue equations in the matrix language.  Consider\footnote{One calculates that $\widetilde{\displaystyle{\not}k} = \displaystyle{\not}k$, $\widetilde{\displaystyle{\not
}k\gamma^5} = \displaystyle{\not}k\gamma^5$ and $\widetilde{p}_1 = \displaystyle{\not}n_1 + i\displaystyle{\not}n_2\gamma^{5}$, $\widetilde{p}_2 = \displaystyle{\not}n_2 - i\displaystyle{\not}n_1\gamma^{5}$.  This means that for self dual fields $X^{\mu}$ one must set $c$ and $d$ to zero.  Both the helicity states $\displaystyle{\not}p_j$ require a mixture of self dual and anti selfdual fields.}
$$\alpha_{\vec{k}} = \alpha  1 + \beta \gamma^5 + \delta_{\alpha \beta} \gamma^{\alpha} \gamma^{\beta}$$ where $\delta_{\alpha \beta}$ is antisymmetric.              Insisting upon $\displaystyle{\not}k \alpha_{\vec{k}} = 0$  leads to
$$\alpha_{\vec{k}} = e_{\vec{k}}\displaystyle{\not}k\displaystyle{\not}l + f_{\vec{k}}(\gamma^5 - i\displaystyle{\not}n_1\displaystyle{\not}n_2) + g_{\vec{k}}\displaystyle{\not}k\displaystyle{\not}n_1 + h_{\vec{k}}\displaystyle{\not}k\displaystyle{\not}n_2$$ where the $\displaystyle{\not}n_j$ are as before and $l$ is the unique null vector satisfying $l.n_j = 0$ and $l.k= -1$.  While the $n_j$ are physical again, the first two are not (especially the $l$ polarization does not appear in standard physics).  To understand better what is going on, let us calculate as before the (anti) commutation relations with the helicity operator.  Since $l=\frac{1}{2}(1,-1,0,0)$ one has
\begin{eqnarray*}
\{ S, \displaystyle{\not}k\displaystyle{\not}n_1 \} & = & 0 \\*
\{ S, \displaystyle{\not}k\displaystyle{\not}n_2 \} & = & 0 \\*
\{ S, \displaystyle{\not}k\displaystyle{\not}l \} & = & \left( \gamma^5 - i\displaystyle{\not}n_1\displaystyle{\not}n_2 \right)\\*
\{ S, \gamma^5 - i\displaystyle{\not}n_1\displaystyle{\not}n_2 \} & = & \displaystyle{\not}k\displaystyle{\not}l   
\end{eqnarray*}  
and the reader can calculate that also all commutators agree as previously.
At this moment, it is useful to tell that all (including the formulae above) expressions one computes do not depend upon the choice of $\displaystyle{\not}n_j$; that is, there is a kind of local gauge invariance on momentum space.  This fact greatly simplifies our computations and the reader can compute that under a rotation $\displaystyle{\not}n'_1 = \cos\psi \displaystyle{\not}n_1 - \sin\psi \displaystyle{\not}n_2$, $\displaystyle{\not}n'_2 = \sin\psi \displaystyle{\not}n_1 + \cos\psi \displaystyle{\not}n_2$, the $\displaystyle{\not}p_j$ transform likewise.  This in turn induces a rotation between the operators $c_{\vec{k}}$ and $d_{\vec{k}}$ implying that both need to satisfy the same type of statistics (and no parastatistics here is possible).  The previous formulae indicate clearly that $\displaystyle{\not}p_j$ and $\displaystyle{\not}k\displaystyle{\not}n_j$ belong together pairwise\footnote{The reader notices that there is nevertheless a symmetry breaking between the odd and even sector respectively.  That is, the notion of (anti)self-duality $\widetilde{X} = X$ on the even sector puts  $e_{\vec{k}}$ and $f_{\vec{k}}$ to zero: therefore different species survive in different sectors.}.    Further computations will reveal that -at the level of the Hamiltonian- the $\displaystyle{\not}p_j$ and $\displaystyle{\not}k$, $\displaystyle{\not}k\gamma^5$ decouple.  Therefore, one might be tempted to simply ignore this sector and proceed by quantizing the ``particles'' determined by the $\displaystyle{\not}p_j$.  However, this is not correct since under a ``translation'' $S(\alpha,\beta)$, the $\displaystyle{\not}p_j$ transform as
\begin{eqnarray*} \displaystyle{\not}p_1 & \Rightarrow & \displaystyle{\not}p_1 + \alpha \displaystyle{\not}k - i \beta \displaystyle{\not}k\gamma^{5} \\*
\displaystyle{\not}p_2 & \Rightarrow & \displaystyle{\not}p_2 + \beta\displaystyle{\not}k + i \alpha\displaystyle{\not}k\gamma^{5}
\end{eqnarray*} while the $\displaystyle{\not}k$ and $\displaystyle{\not}k\gamma^5$ remain invariant under the entire little group.  Likewise, in the even sector, $\displaystyle{\not}k\displaystyle{\not}l$ transforms as
$$\displaystyle{\not}k\displaystyle{\not}l \Rightarrow \displaystyle{\not}k\displaystyle{\not}l + \alpha \displaystyle{\not}k\displaystyle{\not}n_1 + \beta\displaystyle{\not}k\displaystyle{\not}n_2$$
and $$(\gamma^{5} - i\displaystyle{\not}n_1\displaystyle{\not}n_2) \Rightarrow (\gamma^{5} - i\displaystyle{\not}n_1\displaystyle{\not}n_2) - i\alpha\displaystyle{\not}k\displaystyle{\not}n_2 + i\beta\displaystyle{\not}k\displaystyle{\not}n_1$$
while the other two ``states'' remain invariant under $S(\alpha,\beta)$ but not under $R(\theta)$.  This is another asymmetry between the even and odd sector; the attentive reader must have noticed already that in order for the helicity relations to remain valid, the helicity operator itself has to transform properly too.  Let us examine further properties of the first pairs.  That is, it is useful to calculate the following expressions 
$$ \textrm{Tr} \, \widetilde{\left(c_{\vec{k}} \displaystyle{\not}p_1 + g_{\vec{k}}\displaystyle{\not}k\displaystyle{\not}n_1\right)}k_j\gamma^j \left(c_{\vec{k}} \displaystyle{\not}p_1 + g_{\vec{k}}\displaystyle{\not}k\displaystyle{\not}n_1\right)$$ and
$$ \textrm{Tr} \, \widetilde{\left(d_{\vec{k}} \displaystyle{\not}p_2 + h_{\vec{k}}\displaystyle{\not}k\displaystyle{\not}n_2\right)}k_j\gamma^j \left(d_{\vec{k}} \displaystyle{\not}p_2 + h_{\vec{k}}\displaystyle{\not}k\displaystyle{\not}n_2\right)$$ as well as the cross terms between both.  The first expression is calculated to be $$8 k^2 \left( c^{\dag}_{\vec{k}}g_{\vec{k}} + g^{\dag}_{\vec{k}}c_{\vec{k}} \right)$$ and likewise the second term yields $$ 8k^2 \left( d^{\dag}_{\vec{k}}h_{\vec{k}} + h^{\dag}_{\vec{k}}d_{\vec{k}} \right).$$
A straightforward calculation yields that the cross term
$$ \textrm{Tr} \,  \widetilde{\left( d_{\vec{k}} \displaystyle{\not}p_2 + h_{\vec{k}}\displaystyle{\not}k\displaystyle{\not}n_2 \right)}k_j\gamma^j \left(c_{\vec{k}} \displaystyle{\not}p_1 + g_{\vec{k}}\displaystyle{\not}k\displaystyle{\not}n_1\right)$$ 
vanish identically.  Concerning the second sector, one computes  
$$\textrm{Tr} \, \displaystyle{\not}k k_j \gamma^{j}\displaystyle{\not}k \displaystyle{\not}l$$ and $$\textrm{Tr} \, \displaystyle{\not}k k_j \gamma^{j}\left( \gamma^5 - i \displaystyle{\not}n_1 \displaystyle{\not}n_2 \right)$$ and the former equals $-8k^2$ while the latter vanishes as well as all traces with respect to the other ``sector''; replacing $\displaystyle{\not}k$ by $   \displaystyle{\not}k\gamma^5$ and $\displaystyle{\not}k\displaystyle{\not}l$ by $\gamma^5 - i\displaystyle{\not}n_1\displaystyle{\not}n_2$ gives the same result.  Therefore, the relevant expressions are
$$-8k^2\left(a^{\dag}_{\vec{k}}e_{\vec{k}} + e^{\dag}_{\vec{k}}a_{\vec{k}}\right)$$ and
$$-8k^2\left(b^{\dag}_{\vec{k}}f_{\vec{k}} + f^{\dag}_{\vec{k}}b_{\vec{k}}\right).$$
Since 
$$8 k^2 \left( c^{\dag}_{\vec{k}}g_{\vec{k}} + g^{\dag}_{\vec{k}}c_{\vec{k}} \right) = 8 k^2 \left( \frac{1}{2}(c_{\vec{k}} + g_{\vec{k}})^{\dag}(c_{\vec{k}} + g_{\vec{k}}) - \frac{1}{2}(c_{\vec{k}} - g_{\vec{k}})^{\dag}(c_{\vec{k}} - g_{\vec{k}}) \right)$$ it is clear that $(c_{\vec{k}} - g_{\vec{k}})$ cannot be an annihilation operator (otherwise the Heisenberg equations would not hold) - unless we introduce Grassmann numbers.  Hence, if we do not extend the mathematical formalism, negative energies are unavoidable in this formalism; moreover, the particles corresponding to $(c_{\vec{k}} - g_{\vec{k}})$ must be bosons and therefore violating the usual spin statistics connection.  The same comments apply to 
$$-8k^2\left(a^{\dag}_{\vec{k}}e_{\vec{k}} + e^{\dag}_{\vec{k}}a_{\vec{k}}\right) = -8k^2\left(\frac{1}{2}(a_{\vec{k}} + e_{\vec{k}})^{\dag}(a_{\vec{k}} + e_{\vec{k}}) - \frac{1}{2}(a_{\vec{k}} - e_{\vec{k}})^{\dag}(a_{\vec{k}} - e_{\vec{k}}) \right)$$ but now $(a_{\vec{k}} + e_{\vec{k}})$ should be a bosonic creation operator corresponding to a negative energy particle.  Hence, in front of $e^{-i(kt - \vec{k}.\vec{x})}$ one obtains four positive energy particles and four negative energy bosons; more precisely, the term looks like
\begin{eqnarray*} \left(\displaystyle{\not}p_1 + \displaystyle{\not}k\displaystyle{\not}n_1\right)\frac{\left(c_{\vec{k}} + g_{\vec{k}} \right)}{2} + \left(\displaystyle{\not}p_1 - \displaystyle{\not}k\displaystyle{\not}n_1\right)\frac{\left(c_{\vec{k}} - g_{\vec{k}} \right)}{2} + \left(\displaystyle{\not}p_2 + \displaystyle{\not}k\displaystyle{\not}n_2\right)\frac{\left(d_{\vec{k}} + h_{\vec{k}} \right)}{2} \\*
+ \left(\displaystyle{\not}p_2 - \displaystyle{\not}k\displaystyle{\not}n_2\right)\frac{\left(d_{\vec{k}} - h_{\vec{k}} \right)}{2} + \left(\displaystyle{\not}k + \displaystyle{\not}k\displaystyle{\not}l\right)\frac{\left(a_{\vec{k}} + e_{\vec{k}} \right)}{2} + \left(\displaystyle{\not}k - \displaystyle{\not}k\displaystyle{\not}l\right)\frac{\left(a_{\vec{k}} - e_{\vec{k}} \right)}{2} \\*
\left(\displaystyle{\not}k\gamma^5 + \gamma^5 - i\displaystyle{\not}n_1\displaystyle{\not}n_2 \right)\frac{\left(b_{\vec{k}} + f_{\vec{k}} \right)}{2} + \left(\displaystyle{\not}k\gamma^5 - \gamma^5 + i\displaystyle{\not}n_1\displaystyle{\not}n_2 \right)\frac{\left(b_{\vec{k}} - f_{\vec{k}} \right)}{2}. 
\end{eqnarray*} From the transformation properties of the operators under the action of a ``translation'' $S(\alpha,\beta)$ it\footnote{The operators transform as
\begin{eqnarray*}
\frac{\left( c_{\vec{k}} + g_{\vec{k}} \right)}{2} & \Rightarrow & \frac{\left( c_{\vec{k}} + g_{\vec{k}} \right)}{2} + \alpha \left( \frac{\left( a_{\vec{k}} + e_{\vec{k}} \right)}{4} - \frac{\left( a_{\vec{k}} - e_{\vec{k}} \right)}{4} \right) + i\beta \left( \frac{\left(b_{\vec{k}} + f_{\vec{k}} \right)}{4} - \frac{\left( b_{\vec{k}} - f_{\vec{k}} \right)}{4}   \right) \\* 
\frac{\left(c_{\vec{k}} - g_{\vec{k}}\right)}{2} & \Rightarrow & \frac{\left(c_{\vec{k}} - g_{\vec{k}}\right)}{2} - \alpha \left(\frac{\left(a_{\vec{k}} + e_{\vec{k}} \right)}{4} - \frac{\left(a_{\vec{k}} - e_{\vec{k}} \right)}{4}\right) - i\beta \left( \frac{\left(b_{\vec{k}} + f_{\vec{k}} \right)}{4} - \frac{\left(b_{\vec{k}} - f_{\vec{k}} \right)}{4}   \right) \\*
\frac{\left( d_{\vec{k}} + h_{\vec{k}} \right)}{2} & \Rightarrow & \frac{\left( d_{\vec{k}} + h_{\vec{k}} \right)}{2} + \beta \left( \frac{\left( a_{\vec{k}} + e_{\vec{k}} \right)}{4} - \frac{\left( a_{\vec{k}} - e_{\vec{k}} \right)}{4} \right) - i\alpha \left( \frac{\left(b_{\vec{k}} + f_{\vec{k}} \right)}{4} - \frac{\left( b_{\vec{k}} - f_{\vec{k}} \right)}{4} \right) \\*
\frac{\left( d_{\vec{k}} - h_{\vec{k}} \right)}{2} & \Rightarrow & \frac{\left( d_{\vec{k}} - h_{\vec{k}} \right)}{2} - \beta \left( \frac{\left( a_{\vec{k}} + e_{\vec{k}} \right)}{4} - \frac{\left( a_{\vec{k}} - e_{\vec{k}} \right)}{4} \right) + i \alpha \left( \frac{\left(b_{\vec{k}} + f_{\vec{k}} \right)}{4} - \frac{\left( b_{\vec{k}} - f_{\vec{k}} \right)}{4} \right) \\*
\frac{\left( a_{\vec{k}} + e_{\vec{k}} \right)}{2} & \Rightarrow & \frac{\left( a_{\vec{k}} + e_{\vec{k}} \right)}{2} + \alpha \left( \frac{\left( c_{\vec{k}} + g_{\vec{k}} \right)}{4} + \frac{\left( c_{\vec{k}} - g_{\vec{k}} \right)}{4} \right) + \beta \left( \frac{\left(d_{\vec{k}} + h_{\vec{k}} \right)}{4} + \frac{\left( d_{\vec{k}} - h_{\vec{k}} \right)}{4} \right) \\*
\frac{\left( a_{\vec{k}} - e_{\vec{k}} \right)}{2} & \Rightarrow & \frac{\left( a_{\vec{k}} - e_{\vec{k}} \right)}{2} + \alpha \left( \frac{\left( c_{\vec{k}} + g_{\vec{k}} \right)}{4} + \frac{\left( c_{\vec{k}} - g_{\vec{k}} \right)}{4} \right) + \beta \left( \frac{\left(d_{\vec{k}} + h_{\vec{k}} \right)}{4} + \frac{\left( d_{\vec{k}} - h_{\vec{k}} \right)}{4} \right) \\*
\frac{\left( b_{\vec{k}} + f_{\vec{k}} \right)}{2} & \Rightarrow & \frac{\left( b_{\vec{k}} + f_{\vec{k}} \right)}{2} - i\beta \left( \frac{\left( c_{\vec{k}} + g_{\vec{k}} \right)}{4} + \frac{\left( c_{\vec{k}} - g_{\vec{k}} \right)}{4} \right) + i \alpha \left( \frac{\left(d_{\vec{k}} + h_{\vec{k}} \right)}{4} + \frac{\left( d_{\vec{k}} - h_{\vec{k}} \right)}{4} \right) \\*
\frac{\left( b_{\vec{k}} - f_{\vec{k}} \right)}{2} & \Rightarrow & \frac{\left( b_{\vec{k}} - f_{\vec{k}} \right)}{2} - i\beta \left( \frac{\left( c_{\vec{k}} + g_{\vec{k}} \right)}{4} + \frac{\left( c_{\vec{k}} - g_{\vec{k}} \right)}{4} \right) + i \alpha \left( \frac{\left(d_{\vec{k}} + h_{\vec{k}} \right)}{4} + \frac{\left( d_{\vec{k}} - h_{\vec{k}} \right)}{4} \right).
\end{eqnarray*}}
follows that all particles must be bosons which was to be expected on more general grounds.  Hence, a ``proof'' of the traditional spin-statistics relation has to rely upon positive energies again.  When writing a first draft of this book, I thought that this required the machinery of Grassmann numbers.  However, as we shall see later on, the introduction of Grassmann numbers implies the concept of indefinite ``Hilbert'' spaces.  Hence, we shall introduce them right away at this stage since the latter cures the negative energy problem of our theory; it won't provide us a spin-statistics theorem though since positive probabilities were an assumption behind the traditional spin-statistics theorem too.  However, the exchange of negative energies for negative probabilities is not a mathematical isomorphism since the statistics can now be Bose as well as Fermi.  There are further thoughts one could explore here and one of them concerns the role of the Clifford numbers.  Our field operators have Clifford coefficients so should we interpret these as (a) numbers occurring at intermediate stages of the calculation implying that only expressions of the kind $\textrm{Tr}\left(X\gamma^{\alpha} \right)$ are to be evaluated on ``Hilbert'' space or (b) as a ring with respect to which to define the notion of ``Hilbert'' modules or even ``Hilbert'' star algebra's?  The latter would give physical meaning to the Clifford superpostion of two ``states'' and is a drastic extension of quantum mechanics far beyond the scope of quaternionic quantum mechanics explored by Finkelstein.  Getting the physics of (b) right is highly non-trivial and similar ``problems'' regarding the tensor product construction of Clifford modules appear here.  A further thought concerns consistency demands on a sensible interpretation of indefinite ``Hilbert'' modules: all these issues are postponed to the next chapter where some of the reader's worries shall be answered in reasonable detail.  However, for now, we shall take the pragmatic attitude of (a) and not worry about interpretation too much.  All calculations to be done are nearly isomorphic to those including Grassmann numbers and the reader may therefore verify them by himself.  The point is that there exist four particles of helicity $\frac{1}{2}$ created by $a^{\dag}_{\vec{k}}, c^{\dag}_{\vec{k}}, f^{\dag}_{\vec{k}}, h^{\dag}_{\vec{k}}$ where -assuming anti-commutation relations- one concludes that $\{a_{\vec{k}}, a^{\dag}_{\vec{l}}\} = - \{c_{\vec{k}}, c^{\dag}_{\vec{l}}\} = - \{f_{\vec{k}}, f^{\dag}_{\vec{l}}\} = \{h_{\vec{k}}, h^{\dag}_{\vec{l}}\} =  \delta( \vec{k} - \vec{l})$.  Moreover, little transformations $S(\alpha,\beta)$ mix positive and negative probability creation operators.  The latter clearly indicate that one should view $a^{\dag}_{\vec{k}}, c^{\dag}_{\vec{k}}$ and $f^{\dag}_{\vec{k}}, h^{\dag}_{\vec{k}}$ as pairs of ``mirror particles''.  The vacuum state $| 0 \rangle$ is now fixed by defining that it vanishes when acted upon by any annihilation operator (of positive and negative norm) and the one particle Hilbert space $\mathcal{H}$ is defined by applying the canonical positive norm creation operators to $|0 \rangle$.  However, the physical one particle states $\Psi$ with respect to the observer do not reside in $\mathcal{H}$, since a generic interaction term (preserving the particle number) would cause any state in $\mathcal{H}$ to leave it.  Moreover, given the canonical pairs of normed mirror particles $|\vec{k}, \sigma, + \rangle$ and $|\vec{k}, \sigma, - \rangle$ where $\sigma$ is a degeneracy label, $\Psi$ is constrained to satisfy $$\langle \Psi |\vec{k}, \sigma, + \rangle \langle \vec{k}, \sigma, + | \Psi \rangle  - \langle \Psi |\vec{k}, \sigma, - \rangle \langle \vec{k}, \sigma, - | \Psi \rangle > 0$$ implying that the set of physical states $\Psi$ is not linear\footnote{Note that Parsival's identity for indefinite ``Hilbert'' spaces becomes $$| \Psi \rangle = \sum_{\alpha} | + , \alpha \rangle \langle + , \alpha | \Psi \rangle - | - , \alpha \rangle \langle - , \alpha | \Psi \rangle.$$} (and not even convex).  Clearly, going over to another inertial frame preserves the vacuum state (since no creation and annihilation operators are mixed) but changes the one particle Hilbert space as well as the notion of physical states.  This construction can be extended to construct the observers (non-linear) ``Fock-space'' while the most important feature of this quantization is that the Poincar\'e algebra is unbroken and causality holds in an appropriate sense.  \\* \\*       
As mentioned previously, one might contemplate avoiding indefinite ``Hilbert'' spaces and introduce Grassmann numbers instead.  However, indefinite ``Hilbert'' spaces will show up automatically and we investigate all details in this framework.  It is quite something that the Grassmann numbers enter already at this stage where they are absolutely of primordial importance while most people tend to regard them merely as technical tools facilitating computations in supersymmetric theories.  Due to the ``scalar'' transformation law for the Clifford field, we are not obliged to introduce a Majorana superspinor $\theta^{A}$ where $A:0 \ldots 3$ but it suffices to work with merely two Grassmann numbers $\theta, \theta^{*}$.  Supposing that $X$ is not merely Clifford but also Grassmann valued, we arrive at the following action principle
$$S_{6} = i \, \textrm{Tr} \, \int d\theta d\theta^{*}d^{4}x \, \widetilde{X}(x,\theta,\theta^{*})\gamma^{\alpha}\partial_{\alpha}X(x,\theta,\theta^{*})$$ where the reversion has been extended on the Grassmann numbers by $\widetilde{\theta} = \theta^{*}$.  Note that the action has many global symmetries such as $X \Rightarrow e^{ia 1 + ib \gamma^5}X$, a double $U(1)$ group.  More precisely, a pure $U(1)$ and a ``twisted'' $U(1)$ since $\gamma^5$ anti-commutes with the odd algebra elements.  Likewise, one has the invariance $X \Rightarrow Xe^{b\gamma^5}$ which is $\mathbb{R}$ where the $\gamma^5$ here acts like a ``twisted'' scale factor.  Moreover, on solution space, there is a local gauge covariance in momentum space of the $\theta$ sector $\theta \Rightarrow \alpha \theta + \beta \theta^{*}$ preserving all algebraic relations which implies the condition $|\alpha|^2 - |\beta|^2 = 1$.  Hence, the group structure is $U(1) \times U(1) \times SO(1,1)$ or $\alpha = e^{i\psi}\cosh\gamma$ and $\beta=e^{i\phi}\sinh\gamma$.  We shall shortly see that another discrete symmetry $\theta \Rightarrow \theta^{*}$ will be broken by the requirement of positive energies; if we simply would have eliminated the negative energy solutions at the previous stage -which amounts to a breaking of the little group to the helicity subgroup- we would effectively violate causality in the sense that a preferred frame would emerge\footnote{Within this preferred frame, the standard anti-commutation relations between field operators do hold.}.  Even with the insertion of the $\theta$ numbers, allowing for negative energies brings us back to bosonic spin-$\frac{1}{2}$ particles.  This can be easily seen as follows; consider a linear combination of the kind $\theta a + \theta^{*}b$, then performing a boost transforms the operators
\begin{eqnarray*}
a & \Rightarrow & \cosh\gamma a + \sinh\gamma b \\*
b & \Rightarrow & \sinh\gamma a  + \cosh\gamma b
\end{eqnarray*}      
implying that $a$ and $b$ must be bosonic.  Moreover, there is an asymmetry in the sense that one must be a creation operator and the other an annihilation operator.  Calculating the Hamiltonian explicitly would lead one to the same conclusions\footnote{Notice that we encountered already another $U(1) \times U(1)$ as a transformation group on the Clifford valued fields.}.  We will assume here that the $\theta$ sector is $k$ independent which considerably simplifies the analysis; however, the reader may have to insert derivatives of the $\theta$ sector with respect to $k^{j}$ wherever necessary. \\* \\*
In general, it is advantageous to introduce the helicity states $h_{\pm \, \vec{k}} = \frac{1}{\sqrt{2}}\left(q^{1}_{\vec{k}} \mp iq^{2}_{\vec{k}}\right)$ and $s_{\pm \, \vec{k}} = \frac{1}{\sqrt{2}}\left(r^{1}_{\vec{k}} \mp ir^{2}_{\vec{k}}\right)$ where $q^j_{\vec{k}} = \displaystyle{\not}p_j + \displaystyle{\not}k \displaystyle{\not}n_j$ and $r^{j} = \displaystyle{\not}p_j - \displaystyle{\not}k\displaystyle{\not}n_j$.  Under local ``helicity'' transformations on momentum space, $h_{\pm \, \vec{k}}$ transforms as $e^{\mp i \theta}h_{\pm \, \vec{k}}$ and likewise for $s_{\pm \, \vec{k}}$.  The remaining ``states'' are all invariant under the ``helicity'' transformations;  nevertheless, further calculations reveal that one should consider $$t_{\pm \, \vec{k}} = \frac{1}{\sqrt{2}}\left(\displaystyle{\not}k + \displaystyle{\not}k\displaystyle{\not}l \mp (\displaystyle{\not}k\gamma^5 + \gamma^5 - i\displaystyle{\not}n_1\displaystyle{\not}n_2) \right)$$ and similarly $$u_{\pm \, \vec{k}} = \frac{1}{\sqrt{2}}\left(\displaystyle{\not}k - \displaystyle{\not}k\displaystyle{\not}l \mp (\displaystyle{\not}k\gamma^5 - \gamma^5 + i\displaystyle{\not}n_1\displaystyle{\not}n_2) \right).$$  Renaming the operators then leads to
\begin{eqnarray*} h_{+ \, \vec{k}}a_{\vec{k}} + h_{- \, \vec{k}}b_{\vec{k}} + s_{+ \, \vec{k}}c_{\vec{k}} +  s_{- \, \vec{k}}d_{\vec{k}} 
+ t_{+ \, \vec{k}}e_{\vec{k}} + t_{- \, \vec{k}}f_{\vec{k}} 
+ u_{+ \, \vec{k}}g_{\vec{k}} + u_{- \, \vec{k}}h_{\vec{k}}
\end{eqnarray*}
and the reader can work out the transformation laws under translations $S(\alpha,\beta)$.  Hence, we may say that   
\begin{eqnarray*} X(t,\vec{x}) & = & \int \frac{d^3 \vec{k}}{4 \sqrt{k}} \,  e^{-i(kt -\vec{k}.\vec{x})} \left( h_{+ \, \vec{k}}a_{\vec{k}} + h_{- \, \vec{k}}b_{\vec{k}} + s_{+ \, \vec{k}}c_{\vec{k}} +  s_{- \, \vec{k}}d_{\vec{k}} 
+ t_{+ \, \vec{k}}e_{\vec{k}} + t_{- \, \vec{k}}f_{\vec{k}} 
+ u_{+ \, \vec{k}}g_{\vec{k}} + u_{- \, \vec{k}}h_{\vec{k}} \right) + \\* & & e^{i(kt - \vec{k}.\vec{x})}\left( h_{+ \, \vec{k}}a'^{\dag}_{\vec{k}} + h_{- \, \vec{k}}b'^{\dag}_{\vec{k}} + s_{+ \, \vec{k}}c'^{\dag}_{\vec{k}} +  s_{- \, \vec{k}}d'^{\dag}_{\vec{k}} 
+ t_{+ \, \vec{k}}e'^{\dag}_{\vec{k}} + t_{- \, \vec{k}}f'^{\dag}_{\vec{k}} 
+ u_{+ \, \vec{k}}g'^{\dag}_{\vec{k}} + u_{- \, \vec{k}}h'^{\dag}_{\vec{k}} \right). \end{eqnarray*} 
Since,  
$$H = - i \textrm{Tr} \, \int d\theta d\theta^{*} d^3 \vec{x} \widetilde{X} \gamma^{j} \partial_j X$$  a straightforward calculation yields     
\begin{eqnarray*}  H & = & \int d\theta d\theta^{*} \int d^{3} \vec{k} k \left( a^{\dag}_{\vec{k}}a_{\vec{k}} + b^{\dag}_{\vec{k}}b_{\vec{k}} - c^{\dag}_{\vec{k}}c_{\vec{k}} - d^{\dag}_{\vec{k}}d_{\vec{k}} - e^{\dag}_{\vec{k}}e_{\vec{k}} - f^{\dag}_{\vec{k}}f_{\vec{k}} + g^{\dag}_{\vec{k}}g_{\vec{k}} + h^{\dag}_{\vec{k}}h_{\vec{k}} \right) \\*
& + & \int d\theta d\theta^{*} \int d^{3} \vec{k} k \left( - a'_{\vec{k}}a'^{\dag}_{\vec{k}} - b'_{\vec{k}}b'^{\dag}_{\vec{k}} + c'_{\vec{k}}c'^{\dag}_{\vec{k}} + d'_{\vec{k}}d'^{\dag}_{\vec{k}} + e'_{\vec{k}}e'^{\dag}_{\vec{k}} + f'_{\vec{k}}f'^{\dag}_{\vec{k}} - g'_{\vec{k}}g'^{\dag}_{\vec{k}} - h'_{\vec{k}}h'^{\dag}_{\vec{k}} \right). \end{eqnarray*}  
We now come to the conclusion\footnote{The transformation laws are given by
\begin{eqnarray*}
a_{\vec{k}} & \Rightarrow & a_{\vec{k}} + \frac{(\alpha + i \beta)}{2}(f_{\vec{k}} - h_{\vec{k}}) \\*
b_{\vec{k}} & \Rightarrow & b_{\vec{k}} + \frac{(\alpha - i\beta)}{2}(e_{\vec{k}} - g_{\vec{k}}) \\*
c_{\vec{k}} & \Rightarrow & c_{\vec{k}} - \frac{(\alpha + i \beta)}{2}(f_{\vec{k}} - h_{\vec{k}}) \\*
d_{\vec{k}} & \Rightarrow & d_{\vec{k}} - \frac{(\alpha - i\beta)}{2}(e_{\vec{k}} - g_{\vec{k}}) \\*
e_{\vec{k}} & \Rightarrow & e_{\vec{k}} + \frac{(\alpha + i \beta)}{2}(b_{\vec{k}} + d_{\vec{k}}) \\*
f_{\vec{k}} & \Rightarrow & f_{\vec{k}} + \frac{(\alpha - i\beta)}{2}(a_{\vec{k}} + c_{\vec{k}}) \\*
g_{\vec{k}} & \Rightarrow & g_{\vec{k}} + \frac{(\alpha + i \beta)}{2}(b_{\vec{k}} + d_{\vec{k}}) \\*
h_{\vec{k}} & \Rightarrow & h_{\vec{k}} + \frac{(\alpha - i\beta)}{2}(a_{\vec{k}} + c_{\vec{k}})
\end{eqnarray*}
indicating the kind of fermionic prefactor.  Hence, there is a decoupling between $a_{\vec{k}}, c_{\vec{k}}, f_{\vec{k}}, h_{\vec{k}}$ and $b_{\vec{k}}, d_{\vec{k}}, e_{\vec{k}}, g_{\vec{k}}$.} that insisting upon positive energies \emph{requires} one to include negative norm states; moreover, under Lorentz transformations, the positive norm Hilbert space transforms too.  Since the claim is rather unusual -in the sense that such construction is mandatory- we present a full proof.  The beauty of working with Grassmann numbers is that it unifies commutation and anti-commutation relations and reintroduces the commutator as the fundamental bracket.  Now, under $S(\alpha,\beta)$ the usual commutation relations need to be preserved: that is,
\begin{eqnarray*}
\left[ f_{\vec{k}}, f^{\dag}_{\vec{l}} \right] & = & - \left[ h_{\vec{k}}, h^{\dag}_{\vec{l}} \right] \\
\left[ e_{\vec{k}}, e^{\dag}_{\vec{l}} \right] & = & - \left[ g_{\vec{k}}, g^{\dag}_{\vec{l}} \right] \\*
\left[ b_{\vec{k}}, b^{\dag}_{\vec{l}} \right] & = & - \left[ d_{\vec{k}}, d^{\dag}_{\vec{l}} \right] \\*
\left[ a_{\vec{k}}, a^{\dag}_{\vec{l}} \right] & = & - \left[ c_{\vec{k}}, c^{\dag}_{\vec{l}} \right] \\*
\left[ a_{\vec{k}}, a^{\dag}_{\vec{l}} \right] & = & - \left[ f_{\vec{k}}, f^{\dag}_{\vec{l}} \right] \\*
\left[ e_{\vec{k}}, e^{\dag}_{\vec{l}} \right] & = & - \left[ b_{\vec{k}}, b^{\dag}_{\vec{l}} \right].
\end{eqnarray*}
Taking into account our preconsiderations we may without limitation of generality write that 
$a_{\vec{k}} = \left(\alpha 1 + \frac{1}{2 \overline{\alpha}}(1 + ir) \theta^{*}\theta \right)a^{1}_{\vec{k}} + \left(\cosh\phi \, \theta + e^{i\psi}\sinh\phi \, \theta^{*} \right)a^{2}_{\vec{k}}$ where $r$ is any real number and $$\int d\theta d\theta^{*} a^{\dag}_{\vec{k}}a_{\vec{k}} = a^{1 \, \dag}_{\vec{k}}a^{1}_{\vec{k}} + a^{2 \, \dag}_{\vec{k}}a^{2}_{\vec{k}}.$$  Moreover, causality requires that the commutators $\left[a_{\vec{k}}, a_{\vec{l}}\right]$ and $\left[a_{\vec{k}}, a^{\dag}_{\vec{l}}\right]$ have to vanish for $\vec{k} \neq \vec{l}$ implying that $a^{1}_{\vec{k}}$ is bosonic and $a^{2}_{\vec{l}}$ is fermionic (and the usual commutation relations hold between both species).  The commutator
$$\left[a_{\vec{k}} , a^{\dag}_{\vec{l}} \right] = \delta^3 (\,\vec{k} - \vec{l}\,)|\alpha|^2$$ and therefore the fermionic $- \delta^3(\,\vec{k} - \vec{l}\,)\theta^{*}\theta$ is cancelled by the boson.  We can write down that $$c_{\vec{k}} = \left(\alpha e^{i\zeta} - \frac{1}{2\overline{\alpha}e^{-i\zeta}}(1 + is)\theta^{*}\theta\right)c^{1}_{\vec{k}} + \left(\sinh\eta \, \theta + e^{i\kappa}\cosh\eta \, \theta^{*}\right)c^{2}_{\vec{k}}$$ and likewise one can calculate that
$$ -\int d\theta d\theta^{*} c^{\dag}_{\vec{k}}c_{\vec{k}} = c^{1 \, \dag}_{\vec{k}}c^{1}_{\vec{k}} + c^{2 \, \dag}_{\vec{k}}c^{2}_{\vec{k}}$$ 
and
$$\left[c_{\vec{k}} , c^{\dag}_{\vec{l}} \right] = \delta^3 (\,\vec{k} - \vec{l}\,)|\alpha|^2.$$
Obviously, it is impossible to get $- \delta^3 (\,\vec{k} - \vec{l}\,)|\alpha|^2$ and therefore $\alpha = 0$ implying that the bosonic sector has to vanish.  This is the proof of the spin-statistics theorem and indeed, the reader may verify that the fermionic part only satisfies the proper expressions.  The reader must verify that this result crucially depends upon the assumption of positive energies and probabilities and that it falls apart when one condition fails.  Now, we will prove that negative probabilities are unavoidable: one may write down that
\begin{eqnarray*}    
a_{\vec{k}} & = & \left(\cosh\phi \, \theta + e^{i\psi}\sinh\phi \, \theta^{*} \right)\widetilde{a}_{\vec{k}} \\*
h_{\vec{k}} & = & \left(\cosh\phi' \, \theta + e^{i\psi'}\sinh\phi' \, \theta^{*} \right)\widetilde{h}_{\vec{k}} \\*
f_{\vec{k}} & = & \left(\sinh\phi'' \, \theta + e^{i\psi''}\cosh\phi'' \, \theta^{*} \right)\widetilde{f}_{\vec{k}}
\end{eqnarray*}
and under $S(\alpha,\beta)$, $a_{\vec{k}}$ becomes
\begin{eqnarray*}
&& \left(\cosh\phi \, \theta + e^{i\psi}\sinh\phi \, \theta^{*} \right)\widetilde{a}_{\vec{k}} + \\* && \frac{(\alpha + i\beta)}{2}\left( \left(\sinh\phi'' \, \theta + e^{i\psi''}\cosh\phi'' \, \theta^{*} \right)\widetilde{f}_{\vec{k}} - \left(\cosh\phi' \, \theta + e^{i\psi'}\sinh\phi' \, \theta^{*} \right)\widetilde{h}_{\vec{k}}\right).
\end{eqnarray*}
There are two different ways to interpret this result but both lead to the same conclusion: (a) $a_{\vec{k}}$ is \emph{not} of the mandatory form anymore for corresponding to a positive (or negative as a matter of fact) energy particle since the Grassmann numbers mingle with the operators on Hilbert space (b) the expectation value $$ \langle 0 | \widetilde{f}_{\vec{k}} \int d\theta d\theta^{*} a^{\dag}_{\vec{k}}a_{\vec{k}} \, \widetilde{f}^{\dag}_{\vec{k}} | 0 \rangle$$ turns out to be negative.  As a response to (a), one must first realize that the $\theta$  numbers cannot be represented as operators on Hilbert space\footnote{This is quite obvious since $\theta^{*}\theta$ and $\theta\theta^{*}$ are both positive operators.  Demanding them to be opposite implies that $\theta = 0$.} and they require indefinite norm spaces.  The statement of (b) can easily be verified by noticing that  
$$\int d\theta d\theta^{*} a^{\dag}_{\vec{k}}a_{\vec{k}} = \widetilde{a}^{\dag}_{\vec{k}}\widetilde{a}_{\vec{k}} + \frac{(\alpha^2 + \beta^2)}{4}\left( \widetilde{h}^{\dag}_{\vec{k}}\widetilde{h}_{\vec{k}} - \widetilde{f}^{\dag}_{\vec{k}}\widetilde{f}_{\vec{k}} \right) + \, \textrm{cross terms} $$
and therefore one has to redefine the vacuum state and hence the entire Hilbert space.  However, such Hilbert space can easily be found in an indefinite norm space by limiting to the sub-Hilbert space on which the redefined operator $\int d\theta d\theta^{*} \, a^{\dag}_{\vec{k}}a_{\vec{k}}$ is positive definite.  Unfortunately, this viewpoint implies even more serious difficulties such as the breakdown of Lorentz invariance.  Indeed, even if one would manage to redefine $\widetilde{a}_{\vec{k}}$, it would not commute anymore with $\theta^{*}, \theta$ which effectively singles out a preferred frame.  There are two inequivalent ways of dealing with this: (a) the introduction of negative probabilities or (b) a new gauge invariance restoring Lorentz covariance.  In the first case, one should write  
\begin{eqnarray*}    
a_{\vec{k} \, \phi \, \psi} & = & \left(\cosh\phi \, \theta + e^{i\psi}\sinh\phi \, \theta^{*} \right)\widetilde{a}_{\vec{k} \, \phi \, \psi} \\*
h_{\vec{k} \, \phi \, \psi} & = & \left(\cosh\phi \, \theta + e^{i\psi}\sinh\phi \, \theta^{*} \right)\widetilde{h}_{\vec{k} \, \phi \, \psi} \\*
f_{\vec{k} \, \phi \, \psi} & = &\left(\cosh\phi \, \theta + e^{i\psi}\sinh\phi \, \theta^{*} \right)\widetilde{f}_{\vec{k} \, \phi \, \psi}
\end{eqnarray*}
where 
$$\{\widetilde{f}_{\vec{k} \, \phi \, \psi},\widetilde{f}^{\dag}_{\vec{l} \, \phi' \, \psi'} \} = - \delta(\vec{k} - \vec{l}).$$
In this context, the Grassmann numbers haven't done a great deal with respect to the original Clifford theory we started from: at best they have introduced a global symmetry $U(1) \times U(1) \times SO(1,1)$ which might be promoted to a new kind of local gauge invariance.  Strictly speaking, the above results only imply that the Grassmann coefficients in front of little group ``families'' have to be the same, leaving four independent choices.  At this moment, we make the simplification that all these four independent coefficients are the same which only affects the computation of the Lorentz group\footnote{The translation group as well as considerations towards causality are immune to this.}.  The reader is invited to make the minute changes in the formulae.  Again, we have no spin statistics theorem yet: both options are allowed for.  In the computation of the momentum and angular momentum below, we work with this convention and it is quite obvious which operators generate negative norm states and which don't.  Indefinite norm spaces provide a unified framework in which the notion of ``change of observer'' (which is what $S(\alpha,\beta)$ really does) is accompanied by a (not necessarily unitary) transformation between both Hilbert spaces seen as positive definite ``slices'' in the indefinite norm space.  It would be a good exercise to understand the Unruh effect in this way.  \\* \\*
We shall first follow the conventional path and verify whether all results agree with those in the literature \cite{Weinberg}.  Notice also that the Hamiltonian still contains an infinite renormalization constant\footnote{It is here that negative energy particles or the idea of supersymmetry might become useful.} which equals $-8 \delta(0)\int d^3\vec{k}\, k$; these infinities shall persist in the entire algebra and we shall ``ignore'' them in writing out our results (the reader is invited to write them down explicitly).  The classical momentum currents are given by
$$P_{\alpha}^{\beta} = - i \widetilde{X}\gamma^{\beta} \partial_{\alpha} X$$ and the quantized total momentum is computed to be
\begin{eqnarray*} P_{j} & = & \int d^{3} \vec{k} k_j \left( \widetilde{a}^{\dag}_{\vec{k}}\widetilde{a}_{\vec{k}} + \widetilde{b}^{\dag}_{\vec{k}}\widetilde{b}_{\vec{k}} - \widetilde{c}^{\dag}_{\vec{k}}\widetilde{c}_{\vec{k}} - \widetilde{d}^{\dag}_{\vec{k}}\widetilde{d}_{\vec{k}} - \widetilde{e}^{\dag}_{\vec{k}}\widetilde{e}_{\vec{k}} - \widetilde{f}^{\dag}_{\vec{k}}\widetilde{f}_{\vec{k}} + \widetilde{g}^{\dag}_{\vec{k}}\widetilde{g}_{\vec{k}} + \widetilde{h}^{\dag}_{\vec{k}}\widetilde{h}_{\vec{k}} \right) \\*
& + & \int d^{3} \vec{k} k_j \left( - \widetilde{a}'^{\dag}_{\vec{k}}\widetilde{a}'_{\vec{k}} - \widetilde{b}'^{\dag}_{\vec{k}}\widetilde{b}'_{\vec{k}} + \widetilde{c}'^{\dag}_{\vec{k}}\widetilde{c}'_{\vec{k}} + \widetilde{d}'^{\dag}_{\vec{k}}\widetilde{d}'_{\vec{k}} + \widetilde{e}'^{\dag}_{\vec{k}}\widetilde{e}'_{\vec{k}} + \widetilde{f}'^{\dag}_{\vec{k}}\widetilde{f}'_{\vec{k}} - \widetilde{g}'^{\dag}_{\vec{k}}\widetilde{g}'_{\vec{k}} - \widetilde{h}'^{\dag}_{\vec{k}}\widetilde{h}'_{\vec{k}} \right).\end{eqnarray*}  One may compute the total spin operator
 \begin{eqnarray*} S^{rs} & = & 2i \int d^{3}\vec{k} \left( \widetilde{a}^{\dag}_{\vec{k}} k^{[ r} \partial^{s ]} \widetilde{a}_{\vec{k}} + \widetilde{b}^{\dag}_{\vec{k}} k^{[ r} \partial^{s ]} \widetilde{b}_{\vec{k}} - \widetilde{c}^{\dag}_{\vec{k}} k^{[ r} \partial^{s ]} \widetilde{c}_{\vec{k}} - \ldots + \widetilde{h}^{\dag}_{\vec{k}} k^{[ r} \partial^{s ]} \widetilde{h}_{\vec{k}} \right) \\* 
& - & 2i \int d^{3}\vec{k} \left( \widetilde{a}'_{\vec{k}} k^{[ r} \partial^{s ]} \widetilde{a}'^{\dag}_{\vec{k}} + \widetilde{b}'_{\vec{k}} k^{[ r} \partial^{s ]} \widetilde{b}'^{\dag}_{\vec{k}} - \widetilde{c}'_{\vec{k}} k^{[ r} \partial^{s ]} \widetilde{c}'^{\dag}_{\vec{k}} - \ldots + \widetilde{h}'_{\vec{k}} k^{[ r} \partial^{s ]} \widetilde{h}'^{\dag}_{\vec{k}} \right) \\*
& + &\frac{i}{8} \int \frac{d^{3} \vec{k}}{k} \, ik^{[r}\textrm{Tr}\left(\gamma^{5}\gamma^{0}\gamma^{s]}\displaystyle{\not}n_1
\displaystyle{\not}n_2 \right) + ik^{[r}\textrm{Tr}\left(\gamma^{0}\gamma^{s]}\displaystyle{\not}n_1\displaystyle{\not}n_2
\right) + 2ik \textrm{Tr}\left(k^{[r}\partial^{s]}\displaystyle{\not}n_1\displaystyle{\not}n_2 \right) + \\* & & \frac{1}{2}\textrm{Tr}\left(\gamma^{5}\left[\gamma^{0}, \displaystyle{\not}k \right]k^{[r}\partial^{s]}(\displaystyle{\not}n_1 - i\displaystyle{\not}n_2)(\displaystyle{\not}n_1 + i\displaystyle{\not}n_2)\right) \, \, \left( \widetilde{a}^{\dag}_{\vec{k}}\widetilde{a}_{\vec{k}} -  \widetilde{c}^{\dag}_{\vec{k}}\widetilde{c}_{\vec{k}} \right) \\*
& + &\frac{i}{8} \int \frac{d^{3} \vec{k}}{k} \, ik^{[r}\textrm{Tr}\left(\gamma^{5}\gamma^{0}\gamma^{s]}\displaystyle{\not}n_1
\displaystyle{\not}n_2 \right) - ik^{[r}\textrm{Tr}\left(\gamma^{0}\gamma^{s]}\displaystyle{\not}n_1\displaystyle{\not}n_2
\right) - 2ik \textrm{Tr}\left(k^{[r}\partial^{s]}\displaystyle{\not}n_1\displaystyle{\not}n_2 \right) - \\* & &  \frac{1}{2}\textrm{Tr}\left(\gamma^{5}\left[\gamma^{0}, \displaystyle{\not}k \right]k^{[r}\partial^{s]}(\displaystyle{\not}n_1 + i\displaystyle{\not}n_2)(\displaystyle{\not}n_1 - i\displaystyle{\not}n_2)\right) \, \, \left( \widetilde{b}^{\dag}_{\vec{k}}\widetilde{b}_{\vec{k}} -  \widetilde{d}^{\dag}_{\vec{k}}\widetilde{d}_{\vec{k}} \right) \\*
& - & \frac{i}{8}\int \frac{d^3 \vec{k}}{k}\, \, i k^{[r}\textrm{Tr}\left(\gamma^5\gamma^0\gamma^{s]}\displaystyle{\not}n_1\displaystyle{\not}n_2 \right) + ik^{[r}\textrm{Tr}\left(\gamma^{0}\gamma^{s]}\displaystyle{\not}n_1\displaystyle{\not}n_2 \right) + \\* & & \frac{1}{2}\textrm{Tr}\left( \left[ \gamma^{0}, \displaystyle{\not}k \right]k^{[r}\partial^{s]}(\displaystyle{\not}n_1 - i\displaystyle{\not}n_2)(\displaystyle{\not}n_1 + i\displaystyle{\not}n_2) \right) \,\, \left( \widetilde{a}^{\dag}_{\vec{k}}\widetilde{c}_{\vec{k}} - \widetilde{c}^{\dag}_{\vec{k}}\widetilde{a}_{\vec{k}} \right) \\*
& - & \frac{i}{8} \int \frac{d^3 \vec{k}}{k}\, \, i k^{[r}\textrm{Tr}\left(\gamma^5\gamma^0\gamma^{s]}\displaystyle{\not}n_1\displaystyle{\not}n_2 \right) - ik^{[r}\textrm{Tr}\left(\gamma^{0}\gamma^{s]}\displaystyle{\not}n_1\displaystyle{\not}n_2 \right) + \\* & & \frac{1}{2}\textrm{Tr}\left( \left[ \gamma^{0}, \displaystyle{\not}k \right]k^{[r}\partial^{s]}(\displaystyle{\not}n_1 + i\displaystyle{\not}n_2)(\displaystyle{\not}n_1 - i\displaystyle{\not}n_2)\right) \, \, \left( \widetilde{b}^{\dag}_{\vec{k}}\widetilde{d}_{\vec{k}} - \widetilde{d}^{\dag}_{\vec{k}}\widetilde{b}_{\vec{k}} \right) \\*
& + & \, \textrm{other cross terms of the angular momentum} 
\end{eqnarray*}
\begin{eqnarray*}
& + & \frac{1}{8}\int \frac{d^{3}\vec{k}}{k} \, \textrm{Tr}\left(\gamma^{0}\gamma^{r}\gamma^{s}\displaystyle{\not}k\displaystyle{\not}n_2 \displaystyle{\not}n_1\right) \left( \widetilde{a}^{\dag}_{\vec{k}}\widetilde{a}_{\vec{k}} - \widetilde{b}^{\dag}_{\vec{k}}\widetilde{b}_{\vec{k}} - \widetilde{c}^{\dag}_{\vec{k}}\widetilde{c}_{\vec{k}} + \widetilde{d}^{\dag}_{\vec{k}}\widetilde{d}_{\vec{k}}\right) \\*
& + & \frac{1}{8}\int \frac{d^{3}\vec{k}}{k} \, \textrm{Tr}\left(\gamma^{0}\gamma^{r}\gamma^{s}\displaystyle{\not}k\displaystyle{\not}n_2 \displaystyle{\not}n_1\right) \left( \widetilde{e}^{\dag}_{\vec{k}}\widetilde{e}_{\vec{k}} - \widetilde{f}^{\dag}_{\vec{k}}\widetilde{f}_{\vec{k}} - \widetilde{g}^{\dag}_{\vec{k}}\widetilde{g}_{\vec{k}} + \widetilde{h}^{\dag}_{\vec{k}}\widetilde{h}_{\vec{k}}\right) \\*
&+ &\frac{i}{64} \int \frac{d^3 \vec{k}}{k} \textrm{Tr}\left( \left(\displaystyle{\not}\widetilde{p}_1 + i\displaystyle{\not}\widetilde{p}_2\right)\gamma^0\gamma^r\gamma^s\displaystyle{\not}k
\displaystyle{\not}l \right)\left(\widetilde{a}^{\dag}_{\vec{k}}\widetilde{e}_{\vec{k}} + \widetilde{a}^{\dag}_{\vec{k}}\widetilde{f}_{\vec{k}} - \widetilde{a}^{\dag}_{\vec{k}}\widetilde{g}_{\vec{k}} - \widetilde{a}^{\dag}_{\vec{k}}\widetilde{h}_{\vec{k}}\right) \\* & + &  \frac{i}{64} \int \frac{d^3 \vec{k}}{k} \textrm{Tr}\left( \left(\displaystyle{\not}\widetilde{p}_1 + i\displaystyle{\not}\widetilde{p}_2\right)\gamma^0\gamma^r\gamma^s\displaystyle{\not}k
\displaystyle{\not}l \right)\left(- \widetilde{e}^{\dag}_{\vec{k}}\widetilde{b}_{\vec{k}} - \widetilde{f}^{\dag}_{\vec{k}}\widetilde{b}_{\vec{k}} + \widetilde{g}^{\dag}_{\vec{k}}\widetilde{b}_{\vec{k}} + \widetilde{h}^{\dag}_{\vec{k}}\widetilde{b}_{\vec{k}} \right) \\*
& + &  \frac{i}{64} \int \frac{d^3 \vec{k}}{k} \textrm{Tr}\left( \left(\displaystyle{\not}\widetilde{p}_1 + i\displaystyle{\not}\widetilde{p}_2\right)\gamma^0\gamma^r\gamma^s\displaystyle{\not}k
\displaystyle{\not}l \right)\left( \widetilde{c}^{\dag}_{\vec{k}}\widetilde{e}_{\vec{k}} + \widetilde{c}^{\dag}_{\vec{k}}\widetilde{f}_{\vec{k}} - \widetilde{c}^{\dag}_{\vec{k}}\widetilde{g}_{\vec{k}} - \widetilde{c}^{\dag}_{\vec{k}}\widetilde{h}_{\vec{k}} \right) \\*
& + &  \frac{i}{64} \int \frac{d^3 \vec{k}}{k} \textrm{Tr}\left( \left(\displaystyle{\not}\widetilde{p}_1 + i\displaystyle{\not}\widetilde{p}_2\right)\gamma^0\gamma^r\gamma^s\displaystyle{\not}k
\displaystyle{\not}l \right)\left( - \widetilde{e}^{\dag}_{\vec{k}}\widetilde{d}_{\vec{k}} - \widetilde{f}^{\dag}_{\vec{k}}\widetilde{d}_{\vec{k}} + \widetilde{g}^{\dag}_{\vec{k}}\widetilde{d}_{\vec{k}} + \widetilde{h}^{\dag}_{\vec{k}}\widetilde{d}_{\vec{k}} \right) \\*
& + &  \frac{i}{64} \int \frac{d^3 \vec{k}}{k} \textrm{Tr}\left( \left(\displaystyle{\not}\widetilde{p}_1 - i\displaystyle{\not}\widetilde{p}_2\right)\gamma^0\gamma^r\gamma^s\displaystyle{\not}k
\displaystyle{\not}l \right)\left( \widetilde{b}^{\dag}_{\vec{k}}\widetilde{e}_{\vec{k}} + \widetilde{b}^{\dag}_{\vec{k}}\widetilde{f}_{\vec{k}} - \widetilde{b}^{\dag}_{\vec{k}}\widetilde{g}_{\vec{k}} - \widetilde{b}^{\dag}_{\vec{k}}\widetilde{h}_{\vec{k}} \right) \\*
& + & \, \textrm{similar expressions as before and likewise for the primed particles (mind the sign here)} 
\end{eqnarray*}
from the spin currents
$$S^{\alpha \beta}_{\kappa} = - i \left( \widetilde{X} \gamma_{\kappa} \left( x^{\alpha} \partial^{\beta} - x^{\beta}\partial^{\alpha} \right)X + \frac{1}{2} \widetilde{X} \gamma_{\kappa} \gamma^{\alpha}\gamma^{\beta} X \right).$$
It is left as an exercise to the reader to prove that $$i\textrm{Tr}\left(\gamma^5\gamma^0\gamma^r\gamma^s\displaystyle{\not}k\right) = \textrm{Tr}\left(\gamma^{0}\gamma^{r}\gamma^{s}\displaystyle{\not}k\displaystyle{\not}n_2 \displaystyle{\not}n_1\right)$$ which is necessary to obtain the above expressions.  The reader notices that the first part of the angular momentum is identical to the bosonic theory and therefore the Poincar\'e algebra is satisfied if one forgets about the latter parts.  The new part of the angular momentum however is mandatory because the first one is not invariant under local little transformations on momentum space while the combined expression has this invariance.  We now comment briefly upon the second option (b) to rescue Lorentz covariance (and the spin statistics theorem) through a new gauge principle.  Under a local (on momentum space) $S(\alpha,\beta)$, $X$ transforms as
$$X(x,\theta,\theta^{*}) \Rightarrow X(x,\theta,\theta^{*}) + \gamma^{\alpha}\partial_{\alpha}\Omega(\alpha,\beta,x,\theta,\theta^{*}) $$ where $\Omega$ satisfies the Klein-Gordon equation as well as $$\gamma^{\alpha}\partial_{\alpha}\partial_{\beta} \Omega(\alpha,\beta,x, \theta,\theta^{*})\gamma^{\beta} = 0.$$  It is clear that we need a new action principle to have this kind of gauge invariance; it might be useful to construct it from ``Yang Mills'' terms of the kind 
$$\gamma^{\alpha}\partial_{\alpha}X - \partial_{\alpha}X \gamma^{\alpha}$$ where $\Omega(x,\theta,\theta^{*})$ is required to commute with the Dirac operator.  \\* \\*  Suppose we consider a particle with momentum $k^{\mu} = k(1,1,0,0)$ and the ``standard'' complementary vectors, then the angular momentum part of $S^{23}$ vanishes and one calculates that
\begin{eqnarray*}
\textrm{Tr}\left(\gamma^{0}\gamma^{2}\gamma^{3}\displaystyle{\not}k\displaystyle{\not}n_2
\displaystyle{\not}n_1\right) & = & 4k\\*
i\textrm{Tr}\left(\gamma^{5}\gamma^{0}\gamma^{2}\gamma^{3}\displaystyle{\not}k\right)  & = & 4k \\*
\textrm{Tr}\left( \left( \displaystyle{\not}\widetilde{p}_1 \pm i\displaystyle{\not}\widetilde{p}_2 \right)\gamma^{0}\gamma^{2}\gamma^{3}\displaystyle{\not}k\displaystyle{\not}l\right)
& = & 0 
\end{eqnarray*} which implies the standard expressions such as $$S^{23}\widetilde{a}^{\dag}_{\vec{k}}|0 \rangle = \frac{1}{2}\widetilde{a}^{\dag}_{\vec{k}}|0 \rangle $$ and $$S^{23}\widetilde{b}^{\dag}_{\vec{k}}|0 \rangle = -\frac{1}{2}\widetilde{b}^{\dag}_{\vec{k}}|0 \rangle.$$  One may verify that all particles have helicity $\pm \frac{1}{2}$ which confirms our previous calculation of left multiplication.  The reader is invited to study the remaining spin operators $S^{0r}$:   
\begin{eqnarray*} S^{0r} & = & i \int d^{3}\vec{k} k \left( \partial^{r}\widetilde{a}^{\dag}_{\vec{k}}\widetilde{a}_{\vec{k}} +  \partial^{r} \widetilde{b}_{\vec{k}}\widetilde{b}^{\dag}_{\vec{k}} - \partial^{r}\widetilde{c}^{\dag}_{\vec{k}}\widetilde{c}_{\vec{k}} - \ldots \partial^{r}\widetilde{g}^{\dag}_{\vec{k}}\widetilde{g}_{\vec{k}} +
\partial^{r}\widetilde{h}^{\dag}_{\vec{k}}\widetilde{h}_{\vec{k}} \right) + \ldots. \end{eqnarray*}  
To finish the ``isomorphism'' with the traditional results, we aim to verify if causality holds.  The latter issue is however ``broader'' than it is in the standard formulation.  For example, the expressions  
$$\left[ X(t,\vec{x}), \widetilde{X}(s,\vec{y})\gamma^{0} \right] $$
are fundamentally acausal no matter what relations one imposes upon the operators; this can be easily seen by considering the term $$a_{\vec{k}} a^{\dag}_{\vec{l}} h_{+ \,\vec{k}}\widetilde{h}_{+ \, \vec{l}}\gamma^{0} - a^{\dag}_{\vec{l}} a_{\vec{k}} \widetilde{h}_{+ \, \vec{l}} \gamma^{0} h_{+ \, \vec{k}}.$$  To understand how $h_{+ \,\vec{k}}\widetilde{h}_{+ \, \vec{l}} \gamma^{0}$ looks like, one can without loss of generality perform a spin transformation such that $k^{\mu} = k(1,1,0,0)$, $l^{\mu} = l(1,\cos\theta,\sin\theta,0)$ and with respect to $l$, $n_1 = (0,- \sin\theta, \cos\theta, 0)$ and $n_2 = (0,0,0,1)$ (recall that we could freely choose the $n_j$).  Hence, the first terms of the latter expression are computed to be 
$$h_{+ \,\vec{k}}\widetilde{h}_{+ \, \vec{l}}\gamma^{0} = (k+l)(1 + \cos\theta)1 + 0 \gamma^0  - \gamma^{0} \gamma^{1} (k + l)( 1 + \cos\theta) +  \ldots $$
while
$$\widetilde{h}_{+ \, \vec{l}} \gamma^{0} h_{+ \, \vec{k}} = (k+l)(1 + \cos\theta)1 + \frac{1}{2}\gamma^{0}\left(1 + \cos\theta \right)(kl - 2) + \ldots$$
Hence, both ``numbers'' neither commute, nor anti-commute and therefore it is impossible to eliminate the terms with $\displaystyle{\not}k \neq \displaystyle{\not}l$.  As mentioned in the previous chapter, we need our metric to be Clifford valued; therefore, we know already that Minkowski causality won't be satisfied.  One can now be more conventional and consider Lorentz (multi) vectors of the kind $\textrm{Tr}\left(X\gamma^{\alpha}\right)$, $\textrm{Tr}\left(X\gamma^{\alpha}\gamma^{\beta}\right)$ and similar expressions involving $\widetilde{X}$.  In computing these expressions explicitly, it is good to remind a ``trick'' developed first by Wigner and explained more recently by Weinberg \cite{Weinberg}.  Actually, we might have used it already when computing the expressions $S^{rs}$ and $S^{or}$. \\* \\*
For every future pointing null vector $k^{\alpha}$ we can choose a representative Lorentz transformation $L(k)$ such that $$k^{\alpha} = L(k)^{\alpha}_{\,\, \beta} N^{\beta}$$ where $N^{\beta}$ is the ``representative'' null vector defined by $N^{\beta} = (1,1,0,0)$.  Writing $k^{\alpha}$ as $k(1,\cos\theta,\sin \theta \cos\psi, \sin\theta\sin\psi)$ defines the matrices
 \[ \left( \begin{array}{cccc}
\frac{k^2 + 1}{2k} & \frac{k^2 -1}{2k} & 0 & 0 \\
\frac{k^2 -1}{2k} & \frac{k^2 + 1}{2k} & 0 & 0 \\
0 & 0 & 1 & 0 \\
0 & 0 & 0 & 1 \\ \end{array} \right)\] 
which boosts $(1,1,0,0)$ into $(k,k,0,0)$ and the rotation 
 \[ \left( \begin{array}{cccc}
1 & 0 & 0 & 0 \\
0 & \cos\theta & - \sin\theta & 0 \\
0 & \cos\psi\sin\theta & \cos\psi\cos\theta & -\sin\psi \\
0 & \sin\psi\sin\theta & \sin\psi\cos\theta & \cos\psi \\ \end{array} \right)\]
which rotates $(1,1,0,0)$ into $(1,\cos\theta,\sin\theta\cos\psi,\sin\theta\sin\psi)$.   Therefore $L(k)$ equals  
 \[ \left( \begin{array}{cccc}
\frac{k^2 + 1}{2k} & \frac{k^2 - 1}{2k} & 0 & 0 \\
\cos\theta \frac{k^2 - 1}{2k} & \cos\theta \frac{k^2 + 1}{2k} & -\sin\theta & 0 \\
\sin\theta\cos\psi \frac{k^2 - 1}{2k} & \cos\psi\sin\theta\frac{k^2 + 1}{2k} & \cos\psi\cos\theta & -\sin\psi \\
\sin\theta\sin\psi\frac{k^2 - 1}{2k} & \sin\psi\sin\theta\frac{k^2 + 1}{2k} & \sin\psi\cos\theta & \cos\psi \\ \end{array} \right).\]
Defining $n_{1} = L(k)(0,0,1,0)$ and $n_2 = L(k)(0,0,0,1)$, $l = L(k)(1,-1,0,0)$ (notice here that $l.k = - 2$) then the above matrix can also be written as
 \[ \left( \begin{array}{cccc}
\frac{1}{2}(k^0 + l^0) & \frac{1}{2}(k^0 - l^0) & n_1^0 & n_2^0 \\
\frac{1}{2}(k^1 + l^1) & \frac{1}{2}(k^1 - l^1) & n_1^1 & n_2^1 \\
\frac{1}{2}(k^2 + l^2) & \frac{1}{2}(k^2 - l^2) & n_1^2 & n_2^2 \\
\frac{1}{2}(k^3 + l^3) & \frac{1}{2}(k^3 - l^3) & n_1^3 & n_2^3 \\ \end{array} \right).\]        
Let us now calculate the fundamental traces: obviously, $\textrm{Tr}\left( q^{j}_{\vec{k}}\right) = 0$ and $$\textrm{Tr} \left(q^{j}_{\vec{k}} \gamma^{\alpha} \ldots \gamma^{\beta} \right) = \textrm{Tr} \left(U(k)^{-1} q^{j}_{\vec{k}} \gamma^{\alpha} \ldots \gamma^{\beta} U(k) \right) = L(k)^{\alpha}_{\,\, \kappa} \ldots L(k)^{\beta}_{\,\, \delta} \textrm{Tr} \left(q^{j}_{(1,0,0)} \gamma^{\kappa} \ldots \gamma^{\delta} \right) $$ where $U(k)$ is the spin transformation generating $L(k)$.  The latter expressions (with one gamma matrix) are computed to be
$$\textrm{Tr}\left(q^{j}_{\vec{k}} \gamma^{\alpha}\right) = 4n_j^{\alpha}.$$
Likewise, all other formulae are given by
\begin{eqnarray*}
\textrm{Tr}\left(q^{j}_{\vec{k}} \gamma^{\alpha} \gamma^{\beta}\right) & = & 8 n_j^{[ \alpha}k^{\beta]} \\*
\textrm{Tr}\left(q^{j}_{\vec{k}} \gamma^{\alpha} \gamma^{\beta} \gamma^{\kappa} \right) & = & 12 n_j^{[\alpha}k^{\beta}l^{\kappa]}\\*
\textrm{Tr}\left(q^{j}_{\vec{k}} \gamma^5\right) & = & 0. 
\end{eqnarray*}
From these, one can compute all possible traces with as many $q^{j}_{\vec{k}}$, $\widetilde{q}^{j}_{\vec{k}}$ one wants by using the identity operation
\begin{eqnarray*}
4 A & = &  1 \textrm{Tr}\left( A \right) + \gamma^{\alpha} \, \textrm{Tr}\left(\gamma_{\alpha} A \right) + \sum_{\alpha < \beta} \gamma^{\alpha}\gamma^{\beta} \, \textrm{Tr}\left( \gamma_{\beta} \gamma_{\alpha} A \right) \\*
& + & \sum_{\alpha < \beta < \kappa} \gamma^{\alpha}\gamma^{\beta}\gamma^{\kappa} \, \textrm{Tr}\left( \gamma_{\kappa}\gamma_{\beta}\gamma_{\alpha} A \right) + \gamma^5 \, \textrm{Tr}\left( \gamma^5 A \right).
\end{eqnarray*}
All remaining nonzero expression we need are given by 
\begin{eqnarray*}
\textrm{Tr}\left(\displaystyle{\not}k\gamma^{\alpha}\right) & = & 4k^{\alpha} \\*
\textrm{Tr}\left(\displaystyle{\not}k\displaystyle{\not}l\right) & = & - 8 \\*
\textrm{Tr}\left(\displaystyle{\not}k\displaystyle{\not}l\gamma^{\alpha}\gamma^{\beta}\right)
& = & 8l^{[\alpha}k^{\beta]} \\*
\textrm{Tr}\left(\gamma^{5}\gamma^{5}\right) = 4 \\*
\textrm{Tr}\left(\displaystyle{\not}k\gamma^{5}\gamma^{\alpha}\gamma^{\beta}\gamma^{\kappa}\right)
& = & -24ik^{[\alpha}n_1^{\beta}n_2^{\kappa]} \\*
\textrm{Tr}\left(\displaystyle{\not}n_1\displaystyle{\not}n_2\gamma^{\alpha}\gamma^{\beta}\right) 
& = &  8n_{2}^{[\alpha}n_1^{\beta]}.
\end{eqnarray*}
\\*
Returning to the causality question, one notices that 
\begin{eqnarray*} \textrm{Tr} \left( X(t,\vec{x}) \gamma^{\alpha} \right) & = & \int \frac{d^3\vec{k}}{\sqrt{2k}} \,\, a_{\vec{k}}\left( n_1^{\alpha} - in_2^{\alpha} \right) + b_{\vec{k}}\left( n_1^{\alpha} + in_2^{\alpha} \right) + c_{\vec{k}}\left( n_1^{\alpha} - in_2^{\alpha} \right) + d_{\vec{k}}\left( n_1^{\alpha} + in_2^{\alpha} \right) \\* & + & k^{\alpha}e_{\vec{k}} + k^{\alpha}f_{\vec{k}} + k^{\alpha}g_{\vec{k}} + k^{\alpha}h_{\vec{k}} \,\, e^{-i(kt - \vec{k}.\vec{x})} +  e^{i(kt - \vec{k}.\vec{x})} \ldots \end{eqnarray*} therefore 
\begin{eqnarray*} \textrm{Tr} \left( \widetilde{X}(s,\vec{y}) \gamma^{\beta} \right) & = & \int \frac{d^3\vec{k}}{\sqrt{2k}} \,\, a^{\dag}_{\vec{k}}\left( n_1^{\beta} + in_2^{\beta} \right) + b^{\dag}_{\vec{k}}\left( n_1^{\beta} - in_2^{\beta} \right) + c^{\dag}_{\vec{k}}\left( n_1^{\beta} + in_2^{\beta} \right) + d^{\dag}_{\vec{k}}\left( n_1^{\beta} - in_2^{\beta} \right) \\* & + & k^{\beta}e^{\dag}_{\vec{k}} + k^{\beta}f^{\dag}_{\vec{k}} + k^{\beta}g^{\dag}_{\vec{k}} + k^{\beta}h^{\dag}_{\vec{k}} \,\, e^{i(ks - \vec{k}.\vec{y})} +  e^{-i(ks - \vec{k}.\vec{y})} \ldots. \end{eqnarray*}
Hence,
\begin{eqnarray*}
\left[ \textrm{Tr} \left( X(t,\vec{x}) \gamma^{\alpha} \right), \textrm{Tr} \left( \widetilde{X}(s,\vec{y}) \gamma^{\beta} \right) \right] & = & \int \frac{d^3\vec{k}}{2k} \, e^{-i(k(t-s) - \vec{k}.(\vec{x} -\vec{y}))} \theta^{*}\theta \,\, \left( n_1^{\alpha} - i n_2^{\alpha} \right)\left(n_1^{\beta} + in_2^{\beta}\right)(-1 + 1) \\* & + & \left( n_1^{\alpha} + i n_2^{\alpha} \right)\left(n_1^{\beta} - in_2^{\beta}\right)(-1 + 1) + k^{\alpha}k^{\beta}(1 + 1 - 1 - 1) + \\* & & e^{i(k(t-s) - \vec{k}.(\vec{x} -\vec{y}))}\ldots = 0. \\*
\end{eqnarray*}
where the negative probability particles cancel out the amplitudes generated by the positive probability particles.  Other expressions are less trivial; indeed, given
\begin{eqnarray*}
\textrm{Tr} \left( \widetilde{X}(s,\vec{y})\gamma^{\beta}\gamma^{\kappa}\right) & = & \int \frac{d^3 \vec{k}}{\sqrt{2k}} \,\, -2(n_1^{[\beta} + in_2^{[\beta})k^{\kappa]}a^{\dag}_{\vec{k}} -2(n_1^{[\beta} - in_2^{[\beta})k^{\kappa]}b^{\dag}_{\vec{k}} + 2(n_1^{[\beta} + in_2^{[\beta})k^{\kappa]}c^{\dag}_{\vec{k}} \\*
& + & 2(n_1^{[\beta} - in_2^{[\beta})k^{\kappa]}d^{\dag}_{\vec{k}} + \left(- 2l^{[\beta}k^{\kappa]} + 2in_2^{[\beta}n_1^{\kappa]}\right)e^{\dag}_{\vec{k}} + \left(- 2l^{[\beta}k^{\kappa]} - 2in_2^{[\beta}n_1^{\kappa]}\right)f^{\dag}_{\vec{k}} \\*
&+& \left(2l^{[\beta}k^{\kappa]} - 2in_2^{[\beta}n_1^{\kappa]}\right)g^{\dag}_{\vec{k}} + \left(2l^{[\beta}k^{\kappa]}  + 2in_2^{[\beta}n_1^{\kappa]}\right)h^{\dag}_{\vec{k}} \,\, e^{i(ks - \vec{k}.\vec{y})} + \ldots
\end{eqnarray*}
where $l.k = -1$ again, one calculates that
\begin{eqnarray*}
\left[ \textrm{Tr} \left(\widetilde{X}(s,\vec{y})\gamma^{\beta}\gamma^{\kappa}\right), \textrm{Tr} \left(X(t,\vec{x})\gamma^{\alpha}\right) \right] & = & \int \frac{d^3 \vec{k}}{2k}\theta^{*}\theta \\* & & e^{-i(k(t-s) - \vec{k}.(\vec{x} - \vec{y}))} \left( -8n_1^{[\beta}k^{\kappa]}n_1^{\alpha} -8n_2^{[\beta}k^{\kappa]}n_2^{\alpha} + 8l^{[\beta}k^{\kappa]}k^{\alpha} \right) \\*
& + & e^{i(k(t-s) - \vec{k}.(\vec{x} - \vec{y}))} \left( -8n_1^{[\beta}k^{\kappa]}n_1^{\alpha} -8n_2^{[\beta}k^{\kappa]}n_2^{\alpha} + 8l^{[\beta}k^{\kappa]}k^{\alpha} \right) \\*
& = & -4 \int \frac{d^3 \vec{k}}{k}\theta^{*}\theta\eta^{\alpha \,[\beta}k^{\kappa ]} \left( e^{-i(k(t-s) - \vec{k}.(\vec{x} - \vec{y}))} + e^{i(k(t-s) - \vec{k}.(\vec{x} - \vec{y}))} \right)
\end{eqnarray*}
and the reader may check that this expression is manifestly gauge invariant as it should.  This can be rewritten in a more conventional form as
\begin{eqnarray*}
\left[ \textrm{Tr} \left(\widetilde{X}(s,\vec{y})\gamma^{\beta}\gamma^{\kappa}\right), \textrm{Tr} \left(X(t,\vec{x})\gamma^{\alpha}\right) \right] & = & \theta^{*}\theta 4i\eta^{\alpha \, [\beta}\partial^{\kappa]} \int \frac{d^3 \vec{k}}{k} \left(e^{-i(k(t-s) - \vec{k}.(\vec{x} - \vec{y}))} - e^{i(k(t-s) - \vec{k}.(\vec{x} - \vec{y}))}\right)
\end{eqnarray*}    
and it is well know that the latter function is causal \cite{Peskin} \cite{Weinberg}.  It is important to realize that \emph{all} particles contribute to this result which is physically very different from the standard Dirac point of view where the necessary cancellations happen for each species separately.  Of course this is due to our more elaborate notion of Lorentz invariance which mixes different particle species.  The reader is invited to verify the remaining causality expressions\footnote{In the above causality calculations, one should simply drop $\theta^{*}\theta$ for obtaining the correct results in the Clifford theory (and replace the commutator by the anti-commutator).}.  In order to fully grasp the physical implications of this theory, we calculate the three conserved charges (instead of two); the action of the discrete symmetries $C$, $P$ and $T$ is postponed for future research.  The conserved currents are given by
$$Q^{\alpha} = \textrm{Tr}\left( \widetilde{X}(x)\gamma^{\alpha}X(x) \right)$$
$$K^{\alpha} = \textrm{Tr}\left( \widetilde{X}(x)\gamma^{\alpha}\gamma^{5}X(x) \right)$$
and
$$L^{\alpha} = \textrm{Tr}\left( \widetilde{X}(x)\gamma^{\alpha}X(x)\gamma^{5}\right).$$ The corresponding quantized electric charge is 
\begin{eqnarray*}
Q & = & \int d^{3} \vec{k} \left( \widetilde{a}^{\dag}_{\vec{k}}\widetilde{a}_{\vec{k}} + \widetilde{b}^{\dag}_{\vec{k}}\widetilde{b}_{\vec{k}} - \widetilde{c}^{\dag}_{\vec{k}}\widetilde{c}_{\vec{k}} - \widetilde{d}^{\dag}_{\vec{k}}\widetilde{d}_{\vec{k}} - \widetilde{e}^{\dag}_{\vec{k}}\widetilde{e}_{\vec{k}} - \widetilde{f}^{\dag}_{\vec{k}}\widetilde{f}_{\vec{k}} + \widetilde{g}^{\dag}_{\vec{k}}\widetilde{g}_{\vec{k}} + \widetilde{h}^{\dag}_{\vec{k}}\widetilde{h}_{\vec{k}} \right) \\*
& + & \int d^{3} \vec{k} \left( \widetilde{a}'^{\dag}_{\vec{k}}\widetilde{a}'_{\vec{k}} + \widetilde{b}'^{\dag}_{\vec{k}}\widetilde{b}'_{\vec{k}} - \widetilde{c}'^{\dag}_{\vec{k}}\widetilde{c}'_{\vec{k}} - \widetilde{d}'^{\dag}_{\vec{k}}\widetilde{d}'_{\vec{k}} - \widetilde{e}'^{\dag}_{\vec{k}}\widetilde{e}'_{\vec{k}} - \widetilde{f}'^{\dag}_{\vec{k}}\widetilde{f}'_{\vec{k}} + \widetilde{g}'^{\dag}_{\vec{k}}\widetilde{g}'_{\vec{k}} + \widetilde{h}'^{\dag}_{\vec{k}}\widetilde{h}'_{\vec{k}} \right)
\end{eqnarray*}
so that all unprimed (primed) operators have charge $1$ (-$1$).  Note that we already knew how to distinguish $\widetilde{a}^{\dag}_{\vec{k}}$ from $\widetilde{h}^{\dag}_{\vec{k}}$ physically while both are positive norm particles with helicity $\frac{1}{2}$, four momentum $k$ and electric charge $+1$.  Indeed, the distinction between both of them was already clear in the calculation of $S^{rs}$ where different couplings with other particles occurred due to the different transformation laws under the little group.  The following calculation  
\begin{eqnarray*}
K & = & \int d^3 \vec{k} \,\left( -\widetilde{a}^{\dag}_{\vec{k}}\widetilde{a}_{\vec{k}} + \widetilde{b}^{\dag}_{\vec{k}}\widetilde{b}_{\vec{k}} + \widetilde{c}^{\dag}_{\vec{k}}\widetilde{c}_{\vec{k}} - \widetilde{d}^{\dag}_{\vec{k}}\widetilde{d}_{\vec{k}} - \widetilde{e}^{\dag}_{\vec{k}}\widetilde{e}_{\vec{k}} + 
\widetilde{f}^{\dag}_{\vec{k}}\widetilde{f}_{\vec{k}} +
\widetilde{g}^{\dag}_{\vec{k}}\widetilde{g}_{\vec{k}} 
- \widetilde{h}^{\dag}_{\vec{k}}\widetilde{h}_{\vec{k}}\right) \\*
& + & \int d^3 \vec{k} \,\left( \widetilde{a}'^{\dag}_{\vec{k}}\widetilde{a}'_{\vec{k}} - \widetilde{b}'^{\dag}_{\vec{k}}\widetilde{b}'_{\vec{k}} - \widetilde{c}'^{\dag}_{\vec{k}}\widetilde{c}'_{\vec{k}} + \widetilde{d}'^{\dag}_{\vec{k}}\widetilde{d}'_{\vec{k}} + \widetilde{e}'^{\dag}_{\vec{k}}\widetilde{e}'_{\vec{k}} - 
\widetilde{f}'^{\dag}_{\vec{k}}\widetilde{f}'_{\vec{k}} -
\widetilde{g}'^{\dag}_{\vec{k}}\widetilde{g}'_{\vec{k}} +
 \widetilde{h}'^{\dag}_{\vec{k}}\widetilde{h}'_{\vec{k}}\right)
\end{eqnarray*} reveals that $K$ does not distinguish them and likewise for $L$ (the reader is invited to calculate that expression).  Hence, we obtain the following interpretation:  $\widetilde{a}^{\dag}_{\vec{k}}$ creates a particle of helicity $\frac{1}{2}$ with four momentum $k$, $Q$ charge $+1$, $K$ charge $-1$ and positive norm.  $\widetilde{c}^{\dag}_{\vec{k}}$ creates it's mirror particle of negative norm; by convention\footnote{Because we have equaled the Grassmann coefficients over different ``families''.  In the full Grassmann-Clifford theory, there is the possibility for a dynamical symmetry breaking between particles and anti-particles.  This suggests a possible mechanism for baryogenesis.} $\widetilde{c}'^{\dag}_{\vec{k}}$ creates the anti-particle of opposite helicity, $Q$ charge and identical $K$ charge.  Finally, $\widetilde{a}'^{\dag}_{\vec{k}}$ creates the negative norm anti-particle.  $\widetilde{b}^{\dag}_{\vec{k}}$ creates the same particle as $\widetilde{a}^{\dag}_{\vec{k}}$ but then with helicity $-\frac{1}{2}$; the reader may complete the list of corresponding operators.  $\widetilde{h}^{\dag}_{\vec{k}}$ creates a particle of the same species than $\widetilde{a}^{\dag}_{\vec{k}}$ but is physically distinct due to the transformation properties under the little group - again it is easy to find the corresponding mirror (negative norm) anti-particle.  
\\* \\*
Let us do a simple counting exercise: in the massless theory we have 16 operators which can be seen to correspond essentially to two different particles (with two helicity states each).  There is the symmetry positive norm versus negative norm and particle versus anti-particle which provides the number $2.2.2.2 = 16$.  Hence, we might need two different masses, but the mass action $S_4$ we have given so far contains only one.  This is because we have missed a slight extension due to a preoccupation with the standard Dirac theory (and actually, in the Dirac theory such extension is also possible).  Indeed, we can write down
$$S_8 =  \textrm{Tr} \, \int d^4x \widetilde{X}(x)\left( m 1 + n i\gamma^{5} \right) X(x)$$
$$S_9 =  \textrm{Tr} \, \int d^4x \left( m 1 + n i\gamma^{5} \right)\widetilde{X}(x) X(x).$$  In principle, we should study all possible action principles\footnote{The reader may wish to discover that more mass terms exist than I have written down here.} and investigate the physical differences in detail.  Such study however would fill an entire book and it would not reveal the structure of the deeper ideas we want to study from now on.  Hence, we proceed by making the first necessary extension of Quantum Field Theory.             
\chapter{Quantum Field Theory on indefinite Hilbert modules}
The lack of a spin statistics theorem from positive energy, causality, statistics, cluster decomposition, locality and Poincar\'e covariance on Nevanlinna spaces leads one to consider the reverse relation: positive energy, Poincar\'e covariance, locality, cluster decomposition and spin-statistics leads to causality.  This is also an outcome of the work in section six and we investigate whether this relation persists in Clifford-Nevanlinna modules.  The philosophical implications of this work are immense, statistics is more fundamental than causality is.  As we shall learn from the next chapter, we arrive at some form of the holographic principle from which causality automatically follows in \emph{flat} spacetime, not in curved one.  The content in this chapter is presented in the following order: (a) first we define generalized Nevanlinna spaces (b) second, we treat finite dimensional Clifford-Nevanlinna modules and study some spectral properties of Hermitian operators (c) we make some comments about statistics and appropriate probability interpretations.  \\* \\*
What follows could be done in full generality for any associative, unital, involutive algebra but we shall focus on the physical case and leave the rather trivial extension to mathematicians.  Denote by $\mathcal{C}_{(1,3)}$ the complex Clifford algebra with generators given by $\gamma^a$ and let $\mathcal{H}$ be a general $\mathcal{C}_{(1,3)}$ bi-module.  $\mathcal{C}_{(1,3)}$ is equipped with a sesquilinear indefinite ``scalar product'' $B$ defined as $B(v,w) = \widetilde{v}w$.  We furthermore demand that $\mathcal{H}$ is equipped with a ``scalar product'' satisfying the following properties:
\begin{eqnarray*}
\widetilde{\langle  \Psi | \Phi \rangle} & = & \langle \Phi | \Psi \rangle \\*
\langle a \Psi | \Phi \rangle & = & \langle \Psi | \widetilde{a} \Phi \rangle \\*
\langle \Psi + \zeta | \Phi \rangle & = &  \langle \Psi | \Phi \rangle + \langle \zeta | \Phi \rangle \\*
\langle \Psi a | \Phi \rangle & = & \widetilde{a} \langle \Psi | \Phi \rangle.
\end{eqnarray*} 
Note moreover that the Clifford numbers have an active operational meaning in the following sense; suppose $\Psi, \Phi$ are two orthogonal states, then $a \Phi$ and $\Psi$ are not orthogonal anymore.  All properties above are independent from one and another and the second one is rather restrictive and not mandatory in quaternion quantum mechanics.  One might consider dropping it, but I have good physical reasons -which I shall elaborate on later- to include it.  To understand what it means, consider the following finite dimensional representation: take $\mathcal{C}_{(1,3)}^n$ as $\mathcal{C}_{(1,3)}$ bi-module and define $\dag$ as the composition of the standard vector transposition and the Clifford conjugate.  Moreover, define
$$\langle \Phi | \Psi \rangle = \Phi^{\dag} R \Psi $$ where $R^{\dag}= R$.  Then, the second requirement holds if and only if all matrix elements of $R$ are scalar (more general, belong to the center of the algebra).  The reader might have fun generalizing the trace functional and determinant function and see that none of standard transformation properties associated to a change of basis hold anymore.  Also, the interpretation of probability gets seriously extended here: not only are negative probabilities allowed for, but the latter are deduced from the Clifford numbers (by considering the part associated to the algebraic unit only).  Indeed, this is the crucial difference with the division algebra's where an involution $\bar{a}$ exists such that $\bar{a}a$ is a positive real number.  Before we proceed, let me mention again that we could stay closer to the quaternion case and only treat the \emph{real} Clifford algebra instead of the complex one; this would put the imaginary unit outside the center and the standard candidate is $\gamma^{0}\gamma^{1}\gamma^{2}\gamma^{3}$ which anti-commutes with all the odd elements and commutes with the even ones.  One might proceed in this way keeping in mind that even complex linearity will be broken in quantum gravity which I choose not to do.  In contrast to the quaternion case, there is no technical advantage in doing so and therefore we work further in the complex case.  We first study some details regarding the Clifford algebra $\mathcal{C}_{(1,3)}$ where a general number can be written as
$$\textbf{v} = v + v_a \gamma^a + v_{ab} \gamma^{ab} + v'^{a} \gamma^a\gamma^5 + v' \gamma^5$$ and therefore
$$\widetilde{\textbf{v}} = \overline{v} + \overline{v}_a \gamma^a - \overline{v}_{ab} \gamma^{ab} + \overline{v'}_{a} \gamma^{a} \gamma^5 - \overline{v'} \gamma^5.$$
Hence, the scalar product $\widetilde{\textbf{v}}\textbf{w}$ reads
$$\widetilde{\textbf{v}}\textbf{w} = \overline{v}w + \overline{v}_a w^a + 2 \overline{v}_{ab}w^{ab} - \overline{v'}_a w'^a - \overline{v'}w' + \, \textrm{non scalar terms}$$ and the scalar part is an ordinary complex sequilinear form of signature $(8,8)$.  This allows for a splitting of $\mathcal{C}_{(1,3)}$ in two eight dimensional ``Euclidean'' spaces
$$\mathcal{C}_{(1,3)} = V_{+} \oplus V_{-}$$ where the scalar product constrained to either is positive, respectively negative.  In general, some elements of $\mathcal{C}_{(1,3)}$ may not be invertible and wherever appropriate we use the Moore-Penrose pseudo inverse\footnote{Again, D. Constales is gratefully acknowledged for useful conversations in this regard.}.  Of course, what the interpretation is concerned, only the scalar part is important meaning that we have to take the complex viewpoint enunciated above.  The main difference with the complex numbers is that the Clifford numbers also have an \emph{operational} significance which will become clear in the next chapter.    \\* \\*
Sometimes, weaking some conditions gives an entirely different perspective on the matter; old concepts do not make sense anymore and we have to look for a more universal way of thinking about matters.  The generalization of quantum mechanics which we are to set up is such a turning point and I am definitely unhappy with current work performed on these issues for several reasons which I will explain first.  The quaternion case was very close to the standard Hilbert space formalism since $||v||^2 = \langle v | v \rangle > 0$ for $v \neq 0$ and the quaternions themselves induce a positive definite metric.  This implies that $||v||$ is a norm and therefore defines a topology and one can proceed by demanding that $\mathcal{H}$ is a complete module with respect to this norm.  However, we have no such luck here since neither of both properties are satisfied in the generic case.  The second issue I want to address is the way indefinite norm spaces are treated so far in the literature: the original definitions of Krein \cite{Krein1} \cite{Krein2} and Jadczyk \cite{Jadczyk} broke Lorentz invariance in the sense that that the class of splittings in two genuine Hilbert spaces does not carry all \emph{unbounded} unitary operators.  This is a problem of self-reference where the notion of appropriate splittings fixes the notion of boundedness and vice versa.  Jadczyk told me that my problem was that I didn't want to break Lorentz invariance and at the same time I would.  Of course, this remark is very true in some sense, but in another sense it is not and we shall set up a scheme here where topology of observations is observer dependent.  Such strategy removes the absolutism from Hilbert space installed by Hilbert and Von Neumann and necessitates a more physically inspired definition in terms of transformation groups applied to bases.  We shall also assume this line of thought when developping the new geometry in the next chapter.  Having said this, let me give an equivalent relativist definition of Hilbert space and generalize this to spaces with indefinite norm.  It is well known that any Hilbert space $\mathcal{H}$ posesses an orthonormal basis $(e_i)_{i \in I}$ where $I$ is a general index set and where every vector in $\mathcal{H}$ can be written as $v = \sum_{i \in I} v^i e_i$ where at most an $\aleph_0$ of the $v^i$ are different from zero since $\sum_{i \in I} |v^i|^2 < \infty$.  It is well known that the scalar product $$ \langle v | w \rangle = \sum_{i \in I} \overline{v^i}w^i$$ by Parsival's theorem and any two orthonormal bases are connected by unitary operators $U$, where $$e'_i = U^j_{\,\, \, i} e_j.$$ Moreover, the following conditions are met
\begin{eqnarray*}
U^{j}_{\,\, \, i} & = & \langle e_j | U e_i \rangle \\*
e_i & = & \overline{U}^{i}_{\,\,\, j} e'_j. 
\end{eqnarray*}     
where it is understood that 
\begin{eqnarray*}
\sum_{i \in I} \overline{U}^k_{\,\,\, i} U^{k}_{\,\,\, j} & = & \delta_{ij} \\*
\sum_{i \in I} \overline{U}^{i}_{\,\,\, k}U^{j}_{\,\,\, k} & = & \delta^{ij}.
\end{eqnarray*}  
It is of crucial importance here that the \emph{order} in which these sums are taken is completely irrelevant due to the Cauchy-Schwartz inequality. \\* \\* Therefore, an equivalent definition of Hilbert space would go as follows: consider $\mathcal{H}$ to be a linear space over $\mathbb{C}$ and let $\mathcal{N}$ and $I$ be \emph{maximal} index sets such that the following properties are satisfied
\begin{itemize}
\item for any $i \in I$, a unique basis $(e^i_{j})_{j \in \mathcal{N}}$ exists in the sense that any $v \in \mathcal{H}$ can be written as a unique $\aleph_0$ formal sum in the basis elements,
\item for any $i \in I$ and $v, w \in \mathcal{H}$, the sum $\sum_{j \in \mathcal{N}} \overline{v}_j w^j$ exists in $\mathbb{C}$,
\item for any $i$ and $j$, there exists a matrix $U$ such that
\begin{eqnarray*}
e^j_r & = & U^s_{\,\,\, r} e^i_s \\*
e^i_r & = & \overline{U}^r_{\,\,\, s}e^j_s \\*
\delta_{ij} & = &  \sum_{i \in I} \overline{U}^k_{\,\,\, i} U^{k}_{\,\,\, j} \\*
\delta^{ij} & = & \sum_{i \in I} \overline{U}^{i}_{\,\,\, k}U^{j}_{\,\,\, k}, 
\end{eqnarray*}
\item for any $i \in I$, $\mathcal{H}$ is complete in its norm topology induced by its scalar product.
\end{itemize}
The latter is clearly a ``relational'' definition of Hilbert space and it allows us to generalize this construction in an important way; the point is of course that for Hilbert spaces the index set $I$ is completely abundant and the third bullet from the above definition can be entirely dropped.  However, such ``observer independence'' is precisely what we want to tackle which is the reason why I added this axiom to the definition.  From the above definition, one can deduce that for any $i,j,k \in I$ the matrices $U(j,i)$ have a unique extension to a bounded linear operator and moreover $U(k,j)U(j,i) = U(k,i)$.  In either, the latter form a non-abelian group.  Now, for indefinite spaces, all those axioms are in need for modification and we shall first introduce some novel concepts prior to giving the full blown definition in the more general case. \\* \\* 
The most obvious requirement which is going to fall is that $\mathcal{N}$ must be a mere index set.  Indeed, in order for the sums in the second bullet to be well defined for indefinite signature, one needs an order to perform them (unless one demands the sums of positive and negative numbers both to be finite such as Krein and Jadczyk did).  That is, $\mathcal{N}$ needs to be a directed set, which is a partially ordered set $(\mathcal{N}, \leq)$ satisfying the property that for any $a,b \in \mathcal{N}$, there exists a $c \in \mathcal{N}$ so that $a,b \leq c$.
However, this is not sufficient yet and at this point we make a not so obvious generalization which is rooted in the use of continuous bases in Quantum Field Theory.  That is, we give a more intrinsic definition which contains the notion of rigged Hilbert space as a special case. In other words, either we work with bound states or with a continuous spectrum and our formalism should include both; that is, we give an intrisically distributional formulation of the theory.  Therefore, we should make a special kind of measure space $(\mathcal{N}, \Sigma, \mu)$ of $\mathcal{N}$; that is $\mathcal{N}$ is equipped with a topology, $\Sigma$ is its Borel $\sigma$-algebra and $\mu$ a Lebesgue measure.  In order to avoid infinite ambiguities in the order of the ``sum'' we demand that all measurable anti-chains $S$ satisfy $\mu(S) < \infty$.  An anti-chain $S$ is a set of elements $a \in \mathcal{N}$ which are unrelated to one and another.  Since our measure space has a supplementary structure, we will define the notion of balanced integral straight after we complete our definition of Nevanlinna space.  Another useful concept is the one of a kroup which is a weakening of a group complementary to a semi-group and groupoid.  In particular a kroup $G$ is defined by a unitary relation $\star$ satisfying the following properties:
\begin{itemize}
\item if $a \star b$ is well defined for $a,b \in G$ then $a \star b \in G$,
\item there exists a unit element $e \in G$ so that for all $a \in G$, $a \star e = e \star a = a$,
\item every $a \in G$ is invertible, that is there exists a (not necessarily unique) $a^{-1}$ such that $a \star a^{-1} = e = a^{-1} \star a$.
\end{itemize}
In particular, $\star$ is not associative.  Finally, we define the following symbols: $u$ for undefined and $e^{i \theta}\infty$ for a (half) ray in the complex plane with $\theta \in \left[0, 2 \pi \right]$.  All this leads to the following:  let $(\mathcal{N},\Sigma, \mu)$ be as before, $I$ an index set, $(e_j^i)_{j \in \mathcal{N}, i \in I}$ be a matrix of vectors in a linear space $V$ and $R \subset I \times I$ be a symmetric relation.  For all $i \in I$, $(e^i_j)_{j \in \mathcal{N}}$ is the basis of a ``Krein space'' $\mathcal{K}^{i} \subset V$ which is constructed as follows: there exists a measurable function $\chi^{i} : \mathcal{N} \rightarrow \{ - 1,1 \}$ such that for measurable functions $f,g$ the  \emph{formal} vectors $\vec{f}^i = \int_{\oplus} d \mu(x) f(x) e^i_x$ and $\vec{g}^i = \int_{\oplus} d\mu(x) g(x) e^i_x$ obey the following defined scalar product
$$\langle \vec{f}^i | \vec{g}^i \rangle_i = \int d\mu(x) \overline{f}^{i}(x) \chi^i(x) g^{i}(x).$$
Likewise, we allow for the distributional scalar products where 
$\langle e^i_x | e^i_y \rangle$ is defined as 
$$\int d\mu(x) \langle e^i_x | e^i_y \rangle_i h(x) = h(y)\chi^i(y)$$ for all continuous functions $h$.  The scalar product between a ``distributional'' vector and regular vector is defined in the obvious way:
$$\langle e^i_x | \vec{f}^i \rangle_i = f(x) \chi^i(x).$$
  As is usual for Krein spaces, we \emph{define} $\mathcal{K}^i$ by splitting it into two pieces $\mathcal{K}^i = \mathcal{K}^i_{+} \oplus \mathcal{K}^{i}_{-}$ where both spaces are ``spanned'' by the $e^i_j$ corresponding to $\chi^i$ equal to plus one and minus one respectively.  Each $\mathcal{K}^{i}$ is then defined as a ``rigged Hilbert space'' in the following way:  denote by $\Phi^i_j$, where $j \in \{ 1 , - 1\}$ the algebra of continuous functions on $\mathcal{N}$ with compact support in the subset $(\chi^i)^{-1}(j)$ and $\mathcal{H}^i_{j}$ the Hilbert space generated by the vectors $\vec{f}^i_{j}$ where $f^i_j \in \Phi^i_j$.  Then, as usual, we have the inclusion 
$$\Phi^i_j \subset \mathcal{H}^i_j \subset (\Phi^{\star})^{i}_j$$ where 
$(\Phi^{\star})^{i}_j$ is the \emph{ordered} algebraic dual of $\Phi^i_j$; $\mathcal{K}^i_j$ is then defined as $(\Phi^{\star})^{i}_j$.  The ordered algebraic dual is defined as the linear space consisting out of elements of the form
$$\int_{\oplus} d\mu(x) f(x)^i_j e^i_x  + \sum_{<} c^{i \, k}_j e^i_k$$ where at most an $\aleph_0$ of the $c^{i \, k}_j$ are different from zero and the series is absolutely summable on intersections with measurable compacta.  Likewise, $f^i_j \in L^{1 \,\, i}_{\textrm{loc} \, j}(\mathcal{N},\mu)$ where the latter is the space of all locally (meaning on all compact measurable sets) absolute integrable functions with support in $(\chi^i)^{-1}(j)$ (the ordering does not matter in their definition, but it becomes important in the scalar product).  The distributional scalar product
of two elements in the algebraic dual may be regarded as a bi-distribution in the following way.  Consider $h^i_{j}$ a continuous function of compact support on $(\chi^i)^{-1}(j) \times (\chi^i)^{-1}(j)$ and let $\int_{\oplus} d\mu(x) f(x)^i_{j \, r} e^i_x  + \sum_{<} c^{i \, k}_{j \, r} e^i_k$ be two distributions, where $r = 1,2$, then the scalar product
$$
\langle \int_{\oplus} d\mu(x) f(x)^i_{j \, 1} e^i_x  + \sum_{<} c^{i \, k}_{j \, 1} e^i_k | 
\int_{\oplus} d\mu(x) f(x)^i_{j \, 2} e^i_x  + \sum_{<} c^{i \, k}_{j \, 2} e^i_k \rangle_i (h^i_{j})$$ equals
\begin{eqnarray*}
\int d \mu(x) \sum_k \overline{c}^{i \, k}_{j \, 1} \overline{h}^i_{j}(k,x) \sum_{l} c^{i \, l}_{j \, 2} h^i_j(l,x) + \int d\mu(x) \int d\mu(y) \int d\mu(z) \overline{f}^i_{j \, 1}(y) f^i_{j \, 2}(z) \overline{h}^i_j(y,x) h^i_j(z,x) & + &  \\*
\int d\mu(x) \int d\mu(y) \overline{f}^i_{j \, 1}(y) \overline{h}^i_j(y,x) \sum_k c^{i \, k}_{j \, 2} h^i_j(k,x) + 
\int d\mu(x) \int d\mu(y) \sum_k \overline{c}^{i \, k}_{j \, 1} \overline{h}^i_j(k,x) f^i_{j \, 2}(y) h^i_j(y,x). & &
\end{eqnarray*} 
One can generalize this to include limits of continuous functions by ordering all sub-integrals and sums in $\mathcal{N}$ and allowing to swith the order of the integrals and sums in which case the answer might be $u$ (if not all integrals exist and are equal) or $e^{i\theta}\infty$ (in case they exist, are equal and infinity); we shall come back to this later on.  It is now clear how the imbedding of $\mathcal{H}^i_j$ into $\mathcal{K}^i_j$ works.  Notice that we have made the scalar product positive definite by ignoring the sign of the ``norms'' of the basis vectors.  One could now think that this setting is large enough and that all $\mathcal{K}^i$ are equal; however, this is not the case and $V$ is much larger than the individual $\mathcal{K}^i$ are.    \\* \\* 
At this point, we stress that by changing the ``reference frame'' in $V$, we allow for signature changes of the $e^i_k$. Clearly, the spaces $\mathcal{K}^i$ are not necessarily isomorphic as vector spaces for different splits of $\mathcal{N}$.  Full isomorphy is for example guaranteed when two splittings can be connected by a homeomorphism $\phi : \mathcal{N} \rightarrow \mathcal{N}$, that is $\chi^i(x) = \chi^j(\phi(x))$, such that the Radon-Nikodym derivatives $| \frac{d \phi_{*} \mu(x)}{d \mu(x)}|$ and 
$| \frac{d \phi^{*} \mu(x)}{d \mu(x)}|$ exist and are bounded on $\mathcal{N}$.  The evaluation however of some states on (generalized) functions outside $\Phi$ may differ due to different orders of integration.  So, many of these spaces are equivalent in a weak sense but no longer in a stronger sense when evaluation occurs with respect to a ``harder'' algebra of ``test'' functions.  Obviously, it is the latter case which is of interest and which shall be studied further on.  We now consider unbounded unitary relationships between two different reference frames;  the only requirement these unitary matrices have to satisfy is that the basis vectors belong to the appriopriate distributional spaces.  In general, for all $i,j \in I$ such that $(i,j) \in R$, we study transformations of the kind 
\begin{eqnarray*}
e^i_x & = &  \int_{\oplus} U^{(ij)}(x,y) e^j_y d\mu(y) + \sum_{<} D^{(ij)}(x,k)e^j_k \\*
\langle e^i_x | e^i_y \rangle_i & = & \int d\mu(z) \overline{U^{(ij)}}(x,z)\chi^j(z) U^{(ij)}(y,z) + \sum_{<} \overline{U^{(ij)}}(x,k)D^{(ij)}(y,k)\chi^j(k) + \\* & & \sum_{<} \overline{D^{(ij)}}(x,k) \chi^j(k) U^{(ij)}(y,k) + \sum_{<} \overline{D^{(ij}}(x,k) D^{(ij)}(y,l) \langle e^j_k | e^j_l \rangle_j  
\end{eqnarray*} an likewise for the inverse transformation.   
 \\* \\*
The product of such unitary operators is not unambiguously defined; that is, its interpretation depends upon the function algebra one considers.  For example, when summing the whole infinite series, it is not associative: that is $A(Bv) \neq (AB)v$ and $A(BC) \neq (AB)C$ for some $v \in V$ and $A,B,C$ linear operators with some domain in $V$.  Indeed, for $\mathcal{N} = \mathbb{N}$ and $v = \sum_{n = 0}^{\infty} v^n e^i_n$, one could have that $$\sum_{n = 0}^{\infty} A_n^m \left( \sum_{p = 0}^{\infty} B_p^n v^p \right) \neq \sum_{p = 0}^{\infty} \left( \sum_{n = 0}^{\infty} A_n^m B_p^n \right) v^p.$$  This suggests one to take seriously the possibility of non-associative number ``rings'' but in this book, we won't go that far and leave this possibility open for future investigation.  Even if the product $U^{(jk)} U^{(ij)}$ is well defined in some sense, it is not necessarily a unitary transformation regardless of whether some $U^{(ik)}$ exists.  Therefore, vectors do not have an absolute meaning but one which is relative to a basis and the interpretation may become path dependent; this is a well known phenomenon in translations where translating a text directly from english to dutch is likely to give a different result than first passing through german.  Therefore, the addition in $V$ is and will remain \emph{formal} forever, its value is undefined a priori (but the sum is still formally associative for finite linear combinations) and only gets meaning relative to a reference frame.  Therefore, the unitary relationships between different reference frames are merely subjective connotations, they reflect how one frame ``perceives'' the other within its own comfort zone.  There is one supplementary condition (connectedness) the relation $R$ has to satisfy: for any $i,j \in I$ there exists a finite sequence $i = i_1, i_2, \ldots i_{n-1}, i_n = j$ such that $(i_k,i_{k+1}) \in R$ for all $k = 1 \ldots n-1$.  As said before, such chains may have nontrivial homology.  Summarizing:     
\\*  \\*
Let $I$, $\left( \mathcal{N}, \Sigma, \mu \right)$ and $R$ be as before, $(e^i_x)_{x \in \mathcal{N}}$ be formal ``bases'' for any $i \in I$ and $\chi^i$ measurable functions on $\mathcal{N}$ which take value in $\{ -1,1 \}$.  Denote by $\mathcal{K}^i$ the rigged Nevanlinna space defined by $\chi^i$ and $(e^i_x)_{x \in \mathcal{N}}$.  Consider $V$ the formal vector space spanned by all $\mathcal{K}^i$ and for any $(i,j) \in R$ we have a unitary operator and its inverse which is defined in $\mathcal{K}^i$ and $\mathcal{K}^j$ respectively.  The set of unitary operators forms a kroup and this entire structure is defined as a relational Nevanlinna space. \\* \\*
This definition can be expanded further if one adds more structure which is indeed the case following the axiomatic approach in the next chapter.  That is, we may assume that $\mathcal{N}$ has a subnet with the structure of a finite dimensional manifold.  In that case, the whole differential calculus may be imported and the definition of $\mathcal{K}^{i}$ may be generalized by choosing $\Phi^i$ to be a suitable class of Schwartz functions.  We shall not formalize these ideas here but this specification shall be used in the computations in sections nine and ten; where we shall not only work with distributional unitary operators but also with derivatives of them.  Before we proceed to relational Clifford-Nevanlinna modules, let us define ordered Lebesgue integration; a complete treatment of this subject is postponed for future work.  The key idea behind Lebesgue integration is to split up a measurable function in a positive and negative part and define separately the integral for the positive part by making succesive inferior approximations with simple functions.  Hence, the very idea of absolute convergence is build into the foundations while we know this to be a too strong requirement; this implies that a simple supremum or infimum won't suffice anymore.  Before we proceed, some definitions are in place; for a set $A \subset \mathcal{N}$, define
\begin{itemize} 
\item $J^{+}(A)$ as the set of all $x \in \mathcal{N}$ to the future of some point $y \in A$,
\item $\mathcal{J}^{+}(A)$ as the set of all $x \in \mathcal{N}$ to the future of all $y \in A$.
\end{itemize}
Now, it is left as an easy exercise to the reader that the relation $A \prec B$ iff $B \cap \mathcal{J}^{+}(A) \neq \emptyset$ constitutes a partial order.  Consider a countable set of disjoint measurable subsets $A_n$ and let $(\{A_n | n \in N \}, \prec)$ be the induced net which we assume to be past and spatially finite.  Obviously, for a simple function  
$$f = \sum_{n} a_n \chi_{A_n}$$ associated to a spatially finite partition we define
$$\int d\mu(x) f(x) = \sum_{<} a_n \, \mu(A_n)$$ where the last sum is defined by uniquely foliating the infinite poset as follows.  Define the $n$'th layer as the set of elements which have a distance of $n$ to the set of minimal elements (here the distance between $A_k$ and $A_l$ equals the length of the maximal chain between $k$ and $l$).  Then, the sum first goes over the zeroth, first, second, third ... layer and the order in which one sums within a certain layer is unimportant.  Clearly, this kind of partitions are sufficient for simple functions, however, for defining limiting procedures they are not suitable.  Indeed, consider the sequence $1,-1,\frac{1}{2}, - \frac{1}{2}, \frac{1}{3}, - \frac{1}{3}, \ldots$ and take $N > 0$; the partitions defined by taking every odd index as a singleton and likewise for the even numbers smaller than $2N$, complemented with one set containing all even indices bigger than $2N$, obviously satisfy the above requirements and the limit of the ordered sums equals infinity.  On the other hand, the obvious partitions, which consist of grouping the numbers pairwise for sufficiently large $2N$, give zero as result in the limit for $N$ to infinity.  From this example, it is clear that the partitions have to respect the order as much as possible, that is we have to make them ``causally optimal'' in some way.  However, we are not home yet because the set $\mathcal{A}$ of spatially and past finite partitions does not form a net; therefore, it is impossible to define a limiting procedure which is more sophisticated than taking a limsup\footnote{The reader may wish to construct a discrete net with a probability measure and two spatially and past finite partitions such that their intersection is not spatially finite anymore.}.  Hence, we construct a directed filter of $\mathcal{A}$ which resembles the time slicings in general relativistic theories implying that some extra structure on $\mathcal{N}$ is needed.  Closer inspection reveals that we need a measurable fibration of inextendible causal curves $\gamma$ as well as a continuous, measure preserving\footnote{Measure preserving means that $\mu(T^{-1}(\left[a,b \right]) = b - a$ for the \textit{smallest} interval on which $T^{-1}$ remains constant.}, time function\footnote{$T$ is a time function if and only if for any $x < y$ we have that $T(x) < T(y)$.} $T : \mathcal{N} \rightarrow \mathbb{R}$.  By a fibration of causal curves we mean a surjective mapping from a topological space $\gamma : \Sigma \times \mathbb{R} \rightarrow \mathcal{N}$ which is measurable and satisfies (a) $\gamma(x,s)$ is not in the future of $\gamma(x,t)$ if $s < t$ and (b) $\gamma(x,t) \neq \gamma(y,s)$ forall $x \neq y \in \Sigma$.  This is a very weak condition since the hypersurfaces of constant $t$ are not required to be anti-chains which allows for genuine topology change of $\mathcal{N}$.  Any choice of $T, \gamma$ leads to violations of Lorentz covariance unless sampling with continuous functions of compact support occurs.  More in particular we will define \textit{bricks} $\mathcal{B}$ as follows: $\mathcal{B}$ is a brick if and only if there exist $a < b \in \mathbb{R} \cup \{- \infty , + \infty\}$ such that if $x \in \mathcal{B} \subset T^{-1}(\left[a,b \right])$ then the intersection of the unique causal curve $\gamma(z,s)$, for some unique $z \in \Sigma$, which contains $x$ with $T^{-1}((a,b))$ is entirely contained in $\mathcal{B}$.  The \emph{optimal} numbers $a < b$, that means the largest $a$ and smallest $b$, associated to $\mathcal{B}$ are called the boundaries of the brick; it is furthermore easy to see that the intersection of two bricks is again a brick.  The kind of partitions we consider now are restricted by sequences $(a_n)_{n \in N}$ where $a_{0} = - \infty$, $a_n < a_m$ for $n <  m$ and $\lim_{n \rightarrow \infty} a_n = + \infty$.  Indeed, we consider countable partitions of $\mathcal{N}$ up to a subset of measure zero by Bricks which are bricks \emph{up to measure zero} and whose boundaries are determined by $a_n < a_{n+1}$ or $a_n = a_n$ and have the property that only a finite number of Bricks have the same boundaries.  An absolute bound in this case is not required since the order on $R$ defines a coarse way of taking the limit, the internal fluctuations within each layer, how large they may be, are completely irrelevant.  The set $\mathcal{B}$ of such countable partitions can be made into a directed set by means the operation $\subset$; indeed the intersection of two such partitions again belongs $\mathcal{B}$.  Again we define simple functions over partitions of $\mathcal{B}$ and the integral is given by the ordered sum where the order goes in increasing $n$ and one does not break up the layers; if the ordered sum does not exist, the answer is $u$, meaning undefined.  We are now left to determine the unique integral for more general measurable functions $f$; to every element $(A_{n})_n$ in $\mathcal{B}$, we attach a unique simple function by considering the infimum $r_n$ over all $x \in A_n$ of $|f(x)|$.
If there exist sequences for wich $f$ converges to $- r_n$ and $r_n$ then the step function assumes the value $0$ on $A_n$, otherwise the value is given by $r_n$ or $- r_n$.  The limit of the integrals of these step functions over $\mathcal{B}$, if it exists, is called the ordered integral of $f$.  This construction is far more general than the Lebesgue construction in the sense that it can deal with unbounded functions properly, on the other hand it is different due to the special nature of the partitions and the fact that we ask the \emph{limit} to exist which is what balancing is really about.  To understand why it is important that we consider partitions up to measure zero, consider the function $f: \left[0,1 \right] \rightarrow \mathbb{R}$ which maps every rational element to one and every irrational number to minus one.  Then, the Riemann integral does not exist, the Lebesgue integral is minus one and the ordered integral is minus one for the natural time function and trivial fibration.  However, if we wouldn't have added the adjective up to measure zero, then the ordered integral is equal to zero and the reason why such subtle difference exists is because our first class of partitions is finer than the second one\footnote{The definition of directed limit is the following: $a$ is the directed limit of the real function $f$ over the directed set $\mathcal{N}$ if and only if for any $\epsilon > 0$, there exists an $x \in \mathcal{N}$, such that for all $y > x$ one has that  $|f(y) - a| < \epsilon$.  It is easy to see that the limit, if it exists, is unique.}.  The Lebesgue integral however is very unnatural and quite limited in many other ways; for example, the whole integral for more general algebras has to be derived from mapping this algebra to the real numbers.  This is not so for our construction if one puts some further technical restricitions (and again this has everything to do with the notion of balancing) and we leave such natural extensions for future work.  Likewise, further mathematical investigation of this notion of integrability is postponed for work to come.       
\\* \\*  
There is a whole new functional analysis to be developed around the topic of relational Nevanlinna spaces; in particular continuity becomes a relative property and therefore many subtleties arise further on.  We shall, in what follows, only create those concepts which are strictly necessary but a huge remainder is left open.  Therefore, prior to turning to the issue of Clifford-Nevanlinna modules, let us generalize the notion of continuity a bit further.  In standard Hilbert space analysis, continuity is defined with respect to the Hilbert space norm and in Krein space the same concept holds where the norm is defined through the preferred splitting (and changing the minus to plus).  Here, the access a local reference frame $(e^i_x)_{x \in \mathcal{N}}$ has, is to a distributional Nevanlinna space $\mathcal{K}^i$ which is most properly regarded as a locally convex space.  Indeed, on $\mathcal{K}^i_{+}$, one can use continuous bifunctions $h^i_+$ of bounded support on $(\chi^{i})^{-1}(1) \times (\chi^{i})^{-1}(1)$ to define seminorms $p_{h^i_+}$ by using the mapping we constructed previously.  The norm topology on the regular Nevanlinna space $\mathcal{H}^i$ is then recovered by extending the possible functions $h^i_+$ to the square integrable functions and insist upon uniform convergence.  Indeed, this is necessary (and sufficient) as the following easy example demonstrates : consider the series $(x^{\infty}_n)_{n \in N_0} = (\frac{1}{n})_{n \in N_0}$ and $x^m_n = \frac{1}{n} + \delta^m_n$, then $x^m$ converges to $x^{\infty}$ in the orginal locally convex topology but not in the norm topology on $l_2(N_0)$.  Now, we investigate the topological dual $(\mathcal{K}^{i}_{+})^{\star}$, that is the vector space of continuous linear functionals $\phi^i_{+} : \mathcal{K}^{i}_{+} \rightarrow \mathbb{C}$.  As is well known, $\phi^i_{+}$ is continuous if and only if there exists a positive number $M$ and a finite number of seminorms $p^i_k$, $k : 1 \ldots n$ such that
$$ | \phi^i_{+}(x) | \leq M \left( p^i_1(x) + p^i_2(x) + \ldots + p^i_n(x) \right)$$
for all $x \in \mathcal{K}^i_{+}$.  A similar construction holds for $\mathcal{K}^{i}_{-}$ and one has a canonical generalization to $\mathcal{K}^i$ of this concept.  Clearly, $\mathcal{K}^i_{+}$ is not complete in the locally convex topology for the same reason as delta distributions can be limits of square integrable functions.  A linear superfunctional: $\phi^i : \mathcal{K}^i \rightarrow \mathbb{C}$ is a linear superfunctional relative to $(e^i_x)_{x \in \mathcal{N}}$ if and only if $\phi^i$ is linear and $\phi^i(v^i)$ is a bi-distribution on $\mathcal{N}$ for all $v^i \in \mathcal{K}^i$.  For any complex valued continuous function $h$ of compact support on $\mathcal{N} \times \mathcal{N}$, define $\phi^i_{h} : \mathcal{K}^i \rightarrow \mathbb{C}$ as the linear functional 
$$\phi^i_h(v^i) = \phi^i(v^i_{\overline{h}})$$ where $v^i_{\overline{h}} \in \mathcal{H}^i$ is defined by the convolution product as before.  Now, we say that $\phi^i$ is continuous with respect to $(e^i_x)_{x \in \mathcal{N}}$ if and only if $\phi^i_{\overline{h} \star h}: \mathcal{K}^i \rightarrow \mathbb{C}$ is continuous for all $h$ and $\star$ denotes convolution here.  Now, linear (super)functionals on the universal relational Nevanlinna space $V$ clearly have to satisfy $U$ compatibility; that is, for all $i,j \in I$ such that $(i,j) \in R$ we demand that 
$$ \phi(v^i) = \phi(v^j)$$ for whenever $v^i$ and $v^j$ can be related by $U(i,j)$.  Likewise, one can define continuity at this level by demanding that the (super)functional is continuous with respect to all reference frames $(e^i_x)_{x \in \mathcal{N}}$.  This is as complicated life gets for now and we investigate these concepts a bit further. \\* \\*
As is well known, the Hahn-Banach theorem applies for general locally convex spaces and the issues one should address here are completeness and the Riesz representation theorem for (super)continuous functionals relative to $(e^i_x)_{x \in \mathcal{N}}$.  In particular the Riesz representation theorem for a supercontinuous functional $\phi^i$ concerns whether there exists a distribution $x^i \in \mathcal{K}^i$ such that 
$$\phi^i(v^i)(h) = \phi^i_{\overline{h} \star h}(v^i) = \langle x^i | v^i \rangle_i (\overline{h} \star h) = \langle x^i_h | v^i_h \rangle_i = \phi^i_h(v^i_h)$$ for all suitable $h$.  I believe all these questions can be answered in the positive modulo some tiny technical details and mathematicians should further investigate these issues.  We will simply work with continuous functionals and operators such that the Riesz representation automatically holds.  One has two possibilities now to define the adjoint of a distributional operator; either one constructs it directly using the preferred basis, or one tries the usual way via Hahn-Banach and the Riesz representation theorem.  We will assume the first strategy here which just boils down to defining the adjoint of a generalized matrix and leave the second road for future mathematical work.  That is, we consider operators $A$ defined by their matrix expressions
\begin{eqnarray*}
A(e^i_x) & = & \int_{\oplus} d \mu(y) A(x,y) e^i_y +  \sum_{k} A'(x,k) e^i_k 
\end{eqnarray*}
where $A(e^i_x)$ is in $\mathcal{K}^i$ and we demand that 
\begin{eqnarray*}
A^{\dag}(e^i_x) & = & \int_{\oplus} d \mu(y) \chi^i(y) \chi^i(x) \overline{A}(y,x) e^i_y  + \sum_k \chi^i(k) \chi^i(x) \overline{A}'(k,x) e^i_k
\end{eqnarray*}
is in $\mathcal{K}^i$ as well.  Obviously $A^{\dag}$ is called the adjoint of $A$; the definition of a self-adjoint operator (relative to an observer) is easier since all it requires is that $A^{\dag} = A$.  However, it does of course not hold that if $\langle v^i | A w^i \rangle_i$ is well defined as a bi-distribution that therefore $\langle A v^i | w^i \rangle_i$ enjoys these properties.  And even if it were well defined, both distributions are not necessarily equal to one and another.  A further mathematical task is to develop the spectral theory of such operators; obviously the spectrum of self adjoint operators can become complex as is well known to be the case for Nevanlinna space theory. 
  \\* \\*
We turn now to the subject of finite dimensional Clifford-Nevanlinna modules which we started at the first page of this chapter.  To start with, denote by $\mathcal{H}$ a bi-module with a scalar product satisfying our four requirements and construct the indefinite complex vectorspace induced by $\mathcal{S} \, \langle w | v \rangle$.  Clearly, on $(\mathcal{H}, \mathcal{S} \, \langle \, | \, \rangle)$ as a complex Nevanlinna space the Riesz representation theorem holds meaning that complex-linear functionals $\phi$ are all of the form
$$ \phi(v) = \mathcal{S} \, \langle w | v \rangle$$ for some unique $w \in \mathcal{H}$ where $\mathcal{S}$ denotes the unital part.  This allows one to define, as usual, the adjoint $A^{\dag}$ of an operator $A$.  As in the quaternion case \cite{Finkelstein}, one may define right linear and right colinear operators $A$:
\begin{itemize} 
\item $A$ is right linear if and only if $A( v + w) = Av + Aw$ and $A (vb) = (Av)b$ for all $b \in \mathcal{C}_{(1,3)}$.
\item $A$ is right colinear if and only if the second condition gets modified to
$A(vb) = (Av) \kappa_A(b)$ for some automorphism $\kappa$ of $\mathcal{C}_{(1,3)}$.
\end{itemize} 
For the quaternions, right linearity of $A$ automatically implies that $\langle w, Av \rangle = \langle A^{\dag}w | v \rangle$ simply because for any $r \in \mathbb{H}$ there exists an invertible $q \in \mathbb{H}$ such that $rq = \alpha 1$ with $\alpha \in \mathbb{C}$.  Consider $v,w$ such that $\langle v | Aw \rangle$ is not invertible in $\mathcal{C}_{(1,3)}$, then one can find a series of perturbations $\delta v, \delta w, \epsilon(\delta v ,\delta w)$ where the last one is a complex number such that
$$\langle v + \delta v | (A + \epsilon 1)(w + \delta w) \rangle$$ is invertible in $\mathcal{C}_{(1,3)}$.  Hence by continuity, we arrive at the result that 
$$\langle v | Aw \rangle = \langle A^{\dag}v | w \rangle$$ where we have used the property that $\langle A^{\dag} v | w \rangle$ is a scalar if and only if $\langle v | A w \rangle$ is.  By the same logic, we discover that the adjoint of a right colinear operator $A$ satisfies
$$\kappa^{-1}_{A} \left( \langle v | A w \rangle \right) = \langle A^{\dag} v | w \rangle$$ and moreover $A^{\dag}$ is right colinear with automorphism given by $\widetilde{\kappa}^{-1}_{A}$.  For inner automorphisms $\kappa_A(v) = qvq^{-1}$ where $q = \widetilde{q}$, the latter expression equals $\kappa_A$.  Likewise, we can define unitary and co-unitary operators; the former is a right linear operator $U$ satisfying
$$ U U^{\dag} = 1 = U^{\dag} U$$ while the latter is a right colinear operator obeying
$$\langle Uv | Uw \rangle = \widetilde{q}^{-1}\langle v | w \rangle q^{-1}.$$
Due to the second property of the scalar product, the adjoint of any continuous right linear operator satisfies
\begin{eqnarray*} (a U)^{\dag} & = & U^{\dag} \widetilde{a} \\*
(U a)^{\dag} & = & \widetilde{a} U^{\dag}. \end{eqnarray*}
Two important classes of automorphisms are the inner ones and the complex conjugation; that is 
$$\kappa_{A}(v) = qvq^{-1}$$ for some invertible element $q \in \mathcal{C}_{(1,3)}$ or 
$$\kappa_{A}(v) = \overline{v}.$$  From the inner automorphisms, one may still restrict to the subgroup of the real ones (meaning that they map real elements to real elements): that is, $q^{-1} = \widetilde{q}$ which is nothing but the pin group.  This leads to an extension of Wigner's theorem, where ``appropriate'' symmetry transformations either are complex unitary or anti-unitary and moreover a twist is allowed for by an element of the pin-group\footnote{The non-real conjugations are excluded since we basically restrict to co-Hermitian operators.}. \\* \\* We investigate now the status of the spectral decomposition theorem which is somewhat more complicated than is the case for ordinary Euclidean or indefinite norm spaces.  However, in real quaternionic quantum mechanics the situation is \emph{not} more complex because every quaternion module is free meaning it has a basis and the quaternions themselves are a division algebra.  It is important to stress that another property of modules which is often blindly accepted is responsible for this result which is irreducibility of the scalar multiplication: that is, $1.v = v$ for all vectors $v$.  I am not sure that this assumption should hold: the module is to be thought of as all what \emph{is} while the ring represents all that we can \emph{access}.  It is not so that we should be able to access all that is and indeed this stance of hidden variables at the quantum level immediately leads to torsion\footnote{A torsion element $v$ of a module over a ring $R$ without zero divisors is defined by the property that there exists a nonzero $r \in R$ such that $rv = 0$.} ``modules''.  We call such weaker modules wmods and it might be interesting to study spectral theory on such objects.  For standard Clifford bi-modules, the situation should not be overly complex since the Clifford algebras are unital and those elements which break the division property have Lebesgue measure zero.  Also, in real quaternionic quantum mechanics, the status of the number $i$ changes; it becomes contextual meaning it depends upon the normal operators considered.  I am not sure whether this is something deep or not; at least I feel the status of this issue has not been sufficiently clarified.  Therefore, for ordinary Clifford modules (with a real or complex Clifford algebra) the situation should be rather close to quantum theory on standard complex Nevanlinna spaces.  The only distinction is that real eigenvalues are not in the center of the algebra anymore as is the case for the real quaternions, this will complicate matters slightly.  Therefore, we are only starting to wander around in a magical land of mathematical possibilities which has hardly been considered by the physics community so far.  As an alternative road to quantum gravity, one may consider algebras with deformed product and sum structures where the latter is defined as $$X \oplus Y = X + Y + \epsilon \nabla_{X}Y - \frac{\epsilon}{2}\left[ X, Y \right]$$ for vectorfields $X$ and $Y$.  This sum is neither associative, nor commutative and therefore has some nonzero curvature and torsion.  This construction might pose a natural generalization of geometry: indeed parallel transport would mean that $X \oplus X = 2X$ and one would be able to define geodesics purely based upon algebraic properties.  Dynamics could be expressed as a constraint on a particular sum over all possible free vectors and would acquire in this way a direct operational status.  One could define tensors purely based upon associators and commutators and perform a generalized investigation of the equivalence problem.  \\*\\*
We now turn to the construction of a satisfactory spectral theorem for Clifford-Nevanlinna modules which is a rather complicated issue; we will work in a very pedestrian way towards the strongest possible kind of statement one can extract and learn that the standard theorem cannot be upheld by any reasonable standards.  Therefore, it is opportune to ask oneself at this moment what kind of result one really wants from the physical point of view.  Obviously, the standard theorem allows for a straightforward implementation of the Born rule but the latter does not necessitate the former.  Indeed, consider a matrix of the following form expressed in a standard orthonormal basis
$$ 
A = \left(  
\begin{array}{cccc}
2 & 1 & 0 & 0 \\*
0 & 2 & 0 & 0 \\*
0 & 0 & 3  & 0 \\*
0 & 0 & 0 & 3 	
\end{array}
\right)
$$ then $A$ has a Jordan decomposition in which the generalized eigenspaces are orthonormal to one and another which is sufficient to have a straightforward Born rule interpretation.  Actually, it would be already sufficient for $A$ to have such decomposition up to an arbitrary approximation $\epsilon > 0$ in some sense.  $A$ above obeys $A^{\dag} = UAU^{\dag}$ for some unitary transformation $U$ commuting with the Jordan projectors of $A$; indeed, an adequate spectral theorem can be formulated for these twisted Hermitian transformations.  It is easy to prove that all eigenvalues of $A$ are real and that the generalized eigenspaces are perpendicular to one and another. \\* \\* 
Consider $A$ to be a hermitian operator on a finite dimensional Clifford-Nevanlinna module, then we define the right (left) spectrum $\sigma^R(A)$ ($\sigma^L(A)$) as the set of Clifford numbers $\alpha$ such that there exists a vector $v$ so that $Av = v \alpha$ ($Av = \alpha v$).  The first thing one could try now is to define orthogonality as a linear concept: that is, by demanding that $\textrm{Tr} \left( \langle v | w \rangle \right) = 0$.  This however does not work since the natural condition for different eigenvectors corresponding to distinct eigenvalues is that $\langle v | w \rangle$ is not invertible.  Therefore, orthogonality presents itself as a \emph{nonlinear} concept over $\mathbb{R}$ or $\mathbb{C}$; that is, the natural operation is the determinant and not the trace.  Consider then first those eigenvalues $\alpha$ of the right spectrum which are invertible \textit{and} correspond to an eigenvector $v$ such that $\langle v | v \rangle$ is invertible.  Then, hermiticity implies that $$\widetilde{\alpha} = \langle v | v \rangle \alpha \langle v | v \rangle^{-1}$$ meaning $\widetilde{\alpha} \sim \alpha$ and moreover, $\widetilde{\alpha} = \alpha$ if and only if $\alpha$ commutes with $\langle v | v \rangle$.  From the definition of $\, \widetilde{} \,$ it follows that $\overline{\textrm{det}(\alpha)} = \textrm{det}(\widetilde{\alpha})$ and therefore $\textrm{det}(\alpha)$ is a real number.  Also, consider two inequivalent such eigenvalues, then an easy calculation reveals the scalar product between any respective eigenvectors satisfies $\textrm{det} \left( \langle v | w \rangle \right) = 0$.  For complex Clifford algebras, one can formulate a \emph{degenerate} Gram-Schmidt procedure where the degeneracy comes from the nonlinearity.  Indeed, suppose $v,w$ are two invertible eigenvectors corresponding to the eigenvalue $\alpha$ and let $x$ be a complex number which solves the eigenvalue problem
$$\textrm{det} \left( \langle w | v \rangle \langle v | v \rangle^{-1} - x 1 \right) = 0$$ then $w - vx$ is perpendicular to $v$.  It is easy to generalize this to multiple vectors.  Moreover, $vq$ is an eigenvector with invertible norm corresponding to the eigenvalue $q^{-1} \alpha q$ for all invertible $q$.  However, this is entirely consistent since we classified the eigenvalues according to their determinant.  This implies the nice property that we may have a continuum of different ontological eigenvalues which correspond to one empirical eigenvalue and a finite dimensional right module of eigenvectors.  Let us now treat those invertible eigenvalues $\beta$ which have no eigenvector $w$ such that $\langle w | w \rangle$ is invertible.  Then, one cannot conclude that the determinant of $\beta$ is a real number (as is usual in Nevanlinna space) and for $\alpha$ and $v$ of the first kind and $\beta$, $w$ of the second kind it immediately follows that $\textrm{det} \left( \langle w | v \rangle \right) = 0$.  Note that no Gram-Schmidt procedure can be set up amongst null vectors which applies to eigenvalues of both types.  The case when the eigenvalue $\alpha$ is not invertible, that is $\textrm{det}(\alpha) = 0$, is easily treated.  Indeed, all eigenvectors $v$ in cases one and two are perpendicular to all eigenvectors $w$ of $\alpha$ and the Gram-Schmidt procedure is applicable as before.        
\\* \\* 
Let us mention some general facts about the Clifford conjugation:
\begin{itemize}
\item $\widetilde{a} = a$ does not imply that the spectrum is real, for example consider $\widetilde{\gamma}^0 = \gamma^0$ but $(\gamma^0)^{\dag}= - \gamma^{0}$ so the spectrum is $\pm i$  
\item  in fact $\widetilde{a} = a$ does not imply either that $a$ is diagonizable; for example, take $i\left( \gamma^0 \gamma^2 + \gamma^2 \gamma^3 \right)$
\item both conclusions also hold for $b = \widetilde{a}a$; actually $\gamma^0 = \left( \frac{1}{\sqrt{2}} \left( 1 + \gamma^0 \right) \right)^2$ and $i\left( \gamma^0 \gamma^2 + \gamma^2 \gamma^3 \right)$ can be written in this form.
\end{itemize}  Let us first study a simple example in full detail and see what we can learn from that; that is, consider the matrix
$$A = \left( \begin{array}{cc} 0 & a \\
\widetilde{a} & 0 \end{array}  \right)$$ where $a$ is invertible.  Then, the eigenvalue equations become
\begin{eqnarray*}
x z - ay & = & 0 \\*
\widetilde{a}x - yz & = & 0 
\end{eqnarray*}
implying that 
\begin{eqnarray*}
x z^2 - a \widetilde{a}x & = & 0 \\*
ay - xz & = & 0.
\end{eqnarray*}
Now, we have to distinguish two cases: (a) $x$ is invertible and (b) $x$ is not.  If the former holds, then we can perform a gauge transformation $x \rightarrow 1$ such that 
\begin{eqnarray*} z^2 & = & a\widetilde{a} \\*
y & = & a^{-1}z.
\end{eqnarray*}
These equations have a solution if and only if the Jordan type of $a\widetilde{a}$ is in the range of the mapping $z \rightarrow z^2$ which is always the case for \textit{invertible} $a$.  Therefore, the residual ambiguities are given by considering all possible roots of $a\widetilde{a}$ which form a disjoint union of manifolds of different dimension depending upon the commutator of one particular root (in case of multiple eigenvalues of $a\widetilde{a}$).  It is clear that in this case, the eigenvectors of $A$ span the entire module (even if we haven't considered all possible eigenvectors yet) which may be proven by noticing that $1 = q + (1 - q)$ allowing for the second component to be $y' + (y - y')q$ for all solutions $y,y'$ and $q \in \mathcal{C}_{(1,3)}$.  Now, suppose that $a$ is not invertible and the orbit of $a\widetilde{a}$ is not in the range of $z \rightarrow z^2$, then $x$ cannot be invertible and we have to consider the second case (b).  That is, $x z^2 = a \widetilde{a}x$ implies that $z^2$ must map the nucleus of $x$ into the nucleus of $x$ and $a \widetilde{a}$ must map the image of $x$ into the image of $x$; therefore, the image of $x$ must be an invariant subspace of $a\widetilde{a}$.  Actually, this is all information one can get out of the first equation; $x : W/\textrm{Ker}(x) \rightarrow \textrm{Im}(x)$ is invertible and
$$x z^2 x^{-1} = a \widetilde{a}$$ as an equation on $\textrm{Im}(x)$.  Since $x$ only has to satisfy that $\textrm{Im}(x)$ is invariant under $a \widetilde{a}$, $z^2$ is uniquely determined on $W/\textrm{Ker}(x)$ but all the other degrees of freedom of $z$ are completely free.  The remaining question is whether such $z$ exists and if so whether the eigenvectors span the entire module.  Clearly, $\textrm{Im}(x)$ must be an invariant subspace of $a \widetilde{a}$ such that the restriction to it has a Jordan type which is in the image of $z \rightarrow z^2$; this a necessary and sufficient condition for $z$ to exist.  Obviously, the continuum of eigenvectors do not constitute a \emph{right} basis as can be seen from the following example.  Pick $$a = \frac{1}{\sqrt{8}} \left( - i \gamma^0 \gamma^2 - i \gamma^2 \gamma^3 - \gamma^0 \gamma^3 + 1 \right)$$ then
$$a \widetilde{a} = - \frac{i}{2} \left( \gamma^0 \gamma^2 + \gamma^2 \gamma^3 \right)$$
and the invariant subspaces of $a\widetilde{a}$ which are in the image of $z^2$ are all subspaces of $\textrm{Span}\{e_1 , e_4 \}$.  Hence $\textrm{Im}(x) \subset \textrm{Span}\{ e_1 , e_4 \}$ meaning that vectors of the kind $xq$ can never generate the entire Clifford algebra.  One could introduce the notions of approximate eigenvalues and eigenvectors at this point and see where this leads to; the rationale behind this is that every Jordan type is in the closure of the image of $z \rightarrow z^2$.  So far, we have only discussed whether the eigenvectors of $A$ span the two dimensional module; before we investigate a more general case let us see to what extend we can extract a spectral theorem.  Therefore consider the example where $A$ is given by 
$$ A =  
\left( \begin{array}{cc}
0 & \gamma^1 \\*
\gamma^1 & 0	
\end{array} \right)
$$ then the solutions of $z^2 = 1$ are divided into five Jordan classes, where the class with $n$ times the eigenvalue $-1$ has $\frac{4!}{(4 - n)! n!}$ members; one notices that all Jordan matrices commute as they should.  $z = \pm 1$ are special since they are in the center of the algebra and one could expect to obtain an ordinary spectral decomposition.  Indeed, the normalized eigenvectors are $$ v_{\pm} =
\frac{1}{\sqrt{2}} \left( \begin{array}{c}
1 \\*
\pm \gamma^1	
\end{array} \right) $$ respectively and moreover
$$ A = \frac{1}{2} \left( 
\begin{array}{cc}
1 & \gamma^1 \\*
\gamma^1 & 1	
\end{array}
  \right) - \frac{1}{2} \left( 
\begin{array}{cc}
1 & - \gamma^1 \\*
- \gamma^1 & 1	
\end{array}
 \right). $$  Now, we examine the other ``spectral decompositions'' using the fact that the natural diagonal projection operators are given by $1, - i \gamma^1 \gamma^2, \gamma^0 \gamma^3, \gamma^5$; in particular, $\widetilde{}$ induces a permutation within the roots of equal Jordan class which simply consists in swapping the diagonal $2 \times 2$ blocks.  For example,
\begin{eqnarray*}
\widetilde{\left( 
\begin{array}{cccc}
-1 & 0 & 0 & 0 \\*
0 & 1 & 0 & 0 \\*
0 & 0 & 1 & 0 \\*
0 & 0 & 0 & 1	
\end{array}
   \right)} & = & \left( 
\begin{array}{cccc}
1 & 0 & 0 & 0 \\*
0 & 1 & 0 & 0 \\*
0 & 0 & -1 & 0 \\*
0 & 0 & 0 & 1	
\end{array}
   \right) \\*
\widetilde{\left( 
\begin{array}{cccc}
1 & 0 & 0 & 0 \\*
0 & -1 & 0 & 0 \\*
0 & 0 & 1 & 0 \\*
0 & 0 & 0 & 1	
\end{array}
   \right)} & = & \left( 
\begin{array}{cccc}
1 & 0 & 0 & 0 \\*
0 & 1 & 0 & 0 \\*
0 & 0 & 1 & 0 \\*
0 & 0 & 0 & -1	
\end{array}
   \right) \\*     
\end{eqnarray*} 
therefore we have $4$ real roots and $6$ pairs of conjugated ones.  For this type of matrices $A$ in two Clifford dimensions, the spectrum has a peculiar property: that is, if $z$ belongs to the spectrum, then $- z$ also and the scalar product between the associated eigenvectors is exactly zero allowing for ``standard'' spectral decompositions.  The above decomposition could be classified as of type I, that is the projection operators are orthogonal and Hermitian and moreover, the eigenvalues are self adjoint and commute with the latter.  We now discuss a decomposition of type II, where the projection operators are orthogonal and Hermitian, the eigenvalues are self adjoint but do not commute with the relevant operator.  Such decomposition is provided by the eigenvalues 
\begin{eqnarray*}
z_1 & = & \left( 
\begin{array}{cccc}
- 1 & 0 & 0 & 0 \\*
0 & 1 & 0 & 0 \\*
0 & 0 & -1 & 0 \\*
0 & 0 & 0 & 1
\end{array}
  \right)  \\* 
z_2 & = & \left( 
\begin{array}{cccc}
1 & 0 & 0 & 0 \\*
0 & - 1 & 0 & 0 \\*
0 & 0 & 1 & 0 \\*
0 & 0 & 0 & -1
\end{array}
  \right) 
\end{eqnarray*} and 
$$A = \left( 
\begin{array}{cc}
z_1 & 0 \\*
0 & - z_1	
\end{array}
  \right) \frac{1}{2} \left( 
\begin{array}{cc}
1 & z_1 \gamma^1 \\*
\gamma^1 z_1 & 1	
\end{array}
  \right) + \left( 
\begin{array}{cc}
z_2 & 0 \\*
0 & - z_2	
\end{array}
  \right) \frac{1}{2} \left( 
\begin{array}{cc}
1 & z_2 \gamma^1 \\*
\gamma^1 z_2 & 1	
\end{array}
  \right)$$ where the diagonal matrix $\left( 
\begin{array}{cc}
z_1 & 0 \\*
0 & - z_1	
\end{array}
  \right)$ originates from the fact that $\gamma^1$ anticommutes with $z_1$.  We now arrive at a type III decomposition which also occurs in Nevanlinna spaces; that is, the eigenvectors corresponding to $z$ and $\widetilde{z} = -z$ have zero norm and their scalar product equals one.  Moreover, the eigenvalues and eigenvectors commute; this is the case for 
$$z = \left( 
\begin{array}{cccc}
- 1 & 0 & 0 & 0 \\*
0 & 1 & 0 & 0 \\*
0 & 0 & 1 & 0 \\*
0 & 0 & 0 & -1 	
\end{array}
   \right)$$ where the eigenvectors are given by 
   $$ v_{\pm} = \frac{1}{\sqrt{2}} \left( 
\begin{array}{c}
1 \\*
\pm \gamma^1 z	
\end{array}
   \right).$$  The spectral decomposition is then
 $$A = z \frac{1}{2} \left( 
\begin{array}{cc}
1 & z \gamma^1 \\*
\gamma^1 z & 1 
\end{array}
  \right) - z \frac{1}{2} \left( 
\begin{array}{cc}
1 & - z \gamma^1 \\*
- \gamma^1 z & 1 
\end{array}
  \right)$$ and clearly, the projection operators are not Hermitian.  While these types of decompositions are still pretty close to Nevanlinna space, type IV radically departs from it.  Here, the basis eigenvectors corresponding to $z$ and $-z$ have weakly zero norm as well does the scalar product between them; consider
 $$z =  \left(
\begin{array}{cccc}
- 1 & 0 & 0 & 0 \\*
0 & 1 & 0 & 0 \\*
0 & 0 & 1 & 0 \\*
0 & 0 & 0 & 1	
\end{array} \right)
  $$ then the corresponding eigenvectors have norm $$1 + \widetilde{z}z = \left( 
\begin{array}{cccc}
0 & 0 & 0 & 0 \\*
0 & 2 & 0 & 0 \\*
0 & 0 & 0 & 0 \\*
0 & 0 & 0 & 2 	
\end{array}
  \right)$$ and the inner product is given by $$1 - \widetilde{z}z = \left( 
\begin{array}{cccc}
2 & 0 & 0 & 0 \\*
0 & 0 & 0 & 0 \\*
0 & 0 & 2 & 0 \\*
0 & 0 & 0 & 0	
\end{array}
  \right).$$  The most conventional thing to do now is to remark that the eigenvector corresponding to $ - \widetilde{z}$ is perpendicular to $v_{+}$ and has scalar product $2$ with $v_{-}$ and symmetrically for the eigenvector corresponding to $\widetilde{z}$.  This allows one construct the non Hermitian, orthogonal projection operators
 $$P_{\pm} = \frac{1}{2} \left( 
\begin{array}{cc}
1 & \pm z \gamma^1 \\*
\pm \gamma^1 z & 1
\end{array}
   \right) $$ and the spectral decomposition looks like $$A = \left( 
\begin{array}{cc}
z & 0 \\*
0 & \gamma^1 z \gamma^1	
\end{array}
 \right) \frac{1}{2} \left( 
\begin{array}{cc}
1 &  z \gamma^1 \\*  \gamma^1 z & 1
\end{array}
   \right) -  \left( 
\begin{array}{cc}
z & 0 \\*
0 & \gamma^1 z \gamma^1	
\end{array}
 \right) \frac{1}{2} \left( 
\begin{array}{cc}
1 & - z \gamma^1 \\*  - \gamma^1 z & 1
\end{array}
   \right)$$ as the reader may easily verify.  This decomposition is however quite strange, as it relies upon the existence of a complementary basis; however, there is nothing we can do about it since approximate decompositions have ill defined asymptotic behavior (indeed, normally one would expect the eigenvalues \emph{and} eigenvectors to blow up).  The natural strategy would be to make central extensions of the Hermitian generalized projection operators such as
   $$P_{+} = \left( 
\begin{array}{cc}
1 & \widetilde{z}\gamma^1 \\*
\gamma^1 z & \gamma^1 z \widetilde{z} \gamma^1	
\end{array}
   \right).$$  As the reader may easily verify $P_{+} v_{\pm} = v_{+} (1 \pm \widetilde{z}z)$, and as we know $1 \pm \widetilde{z}z$ is nonvanishing and singular.  Picking any $\epsilon > 0$ and replacing $P_{+}$ by $P_{+} + \epsilon 1$ leads to $P_{+} v_{+} = v_{+} ((1 + \epsilon)1 + \widetilde{z}z)$ and $P_{+}v_{-} = v_{+}(1 - \widetilde{z}z) + \epsilon v_{-}$.  Inverting $((1 + \epsilon)1 + \widetilde{z}z)$ leads one to consider the matrices
 \begin{eqnarray*}
 ((1 + \epsilon)1 - \widetilde{z}z)((1 + \epsilon)1 + \widetilde{z}z)^{-1} & = & \left( 
\begin{array}{cccc}
\frac{2 + \epsilon}{\epsilon} & 0 & 0 & 0 \\*
0 & \frac{\epsilon}{2 + \epsilon} & 0 & 0 \\*
0 & 0 & \frac{2 + \epsilon}{\epsilon} & 0 \\*
0 & 0 & 0 & \frac{\epsilon}{2 + \epsilon}	
\end{array}
  \right) \\*
 ((1 - \epsilon)1 - \widetilde{z}z)((1 + \epsilon)1 + \widetilde{z}z)^{-1} & = & \left( 
\begin{array}{cccc}
\frac{2 - \epsilon}{\epsilon} & 0 & 0 & 0 \\*
0 & - \frac{\epsilon}{2 + \epsilon} & 0 & 0 \\*
0 & 0 & \frac{2 - \epsilon}{\epsilon} & 0 \\*
0 & 0 & 0 & - \frac{\epsilon}{2 + \epsilon}	
\end{array}
  \right) 
 \end{eqnarray*} which are both singular in the limit $\epsilon \rightarrow 0$ while all other expressions remain finite.              
\\* \\*
We now adress type V decompositions which depart from type III to the extend that the eigenvalues do not commute with the eigenvectors.  This type is another generalization away from Nevanlinna space and is described in our example by 
$$z = \left( 
\begin{array}{cccc}
- 1 & 0 & 0 & 0 \\*
0 & - 1 & 0 & 0 \\*
0 & 0 & 1 & 0 \\*
0 & 0 & 0 & 1	
\end{array}
  \right)$$ where $\widetilde{z} = - z$.  In this way of splitting things up $$A = \left( 
\begin{array}{cc}
z & 0 \\*
0 & - z	
\end{array}
 \right) \frac{1}{2} \left( 
\begin{array}{cc}
1 & z \gamma^1 \\*
\gamma^1 z & 1	
\end{array}
  \right) - \left( 
\begin{array}{cc}
z & 0 \\*
0 & - z	
\end{array}
 \right) \frac{1}{2} \left( 
\begin{array}{cc}
1 & - z \gamma^1 \\*
- \gamma^1 z & 1	
\end{array}
  \right)$$ as the reader may verify since $z$ anticommutes with $\gamma^1$.  All the remaining pairs $z, - z$ give rise to a type IV decomposition; we now verify whether other types are also possible by combining different eigenvalues.  It is easily seen that the answer is no, since no other pair can form a basis.  Of course, we know that this is not the end of the story as we still need to include the case where $a\widetilde{a}$ is not in the image of $z^2$.  
\\* \\*
To consider these type VI decompositions, let us return to the previous example where $$a = \frac{1}{\sqrt{8}} \left( - i \gamma^{0} \gamma^{2} - i \gamma^2 \gamma^3 - \gamma^{0} \gamma^3 + 1 \right)$$ such that 
$$a \widetilde{a} = - \frac{i}{2} \left( \gamma^2 \gamma^2 + \gamma^2 \gamma^3 \right).$$  The equation $z^2 = a \widetilde{a}$ has four approximate eigenvalues given by
$$z(\epsilon_1, \epsilon_2) = \left( 
\begin{array}{cccc}
\frac{\epsilon_1}{N} & \frac{N \epsilon_1}{2} & 0 & 0 \\*
0 & \frac{\epsilon_1}{N} & 0 & 0 \\*
0 & 0 & \frac{\epsilon_2}{N} & \frac{N \epsilon_2}{2} \\*
0 & 0 & 0 & \frac{\epsilon_2}{N}	 
\end{array}
  \right)$$ where the $\epsilon_j = \pm 1$.  Putting these roots in Jordan normal form is an operation which is unique up to two parameters $\lambda, \mu$ and the corresponding $x$ coordinates of the eigenvectors shift from unity to
$$x(\epsilon_1, \epsilon_2) = \left( 
\begin{array}{cccc}
\lambda & 0 & 0 & 0 \\*
0 & \frac{2 \lambda}{N} & 0 & 0 \\*
0 & 0 & 0 & \frac{2 \mu}{N} \\*
0 & 0 & \mu	& 0 
\end{array}
   \right)$$ and for convenience we will work with $\lambda = \mu = 1$.  Using that 
 $$a = \frac{1}{\sqrt{2}} \left( 
\begin{array}{cccc}
1 & 1 & 0 & 0 \\*
0 & 0 & 0 & 0 \\*
0 & 0 & 0 & 0 \\*
0 & 0 & 1 & 1	
\end{array}
  \right)$$ and 
 $$\widetilde{a} = \frac{1}{\sqrt{2}} \left( 
\begin{array}{cccc}
0 & 1 & 0 & 0 \\*
0 & 1 & 0 & 0 \\*
0 & 0 & 1 & 0 \\*
0 & 0 & 1 & 0	
\end{array}
  \right)$$ we try to solve the system of equations 
\begin{eqnarray*}
ay(\epsilon_1, \epsilon_2) & = & x(\epsilon_1, \epsilon_2)z(\epsilon_1 , \epsilon_2) \\*
\widetilde{a} x(\epsilon_1 , \epsilon_2) & = & y(\epsilon_1 , \epsilon_2) z(\epsilon_1 , \epsilon_2)
\end{eqnarray*} as faithful as possible.  Imposing the first equation leads to 
$$y(\epsilon_1, \epsilon_2) = \left( 
\begin{array}{cccc}
\alpha(\epsilon_j) & y_1 + \frac{N \epsilon_1}{2 \sqrt{2}}& y_2 & y_3 \\*
\beta(\epsilon_j) & - y_1 + \frac{N \epsilon_1}{2 \sqrt{2}}& - y_2 & - y_3 \\*
y_4 & y_5 & \gamma(\epsilon_j) & y_6  + \frac{N \epsilon_2}{2 \sqrt{2}}\\*
- y_4 & - y_5 & \delta(\epsilon_j) & - y_6 + \frac{N \epsilon_1}{2 \sqrt{2}}
\end{array}
  \right)$$ where 
  \begin{eqnarray*}
  \alpha(\epsilon_j) + \beta(\epsilon_j) & = & \epsilon_1 \frac{\sqrt{2}}{N} \\*
  \gamma(\epsilon_j) + \delta(\epsilon_j) & = & \epsilon_2 \frac{\sqrt{2}}{N}
  \end{eqnarray*} and an error of the order $\frac{1}{N^2}$ has been made.  Working out the second equation implies that an approximation error of the order unity has to be made and that optimally $y_2 = y_3 = y_4 = y_5 = 0$.   That is, while the approximate roots of the eigenvalue equation diverge with increasing accuracy, the error margin in the eigenvector equations remains stable and of order unity.  Obviously, better approximations can be found which result for example from perturbing $a$ such that the eigenvalue and eigenvector equations have exact solutions.        
\\* \\*
Let us now treat the general case of $2 \times 2$ Hermitian matrices.  Therefore, we define the non-commutative trace and determinant; that is, consider
$$ A = \left( \begin{array}{cc}
b & a \\
\widetilde{a} & c
\end{array}
\right)
$$ where we automatically assume $a$ to be invertible and $b,c$ to be real, that is $\widetilde{b}=b$ and $\widetilde{c}=c$.  It is easy to compute that the eigenvalue polynomial becomes
$$x z^2 - \left( aca^{-1} + b \right) xz + \left( aca^{-1}b - a\widetilde{a} \right)x = 0$$
leading to a generalized definition of trace and determinant
$$\textrm{Tr}\left( A \right) = b + aca^{-1}$$ and
$$\textrm{Det}\left( A \right) = aca^{-1}b - a\widetilde{a}.$$  As before, we limit ourselves to the case of invertible $x$ which can be put equal to unity by means of a gauge transformation.  Hence, we have to solve a quadratic equation of the kind
$$z^2 - \textrm{Tr}\left( A \right)z + \textrm{Det}\left( A \right) = 0$$ where in this case all factors are non-commutative except for the quaternions, complex and real numbers.  Clearly, the roots of this equation do not necessarily belong to the complex algebra generated by $\textrm{Tr}\left( A \right)$, $\textrm{Det}\left( A \right)$ and unity.  Indeed, let us gain more insight into the solution space of a general quadratic equation 
$$z^2 - \alpha z + \beta = 0.$$  Going over to the variable $w = z - \frac{\alpha}{2}$, the noncommutative structure gets accentuated; indeed, the equation becomes
$$w^2 - \left[ \frac{\alpha}{2} , w \right] + \beta - \frac{\alpha^2}{4} = 0.$$
It is natural to look first for those solutions $w$ satisfying $\left[ w , \alpha  \right] = 0$, implying that $\left[ \alpha , \beta \right] = 0$.  One can easily find commuting $\alpha$ and $\beta$ such that $\frac{\alpha^2}{4} - \beta$ is not in the image of $w^2$ and hence this solution class may be empty.  On the other hand, picking $\alpha = 1$ may lead to a solution space which is continuous, discrete or empty depending upon $\beta$.  The second solution class consists of those $w$ for which $\left[ w , \alpha \right] \neq 0$ and what happens here again depends upon $\alpha$ and $\beta$.  The above example can be easily generalized to the case where $a, b, c$ are invertible but $\textrm{Det}(A)$ is not; specifically choose
$$a = \frac{i}{4} \gamma^0 \gamma^2 + \frac{i}{4} \gamma^2 \gamma^3 + \gamma^0 \gamma^3$$
then $a \widetilde{a} = - 1 - \frac{i}{2} \gamma^0 \gamma^2 - \frac{i}{2} \gamma^2 \gamma^3$ and with $b = 1 = - c$ one arrives at the equation
$$z^2 + \frac{i}{2} \left( \gamma^0 \gamma^2 + \gamma^2 \gamma^3 \right) = 0$$ which again has no roots.   The eigenvector equations decouple if and only if $a$ is invertible or zero in which case there exists a trivial right basis of eigenvectors.  In the former case, we have again a myriad of possibilities containing approximate decompositions; if $a$ is not invertible, we have an extension of a former example we studied.  Note that the basis in which we have written $A$ is an abelian one in the sense that all vectors commute with $\mathcal{C}_{(1,3)}$.  Summarizing, we have the following situation : 
\begin{itemize}
\item only the determinant of the eigenvalues is important for an interpretation and eigenspace classification; it is possible that within such class of eigenvalues, no orthogonal basis of eigenvectors can be constructed,
\item the solution space of the eigenvalue polynomial equation, assuming eigenvectors with invertible components, may be empty, discrete or continuous,
\item there may not exist a right basis of eigenvectors.
\end{itemize}
Since we work with a nonabelian ring as a substitute for the complex numbers, one may wonder wether the concept of a basis is still unique.  Clearly, one can define several notions of a linearly independent or generating set of vectors $v_{i}$ where $i: 1 \ldots n$.  We call $\{ v_{i} | i: 1 \ldots n \}$ right independent if and only if 
$\sum_{i} v_{i} \lambda_i = 0$ implies that $\lambda_i = 0$.  Likewise, one can define left and mixed linear independence.  Similarly, one can define $\{ v_{i} | i: 1 \ldots n \}$ to be right generating if and only if every vector $v$ in the module can be written in the form 
$$v = \sum_i v_i \lambda_i$$ for some $\lambda_i$ and likewise for left and mixed generating.  A set of vectors which is left (right/mixed) independent and generating is called a left (right/mixed) basis.  Now, Clifford algebras are very special in the sense that they can all be reduced to (direct sums of) matrix algebras over the three real division algebras $\mathbb{R}, \mathbb{C}$ and $\mathbb{H}$ in the Euclidean case while the hyperbolic situation is only slightly more complicated.  Indeed, $\mathcal{C}_{(1,n)} = \mathcal{C}_{(0,n-1)} \otimes_{\mathbb{R}} \mathcal{C}_{(1,1)}$ by means of the isomorphism $e_i \rightarrow e_i \otimes {e'}_1 {e'}_2$ and ${e'}_i \rightarrow 1 \otimes {e'}_i$.  Since $\mathcal{C}_{(1,1)} \sim \mathbb{R}(2)$, we have that in the \emph{real} case
$\mathcal{C}_{(1,3)} \sim \mathbb{R}(2) \otimes_{\mathbb{R}} \mathbb{R}(2) = \mathbb{R}(4)$ for signature $(- + + + )$ while for signature $(- - - +)$ it is $$\mathcal{C}_{(3,1)} = \mathbb{H} \otimes_{\mathbb{R}} \mathbb{R}(2) = \mathbb{H}(2).$$  In the complex case, there is only one option of course which does not depend upon the signature, it is given by 
$$\mathcal{C}_{(1,3)} = \mathcal{C}_{(3,1)} = \mathbb{C}(4).$$  The assymetry in the real case is not present in the Euclidean theory in four dimensions since there one has that
$$\mathcal{C}_{(0,4)} = \mathcal{C}_{(4,0)} = \mathbb{H}(2).$$  These considerations suggest that one should really consider the complex Clifford algebra since there ought to be no physical difference between $(- + + +)$ and $(- - - +)$.  A right basis is not necessarily a left basis since the condition for a right basis is that the $4n \times 4n$ matrix
$$ \left(
\begin{array}{ccc}
v_1^1 & \ldots & v_n^1 \\*
\vdots & & \vdots \\*
v_1^n & \ldots & v_n^n	
\end{array} \right) $$ is invertible, while the condition for a left basis is that
$$ \left(
\begin{array}{ccc}
v_1^1 & \ldots & v_1^n \\*
\vdots & & \vdots \\*
v_n^1 & \ldots & v_n^n	
\end{array} \right) $$
is invertible.  Obviously, self adjoint left bases with respect to a trace preserving involution constitute a right basis and vice versa.  The natural bases to consider are the right bases and the previous considerations imply that any basis has precisely $n$ elements and therefore the concept of dimension of this particular Clifford module is well defined. \\* \\*
Let us finish the technical part by defining a natural class of operators $A$ on Nevanlinna space which deserve further study:
\begin{itemize}
\item $A$ is timelike consistent if and only if $\langle Av | Av \rangle < 0$ for all $\langle v | v \rangle < 0$,
\item $A$ is timelike swapping if and only if $\langle Av | A v \rangle > 0$ for all $\langle v | v \rangle < 0$,
\item $A$ is spacelike consistent if and only if $\langle Av | Av \rangle > 0$ for all $\langle v | v \rangle > 0$,
\item $A$ is spacelike swapping if and only if $\langle Av | Av \rangle < 0$ for all $\langle v | v \rangle > 0$.
\end{itemize} These definitions are motivated by the fact that the regions $\langle v | v \rangle > 0$ and $\langle w | w \rangle < 0$ are path connected and we do not wish $\langle Av | Av \rangle$ to become zero anywhere.  Finally, we call $A$ balanced if and only if it is timelike and spacelike consistent.  In chapter eight, it will become clear that the relevant dynamics is no longer given by a unitary one parameter group, but by a set of operators $U(x)$ which depends upon the space-time coordinates and the question one may ask is under what conditions $U(x)$ can be written as
$$U(x) = e^{i H(x)}$$ where $H(x)$ is Hermitian.  For ordinary Hilbert spaces, this is a general property because of the definition of the logarithm.  Also, in standard Krein spaces this result applies but additional subtleties occur here.  Suppose, one has a conjugated pair of null eigenvectors $| v \rangle$ and $| w \rangle$, then standard results imply that their respective eigenvalues should be $\lambda$ and $\overline{\lambda}^{-1}$.  This is a very important fact, since for $\lambda = r e^{i \theta}$ the spectral decomposition of $U$ would look like
$$ U = \lambda | v \rangle \langle w | + \overline{\lambda}^{-1} | w \rangle \langle v | + \ldots $$ and the respective projection operators are not Hermitian at all.  Hence, $ - i \, ln(U)$ can be written as
$$- i \, ln(U ) = - i( ln(r) + i \theta ) | v \rangle \langle w | + i( ln(r) - i \theta ) | w \rangle \langle v | + \ldots$$
and this \emph{entire} expression is Hermitian, although the separate parts are not.  How this result generalizes to our definition of infinite dimensional Nevanlinna spaces and Clifford-Nevanlinna modules is open for future investigation.  In the remainder of the book, we merely investigate the implications of a unitary potential generated by Hermitian operators. \\* \\*  We now make a few comments on the questions of statistics: that is, assuming spin-statistics, Poincar\'e covariance and a well defined tensor product, we derive the consequences.  In other words, one should generalize the work done in chapter five to the context of Clifford-Nevanlinna modules.  The first thing to do is to treat representation theory of the Poincar\'e group in terms of right linear unitary operators on infinite dimensional Clifford-Nevanlinna modules; this determines the one particle states and appropriate quantum numbers for the creation and annihilation operators.  Second, given the results of chapters eight and eleven, we must allow for particle operators to mix with copies of the same species; likewise, it might be possible for different modes to ``interact'' in the fundamental relations mixing the creation and annihilation operators (which would be impossible by the way for massless particles since the little group is not compact in that case).  These are all genuine possibilities which need to be studied.  Finally, we comment upon the issue of the probability interpretation, which is more elaborate than the standard Born rule.  Some of the reasons herefore have been mentioned already in the context of the spectral theorem on Clifford-Nevanlinna modules while other criterea are treated in the next section.  Indeed, the Born rule is rather simplistic and, as Hilbert space itself, absolutist; our task will consist into making the entire interpretation dynamical and relational (but not in Rovelli's equally simplistic sense).  As became very clear now, the objective state of the universe $\Psi \in V$ might suffer from three unrelated ``problems'' : (a) the norm of the state might become undefined, infinite or different from unity in some particular reference frame or (b) some components of $\Psi$ might not be well defined or infinite with respect to a reference frame and (c) there is the issue of negative probabilities in the intermediate stages of the calculation.  However, as we will argue later on, none of these issues is a real physical problem and part of the answer is developed in chapter eight, while the full interpretation remains to be given in chapter eleven.  The point is that we have to go to an open interpretation, open because (a) the new quantum theory itself is \emph{physically} open (in contrast to the standard one) and (b) the interpretation is relational and all macroscopic observational systems in nature are open anyway.                      
\chapter{The relativization of advanced Quantum Field Theory}
This chapter is the theoretical culmination of all previous results and I shall ``axiomatize'' a new quantum-gravity-matter theory.  All axioms are physically motivated and wherever any ambiguity might arise, all possible extensions the author is currently aware of are mentioned.  They might become important at a later stage, albeit I see no reason whatsoever at this point why this should be the case.
The reader should not expect a theory in which all mathematical details are \emph{specified} by which I certainly do not intend to say that the construction is not rigorous.  On the contrary, it certainly is, but what one does not know one should remain silent about and certainly I want to avoid making the mistake of overspecifying the theory.  There is only one way to make progress and that is by dismissing thoughts which lead either to logical contradictions or to physical nonsense; therefore, one must be brutal and pave new ways which lead to a better theory.  In that vein, the practical physicist who is happy with magic and deceit as long as he can make the numbers all right has to give in here: if one speaks about a fundamental theory, the latter has to be nonperturbatively well defined from the very beginning and have a clear ontology as well.  This leads to the uncomfortable situation that I will have to dismiss some ingrained prejudices which very few are willing to give up and which even fewer would know how to replace with a better and deeper principle.  Let me remind the pragmatic physicist that Einstein's laws are also practically irrelevant with respect to Newtonian gravity in many cases; but they impose a completely new way of thinking about the world.   Of course, some people might wish to try to solve these problems in a different way, but given the generality of the laws below and the little room which is left to change some details, I seriously doubt that such enterprise will lead to succes.  This chapter is mainly technical but the philosophical issues treated in chapter three are addressed as well: for example, we will set up a construction for creating a living quantum being in a fully dynamical way inside a dynamical universe.  In short, the theory constructed here is extremely ambitious: it does not only want to avoid technical tools such as the renormalization group method but it also claims to address long standing conceptual issues in quantum mechanics.  \\* \\*   
The first point is rather essential and concerns the role of a symmetry in Quantum Field Theory; this issue might be called somewhat philosophical but it is of absolute physical importance.  For example, why do we quantize the Lorentz transformations in Quantum Field Theory on a flat space-time besides the fact that canonical quantization of a Lorentz invariant classical action gives them for free?  The latter argument is not a good one, we should look for a physical reason and not simply accept a particular procedure (which I am about to dismiss totally).  Well, a good suggestion would be that a Lorentz boost changes the ``object-subject'' relation; in either, it modifies the way ``observers'' look at the system.  Now, a relativist might object and say ``hey, this unitary transformation is against the principle of general covariance'' or a form of ``observer independence''.  Well, my answer to the first remark is definetely no and my reply to the second objection is somewhat more complicated.  It depends of course upon what the unitary transformation does: if it is a trivial Bogoliubov transformation like an ordinary Lorentz transformation is, then nothing really changes to the physics.  However, if this unitary matrix corresponds to a nontrivial Bogliubov transformation such as people \emph{believe} to be occuring for a generic accelerated observer, then I am afraid that my answer might deviate from current consensus.  I shall explain this in full detail in the next chapter, since these matters are not as simple as people usually think they are.  Let me mention upfront that a modification to two of the axioms below allows for different answers regarding this question\footnote{This option was pointed out to me by Rafael Sorkin.}.  Hence, the issue is a deep one and has to do with the distinction between ultralocal and quasi-local particle notions, or equivalently, $TM$ or $TTM$ or higher jet bundles as the basic arena for ``space-time'' physics. Within the philosophy of ultralocal particle notions, an appropriate form of Einstein's principle of ``observer independence'' is correct and the vacuum state does not offer resistance to accelerated, idealized observers.  This puts away the modern ``quantum eather'' and personally, I am convinced that this is the correct answer.  This is the main reason why I did not generalize neither the geometry, nor the quantum dynamics to higher jet bundles; another argument consists in first studying the ``easier'' theory before one would consider something even more complicated.  Such extension however, would be rather straightforward and it is left as a future, perhaps academic, exercise to investigate its implications.  Let me stress that my position is not in conflict with the Casimir effect which is often erroneously regarded as evidence for vacuum fluctuations (which are thought of as being responsible for the ``resistance'').  The former has a perfectly reasonable explanation without such representational artifacts as has been pointed out by many authors, such as Barut and Jaffe, the former even outside the context of Quantum Field Theory.  However, I cannot close this issue on observational grounds and therefore I wished to point out this alternative.  \\* \\*      
Of course, there exist schools which look for alternative quantization \textit{procedures} without really changing any of the basic tenets of quantum physics such as is the case for the LQG community and the polymer quantization.  However, the latter is probably unphysical; indeed, as was explicitly shown by Helling, this quantization method even fails in case of the harmonic oscillator \cite{Helling} where the standard procedure has been tested.  Furthermore, the theoretical input behind the (generalized) Fock representation developed in chapter seven, is simply much deeper physically as well as mathematically than the simple switch between unbounded Heisenberg operators and the ``cleaner'' Wigner picture advocated more than seventy years ago.  Indeed, as will become clear below, the true quantum theory which has a formulation on equal footing with that of general relativity requires even wilder operators than even Heisenberg dreamt of and the Wigner representation is not of much use anymore.  This was already clear in the previous chapter where we had to introduce such exotic structures as relational Clifford-Nevenlinna modules and kroup structures which are far beyond the usual Stone view upon unitary evolution.  What about other symmetries such as gauge invariance and general covariance, you may ask?  Well, gauge symmetry was mainly invented due to a representation problem for massless spin one particles within the context of (free) quantum fields, a problem which is entirely absent in the Nevanlinna space quantization.  Therefore, my suggestion is to forget about gauge symmetry, general covariance however has a somewhat different status: physically it doesn't \emph{do} anything at all (in the sense that it does not alter the subject-object relation), therefore it should not be a quantum symmetry - a conclusion we have reached before in chapter three through a very different line of argumentation.  This means that Dirac quantization of gravity is the wrong thing to do, which invalidates the LQG program from the start.  The conclusion we shall come to here is that gravity has a classical as well as quantum aspect just as this occurs in string theory; but in contrast to string theory this result emerges from a by far superior Quantum Theory which encapsulates some very crucial ideas of Einstein such as manifest classical background independence on space-time as well as a truly local formulation and ontology.  The issue of causality, for example, will indeed be a dynamical one in this theory as explained before in chapter seven; but, it shall be still classical.  How this is commensurable with the quantum world, the Heisenberg uncertainty principle and the detection of gravitons (as quantum particles) is not really stranger than the usual ideas behind semiclassical gravity. \\* \\*
My general problems with Quantum Field Theory do not only originate from the lack of local Lorentz covariance but also from the $S$-matrix \textit{philosophy}.  I mean, there is nothing wrong with an $S$-matrix as such, but it really does not allow you to compute physical amplitudes (albeit there are good reasons why the computed numbers are excellent approximations).  Indeed, the lack of a particle notion for observers within the universe is not just a technical point but a deep physical one; it is rooted in the intrinsic lack of a non-perturbative formulation of Quantum Field Theory from the point of view of Fock space. I emphasized intrinsic since there is nothing you can do about it in the current formalism.  It is constructed in such a way that it splits a theory in a free part and an interaction and the entire idea of quantum fields is supported on that premise (see \cite{Weinberg}).  Quantum Field Theory as it stands is therefore by definition a perturbative game: so what can we do about it?  Do we really have to believe in sloppy path integrals as the only way out, or is the objection of a deeper physical nature and should we rethink Quantum Field Theory from the start in the generalized Fock space formalism of chapter seven?  I think the latter suggestion is clearly the correct one and I shall start to construct a mathematical formalism which includes (a classical and quantum form of) gravity automatically - it is an essential ingredient to make things work out.  But let me first give you a physical reason why it is deeply wrong and this concerns the definition of an observer; if you would like to think (as almost everybody does) that an observer is connected to a coordinate system in space-time which has the usual foliation properties with respect to the background metric, then you must come to the conclusion that a minute local change in this coordinate system is going to influence the particle notion everywhere in the universe.  Therefore, the notion of an observer's reference frame should be one of the tangent bundle and not of spacetime.  This implies that we must define local Poincar\'e groups (or energy momentum tensors as a matter of fact) on the tangent bundle and that dynamics consists in studying unitary equivalences between the different Poincar\'e groups (defining distinct particle notions).  If one would like to think in terms of an action principle (which one should not since such view leads to the wrong type of theory), then I would say that in such action principle one integrates over $TM$ and\footnote{Or, over higher Jet bundles if you believe in quasi-local particle notions.} not over $M$.  It is a background basis $e_a$ of four vector fields determining a Lorentz structure on $TM$ which determines preffered coordinate systems on the tangent bundle.  On $M$ itself, everything is as covariant as it can be.  This also gives the first indication why gravity has two faces instead of one: (a) you have massless spin two particles on $TM$, expressed in terms of the $e_a$ which really are gravitons and (b) the vierbein $e^{a}_{\mu}$ cannot be an operator since it doesn't directly define particles (the latter should not depend upon a change of coordinate system since this has no operational meaning).  Notice also that by construction an arrow of time is present given by $e_0$ which shall also be fixed dynamically.  Let us work out the full quantum Einstein equations, the equations of matter (including gravitons) and the equations determining a dynamical reference frame.  The latter is badly needed since otherwise the theory would not be predictive.  \\*\\*
To conclude this prelude, we take a radical particle perspective and have mainly a physics of relation and less one of propagation; therefore, fields are just to be thought of as ``hidden variables'' which approximate reality to scales at least shorter than $10^{-20}$ meters or so.  Taking the lessons of the previous chapter into account, we start with a Clifford-Nevanlinna module $\mathcal{K}$ associated to a local reference frame, which is to be thought of as the universal particle space which goes beyond space and time and is certainly not associated to some foliation.  The much bigger relational Clifford-Nevanlinna module $V$ is, as stressed in the previous chapter, a dynamical output.  We also attach Fock spaces $\mathcal{F}(e_a(x))$ to an ``observer'' given by $e_{a}(x)$ and the former are determined by the dynamics and initial conditions upon a global ($x$ independent) unitary transformation.  $\mathcal{K}$ is itself a Fock space generated by creation operators (of positive or negative norm) satisfying an appropriate form of statistics applied to a cyclic vacuum state.  The reader might object from the start that local particle notions appear to brutally violate the Heisenberg uncertainty principle.  Well, this is not true at all: the ``spatial'' support for a ``particle'' is determined by the dynamical relations between distinct local Fock spaces as well as the boundary conditions for the universe.  Hence, the Heisenberg principle is dynamical in nature and the one we are aware of is the \emph{unique} flat spacetime limit of the more general construction here.  In the context of chapter seven, we will not allow for signature changes and therefore all $\mathcal{K}$ are isomorphic within $V$; the first AXIOM determines the structure of $\mathcal{K}$.
\begin{itemize}
\item AXIOM 0 : All \textit{local} one particle Nevanlinna modules $\mathcal{H}(e_a(x))$ are second countable and unitarily equivalent to the local model space $\mathcal{L}$.  We do not dispose of an a priori notion of spatiality in $M$; however we have one in $TM$ and the information contained in any one particle Nevanlinna module generating the \textit{universal} Fock space $\mathcal{K}$ is unitary equivalent to $\oplus_{i=1}^{\infty} \mathcal{L}$ which still has cardinality $\aleph_{0}$.  However, the Fock space construction to $\mathcal{K}$ is now more complicated since (a) we genuinely have to describe states with an infinite number of particles and (b) we allow for more complicated forms of statistics.  We use here the Guichardet construction implying that $\mathcal{K}$ has cardinality $\aleph_0^{\aleph_0}$ in contrast to what is usually supposed in Quantum Field Theory\footnote{For Nevanlinna spaces, the Guichardet construction is somewhat more delicate in the sense that in each tensor product only a finite number of negative norm states can occur.  We have treated this already in chapter seven.}.  This fixes the information in the Nevanlinna modules attached to a particular basis and in particular no signature change is allowed for as is the case for the general definition in chapter seven.  The full relational Nevanlinna module however is a dynamical object and not fixed a priori; such as is the case for the spacetime manifold in general relativity. 
\end{itemize}
This axiom is directly addressing Haag's theorem; we have no interaction picture and hence no problem of mapping the physical Hilbert space into the free particle Fock space.  Indeed, in order to implement interactions, one must allow local particle Fock spaces to ``rotate'' into a much bigger universal space, which is precisely what we shall do in the following.
\begin{itemize}
\item AXIOM I : Manifold structure.  As said in the introduction, we regard $TM$ as a manifold, however the geometrical construction here is somewhat different from the standard textbook one \cite{Antonelli}.  Jadczyk originally pointed out to me that the construction below might be a generalization of Finsler geometry, but this is not the case\footnote{I thank Arkadiusz for pushing me to make my notation more intrinsic to facilitate comparison with results in the literature.  He read the original draft in which everything was done in a manifestly coordinate dependent way.}.  The standard view on $TM$ is that ordinary coordinate systems $(x^{\mu})$ get lifted to $TM$ by means of the canonical basis $\partial_{\mu}$; hence every vectorfield $V(x)$ in $TM$ gives rise to natural coordinates $(x^{\mu}, v^{\nu})$.  This point of view is entirely kinematical and only lifts coordinate transformations on $M$ to $TM$.  Such coordinate systems do not have any physical meaning and it is not wise in general to couple the transformation laws of the base space to that of tangent space.  As stressed in the introduction of this chapter, we want the coordinates on tangent space to have physical meaning.  Hence, their very definition must be coupled to dynamical objects on $TM$ as vectorbundle.  The obvious candidate is the vierbein $e_a(x)$ and every vectorfield $V(x) = v^a(x) e_a$ gives rise to coordinates $(x^{\mu}, v^a)$.  Jadczyk pointed out to me that such construction had been made for more general Lie groups in Munteanu \cite{Munteanu}.  The local coordinate transformations on $M$ do not propagate to $TM$, since one simply has $(x'^{\mu},v^a)$.  Under a local Poincar\'e transformation $(\Lambda(x),w(x))$, however, the coordinates on the tangent bundle transform as $$(x,v^{a}) \Rightarrow (x,\Lambda(x)^{a}_{\,\, b}v^{b} + w^{a}(x))$$ since $e_a(x) \Rightarrow \Lambda(x)_{a}^{\,\, b}(x)e_b(x)$.  Therefore, the partial derivatives mix as follows
\begin{eqnarray*}
\partial'_{\mu} & = & \partial_{\mu} + v^{c} \Lambda^{b}_{\,\,c}(x) \partial_{\mu}\Lambda^{\,\,\, a}_{b}(x) \partial_{a} - \Lambda^{\,\,\,a}_{b}(x) \partial_{\mu} w^b(x) \partial_{a} \\*
\partial'_{a} & = & \Lambda_{a}^{\,\,b}(x) \partial_{b} 
\end{eqnarray*}
and the differential forms transform as
\begin{eqnarray*}
dx'^{\mu} & = & dx^{\mu} \\*
dv'^{a} & = & \partial_{\mu} \Lambda^{a}_{\,\,b}(x)v^{b} dx^{\mu} + \partial_{\mu}w^a(x)dx^{\mu} + \Lambda^{a}_{\,\,b}(x) dv^{b}. 
\end{eqnarray*}
This indicates that we better use a distinct notation for the tensors which transform with respect to $e_a$ and the tensors defined by $\partial_{\mu}, \partial_{a}$ even if the basis elements $e_a$ and $\partial_a$ transform identically.  A lesson is that we cannot simply consider $\partial_{\mu}$ separately in the context of $TM$ and where necessary, whe shall use primed indices $a'$ to denote the distinction while unprimed indices always transform with respect to the local Lorentz group.  The invariant tensors are given by $\delta^{a}_{b}$, $\delta^{A}_{B}$ while $\delta^{\mu}_{\nu}$ and $\delta^{a'}_{b'}$ viewed as tensors on $TM$ (the other coordinates are vanishing) are not invariant at all.  Here, $A$ is a shorthand notation for $A = (\mu, a')$, in either it is the natural index on $TM$ where $x^A = (x^{\mu},v^a)$.  Since this is a new geometry and the notation might be a bit unusual to the reader, let me make these statements more explicit.  I presume that the claim for $\delta^{a}_{b}$ is quite obvious since
$$\delta'^{a}_{b} = \Lambda^{a}_{\,\,\, c}(x) \Lambda_{b}^{\,\,\, d}(x) \delta^c_d = \delta^a_b$$ under the action of local Poincar\'e transformations.  Since the case of $\delta^A_B$ is standard in all textbooks on geometry, let me move to 
$\delta^{\mu}_{\nu}$.  The latter is a tensor defined with respect to a prefferred coordinate system $(x^{\mu},v^a)$ and we have to investigate its transformation behavior under local Poincar\'e transformations; an easy computation reveals that
\begin{eqnarray*}
\delta'^{\mu}_{\nu} & = & \delta^{\mu}_{\nu} - \left(v^{c}\partial_{\mu} \Lambda^{b}_{\,\, c}(x) + \partial_{\mu} w^{b}(x)\right)\Lambda^{\,\, \,a}_{b}(x) \delta^{\mu}_{a'} \\*
& = & \delta^{\mu}_{\nu}
\end{eqnarray*} and 
\begin{eqnarray*}
\delta'^{a'}_{\mu} & = & \left( \partial_{\mu} \Lambda^a_{\,\,\,b}(x)v^b + \partial_{\mu} w^a(x) \right).   
\end{eqnarray*}
All other type of coefficients are computed to vanish and the reader is invited to repeat this exercise for $\delta^{a'}_{b'}$.  In the future, we shall make use of $\delta^{\mu}_{\nu}$ as if it were an invariant tensor, which is justified by the fact that the (coordinate dependent) ``projection'' on the $\mu$ indices is.  The reader who thinks that this is a fluffy concept might enjoy the following definition.  We call an object $T^{\alpha_1 \ldots \alpha_r}_{\,\,\,\, \,\, \beta_1 \ldots \beta_s}$ where $\alpha_j, \beta_k \in \{ A, \mu , a' \}$ a partial tensor if the object transforms consistently within the limitation of its indices.  That is, the $\mu$ indices only feel coordinate transformations on $M$, the $a'$ indices transform only under local Lorentz transformations and finally, the $A$ index undergoes the whole transformation group.  In this language, $\delta^{\mu}_{\nu}$ is a partial tensor since $\delta^{\mu}_{a'}$ cannot become nonzero under the full transformation group.  However, it is not a tensor either since the above computation reveals that $\delta^{a'}_{\mu}$ becomes nonzero in different coordinate systems.  Later on, we shall still define physical tensors and write down the relationship between the latter and full or partial tensors on $TTM$.  Now, we do something which is rather similar to what happens in Finsler geometry, we aim to define horizontal subbundles $H_{(x,v^a)}TM$ of $TTM$ over $TM$.  Therefore, we need to introduce a new object $\mathcal{A}_{\mu}^{B}$ which compensates for the action of the local Poincar\'e group on $\partial_{\mu}$.  The latter transforms as 
$$\mathcal{A'}_{\mu}^{B}(x, v'^{b}) = \frac{\partial x'^B}{\partial x^C}\mathcal{A}_{\mu}^{C}(x, v^{b}) - \left(v^{c}\partial_{\mu} \Lambda^{b}_{\,\, c}(x) + \partial_{\mu} w^{b}(x)\right)\Lambda^{\,\, \,a}_{b}(x) \delta_{a}^{C} \frac{\partial x'^B}{\partial x^C}$$ under local Poincar\'e transformations.  Under spacetime transformations, it transforms covariantly in the $\mu$ and $B$ index.  The reader may wish to verify that all this is consistent since $A_{a'}^{B}(x,v^c)$ is defined to be zero and therefore $\mathcal{A}_{\mu}^{B}$ transforms as a partial tensor in the $\mu$ index.  This new type of ``gauge'' theory (which mixes up spacetime and the tangent space) is studied right after all axioms are given.  We change therefore our entire point of view since the relation $$\partial_{\mu} =
e^{a}_{\mu}(x) e_{a}$$ does not behave well under local Lorentz transformations and we want to extend the vierbein to $e^a_{\mu}(x,v^{b})$ so that it lives in $TM$ as a manifold\footnote{The role of the translations might be a bit confusing here since here since all coordinate systems defined so far started from a preferred origin.  However, there is no contradiction since the coordinate definition of the origin just shifts too.}.  Therefore, we define a set of ``gauge'' operators $$\mathcal{D}_{\mu}(x,v^a) = \partial_{\mu} - \mathcal{A}_{\mu}^{B}(x,v^{a})\partial_B$$ which at each point $(x,v^{a})$ span a linear space $H_{(x,v^a)}TM$ isomorphic to $TM_x$ by sending
$$ W^{\mu}(x,v^a) \mathcal{D}_{\mu}(x,v^a) \Rightarrow W^{\mu}(x,v^a) \partial_{\mu}.$$  The latter map is the formal definition of the bundle projection $\tau$ so that we get a formal triple $(H TM, \tau , TM)$.  
The $W^{\mu}(x,v^a)$ transform as scalars under local Lorentz transformations and as ordinary vectors under spacetime $M$ coordinate transformations.  Likewise, one has a vertical subbundle $VTM$ spanned by the $\partial_{a'}$ which gets projected to the zero vector in $TM$.  Later on, we will formulate the necessary condition so that this construction is promoted to a nonlinear connection in the standard Finsler sense.  The original tetrad $e_a(x)$ which does not depend upon $v^a$, but which defines $v^a$, is then to be associated to $e^{a}_{\mu}(x,0)$ by $$e^{a}_{\mu}(x,0) e_a(x) = \partial_{\mu}$$ but it can be redefined as an element of $H_{(x,v^a)}TM$ by the formula
$$e_a(x,v^b) = e_{a}^{\mu}(x,v^b) \mathcal{D}_{\mu}(x,v^b).$$  A constraint invariant under local Lorentz transformations is that $$\mathcal{A}_{\mu}^{B}(x,0) = 0.$$  This implies that the origin of $TM_x$ is an invariant point while the rest of tangent space dynamically positions itself in $H_{(x,v^a)}TM$.  The whole construction depends upon the preferred origin of the tangent space at $x$ but this is entirely physical since the observers still reside there.  This means that in general, the translation degrees of freedom are irrelevant and we ignore them from now.  The reason why I included them in the geometry anyway is motivated by the following : (a) there is nothing wrong with having a pointed $P$ affine space with translation symmetry, it just means you have two preferred points, $P$ and the origin of your coordinate system and there exists exactly one coordinate system in which both agree (b) the translation symmetry is a symmetry of the free theory living in $VTM$ and one might impose that the gravitational theory also obeys it (c) the translation symmetry has to be broken at some point of course since the vielbein is a dynamical entity living on $TM$ and the projection from $TM$ to $M$ is only in the initial conditions, this is of course accomplished by the quantum interactions as will become clear in AXIOM V.  But again, let me stress that we could have broken translation invariance already at the level of the gravitational theory and nothing in what follows would be influenced by this; only the transformation laws for $\mathcal{A}_{\mu}^{B}$ would change. 
\end{itemize}
At this point, it is good to further develop the differential geometry of this construction since we shall need it for axioms VII and VIII where the equations of motion are constructed.  First, let me point out some direct physical implications before I come to the mathematics.  It has been conjectured by as well string theorists as LQG practitioners that smooth geometry must break down around the Planck scale, but both approaches did mean very different statements by this.  LQG postulates that geometry is a quantum observable, which is deeply wrong, and that the breakdown of smooth geometry occurs due to non-commutativity of the spin connection and vielbein variables.  String theory, on the other hand, kept an effective classical geometry and ``imagines'' itself the breakdown of the latter at the Planck scale because the perturbation series determining the background geometry becomes ill defined.  However, they did not propose as yet what the correct ``high energy'' geometry should look like and how this breakdown should be mathematically imagined.  Indeed, as we learn here, there is a breakdown of Riemannian geometry, but the idea of a classical manifold is as good as it ever was; an extension to nonabelian manifolds is presented in chapter eleven.  Riemannian geometry does precisely break down due to the ``gauge'' field $\mathcal{A}_{\mu}^{B}(x,v^a)$ which is of crucial physical importance to make the number of geometric degrees of freedom equal to the physical matter degrees of freedom.  How should we interpret this breakdown of Riemannian geometry physically?  In my view, it means that smooth space evaporates : indeed, every ``infinitesimal'' observer has it's own rest space but those rest spaces simply don't satisfy Frobenius' theorem and therefore smooth space simply is an illusion of our mind, something which was conjectured a while ago in the excellent paper of Aerts \cite{Aerts}. \\*\\*
The transformation law for the ``gauge'' potential under local Lorentz transformations can be further simplified to
$$\mathcal{A'}_{\mu}^{B}(x,v'^a) = \frac{\partial x'^B}{\partial x^C}\mathcal{A}_{\mu}^{C}(x, v^{b}) - \partial_{\mu}\Lambda^{B}_{\,\,c}(x) v^c.$$ Before we proceed, let us further tell something about ordinary gauge theory; the gauge law forces us to introduce a new addition law $\oplus$ satisfying
$$\alpha_{\mu}(G) \oplus \alpha_{\mu}(H) = \alpha_{\mu}(G) + G \alpha_{\mu}(H) G^{-1}$$ where
$$\alpha_{\mu}(G) = \partial_{\mu}G G^{-1}.$$
This is necessary to make the action $\delta_{\mu} :G \Rightarrow \delta_{\mu}(G)$  where 
$$\delta_{\mu}(G):  A_{\mu} \Rightarrow G A_{\mu}G^{-1} + \alpha_{\mu}(G)$$ into a regular group action.  The reader is invited to find out that $\oplus$ is non-commutative, has a unit element $\alpha_{\mu}(1)$, and $\alpha_{\mu}(G)$ has as inverse $\alpha_{\mu}(G^{-1})$.  Morever, there is a canonical way to define mulitple $\oplus$ sums by 
$$\alpha_{\mu}(G) \oplus \alpha_{\mu}(H) \oplus \alpha_{\mu}(K) = \alpha_{\mu}(G) + G\alpha_{\mu}(H) G^{-1} + GH\alpha_{\mu}(K)H^{-1}G^{-1}$$ and it is easy to check that this operation is associative.  Hence, we have a group structure and $\alpha_{\mu}$ is a group homomorphism.  The reader should notice that $\delta^{A}_{a}$ is a well defined invariant tensor and that $\Lambda^{B}_{\,\,a}(x)$ transforms as a scalar under $M$ coordinate transformations (and therefore everything is well defined).  Replacing $\Lambda(x)$ by $\Lambda(x)\Gamma(x)$ transforms the gauge term as 
$$ -\partial_{\mu}\Lambda^{B}_{\,\,b}(x) \Gamma^{b}_{\,\,c}(x)v^{c} - \Lambda^{B}_{\,\, C}(x) \partial_{\mu}\Gamma^{C}_{\,\,d}(x)v^d$$
where $\Lambda^{B}_{\,\,C}(x)$ has not an invariant meaning but the product with $\partial_{\mu}\Gamma^{C}_{\,\,d}(x)$ has.  The extra twist here is of course the dependence of the gauge term upon the $v^a$, but the transformation law as written there is completely logical and gives rise to the sum
$$\left( \alpha_{\mu}(\Lambda(x), \Gamma^{a}_{\,\, b}(x)v^b) \oplus \alpha_{\mu}(\Gamma(x),v^b) \right)^{B} = 
\alpha_{\mu}^{B}(\Lambda(x), \Gamma^{a}_{\,\, b}(x)v^b) + \Lambda^{B}_{\,\, C}(x)\alpha_{\mu}^{C}(\Gamma(x), v^b)$$ where all symbols have their obvious meaning.   As before $\oplus$ has the correct properties with respect to $1$ and $\Lambda^{-1}$, also it is non-commutative and the sum has a clear associative extension.  Therefore, $\alpha_{\mu}(\Lambda(x)\Gamma(x), v^a) = \alpha_{\mu}(\Lambda(x), \Gamma^{a}_{\,\,b}(x)v^b) \oplus \alpha_{\mu}(\Gamma(x), v^a)$ which extends to a homomorphism from the semi-direct product group $SO(1,3) \times \mathbb{R}^4$ to the gauge group by representing the translational part trivially.  A small calculation reveals that this group structure makes $\delta_{\mu}$ into a left action as before.  The question now is how we generate local Poincar\'e invariant ``tensors'' from the gauge potential $\mathcal{A}_{\mu}^{B}(x,v^a)$?  The answer is the usual one, we calculate the commutators of the ``covariant'' derivatives
$$\left[ \mathcal{D}_{\mu}(x,v^a) , \mathcal{D}_{\nu}(x,v^a) \right] = - 2\left( \partial_{[\mu}A^{B}_{\nu]}(x,v^c) - A^{C}_{[\mu}(x,v^c) \partial_{|C|}A^{B}_{\nu]}(x,v^c) \right) \partial_{B}$$    
from which we learn that the field strength
$$ F_{\mu \nu}^{B}(x,v^c) = \partial_{[\mu}A^{B}_{\nu]}(x,v^c) - A^{C}_{[\mu}(x,v^c)\partial_{|C|}A^{B}_{\nu]}(x,v^c)$$ transforms as 
$$F_{\mu \nu}^{'B}(x,\Lambda^{a}_{\,\, b}(x)v^b + w^a(x)) = \frac{\partial x'^B}{\partial x^C} F^{C}_{\mu \nu}(x,v^b)$$ under local Poincar\'e transformations.  Under general coordinate transformations however a gauge term develops as a small calculation reveals
$$F'_{\mu \nu}(x,v^b) = \frac{\partial x^{\alpha}}{\partial x'^{\mu}}\frac{\partial x^{\beta}}{\partial x'^{\nu}}F_{\alpha \beta}(x,v^b) - \mathcal{D'}_{[ \mu}(x(x'),v^a) \left( \frac{\partial x^{\alpha}}{\partial x'^{\nu ]}}\right) \mathcal{D}_{\alpha}(x,v^a)$$ and the reader is invited to write this transformation law out in the somewhat messy basis $\partial_{B}$.  The coordinate invariant vector-fields which respect $M$ as a base manifold are given by $$W^{\mu}(x,v^a) \mathcal{D}_{\mu}(x,v^a)$$ as well as $$V^{a}(x,v^b) \partial_{a}.$$  They span the entire $TTM_{(x,v^a)}$ if and only if the matrix given by $\left( \delta_{\mu}^{\nu} - \mathcal{A}_{\mu}^{\nu}(x,v^a) \right)$ is regular.  Hence, we may bring 
$$\left[ \boldmath{W}(x,v^a) , \boldmath{V}(x,v^a) \right]$$ back in $H_{(x,v^a)}TM \oplus V_{(x,v^a)}TM$ for $\boldmath{W}(x,v^a),\boldmath{V}(x,v^a) \in H_{(x,v^a)}TM \oplus V_{(x,v^a)}TM$ but this transformation will have a complicated rational dependence upon the gauge field.  Again, one might impose the invariant statement that the origin is an exception to this by requiring $F_{\mu \nu}^{B}(x,0) = 0$.  It is important that I make one point clear and comment upon the notation I shall use; we have at this moment two $\mu$'s and $a$'s, one set mixes and transforms according to the ordinary basis $\partial_{A}$ and the other doesn't mix and transforms according to our new physical basis.  We shall not distinguish between them notationally and from now on we shall mostly rely upon the second concept.  However, to make sure the reader understands everything is consistent, let us start from a vector in the unphysical basis
$$ W(x,v^a) = W^{\mu}(x,v^a)\partial_{\mu} + W^{a}(x,v^a) \partial_a $$ and denote as a shorthand $B(x,v^a) = \left(1 - \mathcal{A}(x,v^a) \right)^{-1}$ where the reader may want to check that $B$ is local Lorentz invariant and transforms as a matrix under coordinate transformations of $M$.  Then, we obtain the following decomposition 
$$W(x,v^a) = W^{\mu}(x,v^a) B^{\nu}_{\mu}(x,v^a)\left( \partial_{\nu} - 
\mathcal{A}_{\nu}^{B}(x,v^a) \partial_{B} \right) + \left( W^{b}(x,v^a) + W^{\mu}(x,v^a)B^{\nu}_{\mu}(x,v^a) \mathcal{A}_{\nu}^{b}(x,v^a)\right) \partial_b$$ and we must verify that both coefficients now transform in the new way.  For the first coefficient, this is trivial, so we have to check it only for the second one.  Indeed, the latter transforms as
\begin{eqnarray*} W^{\mu}(x,v^a)\partial_{\mu}\Lambda^{b}_{\,\, c}(x)v^c + \Lambda^{b}_{\,\, c}(x)W^{c}(x,v^a) + \partial_{\nu}\Lambda^{b}_{\,\, c}(x)v^{c}W^{\mu}(x,v^a) B^{\nu}_{\mu}(x,v^a) \mathcal{A}^{\mu}_{\nu}(x,v^a) \\* + W^{\mu}(x,v^a)B^{\nu}_{\mu}(x,v^a) \Lambda^{b}_{\,\, c}(x) \mathcal{A}_{\nu}^{c}(x,v^a) - W^{\mu}(x,v^a) B^{\nu}_{\mu}(x,v^a) \partial_{\nu}\Lambda^{b}_{\,\,c}(x)v^{c} \end{eqnarray*} which reduces to a Lorentz boost of the original expression as it should.  The connection shall always be defined with respect to the physical basis and we perform a basis transformation such that the $A$ in  
$$F_{\mu \nu}^{A}(x,v^a)$$ are also with respect to the new basis.  In fact, the old basis needs only to be used to solve the equations of motion but does not appear anymore in the construction of the field equations.  To please the formal geometers, in the spirit of Finsler geometry, one may define a nonlinear connection by simply stating that everywhere
$$TTM_{(x,v^a)} = H_{(x,v^a)}TM \oplus V_{(x,v^a)}TM$$
holds.  At this point we define physical tensors $T^{\alpha_1 \ldots \alpha_r}_{\,\,\,\,\,\, \beta_1 \ldots \beta_s}$ where $\alpha_j , \beta_k \in \{ \mu, a \}$ and the latter is required to tranform consistently in all indices, meaning that the tensor evaluated in the complementary indices remains zero.  In other words, it is an object acting upon the separate bundles $HTM$, $VTM$ and their duals.  What we just accomplished is to write physical tensors in terms of partial and full tensors.  Before we proceed, the reader might wonder how we construct a dual basis to the $\mathcal{D}_{\mu}$. Obviously, one imposes that 
\begin{eqnarray*}
\mathcal{D}x^{\mu}(x,v^a) \left( \mathcal{D}_{\nu}(x,v^a) \right) & = & \delta^{\mu}_{\nu} \\*
\mathcal{D}x^{\mu}(x,v^a) \left( \partial_a \right) & = & 0 \\*
\mathcal{D}v^a(x,v^a) \left( \partial_b \right) & = & \delta^{a}_{b} \\*
\mathcal{D}v^a(x,v^a) \left( \mathcal{D}_{\nu}(x,v^a) \right) & = & 0 
\end{eqnarray*}      
where it is clear that $\mathcal{D}v^a \neq dv^a$ but explicitly depends upon $v^a$.  If we solve these equations in $T^{*}TM$, then obviously we all assume the $dx^A$ to work \emph{ultralocal} in contrast to the differential operators while the standard duality map allows one to define a Lie bracket which makes the dual base noncommuting\footnote{More precise, if $\alpha$ is the duality map then $[\alpha(V),\alpha(W)] = \alpha([V,W])$.}.  Clearly, the latter requires a quasilocal action meaning the $dx^A$ act differently on a function when it comes with a $\partial_B$ or a $dx^B$.  Also, the exterior derivative needs a correction due to the presence non-symmetric gauge terms as explained on the following page.  For now, we do not care about these issues and simply define covariant tensors (and we have already used them) by imposing the appropriate transformation properties.  Fine, so how should we define a connection?  Since the latter is defined in a universal way depending on four basic axioms, which should all be satisfied, we have no choice but to define the connection on $TTM$.  This is entirely logical since the ``ghost'' gravitational waves should propagate as well in space-time as tangent space.  Before we proceed, let us reflect on the status of the conservation laws and the reader might want to read at this point AXIOM III.   Those conservation laws are merely constraints on the form of matter present and therefore constrain the geometry.  In total, there are $10$ conservation laws of matter which are not implied by the generalized Einstein-Cartan equations.  Indeed, the Einstein tensor is not even covariantly conserved in Einstein-Cartan theory since the contracted Bianchi identity involves torsion and the Riemann tensor as well.  However, the quantum dynamics automatically preserves the conservation laws kinematically since the unitary transformation maps the respective matter tensors to each other.  Therefore, the conservation laws of matter should be thought of as a single ``initial-condition'' in one spacetime point on the free physics of tangent space (see axiom IV).  This implies that a generalization of the ``Einstein tensor'' is not suited for defining the dynamics and we will resort to the torsion tensor instead.  We now turn to connection theory and see if we can still destillate a non-symmetric equivalent of the Levi-Civita connection; again, the formal geometer might call this a generalized Finsler connection.  In total, we have $8^3 = 512$ independent connection components and we eliminate as much of them as possible in the same spirit as the Levi-Civita construction is constructed.  That is, we first identify those parts of the connection which transform as a tensor and put these equal to zero.  As a first calculation, we determine how $\mathcal{D}_{\mu}(x,v^a) W^{\nu}(x,v^a)$ transforms under coordinate transformations.  The formula are 
$$\frac{\partial x^{\alpha}}{\partial x'^{\mu}} \mathcal{D}_{\alpha}(x,v^a) \left( 
\frac{\partial x'^{\nu}}{\partial x^{\beta}}W^{\beta}(x,v^a)\right)$$ which equals
$$\frac{\partial x^{\alpha}}{\partial x'^{\mu}}\frac{\partial x'^{\nu}}{\partial x^{\beta}}\mathcal{D}_{\alpha}(x,v^a)W^{\beta}(x,v^a) + \frac{\partial x^{\alpha}}{\partial x'^{\mu}}\mathcal{D}_{\alpha}(x,v^a)\left( \frac{\partial x'^{\nu}}{\partial x^{\beta}} \right)W^{\beta}(x,v^a)$$ where the ``gauge'' terms explicitly reads
$$\Delta \Gamma_{\mu \gamma}^{\nu} = \frac{\partial x^{\alpha}}{\partial x'^{\mu}} \frac{\partial^2 x'^{\nu}}{ \partial x^{\alpha} \partial x^{\beta}}\frac{\partial x^{\beta}}{\partial x'^{\gamma}} - \frac{\partial x^{\alpha}}{\partial x'^{\mu}}\mathcal{A}_{\alpha}^{\kappa}(x,v^a)\frac{\partial^2 x'^{\nu}}{ \partial x^{\beta} \partial x^{\kappa}}\frac{\partial x^{\beta}}{\partial x'^{\gamma}}.$$           
Unlike in standard relativity, the ``gauge'' term is not symmetric and therefore the connection must contain symmetric as well as antisymmetric terms which adds a nonzero torsion tensor.  The reason here is the noncommuting basis of partial  differential operators $\mathcal{D}_{\mu}$ and the physical origin of this mathematical construction can be traced back to the quantum mechanical spin on $TM$ which should -on average- be balanced by gravitational spin (implying an extension of Einstein-Cartan theory on $TM$ instead of $M$).  The other ``gauge'' term can be read off from the following calculations 
$$\mathcal{D}_{\mu} \left( \Lambda^{a}_{\,\,b}(x) V^{b}(x,v^c) \right) = \mathcal{D}_{\mu} \left( \Lambda^{a}_{\,\,b}(x) \right) \Lambda_{c}^{\, \,b}(x) V'^{c}(x,v'^{d}) + \Lambda^{a}_{\,\,b}(x)\mathcal{D}_{\mu}V^{b}(x,v^c) $$ resulting in a gauge term
$$\Delta \Gamma^{a}_{\mu c}(x,v^e) = \mathcal{D}_{\mu} \left( \Lambda^{a}_{\,\,b}(x) \right) \Lambda_{c}^{\, \,b}(x)$$  which satisfies
$$\Delta \Gamma^{a}_{\mu c}(x,v^e) = -\eta_{dc} \eta^{ab} \Delta \Gamma^{d}_{\mu b}(x,v^e).$$  The reader may verify that no other gauge terms arise, but for reasons which will come clear later on we do not put $\Gamma^{\nu}_{a \mu}$ to zero.  Hence, we have the following equations:
$$ \Gamma^{c}_{ab}(x,v^d) = \Gamma^{\mu}_{ab}(x,v^d)= \Gamma^{b}_{a \mu}(x,v^d) = \Gamma^{b}_{\mu \nu}(x,v^d) = \Gamma^{\nu}_{\mu b}(x,v^d) = 0$$
and we have to determine the remaining $152$ coefficients since $\eta_{a\{c}\Gamma^{a}_{|\mu| b\}} = 0$ implying that $\nabla_{\mu} \eta_{ab} = 0$.  These degrees of freedom can be uniquely filled up by the following equations
$$\nabla_{A}e^{\mu}_{a}(x, v^c) = 0$$ and 
$$T_{\mu \nu}^{\kappa}(x,v^e) = - 2 \Gamma^{\kappa}_{[\mu \nu]}(x,v^e) + 2 F^{\kappa}_{\mu \nu}(x,v^e) = 0$$ which implies that, in the limit for $\mathcal{A}_{\mu}^{B}$ to zero, $\Gamma^{\nu}_{\mu \lambda}$ reduces to the standard Levi-Civita connection.  These restrictions can be uniquely solved to give
$$\Gamma^{\mu}_{a \nu}(x,v^c) = e^{b}_{\nu}(x,v^c) \partial_a e^{\mu}_b(x,v^c)$$
$$\Gamma^{b}_{\mu a}(x,v^c) = - e^{b}_{\nu}(x,v^c) \left(\mathcal{D}_{\mu}(x,v^c)e^{\nu}_{a}(x,v^c) - \Gamma^{\nu}_{\mu \kappa}(x,v^c) e^{\kappa}_{a}(x,v^c) \right)$$ and finally 
\begin{eqnarray*} \Gamma^{\mu}_{\nu \kappa}(x,v^c) &=& - e^{\mu}_{b}(x,v^c) \mathcal{D}_{(\nu}(x,v^c) e^{b}_{\kappa)}(x,v^c) - e^{\mu}_{a}(x,v^c)e^{b}_{(\kappa}(x,v^c)e^{a \beta}(x,v^c) \mathcal{D}_{\nu)}(x,v^c)e_{b\beta}(x,v^c) \\* & & + \frac{1}{2}e^{\mu}_{a}(x,v^c)e^{a \beta}(x,v^c)\mathcal{D}_{\beta}(x,v^c)\left(e^{b}_{\nu}(x,v^c)e_{b\kappa}(x,v^c) \right) - F^{\mu}_{\kappa \nu}(x,v^c) + F_{\kappa \,\,\,\, \nu}^{\,\,\,\, \mu}(x,v^c) - \\* & & F_{\nu \kappa}^{\,\,\,\,\,\,\,\, \mu}(x,v^c) \end{eqnarray*} where the last tensor is written with respect to the physical basis.  The reader may verify that the last formula is a direct consequence of the Koszul formula and moreover, 
$$\Gamma_{a \mu b}(x,v^c)$$ is antisymmetric in $a$ and $b$ as it should.  The reader notices that $$T_{\mu \nu}^{a}(x,v^c) = 2F_{\mu \nu}^{a}(x,v^c)$$ and therefore nontrivial torsion is present.  Define the Riemann tensor as usual by 
$$R(X,Y)Z = \nabla_{X}\nabla_{Y}Z- \nabla_{Y}\nabla_{X}Z - \nabla_{\left[X,Y \right]}Z$$ and with respect to the coordinate basis this gives
$$R(\mathcal{D}_{A}(x,v^c), \mathcal{D}_{B}(x,v^c)) \mathcal{D}_{C}(x,v^c) = 
\nabla_{A}\nabla_{B}\mathcal{D}_{C} - \nabla_{B} \nabla_{A} \mathcal{D}_C + 2 \nabla_{F_{A B}}\mathcal{D}_{C}$$ where, obviously, $$\left[ \mathcal{D}_A(x,v^c), \mathcal{D}_{B}(x,v^c) \right] = - 2F_{AB}(x,v^c).$$  Before we proceed, the reader may want to explicitly verify that everything works out as it should since after all, we are working in a unusual basis.  For example, let us calculate the commutator between two vectorfields $V(x,v^c)$ and $W(x,v^c)$:
\begin{eqnarray*}
\left[ V(x,v^c), W(x,v^c) \right] & = & \left( V^{\mu}(x,v^c)\mathcal{D}_{\mu}(x,v^c)W^{\nu}(x,v^c) - W^{\mu}(x,v^c)\mathcal{D}_{\mu}(x,v^c)V^{\nu}(x,v^c)\right) \mathcal{D}_{\nu}(x,v^c) - \\* & & 2 V^{\mu}(x,v^c)W^{\nu}(x,v^c)F_{\mu \nu}(x,v^c). \end{eqnarray*}
To verify that this expression is well defined, we calculate the transformation behavior under coordinate transformations; the relevant terms are
\begin{eqnarray*}
\Delta & = & 2 V^{\mu}(x,v^c)W^{\kappa}(x,v^c) \mathcal{D}_{[\mu} \frac{\partial x'^{\nu}}{\partial x^{\kappa]}} \frac{\partial x^{\gamma}}{\partial x'^{\nu}}\mathcal{D}_{\gamma}(x,v^c) + 2V^{\mu}(x,v^c)W^{\kappa}(x,v^c) \mathcal{D}_{[\mu}\frac{\partial x^{\gamma}}{\partial x'^{\nu}} \frac{\partial x'^{\nu}}{\partial x^{\kappa}} \mathcal{D}_{\gamma}(x,v^c) \\*
& = & 0.
\end{eqnarray*}  Unlike $F_{\mu \nu}(x,v^c)$, $F_{a \mu}(x,v^c)$ is a tensor under coordinate transformations, but under local Poincar\'e transformations a gauge term of the kind
$$ \frac{1}{2} \mathcal{D}_{\mu}(x,v^c) \left( \Lambda_{a}^{\,\,b}(x) \right) \partial_b$$ develops.  The coordinate expressions of the curvature tensor are given by
\begin{eqnarray*}
R_{abc}^{\,\,\,\,\,\,\,\,d}(x,v^e) & = & 0 \\*
R_{abc}^{\,\,\,\,\,\,\,\, \mu}(x,v^e) & = & 0 \\*
R_{ab\mu}^{\,\,\,\,\,\,\,\, c}(x,v^e) & = & 0\\*
R_{ab \mu}^{\,\,\,\,\,\,\,\, \nu}(x,v^e) & = & - \partial_{a}\Gamma^{\nu}_{b\mu}(x,v^e) + \partial_{b}\Gamma^{\nu}_{a\mu}(x,v^e) + \Gamma^{\nu}_{a \kappa}(x,v^e) \Gamma^{\kappa}_{b\mu}(x,v^e) - \Gamma^{\nu}_{b \kappa}(x,v^e) \Gamma^{\kappa}_{a \mu}(x,v^e) = 0
\end{eqnarray*}
and the reader may verify that the last equation holds.  This means that the tangent space is flat and curvature can at most live on spacetime or in the ``intermediate'' space.  The remaining expressions are computed to be
\begin{eqnarray*}
R_{a \mu b}^{\,\,\,\,\,\,\,\,\, \nu}(x,v^e) & = & 0 \\*
R_{a \mu b}^{\,\,\,\,\,\,\,\, c}(x,v^e) & = & -\partial_a \Gamma^{c}_{\mu b}(x,v^e) - 2F^{\nu}_{a \mu}(x,v^e) \Gamma^{c}_{\nu b}(x,v^e) \\*
R_{a \mu \nu}^{\,\,\,\,\,\,\,\,\, b}(x,v^e) & = & 0 \\*
R_{a \mu \nu}^{\,\,\,\,\,\,\,\,\, \kappa}(x,v^e) & = & - \partial_a \Gamma^{\kappa}_{\mu \nu}(x,v^e) + \mathcal{D}_{\mu} \Gamma^{\kappa}_{a \nu}(x,v^e) + \Gamma^{\alpha}_{\mu \nu}(x,v^e)\Gamma^{\kappa}_{a \alpha}(x,v^e) - \\*
& & \Gamma^{\alpha}_{a \nu}(x,v^e)\Gamma^{\kappa}_{\mu \alpha}(x,v^e) - 2F^{\alpha}_{a \mu}(x,v^e) \Gamma^{\kappa}_{\alpha \nu}(x,v^e) - 2 F^{b}_{a \mu}(x,v^e) \Gamma^{\kappa}_{b \nu}(x,v^e)  \\* 
R_{\mu \nu a}^{\,\,\,\,\,\,\,\,\, \kappa}(x,v^e) & = & 0 \\*
R_{\mu \nu a}^{\,\,\,\,\,\,\,\,\, b}(x,v^e) & = & - \mathcal{D}_{\mu} \Gamma^{b}_{\nu a}(x,v^e) + \mathcal{D}_{\nu} \Gamma^{b}_{\mu a}(x,v^e) + \Gamma^{c}_{\nu a}(x,v^e) \Gamma^{b}_{\mu c}(x,v^e) \\*
& & - \Gamma^{c}_{\mu a}(x,v^e) \Gamma^{b}_{\nu c}(x,v^e) - 2F^{\kappa}_{\mu \nu}(x,v^e) \Gamma^{b}_{\kappa a}(x,v^e) \\*
R_{\mu \nu \kappa}^{\,\,\,\,\,\,\,\,\, a}(x,v^e) & = & 0 \\*
R_{\mu \nu \kappa}^{\,\,\,\,\,\,\,\,\, \lambda}(x,v^e) & = &  - \mathcal{D}_{\mu} \Gamma^{\lambda}_{\nu \kappa}(x,v^e) + \mathcal{D}_{\nu}\Gamma^{\lambda}_{\mu \kappa}(x,v^e) + \Gamma^{\alpha}_{\nu \kappa}(x,v^e) \Gamma^{\lambda}_{\mu \alpha}(x,v^e) - \\*
& & \Gamma^{\alpha}_{\mu \kappa}(x,v^e) \Gamma^{\lambda}_{\nu \alpha}(x,v^e) - 2F^{a}_{\mu \nu}(x,v^e) \Gamma^{\lambda}_{a \kappa}(x,v^e) - 2F^{\alpha}_{\mu \nu}(x,v^e) \Gamma^{\lambda}_{\alpha \kappa}(x,v^e). 
\end{eqnarray*}                     
Let me make some remarks regarding the remarkable structure of these equations.   The expressions $R_{a \mu b}^{\,\,\,\,\,\,\,\,\, c}(x,v^e)$ and $R_{a \mu \nu}^{\,\,\,\,\,\,\,\,\, \kappa}(x,v^e)$ are all first order in the spacetime derivatives; the spacetime derivatives of the different fields $e^{a}_{\mu}(x,v^c)$ and $\mathcal{A}_{\mu}^{B}(x,v^c)$ decouple but there is some novelty in this type of equation in the sense that it may contain both derivatives of the kind $\partial_t e^{a}_{\mu}(x,v^c)$ as $\partial_t \partial_b e^{a}_{\mu}(x,v^c)$ and to uniquely solve those requires a new view on initial value problems.  I believe the linearized equations to be ultrahyperbolic and shall write them out in full detail later on.  The remaining two expressions $R_{\mu \nu a}^{\,\,\,\,\,\,\,\,\, b}(x,v^c)$ and $R_{\mu \nu \kappa}^{\,\,\,\,\,\,\,\,\, \lambda}(x,v^c)$ are classical second order expressions without the above mentioned curiosity.  A two time and six space formalism seems here the right thing to do since we have a direct sum metric $g_{\mu \nu} \oplus \eta_{ab}$ on $TTM$.  This implies physically that non-local (or better non-causal) correlations in the metric tensor will build up instantaneously, the matter equations of motion of course obey the usual hyperbolic laws (with respect to $g_{\mu \nu}$).  A few years ago, I thought about using a two time formalism (where one time is rolled up on a cylinder) to explain away the Bell inequalities; this formalism can be made entirely consistent by declaring that the ``Kaluza-Klein'' modes cannot be observed implying that no tachyons are measured but non-local correlations nevertheless build up rather quickly.  Since the mystery of quantum mechanics is not its non-locality, but rather the wave particle-duality (which we solve in this theory), I felt that a theory explaining non-local correlations would not suffice by itself.  Moreover, I anticipated several problems with the extra dimension and did not like the ad-hoc character of the addition of one time dimension.  One does not solve nature's problems by merely adding new stuff in order to please your philosophical picture about the world; changes in the formalism will always be subtle and arise from conservative arguments which in an uncompromising manner deal with the difficulties in the existing formalism.  We realized\footnote{I thank S. Nobbenhuis for discussions regarding this idea; he suggested me to take $3$ time dimensions instead of $2$.  I was unaware at that time of the work of Bars.} of course that adding just a compactified time dimension was ugly, therefore we thought about starting from a completely symmetric situation in $6$ dimensions with an ultrahyperbolic metric with signature $(+ + + - - -)$.  The compactification of two time dimensions would be a kind of dynamical symmetry breaking giving rise to our world, I have never worked out the full implications of this picture and I do not advise the reader to do so.  The situation concerning the initial value formulation of such theory is rather more delicate; in contrast to what is said in \cite{Tegmark}, the initial value problem for the linear ultrahyperbolic equation can be well defined if one puts suitable constraints on the initial data \cite{Weinstein}.  A non-local deformation of gravity has been proposed in order to explain away the cosmological constant problem \cite{Nima}; however, in the formalism explained below, there is no high vacuum density and therefore no issue with the cosmological constant.  Nevertheless, such non-local deformation is probably necessary to solve the horizon and flatness problem and therefore it serves as an alternative mechanism for inflation as well as the cosmological ``constant''.  We intend to come back to this in the next chapter in more detail.  \\* \\*  In order to better understand what is the right thing to do, we study now the first and second Bianchi identities.  One easily reads off that 
\begin{eqnarray*}
R_{a\mu \left(bc\right)}(x,v^e) & = & 0 \\*
R_{\mu \nu \left(a b \right)}(x,v^e) & = & 0
\end{eqnarray*} 
while the usual standard Bianchi identity in torsionless Riemannian geometry 
$$R_{[\mu \nu \kappa]}^{\,\,\,\,\,\,\,\,\, \,\, \lambda}(x,v^c) = 0$$ does not hold anymore.  Indeed, an elementary calculation yields 
$$R_{[\mu \nu \kappa]}^{\,\,\,\,\,\,\,\,\, \,\, \lambda}(x,v^c) = - 2 \left( \mathcal{D}_{[\mu}(x,v^c) F_{\nu \kappa]}^{\lambda}(x,v^c) - 2 F^{\alpha}_{[\mu \nu}(x,v^c)F_{\kappa]\alpha}^{\lambda}(x,v^c) + F^{a}_{[\mu \nu}(x,v^c) \Gamma^{\lambda}_{|a| \kappa ]}(x,v^c) \right)$$ and it is the last term on the right hand side which makes this expression nonvanishing (due to the Bianchi identities for $\mathcal{A}_{\mu}^{B}(x,v^c)$ which we work out next).  The reader may also verify that 
$$R_{\mu \nu \left( \kappa \lambda \right)}(x,v^c) \neq 0 \neq R_{a \mu \left( \nu \kappa \right)}(x,v^c).$$  It is helpful to first understand the second Bianchi identities for the field strength $F_{AB}^{C}(x,v^c)$.  Although the latter are not tensors, the Bianchi identies are valid in any ``gauge'' and coordinate system.   The first equality is given by
\begin{eqnarray*}
0 & = & \left[ \mathcal{D}_{\mu}(x,v^c) , \left[ \mathcal{D}_{\nu}(x,v^c) , \mathcal{D}_{\kappa}(x,v^c) \right] \right] + \left[ \mathcal{D}_{\nu}(x,v^c) , \left[ \mathcal{D}_{\kappa}(x,v^c) , \mathcal{D}_{\mu}(x,v^c) \right] \right] + \\*
& & \left[ \mathcal{D}_{\kappa}(x,v^c) , \left[ \mathcal{D}_{\mu}(x,v^c) , \mathcal{D}_{\nu}(x,v^c) \right] \right] \\*
& = & - 2 \left( \mathcal{D}_{[\mu}(x,v^c) F^{\lambda}_{\nu \kappa]}(x,v^c) + 
2F^{\lambda}_{\gamma [\mu}(x,v^c)F^{\gamma}_{\nu \kappa]}(x,v^c) + 2F^{\lambda}_{a [\mu}(x,v^c)F^{a}_{\nu \kappa]}(x,v^c) \right) \mathcal{D}_{\lambda}(x,v^c) - \\*
& & 2\left( \mathcal{D}_{[\mu}(x,v^c)F^{a}_{\nu \kappa]}(x,v^c) + 2F^{a}_{b[\mu}(x,v^c) F^{b}_{\nu \kappa]}(x,v^c) + 2 F^{a}_{\lambda [\mu}(x,v^c)F^{\lambda}_{\nu \kappa]}(x,v^c) \right) \partial_a
\end{eqnarray*}
and the reader notices that writing these equations explicitly in terms of the potential $\mathcal{A}_{\mu}^{B}(x,v^c)$ is not that easy given the presence of $B^{\mu}_{\nu}(x,v^c) = \left(\delta^{\mu}_{\nu} - \mathcal{A}^{\mu}_{\nu}(x,v^c) \right)^{-1}$.  The second equality is given by
\begin{eqnarray*}
0 & = & \left[ \mathcal{D}_{\mu}(x,v^c) , \left[ \mathcal{D}_{\nu}(x,v^c) , \partial_a \right] \right] +  \left[ \mathcal{D}_{\nu}(x,v^c) , \left[ \partial_a , \mathcal{D}_{\mu}(x,v^c) \right] \right] + \left[ \partial_a , \left[ \mathcal{D}_{\mu}(x,v^c) , \mathcal{D}_{\nu}(x,v^c) \right] \right] \\*
& = & - 2 \left( \mathcal{D}_{[\mu}(x,v^c) F^{\kappa}_{\nu a]}(x,v^c) - 2 F^{\alpha}_{[\nu a}(x,v^c) F^{\kappa}_{\mu ] \alpha}(x,v^c) - 2 F^{b}_{[\nu a}(x,v^c)F^{\kappa}_{\mu ] b}(x,v^c) \right) \mathcal{D}_{\kappa}(x,v^c) \\*
& & -2 \left( \mathcal{D}_{[\mu}(x,v^c)F^{b}_{\nu a]}(x,v^c) - 2 F^{d}_{[\nu a}(x,v^c)F^{b}_{\mu ] d}(x,v^c) - 2 F^{\kappa}_{[ \nu a}(x,v^c) F^{b}_{\mu ] \kappa}(x,v^c) \right) \partial_b 
\end{eqnarray*} and finally, the last one equals
\begin{eqnarray*}
0 & = & \left[ \mathcal{D}_{\mu}(x,v^c), \left[\partial_a , \partial_b \right] \right] + \left[ \partial_a, \left[\partial_b , \mathcal{D}_{\mu}(x,v^c) \right] \right] + \left[ \partial_b, \left[\mathcal{D}_{\mu}(x,v^c) , \partial_a \right] \right] \\*
& = & - 2\left( \partial_{[a}F^{\lambda}_{b]\mu}(x,v^c) + 2 F^{\lambda}_{\alpha [a}(x,v^c) F^{\alpha}_{b] \mu}(x,v^c) \right) \mathcal{D}_{\lambda}(x,v^c) \\*
& & - 2\left( \partial_{[a}F^{d}_{b] \mu}(x,v^c) + 2 F^{d}_{\alpha [a}(x,v^c) F^{\alpha}_{b] \mu}(x,v^c) \right) \partial_d
\end{eqnarray*} leading in total to six types of Bianchi identities.  Likewise, we now compute the ``ordinary'' second Bianchi identities; the first one is given by
\begin{eqnarray*}
0 & = & \left[ \nabla_{[ \mu}(x,v^c) , \left[ \nabla_{\nu}(x,v^c) , \nabla_{\kappa]}(x,v^c) \right] \right] W^{\gamma}(x,v^c) \\*
& = &  \left( \nabla_{[\mu}(x,v^c) R_{\nu \kappa]\alpha}^{\,\,\,\,\,\,\,\,\,\, \gamma}(x,v^c) + 2 \nabla_{[ \mu}(x,v^c) \left( F^{a}_{\nu \kappa]}(x,v^c) \Gamma^{\gamma}_{a \alpha}(x,v^c) \right) \right) W^{\alpha}(x,v^c) + \\* & & \left( R_{[\nu \kappa \mu]}^{\,\,\,\,\,\,\,\,\,\, \alpha}(x,v^c) + 2 F^{a}_{[\nu \kappa}(x,v^c) \Gamma^{\alpha}_{|a| \mu]}(x,v^c) \right)\nabla_{\alpha}(x,v^c)W^{\gamma}(x,v^c) - 2\nabla_{[\mu}(x,v^c)F^{a}_{\nu \kappa]}(x,v^c) \partial_a W^{\gamma}(x,v^c) \\* & & + 2 F^{a}_{[\nu \kappa} \left( \partial_a \nabla_{\mu]}(x,v^c) - \nabla_{\mu]}(x,v^c)\partial_a \right)W^{\gamma}(x,v^c)
\end{eqnarray*} and the last term is computed to be
\begin{eqnarray*}
 - 2F^{b}_{[\nu \kappa}(x,v^c)\left( 2 F^{a}_{|b| \mu]}(x,v^c) + \Gamma^{a}_{\mu]b}(x,v^c) \right) \partial_a W^{\gamma}(x,v^c) - 4 F^{b}_{[\nu \kappa}(x,v^c) F^{\alpha}_{|b| \mu]}(x,v^c) \nabla_{\alpha}(x,v^c) W^{\gamma}(x,v^c) - \\*
 2 F^{b}_{[\nu \kappa}(x,v^c) \left( 2F^{\beta}_{|b| \mu]}(x,v^c) \Gamma^{\gamma}_{\beta \alpha}(x,v^c) + \partial_b \Gamma^{\gamma}_{\mu] \alpha}(x,v^c) \right) W^{\alpha}(x,v^c)
\end{eqnarray*} and the reader is advised to explicitly check that the correct transformation laws hold.  Given the above Bianchi identities, two new expressions arise; the first (second) one being a correction to the first (second) Bianchi identity:
\begin{eqnarray*}
0 & = & R_{[\nu \kappa \mu]}^{\,\,\,\,\,\,\,\,\,\, \alpha}(x,v^c) + 2 \left( \Gamma^{\alpha}_{a [\mu}(x,v^c) - 2 F^{\alpha}_{a [\mu}(x,v^c) \right) F^{a}_{\nu \kappa]}(x,v^c) \\*
0 & = & \nabla_{[\mu}(x,v^c) R_{\nu \kappa] \alpha}^{\,\,\,\,\,\,\,\,\,\, \gamma}(x,v^c) + 2 \nabla_{[\mu}(x,v^c) \left( F^{a}_{\nu \kappa]}(x,v^c) \Gamma^{\gamma}_{a \alpha}(x,v^c) \right) - \\* & & 2F^{b}_{[\nu \kappa}(x,v^c) \left( 2F^{\beta}_{|b| \mu]}(x,v^c) \Gamma^{\gamma}_{\beta \alpha}(x,v^c) + \partial_b \Gamma^{\gamma}_{\mu] \alpha}(x,v^c) \right).
\end{eqnarray*}  The reader might verify that our new expression for the first Bianchi identity coincides with the old one by making use of previous identities.  One can rewrite these formulae in a more conventional form; indeed, inspection reveals that
\begin{eqnarray*}
0 & = & R_{[\nu \kappa \mu]}^{\,\,\,\,\,\,\,\,\,\, \alpha}(x,v^c) -  T^{\alpha}_{a [\mu}(x,v^c) T^{a}_{\nu \kappa]}(x,v^c) \\*
0 & = & \nabla_{[\mu}(x,v^c) R_{\nu \kappa] \alpha}^{\,\,\,\,\,\,\,\,\,\, \gamma}(x,v^c) - T^{a}_{[\nu \kappa}(x,v^c)R_{\mu]a \alpha}^{\,\,\,\,\,\,\,\,\,\,\, \gamma}(x,v^c)
\end{eqnarray*}
which is identical to the usual Bianchi identities in Einstein-Cartan theory.  This was to be expected since the latter are more universal than the former: indeed, our connection is a constrained affine connection in $6+2$ dimensions written out in a non-holonomic basis.  To appreciate that this is indeed the fact, one may verify that
\begin{eqnarray*}
0 & = & \left[ \nabla_{[ \mu}(x,v^c) , \left[ \nabla_{\nu}(x,v^c) , \nabla_{\kappa]}(x,v^c) \right] \right] V^{a}(x,v^c)
\end{eqnarray*} leads to exactly one new equality
\begin{eqnarray*}
0 & = & \nabla_{[\kappa}(x,v^c) R_{\mu \nu]b}^{\,\,\,\,\,\,\,\,\,\,\, a}(x,v^c) - 2 \left(
2F^{\alpha}_{d [\kappa}(x,v^c) \Gamma^{a}_{\alpha b}(x,v^c) + \partial_{d} \Gamma^{a}_{[\kappa |b|}(x,v^c) \right)F^{d}_{\mu \nu]}(x,v^c) \\*
& = & \nabla_{[\kappa}(x,v^c) R_{\mu \nu]b}^{\,\,\,\,\,\,\,\,\,\,\, a}(x,v^c) - T^{d}_{[\mu \nu}(x,v^c) R_{\kappa]d b}^{\,\,\,\,\,\,\,\,\,\, a}(x,v^c).
\end{eqnarray*}  Therefore, without any further computation, the remaining second Bianchi identities are given by
\begin{eqnarray*}
0 & = & \nabla_{[a}(x,v^c) R_{\mu \nu] \kappa}^{\,\,\,\,\,\,\,\,\,\, \lambda}(x,v^c) - T^{b}_{[\mu \nu}(x,v^c) R_{a]b \kappa}^{\,\,\,\,\,\,\,\,\,\,\, \lambda}(x,v^c) - T^{\alpha}_{[\mu \nu}(x,v^c) R_{a] \alpha \kappa}^{\,\,\,\,\,\,\,\,\,\,\, \lambda}(x,v^c) \\*
0 & = &  \nabla_{[a}(x,v^c) R_{b \mu] \kappa}^{\,\,\,\,\,\,\,\,\,\,\,\, \lambda}(x,v^c) - T^{\alpha}_{[b \mu}(x,v^c) R_{a]\alpha \kappa}^{\,\,\,\,\,\,\,\,\,\, \lambda}(x,v^c) \\*
0 & = & \nabla_{[a}(x,v^c) R_{\mu \nu] b}^{\,\,\,\,\,\,\,\,\,\,\, d}(x,v^c) - 
T^{\alpha}_{[\mu \nu}(x,v^c) R_{a] \alpha b}^{\,\,\,\,\,\,\,\,\,\, d}(x,v^c)
- T^{e}_{[\mu \nu}(x,v^c) R_{a]e b}^{\,\,\,\,\,\,\,\,\,\,\, d}(x,v^c) \\*
0 & = & \nabla_{[a}(x,v^c) R_{b \mu] d}^{\,\,\,\,\,\,\,\,\,\,\, e}(x,v^c) - T^{\alpha}_{[b \mu}(x,v^c) R_{a] \alpha d}^{\,\,\,\,\,\,\,\,\,\, e}(x,v^c)
\end{eqnarray*}     
and the other four, first Bianchi identities are
\begin{eqnarray*} 
0 & = & R_{[\mu \nu a]}^{\,\,\,\,\,\,\,\,\,\, \alpha}(x,v^c) - T^{\alpha}_{b [ \mu}(x,v^c)T^{b}_{\nu a]}(x,v^c) - \nabla_{[\mu}(x,v^c) T^{\alpha}_{\nu a]}(x,v^c) \\*
0 & = & R_{[\mu a b]}^{\,\,\,\,\,\,\,\,\,\,\, \alpha}(x,v^c) - T^{\alpha}_{\beta [\mu}(x,v^c) T^{\beta}_{ab]}(x,v^c) - \nabla_{[\mu}(x,v^c) T^{\alpha}_{ab]}(x,v^c) \\*
0 & = & R_{[\mu \nu a]}^{\,\,\,\,\,\,\,\,\,\, b}(x,v^c) - T^{b}_{d[\mu}(x,v^c) T^{d}_{\nu a]}(x,v^c) - T^{b}_{\alpha [\mu}(x,v^c) T^{\alpha}_{\nu a]}(x,v^c) - \nabla_{[\mu}(x,v^c)T^{b}_{\nu a]}(x,v^c) \\*
0 & = & R_{[\mu a b]}^{\,\,\,\,\,\,\,\,\,\,\, d}(x,v^c) - T^{d}_{\alpha [\mu}(x,v^c)T^{\alpha}_{ab]}(x,v^c) - \nabla_{[\mu}(x,v^c) T^{d}_{ab]}(x,v^c). 
\end{eqnarray*}
All this means that our geometry is an extremely subtle generalization of Riemannian geometry in four spacetime dimensions.  Indeed, it is wider than ordinary geometry of the vielbein and spin connection in $3+1$ dimensions but is much more constrained than Einstein-Cartan geometry in $6+2$ dimensions.  Indeed, the flatness of tangent space as well as the vanishing of many torsion coefficients show that this is the case.  This gives much hope that the standard problems of ordinary gravity theories in higher dimensions are eliminated and we finish this intermezzo on our constrained affine geometry by stuyding whether one can retrieve the correct conservation laws at second order in the partial derivatives.  \\* \\*
One can now calculate the contracted Bianchi identities in order to generate ``conservation laws''; however, Noether's theorem does not apply to geometries with a nonzero torsion and the resulting equations do not permit to extract the correct conserved tensors.  Indeed, from the second Bianchi identities, one calculates that
\begin{eqnarray*}
0 & = & \partial_a \left(R_{b \mu \kappa}^{\,\,\,\,\,\,\,\,\,\,\,\, \mu}(x,v^c)e^{\kappa a}(x,v^c) - \delta^{a}_{b} R_{d \mu \kappa}^{\,\,\,\,\,\,\,\,\,\,\,\, \mu}(x,v^c) e^{\kappa d}(x,v^c) \right) - T^{\alpha}_{b \mu}(x,v^c) R_{a \alpha}^{\,\,\,\,\,\,\, a \mu}(x,v^c) + \\* & &  T^{\alpha}_{a \mu}(x,v^c) R_{b \alpha}^{\,\,\,\,\,\,\, a \mu}(x,v^c) \\*
0 & = & \partial_a \left( R_{b \mu}^{\,\,\,\,\,\,\, a d}(x,v^c) e^{\mu}_{d}(x,v^c) - \delta^{a}_{b} R_{d \mu}^{\,\,\,\,\,\,\, d f}(x,v^c) e^{\mu}_{f}(x,v^c) \right)
- T^{\alpha}_{b \mu}(x,v^c) e^{\mu}_{f}(x,v^c) R_{a \alpha}^{\,\,\,\,\,\,\,\,a f}(x,v^c) + \\* & & T^{\alpha}_{a \mu}(x,v^c) e^{\mu}_{f}(x,v^c) R_{b \alpha}^{\,\,\,\,\,\,\, a f}(x,v^c)   
\end{eqnarray*}
and the reader is invited to construct the four remaining equations (which involve derivatives $\nabla_{\mu}$).  Taking into account the conservation laws in AXIOM III, we must conclude that such laws which involve the Riemann tensor can only be constructed in a theory containing three partial derivatives or more.  However, it is possible to construct the appropriate conservation laws in second order involving the torsion tensor only.  We postpone this issue until AXIOM VII where we compute that the appropriate Newtononian laws emerge in the limit of infinite speed of light.  These results appear to ressurect the idea of a pure torsion theory of gravity which does not imply that the Riemann curvature vanishes of course.         
\begin{itemize}               
\item AXIOM II : At each point $x$ of the manifold $M$, there exists a basic set of particle creation operators $a^{\dag}_{\vec{k} \, m \, \sigma \, c \, \pm}(e_b(x),x)$ where $\vec{k}$ is the three momentum with respect to $e_j(x)$, $m$ the inertial mass, $\sigma$ the spin, $\pm$ indicates whether it corresponds to a particle of positive or negative norm respectively and $c$ is a natural index labeling one of the $\aleph_{0}$ copies mentioned in AXIOM 0.  Moreover, there exists a unique cyclic, generating vacuum state $|0, e_{a}(x),x \rangle$ on which all creation and annihilation operators act as usual.  By convention,  $\mathcal{F}(e_a(x),x)$ is the \emph{local} Fock space ``generated'' by the application of the operators with $c = 0$ on the vacuum state\footnote{See chapter 10 for further explanation.}.  However, the rest of $\mathcal{K}$ is also \emph{ontologically} available to the local observer by which I mean that he ``knows'' about the existence of the particles with $c \neq 0$ in the universe but is unable to measure them and therefore cannot tell anything about the interactions between them.  This implies an extremely important subtlety wich should be well understood: this ``information'' about the rest of the universe \emph{must} be contained in the energy-momentum and spin tensors evaluated at $(x,v^e)$.  However, the local Poincar\'e algebra only depends upon the restriction of these tensors to $\mathcal{F}(e_b(x),x)$.
\end{itemize}
It might be good to philosophize to a greater extend about the meaning of the above AXIOM which is far from trivial.  Every local observer gets information about the rest of the universe through measurements and a model about the rest of the universe is build on $TM_{x}$.  This model is a free quantum theory of matter which does not preclude the incorporation of classical gravitational effects (without curving $VTM$). The reader should await a deeper discussion of this issue following AXIOM VII.       
\begin{itemize} 
\item AXIOM III :  There exist two local conserved, non-symmetric\footnote{It is normal that $T^{ab}$ is non-symmetric for gravitational theories with spin and in general, the natural energy momentum tensor for free fermionic theories is not symmetric either.  We insist, as in Einstein Cartan theory, that this is a physical effect and it is therefore undesirable to apply the Belinfante Rosenfeld symmetrization procedure.} energy momentum tensors $T_{j}^{ab}(x,v^d, e_g(x))$ on $TM$ and anti-symmetric conserved spin tensors $S^{ab}_{j \, c}(x,v^d, e_g(x))$ where the conservation laws are respectively
$$\partial_a T^{ab}_{j}(x,v^d, e_g(x)) = 0$$
and $$\partial^{c} S^{ab}_{j \, c}(x,v^d , e_g(x)) = 0.$$
All tensors are normal ordered expressions in terms of the creation and annihilation operators of the whole universe and local particle space respectively.  \\* \\* 
$T_{2}^{ab}(x,v^e, e_g(x))$ and $S_{2 \, c}^{ab}(x,v^e, e_g(x))$ are the restrictions of $T_{1}^{ab}(x,v^e, e_g(x))$ and $S_{1 \, c}^{ab}(x,v^e, e_g(x))$ respectively to $\mathcal{F}(e_b(x),x)$.  Moreover, the conserved charges $P^{a}_{j}(x, e_c(x))$ and $S^{ab}_{j}(x, e_c(x))$ are generators of the local and global Poincar\'e algebra respectively and all operators annihilate the vacuum state $|0, e_{a}(x) , x \rangle$.  This picture is not complete as yet since gravitons have not been included\footnote{See the discussions about Weinberg-Witten in the introduction.}; therefore, the true generators are constructed from the latter by including the graviton spin and four momentum.  Nevertheless, the way the conservation of energy-momentum and spin is expressed here is entirely physical: all it says is that on tangent space, gravitons do not destroy conservation laws of matter and do not gravitate.  However, these conservation laws might change from one space-time point to another since particles get redefined.  Of course, when conservation laws would be expressed in terms of the spacetime derivatives, we would be in deep problems; but this is not the case here.      
\item AXIOM IV :  Having a totally consistent particle interpretation requires amongst others the commutatation relations 
$$\left[ P^{a}_1 (x , e_b(x)) , a^{\dag}_{\vec{k} \, m \, \sigma \, c \, \pm}(e_b(x),x) \right] = k^{a} a^{\dag}_{\vec{k} \, m \, \sigma \, c \, \pm}(e_b(x),x)$$ as the reader can easily convince himself of (actually this equality is enforced by the multi-particle states).  Similar expressions should hold for $P_2^{a}(x)$ and creation operators corresponding to $c =0$.  This implies that the ``theory'' on the tangent space $TM_x$ is a free one which enforces the physical statement that any legitimate Quantum Theory must be asymptotically free\footnote{We already know this is the case for the strong interactions, Weinberg has speculated quantum gravity to be asymptotically safe.}.
\item AXIOM V :  Space-time interactions are kinematically determined by unitary relational operators $U(e_a(x),e_b(y),x,y)$ inducing the following conditions:
$$ U(e_a(x),e_b(y),x,y) |0, e_{a}(x),x \rangle = |0, e_{a}(y),y \rangle $$
and 
$$ U(e_a(x),e_b(y),x,y) a^{\dag}_{\vec{k} \, m \, \sigma \, c \, \pm}(e_c(x),x) U^{\dag}(e_a(x), e_b(y),x,y) = a^{\dag}_{\vec{k} \, m \, \sigma \, c \, \pm}(e_c(y),y).$$
Moreover, $U^{\dag}(e_a(x),e_b(y),x,y) = U(e_a(y),e_b(x),y,x)$ and we demand the group law to hold
$$U(e_b(y),e_c(z),y,z)U(e_a(x), e_b(y),x,y) = U(e_a(x),e_c(z),x,z)$$
which can be interpreted as a trivial homology condition.  I guess further generalizations can be constructed by going over to higher homology but these would, in general, introduce a path dependence in the above definitions similar to the one of Weyl gravity and the objections against this theory are rather well known.  More precise, the result of a scattering experiment would depend upon the path an unphysical, ``mental'' observer follows in spacetime.  Nevertheless, a reasonable higher theory would satisfy
$$ U(e_c(z),e_a(x),z,x)U(e_b(y),e_c(z),y,z)U(e_a(x), e_b(y),x,y) = D(e_a(x),x)$$ where $D$ is an element of the ``little group'' which is defined as all unitary operations which leave the vacuum state $| 0, e_a(x),x \rangle$ as well as the creation operators $a^{\dag}_{\vec{k} \, m \, \sigma \, n \, \pm}(e_c(x),x)$ invariant up to a phase\footnote{Therefore, the group law is only required to hold on a generalized projective level.}; this is a reasonable alternative with genuine new physics and one might try to work it out in the future. \\* \\* The mathematical implication of a trivial homology is the existence of a unitary ``potential'' $U(e_a(x),x)$ so that $$U(e_a(x),e_b(y),x,y) = U(e_b(y),y)U^{\dag}(e_a(x),x).$$  This axiom needs some further clarification on the following topics: (a) why a unitary transformation of $\mathcal{K}$ and not of the local one particle spaces $\mathcal{F}(e_b(x),x)$ (b) a deeper understanding of the trivial homology condition.  The answer on (a) is at the same time a comment on all those models in the literature where one has a discrete causal graph and one draws arrows between distinct events and associated to the arrows a non-unitary functor between the local Hilbert spaces (usually one employs completely positive maps).  The point is of course that the evolution from $x$ to $y$ also depends upon on other points $z$; this is a deep consequence of the lack of a background metric.  Concerning (b), one way to think about the unitary tranformations is the following: each reference frame contains a list of particles, first the particles which can be observed in the reference frame at that point and second, the way this particular observer thinks about all other particles in the rest of the universe.  The unitary mapping is nothing but a translation from one list to another and what we think about as scattering is nothing but a different perception of the same thing.  Now, if $U(e_b(x),e_b(y),x,y)$ is a translation from the list constructed by $(e_b(x),x)$ to $(e_b(y),y)$ and likewise $U(e_b(y),e_b(z),y,z)$ a translation from the dictionary of $(e_b(y),y)$ to $(e_b(z),z)$, then the trivial homology condition states that translations are perfect.  In the light of chapter seven this would be far too constrained, we would actually demand here that the unitary relators form a group while they should only form a kroup.  However, as mentioned in AXIOM 0, the relational Nevanlinna module is a dynamical object which should be determined from the unitary relators which again can only be defined starting from a local Nevanlinna module attached to some frame.  It is here that we will launch a principle which is very similar to the way the manifold is constructed in general relativity; that is, local triviality.  This means the following: for any reference point $x_0$ there exists an open environment $\mathcal{O}$ of $x_0$ in which the above group property holds.  That is, the group property of the unitary relators between the different bases is assumed to hold locally and the possible breakdown to a kroup might be a global effect.
\end{itemize}
It appears to me that the local triviality statement is a necessary one and it is possible to weaken it slightly by demanding that a covering of charts with \emph{prefferred} points exists and give up on associativity even locally.  It would entail some nonlocality associated with the points and the charts themselves and I have no idea as yet how to interpret this, but it definetly is a possibility.  Likewise, there is not a single state of the universe, but there is a consistent family $\Psi_{\alpha}$ defined with respect to a covering by coordinate charts $\mathcal{O}_{\alpha}$.  This means that for any $\alpha$ and $x \in \mathcal{O}_{\alpha}$, $\Psi_{\alpha}$ is well defined as an element of $\mathcal{K}(x,e_b(x))$ and moreorover, for any $\alpha, \beta$ such that $\mathcal{O}_{\alpha} \cap \mathcal{O}_{\beta} \neq \emptyset$, $\Psi_{\alpha}$ and $\Psi_{\beta}$ coincide.  Of course, this only holds if local triviality is satisfied; in the more general case mentioned before, the definition becomes path dependent and we exclude such possibility for now.  
\begin{itemize} 
\item AXIOM VI: We need a principle of local Lorentz \emph{covariance} since the  dynamics should be covariant with respect to local changes on $M$ in the local reference frames $e_b(x)$ (hence, we need the notion of a quantum spin connection). 
Let me start by saying something about transformation laws in general: the Lorentz transformations depend upon $e^b(x)$ and therefore we write $U(\Lambda(x),e^b(x))$.  Quantum mechanically, all we require for a unitary transformation $T$ from one reference frame to another is that 
$$T(\Gamma(x), \Lambda^{a}_{\,\, b}(x)e^b(x))T(\Lambda(x),e^b(x)) = T(\Gamma(x) \Lambda (x), e^b(x)).$$  This can lead to some different viewpoints; the traditional one being that $$T(\Gamma(x), \Lambda^{a}_{\,\, b}(x)e^b(x)) = U(\Gamma(x))$$ where we then use the ordinary group law.  However, in a fully active point of view, the momentum operators $P^{a}(x, e^b(x))$ should transform covariantly too, that is 
$$P^{a}(x, \Lambda^{b}_{\,\,c}(x)e^c(x)) = \Lambda^{a}_{\, \, b}(x)P^{b}(x,e_c(x))$$ and calculations in Weinberg \cite{Weinberg} reveal that therefore
$$P^{a}(x, \Lambda^{b}_{\,\,c}(x)e^c(x)) = U^{\dag}(\Lambda(x)) P^{a}(x, e^b(x)) U(\Lambda(x))$$ meaning that $U^{\dag}(\Lambda(x), e^b(x))$ is the correct mapping between both modules and not $U(\Lambda(x), e^b(x))$.  This is entirely consistent since now we that 
$$T(\Gamma(x), \Lambda^{a}_{\,\, b}(x)e^b(x) = U^{\dag}(\Lambda(x), e^b(x)) U^{\dag}(\Gamma(x), e^b(x)) U(\Lambda(x), e^b(x))$$ implying that the potential transforms as $$U'(\Lambda(x)_{b}^{\,\,c}e_c(x),x) = U^{\dag}(\Lambda(x),e_b(x))U(e_b(x),x).$$ The dynamics is required to be invariant under these transformations\footnote{From now on, we will drop the reference to the Lorentz frames in the definition of a Lorentz transformation.}; hence, the quantum connection
$A_{a}(e^b(x),x)$ transforms as
$$A'_{a}(e'_b(x),x) = U^{\dag}(\Lambda(x), e^b(x)) A_{a}(e_b(x),x) U(\Lambda(x), e^b(x)) \, - \, i e^{\mu}_{a}(x) \left( \partial_{\mu} U^{\dag}(\Lambda(x), e^b(x))\right) U(\Lambda(x), e^b(x)).$$
This is the second example in this book where two dynamical variables $e^{\mu}_{a}(x)$ and $U(e_b(x),x)$ enter the gauge term of another dynamical variable.  Moreover, the representation of the Poincar\'e group used here is generated by the global algebra and not the local one; we now come to the dynamics.
\item AXIOM VII:  The only way our local particle notions can couple to spacetime is by means of a vielbein $e_{\mu}^{a}(x,v^a)$ and the classical aspect of gravity is fully contained in this symbol and the differential operator $\mathcal{D}_{\mu}(x,v^a)$.  The dynamical content of the theory consists of four pieces (axioms VII till IX): (a) first, we have to construct the equivalent of the Einstein and Spin tensor and put it equal to the expectation values of the local energy momentum tensor $T^{ab}(x,v^a)$ and spin tensor $S^{ab}_{c}(x,v^a)$ in state of the universe (b) second, we have to construct the matter dynamics for the potential $U(e_b(x),x)$ and universal quantum gauge field $A_{\mu}(e_b(x),x)$ (c) third we derive the equations for the time component $e_{0}$ of the tetrad field associated to a local observer.  As a final task, axioms X, XI, XII consist in giving a fully covariant measurement interpretation, remarks on the construction of a theory of consciousness and speculation about the boundary conditions of the universe.  In this axiom, we proceed with (a).  Let us first convince ourselves that all the physics is in the $e_{\mu}^{a}$ and $\mathcal{A}_{\mu}^{B}$ by a counting of degrees of freedom.  On one side, the energy momentum and spin tensor constitute $40$ degrees of freedom.  On the spacetime side, we have the residual symmetries of classical covariance which eliminates $8$ degrees of freedom; both fields have together $16 + 32 = 48$ degrees of freedom which are reduced to $40$ by the residual $8$ dimensional symmetry group\footnote{This counting of degrees of freedom is more delicate than it usually is since $e^{\mu}_a(x,v^c)$ also depends upon $v^c$ and not only $x$.  However, the number of \emph{observable} degrees of freedom are identical on the gravity and matter side, since those all correspond to $v^c = 0$.  There are therefore extra vacuum modes in the gravitational field away from the origin on tangent space which cannot be gauged away.}.  Before we proceed, let us make some remarks about the physical meaning of an action principle, which I haven't met anywhere else in the literature so far.  Usually, what we do is to variate an action principle with respect to the dynamical fields and solve the equations of motion.  In case of vacuum (non-abelian) gauge theories or Einstein gravity, the functional in terms of the gauge fields one arrives at always corresponds to a quantity which is identically conserved.  Since the contracted second Bianchi identities do not give rise to conservation laws due to the presence of torsion, we have resort to the Belinfante Rosenfeld trick and write total derivatives:
$$G^{ab}(x,v^c)  = \partial_{d} K^{dab}(x,v^c)$$ where $K^{dab}(x,v^c)$ is antisymmetric in $d$ and $a$.  As mentioned previously, we can construct a second order theory by using the torsion tensor only; there are three kinds of nonzero torsion coefficients
\begin{eqnarray*}
T_{1}^{abc}(x,v^e) & = & T^{a}_{\,\,\, \mu d}(x,v^e) \eta^{dc} e^{\mu b}(x,v^e) \\*
T_{2}^{abc}(x,v^e) & = & T^{\nu}_{\,\,\, d \mu}(x,v^e) \eta^{db} e^{a}_{\nu}(x,v^e)
e^{\mu c}(x,v^e) \\*
T_{3}^{abc}(x,v^e) & = & T^{a}_{\,\,\, \mu \nu}(x,v^e)e^{\mu b}(x,v^e) e^{\nu c}(x,v^e)  
\end{eqnarray*} and the gauge field $\mathcal{A}_{\mu}^{B}(x,v^e)$ procures another tensor
\begin{eqnarray*}
T_{4}^{abc}(x,v^e) & = & F^{\nu b}_{\,\,\, \,\,\,\, \mu}(x,v^e)e_{\nu}^{a}(x,v^e) e^{\mu c}(x,v^e). 
\end{eqnarray*}
One can write
\begin{eqnarray*}
G^{ab}(x,v^e) & = &  D^{r \, j} \partial_{d} \left( B_{r \, c}^{ \,\,\,\, [d}(x) T_{j}^{|c| a]b}(x,v^e) \right) + E^{r \, j} \partial_{d} \left( B_{r \, c}^{ \,\,\,\, [d}(x) T_{j}^{a]c b}(x,v^e) \right) + \\* & & F^{r \, j} \partial_{d} \left( B_{r \, c}^{ \,\,\,\, [d}(x) T_{j}^{a] b c}(x,v^e) \right) + \, (a \leftrightarrow b) \, +
 K^{i} \partial_d T_{i}^{b[da]}(x,v^e) + \\* & & L^{i} \partial_{d}T_{i}^{[a d] b}(x,v^e) + M^{i} \partial_{d} T_{i}^{[a | b | d]}(x,v^e) + \, \textrm{higher terms in} \\* & & B_{r \, ab}(x), T^{abc}_{j}(x,v^e) \, \textrm{and} \, V^{a}_{r}(x).
\end{eqnarray*} where in principle $K^{i}, L^{i} \ldots$ could be functions of $x$.  Also, one might split the above expressions in a symmetric and anti-symmetric part coming with different coupling functions which we assume to be constants since otherwise the principle of background independence would be violated.  The reader should not be surprised that, even at this level, we have at most $60$ new coupling constants; the reason is of course the symmetry breaking between the spacetime and bundle coordinates which is the very foundation of our geometry\footnote{What we have done here, is to construct first a tensor $K^{[ca]b}$ and contract then with $\partial_{c}$; such tensor is unique up to Hodge terms of the kind $\epsilon^{cade} \partial_{d} R_{e}^{b}$ where $16$ degrees of freedom in $R^a_b$ are available and sixteen equations are satisfied in contrast to what happens for a three dimensional vectorfield, which is the rotation of another one (here one has three degrees of freedom and one equation).  The question then is how to count the degrees of freedom since in the latter case one would say that vectorfields of the form $\epsilon^{abc}\partial_b A_c$ have two of them (due to the zero divergence) while there are three of them available in $A_a$ and the vanishing of the divergence times rotor is an identity.  The answer here depends upon the meaning of the word local since the divergence condition merely eliminates one coordinate dependency of say $A_1$.  Most people would ignore these and say that such functions are global (of codimension $j$), not very physical and do not correspond to propagating degrees of freedom (for example like a constant in a potential does not matter in Newtonian physics).  However, in a higher dimensional theory, merely a dependency upon some extra dimensions could be eliminated, leaving for complete freedom on ``base space''.  This remaining freedom has to be fixed by putting suitable boundary conditions and the upshot of what follows below is that this is indeed the correct strategy for our theory.  Locally, the Hodge terms of the above type determine all tensors $K^{[ca]b}$ satisfying $\partial_c K^{[ca]b} = 0$ and therefore, on the entire tangent bundle and on that ground, one would have to say that only $16$ local degrees of freedom are remaining in $\partial_c K^{[ca]b}$, this is at least so for all $x$ coordinates and in zero'th order of $v^a$ as well.  As mentioned before, there is more freedom than this on base space, which has to be fixed through boundary conditions and is the rationale of our gravitational theory since we aim to look at gravitons as quantum particles and not as classical, locally propating degrees of freedom.  It is so that locally, all two tensors $T_{ab}$ satisfing $\partial^a T_{ab}$ can be written in this way so that no further constraints occur (this is the Hodge dual of the usual result that, locally, any $A$ satisfying $dA = 0$ can be written as $A= dV$ which obviously remains valid as long as one does not mix $x^{\mu}$ and $v^a$ coordinates).}.  Using the torsion tensor only, the spin tensor is a quadratic expression whose linearization around the physical ``vacuum'' $e^{\mu}_a(x,v^e) = \delta^{\mu}_{a}$ and $\mathcal{A}_{\mu}^{B}(x,v^e) = 0$ vanishes identically.  The only way to remedy this is by means of a vector field $V^{a}_{r}(x)$ which is canonically given in our theory by
$$V^{a}_{r}(x) = \mathcal{S} \, \langle \Psi | P^{a}_{r \, 2}(x, e_c(x)) | \Psi \rangle.$$
Likewise, we have a canonical antisymmetric tensor field
$$B^{ab}_{r}(x) = \mathcal{S} \, \langle \Psi | S^{ab}_{r \, 2}(x, e_c(x)) | \Psi \rangle$$ determined by the local matter distribution.  Hence, we have a vector-tensor-gauge theory of gravity which in principle resembles somewhat the situation of the scalar-vector-tensor theory of gravity constructed by Moffat \cite{Moffat}.  Of course, the linearized equations will vanish again when $V^{a}_{r}(x)$ vanishes but this seems to be more like a philosophical point: should the gravity equations be ``deterministic'' when there are ``holes'' in the universe?  Anyhow, the previous considerations result in
\begin{eqnarray*}
H^{ab}_{c}(x,v^e) & = & R^{ij} \partial^{f} \left( T^{d}_{i \,\,\,[f c]}(x,v^e) T_{jd}^{\,\,\,\,\,\,\, [ab]}(x,v^e) \right) + P^{ij} \partial^{f} \left( T_{i \,\,\,\,\,\,\,[c}^{\,\,\, d[a}(x,v^e) T_{j d \,\,\, f]}^{\,\,\,\,\,\,\,b]}(x,v^e) \right) \\*
& & + Q^{ij} \partial^{f} \left( T_{i \,\,\,\,\,\,\, [c}^{[a |d|}(x,v^e) T_{j \,\,\,\,f]d}^{\,\,\, b]}(x,v^e) \right) + V^{ij} \partial^{f} \left( T_{i \,\,\,\,\,\,\, \,\, [c}^{\,\, [a |d|}(x,v^e) T_{jf] \,\,\,\, d}^{\,\,\,\,\, b]}(x,v^e) \right) \\*
& & + Z^{ij}\partial^f \left( T_{i \,\,\,\,\,\,\, [c}^{\,\,\, [a |d|}(x,v^e) T_{j |d|\,\,\,\, f]}^{\,\,\,\,\,\,\, b]}(x,v^e) \right) + \, \textrm{similar terms} \, + \\*
& & \sum_{j=3}^{4} N^{r \, j} \partial^{d} \left( V_{r[d}(x) T_{|j |c]}^{\,\,\,\, \, \, \, \, \, [ab]}(x,v^e) \right) + \sum_{j=3}^4 O^{r \, j} \partial^{d} \left( V_{r[d}(x) T_{|j |\,\,\, c]}^{\,\,\,\, [a \,\, \, \, b]}(x,v^e) \right) +  \\* 
& & \sum_{j=3}^4 P^{r \, j} \partial^{d} \left( V_{r[d}(x) T_{|j |\,\,\, \,\,\,\, c]}^{\,\,\,\, [a b]}(x,v^e) \right) + \, \textrm{higher terms involving} \\* & & V^{a}_{r}(x), B^{ab}_{r}(x) \, \textrm{and} \, T_{j}^{abc}(x,v^e)
\end{eqnarray*} 
where the reader notices that the last three sums go over the indices $3$ and $4$ only.  The reason is that the linearization of those tensors only should depend upon $\mathcal{A}_{\mu}^{B}(x,v^e)$ and therefore vanish identically in the limit for $\mathcal{A}_{\mu}^{B}$ to zero.  Therefore, the first two equations are 
$$G^{ab}(x,v^e) = \mathcal{S} \, \langle \Psi | T_1^{ab}(x,v^e, e_g(x)) | \Psi \rangle$$
and 
$$ H^{ab}_{c}(x,v^e) = \mathcal{S} \, \langle \Psi | S_{1 \, c}^{ab}(x,v^e, e_g(x)) | \Psi \rangle.$$
We are left to determine eight equations of motion to fully fix the vielbein and gauge field.  One might feel that the prime candidate is given by
$$R_{[\mu | \kappa | \nu ]}^{\,\,\,\,\,\,\,\,\, \,\,\,\,\,\, \kappa}(x,v^e) = 0 $$ but this doesn't work since the latter tensor is only first order in the derivatives of the gauge field.  I believe all physics of the matter sector has been imprinted now\footnote{This does not need to be so however; in what follows one can use the antisymmetric part of the energy momentum tensor as well as higher order scalars.} on the gravitational fields; therefore, the remaining equations should be identically equal to zero.  The remaining natural equations are 
\begin{eqnarray*} 0 & = & \alpha^{1} \, \nabla_a(x,v^e) F^{a}_{ \,\,\,\, \mu \nu}(x,v^e) + \alpha^2 \nabla_{\kappa}(x,v^e) \left( e^{\kappa}_{a}(x,v^e) F^{a}_{\,\,\, \mu \nu}(x,v^e) \right) + \\* & & \alpha^3 \nabla^{\kappa}(x,v^e) \left( F^{a}_{\,\,\,\, \kappa [\mu}(x,v^e) e_{\nu]a}(x,v^e) \right) + \alpha^4 \nabla_{b}(x,v^e) \left(e_{a[\mu}(x,v^e)F^{a}_{\,\,\,\, \nu] \kappa}(x,v^e)e^{\kappa b}(x,v^e) \right) + \\* & &  \beta^1 \nabla_{\kappa}(x,v^e) F^{\kappa}_{\,\,\,\, [\mu |a|}(x,v^e)e^{a}_{\nu]}(x,v^e) + \beta^2 \nabla_{a}(x,v^e) \left( F^{\kappa}_{\,\,\,\, [\mu |b |}(x,v^e) e^{a}_{\kappa}(x,v^e) e^{b}_{\nu]}(x,v^e) \right) + \\*  & & \textrm{similar terms}.  
\end{eqnarray*} Furthermore, we insist that the following traces are covariantly conserved:
\begin{eqnarray*}
0 & =  & \gamma^1 \nabla^{\mu}(x,v^e) \left( F_{\,\,\,\, \mu \nu}^{a}(x,v^e) e^{\nu}_{a}(x,v^e) \right) + \gamma^2 \nabla^{b}(x,v^e) \left( F^{a}_{\,\,\,\, \mu \nu}(x,v^e) e^{\mu}_{b}(x,v^e)e^{\nu}_{a}(x,v^e) \right) \\*
0 & = & \delta^1 \nabla^{\mu}(x,v^e) \left( F_{\, \,\,\, \mu a}^{\nu}(x,v^e) e_{\nu}^{a}(x,v^e) \right) + \delta^2 \nabla^{b}(x,v^e) \left( F^{\nu}_{\,\,\,\, \mu a}(x,v^e)e^{a}_{\nu}(x,v^e) e^{\mu}_{b}(x,v^e) \right) + \\* & & \textrm{all other contractions}.
\end{eqnarray*}  All in all, we have of the order of $100$ free parameters which is a rather modest landscape.  This is a simple consequence of the symmetry breaking which took place on $TM$ and it would be utterly naive to expect a smaller number.  We proceed now by investigating the mathematical structure of these equations as well as the emergence of the correct Newtonian limit.             
\end{itemize}
The most important comment here regards the implications of what it means for the expectation values of the energy momentum tensors and spin tensors to be a total divergence.  Indeed, it automatically implies the existence of singularities (point like, string like or higer dimensional) or very nontrivial asymptotic behavior of the gravitational field in tangent space itself where some (finite) asymptotic boundary term exists at a three sphere at infinity\footnote{We have written $\mathcal{S} \langle \Psi | T_{ab} | \Psi \rangle$ as $\partial^{d} K_{\left[ da \right]b}$.  Because of the assymetry of the $K$ tensor, integrating $T_{0b}$ over a four volume in $v^a$ space allows one to apply the three dimensional version of Green's theorem on the slices of constant $v^0$ since the $v^0$ partial derivative is of no relevance in $\partial^{d} K_{\left[ da \right]b}$.  Hence, the whole four dimensional integral can be reduced to a three dimensional one.}.  It is clear that this very formulation is an implementation of the holographic principle where all information of energy momentum and spin of matter is stored on a two surface\footnote{The precise sense in which we employ the name ``holographic'' here is that the total integrated energy momentum and spin within a certain (four or three) volume on $T_xM$ is determined by some integrated quantity over the boundaries.  Of course, it is not sufficient to specify one screen, all screens are necessary unlike what Maldacena has in mind.  The one screen version however does apply, for certain trivial topologies, in the quantum theory to be developed.}.  Curiously enough, we shall see that the theory of matter obeys precisely the same principle on $M$ instead of on $TM$; this reflects the inner consistency of the theory.  It would be too much to ask to perform a complete study of the initial value problem of the full non-linear equations of motion at this moment; such enterprise would be considerably more difficult than the calculations performed in \cite{Hawking1} and those were already rather involving.  We constrain ourselves here to studying the linearized equations and comment upon the latter issue in this respect.  To reassure ourselves that the above equations are a priori not a pile of crap, we first make a quick calculation regarding the Newtonian limit.  One should not expect to get out linearized Einstein gravity since the equations of motion are ultrahyperbolic (but in which variables?). \\* \\*
Therefore, put $e^{\mu}_{a}(x,v^e) = \delta^{\mu}_{a} + h^{\mu}_{a}(x,v^e)$, where the last term is supposed to be small compared to unity.  It is helpful to introduce the following notation: $$\partial_{\widetilde{b}} = \delta^{\mu}_{b} \partial_{\mu}$$ and we shall in general use the index notation $\widetilde{b}$ expressing tensorial properties with respect to $x^{b}$.  Some elementary calculations reveal that
\begin{eqnarray*}
T_{1}^{abc}(x,v^e) & = & \frac{1}{2} \left( \partial^{\widetilde{b}} h^{\widetilde{a}c}(x,v^e) - \partial^{\widetilde{b}}h^{\widetilde{c}a}(x,v^e)
- \partial^{\widetilde{c}}h^{\widetilde{a}b}(x,v^e) + \partial^{\widetilde{a}}h^{\widetilde{c}b}(x,v^e) \right) + \\* & & \frac{1}{2} \left( - \partial^{\widetilde{c}}h^{\widetilde{b}a}(x,v^e) + \partial^{\widetilde{a}}h^{\widetilde{b}c}(x,v^e) \right) \\*
T_{2}^{abc}(x,v^e) & = & - \partial^{b} h^{\widetilde{a}c}(x,v^e) \\*
T_{3}^{abc}(x,v^e) & = & 0.  
\end{eqnarray*} Making the supplementary assumption that only $h^{\widetilde{0}0}(x,v^e)$ is nonzero, it follows that:
\begin{eqnarray*}
G^{0k}(x,v^e) & = & 0 \\*
G^{kl}(x,v^e) & = & 0 \\*
G^{k0}(x,v^e) & = & \frac{1}{2} \partial^{\widetilde{k}} \partial_{0} h^{\widetilde{0}0}(x,v^e) \left(- K^{1} + L^{1} + 2M^{1} \right) + \frac{1}{2} \partial_{0}\partial^{k} h^{\widetilde{0}0}(x,v^e) \left(L^{2} + K^{2} \right) \\*
G^{00}(x,v^e) & = & \frac{1}{2} \partial_{k}\partial^{\widetilde{k}}h^{\widetilde{0}0}(x,v^e)\left(K^{1} - L^{1} - 2M^{1} \right) - \frac{1}{2} \partial_{k} \partial^{k}h^{\widetilde{0}0}(x,v^e) \left( K^{2} + L^{2} \right).
\end{eqnarray*}  It is an \emph{amazing} feature that the structure of $G^{00}(x,v^e)$ is that of an ultrahyperbolic equation of signature $(3,3)$ in the spatial variables only.  ``Time'' is completely eliminated from these equations and therefore this theory has no gravitational waves which is entirely consistent with our previous remark that gravitons belong to the quantum sector.  Insisting upon the correct Newtonian limit allows us to fix some relations between the constants $K^{i}, L^{i}$ and $M^{i}$.  To accomplish that, we need to propose the correct energy-momentum and spin tensor; the only nonzero component is given by     
\begin{eqnarray*}
T^{00}(x,v^e) & = & m \delta^{3} (x^k + v^k - a^{k}(t)).
\end{eqnarray*}
This formula must be puzzling for the reader at first sight, but I will show now that it has to be like that in the light of AXIOM II.  First of all, the point particle has a trajectory in space given by $a^k(t)$ since observation always takes place at $v = 0$.  Now, for $x^k \neq a^k(t)$ the observer at $x$ with vielbein $\delta^{\mu}_a$ will imagine the particle to be at $v^k = a^k(t) - x^k$ and will ascribe to it the above energy momentum on $TM_{x}$.  Another observation has to be made regarding the use of $a^k (t)$ instead of $a^k(t + v^0)$ since the last formula would allow the observer at $x$ to correctly predict the future trajectory of the point particle, while in the former case the particle is imagined to remain stationary.  The point is that the observer at $x$ cannot predict the future trajectory of the particle since (a) the global gravitational field is unknown to him (b) he is unaware of the interactions of the particle with other particles since he cannot infer anything about the state of the gauge field (because photons do not interact with one and another).  The last reason is also the ``philosophical'' stance behind the assumption of a free theory on $TM_x$: nature basically doesn't give us any information about the interaction field, we can only \emph{retrodict} the existence of it by observing that things don't move on a straight line.  All this might be hard to swallow, it is as if the ``personal reality'' of an observer has an (extremely tiny) influence on the trajectories of planets: this is a further
deepening of what people consider to be quantum nonlocality.  Indeed, this nonlocal gravitational interaction is \emph{not} measurable and there is no conflict with quantum physics as we know it.   It is just that this extra level of classical reality allows for this kind of things to happen.  Anyhow, let us return to the computations and let the philosophy to the philosophers.  \\* \\*
Putting $h^{\widetilde{0}0}(x^k,v^k,t)$, one notices that $G^{\widetilde{k}0}(x,v^e)$ vanishes identically and the remaining equation becomes
$$\frac{1}{2}\left(K^1 - L^1 - 2M^1 \right) \partial_j \partial^{\widetilde{j}}h^{\widetilde{0}0}(x^k,v^k,t) \, - \,\frac{1}{2}\left( K^2 + L^2 \right) \partial_j \partial^j h^{\widetilde{0}0}(x^k,v^k,t) = m c^2 \delta^{3}(x^k + v^k - a^k(t)).$$
Further restriction to $h^{\widetilde{0}0}(x^k + v^k,t)$ implies that, since
$$ \Delta \left( - \frac{1}{4 \pi r} \right) = \delta^{3}(\vec{r}) $$
one obtains
$$K^1 -L^1 - 2M^1 - K^2 - L^2 = \frac{c^4}{2 \pi G}.$$
Indeed, the full solution is given by 
$$h^{\widetilde{0}0}(x^k + v^k,t) = - \frac{Gm}{|x^k + v^k - a^k(t)|c^2}$$ as it should.  How should the above equations be interpreted?  Clearly, we cannot think of them as ``evolution'' equations since $\partial_t$ is entirely absent; therefore, we ought to think about them as ultrahyperbolic boundary value problems.  So, philosophically speaking, we have evolving matter within the universe, but the universe itself is completely ``timeless''.  One notices that there is a functional degree of freedom left in the four boundary value problems; indeed $f^{\widetilde{0}0}(\alpha v^k + x^k, t, v^0)$ where
$$\alpha = \frac{K^2 + L^2}{K^1 - L^1 - 2M^1} \neq 1$$ solves $G^{k0}(x,v^e) = 0 = G^{00}(x,v^e)$.  The only constraint is that this function vanishes on the boundary of the eight dimensional universe and this suffices to put the above function to zero.  Indeed, for $\alpha \neq 0$, choose $v^r = \frac{-1}{\alpha}(x^r - z^r)$ for some $r$ and take the limit for $x^r$ to infinity.  Then, one concludes that $$f^{\widetilde{0}0}(z^k, t,v^0) = 0$$ for all variables\footnote{All this holds of course also when $\alpha = 0$.}.  We conclude that the boundary value problems are well posed and do not appear to need such a subtle treatment as the intial value problem\footnote{I again thank S. Nobbenhuis for useful conversations here, usually with a good glass of wine.  His PhD thesis went about the cosmological constant problem and a particular ansatz consisted indeed in going over from inital value formulations to boundary value problems, see \cite{Nobbenhuis}.}.  We look now for further observational evidence and reconstruct some class of ``black-hole'' solutions\footnote{The reader will find out soon why I have put the term black hole between quotation marks.}.  \\* \\*
Again, for simplicity, we switch off $\mathcal{A}_{\mu}^{B}(x,v^e)$ and choose spherical coordinates on spacetime.  The only nonzero components of the vielbein are assumed to be
\begin{eqnarray*}
e^{0}_{t}(x,v^e) & = & f^{0}(x,v^e) \\*
e^{1}_{r}(x,v^e) & = & f^{1}(x,v^e) \\*
e^{2}_{\theta}(x,v^e) & = & f^{2}(x,v^e) \\*
e^{3}_{\phi}(x,v^e) & = & f^{3}(x,v^e)
\end{eqnarray*} where $x = (t, r,\theta,\psi)$.  Some elementary computations reveal that the nonzero ``torsion'' coefficients are given by
\begin{eqnarray*}
T_{1}^{abc}(x,v^e) & = & \frac{1}{f^{b}f^{c}} \partial^{\widetilde{c}} f^{a} \eta^{ab} - \frac{1}{f^{a}f^{c}} \partial^{\widetilde{a}} f^{b} \eta^{bc} \\*
T_{2}^{abc}(x,v^e) & = & \frac{1}{f^c} \partial^{b} f^{a} \eta^{ac}.
\end{eqnarray*}   
Hence, the full equations of motions are
\begin{eqnarray*}
0 & = & \frac{1}{2} K^1 \left( \partial^b \left( \frac{1}{f^a f^d} \partial^{\widetilde{a}} f^b \right) - \partial_{d} \left( \frac{1}{f^a f^d} \partial^{\widetilde{d}} f^b \right) \eta^{ab} \right) + \\*
& & \frac{1}{2}(K^2 + L^2) \left( \partial_d \left( \frac{1}{f^a} \partial^d f^b  \right) \eta^{ab} - \partial^b \left( \frac{1}{f^b} \partial^a f^b \right) \right) + \\*
& & \frac{1}{2}(L^1 + 2M^{1}) \left( \partial_d \left( \frac{1}{f^d f^b} \partial^{\widetilde{d}} f^a \right) \eta^{ab} - \partial^b \left( \frac{1}{f^a f^b} \partial^{\widetilde{a}} f^b \right) \right).
\end{eqnarray*}
Since the purpose here is not a full classification, we make our live somewhat easy and put 
\begin{eqnarray*}
0 & = & K^2 + L^2 \\*
0 & = & L^1 + 2M^1 
\end{eqnarray*} which reduces our previous restriction to 
$$K^1 = \frac{c^4}{2 \pi G}.$$
Furthermore, we look for time independent solutions, that is $f^a(r,\theta,\phi,v^1,v^2,v^3)$ implying that $G^{k0}(x,v^e) = 0 = G^{0k}(x,v^e)$.  The remaining $10$ equations imply that
\begin{eqnarray*} 
\frac{\partial_r \textrm{ln} \, f^2}{f^1} & = & g(r,\theta,\phi ,v^1 - v^3) + h^{2}_{1}(r, \theta , \phi , v^3) \\*
\frac{\partial_{\phi} \textrm{ln} \, f^2}{f^3} & = & g(r, \theta, \phi , v^1 - v^3) + h^{2}_{3}(r, \theta , \phi, v^1) \\*
\frac{\partial_{\theta} \textrm{ln} \, f^1}{f^2} & = & h(r, \theta , \phi , v^2 - v^3) + h^{1}_{3}(r, \theta , \phi , v^3) \\*
\frac{\partial_{\phi} \textrm{ln} \, f^1}{f^3} & = & h(r, \theta , \phi , v^2 - v^3) + h^{1}_{3}(r, \theta , \phi ,v^2) \\*
\frac{\partial_r \textrm{ln} \, f^3}{f^1} & = & p(r, \theta , \phi , v^1 - v^2) + h^{3}_{1}(r, \theta, \phi, v^2) \\*
\frac{\partial_{\theta} \textrm{ln}\, f^3}{f^2} & = & p(r, \theta , \phi, v^1 - v^2) + h^{3}_{2}(r, \theta , \phi, v^1) \\*
\partial_j \left( \frac{1}{f^0 f^j} \partial^{\widetilde{j}} f^0 \right) & = & 0.
\end{eqnarray*}  We partially fix the boundary conditions by imposing that in all limits where at least one of the $r,v^1,v^2,v^3$ goes to infinity, the functions $f^a$ reduce to 
\begin{eqnarray*}
f^0 & = & 1 \\*
f^1 & = & 1 \\*
f^2 & = & r \\*
f^3 & = & r \sin(\theta) 
\end{eqnarray*}
and their derivatives reduce to the derivatives of the right hand side as well.  Taken together with the previous formulae, this results in 
\begin{eqnarray*}
f^1 & = & f^1(r,v^1,v^2,v^3) \\*
f^2 & = & f^2(r,\theta ,v^1,v^2,v^3) \\*
\partial_{r} \textrm{ln}\, f^2 & = & \frac{f^1}{r} \\*
\partial_r \textrm{ln}\, f^3 & = & \frac{f^1}{r} \\*
\partial_{\theta} \textrm{ln}\, f^3 & = & \frac{f^2 \cot(\theta)}{r} 
\end{eqnarray*} taken together with the constraint equation for $f^0$.  This results furthermore in 
\begin{eqnarray*}
f^{1}(s,v^1,v^2,v^3) & = & 1 \\*
f^{2}(r, \theta , v^1,v^2,v^3) & = & r \widetilde{g}(\theta ,v^1,v^2,v^3) \\*
f^{3}(r, \theta, \phi, v^1,v^2,v^3) & = & r \widetilde{h}(\theta , \phi , v^1,v^2,v^3) \\*
\frac{\partial_{\theta} \widetilde{h}(\theta , \phi , v^1,v^2, v^3)}{\widetilde{h}(\theta , \phi, v^1,v^2,v^3)} & = & \widetilde{g}(\theta , v^1,v^2,v^3) \cot(\theta).
\end{eqnarray*}  This last condition implies
\begin{eqnarray*}
\widetilde{h}(\theta , \phi , v^1,v^2,v^3) = \widetilde{f}(\phi ,v^1,v^2,v^3) e^{\int_{0}^{\theta} \widetilde{g}(\alpha, v^1,v^2,v^3) \cot(\alpha) d\alpha}
\end{eqnarray*} and we further restrict by putting $\widetilde{f}(\phi ,v^1,v^2,v^3) = 1 = \widetilde{g}(\theta ,v^1,v^2,v^3)$ which may be thought of as a partial fixing of the boundary conditions at future and past infinity.  Hence, the constraint equation for $f^{0}$ becomes
\begin{eqnarray*}
\partial_1 \partial_r g + \frac{1}{r} \partial_2 \partial_{\theta} g + \frac{1}{r \sin(\theta)} \partial_3 \partial_{\phi} g = 0
\end{eqnarray*} where $g = \textrm{ln} \, f^{0}$.  Assuming that $g(r, \theta,v^1,v^2,v^3)$ one immediately notices that a class of solutions is given by
$$g(r, \theta ,v^1 ,v^2,v^3) = y(\textrm{ln}(r) + v^2,v^3)z(\theta - v^1,v^3)$$ whenever the appropriate boundary conditions are satisfied.  This leads to a very broad class of solutions which does not contain the Kerr or Schwarzschild spacetimes.  Actually, from these considerations, it is already clear that no conventional black holes can be constructed in \emph{this} way (and therefore, well known black hole thermodynamics fails in nonlocal, non-metric theories of gravity at least when one insists upon asymptotically flat solutions\footnote{The small caveat here is that we also demanded that the derivatives of the $f^j$ reduced to the derivatives of the flat spacetime substitutes towards spacelike infinity.}).  Indeed, all coefficients of the metric apart from (possibly) $f^{0}$ behave completely regular so it is just \emph{impossible} for a semi-permeable event horizon to form.  This does not imply that no dark objects exist in this theory, on the contrary, but either (a) they are not perfectly dark or (b) they are, but they are not permeable from \emph{both} sides.  That is, some curvature scalar blows up at the event horizon.  The last class of black holes would be physically very distinct from what we are used to in Einstein gravity and gives rise to naked singular surfaces.  I do not see any \emph{physical} problem in this, since the big bang appears to be exactly of that nature.  Also, it is clear that the ``no-hair'' conjecture is by no means satisfied under reasonable conditions.  We finish this preliminary study by constructing one particular solution of class (b) and calculate a scalar curvature which blows up at the event horizon.  For example,
\begin{eqnarray*}
f^{0}(r, \theta ,v^1,v^2,v^3) & = & e^{- \frac{b}{(\textrm{ln}(r) + v^2 - a)}e^{- c(\theta - v^1)^2 - d(v^3)^2}} \\*
f^{1} & = & 1 \\*
f^{2}(r) & = & r \\*
f^{3}(r, \theta) & = & r \sin(\theta)
\end{eqnarray*} where $b,c,d > 0$.  Hence, for $\textrm{ln}(r) + v^2 - a = 0$ the spacetime volume vanishes and therefore some Ricci scalar must blow up to infinity.
  Before we show by a simple computation that this is indeed the case, let me stress some interesting qualitative feature of this solution class: gravity is actually stronger than it is in the Newtonian theory.  Indeed, putting $a = \textrm{ln}(r_0)$ with $r_0$ some reference length, $b = GM/(c^2r_0)$ and $c \ll 1$ some very small positive number, then in first order
$$f^0(r, \theta, 0,0,0) \sim 1 - \frac{GM}{c^2r_0 \, \textrm{ln}\left(\frac{r}{r_0} \right)}$$ and therefore the gravitational force is given by 
$$\vec{F} =  - \frac{GM}{r_0 \textrm{ln}^2\left( \frac{r}{r_0} \right)r}\vec{e}_{r}.$$  Therefore, $\vec{F}$ is stronger than the usual gravitational force since $\textrm{ln}^2(x) < x$ for $x > 1$.  On the other hand, on short distances $r \ll r_0$ the divergence in the force becomes much softer than is usually the case.  This makes our theory a prime candidate for a gravitational explanation of ``dark matter'' without ``matter''.  Obviously, we also find the ordinary Newtonian gravitational fields (with tiny angular corrections) as solutions to the equations of motion by putting
$$g(\textrm{ln}(r) + v^2) = \textrm{ln}\left( 1 - \frac{GM}{c^2r_0 e^{\textrm{ln}\left( \frac{r}{r_0} \right) + v^2}} \right).$$  This wide diversity of different axisymmetric solutions allows for plenty of interesting new phenomenology and I conjecture it actually suffices to solve all problems Einstein gravity faces.  As a definite step in that direction, it would be nice to see whether the Pioneer anomaly might get a natural explanation within this framework.  People might naively think that we just destroyed the simplicity of Einstein's theory because we have plenty of more constants and plenty of more solutions under the same conditions.  I sharply disagree with that, a larger symmetry group for our higher dimensional geometry would actually have lead to a drastic increase in the complexity of the initial value problem (such as is faced by string theory) and moreover, one would have to put in by hand that tangent space is flat.  Einstein's theory simply appears to be too limited to explain the observed phenomena and any generalization beyond four dimensions which is not constructed in the way our theory is, is bound to be infinitely more complicated.  These $100$ or so free parameters are an extremly tiny price to pay.  However, it is legitimate to wonder why Newton's law appears to hold so well for most celestial bodies up to some small and large distance scale.  We do not only have to figure out anymore why space is almost flat but also why gravity behaves as $1/r^2$ between distance scales of a millimeter and $10^{15}$ kilometer.  Below a millimeter, it might very well be that gravity switches off gradually to vanish entirely at zero distance: these are legitimate solutions of our constraint equations as the reader may easily verify.  At large distances, gravity would have to be stronger than Newtonian theory predicts, so it was rather obvious that any theory capable of such predictions had to be nonlocal (but in contrast to relativistic MOND theories, it does not require a timelike vectorfield \cite{Bekenstein}).  Moreover, the theory is constructed from fewer and more basic principles than MOND is: it is simply so that the boundary value problem simply \emph{appears} somewhat more complicated than it already was.  Indeed, starting from flat spherical coordinates in three space, one might constrain $\partial_r \left( \textrm{ln} f^0(r, \theta , v^1,0,v^3) \right)$ to be a globally well defined Laurent series in $r$ going to zero for $r$ going to zero and infinity.  At that point, one would need to invent a scaling relation between the different coefficients such that Newtonian gravity applies where it is supposed to do so.  Note that the ultra long distance behavior of pure MOND is \emph{not} allowed in our theory since the boundary conditions for $f^0$ would not be satisfied.  Therefore, there are at least three distance scales in cosmology: (a) a short distance cutoff where Newtonian gravity gets modified (b) a long distance scale where a $1/r$ force takes over and (c) an ultralong distance scale where the latter gets switched off.  Hence, our theory has the same virtues as MOND in explaining the galaxy rotation curves but can be corrected for motions between galaxies where MOND appearantly fails.  This is not just a ``small detail'' but a definite indication that our ideas of promoting $TM$ as the basic arena for physics is the correct thing to do.  We gather further observational and theoretical evidence for that later on.  \\* \\*  For general vielbeins of the above type, the nonzero connection coefficients are computed to be
\begin{eqnarray*}
\Gamma^{\widetilde{0}}_{\,\,\, \widetilde{k} \widetilde{0}} & = & - \frac{1}{f^0} \partial_{\widetilde{k}} f^0 \\*
\Gamma^{\widetilde{j}}_{\,\,\, \widetilde{0}\widetilde{0}} & = & - \frac{1}{\left( f^j \right)^2} \left( \partial_{\widetilde{j}}f^0 \right) f^0 \\*
\Gamma^{\widetilde{j}}_{\,\,\, \widetilde{k} \widetilde{l}} & = & - \delta^j_k \frac{1}{f^j} \partial_{\widetilde{l}}f^j - \delta^j_l \frac{1}{f^j} \partial_{\widetilde{k}}f^j + \frac{1}{\left( f^j \right)^2} \delta_{kl} \left( \partial_{\widetilde{j}} f^k \right) f^k.
 \end{eqnarray*}
In our specific case, this becomes 
\begin{eqnarray*}
\Gamma^{\widetilde{0}}_{\,\,\, \widetilde{1} \widetilde{0}} & = & - \frac{b}{r \left( \textrm{ln}\left( \frac{r}{r_0} \right) + v^2 \right)^2} \\*
\Gamma^{\widetilde{0}}_{\,\,\, \widetilde{2} \widetilde{0}} & = & - \frac{2bc(\theta - v^1)e^{- c(\theta - v^1)^2 - d(v^3)^2}}{\left( \textrm{ln} \left( \frac{r}{r_0}\right) + v^2 \right)} \\*
\Gamma^{\widetilde{1}}_{\,\,\, \widetilde{0}\widetilde{0}} & = & - \frac{b}{r\left( \textrm{ln}\left( \frac{r}{r_0} \right) + v^2 \right)^2}e^{- \frac{2b}{\textrm{ln}\left( \frac{r}{r_0} \right) + v^2} e^{- c(\theta - v^1)^2 - d(v^3)^2}} \end{eqnarray*}
and
\begin{eqnarray*}
\Gamma^{\widetilde{2}}_{\,\,\, \widetilde{0}\widetilde{0}} & = & - \frac{2bc(\theta - v^1)e^{-c(\theta - v^1)^2 - d(v^3)^2}e^{- \frac{2b}{\textrm{ln}\left( \frac{r}{r_0} \right) + v^2} e^{- c(\theta - v^1)^2 - d(v^3)^2}}}{r^2\left( \textrm{ln}\left( \frac{r}{r_0} \right) + v^2 \right)} \\*
\Gamma^{\widetilde{1}}_{\,\,\, \widetilde{2} \widetilde{2}} & = & r \\*
\Gamma^{\widetilde{2}}_{\,\,\, \widetilde{1} \widetilde{2}} & = & - \frac{1}{r} \\*
\Gamma^{\widetilde{1}}_{\,\,\, \widetilde{3} \widetilde{3}} & = & r \sin^2(\theta) \\*
\Gamma^{\widetilde{2}}_{\,\,\, \widetilde{3} \widetilde{3}} & = & \cos(\theta) \sin(\theta) \\*
\Gamma^{\widetilde{3}}_{\,\,\, \widetilde{1} \widetilde{3}} & = & - \frac{1}{r} \\*
\Gamma^{\widetilde{3}}_{\,\,\, \widetilde{2} \widetilde{3}} & = & - \cot(\theta). \\* \end{eqnarray*}
The reader may verify that the Ricci scalar blows up at $v^2 = 0$ and $r= r_0$; in particular, it is easy to see that the term $g^{\widetilde{0}\widetilde{0}}R_{\widetilde{0}\widetilde{1}\widetilde{0}}^{\,\,\,\,\,\,\,\,\,\, \widetilde{1}}$ behaves in such way. \\* \\*  We examine now the interesting mathematical structure of the full linearized equations of motion.  The reader notices that we have put the cosmological term in $G^{ab}(x,v^e)$ to zero as well as other ``constant'' terms on $TM$; the reason herefore is being explained in AXIOM XI.  We now linearize the relevant expressions around $e^{\mu}_{a}(x,v^e) = \delta^{\mu}_{a}$ and $\mathcal{A}_{\mu}^B(x,v^e) = 0$; one calculates that
\begin{eqnarray*}
e^{\mu c}(x,v^e) e^{\nu a}(x,v^e) e^{b}_{\kappa}(x,v^e) \Gamma^{\kappa}_{\mu \nu}(x,v^e) & = & \frac{1}{2} \left( \partial^{\widetilde{c}} h^{\widetilde{b}a} + \partial^{\widetilde{a}}h^{\widetilde{b}c}(x,v^e) + \partial^{\widetilde{c}}h^{\widetilde{a}b}(x,v^e) \right) \\*
&  & + \frac{1}{2}\left( \partial^{\widetilde{a}} h^{\widetilde{c}b}(x,v^e) - \partial^{\widetilde{b}}h^{\widetilde{a}c}(x,v^e) - \partial^{\widetilde{b}}h^{\widetilde{c}a}(x,v^e) \right) \\*
& & - \left( \partial^{[\widetilde{a}} \mathcal{A}^{|\widetilde{b}| \widetilde{c}]}(x,v^e) - \partial^{[\widetilde{b}}\mathcal{A}^{|\widetilde{a}| \widetilde{c}]}(x,v^e) + \partial^{[\widetilde{a}} \mathcal{A}^{|\widetilde{c}| \widetilde{b}]}(x,v^e) \right) 
\end{eqnarray*} where $\mathcal{A}^{\widetilde{a}\widetilde{b}}(x,v^e) = \eta^{b \mu} \delta^{a}_{\nu} \mathcal{A}_{\mu}^{\nu}(x,v^e)$ and we have abused notation slightly by writing down $\eta^{b \mu}$.  Also, 
$$2 F_{\mu d}^a(x,v^e) = - \partial_d \mathcal{A}_{\mu}^a(x,v^e)$$
and therefore  
\begin{eqnarray*}
T_{1}^{abc}(x,v^e) & = & \frac{1}{2}\left( \partial^{\widetilde{b}}h^{\widetilde{a}c}(x,v^e) - \partial^{\widetilde{c}}h^{\widetilde{a}b}(x,v^e) - \partial^{\widetilde{b}}h^{\widetilde{c}a}(x,v^e) \right) \\*
& & \frac{1}{2} \left( - \partial^{\widetilde{c}}h^{\widetilde{b}a}(x,v^e) + \partial^{\widetilde{a}}h^{\widetilde{c}b}(x,v^e) + \partial^{\widetilde{a}}h^{\widetilde{b}c}(x,v^e) \right) \\*
& & + \left( \partial^{[\widetilde{c}}\mathcal{A}^{|\widetilde{a}|\widetilde{b}]}(x,v^e) - \partial^{[\widetilde{a}} \mathcal{A}^{|\widetilde{c}|\widetilde{b}]}(x,v^e) + \partial^{[\widetilde{c}}\mathcal{A}^{|\widetilde{b}|\widetilde{a}]}(x,v^e) \right) \\*
& & - \partial^{c} \mathcal{A}^{a \widetilde{b}}(x,v^e)   
\end{eqnarray*}
where $\mathcal{A}^{a \widetilde{b}}(x,v^e)  = e^{\mu b}(x,v^e) \mathcal{A}_{\mu}^{a}(x,v^e)$.  Furthermore,
\begin{eqnarray*}
T_{2}^{abc}(x,v^e) & = & - \partial^b h^{\widetilde{a}c}(x,v^e) + \partial^b \mathcal{A}^{\widetilde{a}\widetilde{c}}(x,v^e)
\end{eqnarray*} and 
\begin{eqnarray*}
T_{3}^{abc}(x,v^e) & = & 2 \partial^{[\widetilde{b}} \mathcal{A}^{|a| \widetilde{c}]}(x,v^e).
\end{eqnarray*} Finally, 
$$T_{4}^{abc}(x,v^e) = \partial^b \mathcal{A}^{\widetilde{a}\widetilde{c}}(x,v^e)$$
and we are now in position to write out the ``full'' equations of motion, where we ignore for now the $B^{ab}(x)$ field.  $G^{ab}(x,v^e)$ is given by
\begin{eqnarray*}
G^{ab}(x,v^e) & = & \partial_d \partial^{[d}h^{|\widetilde{b}|a]}(x,v^e) \left( - K^2 \right) + \partial_d \partial^{[\widetilde{d}}h^{|\widetilde{b}|a]}(x,v^e)\left( K^1 - L^1 - M^1 \right) + \\*
& & \partial_{d}\partial^{[d}h^{\widetilde{a}]b}(x,v^e) \left( - L^2 \right) + 
 \partial_{d}\partial^{[\widetilde{d}}h^{\widetilde{a}]b}(x,v^e) \left( - M^1 \right) + \\* & &
\partial_d \partial^b h^{[\widetilde{d}a]}(x,v^e) \left( M^2 \right) +  
\partial_d \partial^{\widetilde{b}} h^{[\widetilde{d}a]}(x,v^e) \left( - M^1 \right) + \\* & &
\partial_d \partial^{[d} \mathcal{A}^{|\widetilde{b}|\widetilde{a}]}(x,v^e)\left( K^2 + K^4 \right) + \partial_d \partial^{[d} \mathcal{A}^{\widetilde{a}]\widetilde{b}}(x,v^e) \left( L^2 + L^4 \right) + \\* & & \partial_d \partial^{[\widetilde{d}}\mathcal{A}^{|b|\widetilde{a}]}(x,v^e) \left(2K^3 \right) + \partial_d \partial^{[\widetilde{d}} \mathcal{A}^{a]\widetilde{b}}(x,v^e) \left( L^3 - M^3 \right) + \\* & & \partial_d \partial^{[\widetilde{d}} \mathcal{A}^{|\widetilde{b}|\widetilde{a}]}(x,v^e) \left( - K^1 + M^1 \right) + \partial_d \partial^{[d}\mathcal{A}^{|b| \widetilde{a}]}(x,v^e) \left( K^1 \right) + \\* & & \partial_d \partial^b \mathcal{A}^{[a \widetilde{d}]}(x,v^e) \left( - L^1 \right) + \partial_d \partial^{[\widetilde{d}} \mathcal{A}^{|\widetilde{b}| \widetilde{a}]}(x,v^e) \left( L^1 \right) + \\* & & \partial_{d} \partial^{\widetilde{b}} \mathcal{A}^{[d \widetilde{a}]}(x,v^e) \left( L^3 - M^3 \right) + \partial_d \partial^b \mathcal{A}^{[\widetilde{a} \widetilde{d}]}(x,v^e) \left( M^2 + M^4 \right) + \\* & &  \partial_d \partial^{\widetilde{b}} \mathcal{A}^{[\widetilde{d} \widetilde{a}]}(x,v^e) \left( \frac{M^1}{2}\right) + \partial_d \partial^{[\widetilde{d}} \mathcal{A}^{[\widetilde{a}] \widetilde{b}}(x,v^e) \left( M^1\right) + \partial_d \partial^{[d}\mathcal{A}^{a]\widetilde{b}}(x,v^e) \left( - M^1 \right)  
\end{eqnarray*} and the linearization of $H^{ab}_{c}(x,v^e)$ reads
\begin{eqnarray*}
H^{ab}_{c}(x,v^e) & = &  N^3 \partial^d \left( V_{[d}(x) \partial^{[\widetilde{a}}\mathcal{A}^{|g| \widetilde{b}]}(x,v^e) \eta_{c]g} \right) + N^4 \partial^d \left( V_{[d}(x) \partial^{[a} \mathcal{A}^{|\widetilde{g}| \widetilde{b}]}(x,v^e) \eta_{c]g} \right) + \\*
& & O^3 \partial^d \left( V_{[d}(x)\left( \partial_{\widetilde{c}]}\mathcal{A}^{[a \widetilde{b}]}(x,v^e) + \partial^{[\widetilde{a}}\mathcal{A}^{b]\widetilde{g}}(x,v^e) \eta_{c]g} \right) \right) + O^4 \partial^d \left( V_{[d}(x) \partial_{c]}\mathcal{A}^{[\widetilde{a}\widetilde{b}]}(x,v^e) \right) + \\*
& & P^3 \partial^d \left( V_{[d}(x) \left( \partial^{[\widetilde{b}}\mathcal{A}^{a]\widetilde{g}}(x,v^e) \eta_{c]g} - \partial_{\widetilde{c}]} \mathcal{A}^{[a\widetilde{b}]}(x,v^e) \right) \right) - P^4 \partial^d \left( V_{[d}(x) \partial^{[a} \mathcal{A}^{\widetilde{d}]\widetilde{g}}(x,v^e) \eta_{c]g} \right).
\end{eqnarray*}  The remaining eight equations are given by
\begin{eqnarray*}
0 & = & \alpha^1 \partial_a \partial_{[\mu} \mathcal{A}_{\nu]}^{a}(x,v^e) + \alpha^2 \partial_{\widetilde{a}} \partial_{[\mu} \mathcal{A}_{\nu]}^{a}(x,v^e) + \\*
& & \frac{\alpha^3}{2} \partial^{\kappa} \left(  \partial_{[\kappa}\mathcal{A}_{\mu]}^{a}(x,v^e) \eta_{\nu a} - \partial_{[\kappa} \mathcal{A}_{\nu]}^{a}(x,v^e) \eta_{\mu a} \right) + \frac{\alpha^4}{2} \partial_b
\left( \eta_{a \mu} \partial_{[\nu}\mathcal{A}_{\kappa]}^{a}(x,v^e) \eta^{\kappa b} - \eta_{a \nu} \partial_{[\mu} \mathcal{A}_{\kappa]}^a(x,v^e) \eta^{\kappa b} \right) - \\* & & \beta^1 \partial_{\kappa} \partial_a \mathcal{A}_{[\mu}^{\kappa}(x,v^e) \delta^{a}_{\nu]} - \beta^2 \partial_a \partial_b \mathcal{A}_{[\mu}^{\kappa}(x,v^e) \delta^{a}_{|\kappa|} \delta^{b}_{\nu]} + \ldots   
\end{eqnarray*} and 
\begin{eqnarray*}
0 & = & \gamma^1 \partial^{\mu} \partial_{[\mu}\mathcal{A}^{a}_{\widetilde{a}]}(x,v^e) + \gamma^2 \partial^b \partial_{[\widetilde{b}} \mathcal{A}^{a}_{\widetilde{a}]}(x,v^e) \\* 
0 & = & \delta^{1} \partial^{\mu} \partial_{a} \mathcal{A}_{\mu}^{\widetilde{a}}(x,v^e) + \delta^2 \partial^b \partial_{a} \mathcal{A}_{\widetilde{b}}^{\widetilde{a}}(x,v^e) + \ldots
\end{eqnarray*} We have now prepared the ground for an integrability analysis of the linearized equations; the latter subject is however postponed for future work. 
\begin{itemize}
\item AXIOM VIII  The universal equations of motion for the unitary potential $U(e_b(x),x)$ and Hermitian quantum gauge field $A_{\mu}(e_b(x),x)$ are much easier to write down and it is easy to prove that they necessitate the point of view of -at least- an indefinite Clifford Hilbert module.  Indeed, the most general equation for $U(e_b(x),x)$ must satisfy the following conditions: (a) it transforms covariantly under quantum local Lorentz transformations (b) coordinate invariant (c) preserves the unitarity relationship.  From (c), one derives that the equation must be first order in the derivatives and from (b) one concludes one has to contract the covariant derivative $\nabla_{\mu}$ with the vielbein $e^{\mu}_{a}$.  To make this equation generally covariant, we need the gamma matrices, that is the Clifford algebra.  I am not going to dwell here too much on free Clifford Quantum Field theory, but let me tell this much: the number $\gamma^{a}$ should be seen as the operator $\gamma^{a}1$ and therefore transforms under a Lorentz transformation as $U^{\dag}(\Lambda(x), e_b(x))\gamma^{a}U(\Lambda(x), e_b(x))$ which is a natural consistency demand.  This implies that our unitary operators are complex linear and not Clifford linear: this nonlinearity is a genuine new feature of our formalism.  Indeed, we have given in the previous chapter an interpretation to Clifford superpositions of states and under a Lorentz boost $U(\Lambda(x), e_b(x))$,  a superpostion of the kind $a|\Psi \rangle + b |\Phi \rangle$ transforms as \begin{eqnarray*} U^{\dag}(\Lambda(x), e_b(x)) \left( a|\Psi \rangle + b |\Phi \rangle  \right) &  = & \left( U^{\dag}(\Lambda(x), e_b(x))aU(\Lambda(x), e_b(x))\right) U^{\dag}(\Lambda(x), e_b(x)) |\Psi \rangle + \\* & & \left( U^{\dag}(\Lambda(x), e_b(x))bU(\Lambda(x), e_b(x))\right)U^{\dag}(\Lambda(x), e_b(x)) |\Phi \rangle. \end{eqnarray*}  Hence, this also implies that $\gamma^{a}$ must transform under the unitary operators $U(e_b(x),e_{a}(x_0),x,x_0)$ meaning that we might have a different Dirac equation at different spacetime points.  Indeed, the gamma matrices are not universal and under such unitary transformation, all their algebraic properties remain valid (as well as the commutation relations with the particle creation and annihilation operators).  Let therefore $x_0$ be a spacetime point where $$\gamma^{a}(\Lambda_{b}^{\,\,c}(x_0)e_c(x_0),x_0) = U^{\dag}(\Lambda(x_0),e^b(x_0))\gamma^{a}(e_b(x_0),x_0)U(\Lambda(x_0),e^b(x_0)) = \Lambda^{a}_{\,\, b}(x_0) \gamma^{b}(e_b(x_0),x_0).$$  Denote by $$\gamma^{a}(e_b(x),x) = U(e_b(x),x) U^{\dag}(e_b(x_0),x_0) \gamma^{a}(e_b(x_0),x_0)U(e_b(x_0),x_0) U^{\dag}(e_b(x),x)$$ then, since 
$$U(\Lambda(x),e_b(x)) = U(e_b(x),x)U^{\dag}(e_b(x_0),x_0) U(\Lambda(x),e_b(x_0))U(e_b(x_0),x_0) U^{\dag}(e_b(x),x)$$ one obtains that
$$\gamma^{a}(\Lambda_{b}^{\,\,c}(x)e_c(x), x) = U^{\dag}(\Lambda(x),e_b(x))\gamma^a(e_b(x),x))U(\Lambda(x),e_b(x)) =  \Lambda^{a}_{\,\, b}(x)\gamma^{b}(e_{b}(x),x).$$
This view is much more ``covariant'' than the standard treatment of the gamma matrices in curved spacetime where it is gratuitely assumed that they have to be the same at different spacetime points while it is only their algebra which has to be so\footnote{Contrarian to the standard treatment of Dirac theory \cite{Peskin}, there is no reversion between the spin transformations on the Clifford numbers and the unitary operators working on Hilbert space.  Indeed $U(\Lambda(x_0), e_b(x_0))$ should be thought of as the product $u(\Lambda(x_0), e_b(x_0))U'(\Lambda(x_0), e_b(x_0))$ where $u(\Lambda(x_0), e_b(x_0))$ is the spin transformation satisfying  $\widetilde{u}(\Lambda(x_0), e_b(x_0))\gamma^{a}u(\Lambda(x_0), e_b(x_0)) = \Lambda^{a}_{\,\,b}(x_0) \gamma^{b}$.  In this representation $u$ is a Clifford number and $U'$ is a scalar expression consisting out of ordinary creation and annihilation operators commuting with $u$.  The only thing which is important here is that $U(\Lambda(x), e_b(x)) = u(\Lambda(x), e_b(x))U'(\Lambda(x), e_b(x))$ where both are required to commute and moreover, $U'(\Lambda(x), e_b(x))$ commutes with $\gamma^{a}(e_b(x),x)$.  This is a fully active point of view where frames get mapped to frames and there really is no coordinate transformation anymore in the following sense
$$T'^{a b}(x,v'^{c}, e'_d(x)) = U^{\dag}(\Lambda(x), e_d(x)) T^{ab}(x,v'^c, e_d(x))U(\Lambda(x), e_d(x)) = \Lambda^{a}_{\,\,c}(x)\Lambda^{b}_{\,\,d}(x) T^{cd}(x,v^e, e_d(x)).$$  The same result holds for the translation group
$$T'^{ab}(x,v^{c} + w^{c}(x), e_d(x)) = U^{\dag}(w(x), e_d(x)) T^{ab}(x,v^c + w^c(x), e_d(x)) U(w(x), e_d(x)) = T^{ab}(x,v^c, e_d(x))$$ and obviously all this extends to the whole Poincar\'e group.}.  Therefore, the proper transformation laws of the gamma field under quantum Lorentz transformations must be part of the initial conditions (at any point).  The reader notices that nothing here depends upon the choice of the point $x_0$, it is just highly unusual to write down an equation which makes explict reference to a chosen point in spacetime but which transforms completely covariantly under diffeomorphisms and local Lorentz tranformations.  However, there are two caveats here, one would also like to have a Clifford field $\gamma'^{a}(e_b(x),x)$ satisfying not only the Clifford algebra, but also 
\begin{eqnarray*}
\gamma'^{a}(e_b(x),x)U(e_b(x),x) & = & U(e_b(x),x) \gamma^{a}(e_b(x),x) \\*
\gamma'^{a}(\Lambda_{c}^{\,\,d}(x)e_d(x),x) U^{\dag}(\Lambda(x),e_b(x)) & = & U^{\dag}(\Lambda(x),e_b(x)) \Lambda^{a}_{\,\,b}(x)\gamma'^{b}(e_c(x),x) 
\end{eqnarray*}
two properties which are not satisfied by the ``canonical'' Clifford field $\gamma^{a}(e_b(x),x)$ (we shall prove later that unitarity requires such object\footnote{A kind of left right symmetry if you want.}).  The solution to these conditions obviously is unique and given by
\begin{eqnarray*} \gamma'^{a}(e_b(x),x)& = & U(e_b(x),x) \gamma^{a}(e_b(x),x) U^{\dag}(e_b(x),x) \\ & = & 
U^2(e_b(x),x) U^{\dag}(e_b(x_0),x_0) \gamma^{a}(e_b(x_0),x_0) U(e_b(x_0),x_0) \left(U^{\dag}(e_b(x),x)\right)^2 \end{eqnarray*}  which finishes the dicussion of the required tools.
 \\* \\*
As said before, we subtly break linearity in the sense that our operators $X$ still satisfy $$X\left( |\Psi \rangle + | \Phi \rangle \right) = X|\Psi \rangle + X|\Phi \rangle$$
but in general 
$$X\left( a \Psi \right) \neq a X| \Psi \rangle $$
but the equality remains of course true for complex numbers.  Let us first comment upon the construction of the field strength which is not so trivial as it may look.  Take the simple case of a $U(1)$ abelian gauge field, then the field tensor usually is written down as
$$F_{\mu \nu} = \partial_{\mu}A_{\nu} - \partial_{\nu}A_{\mu}$$ and one may think about replacing the partial derivative by the covariant derivative.  However, in the general case of nonzero torsion, this spoils gauge invariance since one obtains a term proportional to
$$ - iT^{\kappa}_{\mu \nu}(x) \partial_{\kappa} \Omega(x)$$
which cannot be compensated for by any means and therefore    
\begin{eqnarray*} \mathcal{F}_{\mu \nu}(e_b(x),x) & = & \left[ i \partial_{\mu} + A_{\mu}(e_b(x),x) , i \partial_{\nu} + A_{\nu}(e_b(x),x) \right]. 
\end{eqnarray*}
Concerning the matter part, an idea coming from Clifford monogetic functions becomes essential to preserve unitarity.  Taking the flat spacetime Dirac operator $D$, one can write the equations $$DX = XD = 0$$ where in the second equation $D$ acts from the right.  The class of solutions is therefore highly constrained but (for example) one can choose $X$ selfadjoint and satisfying $DX = 0$ or $X = DY$, $X = DZD$ where $Y$ in the first case is a scalar and $Z$ can be anything satisfying $\Box Y = \Box Z = 0$.  The second construction can be generalized to the case where $DX = XD' = 0$ where $D'$ is a different Dirac operator.  A whole solution class is given by $X = DZD'$ where $\Box Z = 0$ and, as a matter of fact, plenty of other solutions do exist\footnote{The author thanks D. Constales for a brief but effective discussion on this topic.}.  Strictly speaking a constraint analysis should be performed but we shall be concerned with that later on\footnote{Of course, everything becomes more complicated when the gamma matrices themselves become space-time dependent, but I feel a generalization of the above simple remarks should be possible.}. \\* \\*          
Keeping this in mind, one can write down the following equations of motion
\begin{eqnarray*}
0 & = & e^{\mu}_{a}(x,0) \left( i \partial_{\mu} + A_{\mu}(e_b(x),x) \right) U(e_b(x),x) \gamma^{a}(e_b(x),x) + \\* & &\Lambda \, U(e_b(x),x) + \sqrt{-P^{a}_1 (x)P_{1 \, a}(x)}U(e_b(x),x) + \ldots \\*
0 & = &  e^{\mu}_{a}(x,0) \left( i \partial_{\mu} + A_{\mu}(e_b(x),x) \right) U(e_b(x),x) \gamma^{a}(e_b(x),x) - \\* & &\gamma'^{a}(e_b(x),x) e^{\mu}_{a}(x,0) \left( i \partial_{\mu} + A_{\mu}(e_b(x),x) \right) U(e_b(x),x) \\*
\end{eqnarray*}
where $\ldots$ in the first equation\footnote{Note that the above equations appear not to be recalibration invariant; that is invariant under transformations $U(e_b(x),x)K$ where $K$ is a unitary matrix.  This is most easily repaired by replacing $U(e_b(x),x)$ by $U(e_b(x),x)C$ where $C$ is a recalibration constant transforming as $K^{\dag}C$.  We ignore this issue in this book and always choose $C$ appropriately.} signifies that other physical terms can and possibly must be added\footnote{Renormalization saves quantum field theory from perturbative infinities, but it is somewhat an ugly procedure.  That is, it would be much nicer to directly express the energy dependence of the \emph{physical} parameters without having to perform a resummation procedure akin to Wilson.  Moreover, it would be desirable to have a formulation of the dynamics such that computations of observable quantities are free of infinities.  In our theory, we potentially adress both issues in the following ways : (a) we do not start from Field Theory, therefore no free parameters exist a priori and solving for the quantum dynamics automatically gives the correct energy dependence of the effective parameters in the local Hamiltonians (b) our quantum theory is by definition asymptotically free and states of the kind $$a^{\dag}_{\vec{k} \, m \, \ldots}(e_b(x),x)a^{\dag}_{\vec{l} \, m \, \, \ldots}(e_b(x),x)| 0 , e_b(x) ,x \rangle$$ have a physical mass squared $$ M^2 = 2m^2 - 2k.l $$ and should have a one particle interpretation at $x$.  The latter is indeed a particle with four momentum $k^a + l^a$ and its norm squared precisely equals $M^2$.}.  The square root $\sqrt{-P^{a}_1(x)P_{1 \, a}(x)}$ is defined as 
$$\sqrt{-P^{a}_1(x)P_{1 \, a}(x)} = \pm \int d^3 \vec{k} \, \gamma^{a}(e_b(x),x) \,k_a a^{\dag}_{\vec{k} \, m \, \sigma \, c \, \pm}(e_b(x),x) a_{\vec{k} \, m \, \sigma \, c \, \pm}(e_b(x),x)$$ and the reader verifies that it has the appropriate transformation properties.  Some examples of additional terms are $\alpha S^{ab}_1(x)S_{1 \, ab}(x)U(e_b(x),x)$ and $$\gamma \mathcal{F}_{\mu \nu}(e_b(x),x) \mathcal{F}^{\mu \nu}(e_b(x),x) U(e_b(x),x).$$
All of these terms are necessary to get the correct physics out but they are irrelevant for the issue of unitarity.  A crucial assumption we have made all the time is
$$ \left[ \gamma^{a}(e_b(x),x), a^{\dag}_{\vec{k} \, m \, \sigma \, c \, \pm}(e_b(x),x) \right] = 0$$ which is automatically satisfied if it is so at $x_0$.  Hence, 
the coefficient $\alpha$ can in general belong to the Clifford algebra generated by the $\gamma^{a}(e_b(x),x)$, $\gamma$ however must be a real number.  
One verifies now that unitarity is preserved under the equations of motion\footnote{Notice the remarkable identities $U^{\dag}(e_a(x),x) \gamma^{b}(e_a(x),x) = \gamma^{b}U^{\dag}(e_a(x),x)$ and
$\gamma^{b}(e_a(x),x)U(e_a(x),x) = U(e_a(x),x) \gamma^{b}$.}; indeed
\begin{eqnarray*}
0 & = & \gamma^{a}(e_b(x),x)e_{a}^{\mu}(x,0) \left( i\partial_{\mu} U^{\dag}(e_b(x),x) - U^{\dag}(e_b(x),x) A_{\mu}(e_b(x),x) \right)U(e_b(x),x) + \\*
& & U^{\dag}(e_b(x),x) e^{\mu}_{a}(x,0) \left(i\partial_{\mu} + A_{\mu}(e_b(x),x) \right) U(e_b(x),x) \gamma^{a}(e_b(x),x) \\*
& = & i e_{a}^{\mu}(x,0) \left(\partial_{\mu} U^{\dag}(e_b(x),x)U(e_b(x),x) + U^{\dag}(e_b(x),x) \partial_{\mu}U(e_b(x),x) \right) \gamma^{a}(e_b(x),x)
\end{eqnarray*}
where in the first step, we have immediately ignored those terms which automatically vanish \footnote{We also avoid here the Coleman-Mandula theorem because we do not work within the context of a global Lorentz covariant theory with a Lorentz covariant scattering matrix.  Even if one would manage to construct a scattering matrix in some sense, one would certainly not be able to give the notion of a global spacetime Lorentz transformation any meaning since the local Lorentz groups actually differ from one spacetime point to another (by a unitary transformation).  Another way the Coleman-Mandula theorem does not apply is by means of AXIOM 0 which foresees for an infinite number of copies of the same particle species.  Effectively, our space-time particle notions depend upon a ``coarse graining'' and therefore the mass spectrum effectively becomes continuous around some discrete values.  This of course does not imply that the spectrum of all particles is not measured to be discrete for all practical purposes, it simply means it is not discrete.  This could even mean that in principle the space-time particle spectrum might come arbitrarily close to the ground state, but such coarse grainings would certainly never apply in laboratory experiments since it requires a strong gravitational field, see Weinberg \cite{Weinberg3}.}.  At the second stage, whe have used the left-right symmetry above as well as the equivalence between the different gamma matrices.  Now, we use that the spatial derivatives $$\partial_{\alpha}\left( U^{\dag}(e_b(x),x)U(e_b(x),x) \right) = 0$$ which is a consequence of the initial conditions.  Therefore, we obtain that
$$ 0 = ie_{a}^{t} \partial_t\left(U^{\dag}(e_b(x),x)U(e_b(x),x)\right)\gamma^{a}(e_b(x),x)$$ implying that $$\partial_t \left(U^{\dag}(e_b(x),x)U(e_b(x),x)\right) = 0$$ since $$\gamma^{a}(e_b(x),x)e^{t}_{a}(e_b(x),x)$$ is invertible.  Likewise, one shows that $$\partial_t\left( U(e_b(x),x)U^{\dag}(e_b(x),x) \right) = 0$$
which needs to be done in modules of cardinality greater or equal than $\aleph_{0}$.  Physically, these equations\footnote{In case the above two equations would turn out to be too restrictive, one could drop the unitarity condition for $U(e_b(x),x)$ and write down some polar decomposition (which is not unique anymore because of the indefinite character of Nevanlinna space) $U(e_b(x),x) = V(e_b(x),x) R(e_b(x),x)$.  In that case, one could even just use the first equation only (and drop the holographic principle) and transport creation and annihilation operators by means of $V(e_b(x),x)$.  This would mean that gravity demands the existence of a quantum mechanical hidden variable $R(e_b(x),x)$ (hidden, because it does not show up in the interpretation) and it remains to be seen whether such point of view is mandatory or not.} reveal an important fact: that is, they contain a version of the holographic principle as their very foundation.  Indeed, the unitary potential is fully specified on any spherical tube $B^3 \times \mathbb{R}$ by restriction to the boundary $S^2$ of the three dimensional ball $B^3$. \\*\\*
We now proceed by the construction of the Yang-Mills equations: well, strictly speaking it is not a Yang-Mills symmetry since the gauge groups at different spacetime points are not identical but unitarily equivalent to one and another (in a fully dynamical way).  Indeed, dynamical bundles were the wet dream of any relativist for many years, it is ``surprising'' to see that these ideas can only come to full life by taking Quantum Theory seriously.  The reader may easily verify that the ``gauge'' equations of motion are given by
$$\left( i \widetilde{\nabla}_{\mu} + A_{\mu}(e_b(x),x)\right) \mathcal{F}^{\mu \nu}(e_b(x),x) + hc = \alpha e^{\nu}_{a}(x) \gamma'^{a}(e_b(x),x) + \beta P^{a}_1(e_b(x),x)e^{\nu}_{a}(x) + \ldots$$
where $\alpha, \beta$ are real numbers, $hc$ denotes Hermitian conjugate, $\widetilde{\nabla}_{\mu}$ the ordinary Levi-Civita connection and $\ldots$ denotes insertion of all other possible operators.  The covariant current is not conserved which is no surprise since it is well known in non-abelian gauge theories that no gauge-covariant conserved current exists.    
\end{itemize}
To reassure the reader that everything works out as it should, let me point out how conventional unitary free quantum field theory can be recovered from this scheme.  In the latter theory, the measurable particles at each point in spacetime are the same which appears to contradict AXIOM 0.  However, it does not, the remaining infinite copies are simply not ``activated'' in the dynamics and should be thought of as the particles constituting the observers and not the system under study.  Hence, we choose a gauge and coordinate system such that $e^{\mu}_{a}(x,v^c) = \delta^{\mu}_a$ and $U(t_0, \vec{x}_0) = 1$ and from now on, we surpress the depency upon the frame in our notation.  We ignore the $A_{\mu}$ field and assume $U(x)$ not to depend upon the Clifford numbers.  Then, one obtains that 
$$\gamma^{a}(x) = \gamma'^{a}(x) = \gamma^{a}$$
and the reader may verify that 
$$U(x) = e^{\pm i \int d^3 \vec{k} \, k_{\mu}(x^{\mu} - x_0^{\mu}) \, a^{\dag}_{\vec{k} \, m \, \sigma \, \pm}\, a_{\vec{k} \, m \, \sigma \, \pm}}$$
\emph{uniquely} solves both equations of motion.  Hence, causality is a prediction of our theory and is not something which has been put in by hand through the initial conditions.  Next, one calculates the $x$ dependency of the creation operators:
$$a^{\dag}_{\vec{k} \, m \, \sigma \, \ \pm}(x) = U(x)a^{\dag}_{\vec{k} \, m \, \sigma \, \ \pm} U^{\dag}(x) = e^{i \,k_{\mu}(x^{\mu} - x_0^{\mu})} a^{\dag}_{\vec{k} \, m \, \sigma \, \pm}.$$
If one separates now time from space, it is possible to \emph{interpret} $$a^{\dag}_{\vec{k} \, m \, \sigma \, \pm}(t) = e^{- i k(t - t_0)}a^{\dag}_{\vec{k} \, m \, \sigma \, \pm}$$ as creating a global particle satisfying the usual equations of motion.  Actually, it is fairly easy to construct the appropriate fields (in the vector representation) from this result by putting 
$$ \psi^{+}_l(x) = \int d^3 \vec{k} \, u_l(\vec{k}, m , \sigma, \pm) \, a^{\dag}_{\vec{k} \, m \, \sigma \,\pm}(x) $$ and demanding the correct properties under \emph{local} Lorentz transformations.  Moreover, it follows directly from the commutation relations on $TM_{x_0}$ that 
\begin{eqnarray*}
0 & = & \left[  T^{ab}(x,0), T^{cd}(y,0) \right] \\*
0 & = & \left[  T^{ab}(x,0), S^{cd}_{e}(y,0) \right] \\*
0 & = & \left[  S^{ab}_{c}(x,0), S^{de}_{f}(y,0) \right] 
\end{eqnarray*} for $(x - y)^2 > 0$ since, for example, $T^{ab}(x,0) = T^{ab}(x_0,x -x_0)$.  The reader should pauze a bit and understand the very nontrivial fact which has happened here, although the mathematics is deceivingly simple.  We have just \emph{derived} that one can recuperate the ordinary spacetime field theoretical picture from the local symmetries of our theory.  That is, the assumption of a unitary potential taken together with the constraint equations do not only uniquely (up to a $x$ independent unitary transformation) lead to the correct field equations of motion, but also imply the causal commutation relations.  This was the great achievement of Weinberg's approach \cite{Weinberg} by means of the \emph{global} spacetime Poincar\'e group: indeed, it is underappreciated by several orders of magnitude that his derivation unifies appearantly two different things (the Heisenberg equations and the commutation relations).  We just managed to generalize this virtue to the setting of quantum gravity without using fields at all.  Since in our language, free quantum fields are unnatural concepts, we shall omit such reference in the future (interacting quantum fields are at best an effective low energy description of the world).  To stress again what really happens here is that somehow, the two novel constraint equations do contain the whole information about the quantum theory apart from boundary conditions.  Hence, causality is a byproduct, a consequence of something even more fundamental.    
 \\* \\*    
At this point, we can give a deeper motivation for AXIOM 0 which also answers our previous comments we made regarding the Heisenberg uncertainty principle.  In free Quantum Field Theory, it is possible to define generalizations of the Newton Wigner position operators:
$$X_{n} =  \mp i \int d^3 \vec{k} \, \partial_{k^{n}} a^{\dag}_{\vec{k} \, m \, \sigma \, \pm} a_{\vec{k} \, m \, \sigma \, \pm}$$
and
\begin{eqnarray*}
X_{n} \, a^{\dag}_{\vec{k} \, m \, \sigma \, \pm} | 0 \rangle & = & - i \partial_{k^n} a^{\dag} _{\vec{k} \, m \, \sigma \, \pm} | 0 \rangle.  
\end{eqnarray*}
Likewise,
\begin{eqnarray*}
0 & = & \left[ P^{m} , P^{n} \right] \\*
0 & = & \left[ X_{n}, X_{m} \right] \\*
i \delta^{m}_{n} & = & \left[ X_{n},P^{m} \right].
\end{eqnarray*}
One notices that we have broken translation invariance of space by taking $\vec{x}_{0}$ as the origin.  Anyhow, I wanted to comment why one needs a countable infinite number of copies of the same particle in the module of the universe $\mathcal{H}$.  The reason is that due to a nonzero gravitational field, any position operator applied to the state $  a^{\dag}_{\vec{k} \, m \, \sigma \, \pm} | 0 \rangle$ will have \emph{finite} width because the momentum uncertainty is nonzero for $P^{a}(x)$ where $x \neq x_0$.  In other words, the gravitational field ``localizes'' particles having definite momenta in some spacetime point $x_0$ and there is some rather interesting numerology one encounters.  Indeed, when figuring out an appropriate relation for a particle standing still with respect to the surface of the earth one makes the following remarks : (a) any formula for $\Delta X$ must go to zero if $\hbar \rightarrow 0$ and go to infinity if $G \rightarrow 0$ (b) the expression must be a Laurent series in the rest mass $m$ of the particle and some effective dimensionless parameter $m'$ due to the gravitational field of the earth and only terms of the form $m m'$ arise (c) it must have the same leading term as the formula for the Compton wavelength.  This leads to
 $$ \Delta X \sim \frac{\hbar}{m m' c}$$
and we need to find out the number $m'$.  Here, we apply Einstein's formula in the following way
$$m' c^2 = \frac{GM}{r}$$ where $M$ is the mass of the earth and $r$ its radius.  Therefore,
$$ \Delta X \sim \frac{\hbar}{mc} \frac{c^{2}r}{G M}$$ and the last factor is roughly equal to $10^{9}$.  This is a rather realistic estimate and some interesting consequences should follow from it.  Indeed, an amazing aspect of this formula is that the contribution from the sun to $\Delta X$ is almost equal than the one originating from earth: indeed,
$$ \frac{r}{M_{\textrm{earth}}} \sim 10^{-18}$$ while $$\frac{r_{\textrm{earth-sun}}}{M_{\textrm{sun}}} \sim 10^{-19}.$$  If this were true, it would be a very grand implementation of Mach's principle indeed\footnote{It is rather strange that the ratios satisfy $\frac{r_{\textrm{\textrm{planet}}}}{r_{\textrm{planet-sun}}} \sim \frac{M_{\textrm{planet}}}{M_{\textrm{sun}}} \sim 10^{-6}$.  It is not a mystery that these two separate ratios are more or less the same (up to a relative factor of at most $10^2$) for all planets but that the ratios of two seemingly unrelated quantities are identical.}.   Another strange thing is that $\Delta X$ is of the order of a millimeter for electrons, which is precisely the scale where we expect gravity to be modified.  To my knowledge, no single particle \emph{interference} experiments have been done on this scale yet; the one of Merli in 1974 had a characteristic width of $23$ microns.  For neutrons, such experiment would be easier since one would only need widths of $10^{-7}$ meters.  We will compute in chapter ten whether this estimate holds. \\* \\*
There is another remarkable fact about our matter equations of motion which is that in a nonvanishing gravitational field, the Clifford numbers are ``turned on'' and the equations of motion for the potential effectively become \emph{non-linear} even in the absence of other ``gauge-type'' interactions.  This by itself is consistent with the gravitationally induced loss of coherence explained above since as mentioned previously, $U$ is not linear anymore on Clifford superpositions of states.  Many people have conjectured a loss of quantum coherence in quantum gravity (see Penrose \cite{Penrose1}) as a potential explanation for the absence of a Schrodinger cat.  This formalism makes this explicit even in the context of quantum theory on a curved background and we should be very careful in studying its consequences.    
\begin{itemize}
\item  AXIOM IX : We now come to the equation of motion for the local reference frames in which consciousness operates.  This is far from easy as we explained in chapter three.  Conciousness does not only operate in terms of local particle notions (which we have) but also in terms of ``quasi-local'' particle states which is a dynamical thing.  Let me give an example that something of the second kind is necessary: suppose you have two observers driving cars which are on collision course and they collide roughly at point $x$, then one could try to determine a (upon local rotations) \emph{center of mass} reference frame comoving with the collision.  This would be determined by the equation  
$$\mathcal{S} \, \langle \Psi | P^{k}_2(x,e_b(x)) | \Psi \rangle = 0$$ 
where $k:1 \ldots 3$, in a small neighborhood of $x$; indeed, $e_0$ corresponds to some frame where $p^{k} + q^{k} = 0$.  One needs to be careful here, because the above equations might have no solution due to the indefinite norm induced by the Clifford algebra; indeed, in general, this vector can be anything whatsoever even if the momentum operator gives eigenvalues constituting a timelike vector pointing to the future.  In such cases, no local reference frames exist and the interpretation is that no observation can take place.  In general, it is impossible to construct a comoving frame with $p^{a}$ starting from the state of the universe $|\Psi \rangle$.  Therefore, it is clear that we can only speak of a personal rest space - that is a reference frame in which the local matter distribution of your body is stationary- if we replace the total state $|\Psi \rangle$ by the state of your body\footnote{The latter is defined from the state of the universe $|\Psi \rangle$ by first determining which particles make up your body, shifting the respective creation operators to the right in front of the local vacuum state, inserting at the left of those a new local vacuum state (making a ``bi state'') and calculating the amplitude of this new second state with respect to the natural Fock basis.  This amplitude is defined by taking the square root of the absolute value of the partial ``norm'' while the phase factor is defined by dropping the creation operators of the rest of the universe and integrating out the corresponding momenta (we assume the wave function to be integrable as well in momentum space).  In a complex Nevanlinna space, this concept is insufficient since one would need to remember an eventual minus sign coming with the above scalar product while in Hilbert space no such issue arises. In a Clifford Nevanlinna module however, we can use the $\gamma^0$ associated to a rest frame for the state of the universe $|\Psi \rangle$ to compensate for this; more specifically the latter amplitude -eventually- multiplied with $\gamma^0$ to the left and to the right with creation operators of your body applied to the local vacuum defines $|\Phi \rangle$.} $|\Phi \rangle$.  Some mutually orthogonal parts of the latter may be entangled with different states within $|\Psi \rangle$; therefore, different human beings define distinct trajectories of preffered three spaces and the trajectories might cross each other just because different bodies have different personal states.  Also, the state of your body must be revised once a particle hits you, or a bycicle drives against your car.  This is an extra level of kinematics and dynamics a material theory of the universe has nothing to say about.  The very reason is that it does not define what entities are, since the latter are merely interpretations of states of particles which one considers as a whole.  Nevertheless, perception is relative to these entities and there is something in this world which apart from being dynamical itself, recognizes the dynamics of shapes even though the fundamental materialistic theory does not know shapes and therefore could not even define what it means that they change.  As elaborated upon, this recognition has to occur from \emph{within} the universe and the quantum physicist simply cannot, on his drawing board, change the definitions by hand; hence, the need for a theory of consciousness.  Let me make a provocative remark, obviously I do not expect the physics of elementary particles in scattering experiments to depend upon such considerations, but it might be that for humans, ants, and all kind of structured living beings it does play a role.  By this I mean that a higher non-local variable constructed from elementary particle interactions might dynamically emerge and have an influence on the elementary particles themselves.  That is, any strong enough form of consciousness might not only be a perceptor and interactor through the collapse of the wave function, it might also slightly change the unitary potential $U(e_a(x),x)$.  Such theories however are for the far future but may be a neccesary step if conventional materialism would not succeed in explaining the formation of complex structures.  Related to this issue is the ``problem'' of the arrow of time; an increase of some local entropy function is not responsible for us remembering the distinction between the past and future.  Living creatures are simply disentropic as was mentioned by Norbert Wiener and Alfred Ubbelohde long time ago; therefore, the psychological difference between the past and present has nothing to do with local entropy.  However, the suggestion we just made is that the psychological arrow of time may influence the local entropic arrow of time which might explain the measured entropy increase of (all) approximately closed systems (so far).  The psychological arrow of time is, as explained in AXIOM X, determined by the holistic view any monad has on the rest of the universe which is as global as it may get.  
\item AXIOM X: AXIOM IX was of course the necessary prelude to get a dynamical measurement theory, but we did not specify yet where in spacetime the collapse of the ``wavefunction'' has to take place.  The type of collapse theory one constructs depends of course on the \emph{physical} assumptions one makes regarding the theory of consciousness.  That is, does consciousness operate according to the physical eigentime associated to the observers personal reference frame (upon local rotations), or is there a dynamical universal time function present according to which consciousness operates?  Picking the former, however, does lead to a deepening of thought experiments such as the twin paradox.  The distinction has to do with where you are in space-time and how spacetime physics works.  Let me make an example, take two observers $1$ and $2$ in Minkowski spacetime, $1$ moves on a geodesic $(t,0)$ and $2$ moves almost on a null geodesic to the left from $1$ and then turns back in the same direction again on almost a null geodesic.  Both start at the origin; after say three seconds on his local clock $1$ sends a message in the direction of $2$ which his worline will cross after, say, half a second on his clock since he turns back to $1$ eventually.  The relativistic interpretation is that he always  \emph{measures} this signal; the question however then is where is the first observer on his worldline when the second one reads his message?  The traditional answer a relativist might give is that this is not a well posed question and a quantum field theorist would say it does not matter since spacelike separated operators do commute.  So in the former vein, one has a worldview which consists of ordered personal experiences and it is by no means necessary that those experiences coincide ``mentally''.  In the second view, one supposes an infinite reproduction of the same spatiotemporal situation and state, how else would a statement that two observables at spacelike separated events commute get any operational meaning if not some time isometry between an infinity of such events were dynamically possible?  Of course, in a realistic universe, this will never happen because spacetime changes and even if one could reproduce the local state exactly, it would be impossible to exclude anisotropies coming from the environment which do not average out in the statistics.  How unplausible this may sound, from a conventional point of view there would be no real contradiction as long as the commutation relations were faithfully represented which requires a different representation than free Fock space due to Haag's theorem.  In quantum gravity however, there is a serious problem since spacelike separated observables do not commute anymore and the classical argument gets destroyed because in quantum mechanics, perception has an active meaning and is more than just a form of being.  Therefore, we would need to conclude that there is some global space-time notion of being.  \\* \\*
This implies that communications of the ``mind'' of which we totally unaware are tachyonic in nature with respect to the classical space-time metric.  It is here that the theory of quantum gravity as constructed above might offer a way out since our gravitational theory has precisely such nonlocal features.  More specific, it is the the notion of time created by the normalized \textsl{global} energy momentum vector $$\mathcal{S} \, \langle \Psi | P^a_1(x,e_b(x)) | \Psi \rangle e_a(x)$$ which determines well defined hypersurfaces of equal time as long as this vector remains timelike\footnote{Note that it is possible to attach a constant to any irreducible component of some action of the universal Poincar\'e group and that we did ignore this possibility in the above formula.}.  In case it does not, the universe has no interpretation anymore.  The local energy momentum vectors cannot serve for this purpose since they might cross.  This does not imply that for ordinary lab situations where observers do not really move with respect to one and another, we cannot get away with the old fashioned hand waving Von-Neumann measurement rule.             
\item AXIOM XI : There is no mathematical rule by which we can put the classical cosmological constant to zero, but there is a physical one.  The reader shall have noticed that in AXIOM VIII, we introduced a constant which I called on dimensional grounds the square root of the cosmological constant.  In this axiom, it shall become clear why this name was right on the spot.  Take a coordinate system $(t, \vec{x},v^a)$ such that the initial hypersurface $\Sigma \times \mathbb{R}^4$ coincides with $t = 0$. Moreover, suppose the initial conditions on $\Sigma \times \mathbb{R}^4$ for our universe are given by
\begin{itemize}
\item $e_{\mu}^{a}((0,\vec{x}),v^b) = \delta^{a}_{\mu}$ and all first derivatives vanish.
\item $\mathcal{A}_{\mu}^{A}((0, \vec{x}), v^b) = 0 = A_{\mu}((0,\vec{x}),e_b(0,\vec{x}))$ as well as all first derivatives.
\item $\alpha = \beta = 0$ in the equations for the quantum gauge current.
\item $| 0, e_{b}(0,\vec{x}) ,(0,\vec{x}) \rangle = | 0 \rangle$
\item $| \Psi \rangle = | 0 \rangle$ 
\end{itemize}   
meaning all observers see the same vacuum state (there is no relative acceleration), all ``gauge'' fields are zero (there are no force fields present), the geometry is trivial and the state of the universe is the vacuum state.  Then, we might \emph{impose} that this will remain so at all times which is equivalent to saying that the special relativistic laws of inertia hold in the limit of zero mass.  In order for this to be valid, the usual cosmological constant must vanish, but also must $\Lambda$ from the laws of motion of the unitary potential.  This does not imply there will not be a ``time dependent'' effective cosmological field which is generated by quantum fluctuations of the matter fields if $|\Psi \rangle \neq |0 \rangle$ but all it says is that the average value must be zero.      
\end{itemize} In the beginning of this chapter I promised to describe how observers could come ``alive'' in a dynamical way, how macroscopic objects could get meaning, in other words how creation of ``concepts'' by living beings might occur.  It is of course a logical possibility that this kind of information is encoded from the beginning into the universe, but do we really believe that?  Are paintings like Mona Lisa of da Vinci, Parsifal of Wagner, Relativity of Einstein ever present in the universe or is it possible to start dreaming about a physics of \emph{macroscopic} creation?  The end of this chapter will distinguish itself by its philosophical nature but likewise, it will demonstrate beyond reasonable doubt that quantum gravity is the \emph{easy} problem.  Physicists like mathematicians always confuse a lack of symbolic rigor with a lack of profundity, well this is not the case and the end of this chapter is certainly much more difficult to write down than the technical exercise we just made.  The theory above is strictly speaking an empty box so far, we have a state of the universe and local particle notions, but we haven't touched yet the subject of observation and creation of the observer.  Logically, when something is created, there must be a creator, but the latter must be non-physical and impose meaning to algebraic combinations of local creation beables and Clifford elements.  That is, there must be a ``mental'' source in the universe wich attributes mental capacities to our material configuration and by this very act, we ourselves become conscious observers and influence the very source which gave us these properties in the beginning by our very act of observation.  So, what are these entities which attribute meaning to material configurations?  As the reader has learned up till now, the state of the universe endows us with a preffered time direction at every point in spacetime and every point has a whole ``window'' to the rest of the universe.  This window describes how the ``local'' observer at the origin percieves the matter distribution in the rest of the universe.  Hence, the origin of \emph{local} perception in $x \in M$ resides on $VTM_x$ and the inertial ``glasses'' are fixed by the foliation determined by $e_0(x)$.  The ``monad'' $VTM_x$ attributes meaning to the state of the universe expressed in the preffered basis of creation operators, Clifford elements and the local vacuum state.  Nearby monads compare their ``interpretation'' and once sufficiently strong correlations are found between those pictures, meaning gets transferred to themselves.  This implies not only that the local observer located near the origin of tangent space becomes conscious, but also the different shapes in $VTM_x$ away from the origin, in either the precieved creatures by the local ``observer'' become conscious themselves outside the framework of conventional spacetime.  However, measurements can only be performed by conscious beings inside spacetime (that is near the origin of $VTM_x$ when they are located at $x$ in $M$).  Elementary particles will in this picture not get any consciousness since they are too ``inconsistent'' as explained in chapter three.  The monads themselves are not necessarily a source of eternal Platonic knowledge, but they could by themselves be dynamically evolving ``learning'' devises in the symbolic language of Fock space.  All this is very complicated and sketchy but I am afraid that the honest way to deal with this problem is going to be at least as complicated as this.  In philosophy, these ideas might be classified under neutral monism, where the principal entities are the state of the universe, the creation beables, the Clifford algebra and the Platonic world of shapes; physical and mental properties then emerge in a mutually interacting way.  As mentioned previously in AXIOM IX, such point of view could be used to explain the origin of life without appealing to any God, antropic principle or special initial conditions.  Our universe would be generic in the sense that creating complex (life) forms is a goal of the dynamics without coming into conflict with observations of the second law of thermodynamics for systems which might reasonably be considered as closed.
It puts a death sentence on pure materialism but also on creationist alternatives. \\* \\*
Let me speculate a bit about the kind of mathematics which is necessary for this.  Clearly, the above ideas imply that the ultimate laws have to be self-referential; the reason why the construction in this chapter is not is due to the mere stupidity of its author.  Indeed, my limited brain decided to first tackle the more mundaine problem of quantum gravity but at least I was clever enough to formulate it in a way which \emph{allows} for such extension.  In essence, I believe the ideas of going over to jet bundles, local beable notions, generalizing away to modules are merely first steps in the correct direction.  They are the first things which current knowledge suggests to us, at least to my limited mind.  If I may think a bit further than I did so far, I must conclude that the next thing to fall is classical logic since self referential laws have to escape Russel's paradox.  The axiom of restricted comprehension in Zermelo-Freankel theory is therefore not the right way to go.  In principle, one has two options: either the ultimate laws should be formulated in terms of the \emph{entire} Platonic universe and since there is no way of knowing, describing or calculating it, our work shall never be complete \emph{or} we find out a dynamical theory which allows us to probe this world of concepts without limitation and without them being present in the theory ``at that time''.  We have to find a theory of knowledge creation and acquisition and we are not even at the beginning of that mountain.  We summarize these thoughts in the following principle
\begin{itemize}
\item AXIOM XII: nature adapts its own laws and boundary conditions so that maximal structure formation occurs within the limitations of a well defined second law.  There is no initial value problem nor landscape issue, the laws have a Darwinian \emph{purpose}.
\end{itemize}  
The implications for our worldview of such principle are grand in the sense that it might dispose of the old fashioned concepts of good and evil as Nietzche anticipated more than a century ago.  Indeed, we did not pose ourselves the question yet \emph{why} the monads are asking the questions they do; supposing that all monads are alike, meaning that given an indentical state of the universe at different space time points (with repect to some vielbein and generalized Fock space variables) the result is an identical \emph{potential} questioning\footnote{Meaning the list of questions is the same as well as the probability for asking them.}, then one would need to conclude that the way we percieve the world depends upon the order in which the neigbouring mondads ask them.  Most likely there is no ergodicity in this process in the sense that if you would have two identical initial states at two different spacelike separated points, the distribution of realized future questions and answers for both ``beings'' will differ if the order in which they are asked is not identical.  This might lead to the dangerous suggestion that somehow we are were all identical ``initially'' but that our differences emerge from simply asking different questions and that the ``good'' ones (that is those who harmonize the best) simply read other persons readings much more efficiently and relatively ask the ``right'' questions.  Probably there is more truth to this than one might suspect at first; however it does deem clear to me that it would grant too much power to the spiritual world in order to explain all our (fishes, humans, apes) differences.  Let me try to formulate some principles about how ``self'', ``environment'' as well as a notion of inherent and sociological sanity might be defined.  As mentioned before, neigbouring monads are communicating with one and another on their perceptions of the state of the universe in order to define a coarse grained self as well as what the rest of the universe is; again one must make the assumption here (which is more or less true) that the potential conclusions which are drawn from these are stochastically identical for locally isogravitational regions of spacetime\footnote{Inequivalent gravitational fields might lead to different processes, but obviously one would expect such effects not to be present in a weak field approximation such as seems to be the case on earth.} whose timelike separation with respect to the universal mental time as defined previously is not too large\footnote{We also formulated the principle of creation of intelligence which does take place however, as it appears, on timescales which are significantly larger than at least 10000 years.}.  Taking the above assumptions into account, one might suggest that the way monads split up the self from the environment is dictated by preservation of the self as is any act or measurement the self has with its environment.  Therefore decisions an identity undertakes with respect to other beings must always be seen in the light of a maximal harmonisation benificial to the identity, irrespective whether the consequential actions are physical or purely gravitational (what most people would call ``mental'').  Therefore the concept of inherent sanity means that the \emph{immediate} or first order response of these actions with respect of the self are beneficial to the self.  However, it may be that the other identities to which these actions pertain might have a different perception even though they might be beneficial for their physical constitution (for example a post operational trauma resulting from succesful medical surgery\footnote{To be consistent here, a person having those feelings must be convinced that the foregoing desease must be because of bad interactions with others which have not been dealt with appropriately by the surgery and again the surgery done by doctors (potentially differing from the harmful group) must be understood as benevolent for the collective society as well.}).  Therefore a person, who might be inherently sane, and who receives negative reactions from his actions (even if completely unjustified because of misreadings of others) might be labelled by ``society'' as insane.  It is obvious that this entire process is relative to society as history has shown on many occasions, a social lunatic in Belgium for example might be a hero in South Africa.  I have used the word intelligence here before by which I really mean the extensiveness of the list of questions someone can ask over some period of time, the coherent character of his answers to these questions as well as the productiveness these factors have with respect to \emph{material} reality.  Of course this sets a rather objective benchmark for what \emph{sanity} is supposed to be since someone who can make trains, microwaves, cure cancer and so on clearly has a larger impact on nature than a person who cries that someone is the devil.  One must remark here that the use of objective ``pertains'' to our previous assumption that consciousness cannot change the laws of physics over some sufficiently long time period so that that there is no feedback between sociology and intelligence.  There remains of course the matter of social revolutions which do take place once in a while, where a person previously labelled as sociologically insane becomes sane; basically such revolutions come with modifications to the answers to as well as to the list of questions but not necessarily with an overall increase of intelligence in some neighborhood of space time.  The way in which it might dispose of (relative) good and evil is that the efficient readers can communicate the readings and needs of different persons to one and another; after all, the way we percieve others solely depends upon our own conspiration theories.  This would imply that the suggestion of dynamically emerging quasi local variables in the dynamical picture as well as the role consciousness plays in all of this is substantially larger than probably almost anyone could imagine up till now.  So, if I am allowed to, I guess society should be based upon recognition and respect and certainly those with a conscious lack of the latter should be penalized.  
\\* \\*
One must add that we made no effort to define as yet an observer; obviously, any monad perceives its vertical space in a passive way and the only question which remains is what criteria must be satisfied in order for this perception to be \emph{active}\footnote{In the sense that it causes a collaps of the wavefunction.}.  Experience dictates that monads won't grant this luxury on ``small'' timescales to electrons for example, otherwise we would not have observed the double slit experiments we made so far.   Therefore, only clever experiments in very large vacuum chambers would be able to tell us whether such thing occurs or not and if it would, one would have to grant some consciousness to electrons.  An even crazier idea would be that while the monad learns, it keeps a bookkeeping of all this information even if it materialistically belongs to the past; it could be that those ``images'' become dynamical variables and somehow would start interacting with the material world (probably gravitationally in first order).  These thoughts would lead one even further astray, \emph{assuming} that our consciousness can only perceive materialistic configurations, in the sense that the theory we have written down is one of humans and not one of monads.  This would imply that further unification could be reached by constructing a theory of mental ``images'' on the tangent bundle of some manifold, where the images again live on the vertical spaces\footnote{One cannot dispose of the manifold notion as some people tried to convince me since the monads still need to be labelled and certainly some topology is needed in order to proceed.}.  The notion of a particle then would be entirely relative depending upon the senses of the observer who is again a collective image of neighbouring monads.      
\chapter{The Unruh and Hawking effect revisited}
I urge the reader at this point to ``psychologically distantiate'' himself from what he \emph{believes} to be true.  Indeed, there is no experimental evidence so far that Hawking or Unruh radiation do exist and moreover, there are the well known theoretical arguments against the Hawking effect by 't Hooft and Susskind which stem from exact preservation of unitarity for the outside observer.  My problems with both effects do not have this origin, but I believe conventional quantum mechanics to fail here in another important way.  Mathematically, the Hawking and Unruh effect are treated in an almost isomorphic manner in the literature by going over to Rindler coordinates in the derivation of the former phenomenon although physically, both effects are very distinct.  This chapter is structured as follows: first I treat the Unruh effect and put forwards my objections against the standard interpretation\footnote{As mentioned previously, local Lorentz covariance on $TM$ suggests that no Unruh effect is observed; this is a \emph{weak} prediction of our theory in the sense that it holds for the natural class of observables with local Lorentz symmetry.}; next, I explain precisely why we observe thermal radiation in the case of Quantum Field Theory of a black hole and make the distinction with the Unruh effect.  Even though the latter could be retrieved by, amongst others, giving up upon ultralocal particles and going over to quasi-local ones, the Hawking effect resists such ``cure''.  Therefore, we are not going to try to repair it by constructing backreaction terms coming from quantum gravity; indeed, the computation I will do is as semiclassical as Hawking's with that difference that we use a better quantum theory.  \\* \\* 
Right, so let me start by giving two objective arguments against the Unruh effect and the textbook interpretation of free Quantum Field Theory in general.  A negative argument, which I have stressed also in the previous chapter, is that Quantum Field Theory cannot in general unambiguously answer the question what happens to a locally accelerated observer (by which I mean that you get into your car or personal spaceship and accelerate for some time).  Indeed, one would not even know a priori how to define particle notions, the correct vacuum state and so on since all those depend upon the foliations one chooses.   One could also entertain the thought that the vacuum state and particle notions are the usual ones and that one merely calculates local observables attached to an accelerated two state detector\footnote{I thank Juan Maldacena, who put forwards this point of view, for a useful conversation here.} carrying a trivial representation of the Poincar\'e algebra.  Apart from the fact that the relevant interaction Hamiltonians have insufficient symmetry properties since they are globally Poincar\'e covariant, but not \emph{locally} Poincar\'e covariant, the local reference frame set by the dector is physically distinguished and therefore particle clicks could occur. However, if the relevant observables merely were the local generators of the Poincar\'e algebra of the field expressed in the vielbein associated to the observer, then nothing would occur.  So, I would say the answer is rather inconclusive with perhaps a slight preference for ``no'', but this may just be my perception of the state of affairs. \\* \\*  
Of course, this is a negative argument since it does not directly contradict the validity of the physical effect; it only says that we cannot know (to the highest standards) if some effect is really there or not.  There is however also a positive argument which, I believe, has to do with the cosmological constant problem: the vacuum simply offers no resistance to acceleration which basically eliminates the picture of virtual vacuum fluctuations as being responsible for the huge predicted cosmological constant.  Indeed, the matter tensors which couple to gravity all have a local Lorentz symmetry and obey the standard flat conservation laws (in contradiction to standard relativity); therefore, it will not ``observe'' particles being created by acceleration in the free limit.  As mentioned previously, one might recover an Unruh like effect if one were to consider physics on higher jet bundles which would only make the gravitational theory more non-local than it already is.  Therefore, it seems there really isn't any convincing physical argument \emph{for} the existence of the Unruh effect and I would be very surprised if someone came up with one some day in the future.  \\* \\*
Now we come to the Hawking effect and a discussion of the physical meaning of the original calculation.  Just to be on the safe side, I do not doubt the mathematical validity of the calculation for both effects: the results have been confirmed through different means over and over again (for example Bisognano and Wichmann derived the KMS condition for the Unruh vacuum -before Unruh published his result- from axiomatic Quantum Field Theory).  What I do not agree with, is the physical interpretation which is what makes physics after all an exciting area to study.  There are two logically independent questions one can ask here: (a) do you think a black hole radiates ? (b) if so, did Hawking reveal the correct mechanism by which this happens ?  The anwers to any of these questions can again be further subdivided in different categories.  To anwer question (a), one must look for evidence coming from different directions which ``prove'' the necessity of black hole radiation.  There are two kinds of theoretical evidence and the second one is the most robust.  The first argument I am aware of comes from general considerations about local metric theories of gravity: indeed several authors have proven that the first and second law of thermodynamics must hold for general Killing bifurcation horizons in such theories.  Now, as we have shown in the previous chapter, these seemingly general results \emph{fail} for non-local non-metric theories of gravity where reasonably dark objects may be constructed nevertheless.  Therefore, this type of nonlocal geometries already carry some ``quantum hair'' and are therefore physically superior.  Hence, are these results really as robust as most people believe (as I did until I found out about the failure of these results for a wider class of gravity theories)?  Should one expect classical black hole solutions to be ``real'' in a deep physical sense?  We shall argue from a different perspective later on that the answer to this question is a definite no.  However, I am sure that all black holes radiate (irrespective of whether they are physical or not) because of the following more robust physical argument: the typical gravitational wavelength of a Schwarzschild black hole associated to $\alpha$ solar masses is of the order $10^3 \alpha$ meters in the neighborhood of the event horizon (measured in the standard Schwarzschild coordinates) while the wavelength producing the highest intensity in the black body radiation spectrum (at Hawking temperature) is given by Wien's law as $10^4 \alpha$ which matches perfectly (since I made some tiny numerical errors).  Therefore,  I believe (a) to be correct but I am not sure that Einsteinian black holes are physical, nor that the radiation spectrum is perfectly thermal; the usual argument here is that it would violate unitarity, but global unitarity such as people believe in does not seem to hold anyway in our approach.  The reason is the same as why global conservation laws such as preservation of energy-momentum or spin do not hold in general relativity.  Therefore, my attitude is to simply wait and see for what the proper calculations reveal to us.  The answer I will construct to (b) however is a resounding no since it directly violates our principle of local particle notions.  Here, the physical distinction with the Unruh effect emerges: in the Hawking effect, the observer firing the particles at the black hole from the asymptotic past and measuring them at the asymptotic future is never supposed to enter the bulk of the universe close to the black hole event horizon.  Indeed, it is simply so that for observers at infinity, inertial eigentime and Killing time coincide which allows to define idealized particle notions corresponding to ordinary plane waves (and we know already that gravitational corrections to this are singular).  Therefore the contrast with the Unruh effect is that the Hawking effect is not due to any local acceleration of an observer.  The point is however that real particles of definite momentum are not defined by those asymtotic plane waves and have a finite (instead of infinite) extend.  Therefore, there is nothing special about this Killing time physically and corrections to the Hawking result are going to be large (in either non analytic) \emph{depending} on how far from the black hole the observer shoots the particles (since our particle notions change from one region to another).  For observers close to infinity, one would not see large deviations from the standard result, for observers close to the horizon, matters may be very different.  Now, in contrast to many researchers, I do not find Hawking's conclusions troublesome: indeed, for black holes of solar mass, particles with wavelength less than say 10 meters are not going to feel the gravitational field near the event horizon and are not going to be scattered at all if one does not take into account backreaction effects of the geometry.  The latter however are extremely tiny and I claim it is very unlikely to restore unitarity by such mechanism.  Clearly, many of the black hole solutions in our new theory behave much like ordinary black holes at an ``effective'' event horizon simply because the gravitational force makes it very hard for light to escape (so that one effectively still has a dark object at the sky).  At the same time there are also black holes which behave very differently from ordinary looking black holes but I conjecture that those are all unstable due to Hawking radiation.       
\chapter{The gravitational Heisenberg uncertainty relations and a more advanced double slit experiment} 
In this chapter we accomplish several things at the same time: (a) first we want to study the Heisenberg uncertainty relations in the easiest curved spacetime (b) in order to reach this goal, we have to set up a perturbative formalism to solve the constraint equations for the unitary potential (c) this allows us to draw preliminary conclusions regarding the holographic principle.  Again, no general mathematical theorems are constructed here, we merely want to show that the theory works for non-trivial examples.  From the previous chapter we learn that the tetrad whose nonzero components are given by
\begin{eqnarray*} 
e^{t}_{0}(t, \vec{x}) & = & 1 + \frac{GM}{r} \\*   
e^{x^j}_{k}(t, \vec{x}) & = & \delta^j_k 
\end{eqnarray*} is the restriction to the origin of tangent space of a solution to the vacuum equations of motion (with somewhat unusual boundary conditions).  As before, we shut off interactions, that is $A_{\mu} = 0$ and therefore the constraint equations read
\begin{eqnarray*}
0 & = & i e^{\mu}_{a}(x) \left( \partial_{\mu}U(x)\right) U(x) \gamma^a U^{\dag}(x) + U(x) \int d^3 \vec{k} \, \gamma^a k_a a^{\dag}_{\vec{k}} a_{\vec{k}} \\*
0 & = & e^{\mu}_{a}(x) \left( \partial_{\mu}U(x) \right) U(x) \gamma^a U^{\dag}(x) - e^{\mu}_{a}(x) U^2(x) \gamma^a \left(U^{\dag}(x) \right)^2 \partial_{\mu}U(x)
\end{eqnarray*} where we surpress the details in the creation operators as well as the $\pm$ signs.  Since the gravitational field has a rotational symmetry around the origin of attraction, it appears obvious to only look for potentials $U(x)$ having $SO(3)$ as a symmetry group\footnote{We do not bother at this point about parity transformations.}.  So, what we should do is write out the most general unitary potential one can think of in perturbation theory around the free one and solve the system order by order in the creation and annihilation operators as well as the gamma matrices.  From chapter $7$ we know we can write $$U(x) = e^{i H(x)}$$ where $H(x)$ is Hermitian.   There are $16$ basic types of Clifford valued expressions which may occur for \emph{spinless} particles\footnote{For particles with spin, we get polarization vectors which transfrom covariantly too and which may serve to define more Clifford invariants.}: indeed, one constructs
$$1, \, \gamma^j x_j, \, \gamma^j k_j, \, \gamma^i \gamma^j k_i x_j$$ and all these terms may still be multiplied by $1, \gamma^0, \gamma^5$ and $\gamma^0 \gamma^5$ all coming with the appropriate factor of $i$ to ensure Hermiticity.  We write down the most general local Hamiltonian and solve for the constraint equations in perturbation theory.  The result is:
\begin{eqnarray*} 
H(x) & = & \int d^3 \vec{k} \, k_a x^a \, a^{\dag}_{\vec{k}}a_{\vec{k}} + \\* & & \sum_{i + j \geq 0} \int d \vec{k}_1 \ldots d \vec{k}_i \int d\vec{l}_1 \ldots d\vec{l}_j a^{\dag}_{\vec{k}_1} \ldots a^{\dag}_{\vec{k}_i} a_{\vec{l}_1} \ldots a_{\vec{l}_j}  \\*
& & \left( \alpha_{ij}^{1}(\vec{k}_r, \vec{l}_s, \vec{x},t) + \gamma^{0}\alpha_{ij}^{2}(\vec{k}_r, \vec{l}_s,\vec{x},t) + i \alpha_{ij}^{3}(\vec{k}_r, \vec{l}_s,\vec{x},t) \gamma^5 + 
\gamma^0 \gamma^5 \alpha_{ij}^{4}(\vec{k}_r, \vec{l}_s,\vec{x},t) \right) + \\*
& & \left( \alpha^{5}_{ij,h}(\vec{k}_r, \vec{l}_s,\vec{x},t) \gamma^h + i \gamma^0 \gamma^h \alpha^{6}_{ij,h}(\vec{k}_r, \vec{l}_s,\vec{x},t) + \gamma^5 \gamma^h \alpha^{7}_{ij,h}(\vec{k}_r, \vec{l}_s,\vec{x},t) + \gamma^0 \gamma^5 \gamma^h \alpha^{8}_{ij,h}(\vec{k}_r, \vec{l}_s,\vec{x},t)\right) \\*
& = & H_0(x) + H_{\textrm{int}}(x) \\*
\end{eqnarray*} where $$\alpha^{r}_{ij,\star}(\vec{k}_1 \ldots \vec{k}_i, \vec{l}_1 \ldots \vec{l}_j,\vec{x},t)^{*} = \alpha^{r}_{ji,\star}(\vec{l}_j \ldots \vec{l}_1, \vec{k}_i \ldots \vec{k}_1,\vec{x},t)$$ contain (higher) distributional parts and all coefficients are $SO(3)$ symmetric since the particles are defined with respect to the observer sitting in $VTM_{0}$ (where the gravitational field happens to be infinite - we shall cure that situation later on by replacing $r$ by $r + l_p$ where $l_p$ is the Planck length).  Likewise, the split in a free theory and interaction is inspired by the symmetry of the problem: we have chosen the natural vielbein upon an $SO(3)$ rotation in the coordinates or vielbein indices separately.  We have surpressed the dependency of the $\alpha^{r}_{ij}$ upon the polarization vectors of the particles in case they have a nonzero spin and all other quantum numbers as well.  Moreover, these functions should have the same symmetry as the particle creation and annihilation operators have.  For simplicity, we assume one species of spinless bosons: fermions or mixed spin and statistics can be treated likewise by considering other Young tableaux and more Clifford invariants.  Therefore, all functions are symmetric in the $\vec{k}_r$ and $\vec{l}_s$ separately and no internal quantum numbers matter apart from the rest mass $m$.  Moreover, the $\alpha_{ij,gh}^{r}$ are antisymmetric in $g$ and $h$ simply for not double counting certain terms.  Until now, we have surpressed another important index in our notation and we shall only reveal this extra level of structure when ambiguities may occur.  That is, according to AXIOM 0, any particle of species $N$ and quantum numbers associated to the Poincar\'e algebra, appears in an infinite number of copies due to localization as we explained at the end of chapter $8$.  Therefore, any creation operator $a^{\dag}_{\vec{k}, n}$ must come with an index $n$ which is a natural number; we shall actually do more than just taking AXIOM 0 as an absolute truth, we will try to prove that the theory gets inconsistent without this assumption.  Likewise, the functions $\alpha_{ij}^{r}$ should be thought of as 
$\alpha_{n_1 \ldots n_i m_1 \ldots m_j}^r$ and a permutation of the particles also involves a permutation of the $n_r, m_s$.  As said before, we shall use this notation where it is important.  We write the equations of motion in perturbation series and collect the terms order by order in the operators and Clifford numbers.  In the computations below, we first ignore, unless explicitly mentioned otherwise, the particle copies.  It is natural to start by computing $U(x) \gamma^a U^{\dag}(x)$, which can be written in perturbation series\footnote{We ignored $\hbar$ here, but every power of  $H(x)$ comes with $\hbar$.} as
$$U(x) \gamma^a U^{\dag}(x) = \gamma^a + i \left[ H(x), \gamma^a \right] - \frac{1}{2} \left[ H(x) , \left[ H(x) , \gamma^a \right] \right] + \ldots$$
In the Dyson picture this second order perturbation equals 
$$\gamma^a + i \left[ H_{\textrm{int}}(x), \gamma^{a} \right] - \frac{1}{2} \left[ H_{\textrm{int}}(x), \left[ H_{\textrm{int}}(x), \gamma^a \right] \right] - \frac{1}{2} \left[ \left[H_{0}(x) , H_{\textrm{int}}(x) \right], \gamma^a \right]  $$ which is exactly the same because $H_{0}(x)$ commutes with $\gamma^a$.  The first commutator is computed to be
\begin{eqnarray*}
\left[ H(x) , \gamma^{0} \right] & = & \sum_{i + j \geq 0} \int d^3 \vec{k}_1 \ldots d^3 \vec{k}_i 
\int d^3 \vec{l}_1 \ldots \vec{l}_j \, a^{\dag}_{\vec{k}_1} \ldots a^{\dag}_{\vec{k}_i} a_{\vec{l}_1} \ldots a_{\vec{l}_j} \\*
& & \left( 2i \gamma^5 \gamma^0 \alpha_{ij}^3(\vec{k}_r, \vec{l}_s, \vec{x},t) + 2\gamma^5 \alpha_{ij}^4(\vec{k}_r,\vec{l}_s, \vec{x},t) \right) + 2 i  \alpha_{ij,r}^{6}(\vec{k}_r, \vec{l}_s, \vec{x},t) \gamma^r  + \\* & & 2 \alpha_{ij,r}^5(\vec{k}_r, \vec{l}_s, \vec{x},t) \gamma^r \gamma^0 \end{eqnarray*}
and the remaining expressions are
\begin{eqnarray*}
\left[ H(x) , \gamma^{s} \right] & = & \sum_{i + j \geq 0} \int d^3 \vec{k}_1 \ldots d^3 \vec{k}_i 
\int d^3 \vec{l}_1 \ldots \vec{l}_j \, a^{\dag}_{\vec{k}_1} \ldots a^{\dag}_{\vec{k}_i} a_{\vec{l}_1} \ldots a_{\vec{l}_j} \\*
& & 2 \gamma^0 \gamma^s \alpha_{ij}^{2}(\vec{k}_r, \vec{l}_p, \vec{x},t) + 2i \gamma^5 \gamma^s \alpha_{ij}^{3}(\vec{k}_r, \vec{l}_p, \vec{x},t) + 2 \gamma^{[k}\gamma^{s]} \alpha^{5}_{ij,k}(\vec{k}_r, \vec{l}_p, \vec{x},t) + \\*
& & 2i \gamma^0 \alpha_{ij,s}^{6}(\vec{k}_r, \vec{l}_p, \vec{x},t) + 2 \gamma^5 \alpha_{ij,s}^{7}(\vec{k}_r, \vec{l}_p, \vec{x},t) + 2 i \epsilon_{0r}^{\,\,\,\,\,\,\, k s}\alpha_{ij,k}^{8}(\vec{k}_r, \vec{l}_p, \vec{x},t) \gamma^{r}. 
\end{eqnarray*}
Obviously, higher commutation relations will contain singular terms but those will be required to vanish and constrain therefore the functions $\alpha_{ij,\star}^r$.  This is not a renormalization program, since there are no ad-hoc free parameters one must adjust: on the contrary, all those consistency requirements limit the number of possible theories.  \emph{That} is what the holographic principle is about; from this point on, we have to activate the natural labels of identical particles which requires the study of new representations within the Wigner scheme \cite{Weinberg} (we postpone the construction of such representations on Clifford-Nevanlinna modules for later work).  This task is by no means simple and albeit the geometrical picture is quite easy, the construction of the representation is quite daunting.  The physical idea is the following: it is clear that gravity will modify drastically the Heisenberg uncertainty principle for ``local'' plane waves and therefore one imagines inertial space in $VTM_{0}$, defined by the coordinates $v^1,v^2$ and $v^3$, to be a regular cubic lattice containing the point $v = 0$.  This lattice is not space, but constitutes the observational framework of this particular observer.  A boosted observer defines a different lattice as does a rotated one, this is not an issue; it would be a problem however is the lattice were ``objectively'' representing spacetime which it does not.  So, one has a fundamental length scale $\lambda$ which should be determined by the particle mass and gravitational field.  Each lattice point has three integer coordinates $n = (n^1,n^2,n^3)$, $n^i \in \mathbb{Z}$ and defines creation and annihilation operators $a^{\dag}_{\vec{k},n}, a_{\vec{k},n}$ which, as explained before, all commute for $n \neq m$ and act on the local vacuum $| 0, e_{b}(x) ,x \rangle$ as usual.  Before we proceed, let me explain how this connects with work by 't Hooft on the black hole information puzzle; in his view, the event horizon is subdivided in elementary plaquettes which each carry a bit of information and in his model incoming radiation leaves a ``print'' on the event horizon such that no information gets lost.  So, after a lecture of him in Utrecht, I asked how this was possible since the observer at infinity has an infinite number of degrees of freedom at his disposal, all (approximately) plane waves which hit the event horizon of the black hole.  Nevertheless the black hole only has a finite information, how does this square?  Where happens the transition from infinite to finite?  The kind of answer 't Hooft would need is that a fundamental scale and therefore a lower uncertainty in momentum space (implying an upper uncertainty in the position) exists for a free particle at infinity, otherwise some information loss is unavoidable.  Since I am sure global information loss will occur, but the introduction of some fundamental scale $\lambda$ is nevertheless a good idea, let me proceed.  Perhaps one would like $\lambda$ to be a \emph{local} dynamical variable in the light of the motivation in chapter eight since the \emph{expected} uncertaintly of a local free particle at $x$ would depend upon various characteristics of the gravitational field at that point.  A specific formulation of such idea is not going to be easy since stable free particles of infinite extend do exist on static spacetimes and one would expect anisotropies in the Ricci tensor to induce a direction dependency (in the vielbein variables) in such scale factor too.  Note that all of this is in the mind where an observer at $x$ would again make some assumptions about the whole of real spacetime based upon local information and one might go even that far as to let those assumptions to vary from observer to observer which would introduce additional uncertainties.  I will not proceed in this way in this book and assume $\lambda$ to be kinematical input.  \\* \\*
Since $\lambda$ is a scale which is \emph{invariant} with respect to local Lorentz boosts (it is here that some idea of double special relativity enters but of course not in the naive way those people think about it), one has to determine how the particle notions at $n' = \Lambda(v)$ depend upon those with respect to $n$, for $n$ in the neighborhood of $v$.  There is no real universal answer to this and plenty of acceptable representations could exist which we shall study now.  We present the crucial extensions regarding the analysis made by Wigner \cite{Weinberg} and give a self-contained treatment here for massive particles with mass $m$ and internal quantum numbers $\sigma$ and copies $n$.  The Clifford numbers are not considered in this analysis and an extension in this direction is postponed for future work.  Let $\Psi_{k, \sigma, n}$ be a one particle state with four momentum $k^{a}$, then one computes that $$P^a U(\Lambda) \Psi_{k,\sigma,n} = \Lambda^{a}_{\,\,\, b}k^b U(\Lambda) \Psi_{k, \sigma, n}$$ and therefore
$$U(\Lambda)\Psi_{k,\sigma,n} = C_{(\sigma',n')(\sigma,n)} \Psi_{\Lambda k, \sigma',n'}.$$  Now, in constructing the representation any further, Weinberg chooses as standard vector $p = (m,0,0,0)$ as well as particular Lorentz transformations $L(k)$ such that $k = L(k)p$.  Next, he defines 
$$\Psi_{k,\sigma} = N(k) U(L(k)) \Psi_{p,\sigma}$$ where $N(k)$ is a normalization factor.  The obvious extension for representations with identical copies is
$$\Psi_{k, \sigma, n} = N(k) E_{n'n}(L(k)) U(L(k)) \Psi_{p, \sigma, n'}$$ and we may proceed by calculating 
$$U(\Lambda) \Psi_{k,\sigma,n} = \frac{N(k)}{N(L(\Lambda k))} D_{(\sigma' n'')(\sigma n')}(W(\Lambda,k)) E_{n' n}(L(k)) E^{-1}_{n''' n''}(L(\Lambda k)) \Psi_{\Lambda k, \sigma', n'''}$$ where $$W(\Lambda , k) = L^{-1}(\Lambda k) \Lambda L(k)$$ is an element of the little group of $p$.  Obviously, $D$ is a representation of the little group and in general, the following scalar product holds
$$\left( \Psi_{p,\sigma,n} , \Psi_{p, \sigma', n'} \right) = \delta_{\sigma ,\sigma'} \delta_{n,n'} \delta^{3}(0).$$
Therefore, $D$ is unitary and we require the following extension to hold
$$\left( \Psi_{k,\sigma,n} , \Psi_{k',\sigma', n'} \right) = \delta^{3}(\vec{k} - \vec{k}') \delta_{\sigma,\sigma'} \delta_{n,n'}$$ which implies that 
$$\delta^{3}(0) |N(k)|^2 E^{*}_{m,n}(L(k)) E_{m,n}(L(k)) = \delta^{3}(\vec{k} - \vec{k}') \delta_{n,n'}.$$
By shifting the renormalization constants and picking the correct phase, we may write that
$$N(k) = \sqrt{\frac{m}{k^0}}$$ and
$$E^{\dag}(L(k))E(L(k)) = 1.$$
So our extension of Wigner's formalism consists in the existence of additional unitary operators which do not depend upon the little group, but solely upon the representative Lorentz boosts $L(k)$.  These do not form a group and therefore $E$ is not a representation either (however, the combined expression obviously is).  Let us first define $L(k)$ as
\begin{eqnarray*}
L(k)^{0}_{\,\,\, 0} & = & \gamma(k) \\*
L(k)^{0}_{\,\,\, j} & = & \widehat{k}_j \sqrt{\gamma(k)^2 - 1} \\*
L(k)^{i}_{\, \,\, j} & = & \delta^i_j + (\gamma(k) - 1)\widehat{k}^i \widehat{k}_j  
\end{eqnarray*} where $\gamma(k) = \frac{\sqrt{|\vec{k}|^2 + m^2}}{m}$ and $\widehat{k}_j = \frac{k_j}{|\vec{k}|}$.  Now, we study a simple class of such representations by putting $E = 1$, which simply means that a local particle basis at $n$ only depends on the local generating state $\Psi_{p,\sigma,n}$.  Next, we suppose that 
$$D_{(\sigma' n')(\sigma n)} = F_{\sigma' \sigma} G_{n' n}$$ which physically means that internal labels do not twist with transformations of space.  We do know what $F$ looks like from the analysis of Wigner, we just don't know yet what $G$ has to be.  Obviously, it must be an infinite dimensional representation of the rotation group which can be written as an infinite direct sum of ordinary $(2j +1)$-dimensional spin representations.  The remainder of this intermezzo consists in finding reasonable constraints upon it.  \\* \\*
So, all we have to do is to consider the geometrical problem of space rotations and the corresponding issue for boosts gets automatically solved.  Actually, AXIOM III puts the first constraint on the construction because we require two Poincar\'e groups, a local one associated to $n = 0$ and a global one defined by all $n$.  So, particles with $n' \neq 0$ cannot mix with particles corresponding to $n = 0$, which implies that $G_{n' 0} = G_{0 n'} = 0$ and $G_{0 0} = 1$.  One may think that  the cristallographic group $\mathcal{C}$ of the cubic lattice plays an important role: the latter is generated by the rotations over $90$ degrees over the main axes as well as the translations.  We already know that translation covariance is broken since the origin of tangent space is distinguished; moreover, it is impossible for the rotations $g \in \mathcal{C}$ to satisfy
$$G_{n' n}(g) = \delta_{n', g(n)} e^{i \theta(n,g)}$$ where
$$\theta(n, hg) = \theta(gn,h) + \theta(n,g)$$  
since this would require a spin one representation of the rotation group which is, obviously, impossible.  Finally, one might think it is natural to demand that
$$G_{n' n}(R) = f(n' - Rn).$$ 
However, one can easily argue that such function does not exist; the group property requirement implies the following simpeler condition:
$$f(n) = \sum_{m} f(n - Tm)f(Tm)$$ which must hold for all $T$.  Writing down a formal Laurent series expansion for $f$ implies that one has precisely the same number of equations in $T$ as one has coefficients in $f$ assuming suitable convergence criteria.  Since these equations must hold for all $T$, the associated coefficients must vanish, leaving for an overdetermined system in terms of the original coefficients.  Hence, the group law cannot be satisfied implying that we must look for another, more general, ansatz.  The latter must be something of the most general form $f(n,R,m)$ since combining $R$ with $m$ does not work.   
\\* \\*
The geometry of the problem suggests to consider the spheres $S_m$ of equal distance $m$ to the orgin, where the latter is defined by means of the lattice metric.  An easy argument shows that the number of points in $S_m$ equals $4 m^2 + 2$ and it is natural to consider this as the dimension of a $2 j + 1$ dimensional representation of $SU(2)$.  Obviously $2j$ is always odd and therefore the rotations mixing up the different copies of particles have all half integer spin.  There exists, however, an alternative construction in which all irreducible components have integer spin; that is the one particle space has a splitting into subspaces spanned by
$$\frac{1}{\sqrt{2}} \left( \Psi_{k, \sigma, n} + \psi_{k,\sigma, -n} \right)$$ and
$$\frac{1}{\sqrt{2}} \left( \Psi_{k, \sigma, n} - \psi_{k,\sigma, -n} \right).$$
Every subspace has dimension $2m^2 + 1$ and therefore $j = m^2$.  The remaining problem now is to find a suitable representation of the ladder operators which leaves us with considering the Fourier transform.  The latter is natural since on $\mathbb{Z}^3$ a canonical order exists.  That is $(n_1,n_2,n_3) < (m_1,m_2,m_3)$ if and only if there exists a $j$ such that $n_j < m_j$ and for all $k < j$ we have that $n_k = m_k$.  Therefore, all points on half of $S_m$ can be labelled from $- m^2$ to $m^2$ and one can define
$$|z, m^2 , r \rangle = \sum_{s = 0}^{2m^2} e^{\frac{2 \pi i r s}{2 m^2 + 1}} |z,s \rangle$$ up to a global (r independent) phase, where $s$ uniquely corresponds to an element of $S_m$.  This representation does not coincide with the defining one for $m = 1$; indeed, the former states are eigenstates of the permutation $P = (12 \ldots 2 m^2 + 1)$ and therefore represent a form of ``cyclic'' statistics, the $r = 0$ state being invariant under the entire permutation group.  Other canonical permutations would be those which merely swap the indices; that is $P = (123)$ and $$U(P) |z,(n_1,n_2,n_3) \rangle = |z,(n_2,n_3, n_1) \rangle.$$  The eigenvalues are of course again all $2m^2 + 1$ roots of unity and a different Fourier transform as before may be set up.  All other representations appear to me to be variations on the same theme.  Obviously, instead of spheres, one may consider (half) spherical shells and construct representations as before.       
\\* \\*   
At first, we have to construct a Dyson perturbation theory based upon the natural split of $H = H_0 + H_{\textrm{int}}$ and not just into $H$.  To that end, we write 
$$U(x) = \left( e^{iH(x)} e^{-iH_0(x)} \right) e^{iH_0(x)}$$ and perturb the expression between brackets up to $n$'th order in $H_0, H_{\textrm{int}}$.  In second and third order, this gives respectively
$$U(x) = \left( 1 + iH_{\textrm{int}}(x) - \frac{1}{2}H_{\textrm{int}}^{2}(x) - \frac{1}{2} \left[ H_{0}, H_{\textrm{int}}(x) \right] \right)e^{iH_0(x)}$$ and
\begin{eqnarray*} U(x) & = & \left( e^{iH_{\textrm{int}}(x)}_{|3} - \frac{1}{2} \left[ H_{0}(x), H_{\textrm{int}}(x) \right] - \frac{i}{6} \left[ H_{0}(x), \left[ H_{0}(x), H_{\textrm{int}}(x) \right] \right] \right)e^{iH_{0}(x)} + \\*
& & \left( \frac{i}{12} \left[ H_{\textrm{int}}(x), \left[ H_{\textrm{int}}(x),H_{0}(x) \right] \right] - \frac{i}{4} \left[H_{0}(x), H_{\textrm{int}}^{2}(x) \right] \right)e^{iH_0(x)}.
\end{eqnarray*}
Therefore, in more generality, with $$U(x) \equiv_{n} \left( 1 + \delta_n U(x) \right)e^{iH_{0}(x)},$$ the equations of motion reduce to
\begin{eqnarray*}
0 & = & - e^{\mu}_{a}(x) \left( 1 + \delta_n U(x) \right) e^{iH_{0}(x)} \int d^3 \vec{k} k_{\mu} a^{\dag}_{\vec{k}} a_{\vec{k}} \left( \delta_n U(x)
 \gamma^a + \gamma^a \delta_n U^{\dag}(x) + \delta_n U(x) \gamma^a \delta_n U^{\dag}(x) \right) + \\*
 & &  i e^{\mu}_a(x) \partial_{\mu} \left( \delta_n U(x) \right) e^{iH_{0}(x)}\left( 1 + \delta_n U(x) \right)\gamma^a \left( 1 + \delta_n U^{\dag}(x) \right)  \end{eqnarray*} and
\begin{eqnarray*}
0 & = & i e^{\mu}_{a}(x) \left(1 + \delta_n U(x) \right) e^{iH_{0}(x)} \int d^{3} \vec{k} k_{\mu} a^{\dag}_{\vec{k}} a_{\vec{k}} \left( \delta_n U(x) \gamma^a + \gamma^a \delta_n U^{\dag}(x) + \delta_n U(x) \gamma^a \delta_n U^{\dag}(x) \right) \\* 
& & + \, e^{\mu}_{a}(x) \partial_{\mu} \left( \delta_n U(x) \right) e^{iH_{0}(x)} \left( 1 + \delta_n U(x) \right) \gamma^a \left( 1 + \delta_n U^{\dag}(x) \right) \\* 
& & - \, e^{\mu}_{a}(x) \left[ \left( \left( 1 + \delta_n U(x) \right)e^{iH_{0}(x)} \right)^2 \gamma^a \left(e^{-iH_{0}(x)} \left( 1 + \delta_n U^{\dag}(x) \right) \right)^2 - \gamma^a \right] \partial_{\mu} \left(  \left(  1 + \delta_n U(x) \right)e^{iH_{0}(x)} \right) - \\*
& & e^{\mu}_a(x) \gamma^a \partial_{\mu} \left( \delta_n U(x) \right)e^{iH_{0}(x)} + i e^{\mu}_{a}(x) \left[ \delta_n U(x), \gamma^{a} \right] e^{iH_{0}(x)} \int d^{3} \vec{k} k_{\mu} a^{\dag}_{\vec{k}} a_{\vec{k}}. 
\end{eqnarray*}
In all intermediate computations, it is useful to notice that the product of 
$$\int d\vec{k}_1 \ldots d\vec{k}_i d\vec{l}_1 \ldots d\vec{l}_j a^{\dag}_{\vec{k}_1} \ldots a^{\dag}_{\vec{k}_i} a_{\vec{l}_1} \ldots a_{\vec{l}_j} \beta_{ij} \left( \vec{k}_g, \vec{l}_h, \vec{x},t \right)$$ with $$\int d\vec{r}_1 \ldots d\vec{r}_p d\vec{s}_1 \ldots d\vec{s}_q a^{\dag}_{\vec{r}_1} \ldots a^{\dag}_{\vec{r}_p} a_{\vec{s}_1} \ldots a_{\vec{s}_q} \beta_{pq} \left( \vec{r}_g, \vec{s}_h, \vec{x},t \right)$$ where the $\beta_{ij}$ are Clifford valued functions, equals
$$\sum_{a = 0}^{\min \{ j,p \}} \int \sum_{j_1 < j_2 < \ldots j_a} \frac{1}{(j - a)!} \sum_{\sigma \in S_j} d \vec{k}_1 \ldots d\vec{k}_i d\vec{r}_1 \ldots d\vec{r}_p d\vec{l}_{\sigma(1)} \ldots \widehat{d\vec{l}_{\sigma(j_1)}} \ldots \widehat{d\vec{l}_{\sigma(j_a)}} \ldots d\vec{l}_{\sigma(j)} d\vec{s}_1 \ldots d\vec{s}_q$$  
$$a^{\dag}_{\vec{k}_1} \ldots a^{\dag}_{\vec{k}_i} a^{\dag}_{\vec{r}_1} \ldots 
\widehat{a^{\dag}_{\vec{r}_{j_1}}} \ldots \widehat{a^{\dag}_{\vec{r}_{j_a}}} \ldots a^{\dag}_{\vec{r}_p} a_{\vec{l}_{\sigma(1)}} \ldots \widehat{a_{\vec{l}_{\sigma(j_1)}}} \ldots \widehat{a_{\vec{l}_{\sigma(j_a)}}} \ldots a_{\vec{l}_{\sigma(j)}} a_{\vec{s}_{1}} \ldots a_{\vec{s}_{q}}$$ 
$$\beta_{ij}(\vec{k}_g, \vec{l}_{\sigma(1)}, \ldots, \vec{r}_{j_1}, \ldots , \vec{r}_{j_a}, \ldots, \vec{l}_{\sigma(j)}, \vec{x},t) \beta_{pq}(\vec{r}_g, \vec{s}_h , \vec{x},t)$$ where we have used permutation symmetry of the creation operators and the $\beta$ functions respectively (since we have put $r_{j_k}$ on the $j_k$'th index where normally it should be on the $\sigma(j_k)$'th index).  This can still be further reduced to 
$$\sum_{a = 0}^{\min \{ j,p \}} a! \left( \begin{array}{c}
j \\ a
\end{array}  \right)\left( \begin{array}{c}
p \\ a
\end{array}  \right) \int d \vec{k}_1 \ldots d\vec{k}_i d\vec{r}_1 \ldots d\vec{r}_p d\vec{l}_{1} \ldots d\vec{l}_{j-a} d\vec{s}_1 \ldots d\vec{s}_q $$
$$a^{\dag}_{\vec{k}_1} \ldots a^{\dag}_{\vec{k}_i} a^{\dag}_{\vec{r}_{a + 1}} 
\ldots a^{\dag}_{\vec{r}_{p}} a_{\vec{l}_{1}} \ldots a_{\vec{l}_{j-a}}a_{\vec{s}_{1}} \ldots a_{\vec{s}_{q}}\beta_{ij}(\vec{k}_g, \vec{l}_1, \ldots, \vec{l}_{j-a},\vec{r}_1, \ldots, \vec{r}_{a}, \vec{x},t) \beta_{pq}(\vec{r}_g, \vec{s}_h , \vec{x},t).$$
The second order commutator 
$$\left[ H(x), \left[ H(x) , \gamma^0 \right] \right]$$ is given by the following lengthy expression:
\begin{eqnarray*}
& & \sum_{p + q \geq 0} \sum_{i + j \geq 0} \sum_{a = 0}^{\min \{ q ,i \}} a! \left( \begin{array}{c} q \\ a \end{array} \right) \left( \begin{array}{c} i \\ a \end{array} \right) \int d \vec{r}_1 \ldots d \vec{r}_p d \vec{s}_1 \ldots d \vec{s}_{q - a} d \vec{k}_1 \ldots d \vec{k}_i d \vec{l}_1 \ldots d \vec{l}_j \\*
& & a^{\dag}_{\vec{r}_1} \ldots a^{\dag}_{\vec{r}_p} a^{\dag}_{\vec{k}_{a+1}} \ldots a^{\dag}_{\vec{k}_i} a_{\vec{s}_1} \ldots a_{\vec{s}_{q - a}} a_{\vec{l}_1} \ldots a_{\vec{l}_j} \\*
& & \left( - 4i \alpha_{pq}^4 \alpha_{ij}^3 + 4i \alpha_{pq}^3 \alpha_{ij}^4 + 4i \alpha_{pq,h}^5 \alpha_{ij}^{6 \,\, ,h} - 4i \alpha_{pq,h}^6 \alpha_{ij}^{5 \,\, ,h} \right) + \\* & & \gamma^0 \left( - 4 \alpha_{pq}^3 \alpha_{ij}^3 + 4 \alpha_{pq}^4 \alpha_{ij}^4
- 4 \alpha_{pq,h}^6 \alpha_{ij}^{6 \,\,\, ,h} + 4 \alpha_{pq,h}^5 \alpha_{ij}^{5 \,\,\, ,h} \right) + \\*
& & i \gamma^5 \left( 4 \alpha_{pq}^{(2}\alpha_{ij}^{3)} - 4i \alpha_{pq}^{[1}\alpha_{ij}^{4]} + 4 \alpha_{pq,h}^{(6}\alpha_{ij}^{7) \,\, ,h} - 4i \alpha_{pq,h}^{[8}\alpha_{ij}^{5] \,\, ,h}  \right) + \\*
& & \gamma^0 \gamma^5 \left( - 4i \alpha_{[pq|}^{1} \alpha_{ij]}^{3} + 4 \alpha_{(pq|}^{2}\alpha_{ij)}^{4} - 4i \alpha_{[pq|,h}^{6}\alpha_{ij]}^{8 \,\, ,h}
- 4 \alpha_{(pq|,h}^{7}\alpha_{ij)}^{5 \,\, ,h} \right)  + \\*
& & \gamma^h \left( 4i \alpha_{[pq|}^{3}\alpha_{ij],h}^{8} - 4 \alpha_{(pq|}^{4}\alpha_{ij),h}^{7} + 4i \alpha_{[pq|}^{1}\alpha_{ij],h}^{6} + 4
\epsilon_{0h}^{\,\,\, rs} \alpha_{(pq|,r}^{6}\alpha_{ij),s}^{8} -  4 \alpha_{(pq|}^{2}\alpha_{ij),h}^{5} - 4i \epsilon_{0h}^{\,\,\, rs} \alpha_{[pq|,r}^{7} \alpha_{ij],s}^{5} \right) + \\*
& & i \gamma^0 \gamma^h \left( 4 \alpha_{(pq|}^{3} \alpha_{ij),h}^{7} - 4i \alpha_{[pq|}^{4}\alpha_{ij],h}^{8} + 4 \alpha_{(pq|}^{2}\alpha_{ij),h}^{6}
- 4i \epsilon_{0h}^{\,\,\, rs} \alpha_{[pq|,r}^{6} \alpha_{ij],s}^{7}  + 4i  \alpha_{[pq|}^{1} \alpha_{ij],h}^{5} - 4 \epsilon_{0h}^{\,\,\,rs} \alpha_{(pq|,r}^8 \alpha_{ij),s}^5 \right) + \\*
& & \gamma^5 \gamma^h \left( - 4 \alpha_{(pq|}^3 \alpha_{ij),h}^6 - 4 \alpha_{(pq|}^3 \alpha_{ij),h}^6  - 8 \alpha_{(pq|,h}^5 \alpha_{ij)}^4 + 4i \epsilon_{0h}^{\,\,\, rs} \alpha_{[pq|,r}^6 \alpha_{ij],s}^6 - 4i \epsilon_{0h}^{\,\,\,\, kr}\alpha_{[pq|,k}^{5}\alpha_{ij],r}^{5} \right) + \\*
& & \gamma^0 \gamma^5 \gamma^h  \left( - 8i \alpha_{[pq|,h}^6 \alpha_{ij]}^4 + 4i \alpha_{[pq|}^3 \alpha_{ij],h}^{5} + 4 \epsilon_{0h}^{\,\,\, rs} \alpha_{(pq|,r}^6 \alpha_{ij),s}^5 - 4 \epsilon_{0h}^{\,\,\, rs} \alpha_{(pq|,r}^5 \alpha_{ij),s}^6 - 4i \alpha_{[pq|,h}^{5}\alpha_{ij]}^{3} \right)
\end{eqnarray*} 
where $\left[ ij | pq \right] = \frac{1}{2} \left( ij * pq - pq * ij \right)$ abstractly and the same for the round bracket.  All equal monomials in the creation and annihilation operators can be gathered by rewriting the sum as
$$\sum_{m + n \geq 0} \,\,\, \sum_{i,q \geq 0; a = \max\{0, i - m, q-n \} \ldots \min\{i,q\}; j = n + a - q; p = m + a - i}$$ resulting in infinite complex quadratic forms in the $\alpha's$ as coefficient structures.  A similar calculation for $$\left[ H(x), \left[ H(x), \gamma^h \right] \right]$$ results in 
\begin{eqnarray*}
& & \sum_{p + q \geq 0} \sum_{i + j \geq 0} \sum_{a = 0}^{\min \{ q ,i \}} a! \left( \begin{array}{c} q \\ a \end{array} \right) \left( \begin{array}{c} i \\ a \end{array} \right) \int d \vec{r}_1 \ldots d \vec{r}_p d \vec{s}_1 \ldots d \vec{s}_{q - a} d \vec{k}_1 \ldots d \vec{k}_i d \vec{l}_1 \ldots d \vec{l}_j \\*
& & a^{\dag}_{\vec{r}_1} \ldots a^{\dag}_{\vec{r}_p} a^{\dag}_{\vec{k}_{a+1}} \ldots a^{\dag}_{\vec{k}_i} a_{\vec{s}_1} \ldots a_{\vec{s}_{q - a}} a_{\vec{l}_1} \ldots a_{\vec{l}_j} \\*
& & \left( 8i \alpha_{[pq|}^{6 \,\,\, ,h} \alpha_{ij]}^{2} - 8i \alpha_{[pq|}^{7 \,\,\, ,h}\alpha_{ij]}^{3} + 8i \epsilon_{0}^{\,\, hkr} \epsilon_{[pq|,k}^{8} \alpha_{ij],r}^{5} \right) + \\*
& & \gamma^{0} \left( - 4 \alpha_{(pq|}^{5 \,\,\, ,h}\alpha_{ij)}^{2} - 4i \alpha_{[pq|}^{8 \,\,\, ,h}\alpha_{ij]}^{3} - 4i \epsilon_{0}^{\,\,\, hrs} \epsilon_{[pq|,r}^{7} \alpha_{ij],s}^{5} + 4i \alpha_{[pq|}^{1} \alpha_{ij]}^{6 \,\,\, ,h} + 4 \alpha_{(pq|}^{4} \alpha_{ij)}^{7 \,\,\, ,h} + 4 \epsilon_{0}^{\,\,\, hrs} \alpha_{[ij|,r}^{8} \alpha_{pq],s}^{6} \right) + \\*
& &  i \gamma^5 \left( 4i \alpha_{[pq|}^{8 \,\,\, ,h} \alpha_{ij]}^{2} - 4 \alpha_{(pq|}^{5 \,\,\, ,h} \alpha_{ij)}^{3} - 4i \epsilon_{0}^{\,\, hrs} \alpha_{[pq|,r}^{6} \alpha_{ij],s}^{5} + 4 \alpha_{(pq|}^{4} \alpha_{ij)}^{6 \,\,\, ,h}  - 4i \alpha_{[pq|}^{1} \alpha_{ij]}^{7 \,\,\, ,h} - 4 \epsilon_{0}^{\,\, hrs} \alpha_{(ij|,r}^{8} \alpha_{pq),s}^{7} \right) + \\*
& & \gamma^0 \gamma^5 \left( - 4 \alpha_{(pq|}^{7 \,\,\, ,h}\alpha_{ij)}^{2} + 4 \alpha_{(pq|}^{6 \,\,\, ,h} \alpha_{ij)}^{3}  + 4i \epsilon_{0}^{\,\, hrs}\alpha_{[pq|,r}^{5} \alpha_{ij],s}^{5} + 4 \alpha_{(pq|}^{3} \alpha_{ij)}^{6 \,\,\, ,h} + 4 \alpha_{(pq|}^{2} \alpha_{ij)}^{7 \,\,\, ,h} - 4i \epsilon_{0}^{\,\, hrs} \alpha_{[ij|,r}^{8} \alpha_{pq],s}^{8} \right) + \\*
& & \gamma^r \left( - 4 \alpha_{(pq|}^{2} \alpha_{ij)}^2 \delta^{h}_{r} + 4i \epsilon_{0r}^{\,\,\,\, sh} \alpha_{[pq|,s}^{7} \alpha_{ij]}^{2} - 4 \alpha_{(pq|}^3 \alpha_{ij)}^3 \delta^{h}_{r} + 4i \epsilon_{0r}^{\,\,\,\, sh} \alpha_{[pq|}^{4} \alpha_{ij],s}^{5} + 4 \alpha_{(pq|,s}^{5} \alpha_{ij)}^{5 \,\,\,\, ,s}\delta^{h}_{r} \right) + \\*
& & \gamma^r \left( - 4 \alpha_{(pq|}^{5 \,\,\, ,h} \alpha_{ij),r}^{5} - 4 \alpha_{(pq|,r}^{6} \alpha_{ij)}^{6 \,\,\, ,h} - 4 \alpha_{(pq|,r}^{7} \alpha_{ij)}^{7 \,\,\, ,h} + 4i \epsilon_{0r}^{\,\,\,\, sh}\alpha_{[pq|}^{1}\alpha_{ij],s}^{8} - 4 \epsilon_{0r}^{\,\,\,\, st} \alpha_{(pq|,s}^{8} \epsilon_{0t}^{\,\,\, \, kh} \alpha_{ij),k}^{8} \right) + \\*
& & i \gamma^{0} \gamma^{r} \left(- 4i \alpha_{[pq|}^1 \alpha_{ij]}^{2} \delta^{h}_{r} + 4 \epsilon_{0r}^{\,\,\,\, sh} \alpha_{(pq|,s}^{8} \alpha_{ij)}^{2} + 4i \epsilon_{0r}^{\,\,\,\, sh} \alpha_{[pq|,s}^{5} \alpha_{ij]}^{3} + 4 \alpha_{(pq|}^{4} \alpha_{ij)}^{3} + 4 \alpha_{(pq|,s}^{6} \alpha_{ij)}^{5 \,\,\,\, s} \delta^{h}_{r} - 4 \alpha_{(pq|}^{6 \,\,\,\, ,h}\alpha_{ij),r}^{5} \right) + \\*
& & i\gamma^0 \gamma^r \left( - 4i \epsilon_{0r}^{\,\,\,\, sh} \alpha_{[pq|}^{3} \alpha_{ij],s}^{5} - 4i \alpha_{[pq|,r}^{5} \alpha_{ij]}^{6 \,\,\,\, h} + 4i \alpha_{[pq|,r}^{8} \alpha_{ij]}^{7 \,\,\,\, ,h} + 4 \epsilon_{0r}^{\,\,\,\, kh} \alpha_{[pq|}^{2}\alpha_{ij],k}^{8} - 4 \epsilon_{0r}^{\,\,\,\, st} \alpha_{[pq|,s}^{7} \epsilon_{0t}^{\,\,\,\, kh}\alpha_{ij],h}^{8} \right) + \\*
& & \gamma^{5}\gamma^{r} \left( 4 \alpha_{(pq|}^{4}\alpha_{ij)}^{2} \delta^{h}_{r} + 4i \epsilon_{0r}^{\,\,\,\, sh} \alpha_{[pq|,s}^{5} \alpha_{ij]}^{2} + 4i \alpha_{[pq|}^{1} \alpha_{ij]}^{3} \delta^{h}_{r} + 4 \epsilon_{0r}^{\,\,\,\, sh} \alpha_{[(pq|,s}^{8}\ \alpha_{ij)}^{3} - 4i \epsilon_{0r}^{\,\,\,\, sh} \alpha_{[pq|}^{2} \alpha_{ij],s}^{5} \right) + \\*
& & \gamma^{5} \gamma^{r}\left( + 4 \alpha_{(pq|,s}^{7} \alpha_{ij)}^{5 \,\,\,\, ,s} \delta^{h}_{r} - 4 \alpha_{(pq|}^{7 \,\,\,\, ,h} \alpha_{ij),r}^{5} - 4i \alpha_{[pq|,r}^{8} \alpha_{ij]}^{6 \,\,\,\, ,h} - 4 \alpha_{(pq|,r}^{5} \alpha_{ij)}^{7 \,\,\,\, ,h} - 4i \alpha_{[pq|}^{3} \epsilon_{0r}^{\,\,\,\, kh} \alpha_{ij],k}^{8} \right) \\*
& & - 4 \gamma^5 \gamma^{r}\epsilon_{0r}^{\,\,\,\, st} \alpha_{(pq|,s}^{6} \epsilon_{0t}^{\,\,\,\, kh} \alpha_{ij),k}^{8} + \gamma^0 \gamma^5 \gamma^r \left( - 4i \alpha_{[pq|}^{3}\alpha_{ij]}^{2} \delta^{h}_{r} - 4 \epsilon_{0r}^{\,\,\,\, sh} \alpha_{(pq|,s}^{6} \alpha_{ij)}^{2} + 4i \alpha_{[pq|}^{2} \alpha_{ij]}^{2} \delta^{h}_{r} + 4 \epsilon_{0r}^{\,\,\,\, sh} \alpha_{(pq|,s}^{7} \alpha_{ij)}^{3} \right) \\*
& & + \gamma^{0}\gamma^5 \gamma^r \left( 4i \alpha_{[pq|}^{1} \epsilon_{0r}^{\,\,\,\, sh}\alpha_{ij],s}^{5} + 4 \alpha_{(pq|,s}^{8} \alpha_{ij)}^{5 \,\,\,\, ,s}\delta^{h}_{r} - 4 \alpha_{(pq|}^{8 \,\,\,\, ,h}\alpha_{ij),r}^5 + 4i \alpha_{[pq|,r}^{7} \alpha_{ij]}^{6 \,\,\,\, ,h} - 4i \alpha_{[pq|,r}^{6} \alpha_{ij]}^{7 \,\,\,\, ,h} \right) \\*
& & + \gamma^{0}\gamma^5 \gamma^r \left( 4i \alpha_{[pq|}^{4} \epsilon_{0r}^{\,\,\,\, sh} \alpha_{ij],s}^{8} - 4 \epsilon_{0r}^{\,\,\,\, st}\alpha_{(pq|,s}^{5} \epsilon_{0t}^{\,\,\,\, kh} \alpha_{ij),k}^{8} \right).
\end{eqnarray*}
Finally, it remains to calculate $H_{0}^2$ and the latter equals
$$H_{0}^{2} = \int d^3 \vec{k} (k_a x^a)^2 a^{\dag}_{\vec{k}} a_{\vec{k}} + \int d^3 \vec{k} d^3 \vec{l} (k_a x^a) (l_a x^a) a^{\dag}_{\vec{k}} a^{\dag}_{\vec{l}}a_{\vec{k}} a_{\vec{l}}.$$ 
The reader may calculate that the following lenghty expression equals $H^2$
\begin{eqnarray*}
& & \sum_{i + j \geq 0} \sum_{p + q \geq 0}\sum_{a = 0}^{\min \{ j,p \}} a! \left( \begin{array}{c}
j \\ a
\end{array}  \right) \left( \begin{array}{c} p \\ a \end{array} \right) \int d \vec{k}_1 \ldots d\vec{k}_i d\vec{r}_1 \ldots d\vec{r}_p d\vec{l}_{1} \ldots d\vec{l}_{j-a} d\vec{s}_1 \ldots d\vec{s}_q \\*
& & a^{\dag}_{\vec{k}_1} \ldots a^{\dag}_{\vec{k}_i} a^{\dag}_{\vec{r}_{a + 1}} 
\ldots a^{\dag}_{\vec{r}_{p}} a_{\vec{l}_{1}} \ldots a_{\vec{l}_{j-a}}a_{\vec{s}_{1}} \ldots a_{\vec{s}_{q}} \\*
& & \sum_{c = 1}^{8} i^{F(c \, \text{mod} \, 4)} \alpha_{ij}^c(\vec{k}_g, \vec{l}_{1}, \ldots, \vec{l}_{j-a}, \vec{r}_{1}, \ldots, \vec{r}_{a}, \vec{x},t) \alpha_{pq}^c(\vec{r}_g, \vec{s}_h , \vec{x},t) + \\*
& & \gamma^{0} \left( 2 \alpha_{(ij|}^1 \alpha_{pq)}^2 - 2i\alpha_{[ij|}^3 \alpha_{pq]}^4 - 2i \alpha_{[ij|,h}^5 \alpha_{pq]}^{6 \,\,\, ,h} - 2\alpha_{(ij|,h}^{7} \alpha_{pq)}^{8 \, \, \, h} \right) \\*
& &  + \, i \gamma^5 \left( 2\alpha_{(ij|}^1 \alpha_{pq)}^3 + 2i \alpha_{[ij|}^2 \alpha_{pq]}^4 + 2i \alpha_{[ij|,h}^5 \alpha_{pq]}^{7 \,\,\, ,h} - 2 \alpha_{(ij|,h}^6 \alpha_{pq)}^{8 \,\,\, ,h} \right) \\*
& & + \, \gamma^0 \gamma^5 \left( 2 \alpha_{(ij|}^1 \alpha_{pq)}^4 + 2 i \alpha_{[ij|}^2 \alpha_{pq]}^3 + 2\alpha_{(ij|,h}^8 \alpha_{pq)}^{5 \,\,\, ,h} - 2i \alpha_{[ij|,h}^6 \alpha_{pq]}^{7 \,\,\,\, ,h} \right) \\*
& & + \gamma^h \left( 2 \alpha_{(ij|}^1 \alpha_{pq),h}^5
 - 2i \alpha_{[ij|}^2 \alpha_{pq],h}^6 + 2i \alpha_{[ij|}^3 \alpha_{pq],h}^7 + 2\alpha_{(ij|}^5 \alpha_{pq),h}^8 \right) + \\*
& & i \gamma^0 \gamma^h \left( 2\alpha^1_{(ij|} \alpha_{pq),h}^6 - 2i \alpha_{[ij|}^2 \alpha_{pq],h}^5 - 2\alpha_{(ij|}^3 \alpha_{pq),h}^8 - 2i \alpha_{[ij|}^4 \alpha_{pq],h}^7 \right) + \\*
& & i \gamma^0 \gamma^h \left( 2\epsilon_{0h}^{\,\,\, rs} \alpha_{(ij|,r}^5 \alpha_{pq),s}^7 + 2i \epsilon_{0h}^{\,\,\, rs} \alpha_{[ij|,r}^6 \alpha_{pq],s}^8  \right) + \\*  
& &  \gamma^5 \gamma^h \left( 2\alpha_{(ij|}^1 \alpha_{pq),h}^7 - 2\alpha_{(ij|}^2 \alpha_{pq),h}^8 + 2i \alpha_{[ij|}^3 \alpha_{pq],h}^5 \right) +  \\*
& & \gamma^5 \gamma^h \left(- 2i \alpha_{[ij|,h}^6 \alpha_{pq]}^4 - 2 \epsilon_{0rsh} \alpha_{(ij|,r}^5 \alpha_{pq),s}^6 + 2i \epsilon_{0rsh} \alpha_{[ij|,r}^7 \alpha_{pq],s}^8 \right) + \\*
& & \gamma^0 \gamma^5 \gamma^h \left( 2\alpha_{(ij|}^1 \alpha_{pq),h}^8 + 2\alpha_{(ij|}^2 \alpha_{pq),h}^7 - 2\alpha_{(ij|}^3 \alpha_{pq),h}^6 \right) +  \\* & & \gamma^0 \gamma^5 \gamma^h \left( 2\alpha_{(ij|}^4 \alpha_{pq),h}^5 + 2i \epsilon_{0hrs} \alpha_{[ij|,r}^5 \alpha_{pq],s}^5 - 2i \epsilon_{0hrs} \alpha_{[ij|,r}^7 \alpha_{pq],s}^7 \right)  \\*
\end{eqnarray*}
where in the above formulae involving $H(x)$ we have redefined $\alpha^{1}_{11}(\vec{k}, \vec{l}, \vec{x},t)$ as $\alpha^{1}_{11}(\vec{k}, \vec{l}, \vec{x};t) + k_a x^a \delta^{3}(\vec{k} - \vec{l})$.  In case some confusion may arise between $H(x)$ and $H_{\textrm{int}}$ we shall use $\widetilde{\alpha}^{1}_{11}$ for $\alpha^{1}_{11}(\vec{k}, \vec{l}, \vec{x};t) + k_a x^a \delta^{3}(\vec{k} - \vec{l})$.  The quadratic polynomial $F(x) = 3x + x^2 \, \textrm{mod} \, 4$ possesses a residual ambiguity since 
$$p(x) = 2 x^2 - 2 x \, \textrm{mod} \, 4$$ vanishes identically on $\mathbb{Z}_4$.
While the coefficients of this central term have a nice interpretation, there is nothing of that kind for some other terms.  For example, the $\gamma^0$ coefficients still can be captured by a function from $\mathbb{Z}_4$ to $\mathbb{Z}_4$ but there is no such thing for the $\gamma^0 \gamma^h$ terms since $\alpha_{ij,h}^8$ couples to two different functions.  Likewise,  
$$U^2(x) \gamma^a (U^{\dag})^2(x) = \gamma^a + 2i \left[ H(x) ,\gamma^a \right] - 2 \left[ H(x) , \left[ H(x) , \gamma^a \right] \right] + \ldots $$ which, again, is the correct expression when perturbing around the vacuum in second order.  The first constraint equation in second order perturbation theory equals 
\begin{eqnarray*}
0 & = & - e^{\mu}_a(x)e^{iH_{0}(x)}\int d^3 \vec{k}k_{\mu}a^{\dag}_{\vec{k}}a_{\vec{k}} \left( i \left[ H_{\textrm{int}}(x), \gamma^a \right] - \frac{1}{2} \{ H_{\textrm{int}}^2(x), \gamma^a \} \right) \\*
& & - e^{\mu}_a(x)e^{iH_{0}(x)}\int d^3 \vec{k}k_{\mu}a^{\dag}_{\vec{k}}a_{\vec{k}} \left( - \frac{1}{2} \left[ \left[ H_{0}(x),H_{\textrm{int}(x)}  \right], \gamma^a \right] + H_{\textrm{int}}(x) \gamma^a H_{\textrm{int}}(x) \right)  \\*
& & - e^{\mu}_a(x)H_{\textrm{int}}(x)e^{iH_{0}(x)} \int d^3 \vec{k} k_{\mu}a^{\dag}_{\vec{k}}a_{\vec{k}} \left[H_{\textrm{int}}(x), \gamma^a \right] \\*
& &  - e^{\mu}_{a}(x) \partial_{\mu}H_{\textrm{int}}(x) e^{iH_{0}(x)}\left( \gamma^{a} + i\left[ H_{\textrm{int}}(x), \gamma^a \right] \right) \\*
& & - \frac{i}{2}e^{\mu}_{a}(x) \left( \{ H_{\textrm{int}}(x), \partial_{\mu}H_{\textrm{int}}(x) \} + \left[ H_{0}(x), \partial_{\mu}H_{\textrm{int}}(x) \right] + \left[ \int d^3 \vec{k} k_{\mu}a^{\dag}_{\vec{k}}a_{\vec{k}}, H_{\textrm{int}}(x) \right] \right)e^{iH_{0}(x)} \gamma^{a}.
\end{eqnarray*} and the second constraint equation becomes
\begin{eqnarray*}
0 & = & -i \, \textrm{previous expression} \, + 2e^{\mu}_{a}(x)\left[ H_{\textrm{int}}(x), \gamma^a \right] \left( \partial_{\mu}H_{\textrm{int}}(x) + \int d^3 \vec{k}k_{\mu} a^{\dag}_{\vec{k}}a_{\vec{k}} \right)e^{iH_{0}}(x) + \\*
& & 2i e^{\mu}_{a}(x) \left( \left[ H_{\textrm{int}}(x) + H_{0}(x), \left[ H_{\textrm{int}}(x), \gamma^a \right] \right] \right)\int d^3 \vec{k} k_{\mu} a^{\dag}_{\vec{k}}a_{\vec{k}}\, e^{iH_{0}(x)} - \\*
& & e^{\mu}_a(x) \gamma^a \left( i \partial_{\mu}H_{\textrm{int}}(x) - \frac{1}{2} \{ H_{\textrm{int}}(x), \partial_{\mu}H_{\textrm{int}}(x) \} - \frac{1}{2} \left[ H_{0}(x), \partial_{\mu}H_{\textrm{int}}(x) \right] - \frac{1}{2} \left[ \int d^3 \vec{k} k_{\mu}a^{\dag}_{\vec{k}}a_{\vec{k}}, H_{\textrm{int}}(x) \right]\right) \\*
& & e^{iH_{0}(x)} + ie^{\mu}_a(x) \left[ iH_{\textrm{int}}(x) - \frac{1}{2}H_{\textrm{int}}^2(x) - \frac{1}{2}\left[ H_{0}(x), H_{\textrm{int}}(x) \right] , \gamma^a  \right]e^{iH_{0}(x)}\int d^3 \vec{k}k_{\mu}a^{\dag}_{\vec{k}}a_{\vec{k}}.
\end{eqnarray*}
The idea now is that for any $n$, we cut off the Hamiltonian at terms corresponding to $2 \leq i + j \leq k(n)$ and write the perturbed equations exacty to this order such that every equation has the same number of independent terms as there are free functions (there is a God given order for doing this).  Next, we study whether the matrix elements of the solutions (if they exist) in the natural basis converge in the limit for $n \rightarrow \infty$ (obviously $\lim_{n \rightarrow \infty} k(n) = \infty$).  If this is the case, then we define the corresponding operator to be a perturbatively stable solution of order $k(n)$.  Obviously, not any solution to the constraint equations is of this kind, but there is no other way for finding solutions than either making clever guesses or constructing some perturbative scheme.  In the former case, all one can hope for are solutions for which the Hamiltonian can be expressed in closed form; but given how daunting this even is for the Einstein equations, there is little hope that this can be achieved given that our equations are an infinity more complex than the latter (apart from the free solution which we have and some trivial extensions thereof).  Moreover, there exist good physical reasons why one would like to restrict to perturbatively stable operators of some order and we continue to further investigate those.  \\* \\*
In order to compute the above expressions, we start by calculating $\left[ H_{0}(x), H_{\textrm{int}}(x) \right]$; the latter is given by
\begin{eqnarray*}
& & \sum_{i + j \geq 0} \sum_{a = 0}^{\min \{i,1 \} } a! \left( \begin{array}{c} 1 \\* a \end{array} \right) \left( \begin{array}{c} i \\ a \end{array} \right)  \int d^3 \vec{r} d^3 \vec{k}_1 \ldots d^3 \vec{k}_i \int d^3 \vec{s}^a \int d^3 \vec{l}_1 \ldots d^3 \vec{l}_j a^{\dag}_{\vec{r}} a^{\dag}_{\vec{k}_1} \ldots a^{\dag}_{\vec{k}_{i-a}}a_{\vec{s}^a} a_{\vec{l}_1} \ldots a_{\vec{l}_{j}} \\*
& & \delta^{3}(\vec{r} - \vec{s}) (r_a x^a)\beta_{ij}(\vec{k}, \vec{l}, \vec{x}, t) - \\*
& & \sum_{i + j \geq 0} \sum_{a = 0}^{\min \{j,1 \} } a! \left( \begin{array}{c} j \\* a \end{array} \right) \left( \begin{array}{c} 1 \\ a \end{array} \right)  \int d^3 \vec{r} d^3 \vec{k}_1 \ldots d^3 \vec{k}_i \int d^3 \vec{s} \int d^3 \vec{l}_1 \ldots d^3 \vec{l}_{j-a} a^{\dag}_{\vec{r}^{a}} a^{\dag}_{\vec{k}_1} \ldots a^{\dag}_{\vec{k}_{i}}a_{\vec{s}^a} a_{\vec{l}_1} \ldots a_{\vec{l}_{j-a}}  \\*
& & \delta^{3}(\vec{r} - \vec{s}) (r_a x^a)\beta_{ij}(\vec{k}, \vec{l}_{1}, \ldots, \vec{l}_{j-a}, \vec{s}^{a}, \vec{x}, t)
\end{eqnarray*}       
where $\int d\vec{s}^a$ equals $\emptyset$ for $a = 1$ and all $\vec{s}$ are replaced by $\vec{k}_{i}$ in the first expression.  Changing of variables and summing up the corresponding series leads to
\begin{eqnarray*}
& & \sum_{i + j \geq 0} \int d^3 \vec{k}_1 \ldots d^3 \vec{k}_i d^3 \vec{l}_1 \ldots d^3 \vec{l}_j a^{\dag}_{\vec{k}_1} \ldots a^{\dag}_{\vec{k}_i} a_{\vec{l}_1} \ldots a_{\vec{l}_j} \left[ \left(i k_{i,a} - j l_{j,a} \right) x^{a} \right] \beta_{ij}(\vec{k}, \vec{l}, \vec{x},t).
\end{eqnarray*}
Remains us to calculate $\{ H_{\textrm{int}}(x), \partial_{\mu}H_{\textrm{int}}(x) \}$ which follows
 immediately from the previous expression for $H_{\textrm{int}}^2(x)$.  The first constraint equation is written out in full component 
form as
\begin{eqnarray*}
0 & = & \left( 1 + \frac{GM}{r} \right) \sum_{p + q \geq 0} \int d^3 \vec{k}_1 \ldots d^3 \vec{k}_{p+1} \int d^3 \vec{l}_1 \ldots d^3 \vec{l}_{q+1} a^{\dag}_{\vec{k}_1} \ldots a^{\dag}_{\vec{k}_{p+1}}a_{\vec{l}_1} \ldots a_{\vec{l}_{q+1}} \\*
& & k_{p+1,0} \delta^{3}(\vec{k}_{p+1} - \vec{l}_{q+1}) e^{i \left( \sum_{j=1}^{p+1}k_{j,a} - \sum_{j=1}^{q+1} l_{j,a} \right)x^a} \left( 2 \gamma^5 \gamma^0 \alpha_{pq}^{3} - 2i \gamma^5 \alpha_{pq}^{4} + 2 \alpha_{pq,r}^{6} \gamma^r - 2i \alpha_{pq,r}^{5} \gamma^r \gamma^0 \right) \\*
& & + \left( 1 + \frac{GM}{r} \right) \sum_{p + q \geq 0} p \int d^3 \vec{k}_1 \ldots d^3 \vec{k}_{p} \int d^3 \vec{l}_1 \ldots d^3 \vec{l}_{q} a^{\dag}_{\vec{k}_1} \ldots a^{\dag}_{\vec{k}_{p}}a_{\vec{l}_1} \ldots a_{\vec{l}_{q}} k_{1,0} \\*
& & e^{i \left( \sum_{j=1}^{p}k_{j,a} - \sum_{j=1}^{q} l_{j,a} \right)x^a} \left( 2 \gamma^5 \gamma^0 \alpha_{pq}^{3} - 2i \gamma^5 \alpha_{pq}^{4} + 2 \alpha_{pq,r}^{6} \gamma^r - 2i \alpha_{pq,r}^{5} \gamma^r \gamma^0 \right) + \\*
& & - \sum_{p + q \geq 0} \int d^3 \vec{k}_1 \ldots d^3 \vec{k}_{p+1} \int d^3 \vec{l}_1 \ldots d^3 \vec{l}_{q+1} a^{\dag}_{\vec{k}_1} \ldots a^{\dag}_{\vec{k}_{p+1}}a_{\vec{l}_1} \ldots a_{\vec{l}_{q+1}} e^{i \left( \sum_{j=1}^{p+1} k_{j,a} - \sum_{j=1}^{q+1} l_{j,a} \right)x^a}\delta (\vec{k}_{p+1} - \vec{l}_{q+1}) \\*
& & k_{p+1,s} \left( 2i \gamma^0 \gamma^s \alpha_{pq}^2 - 2 \gamma^5 \gamma^s \alpha_{pq}^{3} - 2i \gamma^0 \gamma^5 \gamma^{r} \epsilon_{0 \,\,\, \,r}^{\,\, ks}\alpha_{pq,k}^{5} - 2 \gamma^0 \alpha_{pq}^{6 \,\,\, ,s} + 2i \gamma^{5} \alpha_{pq}^{7 \,\,\, ,s} - 2 \epsilon_{0r}^{\,\,\,\, ks} \alpha_{pq,k}^{8} \gamma^{r} \right) \\*
& & - \sum_{p+q \geq 0} p \int d^3 \vec{k}_1 \ldots d^3 \vec{k}_{p} d^3 \vec{l}_1 \ldots d^3 \vec{l}_{q} a^{\dag}_{\vec{k}_1} \ldots a^{\dag}_{\vec{k}_{p}}a_{\vec{l}_1} \ldots a_{\vec{l}_{q}} e^{i \left( \sum_{j=1}^{p} k_{j,a} - \sum_{j=1}^{q} l_{j,a} \right)x^a} \\*
& & k_{1,s} \left( 2i \gamma^0 \gamma^s \alpha_{pq}^2 - 2 \gamma^5 \gamma^s \alpha_{pq}^{3} - 2i \gamma^0 \gamma^5 \gamma^{r} \epsilon_{0 \,\,\, \,r}^{\,\, ks}\alpha_{pq,k}^{5} - 2 \gamma^0 \alpha_{pq}^{6 \,\,\, ,s} + 2i \gamma^{5} \alpha_{pq}^{7 \,\,\, ,s} - 2 \epsilon_{0r}^{\,\,\,\, ks} \alpha_{pq,k}^{8} \gamma^{r} \right) + \\*
& & \frac{1}{2} \left( 1 + \frac{GM}{r} \right) \sum_{p + q \geq 0} \sum_{i + j \geq 0} \sum_{a = 0}^{\min \{ q ,i \}} a! \left( \begin{array}{c} q \\ a \end{array} \right) \left( \begin{array}{c} i \\ a \end{array} \right) \int d \vec{r}_1 \ldots d\vec{r}_{p+1} d \vec{s}_1 \ldots d \vec{s}_{q - a} d \vec{k}_1 \ldots d \vec{k}_i d \vec{l}_1 \ldots d \vec{l}_{j+1} \\*
& & a^{\dag}_{\vec{r}_1} \ldots a^{\dag}_{\vec{r}_{p+1}} a^{\dag}_{\vec{k}_{a+1}} \ldots a^{\dag}_{\vec{k}_i} a_{\vec{s}_1} \ldots a_{\vec{s}_{q - a}} a_{\vec{l}_1} \ldots a_{\vec{l}_{j+1}} e^{i \left( \sum_{t=1}^{p+1}r_{t,b} + \sum_{t=a + 1}^{i} k_{t,b} - \sum_{t=1}^{q-a} s_{t,b} - \sum_{t=1}^{j+1} l_{t,b} \right) x^b} \\*
\end{eqnarray*}
\begin{eqnarray*}
& & \delta^3(\vec{r}_{p+1} - \vec{l}_{j+1}) r_{p+1,0} \left( - 4i \alpha_{pq}^4 \alpha_{ij}^3 + 4i \alpha_{pq}^3 \alpha_{ij}^4 + 4i \alpha_{pq,h}^5 \alpha_{ij}^{6 \,\, ,h} - 4i \alpha_{pq,h}^6 \alpha_{ij}^{5 \,\, ,h} \right) + \\* 
& & \gamma^0 \left( - 4 \alpha_{pq}^3 \alpha_{ij}^3 + 4 \alpha_{pq}^4 \alpha_{ij}^4
- 4 \alpha_{pq,h}^6 \alpha_{ij}^{6 \,\,\, ,h} + 4 \alpha_{pq,h}^5 \alpha_{ij}^{5 \,\,\, ,h} \right) + \\*
& & i \gamma^5 \left( 4 \alpha_{pq}^{(2}\alpha_{ij}^{3)} - 4i \alpha_{pq}^{[1}\alpha_{ij}^{4]} + 4 \alpha_{pq,h}^{(6}\alpha_{ij}^{7) \,\, ,h} - 4i \alpha_{pq,h}^{[8}\alpha_{ij}^{5] \,\, ,h}  \right) + \\*
& & \gamma^0 \gamma^5 \left( - 4i \widetilde{\alpha}_{[pq|}^{1} \alpha_{ij]}^{3} + 4 \alpha_{(pq|}^{2}\alpha_{ij)}^{4} - 4i \alpha_{[pq|,h}^{6}\alpha_{ij]}^{8 \,\, ,h}
- 4 \alpha_{(pq|,h}^{7}\alpha_{ij)}^{5 \,\, ,h} \right)  + \\*
& & \gamma^h \left( 4i \alpha_{[pq|}^{3}\alpha_{ij],h}^{8} - 4 \alpha_{(pq|}^{4}\alpha_{ij),h}^{7} + 4i \widetilde{\alpha}_{[pq|}^{1}\alpha_{ij],h}^{6} + 4
\epsilon_{0h}^{\,\,\, rs} \alpha_{(pq|,r}^{6}\alpha_{ij),s}^{8} -  4 \alpha_{(pq|}^{2}\alpha_{ij),h}^{5} - 4i \epsilon_{0h}^{\,\,\, rs} \alpha_{[pq|,r}^{7} \alpha_{ij],s}^{5} \right) + \\*
& & i \gamma^0 \gamma^h \left( 4 \alpha_{(pq|,h}^{3} \alpha_{ij)}^{7} - 4i \alpha_{[pq|,h}^{4}\alpha_{ij]}^{8} + 4 \alpha_{(pq|}^{2}\alpha_{ij),h}^{6}
- 4i \epsilon_{0h}^{\,\,\, rs} \alpha_{[pq|,r}^{6} \alpha_{ij],s}^{7}  + 4i  \widetilde{\alpha}_{[pq|}^{1} \alpha_{ij],h}^{5} - 4 \epsilon_{0h}^{\,\,\,rs} \alpha_{(pq|,r}^8 \alpha_{ij),s}^5 \right) + \\*
& & \gamma^5 \gamma^h \left( - 8 \alpha_{(pq|}^3 \alpha_{ij),h}^6 - 8 \alpha_{(pq|,h}^5 \alpha_{ij)}^4 + 4i \epsilon_{0h}^{\,\,\, rs} \alpha_{pq,r}^6 \alpha_{ij,s}^6 - 4i\epsilon_{0h}^{\,\,\,\, kr}\alpha_{[pq|,k}^{5}\alpha_{ij],r}^{5} \right) + \\*
& & \gamma^0 \gamma^5 \gamma^h  \left( - 8i \alpha_{[pq|,h}^6 \alpha_{ij]}^4 + 4i \alpha_{[pq|}^3 \alpha_{ij],h}^{5} + 4 \epsilon_{0h}^{\,\,\, rs} \alpha_{(pq|,r}^6 \alpha_{ij),s}^5 - 4 \epsilon_{0h}^{\,\,\, rs} \alpha_{(pq|,r}^5 \alpha_{ij),s}^6 - 4i \alpha_{[pq|,h}^{5}\alpha_{ij]}^{3} \right) + \\*
& & \frac{1}{2} \left( 1 + \frac{GM}{r} \right) \sum_{p + q \geq 0} \sum_{i + j \geq 0} \sum_{a = 0}^{\min \{ q ,i \}} a! \left( \begin{array}{c} q \\ a \end{array} \right) \left( \begin{array}{c} i \\ a \end{array} \right) \int d \vec{r}_1 \ldots d\vec{r}_{p} d \vec{s}_1 \ldots d \vec{s}_{q - a} d \vec{k}_1 \ldots d \vec{k}_i d \vec{l}_1 \ldots d \vec{l}_{j} \\*
& & a^{\dag}_{\vec{r}_1} \ldots a^{\dag}_{\vec{r}_{p}} a^{\dag}_{\vec{k}_{a+1}} \ldots a^{\dag}_{\vec{k}_i} a_{\vec{s}_1} \ldots a_{\vec{s}_{q - a}} a_{\vec{l}_1} \ldots a_{\vec{l}_{j}} e^{i \left( \sum_{t=1}^{p}r_{t,b} + \sum_{t= a+1}^{i} k_{t,b} - \sum_{t=1}^{q-a} s_{t,b} - \sum_{t=1}^{j} l_{t,b} \right) x^b} \\*
& & (pr_{1,0} + (i-a)k_{1,0}) \left( - 4i \alpha_{pq}^4 \alpha_{ij}^3 + 4i \alpha_{pq}^3 \alpha_{ij}^4 + 4i \alpha_{pq,h}^5 \alpha_{ij}^{6 \,\, ,h} - 4i \alpha_{pq,h}^6 \alpha_{ij}^{5 \,\, ,h} \right) + \\* 
& & \gamma^0 \left( - 4 \alpha_{pq}^3 \alpha_{ij}^3 + 4 \alpha_{pq}^4 \alpha_{ij}^4
- 4 \alpha_{pq,h}^6 \alpha_{ij}^{6 \,\,\, ,h} + 4 \alpha_{pq,h}^5 \alpha_{ij}^{5 \,\,\, ,h} \right) + \\*
& & i \gamma^5 \left( 4 \alpha_{pq}^{(2}\alpha_{ij}^{3)} - 4i \alpha_{pq}^{[1}\alpha_{ij}^{4]} + 4 \alpha_{pq,h}^{(6}\alpha_{ij}^{7) \,\, ,h} - 4i \alpha_{pq,h}^{[8}\alpha_{ij}^{5] \,\, ,h}  \right) + \\*
& & \gamma^0 \gamma^5 \left( - 4i \widetilde{\alpha}_{[pq|}^{1} \alpha_{ij]}^{3} + 4 \alpha_{(pq|}^{2}\alpha_{ij)}^{4} - 4i \alpha_{[pq|,h}^{6}\alpha_{ij]}^{8 \,\, ,h}
- 4 \alpha_{(pq|,h}^{7}\alpha_{ij)}^{5 \,\, ,h} \right)  + \\*
& & \gamma^h \left( 4i \alpha_{[pq|}^{3}\alpha_{ij],h}^{8} - 4 \alpha_{(pq|}^{4}\alpha_{ij),h}^{7} + 4i \widetilde{\alpha}_{[pq|}^{1}\alpha_{ij],h}^{6} + 4
\epsilon_{0h}^{\,\,\, rs} \alpha_{(pq|,r}^{6}\alpha_{ij),s}^{8} -  4 \alpha_{(pq|}^{2}\alpha_{ij),h}^{5} - 4i \epsilon_{0h}^{\,\,\, rs} \alpha_{[pq|,r}^{7} \alpha_{ij],s}^{5} \right) + \\*
& & i \gamma^0 \gamma^h \left( 4 \alpha_{(pq|,h}^{3} \alpha_{ij)}^{7} - 4i \alpha_{[pq|,h}^{4}\alpha_{ij]}^{8} + 4 \alpha_{(pq|}^{2}\alpha_{ij),h}^{6}
- 4i \epsilon_{0h}^{\,\,\, rs} \alpha_{[pq|,r}^{6} \alpha_{ij],s}^{7}  + 4i  \widetilde{\alpha}_{[pq|}^{1} \alpha_{ij],h}^{5} - 4 \epsilon_{0h}^{\,\,\,rs} \alpha_{(pq|,r}^8 \alpha_{ij),s}^5 \right) + \\*
& & \gamma^5 \gamma^h \left( - 8 \alpha_{(pq|}^3 \alpha_{ij),h}^6 - 8 \alpha_{(pq|,h}^5 \alpha_{ij)}^4 + 4i \epsilon_{0h}^{\,\,\, rs} \alpha_{pq,r}^6 \alpha_{ij,s}^6 - 4i\epsilon_{0h}^{\,\,\,\, kr}\alpha_{[pq|,k}^{5} \alpha_{ij],r}^{5} \right) + \\*
& & \gamma^0 \gamma^5 \gamma^h  \left( - 8i \alpha_{[pq|,h}^6 \alpha_{ij]}^4 + 4i \alpha_{[pq|}^3 \alpha_{ij],h}^{5} + 4 \epsilon_{0h}^{\,\,\, rs} \alpha_{(pq|,r}^6 \alpha_{ij),s}^5 - 4 \epsilon_{0h}^{\,\,\, rs} \alpha_{(pq|,r}^5 \alpha_{ij),s}^6 - 4i \alpha_{[pq|,h}^{5}\alpha_{ij]}^{3} \right) + \\*
& & \frac{1}{2} \sum_{p + q \geq 0} \sum_{i + j \geq 0} \sum_{a = 0}^{\min \{ q ,i \}} a! \left( \begin{array}{c} q \\ a \end{array} \right) \left( \begin{array}{c} i \\ a \end{array} \right) \int d \vec{r}_1 \ldots d \vec{r}_{p+1} d \vec{s}_1 \ldots d \vec{s}_{q - a} d \vec{k}_1 \ldots d \vec{k}_i d \vec{l}_1 \ldots d \vec{l}_{j+1} \\*
& & a^{\dag}_{\vec{r}_1} \ldots a^{\dag}_{\vec{r}_{p+1}} a^{\dag}_{\vec{k}_{a+1}} \ldots a^{\dag}_{\vec{k}_i} a_{\vec{s}_1} \ldots a_{\vec{s}_{q - a}} a_{\vec{l}_1} \ldots a_{\vec{l}_{j+1}} k_{p+1,h} \\* & & \delta^{3}(\vec{k}_{p+1} - \vec{l}_{j+1}) e^{i \left( \sum_{b = 1}^{p+1}r_{b,c} + \sum_{b = a+ 1}^{i}k_{b,c} - \sum_{b = 1}^{q-a}s_{b,c} - \sum_{b = 1}^{j+1}l_{b,c} \right)x^{c}}\\*
& & \left( 8i \alpha_{[pq|}^{6 \,\,\, ,h} \alpha_{ij]}^{2} - 8i \alpha_{[pq|}^{7 \,\,\, ,h}\alpha_{ij]}^{3} + 8i \epsilon_{0}^{\,\, hkr} \epsilon_{[pq|,k}^{8} \alpha_{ij],r}^{5} \right) + \\*
& & \gamma^{0} \left( - 4 \alpha_{(pq|}^{5 \,\,\, ,h}\alpha_{ij)}^{2} - 4i \alpha_{[pq|}^{8 \,\,\, ,h}\alpha_{ij]}^{3} - 4i \epsilon_{0}^{\,\,\, hrs} \epsilon_{[pq|,r}^{7} \alpha_{ij],s}^{5} + 4i \widetilde{\alpha}_{[pq|}^{1} \alpha_{ij]}^{6 \,\,\, ,h} + 4 \alpha_{(pq|}^{4} \alpha_{ij)}^{7 \,\,\, ,h} + 4 \epsilon_{0}^{\,\,\, hrs} \alpha_{[ij|,r}^{8} \alpha_{pq],s}^{6} \right) + \\*
& &  i \gamma^5 \left( 4i \alpha_{[pq|}^{8 \,\,\, ,h} \alpha_{ij]}^{2} - 4 \alpha_{(pq|}^{5 \,\,\, ,h} \alpha_{ij)}^{3} - 4i \epsilon_{0}^{\,\, hrs} \alpha_{[pq|,r}^{6} \alpha_{ij],s}^{5} + 4 \alpha_{(pq|}^{4} \alpha_{ij)}^{6 \,\,\, ,h}  - 4i \widetilde{\alpha}_{[pq|}^{1} \alpha_{ij]}^{7 \,\,\, ,h} - 4 \epsilon_{0}^{\,\, hrs} \alpha_{(ij|,r}^{8} \alpha_{pq),s}^{7} \right) + 
\end{eqnarray*}
\begin{eqnarray*}
& & \gamma^0 \gamma^5 \left( - 4 \alpha_{(pq|}^{7 \,\,\, ,h}\alpha_{ij)}^{2} + 4 \alpha_{(pq|}^{6 \,\,\, ,h} \alpha_{ij)}^{3}  + 4i \epsilon_{0}^{\,\, hrs}\alpha_{[pq|,r}^{5} \alpha_{ij],s}^{5} + 4 \alpha_{(pq|}^{3} \alpha_{ij)}^{6 \,\,\, ,h} + 4 \alpha_{(pq|}^{2} \alpha_{ij)}^{7 \,\,\, ,h} - 4i \epsilon_{0}^{\,\, hrs} \alpha_{[ij|,r}^{8} \alpha_{pq],s}^{8} \right) + \\*
& & \gamma^r \left( - 4 \alpha_{(pq|}^{2} \alpha_{ij)}^2 \delta^{h}_{r} + 4i \epsilon_{0r}^{\,\,\,\, sh} \alpha_{[pq|,s}^{7} \alpha_{ij]}^{2} - 4 \alpha_{(pq|}^3 \alpha_{ij)}^3 \delta^{h}_{r} + 4i \epsilon_{0r}^{\,\,\,\, sh} \alpha_{[pq|}^{4} \alpha_{ij],s}^{5} + 4 \alpha_{(pq|,s}^{5} \alpha_{ij)}^{5 \,\,\,\, ,s}\delta^{h}_{r} \right) + \\*
& & \gamma^r \left( - 4 \alpha_{(pq|}^{5 \,\,\, ,h} \alpha_{ij),r}^{5} - 4 \alpha_{(pq|,r}^{6} \alpha_{ij)}^{6 \,\,\, ,h} - 4 \alpha_{(pq|,r}^{7} \alpha_{ij)}^{7 \,\,\, ,h} + 4i \epsilon_{0r}^{\,\,\,\, sh}\widetilde{\alpha}_{[pq|}^{1}\alpha_{ij],s}^{8} - 4 \epsilon_{0r}^{\,\,\,\, st} \alpha_{(pq|,s}^{8} \epsilon_{0t}^{\,\,\, \, kh} \alpha_{ij),k}^{8} \right) + \\*
& & i \gamma^{0} \gamma^{r} \left(- 4i \widetilde{\alpha}_{[pq|}^1 \alpha_{ij]}^{2} \delta^{h}_{r} + 4 \epsilon_{0r}^{\,\,\,\, sh} \alpha_{(pq|,s}^{8} \alpha_{ij)}^{2} + 4i \epsilon_{0r}^{\,\,\,\, sh} \alpha_{[pq|,s}^{5} \alpha_{ij]}^{3} + 4 \alpha_{(pq|}^{4} \alpha_{ij)}^{3} + 4 \alpha_{(pq|,s}^{6} \alpha_{ij)}^{5 \,\,\,\, s} \delta^{h}_{r} - 4 \alpha_{(pq|}^{6 \,\,\,\, ,h}\alpha_{ij),r}^{5} \right) + \\*
& & i\gamma^0 \gamma^r \left( - 4i \epsilon_{0r}^{\,\,\,\, sh} \alpha_{[pq|}^{3} \alpha_{ij],s}^{5} - 4i \alpha_{[pq|,r}^{5} \alpha_{ij]}^{6 \,\,\,\, h} + 4i \alpha_{[pq|,r}^{8} \alpha_{ij]}^{7 \,\,\,\, ,h} + 4 \epsilon_{0r}^{\,\,\,\, kh} \alpha_{[pq|}^{2}\alpha_{ij],k}^{8} - 4 \epsilon_{0r}^{\,\,\,\, st} \alpha_{[pq|,s}^{7} \epsilon_{0t}^{\,\,\,\, kh}\alpha_{ij],h}^{8} \right) + \\*
& & \gamma^{5}\gamma^{r} \left( 4 \alpha_{(pq|}^{4}\alpha_{ij)}^{2} \delta^{h}_{r} + 4i \epsilon_{0r}^{\,\,\,\, sh} \alpha_{[pq|,s}^{5} \alpha_{ij]}^{2} + 4i \widetilde{\alpha}_{[pq|}^{1} \alpha_{ij]}^{3} \delta^{h}_{r} + 4 \epsilon_{0r}^{\,\,\,\, sh} \alpha_{[(pq|,s}^{8}\ \alpha_{ij)}^{3} - 4i \epsilon_{0r}^{\,\,\,\, sh} \alpha_{[pq|}^{2} \alpha_{ij],s}^{5} \right) + \\*
& & \gamma^{5} \gamma^{r}\left( + 4 \alpha_{(pq|,s}^{7} \alpha_{ij)}^{5 \,\,\,\, ,s} \delta^{h}_{r} - 4 \alpha_{(pq|}^{7 \,\,\,\, ,h} \alpha_{ij),r}^{5} - 4i \alpha_{[pq|,r}^{8} \alpha_{ij]}^{6 \,\,\,\, ,h} - 4 \alpha_{(pq|,r}^{5} \alpha_{ij)}^{7 \,\,\,\, ,h} - 4i \alpha_{[pq|}^{3} \epsilon_{0r}^{\,\,\,\, kh} \alpha_{ij],k}^{8} \right) \\*
& & - 4 \gamma^5 \gamma^{r}\epsilon_{0r}^{\,\,\,\, st} \alpha_{(pq|,s}^{6} \epsilon_{0t}^{\,\,\,\, kh} \alpha_{ij),k}^{8} + \gamma^0 \gamma^5 \gamma^r \left( - 4i \alpha_{[pq|}^{3}\alpha_{ij]}^{2} \delta^{h}_{r} - 4 \epsilon_{0r}^{\,\,\,\, sh} \alpha_{(pq|,s}^{6} \alpha_{ij)}^{2} + 4i \alpha_{[pq|}^{2} \alpha_{ij]}^{2} \delta^{h}_{r} + 4 \epsilon_{0r}^{\,\,\,\, sh} \alpha_{(pq|,s}^{7} \alpha_{ij)}^{3} \right) \\*
& & + \gamma^{0}\gamma^5 \gamma^r \left( 4i \widetilde{\alpha}_{[pq|}^{1} \epsilon_{0r}^{\,\,\,\, sh}\alpha_{ij],s}^{5} + 4 \alpha_{(pq|,s}^{8} \alpha_{ij)}^{5 \,\,\,\, ,s}\delta^{h}_{r} - 4 \alpha_{(pq|}^{8 \,\,\,\, ,h}\alpha_{ij),r}^5 + 4i \alpha_{[pq|,r}^{7} \alpha_{ij]}^{6 \,\,\,\, ,h} - 4i \alpha_{[pq|,r}^{6} \alpha_{ij]}^{7 \,\,\,\, ,h} \right) \\*
& & + \gamma^{0}\gamma^5 \gamma^r \left( 4i \alpha_{[pq|}^{4} \epsilon_{0r}^{\,\,\,\, sh} \alpha_{ij],s}^{8} - 4 \epsilon_{0r}^{\,\,\,\, st}\alpha_{(pq|,s}^{5} \epsilon_{0t}^{\,\,\,\, kh} \alpha_{ij),k}^{8} \right) + \\*
& & \frac{1}{2} \sum_{p + q \geq 0} \sum_{i + j \geq 0} \sum_{a = 0}^{\min \{ q ,i \}} a! \left( \begin{array}{c} q \\ a \end{array} \right) \left( \begin{array}{c} i \\ a \end{array} \right) \int d \vec{r}_1 \ldots d \vec{r}_{p} d \vec{s}_1 \ldots d \vec{s}_{q - a} d \vec{k}_1 \ldots d \vec{k}_i d \vec{l}_1 \ldots d \vec{l}_{j} \\*
& & a^{\dag}_{\vec{r}_1} \ldots a^{\dag}_{\vec{r}_{p}} a^{\dag}_{\vec{k}_{a+1}} \ldots a^{\dag}_{\vec{k}_i} a_{\vec{s}_1} \ldots a_{\vec{s}_{q - a}} a_{\vec{l}_1} \ldots a_{\vec{l}_{j}} \\* & & \left( p r_{1,h} + (i-a)k_{1,h} \right) e^{i \left( \sum_{d = 1}^{p}r_{d,c} + \sum_{d = a + 1}^{i}k_{d,c} - \sum_{d = 1}^{q-a}s_{d,c} - \sum_{d = 1}^{j}l_{d,c} \right)x^{c}}\\*
& & \left( 8i \alpha_{[pq|}^{6 \,\,\, ,h} \alpha_{ij]}^{2} - 8i \alpha_{[pq|}^{7 \,\,\, ,h}\alpha_{ij]}^{3} + 8i \epsilon_{0}^{\,\, hkr} \epsilon_{[pq|,k}^{8} \alpha_{ij],r}^{5} \right) + \\*
& & \gamma^{0} \left( - 4 \alpha_{(pq|}^{5 \,\,\, ,h}\alpha_{ij)}^{2} - 4i \alpha_{[pq|}^{8 \,\,\, ,h}\alpha_{ij]}^{3} - 4i \epsilon_{0}^{\,\,\, hrs} \epsilon_{[pq|,r}^{7} \alpha_{ij],s}^{5} + 4i \widetilde{\alpha}_{[pq|}^{1} \alpha_{ij]}^{6 \,\,\, ,h} + 4 \alpha_{(pq|}^{4} \alpha_{ij)}^{7 \,\,\, ,h} + 4 \epsilon_{0}^{\,\,\, hrs} \alpha_{[ij|,r}^{8} \alpha_{pq],s}^{6} \right) + \\*
& &  i \gamma^5 \left( 4i \alpha_{[pq|}^{8 \,\,\, ,h} \alpha_{ij]}^{2} - 4 \alpha_{(pq|}^{5 \,\,\, ,h} \alpha_{ij)}^{3} - 4i \epsilon_{0}^{\,\, hrs} \alpha_{[pq|,r}^{6} \alpha_{ij],s}^{5} + 4 \alpha_{(pq|}^{4} \alpha_{ij)}^{6 \,\,\, ,h}  - 4i \widetilde{\alpha}_{[pq|}^{1} \alpha_{ij]}^{7 \,\,\, ,h} - 4 \epsilon_{0}^{\,\, hrs} \alpha_{(ij|,r}^{8} \alpha_{pq),s}^{7} \right) + \\*
& & \gamma^0 \gamma^5 \left( - 4 \alpha_{(pq|}^{7 \,\,\, ,h}\alpha_{ij)}^{2} + 4 \alpha_{(pq|}^{6 \,\,\, ,h} \alpha_{ij)}^{3}  + 4i \epsilon_{0}^{\,\, hrs}\alpha_{[pq|,r}^{5} \alpha_{ij],s}^{5} + 4 \alpha_{(pq|}^{3} \alpha_{ij)}^{6 \,\,\, ,h} + 4 \alpha_{(pq|}^{2} \alpha_{ij)}^{7 \,\,\, ,h} - 4i \epsilon_{0}^{\,\, hrs} \alpha_{[ij|,r}^{8} \alpha_{pq],s}^{8} \right) + \\*
& & \gamma^r \left( - 4 \alpha_{(pq|}^{2} \alpha_{ij)}^2 \delta^{h}_{r} + 4i \epsilon_{0r}^{\,\,\,\, sh} \alpha_{[pq|,s}^{7} \alpha_{ij]}^{2} - 4 \alpha_{(pq|}^3 \alpha_{ij)}^3 \delta^{h}_{r} + 4i \epsilon_{0r}^{\,\,\,\, sh} \alpha_{[pq|}^{4} \alpha_{ij],s}^{5} + 4 \alpha_{(pq|,s}^{5} \alpha_{ij)}^{5 \,\,\,\, ,s}\delta^{h}_{r} \right) + \\*
& & \gamma^r \left( - 4 \alpha_{(pq|}^{5 \,\,\, ,h} \alpha_{ij),r}^{5} - 4 \alpha_{(pq|,r}^{6} \alpha_{ij)}^{6 \,\,\, ,h} - 4 \alpha_{(pq|,r}^{7} \alpha_{ij)}^{7 \,\,\, ,h} + 4i \epsilon_{0r}^{\,\,\,\, sh}\widetilde{\alpha}_{[pq|}^{1}\alpha_{ij],s}^{8} - 4 \epsilon_{0r}^{\,\,\,\, st} \alpha_{(pq|,s}^{8} \epsilon_{0t}^{\,\,\, \, kh} \alpha_{ij),k}^{8} \right) + \\*
& & i \gamma^{0} \gamma^{r} \left(- 4i \widetilde{\alpha}_{[pq|}^1 \alpha_{ij]}^{2} \delta^{h}_{r} + 4 \epsilon_{0r}^{\,\,\,\, sh} \alpha_{(pq|,s}^{8} \alpha_{ij)}^{2} + 4i \epsilon_{0r}^{\,\,\,\, sh} \alpha_{[pq|,s}^{5} \alpha_{ij]}^{3} + 4 \alpha_{(pq|}^{4} \alpha_{ij)}^{3} + 4 \alpha_{(pq|,s}^{6} \alpha_{ij)}^{5 \,\,\,\, s} \delta^{h}_{r} - 4 \alpha_{(pq|}^{6 \,\,\,\, ,h}\alpha_{ij),r}^{5} \right) + \\*
& & i\gamma^0 \gamma^r \left( - 4i \epsilon_{0r}^{\,\,\,\, sh} \alpha_{[pq|}^{3} \alpha_{ij],s}^{5} - 4i \alpha_{[pq|,r}^{5} \alpha_{ij]}^{6 \,\,\,\, h} + 4i \alpha_{[pq|,r}^{8} \alpha_{ij]}^{7 \,\,\,\, ,h} + 4 \epsilon_{0r}^{\,\,\,\, kh} \alpha_{[pq|}^{2}\alpha_{ij],k}^{8} - 4 \epsilon_{0r}^{\,\,\,\, st} \alpha_{[pq|,s}^{7} \epsilon_{0t}^{\,\,\,\, kh}\alpha_{ij],h}^{8} \right) + \\*
& & \gamma^{5}\gamma^{r} \left( 4 \alpha_{(pq|}^{4}\alpha_{ij)}^{2} \delta^{h}_{r} + 4i \epsilon_{0r}^{\,\,\,\, sh} \alpha_{[pq|,s}^{5} \alpha_{ij]}^{2} + 4i \widetilde{\alpha}_{[pq|}^{1} \alpha_{ij]}^{3} \delta^{h}_{r} + 4 \epsilon_{0r}^{\,\,\,\, sh} \alpha_{[(pq|,s}^{8} \alpha_{ij)}^{3} - 4i \epsilon_{0r}^{\,\,\,\, sh} \alpha_{[pq|}^{2} \alpha_{ij],s}^{5} \right) + \\*
& & \gamma^{5} \gamma^{r}\left( + 4 \alpha_{(pq|,s}^{7} \alpha_{ij)}^{5 \,\,\,\, ,s} \delta^{h}_{r} - 4 \alpha_{(pq|}^{7 \,\,\,\, ,h} \alpha_{ij),r}^{5} - 4i \alpha_{[pq|,r}^{8} \alpha_{ij]}^{6 \,\,\,\, ,h} - 4 \alpha_{(pq|,r}^{5} \alpha_{ij)}^{7 \,\,\,\, ,h} - 4i \alpha_{[pq|}^{3} \epsilon_{0r}^{\,\,\,\, kh} \alpha_{ij],k}^{8} \right)
\end{eqnarray*}
\begin{eqnarray*}
& & - 4 \gamma^5 \gamma^{r}\epsilon_{0r}^{\,\,\,\, st} \alpha_{(pq|,s}^{6} \epsilon_{0t}^{\,\,\,\, kh} \alpha_{ij),k}^{8} + \gamma^0 \gamma^5 \gamma^r \left( - 4i \alpha_{[pq|}^{3}\alpha_{ij]}^{2} \delta^{h}_{r} - 4 \epsilon_{0r}^{\,\,\,\, sh} \alpha_{(pq|,s}^{6} \alpha_{ij)}^{2} + 4i \alpha_{[pq|}^{2} \alpha_{ij]}^{2} \delta^{h}_{r} + 4 \epsilon_{0r}^{\,\,\,\, sh} \alpha_{(pq|,s}^{7} \alpha_{ij)}^{3} \right) \\*
& & + \gamma^{0}\gamma^5 \gamma^r \left( 4i \widetilde{\alpha}_{[pq|}^{1} \epsilon_{0r}^{\,\,\,\, sh}\alpha_{ij],s}^{5} + 4 \alpha_{(pq|,s}^{8} \alpha_{ij)}^{5 \,\,\,\, ,s}\delta^{h}_{r} - 4 \alpha_{(pq|}^{8 \,\,\,\, ,h}\alpha_{ij),r}^5 + 4i \alpha_{[pq|,r}^{7} \alpha_{ij]}^{6 \,\,\,\, ,h} - 4i \alpha_{[pq|,r}^{6} \alpha_{ij]}^{7 \,\,\,\, ,h} \right) \\*
& & + \gamma^{0}\gamma^5 \gamma^r \left( 4i \alpha_{[pq|}^{4} \epsilon_{0r}^{\,\,\,\, sh} \alpha_{ij],s}^{8} - 4 \epsilon_{0r}^{\,\,\,\, st}\alpha_{(pq|,s}^{5} \epsilon_{0t}^{\,\,\,\, kh} \alpha_{ij),k}^{8} \right) - \\*
& & \left( 1 + \frac{GM}{r} \right) \sum_{i + j \geq 0} \sum_{p+q \geq 0} \sum_{a = 0}^{\min \{ j, p+1 \}} a! \left( \begin{array}{c} j \\* a \end{array} \right) \left( \begin{array}{c} p + 1 \\* a \end{array} \right) \int d \vec{k}_1 \ldots d \vec{k}_i d \vec{r}_1 \ldots d \vec{r}_{p+1} d\vec{l}_1 \ldots d\vec{l}_{j-a} d\vec{s}_1 \ldots \vec{s}_{q+1} \\*
& & a^{\dag}_{\vec{k}_1} \ldots a^{\dag}_{\vec{k}_i} a^{\dag}_{\vec{r}_{a+1}} \ldots a^{\dag}_{\vec{r}_{p+1}} a_{\vec{l}_1}\ldots a_{\vec{l}_{j-a}} a_{\vec{s}_1} \ldots a_{\vec{s}_{q+1}} \delta(\vec{r}_{p+1} - \vec{s}_{q+1})e^{i \left( \sum_{b=1}^{p+1}r_{b,c} - \sum_{b=1}^{q+1}s_{b,c} \right)x^c}r_{p+1,0} \\*
& & \left(2i \alpha_{ij}^{3} \alpha_{pq}^{4} - 2i \alpha_{ij}^{4}\alpha_{pq}^{3} + 
2i \alpha_{ij,h}^{5} \alpha_{pq}^{6 \,\,\, ,h} - 2i\alpha_{ij,h}^{6} \alpha_{pq}^{5 \,\,\, ,h} \right) + \\*
& & \gamma^{0} \left( -2 \alpha_{ij}^{3}\alpha_{pq}^{3} + 2 \alpha_{ij}^{4} \alpha_{pq}^{4} - 2 \alpha_{ij,h}^{6}\alpha_{pq}^{6 \,\,\, ,h} + 2 \alpha_{ij,h}^{5}\alpha_{pq}^{5 \,\,\, ,h} \right) + \\* 
& & \gamma^r \left( - 2i \alpha_{ij,r}^{8}\alpha_{pq}^{3} - 2\alpha_{ij,r}^{7}\alpha_{pq}^{4} + 2i \alpha_{ij}^{1}\alpha_{pq,r}^{6} - 2 \epsilon_{0r}^{\,\,\,\,ks} \alpha_{ij,k}^{8}\alpha_{pq,s}^{6} + 2\alpha_{ij}^{2} \alpha_{pq,r}^{5} - 2i \epsilon_{0r}^{\,\,\,\, ks} \alpha_{ij,k}^{7} \alpha_{pq,s}^{5} \right) + \\*
& & i \gamma^5 \left( 2 \alpha_{ij}^{2} \alpha_{pq}^{3} - 2i \alpha_{ij}^{1}\alpha_{pq}^{4} + 2 \alpha_{ij,h}^{7} \alpha_{pq}^{6 \,\, ,h} - 2i \alpha_{ij,h}^{8} \alpha_{pq}^{5 \,\, ,h} \right) + \\*
& & \gamma^{0} \gamma^{5} \left( - 2i \alpha_{ij}^{1} \alpha_{pq}^{3} + 2 \alpha_{ij}^{2} \alpha_{pq}^{4} + 2i\alpha_{ij,h}^{8} \alpha_{pq}^{6 \,\, ,h} - 2 \alpha_{ij,h}^{7} \alpha_{pq}^{5 \,\, ,h} \right) + \\*
& & \gamma^{0} \gamma^{h} \left(2i \alpha_{ij,h}^{7}\alpha_{pq}^{3} - 2 \alpha_{ij,h}^{8}\alpha_{pq}^{4} + 2i\alpha_{ij}^{2}\alpha_{pq,h}^{6} + 2 \epsilon_{0h}^{\,\,\,\, ks}\alpha_{ij,k}^{7} \alpha_{pq,s}^{6} - 2 \alpha_{ij}^{1}\alpha_{pq,h}^{5} - 2i \epsilon_{0h}^{\,\,\,\, ks}\alpha_{ij,k}^{8} \alpha_{pq,s}^{5} \right) + \\*
& & \gamma^{5}\gamma^{h}\left( - 2 \alpha_{ij,h}^{6}\alpha_{pq}^{3} - 2 \alpha_{ij,h}^{5}\alpha_{pq}^{4} - 2\alpha_{ij}^{3}\alpha_{pq,h}^{6} - 2i\epsilon_{0h}^{\,\,\,\, ks} \alpha_{ij,k}^{6} \alpha_{pq,s}^{6} - 2 \alpha_{ij}^{4}\alpha_{pq,h}^{5} + 2i \epsilon_{0h}^{\,\,\,\, ks} \alpha_{ij,k}^{5} \alpha_{pq,s}^{5} \right) + \\*
& &  \gamma^{0} \gamma^{5} \gamma^{h}\left( - 2i \alpha_{ij,h}^{5} \alpha_{pq}^{3}
- 2i \alpha_{ij,h}^{6} \alpha_{pq}^{4} + 2i \alpha_{ij}^{4} \alpha_{pq,h}^{6} - 2
\epsilon_{0h}^{\,\,\,\, ks}\alpha_{ij,k}^{5} \alpha_{pq,s}^{6} + 2i \alpha_{ij}^{3} \alpha_{pq,h}^{5} + 2 \epsilon_{0h}^{\,\,\,\, ks}\alpha_{ij,k}^{8}\alpha_{pq,s}^{5}  \right) - \\*
& & \left( 1 + \frac{GM}{r} \right) \sum_{i+j \geq 0} \sum_{p+q \geq 0} p \sum_{a= 0}^{\min \{ j,p \}} a! \left( \begin{array}{c} j \\* a \end{array} \right)\left( \begin{array}{c} p \\* a \end{array} \right) \int d\vec{k}_1 \ldots d\vec{k}_i d\vec{r}_1 \ldots d\vec{r}_{p} d\vec{l}_{1} \ldots d\vec{l}_{j-a} d\vec{s}_{1}\ldots d\vec{s}_{q} \\*
& & a^{\dag}_{\vec{k}_{1}} \ldots a^{\dag}_{\vec{k}_{i}}a^{\dag}_{\vec{r}_{a+1}} \ldots a^{\dag}_{\vec{k}_{p}} a_{\vec{l}_{1}} \ldots a_{\vec{l}_{j-a}} a_{\vec{s}_{1}} \ldots a_{\vec{s}_{q}} e^{i \left( \sum_{b=1}^{p}r_{b,c} - \sum_{b=1}^{q}s_{b,c}\right)x^c} r_{1,0} \\*
& & \left(2i \alpha_{ij}^{3} \alpha_{pq}^{4} - 2i \alpha_{ij}^{4}\alpha_{pq}^{3} + 
2i \alpha_{ij,h}^{5} \alpha_{pq}^{6 \,\,\, ,h} - 2i\alpha_{ij,h}^{6} \alpha_{pq}^{5 \,\,\, ,h} \right) + \\*
& & \gamma^{0} \left( -2 \alpha_{ij}^{3}\alpha_{pq}^{3} + 2 \alpha_{ij}^{4} \alpha_{pq}^{4} - 2 \alpha_{ij,h}^{6}\alpha_{pq}^{6 \,\,\, ,h} + 2 \alpha_{ij,h}^{5}\alpha_{pq}^{5 \,\,\, ,h} \right) + \\*
& & \gamma^r \left( - 2i \alpha_{ij,r}^{8}\alpha_{pq}^{3} - 2\alpha_{ij,r}^{7}\alpha_{pq}^{4} + 2i \alpha_{ij}^{1}\alpha_{pq,r}^{6} - 2 \epsilon_{0r}^{\,\,\,\,ks} \alpha_{ij,k}^{8}\alpha_{pq,s}^{6} + 2\alpha_{ij}^{2} \alpha_{pq,r}^{5} - 2i \epsilon_{0r}^{\,\,\,\, ks} \alpha_{ij,k}^{7} \alpha_{pq,s}^{5} \right) + \\*
& & i \gamma^5 \left( 2 \alpha_{ij}^{2} \alpha_{pq}^{3} - 2i \alpha_{ij}^{1}\alpha_{pq}^{4} + 2 \alpha_{ij,h}^{7} \alpha_{pq}^{6 \,\, ,h} - 2i \alpha_{ij,h}^{8} \alpha_{pq}^{5 \,\, ,h} \right) + \\*
& & \gamma^{0} \gamma^{5} \left( - 2i \alpha_{ij}^{1} \alpha_{pq}^{3} + 2 \alpha_{ij}^{2} \alpha_{pq}^{4} + 2i\alpha_{ij,h}^{8} \alpha_{pq}^{6 \,\, ,h} - 2 \alpha_{ij,h}^{7} \alpha_{pq}^{5 \,\, ,h} \right) + \\*
& & \gamma^{0} \gamma^{h} \left(2i \alpha_{ij,h}^{7}\alpha_{pq}^{3} - 2 \alpha_{ij,h}^{8}\alpha_{pq}^{4} + 2i\alpha_{ij}^{2}\alpha_{pq,h}^{6} + 2 \epsilon_{0h}^{\,\,\,\, ks}\alpha_{ij,k}^{7} \alpha_{pq,s}^{6} - 2 \alpha_{ij}^{1}\alpha_{pq,h}^{5} - 2i \epsilon_{0h}^{\,\,\,\, ks}\alpha_{ij,k}^{8} \alpha_{pq,s}^{5} \right) + \\*
& & \gamma^{5}\gamma^{h}\left( - 2 \alpha_{ij,h}^{6}\alpha_{pq}^{3} - 2 \alpha_{ij,h}^{5}\alpha_{pq}^{4} - 2\alpha_{ij}^{3}\alpha_{pq,h}^{6} - 2i\epsilon_{0h}^{\,\,\,\, ks} \alpha_{ij,k}^{6} \alpha_{pq,s}^{6} - 2 \alpha_{ij}^{4}\alpha_{pq,h}^{5} + 2i \epsilon_{0h}^{\,\,\,\, ks} \alpha_{ij,k}^{5} \alpha_{pq,s}^{5} \right) + \\*
& &  \gamma^{0} \gamma^{5} \gamma^{h}\left( - 2i \alpha_{ij,h}^{5} \alpha_{pq}^{3}
- 2i \alpha_{ij,h}^{6} \alpha_{pq}^{4} + 2i \alpha_{ij}^{4} \alpha_{pq,h}^{6} - 2
\epsilon_{0h}^{\,\,\,\, ks}\alpha_{ij,k}^{5} \alpha_{pq,s}^{6} + 2i \alpha_{ij}^{3} \alpha_{pq,h}^{5} + 2 \epsilon_{0h}^{\,\,\,\, ks}\alpha_{ij,k}^{8}\alpha_{pq,s}^{5}  \right) - \\*
\end{eqnarray*}
\begin{eqnarray*}
& & \sum_{i + j \geq 0} \sum_{p+q \geq 0} \sum_{a = 0}^{\min \{ j, p+1 \}} a! \left( \begin{array}{c} j \\* a \end{array} \right) \left( \begin{array}{c} p + 1 \\* a \end{array} \right) \int d \vec{k}_1 \ldots d \vec{k}_i d \vec{r}_1 \ldots d \vec{r}_{p+1} d\vec{l}_1 \ldots d\vec{l}_{j-a} d\vec{s}_1 \ldots \vec{s}_{q+1} \\*
& & a^{\dag}_{\vec{k}_1} \ldots a^{\dag}_{\vec{k}_i} a^{\dag}_{\vec{r}_{a+1}} \ldots a^{\dag}_{\vec{r}_{p+1}} a_{\vec{l}_1}\ldots a_{\vec{l}_{j-a}} a_{\vec{s}_1} \ldots a_{\vec{s}_{q+1}} \delta(\vec{r}_{p+1} - \vec{s}_{q+1})e^{i \left( \sum_{b=1}^{p+1}r_{b,c} - \sum_{b=1}^{q+1}s_{b,c} \right)x^c}r_{p+1,s} \\*
& & \left( 2i \alpha_{ij}^{6 \,\, ,s}\alpha_{pq}^{2} - 2i \alpha_{ij,s}^{7}\alpha_{pq}^{3} - 2 \alpha_{ij,r}^{8}\epsilon_{0}^{\,\, ksr} \alpha_{pq,k}^{5} - 2i \alpha_{ij}^{2} \alpha_{pq}^{6 \,\, ,s} - 2 \alpha_{ij}^{3} \alpha_{pq}^{7 \,\, ,s} + 2i \alpha_{ij,r}^{5} \epsilon_{0}^{rks} \alpha_{pq,k}^{8} \right) + \\*
& & \gamma^{0} \left( - 2\alpha_{ij}^{5 \,\, ,s}\alpha_{pq}^{2} - 2i \alpha_{ij}^{8 \,\, ,s} \alpha_{pq}^{3} + 2 \alpha_{ij,r}^{7} \epsilon_{0}^{\,\, ksr} \alpha_{pq,k}^{5} + 2i \alpha_{ij}^{1} \alpha_{pq}^{6 \,\, ,s} + 2i \alpha_{ij}^{4} \alpha_{pq}^{7 \,\, ,s} - 2 \alpha_{ij,r}^{6} \epsilon_{0}^{\,\, rks} \alpha_{pq,k}^{8} \right) + \\*
& & \gamma^{5} \left( - 2 \alpha_{ij}^{8 \,\, ,s}\alpha_{pq}^{2} - 2i\alpha_{ij}^{5 \,\, ,s}\alpha_{pq}^{3} + 2i \alpha_{ij,r}^{6}\epsilon_{0}^{\,\, ksr} \alpha_{pq,k}^{5} + 2i \alpha_{ij}^{4}\alpha_{pq}^{6 \,\,\, ,s} + 2i \alpha_{ij}^{1} \alpha_{pq}^{7 \,\,\, ,s} + 2i\alpha_{ij,r}^{7} \epsilon_{0}^{\,\, rks}\alpha_{pq,k}^{8} \right) + \\*
& & \gamma^{0}\gamma^{5} \left( 2 \alpha_{ij}^{7 \,\, ,s}\alpha_{pq}^{2} + 2 \alpha_{ij}^{7 \,\, ,s} \alpha_{pq}^{3} - 2 \epsilon_{0h}^{\,\,\,\, ks}\alpha_{ij,k}^{7}\alpha_{pq}^{3} + 2 \alpha_{ij}^{3}\alpha_{pq}^{6 \,\,\, ,s} + 2i \alpha_{ij}^{2}\alpha_{pq}^{7 \,\,\, ,s} + 2i \alpha_{ij,r}^{8} \epsilon_{0}^{\,\, rks} \alpha_{pq,k}^{8} + 2i \alpha_{ij,h}^{5}\alpha_{pq,k}^{5}\epsilon_{0}^{\,\, hks} \right) + \\*
& & \gamma^{h} \left( - 2\alpha_{ij}^{2}\alpha_{pq}^{2}\delta^{s}_{h} + 2i \epsilon_{0h}^{\,\,\,\, ks} \alpha_{ij,k}^{7}\alpha_{pq}^{2} + 2i \epsilon_{0h}^{\,\,\,\, ks}\alpha_{ij,k}^{6} \alpha_{pq}^{3} - 2 \alpha_{ij}^{3}\alpha_{pq}^{3} \delta^{s}_{h} - 2 \epsilon_{0h}^{\,\,\,\, ks}\alpha_{pq,k}^{5} \alpha_{ij}^{4} \right) + \\*
& &  \gamma^{h} \left( -2i \epsilon_{0h}^{\,\,\,\,kr} \epsilon_{0r}^{\,\,\,\, ls} \alpha_{ij,k}^{5}\alpha_{pq,l}^{5} - 2 \alpha_{ij,h}^{6} \alpha_{pq}^{6 \,\,\, ,s} - 2i \alpha_{ij,h}^{7}\alpha_{pq}^{7 \,\,\, ,s} + 2i \alpha_{ij}^{1}\epsilon_{0h}^{\,\,\,\, ks}\alpha_{pq,k}^{8} - 2 \epsilon_{0h}^{\,\,\,\, kr} \epsilon_{0r}^{\,\,\,\, ls} \alpha_{ij,k}^{8} \alpha_{pq,l}^{8} \right) + \\*
& & \gamma^{0} \gamma^{h} \left( 2 \alpha_{ij}^{1} \alpha_{pq}^{2} \delta^{s}_{h} + 2i \epsilon_{0h}^{\,\,\,\, ks} \alpha_{ij,k}^{8}\alpha_{pq}^{2} - 2 \epsilon_{0h}^{\,\,\,\, ks}\alpha_{ij,k}^{6} \alpha_{pq}^{3} + 2i \alpha_{ij}^{4} \alpha_{pq}^{3} \delta^{s}_{h} + 2\epsilon_{0h}^{\,\,\,\, kr}\alpha_{ij,k}^{6}\epsilon_{0r}^{\,\,\,\, ls}\alpha_{pq,l}^{5} + 2i \alpha_{ij}^{3} \epsilon_{0h}^{\,\,\,\, ks}\alpha_{pq,k}^{5}  \right) + \\*
& & \gamma^{0}\gamma^{h}\left( - 2i \alpha_{ij,h}^{5}\alpha_{pq}^{6 \,\,\, ,s} - 2i \alpha_{ij,h}^{8} \alpha_{pq}^{7 \,\,\, ,s} + 2i \epsilon_{0h}^{\,\,\,\, ks} \alpha_{pq,k}^{8}\alpha_{ij}^{2} + 2 \epsilon_{0h}^{\,\,\,\, kr} \epsilon_{0r}^{\,\,\,\, ls} \alpha_{ij,k}^{7} \alpha_{pq,l}^{8} \right) + \\*
& &  \gamma^{5} \gamma^{h} \left( 2 \alpha_{ij}^{4} \alpha_{pq}^{2} \delta^{s}_{h} + 2i \epsilon_{0h}^{\,\,\,\, ks} \alpha_{ij,k}^{5} \alpha_{pq}^{2}+ 2i \alpha_{ij}^{1}\alpha_{pq}^{3} \delta^{s}_{h} - 2\epsilon_{0h}^{\,\,\,\,ks}\alpha_{pq}^{3}\alpha_{ij,k}^{8} + 2 \alpha_{ij}^{2}\epsilon_{0h}^{\,\,\,\, ks}\alpha_{pq,k}^{5} - 2i \epsilon_{0h}^{\,\,\,\, kr} \epsilon_{0r}^{\,\,\,\, ls}\alpha_{ij,k}^{7}\alpha_{pq,l}^{5} \right) + \\*
& & \gamma^{5}\gamma^{h} \left( -2i \alpha_{ij,h}^{8}\alpha_{pq}^{6 \,\,\, ,s} - 2i \alpha_{ij,h}^{5}\alpha_{pq}^{7 \,\,\, ,s} - 2 \epsilon_{0h}^{\,\,\,\, ks} \alpha_{ij}^{3}\alpha_{pq,k}^{8} + 2i \epsilon_{0h}^{\,\,\,\, kr}\epsilon_{0r}^{\,\,\,\, ls}\alpha_{ij,k}^{6}\alpha_{pq,l}^{8} \right)+ \\*
& & \gamma^{0}\gamma^{5}\gamma^{h} \left( -2i \alpha_{ij}^{3}\alpha_{pq}^{2} \delta^{s}_{h} - 2 \epsilon_{0h}^{\,\,\,\, ks}\alpha_{ij,k}^{6}\alpha_{pq}^{2} + 2i \delta^{s}_{h}\alpha_{ij}^{2}\alpha_{pq}^{3} + 2 \epsilon_{0h}^{\,\,\,\, ks} \alpha_{ij,k}^{7}\alpha_{pq}^{3} - 2 \alpha_{ij}^{1}\epsilon_{0h}^{\,\,\,\, ks}\alpha_{pq,k}^{5} \right) + \\*
& & \gamma^{0}\gamma^{5}\gamma^{h} \left(- 2i \epsilon_{0h}^{\,\,\,\, kr}\epsilon_{0r}^{\,\,\,\, ls}\alpha_{ij,k}^{8} \alpha_{pq,l}^{5} + 2i \alpha_{ij,h}^{7}\alpha_{pq}^{6 \,\,\, ,s} + 2 \alpha_{ij,h}^{6}\alpha_{pq}^{7 \,\,\, ,s} + 2i \alpha_{ij}^{4}\epsilon_{0h}^{\,\,\,\, ks}\alpha_{pq,k}^{8} - 2 \epsilon_{0h}^{\,\,\,\, kr}\alpha_{ij,k}^{5} \epsilon_{0r}^{\,\,\,\, ls}\alpha_{pq,l}^{8} \right) - \\*
& &\sum_{i+j \geq 0} \sum_{p+q \geq 0} p \sum_{a= 0}^{\min \{ j,p \}} a! \left( \begin{array}{c} j \\* a \end{array} \right)\left( \begin{array}{c} p \\* a \end{array} \right) \int d\vec{k}_1 \ldots d\vec{k}_i d\vec{r}_1 \ldots d\vec{r}_{p} d\vec{l}_{1} \ldots d\vec{l}_{j-a} d\vec{s}_{1}\ldots d\vec{s}_{q} \\*
& & a^{\dag}_{\vec{k}_{1}} \ldots a^{\dag}_{\vec{k}_{i}}a^{\dag}_{\vec{r}_{a+1}} \ldots a^{\dag}_{\vec{k}_{p}} a_{\vec{l}_{1}} \ldots a_{\vec{l}_{j-a}} a_{\vec{s}_{1}} \ldots a_{\vec{s}_{q}} e^{i \left( \sum_{b=1}^{p}r_{b,c} - \sum_{b=1}^{q}s_{b,c}\right)x^c} r_{1,s} \\*
& & \left( 2i \alpha_{ij}^{6 \,\, ,s}\alpha_{pq}^{2} - 2i \alpha_{ij,s}^{7}\alpha_{pq}^{3} - 2 \alpha_{ij,r}^{8}\epsilon_{0}^{\,\, ksr} \alpha_{pq,k}^{5} - 2i \alpha_{ij}^{2} \alpha_{pq}^{6 \,\, ,s} - 2 \alpha_{ij}^{3} \alpha_{pq}^{7 \,\, ,s} + 2i \alpha_{ij,r}^{5} \epsilon_{0}^{rks} \alpha_{pq,k}^{8} \right) + \\*
& & \gamma^{0} \left( - 2\alpha_{ij}^{5 \,\, ,s}\alpha_{pq}^{2} - 2i \alpha_{ij}^{8 \,\, ,s} \alpha_{pq}^{3} + 2 \alpha_{ij,r}^{7} \epsilon_{0}^{\,\, ksr} \alpha_{pq,k}^{5} + 2i \alpha_{ij}^{1} \alpha_{pq}^{6 \,\, ,s} + 2i \alpha_{ij}^{4} \alpha_{pq}^{7 \,\, ,s} - 2 \alpha_{ij,r}^{6} \epsilon_{0}^{\,\, rks} \alpha_{pq,k}^{8} \right) + \\*
& & \gamma^{5} \left( - 2 \alpha_{ij}^{8 \,\, ,s}\alpha_{pq}^{2} - 2i\alpha_{ij}^{5 \,\, ,s}\alpha_{pq}^{3} + 2i \alpha_{ij,r}^{6}\epsilon_{0}^{\,\, ksr} \alpha_{pq,k}^{5} + 2i \alpha_{ij}^{4}\alpha_{pq}^{6 \,\,\, ,s} + 2i \alpha_{ij}^{1} \alpha_{pq}^{7 \,\,\, ,s} + 2i\alpha_{ij,r}^{7} \epsilon_{0}^{\,\, rks}\alpha_{pq,k}^{8} \right) + \\*
& & \gamma^{0}\gamma^{5} \left( 2 \alpha_{ij}^{7 \,\, ,s}\alpha_{pq}^{2} + 2 \alpha_{ij}^{6 \,\, ,s} \alpha_{pq}^{3} - 2 \epsilon_{0h}^{\,\,\,\, ks}\alpha_{ij,k}^{7}\alpha_{pq}^{3} + 2 \alpha_{ij}^{3}\alpha_{pq}^{6 \,\,\, ,s} + 2i \alpha_{ij}^{2}\alpha_{pq}^{7 \,\,\, ,s} + 2i \alpha_{ij,r}^{8} \epsilon_{0}^{\,\, rks} \alpha_{pq,k}^{8} + 2i\alpha_{ij,h}^{5}\alpha_{pq,k}^{5}\epsilon_{0}^{\,\, hks} \right) + \\*
& & \gamma^{h} \left( - 2\alpha_{ij}^{2}\alpha_{pq}^{2}\delta^{r}_{h} + 2i \epsilon_{0h}^{\,\,\,\, ks} \alpha_{ij,k}^{7}\alpha_{pq}^{2} + 2i \epsilon_{0h}^{\,\,\,\, ks}\alpha_{ij,k}^{6} \alpha_{pq}^{3} - 2 \alpha_{ij}^{3}\alpha_{pq}^{3} \delta^{s}_{h} - 2 \epsilon_{0h}^{\,\,\,\, ks}\alpha_{pq,k}^{5} \alpha_{ij}^{4} \right) + \\*
& &  \gamma^{h} \left( -2i \epsilon_{0h}^{\,\,\,\,kr} \epsilon_{0r}^{\,\,\,\, ls} \alpha_{ij,k}^{5}\alpha_{pq,l}^{5} - 2 \alpha_{ij,h}^{6} \alpha_{pq}^{6 \,\,\, ,s} - 2i \alpha_{ij,h}^{7}\alpha_{pq}^{7 \,\,\, ,s} + 2i \alpha_{ij}^{1}\epsilon_{0h}^{\,\,\,\, ks}\alpha_{pq,k}^{8} - 2 \epsilon_{0h}^{\,\,\,\, kr} \epsilon_{0r}^{\,\,\,\, ls} \alpha_{ij,k}^{8} \alpha_{pq,l}^{8} \right) + \\*
& & \gamma^{0} \gamma^{h} \left( 2 \alpha_{ij}^{1} \alpha_{pq}^{2} \delta^{s}_{h} + 2i \epsilon_{0h}^{\,\,\,\, ks} \alpha_{ij,k}^{8}\alpha_{pq}^{2} - 2 \epsilon_{0h}^{\,\,\,\, ks}\alpha_{ij,k}^{6} \alpha_{pq}^{3} + 2i \alpha_{ij}^{4} \alpha_{pq}^{3} \delta^{s}_{h} + 2\epsilon_{0h}^{\,\,\,\, kr}\alpha_{ij,k}^{6}\epsilon_{0r}^{\,\,\,\, ls}\alpha_{pq,l}^{5} + 2i \alpha_{ij}^{3} \epsilon_{0h}^{\,\,\,\, ks}\alpha_{pq,k}^{5}  \right) + \\*
& & \gamma^{0}\gamma^{h}\left( - 2i \alpha_{ij,h}^{5}\alpha_{pq}^{6 \,\,\, ,s} - 2i \alpha_{ij,h}^{8} \alpha_{pq}^{7 \,\,\, ,s} + 2i \epsilon_{0h}^{\,\,\,\, ks} \alpha_{pq,k}^{8}\alpha_{ij}^{2} + 2 \epsilon_{0h}^{\,\,\,\, kr} \epsilon_{0r}^{\,\,\,\, ls} \alpha_{ij,k}^{7} \alpha_{pq,l}^{8} \right) + \\*
& &  \gamma^{5} \gamma^{h} \left( 2 \alpha_{ij}^{4} \alpha_{pq}^{2} \delta^{s}_{h} + 2i \epsilon_{0h}^{\,\,\,\, ks} \alpha_{ij,k}^{5} \alpha_{pq}^{2}+ 2i \alpha_{ij}^{1}\alpha_{pq}^{3} \delta^{s}_{h} - 2\epsilon_{0h}^{\,\,\,\,ks}\alpha_{pq}^{3}\alpha_{ij,k}^{8} + 2 \alpha_{ij}^{2}\epsilon_{0h}^{\,\,\,\, ks}\alpha_{pq,k}^{5} - 2i \epsilon_{0h}^{\,\,\,\, kr} \epsilon_{0r}^{\,\,\,\, ls}\alpha_{ij,k}^{7}\alpha_{pq,l}^{5} \right) + \\*
& & \gamma^{5}\gamma^{h} \left( -2i \alpha_{ij,h}^{8}\alpha_{pq}^{6 \,\,\, ,s} - 2i \alpha_{ij,h}^{5}\alpha_{pq}^{7 \,\,\, ,s} - 2 \epsilon_{0h}^{\,\,\,\, ks} \alpha_{ij}^{3}\alpha_{pq,k}^{8} + 2i \epsilon_{0h}^{\,\,\,\, kr}\epsilon_{0r}^{\,\,\,\, ls}\alpha_{ij,k}^{6}\alpha_{pq,l}^{8} \right)+ \\*
& & \gamma^{0}\gamma^{5}\gamma^{h} \left( -2i \alpha_{ij}^{3}\alpha_{pq}^{2} \delta^{s}_{h} - 2 \epsilon_{0h}^{\,\,\,\, ks}\alpha_{ij,k}^{6}\alpha_{pq}^{2} + 2i \delta^{s}_{h}\alpha_{ij}^{2}\alpha_{pq}^{3} + 2 \epsilon_{0h}^{\,\,\,\, ks} \alpha_{ij,k}^{7}\alpha_{pq}^{3} - 2 \alpha_{ij}^{1}\epsilon_{0h}^{\,\,\,\, ks}\alpha_{pq,k}^{5} \right) + \\*
& & \gamma^{0}\gamma^{5}\gamma^{h} \left(- 2i \epsilon_{0h}^{\,\,\,\, kr}\epsilon_{0r}^{\,\,\,\, ls}\alpha_{ij,k}^{8} \alpha_{pq,l}^{5} + 2i \alpha_{ij,h}^{7}\alpha_{pq}^{6 \,\,\, ,s} + 2 \alpha_{ij,h}^{6}\alpha_{pq}^{7 \,\,\, ,s} + 2i \alpha_{ij}^{4}\epsilon_{0h}^{\,\,\,\, ks}\alpha_{pq,k}^{8} - 2 \epsilon_{0h}^{\,\,\,\, kr}\alpha_{ij,k}^{5} \epsilon_{0r}^{\,\,\,\, ls}\alpha_{pq,l}^{8} \right) -
\end{eqnarray*}
\begin{eqnarray*}
& & i \left( 1 + \frac{GM}{r} \right) \sum_{i + j \geq 0} \sum_{p+q \geq 0} \sum_{a = 0}^{\min \{ j, p \}} a! \left( \begin{array}{c} j \\* a \end{array} \right) \left( \begin{array}{c} p \\* a \end{array} \right) \int d \vec{k}_1 \ldots d \vec{k}_i d \vec{r}_1 \ldots d \vec{r}_{p} d\vec{l}_1 \ldots d\vec{l}_{j-a} d\vec{s}_1 \ldots \vec{s}_{q} \\*
& & a^{\dag}_{\vec{k}_1} \ldots a^{\dag}_{\vec{k}_i} a^{\dag}_{\vec{r}_{a+1}} \ldots a^{\dag}_{\vec{r}_{p}} a_{\vec{l}_1}\ldots a_{\vec{l}_{j-a}} a_{\vec{s}_1} \ldots a_{\vec{s}_{q}} e^{i \left( \sum_{b=1}^{p}r_{b,c} - \sum_{b=1}^{q}s_{b,c} \right)x^c} \\*
& & \left(2i \partial_t \alpha_{ij}^{3} \alpha_{pq}^{4} - 2i \partial_t \alpha_{ij}^{4}\alpha_{pq}^{3} + 
2i \partial_t \alpha_{ij,h}^{5} \alpha_{pq}^{6 \,\,\, ,h} - 2i\partial_t \alpha_{ij,h}^{6} \alpha_{pq}^{5 \,\,\, ,h} \right) + \\*
& & \gamma^{0} \left( -2 \partial_t \alpha_{ij}^{3}\alpha_{pq}^{3} + 2 \partial_t \alpha_{ij}^{4} \alpha_{pq}^{4} - 2\partial_t \alpha_{ij,h}^{6}\alpha_{pq}^{6 \,\,\, ,h} + 2\partial_t \alpha_{ij,h}^{5}\alpha_{pq}^{5 \,\,\, ,h} \right) + \\*
& & \gamma^r \left( - 2i \partial_t \alpha_{ij,r}^{8}\alpha_{pq}^{3} - 2\partial_t \alpha_{ij,r}^{7}\alpha_{pq}^{4} + 2i \partial_t \alpha_{ij}^{1}\alpha_{pq,r}^{6} - 2 \epsilon_{0r}^{\,\,\,\,ks} \partial_t \alpha_{ij,k}^{8}\alpha_{pq,s}^{6} + 2\partial_t \alpha_{ij}^{2} \alpha_{pq,r}^{5} - 2i \epsilon_{0r}^{\,\,\,\, ks} \partial_t \alpha_{ij,k}^{7} \alpha_{pq,s}^{5} \right) + \\*
& & i \gamma^5 \left( 2 \partial_t \alpha_{ij}^{2} \alpha_{pq}^{3} - 2i \partial_t \alpha_{ij}^{1}\alpha_{pq}^{4} + 2 \partial_t \alpha_{ij,h}^{7} \alpha_{pq}^{6 \,\, ,h} - 2i\partial_t \alpha_{ij,h}^{8} \alpha_{pq}^{5 \,\, ,h} \right) + \\*
& & \gamma^{0} \gamma^{5} \left( - 2i \partial_t \alpha_{ij}^{1} \alpha_{pq}^{3} + 2\partial_t \alpha_{ij}^{2} \alpha_{pq}^{4} + 2i\partial_t \alpha_{ij,h}^{8} \alpha_{pq}^{6 \,\, ,h} - 2\partial_t \alpha_{ij,h}^{7} \alpha_{pq}^{5 \,\, ,h} \right) + \\*
& & \gamma^{0} \gamma^{h} \left(2i \partial_t \alpha_{ij,h}^{7}\alpha_{pq}^{3} - 2 \partial_t \alpha_{ij,h}^{8}\alpha_{pq}^{4} + 2i\partial_t \alpha_{ij}^{2}\alpha_{pq,h}^{6} + 2 \epsilon_{0h}^{\,\,\,\, ks}\partial_t \alpha_{ij,k}^{7} \alpha_{pq,s}^{6} - 2\partial_t \alpha_{ij}^{1}\alpha_{pq,h}^{5} - 2i \epsilon_{0h}^{\,\,\,\, ks}\partial_t \alpha_{ij,k}^{8} \alpha_{pq,s}^{5} \right) + \\*
& & \gamma^{5}\gamma^{h}\left( - 2\partial_t \alpha_{ij,h}^{6}\alpha_{pq}^{3} - 2 \partial_t \alpha_{ij,h}^{5}\alpha_{pq}^{4} - 2\partial_t \alpha_{ij}^{3}\alpha_{pq,h}^{6} - 2i\epsilon_{0h}^{\,\,\,\, ks}\partial_t \alpha_{ij,k}^{6} \alpha_{pq,s}^{6} - 2\partial_t \alpha_{ij}^{4}\alpha_{pq,h}^{5} + 2i \epsilon_{0h}^{\,\,\,\, ks}\partial_t \alpha_{ij,k}^{5} \alpha_{pq,s}^{5} \right) + \\*
& &  \gamma^{0} \gamma^{5} \gamma^{h}\left( - 2i\partial_t \alpha_{ij,h}^{5} \alpha_{pq}^{3}
- 2i\partial_t \alpha_{ij,h}^{6} \alpha_{pq}^{4} + 2i\partial_t \alpha_{ij}^{4} \alpha_{pq,h}^{6} - 2
\epsilon_{0h}^{\,\,\,\, ks}\partial_t \alpha_{ij,k}^{5} \alpha_{pq,s}^{6} + 2i \partial_t \alpha_{ij}^{3} \alpha_{pq,h}^{5} \right) + \\*
& & 2\gamma^{0} \gamma^{5} \gamma^{h} \epsilon_{0h}^{\,\,\,\, ks}\partial_t \alpha_{ij,k}^{8}\alpha_{pq,s}^{5} - \\*
& & i \sum_{i + j \geq 0} \sum_{p+q \geq 0} \sum_{a = 0}^{\min \{ j, p \}} a! \left( \begin{array}{c} j \\* a \end{array} \right) \left( \begin{array}{c} p \\* a \end{array} \right) \int d \vec{k}_1 \ldots d \vec{k}_i d \vec{r}_1 \ldots d \vec{r}_{p} d\vec{l}_1 \ldots d\vec{l}_{j-a} d\vec{s}_1 \ldots \vec{s}_{q} \\*
& & a^{\dag}_{\vec{k}_1} \ldots a^{\dag}_{\vec{k}_i} a^{\dag}_{\vec{r}_{a+1}} \ldots a^{\dag}_{\vec{r}_{p}} a_{\vec{l}_1}\ldots a_{\vec{l}_{j-a}} a_{\vec{s}_1} \ldots a_{\vec{s}_{q}} e^{i \left( \sum_{b=1}^{p}r_{b,c} - \sum_{b=1}^{q}s_{b,c} \right)x^c} \\*
& & \left( 2i \partial_{s} \alpha_{ij}^{6 \,\, ,s}\alpha_{pq}^{2} - 2i \partial_{s} \alpha_{ij,s}^{7}\alpha_{pq}^{3} - 2 \partial_{s} \alpha_{ij,r}^{8}\epsilon_{0}^{\,\, ksr} \alpha_{pq,k}^{5} - 2i \partial_{s} \alpha_{ij}^{2} \alpha_{pq}^{6 \,\, ,s} - 2 \partial_{s} \alpha_{ij}^{3} \alpha_{pq}^{7 \,\, ,s} + 2i \partial_{s} \alpha_{ij,r}^{5} \epsilon_{0}^{rks} \alpha_{pq,k}^{8} \right) + \\*
& & \gamma^{0} \left( - 2 \partial_{s} \alpha_{ij}^{5 \,\, ,s}\alpha_{pq}^{2} - 2i \partial_{s} \alpha_{ij}^{8 \,\, ,s} \alpha_{pq}^{3} + 2\partial_{s} \alpha_{ij,r}^{7} \epsilon_{0}^{\,\, ksr} \alpha_{pq,k}^{5} + 2i \partial_{s} \alpha_{ij}^{1} \alpha_{pq}^{6 \,\, ,s} + 2i\partial_{s}  \alpha_{ij}^{4} \alpha_{pq}^{7 \,\, ,s} - 2 \partial_{s} \alpha_{ij,r}^{6} \epsilon_{0}^{\,\, rks} \alpha_{pq,k}^{8} \right) + \\*
& & \gamma^{5} \left( - 2 \partial_{s} \alpha_{ij}^{8 \,\, ,s}\alpha_{pq}^{2} - 2i \partial_{s} \alpha_{ij}^{5 \,\, ,s}\alpha_{pq}^{3} + 2i \partial_{s} \alpha_{ij,r}^{6}\epsilon_{0}^{\,\, ksr} \alpha_{pq,k}^{5} + 2i \partial_{s} \alpha_{ij}^{4}\alpha_{pq}^{6 \,\,\, ,s} + 2i \partial_{s} \alpha_{ij}^{1} \alpha_{pq}^{7 \,\,\, ,s} + 2i \partial_{s} \alpha_{ij,r}^{7} \epsilon_{0}^{\,\, rks}\alpha_{pq,k}^{8} \right) \\*
& & + \gamma^{0}\gamma^{5} \left( 2\partial_{s} \alpha_{ij}^{7 \,\, ,s}\alpha_{pq}^{2} + 2\partial_{s} \alpha_{ij}^{7 \,\, ,s} \alpha_{pq}^{3} - 2 \epsilon_{0h}^{\,\,\,\, ks} \partial_{s} \alpha_{ij,k}^{7}\alpha_{pq}^{3} + 2 \partial_{s} \alpha_{ij}^{3}\alpha_{pq}^{6 \,\,\, ,s} + 2i \partial_{s} \alpha_{ij}^{2}\alpha_{pq}^{7 \,\,\, ,s} + 2i \partial_{s} \alpha_{ij,r}^{8} \epsilon_{0}^{\,\, rks} \alpha_{pq,k}^{8} \right) \\*
& &  + 2i \partial_{s} \alpha_{ij,h}^{5}\alpha_{pq,k}^{5}\epsilon_{0}^{\,\, hks}\gamma^{0}\gamma^{5} + \gamma^{h} \left( - 2\partial_{s} \alpha_{ij}^{2}\alpha_{pq}^{2}\delta^{r}_{h} + 2i \epsilon_{0h}^{\,\,\,\, ks} \partial_{s} \alpha_{ij,k}^{7}\alpha_{pq}^{2} + 2i \epsilon_{0h}^{\,\,\,\, ks}\partial_{s} \alpha_{ij,k}^{6} \alpha_{pq}^{3} - 2  \partial_{s}\alpha_{ij}^{3}\alpha_{pq}^{3} \delta^{s}_{h} \right) + \\*
& &  \gamma^{h} \left( - 2 \epsilon_{0h}^{\,\,\,\, ks}\alpha_{pq,k}^{5}\partial_{s} \alpha_{ij}^{4} -2i \epsilon_{0h}^{\,\,\,\,kr} \epsilon_{0r}^{\,\,\,\, ls} \partial_{s} \alpha_{ij,k}^{5}\alpha_{pq,l}^{5} - 2\partial_{s} \alpha_{ij,h}^{6} \alpha_{pq}^{6 \,\,\, ,s} - 2i\partial_{s} \alpha_{ij,h}^{7}\alpha_{pq}^{7 \,\,\, ,s} + 2i\partial_{s} \alpha_{ij}^{1}\epsilon_{0h}^{\,\,\,\, ks}\alpha_{pq,k}^{8} \right)  \\*
& &  - 2 \epsilon_{0h}^{\,\,\,\, kr} \epsilon_{0r}^{\,\,\,\, ls}\partial_{s} \alpha_{ij,k}^{8} \alpha_{pq,l}^{8}\gamma^{h} + \\*
& & \gamma^{0} \gamma^{h} \left( 2\partial_{s} \alpha_{ij}^{1} \alpha_{pq}^{2} \delta^{s}_{h} + 2i \epsilon_{0h}^{\,\,\,\, ks}\partial_{s} \alpha_{ij,k}^{8}\alpha_{pq}^{2} - 2 \epsilon_{0h}^{\,\,\,\, ks}\partial_{s} \alpha_{ij,k}^{6} \alpha_{pq}^{3} + 2i\partial_{s} \alpha_{ij}^{4} \alpha_{pq}^{3} \delta^{s}_{h} + 2\epsilon_{0h}^{\,\,\,\, kr}\partial_{s} \alpha_{ij,k}^{6}\epsilon_{0r}^{\,\,\,\, ls}\alpha_{pq,l}^{5} \right) + \\* 
& &  \gamma^{0}\gamma^{h}\left(2i\partial_{s} \alpha_{ij}^{3} \epsilon_{0h}^{\,\,\,\, ks}\alpha_{pq,k}^{5} - 2i\partial_{s} \alpha_{ij,h}^{5}\alpha_{pq}^{6 \,\,\, ,s} - 2i\partial_{s} \alpha_{ij,h}^{8} \alpha_{pq}^{7 \,\,\, ,s} + 2i \epsilon_{0h}^{\,\,\,\, ks} \alpha_{pq,k}^{8}\partial_{s} \alpha_{ij}^{2} + 2 \epsilon_{0h}^{\,\,\,\, kr} \epsilon_{0r}^{\,\,\,\, ls}\partial_{s} \alpha_{ij,k}^{7} \alpha_{pq,l}^{8} \right) + \\*
& &  \gamma^{5} \gamma^{h} \left( 2\partial_{s} \alpha_{ij}^{4} \alpha_{pq}^{2} \delta^{s}_{h} + 2i \epsilon_{0h}^{\,\,\,\, ks}\partial_{s} \alpha_{ij,k}^{5} \alpha_{pq}^{2}+ 2i\partial_{s} \alpha_{ij}^{1}\alpha_{pq}^{3} \delta^{s}_{h} \right) +  \\*
& & \gamma^{5} \gamma^{h} \left( - 2\epsilon_{0h}^{\,\,\,\,ks}\alpha_{pq}^{3}\partial_{s} \alpha_{ij,k}^{8} + 2 \partial_{s} \alpha_{ij}^{2}\epsilon_{0h}^{\,\,\,\, ks}\alpha_{pq,k}^{5} - 2i \epsilon_{0h}^{\,\,\,\, kr} \epsilon_{0r}^{\,\,\,\, ls}\partial_{s} \alpha_{ij,k}^{7}\alpha_{pq,l}^{5} \right) + \\*
& & \gamma^{5}\gamma^{h} \left( -2i\partial_{s} \alpha_{ij,h}^{8}\alpha_{pq}^{6 \,\,\, ,s} - 2i\partial_{s} \alpha_{ij,h}^{5}\alpha_{pq}^{7 \,\,\, ,s} - 2 \epsilon_{0h}^{\,\,\,\, ks}\partial_{s} \alpha_{ij}^{3}\alpha_{pq,k}^{8} + 2i \epsilon_{0h}^{\,\,\,\, kr}\epsilon_{0r}^{\,\,\,\, ls} \partial_{s} \alpha_{ij,k}^{6}\alpha_{pq,l}^{8} \right) + \\*
& & \gamma^{0}\gamma^{5}\gamma^{h} \left( -2i\partial_{s} \alpha_{ij}^{3}\alpha_{pq}^{2} \delta^{s}_{h} - 2 \epsilon_{0h}^{\,\,\,\, ks}\partial_{s} \alpha_{ij,k}^{6}\alpha_{pq}^{2} + 2i \delta^{s}_{h}\partial_{s} \alpha_{ij}^{2}\alpha_{pq}^{3} + 2 \epsilon_{0h}^{\,\,\,\, ks}\partial_{s} \alpha_{ij,k}^{7}\alpha_{pq}^{3} - 2 \partial_{s} \alpha_{ij}^{1}\epsilon_{0h}^{\,\,\,\, ks}\alpha_{pq,k}^{5} \right) + \\*
& & \gamma^{0}\gamma^{5}\gamma^{h} \left(- 2i \epsilon_{0h}^{\,\,\,\, kr}\epsilon_{0r}^{\,\,\,\, ls}\partial_{s} \alpha_{ij,k}^{8} \alpha_{pq,l}^{5} + 2i\partial_{s} \alpha_{ij,h}^{7}\alpha_{pq}^{6 \,\,\, ,s} + 2 \partial_{s} \alpha_{ij,h}^{6}\alpha_{pq}^{7 \,\,\, ,s} + 2i \partial_{s} \alpha_{ij}^{4}\epsilon_{0h}^{\,\,\,\, ks}\alpha_{pq,k}^{8} \right) \\*
& & - 2\gamma^{0}\gamma^{5}\gamma^{h}  \epsilon_{0h}^{\,\,\,\, kr}\partial_{s} \alpha_{ij,k}^{5} \epsilon_{0r}^{\,\,\,\, ls}\alpha_{pq,l}^{8} - \\*
\end{eqnarray*}
\begin{eqnarray*}
& & \sum_{i + j \geq 0} \int d \vec{k}_1 \ldots d \vec{k}_i \int d\vec{l}_1 \ldots d\vec{l}_j a^{\dag}_{\vec{k}_1} \ldots a^{\dag}_{\vec{k}_i} a_{\vec{l}_1} \ldots a_{\vec{l}_j}  \\*
& & \left( 1 + \frac{GM}{r} \right)\left( \partial_t \alpha_{ij}^{1} + \gamma^{0}\partial_t \alpha_{ij}^{2} + i \partial_t \alpha_{ij}^{3}\gamma^5 + 
\gamma^0 \gamma^5 \partial_t \alpha_{ij}^{4} +\partial_t \alpha^{5}_{ij,h} \gamma^h + i \gamma^0 \gamma^h \partial_t \alpha^{6}_{ij,h} + \gamma^5 \gamma^h \partial_t \alpha^{7}_{ij,h} \right)\gamma^{0} + \\*
& & \left( 1 + \frac{GM}{r} \right) \left( \gamma^0 \gamma^5 \gamma^h \partial_t \alpha^{8}_{ij,h} \right) \gamma^{0} + \left( \partial_s \alpha_{ij}^{1} + \gamma^{0}\partial_s \alpha_{ij}^{2} + i \partial_s \alpha_{ij}^{3}\gamma^5 + 
\gamma^0 \gamma^5 \partial_s \alpha_{ij}^{4}\right) \gamma^{s} + \\*
& & \left( \partial_s \alpha^{5}_{ij,h} \gamma^h + i \gamma^0 \gamma^h \partial_s \alpha^{6}_{ij,h} + \gamma^5 \gamma^h \partial_s \alpha^{7}_{ij,h} + \gamma^0 \gamma^5 \gamma^h \partial_s \alpha^{8}_{ij,h} \right)\gamma^{s} - \\*
& & \frac{i}{2} \left( 1 + \frac{GM}{r} \right) \partial_t \sum_{i + j \geq 0} \sum_{p + q \geq 0}\sum_{a = 0}^{\min \{ j,p \}} a! \left( \begin{array}{c}
j \\ a
\end{array}  \right) \left( \begin{array}{c} p \\ a \end{array} \right) \int d \vec{k}_1 \ldots d\vec{k}_i d\vec{r}_1 \ldots d\vec{r}_p d\vec{l}_{1} \ldots d\vec{l}_{j-a} d\vec{s}_1 \ldots d\vec{s}_q \\*
& & a^{\dag}_{\vec{k}_1} \ldots a^{\dag}_{\vec{k}_i} a^{\dag}_{\vec{r}_{a + 1}} 
\ldots a^{\dag}_{\vec{r}_{p}} a_{\vec{l}_{1}} \ldots a_{\vec{l}_{j-a}}a_{\vec{s}_{1}} \ldots a_{\vec{s}_{q}} \\*
& & \sum_{c = 1}^{8} i^{F(c \, \text{mod} \, 4)} \alpha_{ij}^c(\vec{k}_g, \vec{l}_{1}, \ldots, \vec{l}_{j-a}, \vec{r}_{1}, \ldots, \vec{r}_{a}, \vec{x},t) \alpha_{pq}^c(\vec{r}_g, \vec{s}_h , \vec{x},t) + \\*
& & \gamma^{0} \left( 2 \alpha_{(ij|}^1 \alpha_{pq)}^2 - 2i\alpha_{[ij|}^3 \alpha_{pq]}^4 - 2i \alpha_{[ij|,h}^5 \alpha_{pq]}^{6 \,\,\, ,h} - 2\alpha_{(ij|,h}^{7} \alpha_{pq)}^{8 \, \, \, h} \right) \\*
& &  + \, i \gamma^5 \left( 2\alpha_{(ij|}^1 \alpha_{pq)}^3 + 2i \alpha_{[ij|}^2 \alpha_{pq]}^4 + 2i \alpha_{[ij|,h}^5 \alpha_{pq]}^{7 \,\,\, ,h} - 2 \alpha_{(ij|,h}^6 \alpha_{pq)}^{8 \,\,\, ,h} \right) \\*
& & + \, \gamma^0 \gamma^5 \left( 2 \alpha_{(ij|}^1 \alpha_{pq)}^4 + 2 i \alpha_{[ij|}^2 \alpha_{pq]}^3 + 2\alpha_{(ij|,h}^8 \alpha_{pq)}^{5 \,\,\, ,h} - 2i \alpha_{[ij|,h}^6 \alpha_{pq]}^{7 \,\,\,\, ,h} \right) \\*
& & + \gamma^h \left( 2 \alpha_{(ij|}^1 \alpha_{pq),h}^5
 - 2i \alpha_{[ij|}^2 \alpha_{pq],h}^6 + 2i \alpha_{[ij|}^3 \alpha_{pq],h}^7 + 2\alpha_{(ij|}^5 \alpha_{pq),h}^8 \right) + \\*
& & i \gamma^0 \gamma^h \left( 2\alpha^1_{(ij|} \alpha_{pq),h}^6 - 2i \alpha_{[ij|}^2 \alpha_{pq],h}^5 - 2\alpha_{(ij|}^3 \alpha_{pq),h}^8 - 2i \alpha_{[ij|}^4 \alpha_{pq],h}^7 \right) + \\*
& & i \gamma^0 \gamma^h \left( 2\epsilon_{0h}^{\,\,\, rs} \alpha_{(ij|,r}^5 \alpha_{pq),s}^7 + 2i \epsilon_{0h}^{\,\,\, rs} \alpha_{[ij|,r}^6 \alpha_{pq],s}^8  \right) + \\*  
& &  \gamma^5 \gamma^h \left( 2\alpha_{(ij|}^1 \alpha_{pq),h}^7 - 2\alpha_{(ij|}^2 \alpha_{pq),h}^8 + 2i \alpha_{[ij|}^3 \alpha_{pq],h}^5 \right) +  \\*
& & \gamma^5 \gamma^h \left(- 2i \alpha_{[ij|,h}^6 \alpha_{pq]}^4 - 2 \epsilon_{0rsh} \alpha_{(ij|,r}^5 \alpha_{pq),s}^6 + 2i \epsilon_{0rsh} \alpha_{[ij|,r}^7 \alpha_{pq],s}^8 \right) + \\*
& & \gamma^0 \gamma^5 \gamma^h \left( 2\alpha_{(ij|}^1 \alpha_{pq),h}^8 + 2\alpha_{(ij|}^2 \alpha_{pq),h}^7 - 2\alpha_{(ij|}^3 \alpha_{pq),h}^6 \right) +  \\* & & \gamma^0 \gamma^5 \gamma^h \left( 2\alpha_{(ij|}^4 \alpha_{pq),h}^5 + 2i \epsilon_{0hrs} \alpha_{[ij|,r}^5 \alpha_{pq],s}^5 - 2i \epsilon_{0hrs} \alpha_{[ij|,r}^7 \alpha_{pq],s}^7 \right) \gamma^{0} - \\*
& & \frac{i}{2} \partial_{s} \sum_{i + j \geq 0} \sum_{p + q \geq 0}\sum_{a = 0}^{\min \{ j,p \}} a! \left( \begin{array}{c}
j \\ a
\end{array}  \right) \left( \begin{array}{c} p \\ a \end{array} \right) \int d \vec{k}_1 \ldots d\vec{k}_i d\vec{r}_1 \ldots d\vec{r}_p d\vec{l}_{1} \ldots d\vec{l}_{j-a} d\vec{s}_1 \ldots d\vec{s}_q \\*
& & a^{\dag}_{\vec{k}_1} \ldots a^{\dag}_{\vec{k}_i} a^{\dag}_{\vec{r}_{a + 1}} 
\ldots a^{\dag}_{\vec{r}_{p}} a_{\vec{l}_{1}} \ldots a_{\vec{l}_{j-a}}a_{\vec{s}_{1}} \ldots a_{\vec{s}_{q}} \\*
& & \sum_{c = 1}^{8} i^{F(c \, \text{mod} \, 4)} \alpha_{ij}^c(\vec{k}_g, \vec{l}_{1}, \ldots, \vec{l}_{j-a}, \vec{r}_{1}, \ldots, \vec{r}_{a}, \vec{x},t) \alpha_{pq}^c(\vec{r}_g, \vec{s}_h , \vec{x},t) + \\*
& & \gamma^{0} \left( 2 \alpha_{(ij|}^1 \alpha_{pq)}^2 - 2i\alpha_{[ij|}^3 \alpha_{pq]}^4 - 2i \alpha_{[ij|,h}^5 \alpha_{pq]}^{6 \,\,\, ,h} - 2\alpha_{(ij|,h}^{7} \alpha_{pq)}^{8 \, \, \, h} \right) \\*
& &  + \, i \gamma^5 \left( 2\alpha_{(ij|}^1 \alpha_{pq)}^3 + 2i \alpha_{[ij|}^2 \alpha_{pq]}^4 + 2i \alpha_{[ij|,h}^5 \alpha_{pq]}^{7 \,\,\, ,h} - 2 \alpha_{(ij|,h}^6 \alpha_{pq)}^{8 \,\,\, ,h} \right) \\*
\end{eqnarray*}
\begin{eqnarray*}
& & + \, \gamma^0 \gamma^5 \left( 2 \alpha_{(ij|}^1 \alpha_{pq)}^4 + 2 i \alpha_{[ij|}^2 \alpha_{pq]}^3 + 2\alpha_{(ij|,h}^8 \alpha_{pq)}^{5 \,\,\, ,h} - 2i \alpha_{[ij|,h}^6 \alpha_{pq]}^{7 \,\,\,\, ,h} \right) \\*
& & + \gamma^h \left( 2 \alpha_{(ij|}^1 \alpha_{pq),h}^5
 - 2i \alpha_{[ij|}^2 \alpha_{pq],h}^6 + 2i \alpha_{[ij|}^3 \alpha_{pq],h}^7 + 2\alpha_{(ij|}^5 \alpha_{pq),h}^8 \right) + \\*
& & i \gamma^0 \gamma^h \left( 2\alpha^1_{(ij|} \alpha_{pq),h}^6 - 2i \alpha_{[ij|}^2 \alpha_{pq],h}^5 - 2\alpha_{(ij|}^3 \alpha_{pq),h}^8 - 2i \alpha_{[ij|}^4 \alpha_{pq],h}^7 \right) + \\*
& & i \gamma^0 \gamma^h \left( 2\epsilon_{0h}^{\,\,\, rs} \alpha_{(ij|,r}^5 \alpha_{pq),s}^7 + 2i \epsilon_{0h}^{\,\,\, rs} \alpha_{[ij|,r}^6 \alpha_{pq],s}^8  \right) + \\*  
& &  \gamma^5 \gamma^h \left( 2\alpha_{(ij|}^1 \alpha_{pq),h}^7 - 2\alpha_{(ij|}^2 \alpha_{pq),h}^8 + 2i \alpha_{[ij|}^3 \alpha_{pq],h}^5 \right) +  \\*
& & \gamma^5 \gamma^h \left(- 2i \alpha_{[ij|,h}^6 \alpha_{pq]}^4 - 2 \epsilon_{0rsh} \alpha_{(ij|,r}^5 \alpha_{pq),s}^6 + 2i \epsilon_{0rsh} \alpha_{[ij|,r}^7 \alpha_{pq],s}^8 \right) + \\*
& & \gamma^0 \gamma^5 \gamma^h \left( 2\alpha_{(ij|}^1 \alpha_{pq),h}^8 + 2\alpha_{(ij|}^2 \alpha_{pq),h}^7 - 2\alpha_{(ij|}^3 \alpha_{pq),h}^6 \right) +  \\* & & \gamma^0 \gamma^5 \gamma^h \left( 2\alpha_{(ij|}^4 \alpha_{pq),h}^5 + 2i \epsilon_{0hrs} \alpha_{[ij|,r}^5 \alpha_{pq],s}^5 - 2i \epsilon_{0hrs} \alpha_{[ij|,r}^7 \alpha_{pq],s}^7 \right) \gamma^{s} - \\*
& & \frac{i}{2} \left( 1 + \frac{GM}{r} \right) \partial_t \sum_{i + j \geq 0} \int d^3 \vec{k}_1 \ldots d^3 \vec{k}_i d^3 \vec{l}_1 \ldots d^3 \vec{l}_j a^{\dag}_{\vec{k}_1} \ldots a^{\dag}_{\vec{k}_i} a_{\vec{l}_1} \ldots a_{\vec{l}_j} \left[ \left(i k_{i,a} - j l_{j,a} \right) x^{a} \right] \beta_{ij}(\vec{k}, \vec{l}, \vec{x},t))\gamma^{0} - \\*
& & \frac{i}{2}\partial_s \sum_{i + j \geq 0} \int d^3 \vec{k}_1 \ldots d^3 \vec{k}_i d^3 \vec{l}_1 \ldots d^3 \vec{l}_j a^{\dag}_{\vec{k}_1} \ldots a^{\dag}_{\vec{k}_i} a_{\vec{l}_1} \ldots a_{\vec{l}_j} \left[ \left(i k_{i,a} - j l_{j,a} \right) x^{a} \right] \beta_{ij}(\vec{k}, \vec{l}, \vec{x},t) \gamma^{s}.
\end{eqnarray*}
Before we gather all terms and try to solve the truncated system, we perform a drastic simplification by assuming that all
 Clifford coeficients with $h$-indices vanish.  The reason is that is natural to look for rotationally symmetric Hamiltonians $H(x)$,
 defined with respect to $x = 0$, on rotationally symmetric backgrounds.  Up till now, we have not implemented this symmetry for the 
very simple reason that the above calculations need to be performed anyway on generic backgrounds.  Hence, the resulting system of integro-differential equations is given by:
\begin{eqnarray*}
0 & = & \sum_{m + n \geq 0} \int d \vec{k}_1 \ldots d\vec{k}_{m} d\vec{l}_1 \ldots d\vec{l}_{n} a^{\dag}_{\vec{k}_1} \ldots a^{\dag}_{\vec{k}_{m}}a_{\vec{l}_{1}} \ldots a_{\vec{l}_{n}} \\*
& & -4 i \sum_{i,q \geq 0; a = \max \{0, q-n+1, i-m+1 \} \ldots \min \{ i,q \} ;p = m+a-i-1,j=n-q-1+a} a! \left( \begin{array}{c} q \\* a \end{array} \right) \left( \begin{array}{c} i \\* a \end{array} \right) e^{i \left( \sum_{t=1}^{m}k_{t,b} - \sum_{t=1}^{n}l_{t,b} \right)x^b} \\*
& & \left( 1 + \frac{GM}{r} \right) \int d\vec{k}_{m+1}\ldots d\vec{k}_{m+a} \delta(\vec{k}_{p+1} - \vec{l}_{n}) k_{p+1,0} \alpha_{[pq}^{4}\alpha_{ij]}^{3} \\*
& & - 4i \sum_{i,q \geq 0; a = \max \{ 0, q-n, i-m \} \ldots \min \{ i,q \} ; p = m+a-i, j = n-q+a } a!\left( \begin{array}{c} q \\* a \end{array} \right) \left( \begin{array}{c} i \\* a \end{array} \right) \\*
& & \left( 1 + \frac{GM}{r} \right) e^{i \left( \sum_{t=1}^{m}k_{t,b} - \sum_{t=1}^{n}l_{t,b} \right)x^b}\left( pk_{1,0} + (i-a)k_{m+1,0} \right)\alpha_{[pq}^{4}\alpha_{ij]}^{3} \\*  
\end{eqnarray*}
\begin{eqnarray*}
& & - 4i \left( 1 + \frac{GM}{r} \right)\sum_{i,q \geq 0;a = \max \{ 0, i+1 - m,q+1-n \} \ldots \min \{ i+1,q\}; p = -i-1+a+m, j = n+a-q-1 } 
\int d\vec{k}_{m+1} \ldots d\vec{k}_{m+a} \\*
& & \alpha_{[pq}^{3}\alpha_{ij]}^{4} e^{i\left( \sum_{r = p+1}^{m+a}k_{r,b} - \sum_{r = q-a+1}^{n}l_{r,b} \right)x^{b}} \\*
& &
\left( a! \left(\begin{array}{c} q \\* a \end{array}  \right) \left(\begin{array}{c} i \\* a \end{array}  \right)k_{m,0} \delta (\vec{k}_{m} - \vec{l}_{n})
 +  a! \left(\begin{array}{c} q \\* a \end{array}  \right) \left(\begin{array}{c} i \\* a-1 \end{array}  \right)k_{m+1,0} 
\delta (\vec{k}_{m+1} - \vec{l}_{n}) \right) \\*
& & - 4i\left( 1 + \frac{GM}{r} \right) \sum_{i,q \geq 0; a = \max \{ 0,i-m,q-n \} \ldots \min \{ i,q \}; p = - i+a+m,j=n+a-q} i \int 
d\vec{k}_{m+1} \ldots d\vec{k}_{m+a} \\*
& & \alpha_{[pq}^{3}\alpha_{ij]}^{4}e^{i\left( \sum_{r=p+1}^{m+a}k_{r,b} - \sum_{r=q-a+1}^{n}l_{r,b} \right)x^{b}}\left( a! \left(  
\begin{array}{c} q \\* a  \end{array} \right)\left( \begin{array}{c} i-1 \\* a  \end{array} \right)k_{m,0} +
a! \left( \begin{array}{c} q \\* a  \end{array} \right)\left( \begin{array}{c} i-1 \\* a-1  \end{array} \right)k_{m+1,0}  \right) \\*
& &  + 2\left( 1 + \frac{GM}{r} \right) \sum_{i,q \geq 0;a = \max \{0,i-m,q-n \} \ldots \min \{i,q\};p=-i+a+m,j=n+a-q} a! \left(  
\begin{array}{c} q \\* a  \end{array} \right)\left( \begin{array}{c} i \\* a  \end{array} \right) \\*
& & \int d\vec{k}_{m+1} \ldots d\vec{k}_{m+a}e^{i\left( \sum_{r=p+1}^{m+a}k_{r,b} - \sum_{r=q-a+1}^{n}l_{r,b} \right)x^{b}} \left( 
\partial_t \alpha_{pq}^{3}\alpha_{ij}^{4} - \partial_t \alpha_{pq}^{4} \alpha_{ij}^{3} \right) \\*
& & + \left( 1 + \frac{GM}{r} \right) \partial_t \alpha_{mn}^{2} \\*
\\* 
& & + \frac{i}{2} \left( 1 + \frac{GM}{r} \right) \sum_{i,q \geq 0; a = \max \{ 0,i-m,q-n \} \ldots \min \{ i,q \};p = -i+a+m,j=n+a-q } a! \
\left( \begin{array}{c} q \\* a \end{array} \right)\left( \begin{array}{c} i \\* a \end{array} \right) \\*
& & \int d\vec{k}_{m+1} \ldots d\vec{k}_{m+a} \partial_t \left( 2\alpha_{(pq}^{1}\alpha_{ij)}^{2} - 2i\alpha_{[pq}^{3}\alpha_{ij]}^{4} \right) \\*
& & + \frac{i}{2}\left( 1 + \frac{GM}{r} \right) \partial_t \left((mk_{m,b} - nl_{n,b})x^{b}\alpha_{mn}^{2} \right)
\end{eqnarray*}
for the central component and obviously, no spatial derivatives enter at this stage.  The $\gamma^{0}$ component is given by
\begin{eqnarray*}
0 & = & \sum_{m+n \geq 0} \int d\vec{k}_{1} \ldots d\vec{k}_{m} d\vec{l}_{1} \ldots d\vec{l}_{n}a^{\dag}_{\vec{k}_{1}} \ldots a^{\dag}_{\vec{k}_{m}}
a_{\vec{l}_{1}} \ldots a_{\vec{l}_{n}} \\*
& & - 4\sum_{i,q \geq 0; a = \max \{0,q-n+1,i-m+1\} \ldots \min \{ i,q \}; p = m+a-i-1,j=n+a-q-1}a! \left( \begin{array}{c} q \\* a \end{array}\right)
\left( \begin{array}{c} i \\* a \end{array}\right) e^{i\left( \sum_{t=1}^{m}k_{t,b} - \sum_{t=1}^{n} l_{t,b} \right)x^{b}} \\*
& & \left( 1 + \frac{GM}{r}  \right) \int d\vec{k}_{m+1} \ldots d\vec{k}_{m+a} \delta(\vec{k}_{p+1} - \vec{l}_{n})k_{p+1,0} (\alpha_{pq}^{3}
\alpha_{ij}^{3} - \alpha_{pq}^{4}\alpha_{ij}^{4}) \\*
& & - 4\sum_{i,q \geq 0;a = \max \{ 0,q-n,i-m \} \ldots \min \{ i,q \}; p = m+a-i,j=n-q+a}a! \left( \begin{array}{c} q \\* a \end{array} \right)
\left( \begin{array}{c} i \\* a \end{array} \right) \\*
& & \left( 1 + \frac{GM}{r} \right) e^{i\left( \sum_{t=1}^{m}k_{t,b} - \sum_{t=1}^{n}l_{t,b} \right)x^{b}} \int d\vec{k}_{m+1} \ldots d\vec{k}_{m+a}
(pk_{1,0} + (i-a)k_{p+1,0})(\alpha_{pq}^{3} \alpha_{ij}^{3} - \alpha_{pq}^{4}\alpha_{ij}^{4}) \\*
& & + 2\left( 1 + \frac{GM}{r} \right) \sum_{i,q \geq 0; a= \max \{ 0,i+1-m,q+1-n \} \ldots \min \{ i+1,q \};p = m+a-i-1,j=n+a-q-1} \int
d\vec{k}_{m+1} \ldots d\vec{k}_{m+a}  \\*
\end{eqnarray*}
\begin{eqnarray*}
& & e^{i\left( \sum_{r=p+1}^{m+a}k_{r,b} - \sum_{r=q-a+1}^{n} l_{r,b} \right)x^{b}} (\alpha_{pq}^{3} \alpha_{ij}^{3} - 
\alpha_{pq}^{4} \alpha_{ij}^{4} ) \\*
& & \left( a! \left( \begin{array}{c} q \\* a \end{array} \right)\left( \begin{array}{c} i \\* a \end{array} \right)k_{m,0}\delta(
\vec{k}_{m} - \vec{l}_{n}) +  a! \left( \begin{array}{c} q \\* a \end{array} \right)\left( \begin{array}{c} i \\* a-1 \end{array} \right)
k_{m+1,0}\delta(\vec{k}_{m+1} - \vec{l}_{n}) \right) \\*
& & + 2\left( 1 + \frac{GM}{r} \right) \sum_{i,q \geq 0;a = \max \{ 0,i-m,q-n \} \ldots \min \{ i,q \};p=m+a-i,j=n+a-q} i\int d\vec{k}_{m+1}
\ldots d\vec{k}_{m+a} \\*
& & e^{i\left( \sum_{r=p+1}^{m+a}k_{r,b} - \sum_{r=q+1-a}^{n}l_{r,b} \right)x^{b}}(\alpha_{pq}^{3}\alpha_{ij}^{3} - \alpha_{pq}^{4}\alpha_{ij}^{4}) \\*
& & \left( a! \left( \begin{array}{c} q \\* a  \end{array} \right)\left( \begin{array}{c} i-1 \\* a  \end{array} \right)k_{m,0} 
+ a! \left( \begin{array}{c} q \\* a  \end{array} \right)\left( \begin{array}{c} i-1 \\* a-1  \end{array} \right)k_{m+1,0} \right) \\*
& & + 2i\left( 1 + \frac{GM}{r} \right) \sum_{i,q \geq 0;a = \max \{ 0,i-m,q-n \} \ldots \min \{ i,q \}; p = a+m-i,j=n+a-q}a! \left( 
\begin{array}{c} q \\* a \end{array} \right)\left( \begin{array}{c} i \\* a \end{array} \right) \\*
& & \int d\vec{k}_{m+1} \ldots d\vec{k}_{m+a} e^{i\left( \sum_{r=p+1}^{m+a}k_{r,b} - \sum_{r=q+1-a}^{n}l_{r,b}\right)x^{b}} (\partial_t 
\alpha_{pq}^{3}\alpha_{ij}^{3} - \partial_t \alpha_{pq}^{4}\alpha_{ij}^{4}) \\*
& & + \left( 1 + \frac{GM}{r} \right) \partial_t \alpha_{mn}^{1} \\*
& & - \frac{i}{2} \left( 1 + \frac{GM}{r} \right) \sum_{i,q \geq 0; a = \max \{ 0,i-m,q-n \} \ldots \min \{ i,q \}; p = m+a-i,j=n+a-q} a!
\left( \begin{array}{c} q \\* a \end{array}  \right)\left( \begin{array}{c} i \\* a \end{array}  \right) \\*
& & \int d\vec{k}_{m+1} \ldots d\vec{k}_{m+a} \partial_t \left( \alpha_{pq}^{1}\alpha_{ij}^{1} - \alpha_{pq}^{2}\alpha_{ij}^{2} 
- \alpha_{pq}^{3}\alpha_{ij}^{3} +  \alpha_{pq}^{4}\alpha_{ij}^{4}  \right) \\*
& & - \frac{i}{2} \left( 1 + \frac{GM}{r} \right) \partial_t ((mk_{m,b} - nl_{n,b})x^{b}\alpha_{mn}^{1}) \\*
\end{eqnarray*}
which has a rather similar structure as the previous one.  The $\gamma^{5}$ equation is given by
\begin{eqnarray*}
0 & = & \sum_{m+n \geq 0} \int d\vec{k}_{1} \ldots d\vec{k}_{m} d\vec{l}_{1} \ldots d\vec{l}_{n}a^{\dag}_{\vec{k}_{1}} \ldots
 a^{\dag}_{\vec{k}_{m}}a_{\vec{l}_{1}} \ldots a_{\vec{l}_{n}} \\*
& & - 2i\chi(m-1) \chi(n-1) \left( 1 + \frac{GM}{r} \right)k_{m,0}\delta(\vec{k}_{m} - \vec{l}_{n})e^{i\left( \sum_{t=1}^{m}k_{t,b} - \sum_{t=1}^{n}
l_{t,b} \right)x^{b}} \alpha_{(m-1)(n-1)}^{4} \\*
& & -2i\left( 1 + \frac{GM}{r} \right) mk_{1,0}e^{i\left( \sum_{t=1}^{m}k_{t,b} - \sum_{t=1}^{n}l_{t,b} \right)x^{b}} \alpha_{mn}^{4} \\*
& & + 2i\left( 1 + \frac{GM}{r} \right) \sum_{i,q \geq 0;a = \max \{ 0,q-n+1,i-m+1 \} \ldots \min \{ i,q \};p=m+a-i-1,j=n+a-q-1} a! 
\left( \begin{array}{c} q \\* a \end{array}  \right)\left( \begin{array}{c} i \\* a \end{array}  \right) \\*
& & e^{i\left( \sum_{t=1}^{m}k_{t,b} - \sum_{t=1}^{n}l_{t,b} \right)x^{b}} \delta(\vec{k}_{p+1} - \vec{l}_{n}) \int d\vec{k}_{m+1} \ldots d\vec{k}_{m+a}
\left( \alpha_{(pq}^{2} \alpha_{ij)}^{3} - i \widetilde{\alpha}_{[pq}^{1}\alpha_{ij]}^{4} \right) \\*
& & + 2i\left( 1 + \frac{GM}{r} \right) \sum_{i,q \geq 0;a = \max \{ 0,q-n+1,i-m+1 \} \ldots \min \{ i,q \};p = m+a-i,j=n+a-q} a! 
\left( \begin{array}{c} q \\* a \end{array}  \right)\left( \begin{array}{c} i \\* a \end{array}  \right) \\*
\end{eqnarray*}
\begin{eqnarray*}
& & e^{i\left( \sum_{t=1}^{m}k_{t,b} - \sum_{t=1}^{n}l_{t,b}\right)x^{b}} \int d\vec{k}_{m+1} \ldots d\vec{k}_{m+a}(pk_{1,0} + (i-a)k_{p+1,0})
\left( \alpha_{(pq}^{2}\alpha_{ij)}^{3} - i \widetilde{\alpha}_{[pq}^{1}\alpha_{ij]}^{4} \right) \\*                                                 
& & - 2i\left( 1 + \frac{GM}{r} \right) \sum_{i,q \geq 0; a = \max \{ 0; i+1-m,q+1-n \} \ldots \min \{ i,q \};p=m+a-i-1,j=n+a-q-1 }
\int d\vec{k}_{m+1} \ldots d\vec{k}_{m+a} \\*
& &  e^{i\left( \sum_{r=p+1}^{m+a}k_{r,b} - \sum_{r=q-a+1}^{n}l_{r,b} \right)x^{b}} (\alpha_{pq}^{2} \alpha_{ij}^{3} - i \alpha_{pq}^{1} \alpha_{ij}^{4}) \\*
& & \left( a! \left( \begin{array}{c} q \\* a \end{array} \right)\left( \begin{array}{c} i \\* a \end{array} \right)k_{m,0} \delta(\vec{k}_{m}
- \vec{l}_{n}) +  a! \left( \begin{array}{c} q \\* a \end{array} \right)\left( \begin{array}{c} i \\* a-1 \end{array} \right)k_{m+1,0} 
\delta(\vec{k}_{m+1} - \vec{l}_{n}) \right) \\*
& & - 2i \left( 1 + \frac{GM}{r} \right) \sum_{i,q \geq 0; a= \max \{ 0,i-m,q-n \} \ldots \min \{ i+1,q \}; p=m+a-i,j=n+a-q}i \int d\vec{k}_{m+1}
\ldots d\vec{k}_{m+a} \\*
& & e^{i \left( \sum_{r=p+1}^{m+a} k_{r,b} - \sum_{r=q-a+1}^{n} l_{r,b} \right)  x^{b}} (\alpha_{pq}^{2} \alpha_{ij}^{3} - i \alpha_{pq}^{1} \alpha_{ij}^{4}) \\*
& & \left( a! \left( \begin{array}{c} q \\* a \end{array} \right)\left( \begin{array}{c} i-1 \\* a \end{array} \right)k_{m,0} + a!
\left( \begin{array}{c} q \\* a \end{array} \right)\left( \begin{array}{c} i-1 \\* a-1 \end{array} \right)k_{m+1,0} \right) \\*
& & + 2\left( 1 + \frac{GM}{r} \right) \sum_{i,q \geq 0; a= \max \{0,i-m,q-n \} \ldots \min \{ i+1,q \};p=m+a-i,j=n+a-q }a! 
\left( \begin{array}{c} q \\* a \end{array} \right)\left( \begin{array}{c} i \\* a \end{array} \right) \\*
& & \int d\vec{k}_{m+1} \ldots d\vec{k}_{m+a} e^{i\left( \sum_{r=p+1}^{m+a}k_{r,b} - \sum_{r=q-a+1}^{n} l_{r,b} \right)x^{b}}k_{m+1,0}
(\partial_t \alpha_{pq}^{2}\alpha_{ij}^{3} - i \partial_t \alpha_{pq}^{1}\alpha_{ij}^{4}) \\*
& & - \left( 1 + \frac{GM}{r} \right)\partial_t \alpha_{mn}^{4} \\*
& & - i\left( 1 + \frac{GM}{r} \right)\sum_{i,q \geq 0;a= \max \{ 0,i-m,q-n \} \ldots \min \{ i+1,q \};p=m+a-i,j=n+a-q } a!   
\left( \begin{array}{c} q \\* a \end{array} \right)\left( \begin{array}{c} i \\* a \end{array} \right) \\*
& & \int d\vec{k}_{m+1} \ldots d\vec{k}_{m+a} \partial_t \left( \alpha_{(pq}^{1}\alpha_{ij)}^{4} + i \alpha_{[pq}^{2}\alpha_{ij]}^{3} \right) \\*
& & - \frac{i}{2} \left( 1 + \frac{GM}{r} \right) \partial_t \left((mk_{m,b} - nl_{n,b})x^{b}\alpha_{mn}^{4} \right) \\*
\end{eqnarray*}
and the last set of equations which does not depend upon the spatial derivatives, that is the coefficients of $\gamma^{0}\gamma^{5}$ is given by
\begin{eqnarray*}
0 & = & \sum_{m+n \geq 0} \int d\vec{k}_{1} \ldots d\vec{k}_{m}d\vec{l}_{1} \ldots d\vec{l}_{n}a^{\dag}_{\vec{k}_{1}} \ldots a^{\dag}_{\vec{k}_{m}}
a_{\vec{l}_{1}} \ldots a_{\vec{l}_{n}} \\*
& & - 2 \chi(m-1) \chi(n-1) \left( 1 + \frac{GM}{r} \right)k_{m,0}e^{i\left( \sum_{t=1}^{m}k_{t,b} - \sum_{t=1}^{n}l_{t,b} \right)x^{b} }
\alpha_{(m-1)(n-1)}^{3} \delta(\vec{k}_{m} - \vec{l}_{n}) \\*
& & -2\left( 1 + \frac{GM}{r} \right)mk_{1,0}e^{i\left( \sum_{t=1}^{m}k_{t,b} - \sum_{t=1}^{n}l_{t,b}  \right)x^{b}}\alpha_{mn}^{3} \\*
& & -2i\left( 1 + \frac{GM}{r} \right) \sum_{i,q \geq 0;a= \max \{0,q-n+1,i-m+1 \} \ldots \min \{ i,q \}; p = m+a-i-1,j=n+a-q-1}
a! \left( \begin{array}{c} q \\* a \end{array}  \right) \left( \begin{array}{c} i \\* a \end{array}  \right) \\*
& & e^{i \left( \sum_{t=1}^{m}k_{t,b} - \sum_{t=1}^{n}l_{t,b}  \right)x^{b}}\delta(\vec{k}_{p+1} - \vec{l}_{n}) \int d\vec{k}_{m+1} \ldots
d\vec{k}_{m+a} \left( \widetilde{\alpha}_{[pq}^{1}\alpha_{ij]}^{3} + i \alpha_{(pq}^{2}\alpha_{ij)}^{4} \right) \\*
& & - 2i \left( 1 + \frac{GM}{r} \right)\sum_{i,q \geq 0;a = \max \{ 0,q-n+1,i-m+1 \} \ldots \min \{ i,q \};p=m+a-i,j=n+a-q } a!  
\left( \begin{array}{c} q \\* a \end{array}  \right) \left( \begin{array}{c} i \\* a \end{array}  \right) \\*
& & e^{i\left( \sum_{t=1}^{m}k_{t,b} - \sum_{t=1}^{n}l_{t,b} \right)x^{b}} \int d\vec{k}_{m+1} \ldots d\vec{k}_{m+a}(pk_{1,0} + (i-a)k_{p+1,0})
\left( \widetilde{\alpha}_{[pq}^{1}\alpha_{ij]}^{3} + i \alpha_{(pq}^{2}\alpha_{ij)}^{4} \right) \\*
& & + 2i\left( 1 + \frac{GM}{r} \right) \sum_{i,q \geq 0; a = \max \{ 0,i+1-m,q+1,n\} \ldots \min \{ i+1,q \};p=m+a-i-1,j=n+a-q-1} \int d\vec{k}_{m+1}
\ldots d\vec{k}_{m+a} \\*
& & e^{i\left( \sum_{r=p+1}^{m+a}k_{r,b} - \sum_{r=q-a+1}^{n} l_{r,b} \right)x^{b}} (\alpha_{pq}^{1}\alpha_{ij}^{3} + i\alpha_{pq}^{2}\alpha_{ij}^{4}) \\*
& & \left( a! \left( \begin{array}{c} q \\* a \end{array}  \right)\left( \begin{array}{c} i \\* a \end{array}  \right)k_{m,0} \delta(\vec{k}_{m} 
- \vec{l}_{n}) + a! \left( \begin{array}{c} q \\* a \end{array}  \right)\left( \begin{array}{c} i \\* a-1 \end{array}  \right)k_{m+1,0}
 \delta(\vec{k}_{m+1} - \vec{l}_{n}) \right) \\*
& &+ 2i \left( 1 + \frac{GM}{r} \right) \sum_{i,q \geq 0;a = \max \{ 0,i-m,q-n \} \ldots \min \{ i+1,q \};p=m+a-i,j=n+a-q}i \int d\vec{k}_{m+1}
\ldots d\vec{k}_{m+a} \\*
& & e^{i \left( \sum_{r=p+1}^{m+a}k_{r,b} - \sum_{t=q+1-a}^{n}l_{r,b} \right)x^{b}}(\alpha_{pq}^{1}\alpha_{ij}^{3} + i \alpha_{pq}^{2}\alpha_{ij}^{4}) \\*
& & \left( a! \left( \begin{array}{c} q \\* a \end{array}  \right)\left( \begin{array}{c} i-1 \\* a \end{array}  \right)k_{m,0} + 
a! \left( \begin{array}{c} q \\* a \end{array}  \right)\left( \begin{array}{c} i-1 \\* a-1 \end{array}  \right)k_{m+1,0} \right) \\*
& & -2\left( 1 + \frac{GM}{r} \right) \sum_{i,q \geq 0; a= \max \{ 0,i-m,q-n \} \ldots \min \{ i+1,q \};p=m+a-i,j=n+a-q } a! \left(
\begin{array}{c} q \\* a \end{array} \right)\left(\begin{array}{c} i \\* a \end{array} \right) \\*
& & \int d\vec{k}_{m+1} \ldots d\vec{k}_{m+a} e^{i\left( \sum_{r=p+1}^{m+a}k_{r,b} - \sum_{r=q+1-a}^{n}l_{r,b}\right)x^{b}}k_{m+1,0}
(\partial_t \alpha_{pq}^{1}\alpha_{ij}^{3} + i \partial_t \alpha_{pq}^{2}\alpha_{ij}^{4}) \\*
\end{eqnarray*}
\begin{eqnarray*}
& & + i\left( 1 + \frac{GM}{r} \right)\partial_t \alpha_{mn}^{3} \\*
& & + \left( 1 + \frac{GM}{r} \right) \sum_{i,q \geq 0;a= \max \{ 0,i-m,q-n \} \ldots \min \{ \i+1,q \};p = m+a-i,j=n+a-q} a! 
\left( \begin{array}{c} q \\* a \end{array} \right)\left(\begin{array}{c} i \\* a \end{array} \right) \\*
& & \int d\vec{k}_{m+1} \ldots d\vec{k}_{m+a} \partial_t \left( \alpha_{(pq}^{1}\alpha_{ij)}^{3} + i \alpha_{[pq}^{2}\alpha_{ij]}^{4} \right) \\*
& & - \frac{1}{2}\left( 1 + \frac{GM}{r} \right)\partial_t ((mk_{m,b} - nl_{n,b})x^{b}\alpha_{mn}^{3}). \\*
\end{eqnarray*}
Now, we write down the equations which break the spatial isotropy and restrict the spatial derivatives of the $\alpha$ functions; the coefficients
of $\gamma^h$ are given by
\begin{eqnarray*}
0 & = & \sum_{m+n \geq 0} \int d\vec{k}_{1} \ldots d\vec{k}_{m}d\vec{l}_{1} \ldots d\vec{l}_{n}a^{\dag}_{\vec{k}_{1}} \ldots a^{\dag}_{\vec{k}_{m}} 
a_{\vec{l}_{1}} \ldots a_{\vec{l}_{n}} \\*
& & -2 \sum_{i,q \geq 0; a= \max \{ 0,i+1-m,q+1-n\} \ldots \min \{ i,q\};p =m+a-i-1,j=n+a-q-1}a! \left( \begin{array}{c} q \\* a \end{array}  \right)
\left( \begin{array}{c} i \\* a \end{array}  \right) \\*
& & \int d\vec{k}_{m+1} \ldots d\vec{k}_{m+a} \delta(\vec{k}_{p+1} - \vec{l}_{n}) \alpha_{(pq}^{2}\alpha_{ij)}^{2}k_{p+1}^{\,\,\, ,h}
e^{i\left(\sum_{t=1}^{m}k_{t,b} - \sum_{t=1}^{n}l_{t,b} \right)x^{b}} \\*
& & -2\sum_{i,q \geq 0;a= \max \{0,i-m,q-n \} \ldots \min \{ i,q \}; p =m+a-i,j=n+a-q}a! \left( \begin{array}{c} q \\* a \end{array}  \right)
\left( \begin{array}{c} i \\* a \end{array}  \right) \\*
\end{eqnarray*}
\begin{eqnarray*}
& & \int d \vec{k}_{m+1} \ldots d\vec{k}_{m+a}\delta(\vec{k}_{p+1} - \vec{l}_{n})\alpha_{(pq}^{2}\alpha_{ij)}^{2}e^{i\left( \sum_{t=1}^{m} 
k_{t,b} - \sum_{t=1}^{n} l_{t,b} \right)x^{b}} \left( pk_{1}^{\,\,\, ,h} + (i-a)k_{p+1}^{\, \, \, ,h}  \right) \\*
& & + 2\sum_{i,q \geq 0;a = \max \{0,q+1-n,i+1-m  \} \ldots \min \{ i+1,q \};p=m+a-i-1,j=n+a-q-1  } \int d\vec{k}_{m+1} \ldots d\vec{k}_{m+a} \\*
& & e^{i\left( \sum_{t=p+1}^{m+a}k_{t,b} - \sum_{t=q-a+1}^{n}l_{t,b} \right)x^{b}} \\*
& & \left( a! \left( \begin{array}{c} q \\* a \end{array} \right)\left( \begin{array}{c} i \\* a \end{array} \right)k_{m}^{\,\, \, ,h}
\delta(\vec{k}_{m} - \vec{l}_{n}) + a! \left( \begin{array}{c} q \\* a \end{array} \right)\left( \begin{array}{c} i \\* a-1 \end{array} \right)k_{m}^{\,\, \, ,h}
\delta(\vec{k}_{m+1} - \vec{l}_{n}) \right)\alpha_{pq}^{2}\alpha_{ij}^{2} \\*
& & +2 \sum_{i,q \geq 0;a = \max \{ 0,q-n,i-m \} \ldots \min \{ i,q \};p=m+a-i,j=n+a-q  }i \int d\vec{k}_{m+1} \ldots d\vec{k}_{m+a}
e^{i\left( \sum_{t=p+1}^{m+a}k_{t,b} - \sum_{t=q+1-a}^{n}l_{t,b} \right)x^{b}} \\*
& & \left( a! \left( \begin{array}{c} q \\* a  \end{array} \right)\left( \begin{array}{c} i-1 \\* a  \end{array} \right)k_{m}^{\,\,\, ,h}
+ a!\left( \begin{array}{c} q \\* a  \end{array} \right)\left( \begin{array}{c} i-1 \\* a-1  \end{array} \right)k_{m+1}^{\,\,\, ,h} \right)
\alpha_{pq}^{2}\alpha_{ij}^{2} \\*
& & +2i \sum_{i,q \geq 0;a = \max \{ 0,q-n,i-m \} \ldots \min \{ i,q \};p=m+a-i,j=n+a-q }i \int d\vec{k}_{m+1} \ldots d\vec{k}_{m+a}
e^{i\left( \sum_{t=p+1}^{m+a}k_{t,b} - \sum_{t=q+1-a}^{n}l_{t,b} \right)x^{b}} \\*
& & a! \left( \begin{array}{c} q \\* a  \end{array} \right)\left( \begin{array}{c} i \\* a  \end{array} \right)\partial^h \alpha_{pq}^{2}
\alpha_{ij}^{2} \\*
& & - \partial^h \alpha_{mn}^{1} \\*
& & - \frac{i}{2} \sum_{i,q \geq 0; a = \max \{ 0,q-n,i-m \}\ldots \min \{ i,q \};p=m+a-i,j=n+a-q }a! \left( \begin{array}{c} i \\* a  \end{array} \right)
\left( \begin{array}{c} q \\* a  \end{array} \right) \\*
& & \int  d\vec{k}_{m+1} \ldots d\vec{k}_{m+a} \partial^h \left( \alpha_{pq}^{1}\alpha_{ij}^{1} - \alpha_{pq}^{2}\alpha_{ij}^{2}
 - \alpha_{pq}^{3}\alpha_{ij}^{3} + \alpha_{pq}^{4}\alpha_{ij}^{4}  \right) \\*
& & - \frac{i}{2} \partial^h \left( (mk_{m,b} - nl_{n,b})x^{b}\alpha_{mn}^{1} \right). \\*
\end{eqnarray*}
The $\gamma^{0}\gamma^{h}$ set of equations is given by
\begin{eqnarray*}
0 & = & \sum_{m+n \geq 0}\int d\vec{k}_{1} \ldots d\vec{k}_{m}d\vec{l}_{1} \ldots d\vec{l}_{n}a^{\dag}_{\vec{k}_{1}} \ldots a^{\dag}_{\vec{k}_{m}}
a_{\vec{l}_{1}} \ldots a_{\vec{l}_{n}} \\*
& & -2ie^{i\left( \sum_{t=1}^{m}k_{t,b} - \sum_{t=1}^{n}l_{t,b} \right)x^{b}}\delta(\vec{k}_{m} - \vec{l}_{n})\chi(m-1)\chi(n-1)k_{m}^{\,\,
\, ,h} \alpha_{(m-1)(n-1)}^{2} \\*
& & -2ime^{i\left( \sum_{t=1}^{m}k_{t,b} - \sum_{t=1}^{n}l_{t,b} \right)x^{b}}k_{1}^{\,\,\, ,h}\alpha_{mn}^{2} \\*
& & + 2\sum_{i,q \geq 0; a= \max \{ 0,q+1-n,i+1-m \} \ldots \min \{ i,q \}; p=m+a-i-1,j=n+a-q-1} a! \left(  \begin{array}{c} i \\* a  \end{array} \right)
\left(  \begin{array}{c} q \\* a  \end{array} \right) \\*
& & \int d\vec{k}_{m+1} \ldots d\vec{k}_{m+a}\delta(\vec{k}_{p+1} - \vec{l}_{n})e^{i\left( \sum_{t=1}^{m}k_{t,b} - \sum_{t=1}^{n}l_{t,b} 
\right)x^{b}}k_{p+1}^{\,\,\, ,h} \widetilde{\alpha}_{[pq}^{1}\alpha_{ij]}^{2} \\*
& & + 2\sum_{i,q \geq 0; a= \max \{ 0,q-n,i-m \} \ldots \min \{ i,q \}; p=m+a-i,j=n+a-q} a! \left(  \begin{array}{c} i \\* a  \end{array} \right)
\left(  \begin{array}{c} q \\* a  \end{array} \right) \\*
& & \int d\vec{k}_{m+1} \ldots d\vec{k}_{m+a}e^{i\left( \sum_{t=1}^{m}k_{t,b} - \sum_{t=1}^{n}l_{t,b} 
\right)x^{b}}k_{p+1}^{\,\,\, ,h}\left( pk_{1}^{\,\,\, ,h} + (i-a)k_{p+1}^{\,\,\, ,h} \right) \widetilde{\alpha}_{[pq}^{1}\alpha_{ij]}^{2} \\*
\end{eqnarray*}
\begin{eqnarray*}
& & -2 \sum_{i,q \geq 0; a= \max \{ 0,q-n+1,i-m+1 \} \ldots \min \{ i+1,q \};p=m+a-i-1,j=n+a-q-1} \int d\vec{k}_{m+1} \ldots d\vec{k}_{m+a} \\*
& & e^{i\left( \sum_{t=p+1}^{m+a}k_{t,b} - \sum_{t=q-a+1}^{n}l_{t,b} \right)x^{b}} \\*
& & \left( a! \left( \begin{array}{c} q \\* a \end{array} \right)\left( \begin{array}{c} i \\* a  \end{array} \right)k_{m}^{\,\, \, h}
\delta(\vec{k}_{m} - \vec{l}_{n}) + a! \left( \begin{array}{c} q \\* a \end{array} \right)\left( \begin{array}{c} i \\* a-1  \end{array} 
\right)k_{m+1}^{\,\, \, h} \delta(\vec{k}_{m+1} - \vec{l}_{n}) \right)\alpha_{pq}^{1}\alpha_{ij}^{2} \\*
& & -2 \sum_{i,q \geq 0; a= \max \{ 0,q-n,i-m \} \ldots \min \{ i,q \};p=m+a-i,j=n+a-q}i \int d\vec{k}_{m+1} \ldots d\vec{k}_{m+a} 
e^{i\left( \sum_{t=p+1}^{m+a}k_{t,b} - \sum_{t=q-a+1}^{n}l_{t,b} \right)x^{b}} \\*
& & \left( a! \left( \begin{array}{c} q \\* a \end{array} \right)\left( \begin{array}{c} i-1 \\* a  \end{array} \right)k_{m}^{\,\, \, h} 
+ a! \left( \begin{array}{c} q \\* a \end{array} \right)\left( \begin{array}{c} i-1 \\* a-1  \end{array} 
\right)k_{m+1}^{\,\, \, h} \right)\alpha_{pq}^{1}\alpha_{ij}^{2} \\*
& & -2i \sum_{i,q \geq 0;a= \max \{ 0,q-n,i-m \} \ldots \min \{ i,q \};p = m+a-i,j=n+a-q }i \int d\vec{k}_{m+1} \ldots d\vec{k}_{m+a}
e^{i \left( \sum_{t=p+1}^{m+a}k_{t,b} - \sum_{t=q+1-a}^{n}l_{t,b} \right)x^{b}} \\*
& & a! \left( \begin{array}{c} q \\* a \end{array} \right)\left( \begin{array}{c} i \\* a  \end{array} \right) \partial^h \alpha_{pq}^{1}
\alpha_{ij}^{2} \\*
& & - \partial^h \alpha_{mn}^{2} \\*
& & - i \sum_{i,q \geq 0; a= \max \{ 0,q-n,i-m \} \ldots \min \{ i,q \};p=m+a-i,j=n+a-q }i \int d\vec{k}_{m+1} \ldots d\vec{k}_{m+a} \\*
& & a! \left( \begin{array}{c} q \\* a \end{array} \right)\left( \begin{array}{c} i \\* a  \end{array} \right)
\partial^h \left( \alpha_{(pq}^{1}\alpha_{ij)}^{2} - i \alpha_{[pq}^{3} \alpha_{ij]}^{4} \right) \\*
& &  - \frac{i}{2} \partial^h \left( (mk_{m,b} - nl_{n,b})x^{b}\alpha_{mn}^{2} \right). \\*
\end{eqnarray*}
The $\gamma^{5}\gamma^{h}$ equations read
\begin{eqnarray*}
0 & = & \sum_{m+n \geq 0} \int d\vec{k}_{1} \ldots d\vec{k}_{m}d\vec{l}_{1} \ldots d\vec{l}_{n}a^{\dag}_{\vec{k}_{1}} \ldots a^{\dag}_{\vec{k}_{m}}
a_{\vec{l}_{1}} \ldots a_{\vec{l}_{n}} \\*
& & +2e^{i\left( \sum_{t=1}^{m}k_{t,b} - \sum_{t=1}^{n}l_{t,b} \right)x^{b}}\delta(\vec{k}_{m} - \vec{l}_{n})\chi(m-1)\chi(n-1)k_{m}^{\,\,
\, ,h} \alpha_{(m-1)(n-1)}^{3} \\*
& & +2me^{i\left( \sum_{t=1}^{m}k_{t,b} - \sum_{t=1}^{n}l_{t,b} \right)x^{b}}k_{1}^{\,\,\, ,h}\alpha_{mn}^{3} \\*
& & + 2\sum_{i,q \geq 0; a= \max \{ 0,q+1-n,i+1-m \} \ldots \min \{ i,q \}; p=m+a-i-1,j=n+a-q-1} a! \left(  \begin{array}{c} i \\* a  \end{array} \right)
\left(  \begin{array}{c} q \\* a  \end{array} \right) \\*
& & \int d\vec{k}_{m+1} \ldots d\vec{k}_{m+a}\delta(\vec{k}_{p+1} - \vec{l}_{n})e^{i\left( \sum_{t=1}^{m}k_{t,b} - \sum_{t=1}^{n}l_{t,b} 
\right)x^{b}}k_{p+1}^{\,\,\, ,h} \widetilde{\alpha}_{(pq}^{4}\alpha_{ij)}^{2} \\*
& & + 2\sum_{i,q \geq 0; a= \max \{ 0,q-n,i-m \} \ldots \min \{ i,q \}; p=m+a-i,j=n+a-q} a! \left(  \begin{array}{c} i \\* a  \end{array} \right)
\left(  \begin{array}{c} q \\* a  \end{array} \right) \\*
& & \int d\vec{k}_{m+1} \ldots d\vec{k}_{m+a}e^{i\left( \sum_{t=1}^{m}k_{t,b} - \sum_{t=1}^{n}l_{t,b} 
\right)x^{b}}\left( pk_{1}^{\,\,\, ,h} + (i-a)k_{p+1}^{\,\,\, ,h} \right) \widetilde{\alpha}_{(pq}^{4}\alpha_{ij)}^{2} \\*
\end{eqnarray*}
\begin{eqnarray*}
& & -2 \sum_{i,q \geq 0; a= \max \{ 0,q-n+1,i-m+1 \} \ldots \min \{ i+1,q \};p=m+a-i-1,j=n+a-q-1} \int d\vec{k}_{m+1} \ldots d\vec{k}_{m+a} \\*
& & e^{i\left( \sum_{t=p+1}^{m+a}k_{t,b} - \sum_{t=q-a+1}^{n}l_{t,b} \right)x^{b}} \\*
& & \left( a! \left( \begin{array}{c} q \\* a \end{array} \right)\left( \begin{array}{c} i \\* a  \end{array} \right)k_{m}^{\,\, \, h}
\delta(\vec{k}_{m} - \vec{l}_{n}) + a! \left( \begin{array}{c} q \\* a \end{array} \right)\left( \begin{array}{c} i \\* a-1  \end{array} 
\right)k_{m+1}^{\,\, \, h} \delta(\vec{k}_{m+1} - \vec{l}_{n}) \right)\alpha_{pq}^{4}\alpha_{ij}^{2} \\*
& & -2 \sum_{i,q \geq 0; a= \max \{ 0,q-n,i-m \} \ldots \min \{ i,q \};p=m+a-i,j=n+a-q}i \int d\vec{k}_{m+1} \ldots d\vec{k}_{m+a} 
e^{i\left( \sum_{t=p+1}^{m+a}k_{t,b} - \sum_{t=q-a+1}^{n}l_{t,b} \right)x^{b}} \\*
& & \left( a! \left( \begin{array}{c} q \\* a \end{array} \right)\left( \begin{array}{c} i-1 \\* a  \end{array} \right)k_{m}^{\,\, \, h} + 
a! \left( \begin{array}{c} q \\* a \end{array} \right)\left( \begin{array}{c} i-1 \\* a-1  \end{array} 
\right)k_{m+1}^{\,\, \, h} \right)\alpha_{pq}^{4}\alpha_{ij}^{2} \\*
& & -2i \sum_{i,q \geq 0;a= \max \{ 0,q-n,i-m \} \ldots \min \{ i,q \};p = m+a-i,j=n+a-q }i \int d\vec{k}_{m+1} \ldots d\vec{k}_{m+a}
e^{i \left( \sum_{t=p+1}^{m+a}k_{t,b} - \sum_{t=q+1-a}^{n}l_{t,b} \right)x^{b}} \\*
& & a! \left( \begin{array}{c} q \\* a \end{array} \right)\left( \begin{array}{c} i \\* a  \end{array} \right) \partial^h \alpha_{pq}^{4}
\alpha_{ij}^{2} \\*
& & - i \partial^h \alpha_{mn}^{3} \\*
& & + \sum_{i,q \geq 0; a= \max \{ 0,q-n,i-m \} \ldots \min \{ i,q \};p=m+a-i,j=n+a-q }i \int d\vec{k}_{m+1} \ldots d\vec{k}_{m+a} \\*
& & a! \left( \begin{array}{c} q \\* a \end{array} \right)\left( \begin{array}{c} i \\* a  \end{array} \right)
\partial^h \left( \alpha_{(pq}^{1}\alpha_{ij)}^{3} + i \alpha_{[pq}^{2} \alpha_{ij]}^{4} \right) \\*
& &  + \frac{1}{2} \partial^h \left( (mk_{m,b} - nl_{n,b})x^{b}\alpha_{mn}^{3} \right). \\*
\end{eqnarray*}
Finally, the $\gamma^{0}\gamma^{5}\gamma^{h}$ equations are
\begin{eqnarray*}
0 & = & \sum_{m+n \geq 0} \int d\vec{k}_{1} \ldots d\vec{k}_{m}d\vec{l}_{1} \ldots d\vec{l}_{n}a^{\dag}_{\vec{k}_{1}} \ldots a^{\dag}_{\vec{k}_{m}}
a_{\vec{l}_{1}} \ldots a_{\vec{l}_{n}} \\*
& & - 2i\sum_{i,q \geq 0; a= \max \{ 0,q+1-n,i+1-m \} \ldots \min \{ i,q \}; p=m+a-i-1,j=n+a-q-1} a! \left(  \begin{array}{c} i \\* a  \end{array} \right)
\left(  \begin{array}{c} q \\* a  \end{array} \right) \\*
& & \int d\vec{k}_{m+1} \ldots d\vec{k}_{m+a}\delta(\vec{k}_{p+1} - \vec{l}_{n})e^{i\left( \sum_{t=1}^{m}k_{t,b} - \sum_{t=1}^{n}l_{t,b} 
\right)x^{b}}k_{p+1}^{\,\,\, ,h} \widetilde{\alpha}_{[pq}^{3}\alpha_{ij]}^{2} \\*
& & - 2i\sum_{i,q \geq 0; a= \max \{ 0,q-n,i-m \} \ldots \min \{ i,q \}; p=m+a-i,j=n+a-q} a! \left(  \begin{array}{c} i \\* a  \end{array} \right)
\left(  \begin{array}{c} q \\* a  \end{array} \right) \\*
& & \int d\vec{k}_{m+1} \ldots d\vec{k}_{m+a}e^{i\left( \sum_{t=1}^{m}k_{t,b} - \sum_{t=1}^{n}l_{t,b} 
\right)x^{b}}\left( pk_{1}^{\,\,\, ,h} + (i-a)k_{p+1}^{\,\,\, ,h} \right) \widetilde{\alpha}_{[pq}^{3}\alpha_{ij]}^{2} \\*
& &  + 2i\sum_{i,q \geq 0; a= \max \{ 0,q+1-n,i+1-m \} \ldots \min \{ i+1,q \}; p=m+a-i-1,j=n+a-q-1} \int d\vec{k}_{m+1} 
\ldots d\vec{k}_{m+a} \\*
& & e^{i\left( \sum_{t=p+1}^{m+a}k_{t,b} - \sum_{t=q+1-a}^{n}l_{t,b} \right)x^{b}} \\*
& &  \left( a! \left( \begin{array}{c} q \\* a \end{array} \right)\left( \begin{array}{c} i \\* a  \end{array} \right) k_{m}^{\,\,\ ,h}
\delta(\vec{k}_{m} - \vec{l}_{n}) + a! \left( \begin{array}{c} q \\* a \end{array} \right)\left( \begin{array}{c} i \\* a-1  \end{array}
 \right) k_{m+1}^{\,\,\ ,h} \delta(\vec{k}_{m+1} - \vec{l}_{n}) \right) \widetilde{\alpha}_{pq}^{3}\alpha_{ij}^{2} \\*
\end{eqnarray*}
\begin{eqnarray*}
& & +2i \sum_{i,q \geq 0; a= \max \{ 0,q-n,i-m \} \ldots \min \{ i,q \};p=m+a-i,j=n+a-q} i \int d\vec{k}_{m+1}
 \ldots d\vec{k}_{m+a} e^{i\left( \sum_{t=p+1}^{m+a}k_{t,b} - \sum_{t=q-a+1}^{n}l_{t,b} \right)x^{b}} \\*
& & \left( a! \left( \begin{array}{c} q \\* a \end{array} \right)\left( \begin{array}{c} i-1 \\* a  \end{array} \right)k_{m}^{\,\, \, h}
+ a! \left( \begin{array}{c} q \\* a \end{array} \right)\left( \begin{array}{c} i-1 \\* a-1  \end{array} 
\right)k_{m+1}^{\,\, \, h}\right)\alpha_{pq}^{3}\alpha_{ij}^{2} \\*
& & -2 \sum_{i,q \geq 0; a= \max \{ 0,q-n,i-m \} \ldots \min \{ i,q \};p=m+a-i,j=n+a-q}i \int d\vec{k}_{m+1} \ldots d\vec{k}_{m+a} 
e^{i\left( \sum_{t=p+1}^{m+a}k_{t,b} - \sum_{t=q-a+1}^{n}l_{t,b} \right)x^{b}} \\*
& &  a! \left( \begin{array}{c} q \\* a \end{array} \right)\left( \begin{array}{c} i \\* a  \end{array} \right) \partial^h \alpha_{pq}^{3}
\alpha_{ij}^{2} \\*
& & -  \partial^h \alpha_{mn}^{4} \\*
& & + i \sum_{i,q \geq 0; a= \max \{ 0,q-n,i-m \} \ldots \min \{ i,q \};p=m+a-i,j=n+a-q }i \int d\vec{k}_{m+1} \ldots d\vec{k}_{m+a} \\*
& & a! \left( \begin{array}{c} q \\* a \end{array} \right)\left( \begin{array}{c} i \\* a  \end{array} \right)
\partial^h \left( \alpha_{(pq}^{1}\alpha_{ij)}^{4} + i \alpha_{[pq}^{2} \alpha_{ij]}^{3} \right) \\*
& &  + \frac{i}{2} \partial^h \left( (mk_{m,b} - nl_{n,b})x^{b}\alpha_{mn}^{4} \right). \\* 
\end{eqnarray*}
meaning that the gravitational potential factorizes from the first equation in second order perturbation theory
 (even if it were time dependent),
\emph{assuming} that local Hamiltonians with such symmetry exist.  A similar effect takes place in perturbation theory for the second 
constraint equation and this result is clearly independent of the order of perturbation.  Therefore, it would be desirable that no solutions
of this system exist which would automatically imply that the quantum theory needs to have rotational momentum which is what one would expect
 physically.  Indeed, our equations resemble the ordinary spin-$\frac{1}{2}$ Dirac equation so that angular momentum always has to be present, in
contrast to the spin less Schroedinger equation for a particle in a three dimensional harmonic potential.  We now determine $k(2)$ to be
$1$, meaning that we cut off the interaction Hamiltonian at $i,j \leq 1$; therefore the first constraint equation determines $32$ scalar
equations in $16$ variables.  The notation we use is $(\star,m,n)$ where $\star$ is an element of the Clifford basis.  Hence, $(1,0,0)$
is determined by
\begin{eqnarray*}
 0 & = & \int d\vec{k}e^{ik_{a}x^{a}} \left( -4i\alpha_{[01}^{3}\alpha_{10]}^{4}k_{0} + 2\partial_t \alpha_{01}^{3}\alpha_{10}^{4} - 
2\partial_t \alpha_{01}^{4} \alpha_{10}^{3}  \right) + \partial_t \alpha_{00}^{3}\alpha_{00}^{4} - \partial_t \alpha_{00}^{4}\alpha_{00}^{3}
+ \partial_t \alpha_{00}^{2} \\*
& & + i\partial_t \left( \alpha_{00}^{1} \alpha_{00}^{2} \right) + i \int d\vec{k} \partial_t \left( \alpha_{(01}^{1}\alpha_{10)}^{2} - i 
\alpha_{[10}^{3}\alpha_{10]}^{4}  \right). 
\end{eqnarray*}
$(1,1,0)$ results in
\begin{eqnarray*}
 0 & = & -4ie^{ik_{b}x^{b}} \int d\vec{k}_{1}k_{0} \alpha_{[11}^{4}\alpha_{10]}^{3} - 4i\int d\vec{k}_{1} \alpha_{[11}^{3}\alpha_{10]}^{4}
e^{ik_{1,b}x^{b}}k_{1,0} - 4i\alpha_{[00}^{3}\alpha_{10]}^{4}e^{ik_{b}x^{b}}k_{0} \\*
& & +2\left( \partial_t \alpha_{10}^{3}\alpha_{00}^{4} - \partial_t \alpha_{10}^{4} \alpha_{00}^{3} \right) + 2\int d\vec{k}_{1}e^{ik_{1,b}x^{b}}
\left( \partial_t \alpha_{11}^{3}\alpha_{10}^{4} - \partial_t \alpha_{11}^{4}\alpha_{10}^{3}  \right) \\*
& & + 2e^{ik_{b}x^{b}} \left( \partial_t \alpha_{00}^{3} \alpha_{10}^{4} - \partial_t \alpha_{00}^{4} \alpha_{10}^{3} \right) + \partial_t \alpha_{10}^{2}
+ 2i \partial_t \left( \alpha_{(10}^{1}\alpha_{00)}^{2} \right) \\*
& & + i\int d\vec{k}_{1} \partial_t \left(\alpha_{(11}^{1}\alpha_{10)}^{2} - i \alpha_{[11}^{3} \alpha_{10]}^{4}  \right) + \frac{i}{2} 
\partial_t \left( k_bx^{b} \alpha_{10}^{2} \right)
\end{eqnarray*}
while $(1,0,1)$ reads
\begin{eqnarray*}
0 & = & -4i\alpha_{[01}^{3}\alpha_{00]}^{4}l_{0} - 4i\int d\vec{k}_{1}\alpha_{[01}^{3}\alpha_{11]}^{4}e^{i\left( k_{1,b} - l_{b} \right)x^{b}}
k_{1,0} + 2\left( \partial_t \alpha_{01}^{3}\alpha_{00}^{4} - \partial_t \alpha_{01}^{4} \alpha_{00}^{3} \right) \\*
& & +2e^{-il_{b}x^{b}}\left( \partial_t \alpha_{00}^{3} \alpha_{01}^{4} - \partial_t \alpha_{00}^{4}\alpha_{01}^{3} \right) + 2
\int d\vec{k}_{1} e^{i\left( k_{1,b} - l_{b} \right)x^{b}} \left( \partial_t \alpha_{01}^{3} \alpha_{11}^{4} - \partial_t \alpha_{01}^{4}
\alpha_{11}^{3} \right) \\*
& & + \partial_t \alpha_{01}^{2} + 2i\partial_t \left( \alpha_{(01}^{1}\alpha_{00)}^{2}  \right) + i \int d\vec{k}_{1} \partial_t 
\left( \alpha_{(01}^{1}\alpha_{11)}^{2} - i\alpha_{[01}^{3}\alpha_{11]}^{4} \right) - \frac{i}{2} \partial_t \left( l_b x^{b} \alpha_{01}^{2}\right)
\end{eqnarray*}
and finally $(1,1,1)$ equals
\begin{eqnarray*}
0 & = & -4i\delta(\vec{k} - \vec{l})k_{0} \int d\vec{k}_{1} \alpha_{[01}^{3} \alpha_{10]}^{3} - 4ik_{0}\delta(\vec{k} - \vec{l})\int d\vec{k}_{1}
\alpha_{[01}^{3}\alpha_{10]}^{4}e^{ik_{1,b}x^{b}} \\*
& & - 4i\alpha_{[11}^{3}\alpha_{00]}^{4}l_{0} - 4i\alpha_{[00}^{3}\alpha_{11]}^{4}e^{i\left( k_{b} - l_{b} \right)x^{b}}k_{0} - 4i\alpha_{[01}^{3}
\alpha_{10]}^{4}e^{ik_{b}x^{b}}k_{0} \\*
& & +2e^{i(k_{b} - l_{b})x^{b}}\left( \partial_t \alpha_{00}^{3} \alpha_{11}^{4} - \partial_t \alpha_{00}^{4} \alpha_{11}^{3} \right) + 
2\left(  \partial_t \alpha_{11}^{3} \alpha_{00}^{4} - \partial_t \alpha_{11}^{4} \alpha_{00}^{3} \right) \\*
& & + 2\partial_t \alpha_{11}^{2}  + 2e^{-il_{b}x^{b}} \left( \partial_t \alpha_{10}^{3} \alpha_{01}^{4} - \partial_t \alpha_{10}^{4}
\alpha_{01}^{3} \right) + 2e^{ik_{b}x^{b}} \left( \partial_t \alpha_{01}^{3}\alpha_{10}^{4} - \partial_t \alpha_{01}^{4}\alpha_{10}^{3} \right) \\*
& & +2\int d\vec{k}_{1}e^{i(k_{1,b} - l_{b})x^{b}}\left( \partial_t \alpha_{11}^{3}\alpha_{11}^{4} - \partial_t \alpha_{11}^{4}\partial_{11}^{3} \right)
+ 2i\partial_t \left( \alpha_{(00}^{1}\alpha_{11)}^{2} \right) + 2i\partial_t \left( \alpha_{(01}^{1} \alpha_{10)}^{2}  \right) \\*
& & +i \int d\vec{k}_{1} \partial_t \left( \alpha_{(11}^{1} \alpha_{11)}^{2} \right)  + \frac{i}{2} \partial_t \left( (k_{b} - l_{b})x^{b}
\alpha_{11}^{2} \right).
\end{eqnarray*}
A similar series for $\gamma^{0}$ is computed; $(\gamma^{0},0,0)$ is given by
\begin{eqnarray*}
 0 & = & 2\int d\vec{k}_{1}\left( \alpha_{01}^{3}\alpha_{10}^{3} - \alpha_{01}^{4}\alpha_{10}^{4}  \right)e^{ik_{1,b}x^{b}}k_{1,0} + 
2i\left( \partial_t \alpha_{00}^{3}\alpha_{00}^{3} - \partial_t \alpha_{00}^{4} \alpha_{00}^{4} \right)  \\*
& & + 2i\int d\vec{k}_{1}e^{ik_{1,b}x^{b}}\left( \partial_t \alpha_{01}^{3}\alpha_{10}^{3} - \partial_{t} \alpha_{01}^{4} \alpha_{10}^{4} \right)
+\partial_t \alpha_{00}^{1} - \frac{i}{2}\partial_t \left( \left( \alpha_{00}^{1} \right)^{2} - \left( \alpha_{00}^{2} \right)^{2} -
 \left( \alpha_{00}^{3} \right)^{2} + \left( \alpha_{00}^{4} \right)^{2} \right) \\*
& & - \frac{i}{2}\int d\vec{k}_{1}\partial_t \left( \alpha_{(01}^{1}\alpha_{10)}^{1} - \alpha_{(01}^{2}\alpha_{10)}^{2} - 
\alpha_{(01}^{3}\alpha_{10)}^{3} + \alpha_{(01}^{4}\alpha_{10)}^{4} \right)
\end{eqnarray*}
while $(\gamma^{0},1,0)$ reads
\begin{eqnarray*}
0 & = & - 8e^{ik_{b}x^{b}}k_{0}\left( \alpha_{10}^{3}\alpha_{00}^{3} - \alpha_{10}^{4}\alpha_{00}^{4} \right) + 4e^{ik_{b}x^{b}}k_{0}
\int d\vec{k}_{1} \left( \alpha_{11}^{3}\alpha_{10}^{3} - \alpha_{11}^{4}\alpha_{10}^{4} \right) \\*
& & +2\int d\vec{k}_{1}e^{ik_{1,b}x^{b}}\left( \alpha_{11}^{3}\alpha_{10}^{3} - \alpha_{11}^{4}\alpha_{10}^{4} \right)k_{1,0} + 2e^{ik_{b}
x^{b}}k_{0}\left( \alpha_{00}^{3}\alpha_{10}^{3} - \alpha_{00}^{4}\alpha_{10}^{4} \right) \\*
& & + 2i\left( \partial_t \alpha_{10}^{3} \alpha_{00}^{3} - \partial_t \alpha_{10}^{4} \alpha_{00}^{4} \right) + 2i\int d\vec{k}_{1}e^{ik_{1,b}x^{b}}
\left( \partial_t \alpha_{11}^{3} \alpha_{10}^{3} -  \partial_t \alpha_{11}^{4} \alpha_{10}^{4}  \right) + 2ie^{ik_{b}x^{b}}
\left( \partial_t \alpha_{00}^{3} \alpha_{10}^{3} - \partial_t \alpha_{00}^{4} \alpha_{10}^{4} \right) \\*
& & + \partial_t \alpha_{10}^{1} - i\partial_t \left( \alpha_{10}^{1} \alpha_{00}^{1} - \alpha_{10}^{2} \alpha_{00}^{2} 
- \alpha_{10}^{3} \alpha_{00}^{3} + \alpha_{10}^{4} \alpha_{00}^{4}  \right) - \frac{i}{2} \int d\vec{k}_{1} \partial_t \left(
 \alpha_{11}^{1} \alpha_{10}^{1} - \alpha_{11}^{2} \alpha_{10}^{2} 
- \alpha_{11}^{3} \alpha_{10}^{3} + \alpha_{11}^{4} \alpha_{10}^{4}  \right) \\*
& & -\frac{i}{2} \partial_t \left( k_{b}x^{b}\alpha_{10}^{1} \right) \\* 
\end{eqnarray*}
and $(\gamma^{0},0,1)$ produces
\begin{eqnarray*}
0 & = & 2l_{0}\left( \alpha_{01}^{3}\alpha_{00}^{3} -  \alpha_{01}^{4}\alpha_{00}^{4} \right) + 2\int d\vec{k}_{1}e^{i(k_{1,b} - l_{n})x^{b}}
k_{1,0}( \alpha_{01}^{3}\alpha_{11}^{3} - \alpha_{01}^{4}\alpha_{11}^{4}) \\*
& & + 2ie^{-il_{b}x^{b}}\partial_t \left( \alpha_{00}^{3}\alpha_{01}^{3} 
- \alpha_{00}^{4}\alpha_{01}^{4} \right) + 2i \int d\vec{k}_{1}e^{i(k_{1,b} - l_{b})x^{b}}\left( \partial_t \alpha_{01}^{3}\alpha_{11}^{3} 
- \partial_t \alpha_{01}^{4}\alpha_{11}^{4} \right) \\*
& & + \partial_t \alpha_{01}^{1} - i\partial_t \left( \alpha_{00}^{1}\alpha_{01}^{1} - \alpha_{00}^{2}\alpha_{01}^{2} - 
\alpha_{00}^{3}\alpha_{01}^{3} + \alpha_{00}^{4}\alpha_{01}^{4}  \right) \\*
& & - \frac{i}{2} \int d\vec{k}_{1} \partial_t \left( \alpha_{01}^{1}\alpha_{11}^{1} - \alpha_{01}^{2}\alpha_{11}^{2} 
- \alpha_{01}^{3}\alpha_{11}^{3} + \alpha_{01}^{4}\alpha_{11}^{4}\right) \\*
& & + \frac{i}{2} \partial_t \left( l_{b}x^{b}\alpha_{01}^{1} \right) \\* 
\end{eqnarray*}
and finally, $(\gamma^{0},1,1)$ gives
\begin{eqnarray*}
0 & = & -4k_{0}\delta(\vec{k} - \vec{l})\left( \left( \alpha_{00}^{3} \right)^{2} - \left( \alpha_{00}^{4} \right)^{2}  \right) - 4\delta(\vec{k} 
- \vec{l})k_{0} \int d\vec{k}_{1} \left( \alpha_{01}^{3} \alpha_{01}^{3} - \alpha_{01}^{4} \alpha_{01}^{4} \right) \\*
& & - 8e^{i(k_{b} - l_{b})x^{b}}k_{0} \left( \alpha_{00}^{3}\alpha_{11}^{3} - \alpha_{00}^{4}\alpha_{11}^{4} \right) - 8e^{i(k_{b} - l_{b})x^{b}}
k_{0} \left( \alpha_{10}^{3} \alpha_{01}^{3} - \alpha_{10}^{4} \alpha_{01}^{4}  \right) \\*
& & -4e^{i(k_{b} - l_{b})x^{b}}k_{0}\int d\vec{k}_{1} \left( \left( \alpha_{11}^{3} \right)^{2} - \left( \alpha_{11}^{4} \right)^{2} \right)
+ 2k_{0}\delta(\vec{k} - \vec{l})\left( \left( \alpha_{00}^{3} \right)^{2} - \left( \alpha_{00}^{4} \right)^{2} \right) \\*
& & +2k_{0}\delta(\vec{k} - \vec{l}) \int d\vec{k}_{1}e^{ik_{1,b}x^{b}}\left( \alpha_{01}^{3}\alpha_{10}^{3} - \alpha_{01}^{4}\alpha_{10}^{4}  \right)
 + 2l_{0}\left( \alpha_{11}^{3}\alpha_{00}^{3} -  \alpha_{11}^{4}\alpha_{00}^{4} \right) \\*
& & +2e^{i(k_{b} - l_{b})x^{b}}k_{0} \left( \alpha_{00}^{3}\alpha_{11}^{3} - \alpha_{00}^{4}\alpha_{11}^{4} \right) + 2e^{ik_{b}x^{b}}k_{0}
\left( \alpha_{01}^{3}\alpha_{10}^{3} -  \alpha_{01}^{4}\alpha_{10}^{4}  \right) \\*
& & +2\int d\vec{k}_{1}e^{i(k_{1,b} - l_{b})x^{b}}k_{1,0}\left( \left( \alpha_{11}^{3} \right)^{2}  - \left( \alpha_{11}^{3} \right)^{2} \right)
+ \partial_t \alpha_{11}^{1} - \frac{i}{2}\partial_t ((k_{b} - l_{b})x^{b}\alpha_{11}^{1}) + \\*
& & 2ie^{i(k_{b} - l_{b})x^{b}} \left( \partial_t \alpha_{00}^{3}\alpha_{11}^{3} - \partial_t \alpha_{00}^{4}\alpha_{11}^{4} \right) + 2i
\left( \partial_t \alpha_{11}^{3}\alpha_{00}^{3} - \partial_t \alpha_{11}^{4}\alpha_{00}^{4} \right) + 2ie^{-il_{b}x^{b}} \left( 
\partial_t \alpha_{10}^{3} \alpha_{01}^{3} - \partial_t \alpha_{10}^{4} \alpha_{01}^{4} \right) \\*
& & + 2ie^{ik_{b}x^{b}} \left( \partial_t \alpha_{01}^{3} \alpha_{10}^{3} - \partial_t \alpha_{01}^{4} \alpha_{10}^{4} \right) + 2i
\int d\vec{k}_{1}e^{i(k_{1,b} - l_{b})x^{b}} \left( \alpha_{01}^{1} \alpha_{10}^{1} - \alpha_{01}^{2} \alpha_{10}^{2}
 - \alpha_{01}^{3} \alpha_{10}^{3} + \alpha_{01}^{4} \alpha_{10}^{4}  \right) \\*
& &  -i\partial_t \left( \alpha_{00}^{1}\alpha_{11}^{1} - \alpha_{00}^{2}\alpha_{11}^{2} - \alpha_{00}^{3}\alpha_{11}^{3} 
+ \alpha_{00}^{4}\alpha_{11}^{4}  \right) - i\partial_t \left( \alpha_{10}^{1}\alpha_{01}^{1} - \alpha_{10}^{2}\alpha_{01}^{2} 
- \alpha_{10}^{3}\alpha_{01}^{3} + alpha_{10}^{4}\alpha_{01}^{4} \right) \\*
& & - \frac{i}{2} \int d\vec{k}_{1}\partial_t \left( \left( \alpha_{11}^{1}  \right)^{2} - \left( \alpha_{11}^{2}  \right)^{2}
- \left( \alpha_{11}^{3}  \right)^{2} +  \left( \alpha_{11}^{4}  \right)^{2}  \right).
\end{eqnarray*}
We now study the $\gamma^{5}$ series; $(\gamma^5,0,0)$ is determined by
\begin{eqnarray*}
0 & = & -2i\int d\vec{k}_{1}e^{ik_{1,b}x^{b}}k_{1,0} \left( \alpha_{01}^{2}\alpha_{10}^{3} - i\alpha_{01}^{1}\alpha_{10}^{4} \right) +
2 \int d\vec{k}_{1}k_{1,0}e^{ik_{1,b}x^{b}} \left( \partial_t \alpha_{01}^{2} \alpha_{10}^{3} - i  \partial_t \alpha_{01}^{1} \alpha_{10}^{4} \right) \\*
& & - i\partial_t \left( \alpha_{00}^{1} \alpha_{00}^{4} \right) - i\int d\vec{k}_{1} \partial_t \left( \alpha_{(01}^{1}\alpha_{10)}^{4}
 + i \alpha_{[01}^{2}\alpha_{10]}^{3} \right) - \partial_t \alpha_{00}^{4} 
\end{eqnarray*}
$(\gamma^5,1,0)$ reads
\begin{eqnarray*}
 0 & = & -2ik_{0}e^{ik_{b}x^{b}}\alpha_{10}^{4} - 2ie^{ik_{b}x^{b}}k_{0} \left( \alpha_{00}^{2}\alpha_{10}^{3} - i\alpha_{00}^{1}\alpha_{10}^{4}
 \right) \\*
& & - 2i\int d\vec{k}_{1}e^{ik_{1,b}x^{b}}k_{1,0} \left( \alpha_{11}^{2}\alpha_{10}^{3} - i\alpha_{11}^{4}\alpha_{10}^{4} \right) + 2
\int d\vec{k}_{1}e^{ik_{1,b}x^{b}}k_{1,0} \left( \partial_t \alpha_{11}^{2}\alpha_{10}^{3} - i \partial_t \alpha_{11}^{1}\alpha_{10}^{4}
 \right) \\*
& & - \partial_t \alpha_{10}^{4} - 2i \partial_t \left( \alpha_{(10}^{1} \alpha_{00)}^{4} \right) - i \int d\vec{k}_{1} \partial_t 
\left( \alpha_{(11}^{1}\alpha_{10)}^{4} + i\alpha_{(11}^{2}\alpha_{10)}^{3}  \right) - \frac{i}{2}\partial_t (k_b x^{b}\alpha_{10}^{4}).
\end{eqnarray*}
$(\gamma^5,0,1)$ is given by
\begin{eqnarray*}
0 & = & - 2i\int d\vec{k}_{1}e^{i(k_{1,b} - l_{b})x^{b}}k_{1,0}\left( \alpha_{01}^{2}\alpha_{11}^{3} - i\alpha_{01}^{1}\alpha_{11}^{4} 
\right) + 2\int d\vec{k}_{1}e^{i(k_{1,b} - l_{b})x^{b}}k_{1,0} \left( \partial_t \alpha_{01}^{2}\alpha_{11}^{3} - i
\partial_t \alpha_{01}^{1}\alpha_{11}^{4}  \right) \\*
 & & - \partial_t \alpha_{01}^{4} - 2i\partial_t \left( \alpha_{(00}^{1} \alpha_{01)}^{4} \right) - i \int d\vec{k}_{1} \partial_t 
\left( \alpha_{(01}^{1} \alpha_{11)}^{4} + i \alpha_{[01}^{2} \alpha_{11]}^{3}   \right) + \frac{i}{2} \partial_t (l_b x^b \alpha_{01}^{4})
- 2il_{0}\left( \alpha_{01}^{2}\alpha_{00}^{3} - i\alpha_{01}^{1}\alpha_{00}^{4} \right) 
\end{eqnarray*}
and finally, $(\gamma^5,1,1)$ reads
\begin{eqnarray*}
0 & = & -2ik_{0}\delta(\vec{k} - \vec{l})\alpha_{00}^{4} - 2ik_{0}e^{i(k_{b} - l_{b})x^{b}}\alpha_{11}^{4} \\*
& &  +2i\delta(\vec{k} - \vec{l})\left( \alpha_{00}^{2}\alpha_{00}^{3} + \int d\vec{k}_{1} \left( \alpha_{(01}^{2} \alpha_{10)}^{3} 
- i\widetilde{\alpha}_{[01}^{1} \alpha_{10]}^{4} \right)  \right) \\*
& & + 2ie^{i(k_b - l_b)x^b} \left( k_{0} \alpha_{(10}^{2}\alpha_{01)}^{3} + k_{0}\int d\vec{k}_{1} \alpha_{(11}^{2} \alpha_{11)}^{3} \right)
- 2ik_{0} \delta(\vec{k} - \vec{l}) \left( \alpha_{00}^{2} \alpha_{00}^{3} - i \alpha_{00}^{1} \alpha_{00}^{4} \right) \\*
& & -2ik_{0}\delta(\vec{k} - \vec{l}) \int d\vec{k}_{1}e^{ik_{1,b}x^{b}}\left( \alpha_{01}^{2}\alpha_{10}^{3} - i\alpha_{01}^{1} \alpha_{10}^{4} \right)
- 2il_{0}\left( \alpha_{11}^{2}\alpha_{00}^{3} - i\alpha_{11}^{1}\alpha_{00}^{4} \right) \\*
& & -2ie^{i(k_b - l_b)x^{b}}k_{0} \left( \alpha_{00}^{2}\alpha_{10}^{3}- i\alpha_{00}^{1}\alpha_{10}^{4} \right) - 2i\int d\vec{k}_{1}
e^{i(k_{1,b} -l_{b})x^{b}} k_{1,0} \left( \partial_t \alpha_{11}^{2} \alpha_{11}^{3} - i\partial_t \alpha_{11}^{1}\alpha_{11}^{4} \right) \\*
& & -2ik_{0}e^{i(k_b - l_b)x^b} \partial_t \left( \alpha_{(01}^{1}\alpha_{10)}^{4} + \alpha_{00}^{1}\alpha_{00}^{4} \right) - 
i\int d\vec{k}_{1} \partial_t \left( \alpha_{11}^{1} \alpha_{11}^{4} \right) - \frac{i}{2} \partial_t ((k_b - l_b)x^b \alpha_{11}^{4}) \\*
& & - 2ie^{ik_{b}x^{b}}l_{0} \left( \alpha_{01}^{2}\alpha_{10}^{3} - i\alpha_{01}^{1}\alpha_{10}^{4}  \right).
\end{eqnarray*}
The $(\gamma^0 \gamma^5, 0 , 0)$ equation is given by
\begin{eqnarray*}
 0 & = & 2i\int d\vec{k}_{1} e^{ik_{1,b}x^{b}} k_{1,0} \left( \alpha_{01}^{1}\alpha_{10}^{3} + i\alpha_{01}^{2}\alpha_{10}^{4} \right) + i
\partial_t \alpha_{00}^{3} \\*
& & - 2\int d\vec{k}_{1}e^{ik_{1,b}x^{b}}k_{1,0} \left( \partial_t \alpha_{01}^{1} \alpha_{10}^{3} + i \partial_t \alpha_{01}^{2} \alpha_{10}^{4}
 \right) + \partial_t \left( \alpha_{00}^{1}\alpha_{00}^{3} \right) + \int d\vec{k}_{1} \partial_t \left( \alpha_{(01}^{1}\alpha_{10)}^{3} + 
i\alpha_{[01}^{2}\alpha_{10]}^{4} \right) 
\end{eqnarray*}
while $(\gamma^0 \gamma^5, 1,0)$ reads
\begin{eqnarray*}
0 & = & - 2k_{0}e^{ik_b x^b}\alpha_{10}^{3} + 2i \int d\vec{k}_{1}k_{1,0}e^{ik_{1,b}x^{b}}\left( \alpha_{11}^{1}\alpha_{10}^{3} + i 
\alpha_{11}^{2}\alpha_{10}^{4} \right) + 2ik_{0}e^{ik_b x^b} \left( \alpha_{00}^{1} \alpha_{10}^{3} + i \alpha_{00}^{2}\alpha_{10}^{4} \right) \\*
& & -2\int d\vec{k}_{1}e^{ik_{1,b}x^{b}}k_{1,0}\left( \partial_t \alpha_{11}^{1} \alpha_{10}^{3} + i\partial_t \alpha_{11}^{2} \alpha_{10}^{4}  \right)
+ i \partial_t \alpha_{10}^{3} + 2 \partial_t \left( \alpha_{(10}^{1} \alpha_{00)}^{3} \right) \\*
& & + \int d\vec{k}_{1} \partial_t \left( \alpha_{(11}^{1} \alpha_{10)}^{3} \right) - \frac{1}{2} \partial_t \left( k_b x^b \alpha_{10}^{3}  \right). 
\end{eqnarray*}
The $(\gamma^0 \gamma^5,0,1)$ equation is
\begin{eqnarray*}
0 & = & 2il_{0} \left( \alpha_{01}^{1}\alpha_{00}^{3} + i\alpha_{01}^{2}\alpha_{00}^{4} \right) + 2i \int d \vec{k}_{1}e^{i(k_{1,b} - l_b)
x^b}k_{1,0} \left( \alpha_{01}^{1}\alpha_{11}^{3} + i\alpha_{01}^{2}\alpha_{11}^{4} \right) \\*
& &  -2\int d\vec{k}_{1}e^{i(k_{1,b} - l_b )x^b}k_{1,0} \left( \partial_t \alpha_{01}^{1}\alpha_{11}^{3} + i \partial_t \alpha_{01}^{2}\alpha_{11}^{4} 
\right) + i \partial_t \alpha_{01}^{3} + 2 \partial_t \left( \alpha_{(00}^{1}\alpha_{01)}^{3} \right) \\*
& & + \int d\vec{k}_{1} \partial_t \left( \alpha_{(01}^{1} \alpha_{11)}^{3} + i \alpha_{[01}^{2} \alpha_{11]}^{4} \right) + \frac{1}{2}
\partial_t (l_b x^b \alpha_{01}^{3}) \\*
\end{eqnarray*}
and finally, $(\gamma^0 \gamma^5 ,1, 1)$ reads
\begin{eqnarray*}
 0 & = & - 2k_{0}\alpha_{00}^{3}\delta + (\vec{k} - \vec{l}) - 2k_{0}e^{i(k_b - l_b)x^b}\alpha_{11}^{3} - 2i\delta(\vec{k} - \vec{l})
\left( i\alpha_{00}^{2}\alpha_{00}^{4} + \int d\vec{k}_{1}\left( \widetilde{\alpha}_{[10}^{1}\alpha_{01]}^{3} + i \alpha_{(10}^{2}\alpha_{01)}^{4} 
\right)  \right) \\*
& & -2ie^{i(k_b - l_b)x^b}k_{0} \left( \widetilde{\alpha}_{[10}^{1} \alpha_{01]}^{3} + i\alpha_{(10}^{2} \alpha_{01)}^{4} + i\int d\vec{k}_{1}
\alpha_{11}^{2} \alpha_{11}^{4} \right) + \\*
& & +2il_{0}e^{ik_{b}x^{b}} \left( \alpha_{01}^{1}\alpha_{10}^{3} + i \alpha_{01}^{2}\alpha_{10}^{4} \right)  + 2ik_{0}\delta(\vec{k} - \vec{l})
e^{i(k_b - l_b)x^b} \left( \alpha_{00}^{1}\alpha_{00}^{3} + i \alpha_{00}^{2}\alpha_{00}^{4} \right) \\*
& & + 2ik_{0}\int d\vec{k}_{1}e^{ik_{1,b}x^b} \left( \alpha_{01}^{1}\alpha_{10}^{3} + i \alpha_{01}^{2}\alpha_{10}^{4} \right) + 2il_{0}
\left( \alpha_{11}^{1}\alpha_{00}^{3} + i \alpha_{11}^{2}\alpha_{00}^{4} \right) + 2ie^{i(k_b - l_b)x^b}k_{0} \left( \alpha_{00}^{1}\alpha_{11}^{3} + 
i \alpha_{00}^{2}\alpha_{11}^{4} \right) \\*
& & +2ik_{0}e^{ik_{b}x^{b}}\left( \alpha_{01}^{1} \alpha_{10}^{3} +i\alpha_{01}^{2}\alpha_{10}^{4} \right) + 2i\int d\vec{k}_{1}e^{i(k_{1,b}
- l_{b})x^{b}}k_{1,0}\left( \alpha_{11}^{1}\alpha_{11}^{3} + i\alpha_{11}^{2}\alpha_{11}^{4} \right) \\*
& & -2\int d\vec{k}_{1}e^{i(k_{1,b} - l_{b})x^{b}}k_{1,0} \left( \partial_t \alpha_{11}^{1} \alpha_{11}^{3} + i \partial_t \alpha_{11}^{2} \alpha_{11}^{4} \right)
+ i\partial_t \alpha_{11}^{3} - \frac{1}{2} \partial_t ((k_b - l_b)x^b \alpha_{11}^{3}) \\*
& & + \int d\vec{k}_{1} \partial_t (\alpha_{11}^{1}\alpha_{11}^{3}) + 2 \partial_t \left( \alpha_{(00}^{1}\alpha_{11)}^{3} + \alpha_{(10}^{1}
\alpha_{01)} ^{3}\right). 
\end{eqnarray*}
We now calculate the $\gamma^h$ series; $(\gamma^h,0,0)$ reads
\begin{eqnarray*}
0 & = & 2\int d\vec{k}_{1}e^{ik_{1,b}x^{b}}k_{1}^{\,\,\, ,h} \alpha_{01}^{2}\alpha_{10}^{2} + 2i\int d\vec{k}_{1}e^{ik_{1,b}x^{b}}
\partial^h \alpha_{01}^{2} \alpha_{10}^{2} \\*
& & - \partial^h \alpha_{00}^{1} - \frac{i}{2} \partial^h \left( \left( \alpha_{00}^{1} \right)^{2} - \left( \alpha_{00}^{2} \right)^{2} 
- \left( \alpha_{00}^{3} \right)^{2} + \left( \alpha_{00}^{4} \right)^{2} \right) 
\end{eqnarray*}
while $(\gamma^h,1,0)$ reads
\begin{eqnarray*}
0 & = & 2\int d\vec{k}_{1}e^{ik_{1,b}x^{b}}k_{1}^{\,\,\, ,h}\alpha_{11}^{2}\alpha_{10}^{2} + 2e^{ik_{b}x^{b}}k^{h} + 2i\int d\vec{k}_{1}
e^{ik_{1,b}x^{b}} \partial^ h \alpha_{11}^{2}\alpha_{10}^{2} \\*
& &  + 2ie^{ik_{b}x^{b}}\partial^h \alpha_{00}^{2}\alpha_{10}^{2} -\partial^h \alpha_{10}^{1} - i\partial^h \left( \alpha_{10}^{1} \alpha_{00}^{1}
-  \alpha_{10}^{2} \alpha_{00}^{2} -  \alpha_{10}^{3} \alpha_{00}^{3} +  \alpha_{10}^{4} \alpha_{00}^{4} \right) \\*
& & - \frac{i}{2} \int d\vec{k}_{1} \partial^h \left( \alpha_{11}^{1}\alpha_{10}^{1} - \alpha_{11}^{2}\alpha_{10}^{2}
- \alpha_{11}^{3}\alpha_{10}^{3} + \alpha_{11}^{4}\alpha_{10}^{4}  \right) - \frac{i}{2}\partial^h \left( k_b x^b \alpha_{10}^{1} \right).
\end{eqnarray*}
The $(\gamma^h,0,1)$ equation is given by
\begin{eqnarray*}
 0 & = & 2l^h \alpha_{01}^{2}\alpha_{00}^{2} + 2\int d\vec{k}_{1}e^{i(k_{1,b} - l_{b})x^{b}}k_{1}^{\,\,\, ,h}\alpha_{01}^{2}\alpha_{11}^{2}
+2i\int d\vec{k}_{1}e^{i(k_{1,b} - l_b)x^b}\partial^h \alpha_{01}^{1}\alpha_{11}^{2} \\*
& & - \partial^h \alpha_{01}^{1} - i\partial^h \left( \alpha_{00}^{1}\alpha_{01}^{1} - \alpha_{00}^{2}\alpha_{01}^{2} - \alpha_{00}^{3}\alpha_{01}^{3}
+ \alpha_{00}^{4}\alpha_{01}^{4}  \right) + \frac{i}{2}\partial^h(l_bx^b \alpha_{01}^{1}) \\*
& & - \frac{i}{2} \int d\vec{k}_{1} \partial^h \left( \alpha_{01}^{1}\alpha_{11}^{1} - \alpha_{01}^{2}\alpha_{11}^{2} 
- \alpha_{01}^{3}\alpha_{11}^{3}  + \alpha_{01}^{4}\alpha_{11}^{4} \right)
\end{eqnarray*}
and finally, $(\gamma^h,1,1)$ reads
\begin{eqnarray*}
0 & = & -2\delta(\vec{k} - \vec{l})k^h \left( \left( \alpha_{00}^{2} \right)^{2}  + 2\alpha_{00}^{2}\alpha_{11}^{2} + 2\alpha_{10}^{2} \alpha_{01}^{2}
+ \int d\vec{k}_{1} \alpha_{01}^{2}\alpha_{10}^{2} \right)  \\*
& & - 2k^h e^{i(k_b - l_b)x^b} \int d\vec{k}_{1} \alpha_{11}^{2}\alpha_{11}^{2} \delta(\vec{k}_{1} - \vec{l}) + 2k^h \delta(\vec{k} - \vec{l})
\alpha_{00}^{2} + 2k^h \int d\vec{k}_{1} e^{ik_{1,b}x^{b}} \alpha_{01}^{2} \alpha_{10}^{2} \delta(\vec{k} - \vec{l}) + \\*
& & 2l^h e^{ik_{b}x^{b}}\alpha_{01}^{2} \alpha_{10}^{2} + 2l^h \alpha_{00}^{2} \alpha_{11}^{2} + 2e^{i(k_b - l_b)x^b}k^h \alpha_{00}^{2}
\alpha_{11}^{2} + 2e^{ik_b x^b}k^h \alpha_{01}^{2}\alpha_{10}^{2} + \\*
& & 2\int d\vec{k}_{1}e^{i(k_{1,b} - l_{b})x^{b}}k_{1}^{\,\, h} \alpha_{11}^{2}\alpha_{11}^{2} + 2ie^{i(k_b - l_b)x^b}\partial^h \alpha_{00}^{2}
\alpha_{11}^{2} + \\*
& & 2ie^{ik_{b}x^{b}}\partial^h \alpha_{01}^{2}\alpha_{10}^{2} + 2i \int d\vec{k}_{1}e^{i(k_{1,b} - l_{b})x^{b}} \partial^h \alpha_{11}^{2}
\alpha_{11}^{2} - \partial^h \alpha_{11}^{1} - i\partial^h \left( \alpha_{00}^{1}\alpha_{11}^{1} - \alpha_{00}^{2}\alpha_{11}^{2}
- \alpha_{00}^{3}\alpha_{11}^{3} + \alpha_{00}^{4}\alpha_{11}^{4} \right) \\*
& & - \frac{i}{2} \partial^h \left( \alpha_{01}^{1}\alpha_{10}^{1} - \alpha_{01}^{2}\alpha_{10}^{2} - 
\alpha_{01}^{3}\alpha_{10}^{3} + \alpha_{01}^{4}\alpha_{10}^{4} \right) \\*
& & -\frac{i}{2} \int d\vec{k}_{1} \partial^h \left( \left( \alpha_{11}^{1} \right)^{2} - \left( \alpha_{11}^{2} \right)^{2}
- \left( \alpha_{11}^{3} \right)^{2} + \left( \alpha_{11}^{4} \right)^{2}   \right) - \frac{i}{2}\partial^h ((k_{b} - l_{b})x^{b}
\alpha_{11}^{1}).
\end{eqnarray*}
We now calculate the $\gamma^0 \gamma^h$ series; $(\gamma^0 \gamma^h, 0,0)$ reads
\begin{eqnarray*}
0 & = & -2 \int d\vec{k}_{1}e^{ik_{1,b}x^{b}}k_{1}^{\,\, h}\alpha_{01}^{1}\alpha_{10}^{1} - 2i \int d\vec{k}_{1}e^{ik_{1,b}x^{b}} \partial^h
\alpha_{01}^{1} \alpha_{10}^{2} - \partial^h \alpha_{00}^{2} \\*
& & - i\int d\vec{k}_{1} \partial^h \left( \alpha_{(01}^{1}\alpha_{10)}^{2} - i\alpha_{(01}^{3}\alpha_{10)}^{4} \right)
\end{eqnarray*}
while $(\gamma^0 \gamma^h, 1, 0)$ is given by
\begin{eqnarray*}
0 & = & -2ie^{ik_{b}x^{b}}k^h \left( \alpha_{10}^{2} + i \int d\vec{k}_{1} \widetilde{\alpha}_{[11}^{1} \alpha_{10]}^{2} \right)
-2e^{ik_{b}x^{b}}k^{h} \alpha_{00}^{1} \alpha_{10}^{2} - 2\int d\vec{k}_{1}e^{ik_{1,b}x^{b}}k_{1}^{\,\, \, h}\alpha_{11}^{1} \alpha_{10}^{2} \\*
& & -2ie^{ik_{b}x^{b}} \partial^h \alpha_{00}^{1} \alpha_{10}^{2} - 2i \int d\vec{k}_{1}e^{ik_{1,b}x^{b}} \partial^h \alpha_{11}^{1} \alpha_{10}^{2}
 - \partial^h \alpha_{10}^{2} \\*
& & -i \partial^h \left( \alpha_{(00}^{1}\alpha_{10)}^{2} - i \alpha_{[00}^{3}\alpha_{10]}^{4} \right) - i \int d\vec{k}_{1} \partial^h
\left( \alpha_{(11}^{1} \alpha_{10)}^{2} - i \alpha_{[11}^{3}\alpha_{10]}^{4} \right) - \frac{i}{2} \partial^h \left( k_b x^b \alpha_{10}^{2}  \right)
\end{eqnarray*}
and $(\gamma^0 \gamma^h, 0, 1)$ reads
\begin{eqnarray*}
0 & = &  - 2l^h \alpha_{01}^{1}\alpha_{00}^{2} - 2\int d\vec{k}_{1}e^{i(k_{1,b} - l_{b})x^{b}}k_{1}^{\,\, h} \alpha_{01}^{1}\alpha_{11}^{2}
-2i \int d\vec{k}_{1}e^{i(k_{1,b} - l_{b})x^{b}} \partial^h \alpha_{01}^{1} \alpha_{11}^{2} \\*
& & - \partial^h \alpha_{11}^{2} - i \int d\vec{k}_{1}\partial^h \left( \alpha_{(01}^{1}\alpha_{11)}^{2} - i \alpha_{[01}^{3} \alpha_{10]}^{4} \right)
+ \frac{i}{2}\partial^h(l_b x^b \alpha_{01}^{2})
\end{eqnarray*}
while finally $(\gamma^0 \gamma^h, 1,1)$ equals
\begin{eqnarray*}
0 & = & -2i\delta(\vec{k} - \vec{l})k^h \alpha_{00}^{2} - 2ie^{i(k_b - l_b)x^{b}}k^{h} \alpha_{11}^{2} + 2 k^h \delta(\vec{k} - \vec{l})
\int d\vec{k}_{1}\widetilde{\alpha}_{[01}^{1} \alpha_{10]}^{2} \\*
& & +2k^h  e^{i(k_{b} - l_{b})x^{b}} \widetilde{\alpha}_{[00}^{1} \alpha_{11]}^{2} - 2k^h \delta(\vec{k} - \vec{l})\alpha_{00}^{1}\alpha_{00}^{2}
- 2k^h \delta(\vec{k} - \vec{l})\int d \vec{k}_{1}e^{ik_{1,b}x^{b}}\alpha_{01}^{1}\alpha_{10}^{2} \\*
& & -2l^h e^{ik_{b}x^{b}} \alpha_{01}^{1} \alpha_{10}^{2} - 2l^h \int d \vec{k}_{1} \alpha_{11}^{1} \alpha_{01}^{2} - 2e^{i(k_{b} - l_{b})x^{b}}
k^h \alpha_{00}^{1} \alpha_{11}^{2} - 2e^{ik_{b}x^{b}} \alpha_{01}^{1} \alpha_{10}^{2} \\*
& & -2 \int d\vec{k}_{1}e^{i(k_{1,b} - l_{b})x^{b}}k_{1}^{h} \alpha_{11}^{1} \alpha_{11}^{2} - 2ie^{i(k_b - l_b)x^b}\partial^h \alpha_{00}^{1}
\alpha_{11}^{2} - 2e^{ik_{b}x^{b}} \partial^h \alpha_{01}^{2} \alpha_{10}^{2} \\*
& & -2i\int d\vec{k}_{1} e^{i(k_{1,b} - l_{b})x^{b}} \partial^h \alpha_{11}^{1} \alpha_{11}^{2} - \frac{i}{2} \partial^h ((k_b - l_b)x^b \alpha_{11}^{2}).
\end{eqnarray*}
The $\gamma^5 \gamma^h$ series is calcualted as follows; $(\gamma^5 \gamma^h,0 , 0)$ reads
\begin{eqnarray*}
0 & = & -2\int d\vec{k}_{1}e^{ik_{1,b}x^{b}}k_{1}^{\,\, h}\alpha_{01}^{4} \alpha_{10}^{2} - 2i\int d\vec{k}_{1}e^{ik_{1,b}x^{b}}
\partial^h \alpha_{01}^{4}\alpha_{10}^{2} - \partial^h \alpha_{00}^{3} \\*
& & + \int d\vec{k}_{1} \partial^h \left( \alpha_{(01}^{1}\alpha_{10)}^{3} + i \alpha_{[01}^{2}\alpha_{10]}^{4} \right) \\* 
\end{eqnarray*}
while $(\gamma^5 \gamma^h,1,0)$ is given by
\begin{eqnarray*}
0 & = & 2e^{ik_{b}x^{b}}k^h \alpha_{10}^{3} - 2\int d\vec{k}_{1}e^{ik_{1,b}x^{b}}k_{1}^{h}\alpha_{11}^{4}\alpha_{10}^{2} - 2
e^{ik_{b}x^{b}}k^h \alpha_{00}^{4} \alpha_{10}^{2} \\*
& & - 2i\int d\vec{k}_{1}e^{ik_{1,b}x^{b}}\partial^h \alpha_{11}^{4} \alpha_{10}^{2} - 2ie^{ik_{b}x^{b}}\partial^h \alpha_{00}^{4} \alpha_{10}^{2}
- i\partial^h \alpha_{10}^{3} \\*
& & + \int d\vec{k}_{1}\partial^h \left( \alpha_{(11}^{1}\alpha_{10)}^{3} + i \alpha_{[11}^{2}\alpha_{10]}^{4} \right) + 
\partial^h \left( \alpha_{(00}^{1} \alpha_{10)}^{3} + i\alpha_{[00}^{2} \alpha_{10]}^{4} \right) + \frac{1}{2}\partial^h (k_b x^b \alpha_{10}^{3})
\end{eqnarray*}
and $(\gamma^5 \gamma^h, 0, 1)$ equals
\begin{eqnarray*}
0 & = & -2l^h \alpha_{01}^{4} \alpha_{00}^{2} - 2 \int d\vec{k}_{1}e^{i(k_{1,b} - l_{b})x^{b}}k_{1}^{h} \alpha_{01}^{4} \alpha_{11}^{2}
- 2i \int d\vec{k}_{1}e^{i(k_{1,b} - l_{b})x^{b}} \partial^h \alpha_{01}^{4} \alpha_{11}^{2} \\*
& & - \partial^h \alpha_{01}^{3} + \int d\vec{k}_{1} \partial^h \left( \alpha_{(01}^{1} \alpha_{11)}^{3} + i\alpha_{[01}^{2} \alpha_{11]}^{4}  \right)
- \frac{1}{2} \partial^h (l_b x^b \alpha_{01}^{3})
\end{eqnarray*}
and finally, $(\gamma^5 \gamma^h,1,1)$ is given by
\begin{eqnarray*}
0 & = & 2e^{i(k_b - l_b)x^{b}}k^{h}\alpha_{11}^{3} + 2 \delta(\vec{k} - \vec{l})k^h \left( \alpha_{00}^{4}\alpha_{00}^{2} + \int d\vec{k}_{1}
\alpha_{(01}^{4} \alpha_{10)}^{2} \right) \\*
& & + 2e^{i(k_b - l_b)x^b}k^h \alpha_{(10}^{4} \alpha_{01)}^{2} + 2e^{i(k_b - l_b)x^b}k^h\alpha_{(01}^{4}\alpha_{10)}^{2} + 
2\int d\vec{k}_{1}e^{i(k_b - l_b)x^b}k^h \alpha_{11}^{4} \alpha_{11}^{2} \\*
& & -2k^h \delta(\vec{k} - \vec{l})\alpha_{00}^{4} \alpha_{00}^{2} - 2\int d\vec{k}_{1}e^{ik_{1,b}x^b} \alpha_{01}^{4}\alpha_{10}^{2}\delta(\vec{k} - \vec{l})
 - 2l^he^{ik_b x^b}\int d\vec{k}_{1}\alpha_{01}^4 \alpha_{10}^{2} \delta(\vec{k} - \vec{l}) 
\end{eqnarray*}
\begin{eqnarray*}
& & -2l^h\alpha_{11}^{4} \alpha_{00}^{2} - 2e^{i(k_b - l_b)x^{b}}k^h \alpha_{01}^{4}\alpha_{10}^{2} - 2\int d\vec{k}_{1}e^{i(k_{1,b} - l_{b})x^{b}}
k_{1}^{h} \alpha_{11}^{4} \alpha_{11}^{2} \\*
& & -2e^{i(k_b - l_b)x^{b}}k^h \alpha_{00}^{4}\alpha_{11}^{2} - 2ie^{ik_{b}x^{b}}\partial^h \alpha_{01}^{4} \alpha_{10}^2
 - 2i\int d\vec{k}_{1}e^{i(k_{1,b} - l_{b})x^{b}} \partial^h \alpha_{11}^{4} \alpha_{11}^{2} \\*
& & -2ie^{i(k_b - l_b)x^b}\partial^h \alpha_{00}^{4} \alpha_{11}^{2} - \partial^h \alpha_{11}^{3} + \partial^h \left( \alpha_{(01}^{1}\alpha_{10)}^{3}
 + i \alpha_{[01}^{2}\alpha_{10]}^{4} \right) + \int d\vec{k}_{1}\partial^h \left( \alpha_{11}^{1}\alpha_{11}^{3} \right) + \\*
&  & \partial^h \left( \alpha_{(00}^{1}\alpha_{11)}^{3} + i \alpha_{[00}^{2}\alpha_{11]}^{4}  \right) + \frac{1}{2} \partial^h ((k_b - l_b)x^b \alpha_{11}^{3}). 
 \end{eqnarray*}
Finally, the $\gamma^0 \gamma^5 \gamma^h$ series reads: $(\gamma^0 \gamma^5 \gamma^h , 0 ,0)$ equals
\begin{eqnarray*}
0 & = & 2i\int d\vec{k}_{1}e^{ik_{1,b}x^{b}}k_{1}^{h}\alpha_{01}^{3} \alpha_{10}^{2} - 2\int d\vec{k}_{1}e^{ik_{1,b}x^{b}}\partial^{h}
\alpha_{01}^{3} \alpha_{10}^{2} \\*
& &  - \partial^h \alpha_{00}^{4} + i\int d\vec{k}_{1} \partial^h \left( \alpha_{(01}^{1}\alpha_{10)}^{4} + i\alpha_{[01}^{2} \alpha_{10]}^{4} \right)
\end{eqnarray*}
while $(\gamma^0 \gamma^5 \gamma^h, 1, 0)$ is given by
\begin{eqnarray*}
0 & = & -2ie^{ik_{b}x^{b}}k^h \int d\vec{k}_{1} \alpha_{[11}^{3}\alpha_{10]}^{2} + 2i\int d\vec{k}_{1}e^{ik_{1,b}x^{b}}k_{1}^{h}\alpha_{11}^{3}
\alpha_{10}^{2} + 2ie^{ik_{b}x^{b}}k^h \alpha_{10}^{3}\alpha_{00}^{2} - 2\int d\vec{k}_{1}e^{ik_{1,b}x^{b}}\partial^h \alpha_{11}^{3}\alpha_{10}^{2} \\*
& & - 2e^{ik_{b}x^{b}}\partial^h \alpha_{00}^{3} \alpha_{10}^{2} - \partial^h \alpha_{10}^{4} + i\int d\vec{k}_{1}\partial^h
\left( \alpha_{(11}^{1}\alpha_{10)}^{4} + i\alpha_{[11}^{4}\alpha_{10]}^{3} \right) \\*
& &  + i\partial^h \left( \alpha_{(00}^{1}\alpha_{10)}^{4} + i\alpha_{[00}^{2} \alpha_{10]}^{3} \right) + \frac{i}{2}\partial^h \left( k_b x^b \alpha_{10}^{4} \right) \\* 
\end{eqnarray*}
and $(\gamma^0 \gamma^5 \gamma^h, 0 ,1)$ equals
\begin{eqnarray*}
0 & = & 2il^h\alpha_{01}^{3}\alpha_{00}^{2} + 2i\int d\vec{k}_{1}e^{i(k_{1,b} - l_{b})x^{b}}k_{1}^{h}\alpha_{01}^{3}\alpha_{11}^{2}
-2\int d\vec{k}_{1}e^{i(k_{1,b} - l_{b})x^{b}} \partial^h \alpha_{01}^{3} \alpha_{11}^{2} \\*
& & -\partial^h \alpha_{01}^{4} + i\int d\vec{k}_{1} \partial^h \left( \alpha_{(01}^{1} \alpha_{11)}^{4} + i \alpha_{[01}^{2}\alpha_{11]}^{3} \right)
 - \frac{i}{2}\partial^h \left( l_b x^b \alpha_{10}^{4} \right) 
\end{eqnarray*}
while finally, $(\gamma^0 \gamma^5 \gamma^h, 1 , 1)$ reads
\begin{eqnarray*}
0 & = & -2i\delta(\vec{k} - \vec{l}) \int d\vec{k}_{1}k_{1}^{h} \alpha_{[01}^{3}\alpha_{10]}^{2} - 2ie^{i(k_b - l_b)x^b}k^h\alpha_{[10}^{3}
\alpha_{01]}^{2} + 2ik^h \delta(\vec{k} - \vec{l}) + 2ik^h\delta(\vec{k} - \vec{l}) \int d\vec{k}_{1}e^{ik_{1,b}x^{b}} \alpha_{01}^{3}
\alpha_{10}^{2} \\*
& & + 2il^h e^{ik_{b}x^{b}}\int d\vec{k}_{1}\alpha_{01}^{3}\alpha_{10}^{2}  + 2il^h\alpha_{11}^{3}\alpha_{00}^{2} + 2i
\int d\vec{k}_{1}e^{i(k_{1,b} - l_{b})x^{b}}k_{1}^{h} \alpha_{11}^{3}\alpha_{11}^{2} + 2ie^{i(k_b - l_b)x^b}k^h \alpha_{10}^{3}\alpha_{01}^{2} \\*
& & -2\int d\vec{k}_{1}e^{i(k_{1,b} - l_{b})x^{b}}\partial^h \alpha_{11}^{3} \alpha_{11}^{2} - 2e^{i(k_b - l_b)x^b} \partial^h \alpha_{00}^{3}
\alpha_{11}^{2} - \partial^h \alpha_{11}^{4} \\*
& & +i\int d\vec{k}_{1} \partial^h (\alpha_{11}^{1}\alpha_{11}^{4}) +i\partial^h \left( \alpha_{(00}^{1}\alpha_{11)}^{4} + i\alpha_{[00}^{2}
\alpha_{11]}^{3} \right) + \frac{i}{2} \partial^h ((k_b - l_b)x^{b} \alpha_{11}^{4}).  
\end{eqnarray*}
We now investigate whether solutions or not exist; rotational invariance implies the following functional dependencies 
$\alpha^i_{00}(r,t)$, 
$\alpha^i_{01}(r,\vec{x}.\vec{l}, l,t)$, $\alpha^i_{11}(r,\vec{x}.\vec{k}, \vec{x}.\vec{l}, \vec{k}.\vec{l},k,l,t)$ and moreover
$\overline{\alpha}^{i}_{01} = \alpha^i_{10}$, $\overline{\alpha}^{i}_{11}(\vec{x}, \vec{k}, \vec{l},t) = \alpha^{i}_{11}(\vec{x}, \vec{l}, \vec{k},t)$ and
$\overline{\alpha}^i_{00} = \alpha^i_{00}$.  Before we actually start to do any boring work, let us think about a clever method to compute solutions to this
type of equations.  The most natural method is iteration; that is, linearize around the free solution on the origin of space and time, 
compute the (approximate) solutions to these equations if they exist, linearize around this solution and study whether this process converges in a well defined sense.
 There are several problematic aspects with this line of thought which are (a) the fact that one might miss out lots of solutions to the 
full equations because the linearized equations (which constitute an overdetermined system) give an ill posed problem (b) near the singularity it just might be
that iteration does not work or that convergence is way too slow.  To compensate for these shortcomings, one might restrict to meromorphic solutions in suitable 
variables which would be $r, \vec{x}.\vec{k},k$ and $t$ in our case; this would result in an infinite dimensional matrix problem which we can only solve
approximately.  A further remark would be that generically coordinate singularities might show up in solutions to the above equations or that 
we are interested in the regime sufficiently far away from the singularity.  In that case we would better shift the origin and give up rotational
symmetry because some terms in the above equations would clearly be of leading order and solutions are much easier found.  This would lead to the
notion of space-time renormalization which I will work towards at this point.  First of all, in asymptotically flat spacetimes, it is natural to look
for solutions which are asymptotically free; this is clear from the usual $S$-matrix motivation as well as the special geometrical configuration.  
There exist however several notions of asymptotic freeness, depending on how one makes the following idea rigorous; it is natural to expect that for $x$ and $y$ sufficiently 
large, the following holds
$$U(e_b(x), e_b(y),x,y) \sim e^{i \int d\vec{k}k_a(y^a - x^a)a^{\dag}_{\vec{k}}(e_b(x),x) a_{\vec{k}}(e_b(x),x)}$$ for any ``almost flat'' 
vierbein $e_b^{\nu}(x)$, that is
$$\nabla_{\mu}e_{b}^{\nu}(z) \sim 0$$ for any $z$ in some asymptotic region containing $x,y$ spacetime indices $\mu, \nu$ any flat Lorentz index $b$ and 
``almost flat'' coordinate system $x^{\mu}$.  Likewise, there is the notion of asymptotic abelianess, which dictates that for $x,y$ sufficiently large
 $$\left[ H(x,e_b(x)),H(y,e_b(y)) \right] \sim 0$$
with respect to an ``almost'' flat vielbein in an ``almost'' flat coordinate system.  
Obviously, asymptotically free solutions ought to be asymptotically abelian; the latter notion however is much weaker given that the nontrivial commutant of 
some free Hamiltonian $H(x)$ (with possibly a central extension) is algebraically generated by monomials of the kind 
$$\int d\vec{k}_{1} \ldots d\vec{k}_{m} d\vec{l}_{1} \ldots d\vec{l}_{n}a^{\dag}_{\vec{k}_{1}}(x,e_b(x)) \ldots a^{\dag}_{\vec{k}_{m}}(x,e_b(x))
a_{\vec{l}_{1}}(x,e_b(x)) \ldots a_{\vec{l}_{n}}(x,e_b(x))$$ $$ f_{mn}(\vec{k}_{1}, \ldots \vec{k}_{m}, \vec{l}_1, \ldots, \vec{l}_{n},x) \delta\left( \sum_{j=1}^{m}
\vec{k}_j - \sum_{j=1}^{n}\vec{l}_j \right).$$
The ambiguities in the above reside in the notions of ``almost'' and ``limit'', where the latter may be taken over 
lightcones associated to points or worldines or to some asymptotic sections of spacetime associated to a preferred (eg. Killing) observer.
  We shall not be concerned with those details over here and the reader should keep in mind that in our example limits are always
 understood by using the Killing time $t$.  We now come to the notion of spacetime renormalization which is nothing but the effect of 
shifting points on the local Hamiltonians, so we might call it as well spacetime recalibration.  We \textit{assume} that\footnote{That is, we posit by hand
that the group properties hold between the origin, $x$ and $x_0$.}
$$U(x,e_b(x))U^{\dag}(x_0,e_b(x_0)) = U(x_0,x,e_b(x_0),e_b(x))$$ and the trick is now to write
$$U(x_0,x,e_b(x_0),e_b(x)) = \widetilde{U}(x,e_b(x)) = e^{i\widetilde{H}(x)}$$
with respect to some fixed vierbein $e_b(z)$ and henceforth we surpress reference to the latter in the Hamiltonians $\widetilde{H}(x), H(x)$.  This allows us to rewrite local $\widetilde{\alpha}$ functions in terms of local
 $\alpha$ functions and those with respect to the new reference point $x_0$.  Again, the solution to this problem resides in an appropriate perturbation theory
around the free solution; however the latter trick is only meaningful with respect to some point $x_0$ sufficiently far from the singularity.  Therefore,
 we write
$$U(x)U^{\dag}(x_0) = \left( e^{i\widetilde{H}(x)}e^{-i\widetilde{H}_{0}(x)} \right)e^{i\widetilde{H}_{0}(x)}$$ and expand the expression between brackets
in $n$'th order in $\widetilde{H}(x)$ and $\widetilde{H}_{0}(x)$ with respect to the \emph{new} Clifford basis.  Next, one may rewrite the new annihilation/creation operators
and Clifford elements in turn of the old ones by using $H(x_0)$ and retain only terms up till order $n$ in $H(x_0)$.  
An appropriate perturbation scheme for the left hand side could also reside in perturbation around the 
free theory with respect to the orgin (but possibly of an order $m \neq n$) or might consist in a straightforward perturbation expansion in terms of the full Hamiltonians $H(x), H(x_{0})$ depending on which scheme has the best convergence properties.  The latter scheme however is quite unnatural since one would expect a symmetric treatment on both sides, therefore we develop the former\footnote{It is possible to demand that $x,x_0$ are close to one and another in which case a straightforward perturbation expansion on \emph{both} sides is expected to hold.}; an elementary calculation yields
$$\left( e^{iH(x)}e^{-iH_{0}(x)} \right)\left[ e^{iH_{0}(x-x_0)}\left( e^{iH_0(x_0)}e^{-iH(x_0)} \right)e^{-iH_{0}(x-x_{0})} \right]\left( e^{iH_0(x-x_0)}e^{-i\widetilde{H}_0(x)}\right) =$$ $$\sum_{r = 0}^{n} \sum_{k+l = r} 
\frac{i^{k+l}(-1)^{l}}{k!l!}\widetilde{H}(x)^k \widetilde{H}_{0}(x)^{l}.$$
As an example, we put $m=n=2$ and make a cutoff in the matrix elements as before; that is, we neglect ``scattering'' with a high number
of particles; that is, since all expressions between brackets are expected to be small we do perturbation theory in second order in $H,H_0$.  While the right hand side reduces to
$$1 + i\widetilde{H}(x) - i\widetilde{H}_{0}(x) + \widetilde{H}(x)\widetilde{H}_{0}(x) - \frac{1}{2}\widetilde{H}^2(x) - \frac{1}{2}\widetilde{H}_{0}^{2}(x)$$ the left hand side is given by the following lengthy expression
\begin{eqnarray*}
& & \left( 1 + iH(x) - iH_0(x) + H(x)H_0(x) - \frac{1}{2}H^2(x) - \frac{1}{2}H_{0}^2(x) \right) \\*
& & \left[ e^{iH_0(x-x_0)}\left( 1 + iH_0(x_0) - iH(x_0) + H_0(x_0)H(x_0) - \frac{1}{2}H_0^2(x_0) - \frac{1}{2}H^2(x_0) \right)e^{-iH_0(x - x_0)} \right] \\*
& & \left( 1 + iH_0(x-x_0) - i\widetilde{H}_0(x) + H_0(x-x_0)\widetilde{H}_0(x) - \frac{1}{2}\widetilde{H}_{0}^2(x) - \frac{1}{2}H_{0}^2(x-x_0) \right).
\end{eqnarray*}
One should write out both expressions in their respective basis and finally express the basis at $x_0$ in terms of the basis at $0$ in second order perturbation theory in $H_{\textrm{int}}(x)$.  We leave this as a labourious exercise to the reader.  
\chapter{Some sobering final comments and exciting possibilities the future may bring us}
As if the present construction is not enough to digest yet, let me sketch some ideas which might evolve the theory into the future.  Our method has been a healthy mixture of a hands-on conservative approach and a quite general axiomatic one; however, one could decide now to push the axiomatic framework even further by pure thought or to wait again before real trouble would show up with this theory.  Let me indulge in the luxury of making the former exercise for a brief moment; we have already partially dispelled the absolutism from quantum mechanics, there is however still one absolute remnant which is the local Nevanlinna module $\mathcal{K}$ and the free theory on the tangent bundle.  On the other hand, there is an unknown theory of awareness or meaning which has still to be constructed and interferes with the dynamics for macroscopic bodies as speculated previously.  How to construct a theory of genuine creation?  As Godel's theorem seems to suggest, such adventure cannot be grounded within classical logic; actually, I would speculate that it cannot be constructed within the limitations of any fixed framework for rationality (even the quantum one).  Since our equations are grounded within logic by the very definition of equality, logical equivalence and implication we have to ask again the question what it means for two things to be equal to one and another.  A subject which has been touched upon many times by philosophers and logicians as the excellent accounts of Whitehead and Russell demonstrate.  The idea is that logic is dynamical, therefore equations and definitions get a different meaning and as a back reaction logic changes too.  Furthermore, logic depends not only upon the dynamics but also on the ``state of the universe'' (in the sense explained previously) as is clearly demonstrated in everyday life: in court for example, a proof of guilt may consist in the frowning of a respected witness upon the accused while the connotation of evidence has a different meaning in science; indeed, the contextuality of logic is a basic fact of life.  So, the quest is to find out means to write out such theory whereby the meaning of symbols defines itself.  In this quest, we must touch upon the deepest question, that is the one of meaning of things or what this mysterious quality of understanding which we really possess is?  For example, how should one tell to a computer what the quantifiers $\forall$ or $\exists$ mean?  There is no way of doing that !  This implies a human cannot \emph{define} it neither, but clearly we know what it means.  A computer could not even \textit{understand} it in a relational context; for example, take the definition of continuity for functions from $\mathbb{R} \rightarrow \mathbb{R}$.  Give now the computer the function $x \rightarrow x$ and ask whether it is continuous.  I conjecture that the computer will never ever give an answer as quick as a human will on these type of questions because it doesn't know the meaning of \emph{pick} $\epsilon > 0$, \emph{choose} $\delta \leq  \epsilon$, then $|y- x| < \delta$ implies that $|y -x| < \epsilon$.  Indeed, formally a computer will find out an answer to all questions that are true and can be proven to be true in a finite number of lines, but a simple question as the one above in a most primitive language will take perhaps millions of years or it might even be possible that one could formulate finite sentences using new words which by themselves cannot be expressed in a finite number of elementary letters in the primitive language.  We know such things exist; for example functions which have no closed prescription in a finite language.  Perhaps, a human can ``grasp'' infinity and writes down finite projections of it on a sheet of paper; this at least seems to have been suggested by Cantor and Godel.  As an example of this, we have the illusion we can draw \emph{any} curve on a plane in our minds (for almost all, we could never give a function prescription) while a computer clearly could not do that.  The only thing which can help a machine would be to increase its computation power drastically, and here physics might put a limit to what such machine can accomplish without becoming an organic, living, conscious being by itself.  But then, we are in a very paradoxical situation: our human brains are capable of making far quicker more complex thoughts than computers can while we fail by many margins to calculate something as simple as the square root of two up to ten numbers after the first digit as quick as a machine\footnote{I thank Geert Vernaeve for interesting discussions regarding this topic}.  Still, this allows for the possibility that nature is a Turing machine but one which is not based in physical reality; this would lead to a weaker form of the Penrose conclusion that we shall never be able to compute what nature actually does.  Both conclusions, one which takes meaning to be metaphysical and not a mere illusion and the other which takes meaning to be an illusion but one which cannot be computed by any \textit{physical} machine, are sobering remarks to the limited scope mathematics and physics have regarding reality.  In these issues, I side with Godel and \emph{believe} the universe is a living entity with two components : (a) a dynamic symbolic language of meaning and knowledge containing a dynamical logic and (b) actors and states.  States define how ``things are'', actors relate states to the symbolic language of meaning and the total dynamics should relate all three of them.  The reason why I side with Godel comes from another metaphysical consideration that to any machine or computer, you have to tell what it has to do, what the ground rules are.  A really closed system defines itself and that is the very essence of the liar paradox which inspired Godel to do his work. \\* \\*
Only meaning resists definition, it transcends knowledge and all the rest; still it is the driving force behind our actions and the way we think about the world.  Meaning, I believe, is an eternal \textit{self-referring} concept which catches words like: I, You, Survival, Food, Procreation, Love and so on ... which never ever change.  Knowledge is a dynamic \textit{relational} component between entities having a certain meaning; therefore it is only very basic in the sense that it only uses well defined words like (probably) implies, is (more or less) equivalent to and so on.  The creation of new knowledge always involves an interplay    
between meaning and contemporary knowledge; therefore, by the lack of a definition of meaning, we will never ever be able to write down a theory of everything.  Science will always be a game of humans running behind their own tail as well as a vital ingredient in changing our own future and the laws of the universe themselves; therefore the way things are will also change depending on the knowledge we have of it.  There is however hope and the scientific enterprise is certainly not doomed; that is, I think it is fair to say that microscopic physics will satisfy the general  principles we laid out in chapter eight.  Surely, one may increase complexity in the gravitation theory by going over to higher bundles; this will necessarily change the way we think about quantum physics too meaning that the implementation of the principle of Lorentz covariance has to undergo some modifications.  Likewise, we may further generalize quantum physics by allowing for nonassociative ``algebras'' or by introducing some nonlocality by letting the product depend upon the number of factors.  But that doesn't imply we don't understand microscopic phyisics, we do: we know the basic principles and all the rest is merely representation.  Likewise, I have tried to formulate a principle for macroscopic physics; in contrast to the microscopic world, this does not allow for a well-defined theory but we can produce well-defined approximations.  This is good enough and we better learn to live with this limitation.  It is as if we are allowed to understand and grasp the linearization of a reality in which the nonlinear terms involve that what we cannot define: understanding. \\* \\*
On the positive side, a completely new scala of mathematical possibilities is opened and the remainder of this chapter contains the seeds of such constructions.  I have put a lot of emphasis on the principle of locality which is grounded in the continuum and the entire geometric construction turned around the most straightforward universal construction of this idea, that is real $n$ dimensional manifolds and its derived bundle structures.  However, the fashionable idea in these days is that spacetime should display a scale dependent granularity and the only question I am allowed to ask is if there exists a universal geometrical construction allowing for this while canonically lifting the locality notion of $\mathbb{R}^n$ and having no space-time superposition principle.  The answer to this question is yes and the difference lies in the density matrix approach to quantum mechanics versus the state approach.  The problematic aspect of all noncommutative approaches so far is that the diffeomorphism group has no natural place in the formalism and indeed, imposing algebraic relations by hand breaks diffeomorphism invariance of the single algebra.  The answer to this problem is to consider all possible algebras and modelling one manifold on a particular one.  Hence, a diffeomorphism will map one manifold into another and the only fixed manifolds are the abelian and free ones.  Moreover, the abelian continuum spacetimes have the largest symmetry group and therefore they are preffered from the point of view of internal symmetries.  Therefore, any quantum spacetime dynamics should be based upon the fact that a maximal internal symmetry group, as a subgroup of the free diffeomorphisms, determines the only stable ground state.  Hence, we conjecture that the theory developed so far describes the ground state of a much larger one which allows for small scale granularity as quantum fluctuations at sufficiently small scales.  Likewise, the attentive reader must have noticed that the formulation of a covariant quantum mechanics requires \emph{quantum} objects which transform nontrivially under coordinate as well as local Lorentz transformations while we just started out from a scalar section\footnote{That is the local Hamiltonians $H(x)$.}.  Therefore, it would be natural to extend the whole theory to the infinite dimensional tangent bundle
$$T^{\infty} M = \oplus_{k = 0}^{\infty} T^{k} M$$ and consider particular quantum representations of those with respect to the coordinate transformations as well as the internal Lorentz symmetries.   We will present some of the mathematics behind the latter idea first, then come back to the former and finally present some possibly important modifications to the latter in this new context.
\section{Tensorial and spinorial quantum mechanics.}
Let us first start from defining the most general kind of involutive algebra on $T^{\infty} M$; the natural sum of $(H_{\alpha_1 \ldots \alpha_k})_{k=0}^{\infty}$ and $(K_{\alpha_1 \ldots \alpha_k})_{k=0}^{\infty}$ is given by
$$(H_{\alpha_1 \ldots \alpha_k})_{k=0}^{\infty} + (K_{\alpha_1 \ldots \alpha_k})_{k=0}^{\infty} = (H_{\alpha_1 \ldots \alpha_k} + K_{\alpha_1 \ldots \alpha_k})_{k=0}^{\infty}$$ whereas some more general\footnote{The only further generalization consists in allowing for indices to swap in the definition below.  In terms of monoidial category theory, we do incorporate a duality (contraction) but not a swapping operation; such extension being left for future work.} product structure is given by
$$(H_{\alpha_1 \ldots \alpha_k})_{k=0}^{\infty}.(K_{\alpha_1 \ldots \alpha_k})_{k=0}^{\infty} = \left(\sum_{k+l = m + 2r; r \geq 0} \gamma^{m}_{kl\sigma_r} H_{\alpha_1 \ldots \alpha_k} K_{\beta_1 \ldots \beta_l } \prod_{p = 1}^{r} g^{\mu_{p} \nu_{\sigma_{r}(p)} } \right)_{m=0}^{\infty}$$ where the coupling constants $\gamma^{m}_{kl\sigma_r}$ are real, $\gamma^{m}_{0k\sigma_r}= \gamma^{m}_{k0\sigma_r} = \delta^{0}_{r}$ and $\sigma_r$ indicates $r$ contractions in the $k+l = m+2r$ indices by means of the spacetime metric.  The above fixing of those $\gamma$ symbols involving a zero lower index originates from the demand that $(1,0,0,\ldots)$ must be the identity element.  Insisting that the product is associative gives an infinite tower of ``braid'' relations between the $\gamma$ coefficients; more precisely:
$$\sum_{\textrm{equivalent braidings}} \gamma^{n}_{jm\sigma_{\frac{j+m-n}{2}}} \gamma^{m}_{kl\sigma_{\frac{k+l-m}{2}}} = \sum_{\textrm{equivalent braidings}} \gamma^{p}_{jk\sigma_{\frac{j+k-p}{2}}}\gamma^{n}_{pl\sigma_{\frac{p+l-n}{2}}}.$$
Obviously, the tensoralgebra without contractions and therefore no braiding is a straightforward example; another less trivial one consists in setting all symbols with internal contractions to zero and those with $r$ cross contractions\footnote{One notices in this case that the braiding on each side is unique.} equal to $\alpha^r$.  The construction of less trivial examples is a task for the future.  The canonical involution simply is given by complex conjugation as there exists a fairly unique scalar product on $T^{\infty} M$ given by
$$\langle (H_{\alpha_1 \ldots \alpha_k})_{k=0}^{\infty} | (K_{\alpha_1 \ldots \alpha_k})_{k=0}^{\infty} \rangle = \sum_{k=0}^{\infty} \kappa_k \overline{H_{\alpha_1 \ldots \alpha_k}}g^{\alpha_1 \beta_1}\ldots g^{\alpha_k \beta_k}K_{\beta_1 \ldots \beta_k}$$ where $\kappa_k$ are real numbers.  The latter product is obviouly indefinite no matter what the signs of the $\kappa_k$ are and therefore the natural language for tensorial quantum mechanics appears to be the one of a ``Nevanlinna algebra''.  One may now look for representations of different types as we illustrated rigorously in chapter seven depending on whether or not you take the dual point of view.  To illustrate what we mean by this, consider the trivial example without contractions and define the action of $(H_{\alpha_1 \ldots \alpha_k})_{k=0}^{\infty}$ on a vector $v$ as
$$(H_{\alpha_1 \ldots \alpha_k})_{k=0}^{\infty}v = (H_{\alpha_1 \ldots \alpha_k}v)_{k=0}^{\infty}$$ where the expressions on both sides must be understood in the dual way with respect to a section of $T^{\infty}M$.  In that case, the inner product is the standard one in this dual interpretation and clearly this action respects the law that $\left( (H_{\alpha_1 \ldots \alpha_k})_{k=0}^{\infty}.(K_{\alpha_1 \ldots \alpha_k})_{k=0}^{\infty} \right)v = (H_{\alpha_1 \ldots \alpha_k})_{k=0}^{\infty}. \left( (K_{\alpha_1 \ldots \alpha_k})_{k=0}^{\infty}v \right)$ and the adjoint is what it should be.  Abandoning the dual point of view requires one to consider vectors of the type $(v_{\alpha_1 \ldots \alpha_k})_{k=0}^{\infty}$ and one must look for a suitable scalar product as well as action in order to respect the algebraic properties.  In general, it might be conceivable to consider braided actions of the kind
$$(H_{\alpha_1 \ldots \alpha_k})_{k=0}^{\infty} (v_{\beta_1 \ldots \beta_l})_{l=0}^{\infty} = \sum_{k + l =  m + 2r; m,r \geq 0} \lambda^{m}_{kl\sigma_r} H_{\alpha_1 \ldots \alpha_k} v_{\beta_1 \ldots \beta_l}\prod_{p=1}^{r} g^{\mu_{p} \nu_{\sigma_{r}(p)} }$$ so that the following tower of braid relations is satisfied
$$\sum_{\textrm{equivalent braidings}} \lambda^{n}_{jm \sigma_{\frac{j+m-n}{2}}} \lambda^{m}_{kl\sigma_{\frac{k+l-m}{2}}} = \sum_{\textrm{equivalent braidings}} \lambda^{n}_{ml\sigma_{\frac{m+l-n}{2}}} \gamma^{m}_{jk\sigma_{\frac{j+k-m}{2}}}$$ as well as the standard gauge fixing conditions originating from left multiplication with the identity operator.  Again, it might be interesting to find braided actions for the trivial product structure on the sections of $T^{\infty}M$; the scalar product structure is similar to the one above for the ``Nevanlinna algebra'' where the $\kappa_k$ represent this time Hermitian operators squeezed between the remaining vectors.  Therefore the only remaining task is to find quantum representations of the group of coordinate transformations commuting with the quantum local Poincar\'e transformations\footnote{This point may differ for noncommutative manifolds however where one would need to consider twisted representations and we come back to this later on.}.  The straightforward way to construct those would be to construct unitary representations of the sheaf of differentiable local $GL_n(\mathbb{R})$ sections over $M$ and restrict those to the integrable sections.  The former are of course most easily found by considering differentiable distributions $U_x$ of (quasilocal of ultralocal) unitary $GL_n(\mathbb{R})$ representations commuting with the local Poincar\'e algebras.  One may wonder whether one may find a criterion for classical integrability at the quantum level and the answer to this may be yes.  Consider the usual linear representation\footnote{Permutations of colums in a matrix do not preserve the transpose, neither multiplication.} of the permutation group $S_n$ on $GL_n(\mathbb{R})$, then $\sigma_x$ twisted integrability of $U_x$ could be defined by comparing the singular terms in $$\partial_k U_x((E_i + \epsilon 1)M(x)(E_j + \epsilon 1))$$ and $$\partial_j \sigma_x \left( U_x \left( (E_i + \epsilon 1)\sigma_{jk}\left(M(x) \right)(E_j + \epsilon 1) \right)\right)$$ near $\epsilon \sim0$ where $\sigma_{jk}$
is the representation of the permutation $(jk)$, $\partial_k E_i M(x)E_j = \partial_j E_i \sigma_{jk}\left( M(x) \right)E_j$, $\sigma_x$ is a distribution of $\star$ automorphisms and $E_i$ is the projection on the i'th coordinate.  Quantum representations of the above algebra are then most easily found by considering the usual actions by internal automorphisms; obviously this construction can be extended to the whole cotangent bundle.  Everything done in chapter $8$ can be straightforwardly generalized in this extended mathematical setting.  Obviously, it is possible to regard spacetime from the eight dimensional spinor perspective which would lead to spinor bundles and half integer quantal operators.  In terms of creation and annihilation operators, all the above really means that we attach supplementary (half) integer quantum numbers to the standard ones which indicate at which level of the spacetime bundle these operators are working.  Of course, as mentioned previously, the creation and annihilation operators in quantum field theory are scalar with respect to spacetime in the sense that they do not appreciate the higher bundle structure.  We now turn to the construction of noncommutative manifolds prior to dealing with some further generalizations of this construction.    
\section{Topological quantum manifolds}
Basically, the universal complex (or real) algebra in $n$ variables $\widehat{x}_i$ is the free one $\mathcal{F}^{\infty}_n$; we shall 
also be concerned with the free algebra of finite words $\mathcal{F}_n$ which is equipped with a canonical involution $\star$ which
 simply reverses the order of the words and conjugates the complex numbers.  Hence, every generator is Hermitian and therefore has a real
 spectrum if one restricts to $W^{\star}$ algebraic representations.  Besides $\mathcal{F}^{r}_n$, there is the totally commutative
 algebra $\mathcal{C}^{r}_n$ in $n$ variables $x_i$ and we denote by 
$\phi : \mathcal{F}^{r}_n \rightarrow \mathcal{C}^{r}_n : \widehat{x}_i \rightarrow x_i$ the canonical homomorphisms
 where $r \in \{ \emptyset , \infty \}$.  Morever, we adjoin all algebras with an identity element and restrict to unital $\star$ 
homomorphisms.  The idea is to represent $\mathcal{F}_n$ in unital $W^{\star}$ algebras $\mathcal{A}$ equipped with a 
trace functional $\omega_{\mathcal{A}}$.  Therefore
 let $\pi : \mathcal{F}_n \subset \textrm{Dom}(\pi) \subset \mathcal{F}^{\infty}_n \rightarrow \mathcal{A}$ be a unital, maximal, star 
homomorphism (where $\textrm{Dom}(\pi)$ is a subalgebra) with a dense image and denote by $\sigma(i, \pi, \mathcal{A})$ the spectrum 
of $\pi(\widehat{x}_i)$ in $\mathcal{A}$; then it is natural to construct the compact and bounded ``cube''
$$
\mathcal{O}(\pi,\mathcal{A}) = \times_{i = 1}^{n} \sigma(i, \pi, \mathcal{A})\;.
$$
Likewise, one can restrict the variables in $\mathcal{C}_n$ to $\mathcal{O}(\pi, \mathcal{A})$.  Because of the spectral decompositon theorem, for every $n$ vector $\vec{\alpha}$ in the cube, index $i$ and $\epsilon_i > 0$, one has a unique Hermitian spectral operator $P^{\epsilon_i}_{\alpha_i}$ which is by definition a shorthand for 
$$
P^i (( \alpha_i - \epsilon_i, \alpha_i + \epsilon_i ))\;.
$$
The operators have the usual intersection properties.  Hence for every resolution $\vec{\epsilon}$, we may define an event $P^{\vec{\epsilon}}(\vec{\alpha})$ in the algebra $\mathcal{A}$ as the maximal Hermitian projection operator wich is smaller than all $P^{\epsilon_i}_{\alpha_i}$ (notice that this projection operator may become zero if the resolution becomes to high, that is $\epsilon_i$ too small).  Now, it is easy to see that if one were to cover a cube by smaller cubes (arbitrary overlaps are allowed), take the projection operators associated to those and consider the smallest projection operator which majorizes all of these, then, by the superposition principle, the latter is smaller or equal to the projection operator of the full cube.  This is a very quantum mechanical idea where we acknowledge that the whole is more than the sum of its parts and therefore we have to give up the idea of a classical partition.  Hence, for any relative open subset $\mathcal{W}$; there exists a unique smallest projection operator which majorizes all projection operators attached to subcoverings of $\mathcal{W}$ by relative open cubes (a subcovering simply is a set of relative open cubes contained in $\mathcal{W}$).  Hence, there is a natural almost everywhere weakly continuous\footnote{We shall explain this notion later on.} mapping $\kappa_{(\pi,\mathcal{A})}$ from relative open subets $\mathcal{W}$ of $\mathcal{O}(\pi,\mathcal{A})$ to $\mathcal{A}$ given by
$$
\kappa_{(\pi,\mathcal{A})}(\mathcal{W}) = P(\mathcal{W})\;.
$$
For disjoint $\mathcal{W}_j$ one obtains that
$$
P(\mathcal{W}_1)\, P(\mathcal{W}_2) = 0\;,
$$
meaning that the coherence of the theory depends upon the scale you are observing at.  Concretely, if you zoom into the region 
$\mathcal{W}_1$ you will be oblivious to the entanglement with the region $\mathcal{W}_2$; however, looking at both together gives a very
 different picture.  If the \emph{dynamics} itself were scale dependent in this way, then it might explain why we see a local world on
 our scales of observation and above, while the microscopic world would seem to be completely entangled.  This picture would offer a complete relativization of physics where giants would look to us as if we were electrons.  Also, 
$$
P(\mathcal{W}_1) \prec P(\mathcal{W}_2)
$$
of $\mathcal{W}_1 \subset \mathcal{W}_2$, which means that zooming in is a consistent procedure.  Now, we can go on and construct several
 forms of equivalence, going from ultra strong to ultra weak.  Two 
representations $\pi_i: \mathcal{F}_n \subset \textrm{Dom}(\pi_i) \subset \mathcal{F}^{\infty}_n \rightarrow \mathcal{A}_i$ are ultra
 strongly isomorphic if and only if there exists a $W^{\star}$ isomorphism $\gamma : \mathcal{A}_1 \rightarrow \mathcal{A}_2$ such
 that $\pi_2 = \gamma \circ \pi_1$ and $\textrm{Dom}(\pi_1) = \textrm{Dom}(\pi_2)$.  They are called strongly isomorphic it is only 
demanded that $\gamma$ is a unital star isomorphism from $\pi_1(\textrm{Dom}(\pi_1))$ to $\pi_2(\textrm{Dom}(\pi_2))$.  We say, moreover,
 that they are weakly isomorphic when equality is supposed to only hold on $\textrm{Dom}(\pi_1) \cap \textrm{Dom}(\pi_2)$ and finally we
 define them to be ultra weakly equivalent if and only if $\gamma$ is a star isomorphism from
 $\pi_1(\mathcal{F}_n)$ to $\pi_2(\mathcal{F}_n)$ and equality only holds on $\mathcal{F}_n$.  In the case of real manifolds, ultra weak covariance is the only notion which applies and we continue now to investigate it.  Now, we are ready to go over to an atlas construction; a topological space $\mathcal{M}$ is said to be a real, $n$-dimensional, non-commutative manifold if there exists a covering of $\mathcal{M}$ by open sets $\mathcal{V}_{\beta}$, a homeomorphism $\phi_{\beta}$ from $\mathcal{V}_{\beta}$ to a relative open subset of the cube $\mathcal{O}(\pi_{\beta},\mathcal{A}_{\beta})$ associated to some representation $\pi_{\beta}: \textrm{Dom}(\pi_{\beta}) \rightarrow \mathcal{A}_{\beta}$ of the free algebra in n letters.  This homeomorphism canonically lifts to the algebra on the open subsets $\mathcal{W} \subset \mathcal{V}_{\beta}$ by stating that $\widehat{\phi}_{\beta}(\mathcal{W}) = \kappa_{(\pi_{\beta}, \mathcal{A}_{\beta})}(\phi_{\beta}(\mathcal{W}))$.  Hence, a single chart is a tuple $\left( \mathcal{V}_{\beta}, \pi_{\beta}, \mathcal{A}_{\beta}, \phi_{\beta} \right)$ and we proceed now to construct an atlas by demanding compatibility.  Two charts $\mathcal{V}_{\beta_j}$ with some non zero overlap $\mathcal{V}_{\beta_1} \cap \mathcal{V}_{\beta_2} \neq \emptyset$ are said to be compatible if and only if the canonical mapping between the normed subsets
$$
\{ P_{\beta_j}(\widehat{\phi}_{\beta_{j}}(\mathcal{W})) \, | \, \mathcal{W} \subset \mathcal{V}_{\beta_1} \cap \mathcal{V}_{\beta_2} \}
$$
induces a star isomorphism between the normed algebras generated by them; the latter preserves the trace functionals $\omega_{\mathcal{A}_{\beta}}$.  We now proceed by giving some examples. \\* \\*
We start by the most trivial thing and show that ordinary real manifolds have a natural place in this setup.  Let $\mathcal{M}$ be
 an $n$-dimensional real manifold and consider the coordinate chart $(\mathcal{V}, \psi)$.  Define now the Hilbert 
space L$^2(\overline{\psi(\mathcal{V})}, \dd^n x)$ and the multiplication operators $x_i$.  Define $\mathcal{A}$ to be 
the $W^{*}$ algebra generated by the $x_i$, then $\pi : \mathcal{F}_n \rightarrow \mathcal{A} : \widehat{x}_i \rightarrow x_i$ 
has a unique maximal extension.  The spectrum of each of these multiplication operators is continuous and varies between $a_i < b_i$
 and the canonical mapping $\phi$ is given by $\phi(v) = \psi(v)$.  Then, the canonical projectors associated to
 $\mathcal{W} \subset \mathcal{V}$ are given by $P(\mathcal{W}) = \chi_{\phi (\mathcal{W})}$ where the latter is the characteristic 
function on $\mathcal{W}$.   Clearly, a coordinate tranformation induces a $W^{\star}$ algebraic isomorphism between these 
commutative projection operators.  By the same arguments, one sees that any commutative $n$-dimensional measure space is represented 
in this framework; so we are left with presenting a non abelian example.  A very simple example is a double sheeted manifold 
constructed from the Hilbert space L$^2(\mathbb{R}^4, \dd^4 x) \otimes \mathbb{C}^2$ and consider the algebra generated by the
 operators $x^{\mu} \otimes \sigma^{\mu}(x)$ where the $\sigma^{\mu}(x)$ are automorphic to the standard spacetime Pauli algebra
 $(\sigma^{\mu}) = (1, \sigma^i)$.  That is $\sigma^{\mu}(x) = U(x)\, \sigma^{\mu} U^{\dag}(x)$ for $U(x)$ some $2 \times 2$ complex 
unitary matrix.  The whole manifold structure depends upon $U(x)$, since suppose $U(x) = 1$, then the cube is
 $\mathbb{R}^4$ and the set of basic projection operators is given by
\begin{eqnarray*}
P^{\epsilon}_t & = & \chi_{\left[ t - \epsilon , t + \epsilon \right]} \otimes 1 \\
P^{\epsilon}_x & = & \frac{1}{2} \left[ \chi_{\left[ x - \epsilon, x + \epsilon \right]} \otimes |1,1 \rangle \langle 1,1 |  + \chi_{\left[ - x - \epsilon , - x + \epsilon \right]} \otimes | 1,-1 \rangle \langle 1, -1| \right]  \\
P^{\epsilon}_y & = & \frac{1}{2} \left[ \chi_{\left[ y - \epsilon, y + \epsilon \right]} \otimes |i,1 \rangle \langle i,1 |  + \chi_{\left[ - y - \epsilon , - y + \epsilon \right]} \otimes | -i,1 \rangle \langle -i, 1| \right] \\
P^{\epsilon}_z & = & \frac{1}{2} \left[ \chi_{\left[ z - \epsilon, z + \epsilon \right]} \otimes |0,1 \rangle \langle 0,1 |  + \chi_{\left[ - z - \epsilon , - z + \epsilon \right]} \otimes | 1,0 \rangle \langle 1, 0| \right].   
\end{eqnarray*}
Hence, the operators $P^{\epsilon}_{(t,x,y,z)}$ vanish as soon as at least \emph{two} of the spatial coordinates have modulus greater
 or equal to $\epsilon$.  Therefore, if one is far away in two coordinates from the origin, one sees nothing except on the scales of 
the distances to the orgin itself.  If only one coordinate, say $z$ has a modulus greater than $\epsilon$, then the projection operator
 is given by 
\begin{eqnarray*}
P^{\epsilon}_{(t,x,y,z)} &=& \half\, \chi{\vphantom{\big|}}_{\left[ |x| - \epsilon, - |x| + \epsilon \right] \times \left[ |y| - \epsilon, - |y| + \epsilon \right] \times \left[ z - \epsilon, z + \epsilon \right]} \otimes |0,1 \rangle \langle 0,1 |  + \\
& & \half\, \chi{\vphantom{\big|}}_{\left[ |x| - \epsilon, - |x| + \epsilon \right] \times \left[ |y| - \epsilon, - |y| + \epsilon \right] \times \left[ - z - \epsilon , - z + \epsilon \right]} \otimes | 1,0 \rangle \langle 1, 0|\;,
\end{eqnarray*}
and the reader is invited to work out the projection operator for a case in which all spatial coordinates have a modulus smaller
 than $\epsilon$.   Therefore, one obtains an axial structure where any of the coordinate axes (added with time) are priviliged as well as the origin; obviously, these correspond to abelian subalgebra's. \\* \\*
In case $U(x) \neq 1$, several interesting structures may emerge where locally multiple foldings arise (corresponding to many $x$ regions); 
this can be seen as follows.  For $U(x) = e^{i\vec{a}(x).\vec{\sigma}}$ one has that $U(x)\sigma^i U(x)^{\dag}$ induces a rotation of an 
angle $2 ||\vec{a}|| \textrm{mod}
2 \pi$ around the vector $\vec{a}$.  These can be computed exactly, as well as can the eigenvectors (although they are rather ugly 
functions of $\vec{a}$) and the latter are all of the type $v(x)\delta^n(y-x)$ where $v(x) \in \mathbb{C}^{2}$.  The reader may well 
have noticed that we still have to say something about dimension since dimensional collapse is possible; 
indeed any real $n$ dimensional manifold is a $m$ dimensional noncommutative one if and only if $m \geq n$.  On the other hand, 
discrete manifolds do not necessarily have a one dimensional representation due to the algebraic relations 
(so we have some kind of entanglement dimension).
  Therefore, one might be tempted to declare the dimension of a manifold to be 
the minimal one; it is for now a matter of taste whether one allows for collapse or not and we leave this to the discretion of 
the reader.
\section{Canonical Differentiable Structure}
Before we define a differential structure, we have to identify the natural class of functions on a local chart $(V_{\beta}, \pi_{\beta}, \mathcal{A}_{\beta}, \phi_{\beta})$.  The thing is that points and functions are simply unified in the algebraic context; they just are elements of $\mathcal{A}_{\beta}$.  Indeed, a function is nothing than some limit of a finite polynomial in the $\pi_{\beta}(\widehat{x}_i)$ and the natural question is how we should define the function on an open set $\mathcal{W} \subset \mathcal{V}_{\beta}$.  There are two natural candidates for local functions which we call the entangled and unentangled one for obvious reasons.  The former forgets how an element $A \in \mathcal{A}_{\beta}$ arises from the fundamental building blocks and maps $A \rightarrow \widehat{A}$, where the latter is defined as
$$
\widehat{A}(\mathcal{W}) = P_{\beta}(\mathcal{W})A P_{\beta}(\mathcal{W})\;,
$$
and obviously $\widehat{A}$ maps distinct regions to orthogonal operators; moreover, $\widehat{A}$ preserves the order relation in the sense that
$$
\widehat{\widehat{A}(\mathcal{W}_2)}(\mathcal{W}_1) = \widehat{A}(\mathcal{W}_1)
$$
for $\mathcal{W}_1 \subset \mathcal{W}_2$.  However, this transformation does not erase entanglement with regions outside $\mathcal{W}$ as the reader may easily verify and obviously, this ansatz is not a suitable candidate for defining a differential since it does not ``feel'' the order in which the elementary variables occur.  Let us start with finite polynomials in unity and the preferred variables $\pi_{\beta}(\widehat{x}_i)$, then one meets a rarity which might seem to be a lethal problem at first sight but really is nothing but a manifestation of what breaking of entanglement means.  That is let 
$A = Q(1,\pi_{\beta}(\widehat{x}_i))$, where $Q$ is some polynomial of finite degree, then we define 
$$
\widehat{Q}(\mathcal{W}) = Q(P_{\beta}(\mathcal{W}), P_{\beta}(\mathcal{W}) \pi_{\beta}(\widehat{x}_i) P_{\beta}(\mathcal{W}))
$$
as the local unentangled realization of $Q$.  Now, it is possible for two polynomials $Q_1$ and $Q_2$ to determine identical elements in $\mathcal{A}_{\beta}$, but the local realizations
$\widehat{Q}_j$ differ; also, the reader is invited to construct some examples on this.  All this implies that we have to define nets of polynomials and declare equivalence with respect to the resolution one is measuring which removes the absolutism from $\mathcal{A}_{\beta}$; that is,
$$
\widehat{Q}_1 \sim_{\mathcal{W}} \widehat{Q}_2
$$
if and only if $\widehat{Q}_1(\mathcal{W}) = \widehat{Q}_2(\mathcal{W})$.  One verifies moreover that the local unentangled $\widehat{A}$ has the same inclusion and disjoint properties than the entangled one.  Therefore, consider a natural directed net $(Q_i, i \in \mathbb{N})$ of finite polynomials in the fundamental variables $\widehat{x}_i$ and unity, then we say that the domain $\textrm{Dom}((Q_i, i \in \mathbb{N}), (\pi_{\beta}, \mathcal{A}_{\beta}))$ of this net relative to the chart $(\pi_{\beta}, \mathcal{A}_{\beta})$ is given by the set of relative opens $\mathcal{W} \subset \mathcal{O}(\pi_{\beta}, \mathcal{A}_{\beta})$ so that $\widehat{Q}_i(\mathcal{W})$ is a weakly convergent series of operators.  For the general reader, the weak topology on a $W^{\star}$ algebra is the locally convex topology generated by the continuous complex linear functionals $\psi_{\beta} : \mathcal{A}_{\beta} \rightarrow \mathbb{C}$.  Now in order to define continuity and differentiability of such functions, we need to equip the relative open sets with a canonical topology, that is the Vietoris topology which is defined by the relative open subsets $(\mathcal{O}, \mathcal{V})(\mathcal{W})$ where $\overline{\mathcal{V}} \subset \mathcal{W} \subset \overline{\mathcal{W}} \subset \mathcal{O}$ and $(\mathcal{O}, \mathcal{V})(\mathcal{W})$ is the set of all open sets $\mathcal{Z}$ satisfying $\mathcal{V} \subset \mathcal{Z} \subset \mathcal{O}$.
\newtheorem{theo}{Definition} 
\begin{theo} Therefore, the net $(Q_i, i \in \mathbb{N})$ is of bounded variation relative to $(\pi_{\beta}, \mathcal{A}_{\beta})$ in $\mathcal{W} \in \textrm{Dom}((Q_i, i \in \mathbb{N}), (\pi_{\beta}, \mathcal{A}_{\beta}))$ if and only if for every $\epsilon > 0$ and continuous functional $\psi_{\beta}$, there exists an open set containing $\mathcal{W}$ such that for any open $\mathcal{Z}$ contained in it we have that
$$
|\psi_{\beta}((\widehat{Q}_i, i \in \mathbb{N})(\mathcal{W}) - (\widehat{Q}_i, i \in \mathbb{N})(\mathcal{Z}))| < \epsilon\;.
$$ \end{theo}
In order to define directional continuity, partial differential operators and finite difference operators, we need the notion of directional displacement.  Therefore, let $\vec{e}$ be a unit vector in $\mathbb{R}^n$ and $\delta$; then the translation $T_{(\delta \vec{e})}$ canonically lifts as a continuous map to the space of all open sets by the prescription
$$
T_{(\delta \vec{e})}(\mathcal{W}) = \mathcal{W} + \delta \vec{e}\;.
$$
We need also need to lift the translations to homomorphisms between the local algebras $\mathcal{A}^{loc}_{\beta}(\mathcal{W})$ which requires the use of a quantum connection.  Here, $\mathcal{A}^{loc}_{\beta}(\mathcal{W})$ is the $W^{\star}$ subalgebra of $\mathcal{A}_{\beta}$ generated by $P_{\beta}(\mathcal{W})\pi_{\beta}(\widehat{x}_i)P_{\beta}(\mathcal{W})$ and $P_{\beta}(\mathcal{W})$ which is not the same as $P_{\beta}(\mathcal{W})\mathcal{A}_{\beta} P_{\beta}(\mathcal{W})$ (which is also a Von Neumann algebra) as explained before.  The reason why we need a connection is because at some resolution $\epsilon$, $P_{\beta}(\mathcal{W})$ will not majorize, nor commute with the $P^i((\alpha_i - \epsilon, \alpha_i + \epsilon))$ so that the projection operators will not be projection operators anymore but twisted depending upon the region $\mathcal{W}$ and spectral operator at hand.  This does of course not happen in the abelian case where everything remains trivial.  Also, it is generally not so that for $\mathcal{V} \subset \mathcal{W}$ one obtains that $$\mathcal{A}^{loc}_{\beta}(\mathcal{V}) \subset \mathcal{A}^{loc}_{\beta}(\mathcal{W})$$  and the reason is that fine grained projections can add a twist where coarser grained projections do not.  Of course, this inclusion property does hold when we do not cut entanglement, that is
$$P_{\beta}(\mathcal{V}) \mathcal{A}_{\beta} P_{\beta}(\mathcal{V}) \subset P_{\beta}(\mathcal{W}) \mathcal{A}_{\beta} P_{\beta}(\mathcal{W})$$ for $\mathcal{V} \subset \mathcal{W}$.  Let us give some example confirming these facts, consider the following discrete four dimensional quantum manifold
\begin{eqnarray*}
t & = & \left( \begin{array}{cc}
0 & 1 \\*
1 & 0 \\*	
\end{array} \right) \\
x & = & \left( \begin{array}{cc}
0 & \sigma_1 \\*
\sigma_1 & 0 \\*	
\end{array} \right) \\
y & = & \left( \begin{array}{cc}
0 & \sigma_2 \\*
\sigma_2 & 0 \\*	
\end{array} \right) \\
z & = & \left( \begin{array}{cccc}
2 & 0 & 0 & 0 \\*
0 & 1 & 0 & 0 \\*
0 & 0 & 1 & 0 \\*
0 & 0 & 0 & 0	\\*
\end{array} \right).
\end{eqnarray*} A little algebra reveals that $\left[ t, x \right] = \left[t, y \right] = 0$, $\{ x, y \} = 0$ and 
$t^2 = x^2 = y^2 = 1$.  Also, one notices that $y$ and $z$ do not commute nor anticommute.  The spectrum of $t,x,y$ is $\{ -1, 1 \}$ and both eigenspaces have dimension two; for $z$ it clearly is $\{0,1,2\}$ and therefore the cube consists out of $24$ points.  Associate $\mathcal{V}$ to that subset of the cube with arbitrary values for $t,x$ and $y = 1 = z$ and $\mathcal{W}$ to arbitrary values for $t,x,z$ and $y = 1$, then clearly $\mathcal{V} \subset \mathcal{W}$.  One computes that
\begin{eqnarray*}
P(\mathcal{V}) & = & \frac{1}{2}
\left( \begin{array}{cccc}
0 & 0 & 0 & 0 \\*
0 & 1 & - i & 0 \\*
0 & i & 1 & 0 \\*
0 & 0 & 0 & 0 \\*	
\end{array} \right) 
\end{eqnarray*} 
and $P(\mathcal{W}) = \frac{1}{2} \left( 1 + y  \right)$.  We compute $\mathcal{A}^{loc}(\mathcal{W})$ and show that $P(\mathcal{V})$ does not belong to it.  Elementary algebra shows that
\begin{eqnarray*}
P(\mathcal{W})t P(\mathcal{W}) & = & \frac{1}{2} \left( \begin{array}{cc}
\sigma_2 & 1 \\*
1 & \sigma_2 \\*
\end{array} \right) \\
P(\mathcal{W})x P(\mathcal{W}) & = & 0 \\
P(\mathcal{W})y P(\mathcal{W}) & = & P(\mathcal{W}) \\
P(\mathcal{W})z P(\mathcal{W}) & = & P(\mathcal{W}) \\
\end{eqnarray*} even though $P(\mathcal{W})$ does not commute with $z$.  It is now easy to show that $\mathcal{A}^{loc}(\mathcal{W})$ is two dimensional and that $P(\mathcal{V})$ is not in it.  Finally, we compute the dimension of $P(\mathcal{W}) \mathcal{A} P(\mathcal{W})$; the latter is four as can be easily seen by starting from the expression
\begin{eqnarray*} \frac{3}{2}z - \frac{1}{2}z^2 & = & \left( \begin{array}{cccc}
1 & 0 & 0 & 0\\*
0 & 1 & 0 & 0 \\*
0 & 0 & 1 & 0 \\*
0 & 0 & 0 & 0 \\*
\end{array} \right) = \alpha \end{eqnarray*}
and notice that $P(\mathcal{W}) \alpha P(\mathcal{W}) - \frac{1}{2} P(\mathcal{W}) \sim P(\mathcal{V})$.    
 \\* \\*
From the weak continuity of $\kappa$ ``almost everywhere'' one deduces that the local algebra's $\mathcal{A}^{loc}(\mathcal{W})$ almost never jump when we move $\mathcal{W}$ around.  Therefore, what one could call quasilocal algebra's are basically the same as the local ones.  Hence, we define a connection, or parallel transport, $\Gamma_{\beta}(\mathcal{V}, \mathcal{W})$ as a bifunction of two relatively open sets which map to a star homomorphism between the respective local algebras; that is,
$$\Gamma_{\beta}(\mathcal{V}, \mathcal{W}) : \mathcal{A}^{loc}_{\beta}(\mathcal{V}) \rightarrow \mathcal{A}^{loc}_{\beta}(\mathcal{W})$$ where a path dependence is possible in the composition and we could at most look for rules of intersection and inclusion.  For $\mathcal{V} \subset \mathcal{W}$, one has that when a spectral projector $P \prec P(\mathcal{V})$ or $P(\mathcal{V}) P P(\mathcal{V}) = P$ then the same is true for $P(\mathcal{W})$ and we demand $\Gamma(\mathcal{V}, \mathcal{W})$ to preserve these fixpoints.  Other principles of this kind are not possible, it might be that $P$ commutes with $P(\mathcal{V})$ but not with $P(\mathcal{W})$ and vice versa.  We might still ask however for the connection to be optimal which means that the homomorphisms cannot have a smaller kernel.  Therefore, in case the local algebra's are isomorphic, $\Gamma(\mathcal{V}, \mathcal{W})$ is an isomorphism too.  Also, we demand the connection to be unital, meaning that $\Gamma(\mathcal{W}, \mathcal{W})$ is equal to the identity.  There will be two further requirements on the connection which is that the basic functions $\pi_{\beta}(\widehat{x}_i)$ are weakly continuous or differentiable wherever $\kappa_{(\pi_{\beta},\mathcal{A}_{\beta})}$ is in all or some directions $\vec{e}$.  The latter is a huge constraining between the analytical and $W^{\star}$ algebraic aspects of $\mathcal{A}_{\beta}$. \\* \\* 
We have two different notions of continuity and differentiability because $\kappa_{(\pi_{\beta}, \mathcal{A}_{\beta})}$ has a peculiar and natural status within our construction.  First of all, we say that $\kappa_{(\pi_{\beta}, \mathcal{A}_{\beta})}$ is weakly continuous in a point $\mathcal{W}$ in the Vietoris topology when for all $\epsilon > 0$ and continuous functionals $\psi_{\beta}$, there exists an open neighborhood $\mathcal{O}$ in the Vietoris topology such that for any $\mathcal{Z} \in \mathcal{O}$ we have that
$$| \psi_{\beta} \left( \kappa_{(\pi_{\beta}, \mathcal{A}_{\beta})}(\mathcal{W}) - \kappa_{(\pi_{\beta}, \mathcal{A}_{\beta})}(\mathcal{Z}) \right) | < \epsilon.$$  Likewise, we say that $\kappa_{(\pi_{\beta}, \mathcal{A}_{\beta})}$ is continuous in the direction $\vec{e}$ at $\mathcal{W}$ when for any $\epsilon > 0$ and $\psi_{\beta}$, there exists a $\delta > 0$ so that for any $|h| < \delta$ 
$$|\psi_{\beta} \left( \kappa_{(\pi_{\beta}, \mathcal{A}_{\beta})}(\mathcal{W}) - \kappa_{(\pi_{\beta}, \mathcal{A}_{\beta})}(T_{(h\vec{e})}(\mathcal{W})) \right) | < \epsilon.$$
Concerning the notion of weak differentiability of $\kappa_{(\pi_{\beta}, \mathcal{A}_{\beta})}$, there exist several and we have to find out if some of them are equivalent or not.  Let me first start by examining the abelian case in sufficient detail and then generalize to the nonabelian setting.  In the Schrodinger like setting explained before, the projection operators are just characteristic functions and in one dimension, the computations simplify considerably (however, there is no problem generalizing this to higher dimensions as the reader may try to do) while the results are universal.  Naively, one would think we have to calculate the limit of
$$\frac{1}{\delta} \left( \chi_{(a + \delta, b + \delta)} - \chi_{(a,b)} \right)$$ for $0 < \delta \rightarrow 0$.  If one would restrict to the continuous functions as a separating \emph{subalgebra} of the $L^{2}$ functions (at least on a compact measure space), then this limit exists in the weak sense and it is $\delta(b) - \delta(a)$ which is outside the algebra since it is not well defined on the whole Hilbert space.  Now, if again, we would only restrict to the continuous functions, then the limit  
$$\frac{1}{\delta^{1 - \gamma}} \left( \chi_{(a + \delta, b + \delta)} - \chi_{(a,b)} \right)$$ is zero and independent of $\gamma > 0$.  However, if one were to go over to the full Hilbert space, then it is necessary and sufficient that $\gamma > \frac{1}{2}$ in which case the limit is also zero.  Therefore, we say that $\kappa_{(\pi_{\beta}, \mathcal{A}_{\beta})}$ is $\gamma$-weakly differentiable with respect to a separating\footnote{Separating means that for all distinct $A,B \in \mathcal{A}_{\beta}$ there exists a $\psi_{\beta} \in \Psi_{\beta}(\gamma)$ such that $\psi_{\beta}(A) \neq \psi_{\beta}(B)$.} subset $\Psi_{\beta}(\gamma)$ of continuous functionals in the direction $\vec{e}$ at $\mathcal{W}$ if there exists an element $\partial^{\gamma}_{\vec{e}} \, \kappa_{(\pi_{\beta}, \mathcal{A}_{\beta})}(\mathcal{W})$  such that for all $\epsilon > 0$ and $\psi_{\beta} \in \Psi_{\beta}(\gamma)$, there exists a $\delta > 0$ such that for all $0 < h < \delta$ we have that 
$$\left|\psi_{\beta} \left( \frac{1}{h^{1 - \gamma}} \left( \kappa_{(\pi_{\beta}, \mathcal{A}_{\beta})}(T_{(h\vec{e})}(\mathcal{W})) - \kappa_{(\pi_{\beta}, \mathcal{A}_{\beta})}(\mathcal{W}) \right) - \partial^{\gamma}_{\vec{e}} \, \kappa_{(\pi_{\beta}, \mathcal{A}_{\beta})}(\mathcal{W}) \right)  \right| <  \epsilon.$$
Similarly, one could forget about $\Psi_{\beta}(\gamma)$ and demand that $\gamma > \frac{1}{2}$.  This attitude could lead to very different algebra's and we will not even start its investigation in this short paper.  An obvious property is that if $\kappa_{(\pi_{\beta}, \mathcal{A}_{\beta})}$ is differentiable with respect to $(\gamma_1, \Psi_{\beta}(\gamma_1))$, then it is also the case for $(\gamma_2, \Psi_{\beta}(\gamma_1))$ where $\gamma_2 > \gamma_1$ and the differential is exactly zero. \\* \\*
We now turn to continuity and differentiability of nets $(\widehat{Q}_i , i \in \mathbb{N})$ of finite polynomials on their relative domain (with respect to $(\pi_{\beta}, \mathcal{A}_{\beta})$).  Define now $$\widehat{T}_{(\delta \vec{e})}(\mathcal{W}) = \Gamma(\mathcal{W}, T_{(\delta \vec{e})}(\mathcal{W}))$$ then we say that $(\widehat{Q}_i , i \in \mathbb{N})$ differentiable at $\mathcal{W}$ in the interior of its relative domain in the direction of $\vec{e}$ if and only if there exists a unique element $$\partial_{\vec{e}} \,(\widehat{Q}_i, i \in \mathbb{N})(\mathcal{W}) \in \mathcal{A}^{loc}(\mathcal{W})$$ such that for any $\psi_{\beta}$, 
\begin{eqnarray}
& &\psi_{\beta} \left( \partial_{\vec{e}} \, (\widehat{Q}_i, i \in \mathbb{N})(\mathcal{W}) \right) \nonumber\\
& &=\ \lim_{\delta \rightarrow 0}  \frac{1}{\delta} \, \psi_{\beta} \left( \widehat{T}_{- \delta \vec{e}} \left[(\widehat{Q}_i, i \in \mathbb{N})(T_{(\delta \vec{e})}(\mathcal{W})) \right] - (\widehat{Q}_i, i \in \mathbb{N})(\mathcal{W}) \right). \nonumber
\end{eqnarray}
So, the differential operator is only defined if some translates of $\mathcal{W}$ belong to the relative domain of $(\widehat{Q}_i, i \in \mathbb{N})$ for arbitrarily small $\delta$.  Therefore, partial differentials are not defined for directions in which the set at hand is isolated.  Of course, if one looks only at larger scales, then jumps may be accomplished and the difference operators are canonically defined.  One could also resort here to notions of $(\gamma, \Psi_{\beta}(\gamma))$ differentiability, but I see no stringent need to do it at this point. \\* \\* 
Before we give some examples, let us proceed by defining the holonomy groups attached to the connection;  for any $\mathcal{W}$, we define $H(\mathcal{W})$ as the group of homomorphisms from $\mathcal{A}^{loc}(\mathcal{W})$ to itself generated by finite compositions of the kind $$\Gamma(\mathcal{W}_n, \mathcal{W}) \Gamma(\mathcal{W}_{n-1}, \mathcal{W}_n) \ldots \Gamma(\mathcal{W}_{1}, \mathcal{W}_{2})\Gamma(\mathcal{W}, \mathcal{W}_1).$$
We say that a connection is flat when all the holonomy groups are equal to the identity.  Consider as before the trivial example of a real $n$ dimensional manifold, then the translation mappings induce a canonical flat connection on the pairs of opens differing by a translate as follows: every spectral operator $P((\alpha_i - \epsilon, \alpha_i + \epsilon) \cap \mathcal{W}) = P(\mathcal{W}) P^{i}((\alpha_i - \epsilon, \alpha_i + \epsilon))P(\mathcal{W})$ gets mapped to $$P((\alpha_i - \epsilon + \delta e_i, \alpha_i + \epsilon + \delta e_i) \cap T_{(\delta \vec{e})}(\mathcal{W}))$$ provided $\mathcal{W}$ and $\mathcal{W} + \delta \vec{e}$ belong to the cube.  Actually, this is all we need to calculate differentials and so on, but the reader might wish to extend this definition in a canonical way to generic pairs.  For $\mathcal{W}$ of compact closure and real differentiable function $f$ (with uniformly continuous partial derivatives) with $\overline{\mathcal{W}} \subset \textrm{Dom}(f)$ one associates a unique algebra element $\widetilde{f}$ (in the commutative case we do not need the nets).  It is easy to calculate that the new differential
$$
\partial_{\vec{e}} \, \widehat{\widetilde{f}}(\mathcal{W}) = \widetilde{\partial_{\vec{e}} \, f \, \chi_{\mathcal{W}}}\;,
$$
reduces to the old one and that the latter even exists in the norm topology in this case\footnote{The reader is invited to fill in all intermediate steps.}.  \\* \\*
All these results allow us now to obtain a better insight into the nature of noncommutative $n$ dimensional manifolds.  Before we engage in this discussion we still need to solve some questions:  
\begin{itemize} 
\item We have demanded that for overlapping charts the algebra's of local projection operators (with respect to these charts) are isomorphic; how does this algebra relate to the local algebra with respect to that chart?
\item We have seen that for $\mathcal{V} \subset \mathcal{W}$, it does not necessarily hold that $\mathcal{A}^{loc}(\mathcal{V}) \subset \mathcal{A}^{loc}(\mathcal{W})$.  However, does there exist an isomorphism of $\mathcal{A}^{loc}(\mathcal{V})$ into a subalgebra of $\mathcal{A}^{loc}(\mathcal{W})$ ?
\item  Finally, say that $\mathcal{W}$ contains $r$ components with respect to $\mathcal{V}_{\beta}$; does the spectrum of the local algebra $\mathcal{A}^{loc}(\mathcal{W})$ contain at least $r$ components ?  
\end{itemize} 
As a response to the first question, we already know that the algebra of local projection operators is not necessarily contained in the local algebra and the question is whether the inverse holds.  But before we treat these questions in generality, let us see how they are answered in the our previous example.  Concerning the first question, we notice that the only nonzero projection operators (apart from $P(\mathcal{V})$ and $P(\mathcal{W})$) arise from $y = 1$ and $t = \pm 1$; they are given by
\begin{eqnarray*}
P(t = 1 = y) & = & \frac{1}{4}
\left( \begin{array}{cccc}
1 & i & 1 & i \\*
- i & 1 & - i & 1 \\*
1 & i & 1 & i \\*
-i & 1 & -i & 1	
\end{array} \right) \\
P(t = -1 = -y) & = & \frac{1}{4} \left(
\begin{array}{cccc}
1 & -i & -1 & i \\*
i & 1 & - i & -1 \\*
-1 & i & 1 & -i \\*
-i & -1 & i & 1	
\end{array} \right).
\end{eqnarray*}
It is most easily seen that $P(\mathcal{W})tP(\mathcal{W}) = 2 P(t = 1 = y) - P(\mathcal{W})$ which shows that $\mathcal{A}^{loc}(\mathcal{W})$ is a subalgebra of the algebra generated by the local projection operators $P(\mathcal{V})$ with $\mathcal{V} \subseteq \mathcal{W}$.  The second question is answered in the \emph{negative}  since $\mathcal{A}^{loc}(\mathcal{V})$ is generated by $P(\mathcal{V})$ and
\begin{eqnarray*}
P(\mathcal{V}) t P(\mathcal{V}) & = & \frac{1}{2} \left(
\begin{array}{cccc}
1 & 0 & 0 & i \\*
0 & 0 & 0 & 0 \\*
0 & 0 & 0 & 0 \\*
- i & 0 & 0 & 1	
\end{array} \right) 
\end{eqnarray*} and it is easy to verify that this algebra is not isomorphic to $\mathcal{A}^{loc}(\mathcal{W})$.
Therefore, the answer to the second question is inconclusive since in the commutative case $\mathcal{A}^{loc}(\mathcal{V}) \subseteq \mathcal{A}^{loc}(\mathcal{W})$.  Regarding the third issue, $\mathcal{W}$ contains $12$ points and the cube of $\mathcal{A}^{loc}(\mathcal{W})$ contains also $12$ of them\footnote{One calculates that the spectrum of $P(\mathcal{W})tP(\mathcal{W})$ is $\{ - 2, 2, 0\}$ and the projection operator on the zero eigenvalue is $\frac{1}{2}( 1 - y)$.}.  However, all projection operators vanish in the former case while in the latter exactly $3$ of them are nonzero.  Therefore, the question appears to hold on the ontological as well as the empirical level. \\* \\*  Let us start with some mathematical preliminaries.
\newtheorem{theor}{Theorem} 
\begin{theor} Let $P$ and $Q$ be two (noncommuting) Hermitian projection operators then the projection operators $P \wedge Q$ and $P \vee Q$ belong to $\mathcal{M}' \cap \mathcal{M}$, where $\mathcal{M}'$ is the commutant in $\mathcal{A}_{\beta}$ of the Von Neumann algebra $\mathcal{M}$ generated by $P$ and $Q$.  In particular, any Hermitian projection operator which is smaller than $P \wedge Q$ or larger than $P \vee Q$ belongs to $\mathcal{M}'$.
\end{theor} 
\textit{Proof} : Represent $P$ and $Q$ on a Hilbert space $\mathcal{H}$ and consider the smallest closed subspace $\mathcal{H}'$ which is left invariant by both of them.  Then this $\mathcal{H}'$ has $P \vee Q$ as identity operator and we have to show that it is generated by $P$ and $Q$.
For the intersection, the proof is easy: $\frac{1}{2} \left( PQ + QP \right) = P \wedge Q + A$ where $(P \wedge Q)A = 0$, $A^{\star} = A$, $|| A || \leq 1$ but $1$ does not belong to the discrete spectrum, and therefore
$$P \wedge Q = \lim_{n \rightarrow \infty} \left( \frac{1}{2} (PQ + QP) \right)^{n}$$ in the weak sense.  Replacing $Q$ by $Q' = Q - PQ - QP + PQP$, we see that it is zero if and only if $Q = P$; moreover, $PQ' = Q'P = 0$ and $Q'$ as a mapping from $(1 - P)\mathcal{H}'$ to $(1 - P)\mathcal{H}'$ does not contain $0$ in its discrete spectrum.  Otherwise, there would exist a vector $v \in (1-P)\mathcal{H}'$ such that $(1 - P)Qv = 0$ or $Qv = PQv$ which is impossible unless $v$ is in the intersection of both hyperspaces which implies it must be the zero vector.  In the finite dimensional case, it easy to construct polynomials $f_{\alpha}(x)$ with $f_{\alpha}(0) = 0$ such that $$f_{\alpha}(Q') = P_{\alpha}$$ where $\alpha \in \sigma(Q')$ and $P_{\alpha}$ is its spectral operator.  Therefore, one can recuperate the identity $P \vee Q - P$ on $(1 - P)\mathcal{H}'$ in the algebra of $Q'$ only.  In the infinite dimensional case, this technique fails since the polynomials will start to oscillate heavily which has a detrimental effect on the continuous spectrum.   However, if one considers the algebra generated by $1,P,Q$ a similar argument holds due to the Stone Weierstrass and spectral theorem. \\* \\*
Concerning the first question, let us elaborate on whether given a cube $P_1, P_2$ where $P_1 = P + Q$ with $PQ = 0$ and corresponding to distinct discrete eigenvalues, it is true that
$$(P_1 \wedge P_2) P (P_1 \wedge P_2) = \alpha P_1 \wedge P_2 + (1 - \alpha)P \wedge P_2 - \alpha Q \wedge P_2$$
for some $\alpha \in \mathbb{R}$ (actually the reader can check that any linear combination of these operators has to be of this form).  It is easily seen that this statement is false, since consider the orthonormal unit vectors $e_i$, $i : 1 \ldots 5$, and the following subspaces:
\begin{eqnarray*}
\mathcal{P} & = & \textrm{Span} \{ \cos (\theta) e_1 + \sin (\theta) e_2, \cos (\psi)e_3 + \sin (\psi) e_4  \} \\*
\mathcal{Q} & = & \textrm{Span} \{ \sin (\theta)e_1 - \cos (\theta) e_2 , \sin (\psi)e_3 - \cos(\psi) e_4 \} \\*
\mathcal{P}_2 & = & \textrm{Span} \{ e_2, e_3 , e_5 \}.
\end{eqnarray*} Then, one has the following identities: 
\begin{eqnarray*}
P_1 \wedge P_2 & = & | e_2 \rangle \langle e_2 | + | e_3 \rangle \langle e_3 | \\*
P \wedge P_2 & = & 0 \\*
Q \wedge P_2 & = & 0 \\*
PQ & = & 0.
\end{eqnarray*} However, one easily calculates that
\begin{eqnarray*}
(P_1 \wedge P_2) P (P_1 \wedge P_2)  & = & \sin^2(\theta) |e_2 \rangle \langle e_2 | + \cos^2(\psi) | e_3 \rangle \langle e_3 |
\end{eqnarray*} which is not a multiple of $P_1 \wedge P_2$.  Therefore, one has that $P(\mathcal{W}) P^i P(\mathcal{W})$ is in general not in the algebra generated by $P(\mathcal{V})$ where $\mathcal{V} \subseteq \mathcal{W}$.  It is now easy to pick $\pi_{\beta}(\widehat{x}_i) = P + \mu R$ where $R = |e_5 \rangle \langle e_5|$ to conclude that $$(P_1 \wedge P_2) \pi_{\beta}(\widehat{x}_i) (P_1 \wedge P_2)$$ is not in the algebra generated by the $P(\mathcal{V})$.  This shows that $\mathcal{A}^{loc}(\mathcal{W})$ and the $W^{\star}$ algebra  $\mathcal{A}^{open}(\mathcal{W})$ generated by the $P(\mathcal{V})$ where $\mathcal{V} \subseteq \mathcal{W}$ have no relation to one and another.
\begin{theo} We call the chart $(\mathcal{V}_{\beta}, \pi_{\beta}, \mathcal{A}_{\beta} , \phi_{\beta})$ \emph{pointed} when for all $\mathcal{W}$, $$\mathcal{A}^{loc}(\mathcal{W}) \subseteq \mathcal{A}^{open}(\mathcal{W}).$$
\end{theo}  
We now proceed to answer the third question which intuitively means that if you zoom in you see more and more disconnected components.  Now, it is obvious that this property does not even hold in the commutative case where on large scales one may see many isles but on small scales all one sees is one of them.  However, a refinement of the question is nevertheless interesting and one might want to look for manifolds which have only one component on a given scale and where the number of components grows polynomially (or even exponentially) in the inverse scaling $\frac{1}{\lambda}$.  \\*  \\*
We now have obtained a better view on how we should do function theory on a noncommutative topological manifold although we are confronted with an apparent dilemma.  On one side $\mathcal{A}^{open}_{\beta}(\mathcal{W})$ is the natural algebra we should use to compare overlapping charts, but $\mathcal{A}^{loc}_{\beta}(\mathcal{W})$ is the natural algebra for function theory.  What we learned is that they have generically little to do with one and another; therefore, this begs the question of how to even define \emph{algebraic} functions on the entire manifold.  It is here that the (trace) functionals $\omega_{\mathcal{A}_{\beta}}$ come into play in the following sense: let $\mathcal{M}$ be a noncommutative manifold, then $F: \tau(\mathcal{M}) \rightarrow \mathbb{C}$, where $\tau(\mathcal{M})$ is the set of open subsets of $\mathcal{M}$ equipped with the Vietoris topology, is an algebraic function if and only if for any chart $(\mathcal{V}_{\beta}, \pi_{\beta}, \mathcal{A}_{\beta} , \phi_{\beta})$, there exists a net of polynomials $(Q^{\beta}_i , i \in \mathbb{N})$ such that $$F(\mathcal{W}) = \omega_{\mathcal{A}_{\beta}} \left(  \left( \widehat{Q}^{\beta}_{i} , i \in \mathbb{N} \right)(\mathcal{W}) \right).$$  Continuity of $F$ is obviously defined with respect to the Vietoris topology.  We call $F$ nuclear if and only if for any $\mathcal{V}, \mathcal{W}$, one has that 
$$F(\mathcal{V} \cup \mathcal{W}) = F(\mathcal{W}) + F(\mathcal{V}) - F(\mathcal{V} \cap \mathcal{W}).$$  
Obviously, the standard continuous functions on a real $n$ dimensional manifold with a volume element induce nuclear continuous functions
 by putting the trace functional equal to the $n$ dimensional integral.  We can define higher order algebraic functions as follows 
$$F(\mathcal{W}, \mathcal{V}_1, \ldots \mathcal{V}_m) = \omega_{\mathcal{A}_{\beta}} \left( P_{\beta}(\mathcal{V}_1 ) \ldots P_{\beta}(\mathcal{V}_m) \left( \widehat{Q}^{\beta}_i , i \in \mathbb{N} \right)(\mathcal{W}) \right)$$ where
$\mathcal{V}_j  \subset \mathcal{W}$.  
The gluing conditions ensure us that the identity element in $\mathcal{F}_n$ canonically defines a set of (higher order) algebraic 
functions.  One could now study trace abelian local representations of algebraic functions $F$; more specifically, consider any two overlapping
charts $(\mathcal{V}_{\beta_i},\pi_{\beta_i}, \mathcal{A}_{\beta_i},\phi_{\beta_i})$ and consider any open $\mathcal{W} \subset 
\mathcal{V}_{\beta_1} \cap \mathcal{V}_{\beta_2}$.  We demand there to exist a matrix valued function $h^{k}_{l}(\beta_{1},\beta_{2})$
 on the topology such that 
$$\omega_{\beta_1} \left( \widehat{\partial}_{k} \left( \widehat{Q}^{1}_{j} ; j \in \mathbb{N} \right)(\mathcal{W}) \right) = 
\omega_{\beta_2}  \left( \int_{\phi_{\beta_1}(\mathcal{W})} \frac{\partial^l (\phi_{\beta_2} \circ \phi_{\beta_1}^{-1})(x)}{ \partial x^k}
h^{r}_{l}(x) \widehat{\partial}_{r} \left( \widehat{Q}^{2}_{j}; j \in \mathbb{N} \right)(x) \right) $$
where the integral is understood to be taken in some ordered sense by evaluating the functions on (almost everywhere) partitions by open subsets.  For 
standard abelian manifolds and nuclear functions, the standard matrix $h^{k}_{l}$ is given by
$$h^{k}_{l}(\mathcal{W}) = \delta^{k}_{l} \frac{1}{\textrm{Vol}(\mathcal{W})}.$$   
Let us finish by commenting upon the very act of pasting together ``algebraic charts''. 
  We have learned two ways of cutting entanglement, which was by going over to local and open $W^{\star}$ algebra's associated to open 
subsets of $\mathcal{M}$; also, the $W^{\star}$ algebraic framework forces us in the cauldron of relatively open subsets of
 $\mathbb{R}^n$.  This implies that in order to generate a nontrivial topology (with respect to a continuum background) some sort of 
``decoherence'' has to occur.  Indeed, saying that two charts are described by separate $W^{\star}$ algebra's really means that the 
points in both charts do not ``entangle'' in some sense.  Whether or not this is a desirable conclusion remains to be seen.  
\section{Realizations of non-commutative continuum manifolds.}
All our nontrivial examples in the previous section were discrete and it is desirable to look for continuum representations.  The reason we did not study these so far in detail is that it requires nontrivial sets of operators; as is well known, traditional quantum physics is founded upon the Heisenberg commutation relations
$$[ X,P ] = i\hbar 1$$
and it is well known that there exist no invariant subspaces of $P$ with compact support in $x$ space in the usual Schrodinger representation.  Actually, there exist no regular differential operators in $x$ space whose (approximate) eigenfunctions are of compact support.  The realisation of such feature requires boundary conditions (a compactification of space), distributional operators (eg potentials, like the free particle in an infinite potential well) or irregular operators such as
$$e^{- \frac{1}{(x^2 - a^2)^2}}\frac{\partial^2}{\partial x^2} + b$$ where the operator just becomes constant for $|x| > a$.  It is possible to consider analytic operators such as
$$\arctan \left( -i\partial_x \right)$$ so that the entire real line gets compactified to $(-\frac{\pi}{2},\frac{\pi}{2})$ meaning there is an infinite spectral compression while the eigenfunctions remain $e^{ikx}$.  However, this is not a useful way of thinking about it because it still remains impossible to consider subintervals of $(-\frac{\pi}{2},\frac{\pi}{2})$.  To formalize what we are looking for, it is useful to introduce the notion of relatively compact operators of scale $\delta$,  $A$ is relatively compact with respect to $B$ on a scale $\delta$ if and only if for any $\lambda \in \sigma(A)$ there exists a projection operator $p_{\lambda}^{A} \prec p_{[a,b]}^{A}$ and $p_{\lambda}^{A} \prec p_{[c,d]}^{B}$ (where by convention we choose the maximal a,c and minimal b,d) with $a \leq \lambda \leq b$ and $b-a < \delta$.  A direct consequence is the notion of spectral compression on a scale $\delta$
$$R^{\delta}(A,B) = \sup_{\lambda} \frac{b - a}{d - c}.$$
The latter is zero or not well defined for commuting operators with a discrete spectrum; in case both spectra are continuous, the spectral compression could be any element of $\mathbb{R}_{+}^{\infty}$ or ill defined.  For the standard position operators $X,Y$ the spectral compression equals $\infty$.  Finally, we call the pair $(A,B)$ relatively compact on a scale of $\delta$ when both are relatively compact on that scale with respect to one and another.  We already established that regular local differential operators are not relatively compact to $X$; the same however holds for the much wider class of integral operators.  For example, it is instructive to study the operator
$$L_a(f)(x) = \int_{x}^{x+a} f(s)ds $$ where the eigenvalue problem $L_a(f) = bf$ reduces to solving a nonlocal differential equation
$$bf'(x) = f(x+a) - f(x).$$
Another way to put this is that $\frac{b}{a}f'(x) = \frac{f(x+a) - f(x)}{a}$; that is, a multiple of the derivative equals the discrete derivative at scale $a$.  It is easy to see that this equation has no nontrivial solutions of compact support.  Therefore, in order to give an example of an operator relatively compact with respect to $X$, it is best at this point to give the spectral decomposition straight away.  Subdivide the real line by unit intervals and consider the functions $e^{i2\pi n x}_{m}$ which are zero outside $[m,m+1]$ and take the value $e^{i2\pi n x}$ within.  One can now construct two operators, one which attaches to the latter function the eigenvalue $n\pi^{m^2}$ and another wich attributes simply $m^2$.  In the former case, the operator is relatively compact to $X$ on any scale $>0$ but not the other way around while in the latter case they form a relatively compact pair on the same scales.  In both cases the compression is zero on scales smaller than one\footnote{In the latter case one can construct a sequence of projection operators in the eigenspace of $m^2$ whose support becomes arbitrarily small, but the point is of course never reached.}.  We end up with an example of a nonlocal operator $P$ on the half line $x \geq 0$ with as spectrum $\mathbb{R}_{\geq 0}$ which is relatively compact to $X$ on any scale $>0$.  Consider a natural number $n > 0$, a unitary $n \times n$ matrix $U$ and for any point $x \in [0,1)$, the sequence $x,x+1,x+2, \ldots$.  Define the vectors $|x,i,j\rangle = \sum_{k=0}^{n-1} U_{jk}|x + in + k\rangle$ where $i=0 \ldots \infty$ and $j = 0 \ldots n-1$ and attribute the eigenvalue $x+ in + j$; then the corresponding operator defined by the spectral measure $\sum_{i=0}^{\infty} \sum_{j=0}^{n-1} \int_{0}^{1} dx$ is relatively compact on any scale to $X$.  In the light of the previous comments, it is obvious that we need differential operators with memory if we want local operators which both have a continuum spectrum or we need to introduce bundle structures (such as the Clifford bundle); that is, we have to go over to path space.  Assuming that one of our two operators (generalizations to more coordinates can be made) has a uniformly degenerate continuum spectrum\footnote{This condition means that for any $a,b,\delta$ such that $[a,a+\delta]$ and $[b,b+\delta]$ belong to the continuum spectrum, there exists an algebra automorphism mapping $p_{[a,a+\delta]}$ to $p_{[b,b+\delta]}$.} one can assume that $X$ equals the multiplication by $x$ on some $L^2(\mathbb{R} \times \mathcal{M}, dx \wedge \mu)$ where $\mathcal{M}$ is diffeomorphic to $\mathcal{P}(\mathbb{R}) \times \mathcal{N}$, where $\mathcal{P}(\mathbb{R})$ is the path space of $\mathbb{R}$ and $\mathcal{N}$ is a finite dimensional manifold.  Now, it is well known that $\mathcal{P}(\mathbb{R})$, where we restrict to the continuous paths, is a Fr\'echet manifold; basically this means that the local model is a complete locally convex space and not a vector space.  Intuitively, this signifies that one cannot vary the infinite number of coordinates freely which makes a careful definition of a notion of integration necessary, something we come back to in greater detail later on.  In general, let $V$ be a vectorfield, then the free displacement over a parameter range $\lambda$ is given by
$$\Delta^{\lambda}_V \gamma(s) = \exp_V(\lambda)\left[\gamma(s)\right]$$
which defines a curve with the same parameter domain.  One can also consider extensionsof the curve either on the  front or tail:
$$\Gamma^{\lambda,+}_{V} \gamma(s) = \exp_{V}(s-b)\left[ \gamma(b)\right]$$
for $b+\lambda \geq s \geq b$ and $\gamma(s)$ otherwise.  Therefore, the non-abelian semi group\footnote{Not every generator has an inverse.} of difference land is generated by $\Delta^{\lambda}_{V}, \Gamma^{\lambda,+}_{V}$ as well as the operations $T$ and $R^{\lambda}$ where $T\gamma(s) = \gamma(b+a-s)$ and $R^{\lambda}\gamma$ is the restriction of $\gamma$ to $\left[a,a + \lambda\right]$ for\footnote{One can also define the operation $S^{\lambda}$ as the restriction of $\gamma$ to $\left[b - \lambda,b \right]$.  $S^{\lambda} = TR^{\lambda}T$ as an easy calculation reveals; also, $T$ commutes with the displacements $\Delta^{\lambda}_{V}$ and $T\Gamma^{\lambda,+}_{V}T = \Gamma^{\lambda,-}_{V}$.} $\lambda \leq b-a$.  We now study cases where $V$ is dependent upon $\gamma$ itself; in this context, two notions are particularly useful: (a) a curve $\gamma$ and vectorfield $V$ are locally alligned if and only if $\dot{\gamma}(t) = \alpha(t)V(\gamma(t))$ where $\alpha(t)$ is real valued and
$$\exp^{\lambda}_{V} \gamma(s) = \gamma \left( s + \int_{s}^{s + \lambda}\alpha(t)dt \right)$$
whenever the last expression between brackets remains smaller or equal to $b$ and $(b)$ $V$ is called a velocity field if $$\dot{\gamma}(t) = V(\gamma(t))$$ and the associated displacement $$\exp^{\lambda}_{V}\gamma(s)$$ is given by the expression $\gamma(s+\lambda)$ for $a \leq s \leq b - \lambda$.  We now resort to displacements with memory and depending upon a background vectorfield $W$ as well.
\subsection{Differences with memory and a background influence.}
Basically, one can decide to displace a curve in a determinstic or stochastic manner and the latter only requires probability measures on the appropriate infinite dimensional spaces making the relevant objects into stochastic variables.  For now, we will only present some deterministic options and require the presence of a non-degenerate Riemannian or Lorentzian metric $h$ which canonically defines a Levi Civita connection $\nabla$.  That is, we consider vector valued functions $\omega(p,\gamma,h,t)$ defining a vector at $p$ by means of the inverse exponential map $D \exp^{-1}_{h}(p):T\mathcal{M} \rightarrow T_{p}\mathcal{M}$ from a vector at $\gamma(t)$ constructed by means of local quantities $\dot{\gamma}(t), \ddot{\gamma}(t), \ldots$ where $\ddot{\gamma}(t)$ is defined as $\nabla_{\dot{\gamma}(t)} \dot{\gamma}(t)$.  Then, 
$$V(p) = \int_{a}^{b} \omega(p,\gamma,h,t)dt$$ and we give now some examples where $h$ is Lorentzian versus Riemannian.  In case of a time oriented Lorentzian manifold and causal curves, one has to allow for an assymetric treatment of the past and future.  More specifically $\omega(p,\gamma,h,t)$ vanishes for all $t$ suh that $\gamma(t) \notin J^{+}(p) \cup J^{-}(p)$ and its functional form depends upon whether $\gamma(t) \in J^{+}(p)$ or $J^{-}(p)$ respectively.  For an exclusively retarded prescription on flat Minkowski, one may choose
$$V(p) = \int_{a}^{\alpha_{p}} e^{-\mu\left( A(\gamma(t),p)\right)}\dot{\gamma}(t)dt $$
where $\gamma(\alpha_p)$ is the intersection point of $\gamma$ with $J^{-}(p)$ and $A(p,q)$ denotes the Alexandrov set between $p$ and $q$.  In case $h$ is the standard flat Euclidean metric, it is natural to replace the volume of the Alexandrov set by the Euclidean distance between $\gamma(t)$ and $p$.  When allowing for an external perturbation $W$, one could add to the above prescription terms of the kind
$$\alpha(p,\gamma)W(p) + \int_{a}^{b} \dot{\gamma}(s)h(\dot{\gamma}(s),W(\gamma(s)))\beta(\gamma(s),p)ds $$ as well as many other forms. 
\subsection{Construction of operators relatively compact to some position operator.}
Before we proceed, let us define differential operators on smooth functions $\Psi: \mathcal{P}(\mathcal{M}) \rightarrow \mathbb{R}$ by using the basic differences $\Delta$ and $\Gamma$.  Let $V$ be a smooth vectorfield, then
$$\partial_{V} \Psi(\gamma) = \lim_{\lambda \rightarrow 0} \frac{\Psi\left( \Delta^{\lambda}_{V}(\gamma)\right) - \Psi(\gamma)}{\lambda}$$ and using $\Gamma^{\lambda,\pm}_{V}$ one can define $\partial^{\pm}_{V}\Psi$.  One notices that for elementary coordinate functions of endpoints, that is 
$$x^{\mu}_{f}(\gamma)= x^{\mu}(\gamma(b))$$ the derivative 
$$\partial^{+}_{V}x^{\mu}_{f}(\gamma) = V^{\mu}(\gamma(b)) = \partial_{V} x^{\mu}_{f}(\gamma).$$
\subsubsection{Addition rule.}
The derivative associated to a composition of ordinary displacements (not involving the $T$ operator) obviously satisfies a chain rule depending upon the particular scaling limit which has been taken.  Concretely, let  $\alpha(\lambda,\delta)$ be a positive function of $\lambda,\delta$ which takes the value zero if and only if $\delta=\lambda=0$.  Then, the expression
$$\frac{\Psi\left( \Delta^{\delta}_{V}\Gamma^{\lambda,+}_{W}(\gamma) \right) - \Psi\left(\gamma\right)}{\alpha(\delta,\lambda)} = \frac{\delta}{\alpha(\delta,\lambda)}\frac{\Psi\left( \Delta^{\delta}_{V}\Gamma^{\lambda,+}_{W}(\gamma) \right) - \Psi\left(\Gamma^{\lambda,+}_{W}(\gamma)\right)}{\delta} + \frac{\lambda}{\alpha(\delta,\lambda)}\frac{\Psi\left(\Gamma^{\lambda,+}_{W}(\gamma) \right) - \Psi\left(\gamma\right)}{\lambda} $$ and the behaviour of this expression in the double limit depends upon the scaling $\lambda(\delta)$ as well as the function $\alpha$.  In case $\alpha(\lambda,\delta) = \frac{1}{2}(\lambda + \delta)$, the right hand side reduces to 
$$\partial_{V}\Psi(\gamma) + \partial^{+}_{W}\Psi(\gamma)$$ in the scaling limit $\lambda = \delta$.  
\subsection{Coordinatization.}
Let us first construct suitable coordinate systems so that explicit calculations become possible.  That is, let $B\left( \mathcal{M} \right)$ be a countable basis of vectorfields on $\mathcal{M}$ and likewise consider $B\left( \mathcal{O} \right)$ to be a filter of local subbases meaning $B\left( \mathcal{O} \right) \subset B\left( \mathcal{V} \right)$ for $\mathcal{O} \subset \mathcal{V}$.  If $\mathcal{O}$ is a set of $s$ points and $\mathcal{M}$ is $n$-dimensional, then teh dimension of $B\left( \mathcal{O} \right)$ is given by $sn$.  Obviously, it is sufficient to construct charts based around a curve $\gamma$ an open set $\mathcal{O}$ around it and some open interval $\left( - \delta, \delta \right)$; more specifically, consider the subspace of $T\mathcal{O}$ consisting of vectorfields $V$ of sup-norm one on $\mathcal{O}$ (with respect to some Riemannian metric $h$) and study the actions
$$T\left[ \lambda,\delta,V \right](\gamma) = \Delta^{\lambda}_{V}\Gamma^{\delta,+}_{W_{\gamma}}(\gamma)$$ where $W_{\gamma}$ is some non-vanishing background field with $\gamma$ as an integral curve.  Now, while the action above is globally uniquely defined regarding $V$ in the sense that in case 
$$T\left[ \lambda,\delta,V_{1}\right](\gamma) = T\left[ \lambda,\delta,V_{2} \right](\gamma)$$
for all $\lambda,\delta$ and $\gamma$, then $V_1 = V_2$ and ordinary rescalings have been excluded already by using the sup norm.  However, the Gribov problem that for \emph{some} $\delta_i,V_i$ the above expressions might be equal remains and appropriate identifications have to be made.  Obviously, the solution to this problem is to further restrict to vectorfields defined from the maximal $W_{\gamma}$ future extension $\gamma_{\textrm{max}}$ of $\gamma$ by using the exponential map $\exp_{\lambda}^{h}$.  One knows that $\exp_{\lambda}^{h}\left[ V\left( \gamma_{\textrm{max}}(s) \right) \right]$ is a bijection on $(\lambda,s) \in (-d,d) \times \gamma_{\textrm{max}}$ for some suitable $d > 0$ given the unit sup-norm vectorfield $V$.  However, we cannot disgard focal points, otherwise curves were not allowed to curl and have to stay transversal with respect to nonintersecting geodesic bundles.  Therefore, we define multivalued vectorfields $V$ to be consistent if and only if differentiable integral curves are well defined (in the sense that they are well defined whenever $V$ is).  Hence, we work with vectorfields of unit sup-norm on $\gamma_{\textrm{max}}$ which may be geodesically mapped to a multivalued field in case the geodesics cross.  In particular, let $\gamma_{\textrm{max}}:\left[ a,c  \right] \rightarrow \mathcal{M}$ be a (possibly self-intersecting) curve in $\overline{\mathcal{O}}$ where $a<b<c$ and consider as basis (in $L^{2}$ norm) of the function space of continuous functions on $\left[ a,c \right]$ the family $\sin \left( n\pi\frac{s-a}{c-a} \right),\cos \left( n\pi\frac{s-a}{c-a} \right)$ where $n \in \mathbb{N}$.  Moreover, let $E_i(s)$ be some parallel transported basis with respect to $h$, then every (possibly multivalued) orthonormal vectorfield $V$ on $\gamma_{\textrm{max}}$ can be uniquely written as
$$V\left( \gamma_{\textrm{max}}(s) \right) = \sum_{i=1}^{n} \left(a_{i,0} + \sum_{n=1}^{\infty}\left( a_{i,n} \sin \left( n\pi \frac{s-a}{c-a} \right) + b_{i,n} \cos \left( n\pi \frac{s-a}{c-a} \right)  \right)  \right)E_{i}(s)$$
in the $L^{2}$ sense.  Since normalization with respect to the sup-norm is technically akward, it is much easier to consider the map $\exp_{1}^{h}\left( V\left( \gamma_{\textrm{max}}(s) \right) \right)$ wherever the image of the latter is a connected curve\footnote{Taking into account that some pieces may drop off the manifold.} since $\exp_{1}^{h}\left( V(\gamma_{\textrm{max}}(s)) \right) = \exp_{\lambda}^{h}\left( \frac{1}{\lambda}V\left( \gamma_{\textrm{max}}(s) \right) \right)$.   Therefore, the infinite dimensional coordinate charts look like $(\lambda,a_{i,0},a_{i,n},b_{i,n})$ with as chart mapping 
$$\exp_{1}^{h}\left( V^{\lambda}(s) \right)R^{b-a+\lambda}\left( \gamma_{\textrm{max}}\right)$$ where
$$V^{\lambda}(s) = \sum_{i=1}^{n} \left(a_{i,0} + \sum_{n=1}^{\infty}\left( a_{i,n} \sin \left( n\pi \frac{s-a}{b+\lambda -a} \right) + b_{i,n} \cos \left( n\pi \frac{s-a}{b+\lambda-a} \right)  \right)  \right)E_{i}(s)  $$ and the reader may verify that everything is well defined.  Notice that not 
every vectorfield defined in this way has a continuous representant and that therefore holes are to be pinched in this chart; however, the set of vectors with a continuous representant is open in the compact-open topology on $\mathbb{R}^{\infty}$ and everything below has to be understood in this way.  
\subsection{Explicit expressions of operators.}
We now calculate leading terms for the standard differential operators depending upon nonlocal functions of local geometric tensors derived from the metric $h$ and give exact expressions in the Euclidean case.  Let us start with the difference operator $\Delta^{\mu}_{W}$; that is, we calculate
$$\Delta^{\mu}_{W}(p)^{co}(s) = \left( \exp^{h}_{1}\left( \gamma_{\textrm{max}}(s) \right) \right)^{-1}  \left( 
\Delta^{\mu}_{W} \left( \exp^{h}_{1}\left( V^{\lambda}(s) \right) \right) \left( R^{b-a + \lambda} \left( \gamma_{\textrm{max}}(s) \right) \right) \right)$$ where $p=(\lambda,a_{i,0},a_{i,n},b_{i,n})$ and rescale to the entire interval $\left[ a,c \right]$.  More precisely, we calculate the first and second derivatives with respect to $\mu$ and $W$ as far as we can for a general Riemannian metric, the Lorentzian case being somewhat more subtle.  Concerning the $\mu$-derivative, one needs to remember the notion of Fermi transport and calculate the geodesic difference equation.  Let $V$ be a vectorfield, then the Fermi derivative of $W$ along $V$ is defined as 
$$D^{F}_{V}\left(W\right) = \nabla_{V}W - \frac{h(\nabla_{V}W,V)}{h(V,V)}V$$ and coincides with the standard Levi-Civita derivative in case $V$ has geodesic integral curves.  Therefore, let $\gamma(s,t)$ be a one parameter family of geodesic curves in the sense that $$s \rightarrow \gamma(s,t)$$ defines a geodesic in affine parametrization for all $t$.  We wish to write down an evolution equation for the orthogonal part $Z_{+} = Z - \frac{h(Z,V)}{h(V,V)}V$ of the geodesic deviation vector $Z = \partial_t \gamma(s,t)$ and note that $\mathcal{L}_V Z = \left[ V,Z \right] = 0$ by construction.  Taking into account that $D^{F}_{V}Z_{+} = \nabla_{V}Z_{+}$, we arrive at
$$\frac{D^F}{ds} Z_{+} = \nabla_{V} Z_{+}$$
and
$$\frac{D^F}{ds}\frac{D^F}{ds} Z_{+} = \nabla_V \nabla_V Z_{+} = \nabla_V \nabla_Z V = - R(Z,V)V = - R(Z_{+},V)V$$
due to the geodesic equation.  Let us study more in particular the geodesic congruence 
$$\exp^{s}_{tZ + V}(p)$$ where $s \geq 0$, $t \in (-\delta,\delta)$ and $h(Z,V) = 0$ at $t=0$.  Consider an $n-1$ bein $E_{i}(s)$ perpendicular to $V(s) \equiv \frac{d}{ds}\exp^{s}_{V}(p)$ satisfying $\nabla_{V(s)}E_{i}(s) = 0$; then with $Z(s)= Z^{i}(s)E_{i}(s)$ the above equation reduces to
$$\frac{d^2}{ds^2}Z^{i}(s) = - R^{i}_{\,\, njn}(\exp^{s}_{V}(p))Z^{j}(s)h(V,V)$$ with $Z(0) = 0$ and $\frac{D}{ds}Z(0) = Z$.  Last, but not least, we need a method to find excellent approximations to solutions of the geodesic equation; first choose a coordinate system such that the Christoffel symbols $\Gamma^{\alpha}_{\,\, \beta \delta}$
are all smaller in absolute value than some $\epsilon > 0$ and proceed with a Newton-Rhapson iteration scheme.  That is, linearize the geodesic equation for a bundle of geodesics given by integral curves of some vectorfield $V$
$$V^{\mu}\partial_{\mu} V^{\nu} + \Gamma^{\nu}_{\alpha \beta} V^{\alpha}V^{\beta} = 0$$ by means of the equation
$$V_{i}^{\mu}\partial_{\mu} V_{i+1}^{\nu} + \Gamma^{\nu}_{\alpha \beta} V_{i}^{\alpha}V_{i+1}^{\beta} = 0$$ where $i\geq 1$ and at each iteration step fixed initial data are held for some maximal $n-1$ surface $\Sigma$.  For example, in such coordinate system, one can pick $\Sigma$ as $x^n=0$ and consider 
$$V_{1}(x^{1} + W^{1}(x^{1}, \ldots, x^{n-1})x^{n},\ldots, x^{n-1} + W^{n-1}(x^{1}, \ldots, x^{n-1})x^{n},x^{n}W^{n}(x^{1},\ldots, x^{n-1})) = W(x^{1}, \ldots, x^{n-1})$$ where $W$ is a vectorfield on and transversal to $\Sigma$ kept fixed at any iteration step.  We should check that the above procedure converges to a fix point (an obvious fact in one dimension where $V_3 = V_2$).  In that regard, it is easier to use the notation $A_{\alpha} = \partial_{\alpha} + \Gamma_{\alpha}$ and frame the convergence of the Newton-Rhapson procedure within the context of that family of operators; defining $V_{i+1} = V_{i} + \delta V_{i}$ where $\delta V_{i}= 0$ on $\Sigma$, one obtains that $$V_{1}^{\alpha} \Gamma^{\mu}_{\alpha \beta} V_{1}^{\beta} + V_{1}^{\alpha}\partial_{\alpha}\delta V_{1}^{\mu} + V_{1}^{\alpha}\Gamma^{\mu}_{\alpha \beta}\delta V_{1}^{\beta} = 0$$ and it is a matter of uniformly controlling the behaviour of $\delta V_{i}$ in a neighborhood of $\Sigma$ given that the equation $\nabla_{V_i} \delta V_{i} = 0$ is equivalent to $\delta V_{i} = 0$.  It is natural to consider the Hilbert space  $\mathcal{H}$ of vectorfields vanishing on $\Sigma$ and study spectral properties of the operators $A_{i} = V_{i}^{\alpha}A_{\alpha}$; hence, we construct the Green kernels $G_{i \, \beta}^{\alpha}(x,y)$ satisfying
$$A_{i \, \gamma}^{\alpha}G_{i \, \beta}^{\gamma}(x,y) = \sqrt{h}(y)\delta^{n}(x,y)\delta^{\alpha}_{\beta}$$ and with boundary conditions $G_{i}(x,y) = 0$ whenever $x \in \Sigma$.  It is most convenient to go to a $V$ adapted coordinate system $(x^1,\ldots, x^{n-1},x^n)$ where the first $n-1$ digits constitute the standard coordinates of the intersection point of the unique integral curve of $V$ with $\Sigma$ and $x^n$ is the affine parameter vanishing when $x \in \Sigma$.
 Covariance of the above equation then implies $$\frac{\partial}{\partial x^n} G^{\alpha}_{\beta}(x,y) + \Gamma^{\alpha}_{n \gamma}(x)G_{\beta}^{\gamma}(x,y) = \sqrt{h}(y)\delta^{n}(x,y)\delta^{\alpha}_{\beta}$$ and the latter is most easily solved by noticing that for $x^{n} > 0$
$$\frac{\partial}{\partial x^n}\left( \mathcal{T} G(x,y) \, \exp{\left( \int_{0}^{x^n} \Gamma_{n}(x^{1},\ldots,x^{n-1},s) ds \right)} \right) = \mathcal{T} \exp{\left( \int_{0}^{x^n} \Gamma_{n}(x^{1},\ldots,x^{n-1},s) ds \right)} \,\sqrt{h}(y)\delta^{n}(x,y)$$ where the operation $\mathcal{T}$ orders the expression in decreasing values of time.  Consequently, $G(x,y)$ is given by
$$\sqrt{h}(y)\mathcal{T} \int_{0}^{x^{n}}dt \left( \exp{\left( \int_{0}^{t} \Gamma_{n}(x^{1},\ldots,x^{n-1},s) ds \right)} \delta^{n}(x^{1},\ldots, x^{n-1},t,y) \right) \exp{\left( - \int_{0}^{x^n} \Gamma_{n}(x^{1},\ldots,x^{n-1},s) ds \right)}.$$  The above formula has a nice geometric interpretation in the sense that 
$\int G(x,y)f(y)d^{n}y$ only depends on the values of $f$ on the integral curve of $V$ through $x$.  It is clear that, although formal proofs of convergence can be set up, this method is not going to be of much help since calculating integral curves in closed form is usually not possible and therefore one is working with approximations of approximations.
 \\* \\*    
 Another, much more geometrical and direct method, consists in constructing geodesics as an ordered integral from some initial values; that is, choose $\delta$ small enough such that variations of $\Gamma$ become small on scales of $W\delta$ and construct a piecewise linear curve starting on $\Sigma$ at $x_0$ with initial direction $W(x_{0})$ and length of the first linear piece $\delta \sqrt{h(W(x_{0}), W(x_{0}))}$ ending at $x_{1}$.  At $x_{1}$, twist the vector $W(x_{0})$ by an amount of $ - \delta W(x_{0})\Gamma(x_{0})W(x_{0})$ and repeat this procedure $n$ times such that $n\delta$ remains constant in the limiting procedure.  It is very easy to track the evolution of the geodesic deviation in this way by constructing two geodesics with nearly identical initial conditions in phase space and comparing the endpoints at identical parameter lengths.  \\* \\*
 Finally, exact computations are only possible in highly symmetric spaces such as Euclidean space, higher dimensional tori or some $n$-dimensional sphere; we shall treat here to some detail the first case.  Since general exact computations of nonlocal operators are out of reach, we merely compute the derivative $\frac{d}{d\lambda}\Delta^{\lambda}_{W \, |\lambda=0}$; in general, one has to obtain the Fourier coefficients of $$V^{\kappa}(s) + \lambda W\left( R^{b-a + \kappa}(\gamma_{\textrm{max}})(s) + V^{\kappa}(s)\right)$$ with respect to a parallel transported vielbein $E_{i}(s)$.  As an example, let $$\gamma: \left[ 0,\frac{1}{2} \right] \rightarrow \mathbb{R}^2 : s \rightarrow (0,s)$$ and $\gamma_{\textrm{max}}$ be the unique maximal extension to $\left[0,1\right]$.  With $V^{0}(s)$ as before and $W(x,y) = (1,y^2)$, we are left with Fourier decomposing
 $$\left( \sum_{m=1}^{\infty}\left( a_{2,m}\sin(2\pi ms ) + b_{2,m} \cos(2 \pi ms) \right) + a_{2,0} + s \right)^2$$ in order to determine the corresponding vectorfield on $\mathbb{R}^{\infty}$.  Obviously, the reader can immediately construct a few differential operators forming a relatively compact pair on any scale to the endpoint operators
  $$x^{i}_{f}(\gamma) = a_{i,0} + \sum_{m=1}^{\infty} (-1)^{m} b_{i,m}$$ 
  and
  $$x_{f}^{\mu} = \left( \exp_{\sum x^{i}_{f}E_{i}(b+\lambda)}^{1}(\gamma_{\textrm{max}}^{i}(b+\lambda)) \right)^{\mu}$$ on the space of square integrable functions on path space\footnote{We will come back to the accurate description later on.} for the same reason that $\partial_y$ and $x$ are.  More interesting examples are to be constructed later on.   
\section{Another view on constructing solutions of differential equations.} 
There are good reasons to suspect that there is little restriction in standard quantum mechanics concerning its linear nature.  Indeed, it is well known one can write down any classical field theory on a Fock space, using the Heisenberg equations, where the allowed states are coherent states and not just any state\footnote{Indeed, considering the equation $\dot{z} = F(z)$ or better $\frac{d}{dt}e^{iz(t)} = iF(z(t))e^{iz(t)}$ with initial conditions $z(0) = z_0$ one can define the standard bosonic Fock algebra $[a_x,a^{\dag}_y] = \delta(x-y)$ resulting in the operators $H = \int F(x) a^{\dag}_{x}a_{x} dx$ and $K(t) = \int \left( e^{- ix(t;x_0)}a^{\dag}_{x(t;x_0)} + e^{ix(t;x_0)}a_{x(t;x_0)} \right) dx_0$.  Obviously, the equation $\frac{d}{dt}K = 0$ is equivalent to the Heisenberg equations for the operators $a^{\dag}_{x(t;x_0)}$.}.  We just encountered a similar issue in the above, we concluded that we need linear operators which correspond on the real line to a process with memory which comes at the price of adding an infinite number of degrees of freedom in terms of path space.  The natural thing to do however is to directly define a Markov process with memory and see how we can represent this as a linear operator on our infinite dimensional Fr\'echet manifold.  In the beginning however, we adress an even more moderate task which is to study Finslerian random walks and study how they correspond to partial differential equations and vice versa.  Therefore, instead of solving difference equations, we simply count paths with suitable boundary conditions imposed, which is a much easier thing to do.  The other road, which consists in solving the linear problem and restricting to coherent states, does have an easy formulation on a random lattice as we shall illustrate later. \\* \\*
It is well known that the Green's function of the diffusion equation can be constructed as a particular scaling limit of the random walk, just as is the case for the telegraph equations starting from a process with one step memory.  The aim of this paragraph is to show that actually any nonlinear partial differential equation can be constructed in this way by means of nonlinear Finslerian random walks as long as one accepts the concept of negative probability, which merely amounts to saying that there can appear and dissapear drunken sailors either out of nothing or from the sea as it pleases the reader.  As an example consider any continuous functions $f(x), g(x), h(x)$ and spacetime grid $n\Delta t, m\Delta x$, then the walk defined by
\begin{eqnarray*}
u(t + \Delta t,x) & = & \left( f(x) - \frac{\Delta x}{2} g(x) \right) u(t,x - \Delta x) + \left( f(x) + \frac{\Delta x}{2} g(x) \right) u(t, x + \Delta x) \\* 
& & - (f(x) - 1 - h(x)(\Delta x)^2) u(t,x)
\end{eqnarray*} leads in the scaling limit $\Delta t = (\Delta x)^2$ to the Green's function of the differential equation 
$$\partial_t u(t,x) = f(x) \frac{\partial^2}{\partial x^2} u(t,x) + g(x) \frac{\partial}{\partial x} u(t,x) + h(x)u(t,x).$$  It is easily seen that the solution to any nonlinear equation with particular initial values can be constructed in this way by considering a weighted collection of random walkers where the number of walkers scales with $\frac{1}{\Delta x}$.  It remains to ``integrate'' this discretization procedure and write it as a weighted sum over oriented paths starting at the origin and ending up at $m\Delta x$; suffice it to say that in the former case
$$ u(n \Delta t,m \Delta x) = \sum_{\textrm{paths} \, \gamma \, \textrm{of length} \, n} \prod_{j = 1}^{n} w_{(\gamma(j-1), \gamma(j))}$$ where the paths $\gamma$ can stand still and $w_{(k,k+1)} = f((k + 1)\Delta x) - \frac{\Delta x}{2}g((k+1)\Delta x)$, $w_{(k,k)} = - f(k \Delta x) + 1 + h(k \Delta x)(\Delta x)^2$ and $w_{(k,k-1)} = f((k - 1)\Delta x) + \frac{\Delta x}{2}g((k-1)\Delta x)$.  As is well known to physicists, this procedure can be extended to construct solutions of nonlinear equations by means of Feynman diagrams\footnote{This procedure can be carried out from the classical path integral, which requires a doubling of the phase space degrees of freedom, and by making a standard Feynman perturbation expansion.}.  To illustrate a different procedure, let us add a $u^2(t,x)$ instability to the former equation.  Then, it is obvious that $u(n\Delta t, m \Delta x)$ must be written in terms of sums over multipaths\footnote{A multipath ending at $m \Delta x$ is a collection of $k$ paths with $n$ steps and identical (marked) endpoints, the origin and $m\Delta x$.  Every oriented edge comes with a weight $w$ or with $1$ and the reader has some choice in where to put $1$.}, each term coming with its own coupling constant.  Specifically, adding $\kappa u^2(t,x)$ to the random walk and writing down
$$u(n \Delta t, m \Delta x) =  \sum_{\textrm{multipath ending at} \, m\Delta x.} \lambda(\textrm{multipath},n,m,\kappa) \prod_{e_{+}, e_{-}, v} w(e_{+})^{\beta(e_{+})}w(e_{-})^{\gamma(e_{-})}w(v)^{\alpha(v)}$$ the task remains to find out the coupling \emph{functions} $\lambda$ which do depend upon space and time in contrast to the standard timeless $S$ matrix expansion in quantum field theory.
\section{Mass, conformal invariance, extra dimensions and all that}
There is the well known tension in the last century between the universal cover of the Poincar\'e group and the whole conformal group of electromagnetism.  Being minimalist, one could argue that conformal invariance is not a symmetry of nature since our measurements break it; however this leaves the opportunity for spontaneous symmetry breaking due to a vacuum state which is not conformally invariant.  A second, equally sloppy argument has always been that such theory cannot contain massive particles which is not true even in four spacetime dimensions.  Indeed, it is a piece of cake to introduce a conformal mass field $G$ replacing the Higgs as well as a conformal gauge field satisfying conformally invariant equations of motion.  The only valid objection I know of against such theory came from Einstein who pointed out that particle masses become path dependent and therefore, at first sight, one would expect trouble with the Pauli principle; this is philosophically puzzling indeed and does not appear within measurement error here on earth.  Of course, this is still not a leathal objection, it merely poses a fine tuning problem which may have a perfectly satisfying solution.  Still, it appears that we did not get any closer to answering the question why the particle masses are what they are.  There is of course an easy kinematical way out of this problem which constitutes in allowing for extra spacelike dimensions and by compactifying the latter.  However, there are the well known stability issues here and one must make certain one is not exchanging the lack of a scaling symmetry for a myriad of other problems.  It is not my intention to deal with such an approach here since it would force me to redo the entire representation theory of the conformal group.  I will just content myself with writing down the most generic Lagrangian with the full symmetry of electromagnetism with associated mass terms and so on.  Consider the following transformation properties of elementary fields under a local conformal transformation dictated by a positive function $\Omega(x)$:
\begin{eqnarray*}
\gamma^a & \rightarrow & \Omega^{-1} \gamma^a \\*
\Psi & \rightarrow & \Omega^{\alpha} \Psi \\*
G & \rightarrow & \Omega^{-1} G \\*
\mathcal{A}_{\mu} & \rightarrow & \mathcal{A}_{\mu} + G^{-1} \partial_{\mu} G.
\end{eqnarray*} 
Defining a scale covariant $\mathcal{D}_{\mu}^{q}$ partial derivative attached to a charge $q$ by
$$\mathcal{D}_{\mu}^{q} = \partial_{\mu} - q \mathcal{A}_{\mu}$$ we may write down that
\begin{eqnarray*} & & \mathcal{L} = \int d^4 x \, \sqrt{ \textrm{det} \left( g_{\mu \nu} \right) } \\* 
& & \left[ G^{2 \alpha + 3} \overline{\Psi} \gamma^a e_a^{\mu} \mathcal{D}_{\mu}^{\alpha} \Psi + \zeta G^{2 \alpha + 4} \overline{\Psi} \Psi + \chi F_{\mu \nu} F^{\mu \nu} + 
\xi g^{\mu \nu} \mathcal{D}_{\mu}^{-1} G \mathcal{D}_{\nu}^{-1} G + \epsilon G^4 + \lambda G^2 R \right]
\end{eqnarray*}
where $\zeta, \chi, \xi, \epsilon, \lambda$ are dimensionless constants and $R$ is the Ricci scalar for the conformal connection and the potential for $G$ looks analogous to the Mexican hat for the Higgs field.  This Lagrangian can be brought to standard form by the field redefinition
$$\Psi \rightarrow G^{\alpha + \frac{3}{2}} \Psi.$$  As mentioned previously, I leave the further development of these ideas for future work.
\chapter{Conclusions}
This book contained many ideas which came together in one coherent entity and caused a ``philosophical'' revolution of some kind.  Let me further comment on the possible implications of the core principles on which the entire construction is founded.  First of all, there is the principle of locality by which I really mean a \emph{local} and manifestly covariant formulation of the laws of physics.  This core idea allows for a consistent treatment of classical singularities and disposes of strict determinism by allowing for an indeterminate topology amongst other things.  That is, any non-operational theory which would replace the above scheme has to be at least founded on a stochastic dynamics; this remains possible since Bell's theorem does not apply to ``spacetimes'' with multiple time directions.  However, I do not deem this a plausible course to persue since operationalism appears to be deeply ingrained in nature.  Moreover, our theory is not Bell local either since at any point in spacetime, gravitationally distorted information about the whole universe is available to the monad; therefore, any local observer knows about these correlations and only the issue of absence of a ``pictorial'' three dimensional representation remains.  The latter would require an infinite tensorial product of a spinorial construction of spacetime and somehow Plato's shades, which we call reality, would have to be extracted from this infinite dimensional cage.  All this can be circumvented of course when a causal structure would not be present from the beginning and even the local Poincar\'e group would be emergent; a course of action which is doomed from the outset due to the pleitoria of possible non-existent particle statistics which would show up in such theory.  In chapter 11, we have canonically weakened the notion of locality allowing for spacetimes which behave like a heat bath with possibly long range entanglement; this line of research may open promising avenues but a lot of preliminary mathematical groundwork needs to be performed in order to get a sufficient understanding of these objects.  \\* \\*
A second core realization was that causality needs to be sacrified, that is, the effective statistics needs to be the result of a computation and not being put in by hand.  However, the equations of motion for the quantum potential are causal; for example if interactions are switched off, a Cauchy Kowalevskaja type result for coupled scalar fields
$$A_i(\psi_j, \partial_{\mu} \psi_j) \Box \psi_i + B_{i}^{k \nu}(\psi_j , \partial_{\mu}\psi_j) \partial_{\nu} \psi_{k} + C_i(\psi_j) = 0$$ insures that the values of $\psi_j(x)$ are determined by the initial data on the intersection with the past or future lightcone.  Nevertheless, this result has nothing to do with the notion of \emph{quantum} causality in the sense that the commutators of local observables vanish at spacelike separated events and the degree in which those are violated depends upon the boundary data\footnote{This issue becomes even more complex when one would make the vielbein noncommutative or when spacetime itself is noncommutative.}.  The preliminary result regarding the free theory on Minkowski is hopeful in this respect in the sense that there exists a natural criterion for the commutation relations in the gravitational vacuum to be inherited from the free theory on the vertical spaces.  Therefore, from a perturbative point of view, one would expect communication of quantum signals faster than light to generically happen at a speed close to the local speed of light.  I am aware that recently claims towards causality violation have been made, some by sending Mozart's 40'th symphony by $4.7$ times the speed of light.  Let me stress that I did not delve into the details here, but the mainstream claim that this can be perfectly explained due to the fact that group and phase velocities can travel faster than light needs a lot of more detail.  Actually, I was under the impression that Mozart's masterpiece came through with very little distortion which appears to strongly suggest that the front velocity exceeded the velocity of light in vacuum too.  Wether or not this experiment was a true breach of limits of causality, the point is that such a qualitative feature is expected to occur (with generically a pretty low probability) within my knowledge.  This brings one back to the recovery of an approximate ``Quantum Electrodynamics'' from our generalized quantum theory; there are a few issues here, the first one being ``does some approximation to QED exist in the absence of any other interactions (including gravity) on Minkowski?''.   If the answer is yes, then such theory would need to have a global Poincar\'e symmetry if one looks at the equations of motion while the local free Poincar\'e algebras differ from one spacetime point to another.  Given this fact, there is no indication yet that causality is preserved; indeed, most likely there is no well defined number operator in the bulk of spacetime which causes a different ``effective'' statistics than Bose and Fermi.  Now, if one would turn on the gravitational field, which is the case in nature, then for sure global Poincar\'e invariance is broken and generic causality violations will occur.  This might be hard to swallow but I really see no obvious problem, neither from the theory nor from the experimental side. \\* \\*
The second major correction to standard quantum theory resided in the substantial generalization of the Born rule which became apparant in the formulation of the spectral theorem on Clifford-Nevanlinna modules.  Also, it is clear that realistic observables will in general only be defined on a dense \emph{subset} of\footnote{For example, consider particle states $|0\rangle, |1\rangle$ and assume $|1\rangle$ leads to a measurement, then all states $\lambda |0\rangle + \mu|1\rangle$ with $\mu \neq 0$ are in the domain and the image is $\mu | 1 \rangle$. } the full local Clifford Nevanlinna space; the reason for this simply being that no screen is large enough to measure all electrons in the universe.  This subtle violation of the superposition principle does not pose any problems regarding a spectral theorem where the spectral measure will not constitute a resolution of the identity.   
\chapter{Acknowledgements}
In the first place, I am indebted to Rafael Sorkin for many interesting, long conversations about physics and some discussions about this work in particular.  Also, my regards go to Steve Adler for the visit at the IAS at Princeton University as well as some helpful comments.  Furthermore, my gratitude goes towards Stefan Nobbenhuis for the interesting and open minded discussions we had about physics for the last five years; some of the speculative ideas Stefan and I spoke about came back in this work.  Finally, I thank Wilhelm Furtwangler and Kirsten Flagstad for harmonizing my brain again when it got overloaded due to thinking too deeply; that his genius may come back to this world.   


\begin{thebibliography}{99}
\bibitem{'t Hooft}  G. 't Hooft, the Free-Will postulate in quantum mechanics, arXiv:quant-ph/0701097v1
\bibitem{Conway1}  J. Conway and S. Kochen, The strong free will theorem, arXiv:0807.3286 
\bibitem{Dowker} F. Dowker and J. Halliwell, Quantum mechanics of history: The decoherence functional in quantum mechanics, Phys. Rev. D 46, 1580-1609 (1992) 
\bibitem{Conway2}  J. Conway and S. Kochen, Reply to Comments of Bassi, Ghirardi, and Tumulka on the Free Will Theorem, arXiv:quant-ph/0701016  
\bibitem{Sorkin1} R. Sorkin and D. Rideout, A classical sequential growth dynamics for causal sets,  arXiv:gr-qc/9904062
\bibitem{Rideout} D. Rideout, Dynamics of causal sets, arXiv:gr-qc/0212064
\bibitem{Sorkin2}  R. Sorkin, Quantum mechanics as a quantum measure theory,  arXiv:gr-qc/9401003
\bibitem{Sorkin3} R. Sorkin, Quantum measure theory and its interpretation, arXiv:gr-qc/9507057
\bibitem{Sorkin4}  R. Sorkin, To What Type of Logic Does the "Tetralemma" Belong?, arXiv:1003.5735
\bibitem{Sorkin5}  R. Sorkin,  Logic is to the quantum as geometry is to gravity, arXiv:1004.1226
\bibitem{three}  Testing quantum mechanics using a three slit experiment, pirsa.org/pdf/files/366ce8ad-908b-4a2e-9909-33d12070d25d.pdf
\bibitem{Sorkin6} Xavier Martin, Denjoe O'Connor, R.D. Sorkin, The Random Walk in Generalized Quantum Theory, arXiv:gr-qc/0403085
\bibitem{Salmhofer}  M. Salmhofer, Renormalization An Introduction, Springer 
\bibitem{Wald1} R. Wald,  The Thermodynamics of Black Holes, arXiv:gr-qc/9912119
\bibitem{Sorkin7}  R. Sorkin, Ten Theses on Black Hole Entropy, arXiv:hep-th/0504037
\bibitem{'t Hooft1}  G. 't Hooft,  The black hole horizon as a dynamical system, arXiv:gr-qc/0606026
\bibitem{'t Hooft2}  G. 't Hooft, Horizon Operator Approach to Black Hole Quantization, arXiv:gr-qc/9402037
\bibitem{Susskind}  L. Susskind, The Black Hole War: My Battle with Stephen Hawking to Make the World Safe for Quantum Mechanics, American Journal of Physics -- March 2010 -- Volume 78, Issue 3, pp. 318-319 
\bibitem{Gudder}  S. Gudder, Quantum measures and the coevent interpretation, arXiv:1005.2242
\bibitem{d'Espagnat} B. d'Espagnat, Consciousness and the Wigner's friend problem, arXiv:quant-ph/0402121
\bibitem{'t Hooft3} G. 't Hooft, How Does God Play Dice? (Pre-)Determinism at the Planck Scale, arXiv:hep-th/0104219
\bibitem{'t Hooft4} G. 't Hooft, Does God play dice?, http://www.phys.uu.nl/~thooft/gthpub/DiceWorld.pdf
\bibitem{Wald2} R. Wald, General Relativity, The university of Chicago Press
\bibitem{Hawking1} S. Hawking and G. Ellis, The Large Scale structure of space-time, Cambridge monographs on mathematical physics
\bibitem{Beem} J. Beem, P. Ehrlich and K. Easley, Global Lorentzian Geometry, Marcel Dekker
\bibitem{Thiemann} T. Thiemann, Introduction to Modern Canonical Quantum General Relativity, arXiv:gr-qc/0110034
\bibitem{Henneaux}  M. Henneaux and C. Teitelboim, Quantization of Gauge Systems, Princeton university press 
\bibitem{Rovelli1} C. Rovelli, Partial Observables, gr-qc/0110035
\bibitem{Rovelli2} C. Rovelli, GPS observables in general relativity, arXiv:gr-qc/0110003
\bibitem{Szabados1}  L. Szabados, On the roots of the Poincar\'e structure of asymptotically flat spacetimes, 2003 Class. Quantum Grav. 20 2627
\bibitem{Szabados2}  L. Szabados, The Poincar\'e Structure
and the Centre-of-Mass of Asymptotically Flat Spacetimes, Lect. Notes Phys. 692, 157-184 (2006)
\bibitem{Szabados3} L. Szabados, Quasi-Local Energy-Momentum and Angular Momentum in General Relativity, http://blog.relativity.livingreviews.org/tag/quasi-local-angular-momentum/
\bibitem{Anderson} Anderson, Paul R.; and Brill, Dieter R. (1997). "Gravitational Geons Revisited". Phys. Rev. D 56: 4824-4833
\bibitem{Marolf} Louko, Jorma; Mann, Robert B.; and Marolf, Donald (2005). "Geons with spin and charge". Class. Quant. Grav. 22: 1451-1468.
\bibitem{Canfora}  F. Canfora, L. Parisi, G. Vilasi, Spin-1 gravitational waves. Theoretical and experimental aspects, arXiv:gr-qc/0512159
\bibitem{Tumulka} R. Tumulka, A Relativistic Version of the Ghirardi Rimini Weber Model, Journal of Statistical Physics Volume 125, Number 4, 821-840
\bibitem{Penrose1} R. Penrose, The road to reality, Jonathan Cape London
\bibitem{'t Hooft5} G. 't Hooft, The fundamental nature of space and time, in Approaches to Quantum Gravity, Toward a New Understanding of Space, Time and Matter, D. Oriti Ed.,  Cambridge Univ. Press 2009, ISBN 978-0-521-86045-1 (pbk), pp. 13-25.
\bibitem{Westra}  W. Westra, Topology Change and the Emergence of Geometry in Two Dimensional Causal Quantum Gravity, arXiv:0810.0771
\bibitem{'t Hooft6} G. 't Hooft, Unitarity in the Brout-Englert-Higgs Mechanism for Gravity, arXiv:0708.3184
\bibitem{Noldus1} J. Noldus, A Lorentzian Gromov-Hausdoff notion of distance, arXiv:gr-qc/0308074 
\bibitem{Noldus2} J. Noldus, The limit space of a Cauchy sequence of globally hyperbolic spacetimes, arXiv:gr-qc/0308075
\bibitem{Noldus3} J. Noldus and L. Bombelli, The moduli space of isometry classes of globally hyperbolic spacetimes, arXiv:gr-qc/0402049
\bibitem{Lasenby} A. Lasenby, C. Doran and S. Gull, Gravity, gauge theories and geometric algebra, Phil. Trans. R. Soc. Lond. A 15 March 1998 vol. 356 no. 1737 487-582 
\bibitem{Leblond}  J. L\'evy-Leblond, One more derivation of the Lorentz transformation, American Journal of Physics Vol. 44, No. 3, March 1976
\bibitem{Penrose2} R. Penrose, The Large, the Small and the Human Mind, Canto
\bibitem{Witten} M. Green, J. Schwarz and E. Witten, Superstring theory, Volume 1, Cambridge Monographs on mathematical physics.
\bibitem{Georgiev} D. Georgiev, Consciousness operates beyond the timescale for discerning time intervals: implications for Q-mind theories and analysis
of quantum decoherence in brain, http://cogprints.org/3318/1/time.pdf
\bibitem{Wald3} S. Gralla and R. Wald, A Rigorous Derivation of Gravitational Self-force, arXiv:0806.3293
\bibitem{Cooperstock} G. Perry and F. Cooperstock, Stability of Gravitational and Electromagnetic Geons, gr-qc/9810045
\bibitem{Thorne} K. Thorne, The Instability of a Toroidal Magnetic Geon against Gravitational Collapse, Proceedings of the 1st Texas Symposium on Relativistic Astrophysics. Edited by Ivor Robinson, Alfred Schild and E.L. Schucking. Chicago: University of Chicago Press, 1965., p.83
\bibitem{Surya}  R. Sorkin and S. Surya, An Analysis of the Representations of the Mapping Class Group of a Multi-Geon Three-Manifold, arXiv:gr-qc/9605050
\bibitem{Majid}  S. Majid, Beyond supersymmetry and quantum symmetry, hep-th/9212151
\bibitem{Greenberg} O. Greenberg, Theories of violation of statistics, hep-th/0007054
\bibitem{Peskin} M. Peskin and D. Schroeder, An introduction to quantum field theory, Westview Press
\bibitem{Hestenes}  D. Hestenes, Geometry of the Dirac theory, A symposium on the mathematics of physical space-time, Universidad nacional Autonoma de Mexico, $67-96$, $1981$ 
\bibitem{Weinberg} S. Weinberg, The Quantum Theory of Fields I, Cambridge University Press.
\bibitem{Krein1} I.S. Iohvidov an M.G. Krein, Spectral theory of operators in spaces with an indefinite metric. I
\bibitem{Krein2} I.S. Iohvidov an M.G. Krein, Spectral theory of operators in spaces with an indefinite metric. II
\bibitem{Aerts} D. and S. Aerts, Quo Vadis Quantum mechanics, Springer
\bibitem{Helling} R.C. Helling and G. Policastro, String quantization: Fock versus LQG representations, hep-th/0409182
\bibitem{Baez} J.C. Baez and A.D. Lauda, Higher dimensional algebra V: 2-Groups, math/0307200
\bibitem{Weinberg2} S. Weinberg, The quantum theory of fields, Volume II : modern applications, Cambridge University Press 
\bibitem{Weinberg3} S. Weinberg, The quantum theory of fields, Volume III: supersymmetry, Cambridge University Press
\bibitem{Tegmark} M. Tegmark, On the dimensionality of spacetime, Class. Quantum gravity, 14 (1997), L69-L75
\bibitem{Weinstein} W. Craig and S. Weinstein, On determinism and well-posedness in multiple time dimensions, Proc. Royal Society A (2009)465, 3023-3046.
\bibitem{Nima} N. Arkani-Hamed, S. Dimopoulos, G. Dvali, G. Gabadadze, Non-local deformation of gravity and the cosmological constant problem, hep-th/0209277v1.
\bibitem{Jenkins} A. Jenkins, Topics in theoretical particle physics and cosmology beyond the standard model, PhD-thesis, hep-th/0607239
\bibitem{Nobbenhuis} S. Nobbenhuis, The Cosmological Constant Problem: an inspiration for New Physics, gr-qc/0609011
\bibitem{Moffat}  J. Moffat and V.T. Toth, Modified gravity: Cosmology without dark matter or Einstein's cosmological constant, arxiv:0710.0364 
\bibitem{Bekenstein} J.D. Bekenstein, Relativistic gravitation theory for the Mond paradigm, astro-ph/040369
\bibitem{Jadczyk}  A. Jadczyk, Geometry of indefinite metric spaces, Reports on Mathematical physics, Volume 2, 1971
\bibitem{Antonelli} P.L. Antonelli, Handbook of Finsler geometry, Kluwer Academic publishers. 
\bibitem{Munteanu} Gheorghe Munteanu, Complex spaces in Finsler, Lagrange and Hamilton geometries, Kluwer Academic Press.
\bibitem{Finkelstein} D. Finkelstein, J.M. Jauch, S. Schiminovich and D. Speiser, Foundations of quanternion quantum mechanics, Journal of mathematical physics, Volume 3 number 2.
\end{thebibliography}
\end{document}